\documentclass[12pt]{article}

\usepackage[top=1in, bottom=1in, left=1in, right=1in]{geometry}
\usepackage{amsmath}
\usepackage{amssymb}
\usepackage{amsthm}
\usepackage{bm}
\usepackage{mathrsfs}
\usepackage{graphicx}
\usepackage{graphics}
\usepackage{epsfig}
\usepackage{subfigure}
\usepackage{cite}
\usepackage{setspace}
\usepackage{cancel} 
\usepackage{appendix}
\usepackage{color}
\usepackage{wrapfig}
\usepackage{xspace}
\usepackage{epstopdf}
\usepackage{imakeidx}

\usepackage{amsfonts}		
\usepackage{longtable} 
\usepackage{rotating}     

\usepackage{caption}
\usepackage{arydshln}
\usepackage{indentfirst}
\usepackage{float}
\usepackage{url}
\usepackage[colorlinks,citecolor=black,linkcolor=black]{hyperref}
\usepackage{array}
\usepackage{times}
\usepackage{accents}
\usepackage{enumerate}

\numberwithin{equation}{section}

\newcommand{\ignore}[1]{}

\newcommand{\norm}[1]{\left\Vert#1\right\Vert} 
\newcommand{\abs}[1]{\left\vert#1\right\vert} 

\newcommand{\bbm}{\begin{bmatrix}}
\newcommand{\ebm}{\end{bmatrix}}
\newcommand{\bma}[1]{\left[\begin{array}{#1}}
\newcommand{\ema}{\end{array}\right]}

\DeclareMathAlphabet{\mbf}{OT1}{ptm}{b}{n}
\newcommand{\mbs}[1]{{\boldsymbol{#1}}}
\newcommand{\mbc}[1]{ \boldsymbol{\mathcal{#1}} } 

\newcommand{\mbfbar}[1]{{\bar{\mbf{#1}}}}
\newcommand{\mbfhat}[1]{{\hat{\mbf{#1}}}}

\newcommand{\mbftilde}[1]{{\tilde{\mbf{#1}}}}
\newcommand{\mbfubar}[1]{{\underline{\mbf{#1}}}}

\def\dotb{{\raisebox{-0.6ex}{ \kern0.2ex\raisebox{0.8ex}{\tiny $\circ$}}}}
\def\ddota{{\raisebox{-0.6ex}{ \raise0.2ex\hbox{ \LARGE $\cdot\hspace*{-0.2ex}\cdot$}}}}
\def\ddotb{{\raisebox{-0.6ex}{ \kern0.2ex\raisebox{0.8ex}{\tiny $\circ\circ$}}}}

\newcommand{\trans}{{\ensuremath{\mathsf{T}}}} 
\newcommand{\herm}{{\ensuremath{\mathsf{H}}}} %
\newcommand{\frob}{{\ensuremath{\mathsf{F}}}} %
\newcommand{\trace}{ {\ensuremath{\mathrm{tr}}} } 

\newcommand{\onehalf}{\mbox{$\textstyle{\frac{1}{2}}$}}

\newcommand{\beq}{\begin{equation}}
\newcommand{\eeq}{\end{equation}}
\newcommand{\bdis}{\begin{displaymath}}
\newcommand{\edis}{\end{displaymath}}
\newcommand{\beqarray}{\begin{eqnarray}}
\newcommand{\eeqarray}{\end{eqnarray}}
\newcommand{\beqarraynn}{\begin{eqnarray*}}
\newcommand{\eeqarraynn}{\end{eqnarray*}}

 \theoremstyle{definition} 
 \newtheorem{example}{Example}[section]
\newtheorem{theorem}{Theorem}[section]
\newtheorem{lemma}{Lemma}[section]

\newtheorem{SynthMeth}[theorem]{Synthesis Method}

\newcommand{\du}[1]{{_{#1}}}
\newcommand{\dup}[2]{{_{#1 #2}}}

\newcommand{\itemcite}{\item \hspace{-1pt}}

  \newcommand\munderbar[1]{\underaccent{\bar}{#1}}

\theoremstyle{definition} 
\newtheorem{definition}[theorem]{Definition}

\makeindex[intoc]

\begin{document}

\title{\textbf{LMI Properties and Applications in Systems, Stability, and Control Theory}}

\author{Ryan James Caverly$^1$ and James Richard Forbes$^2$
}
\date{%
\small $^1$ \textit{Associate Professor, Department of Aerospace Engineering and Mechanics, University of Minnesota, 110~Union St. SE, Minneapolis, MN 55455, USA}, \texttt{rcaverly@umn.edu}.\\%
$^2$ \textit{Professor, Department of Mechanical Engineering, McGill University, 817~Sherbrooke~St.~West, Montreal, QC, Canada H3A 0C3}, \texttt{james.richard.forbes@mcgill.ca}.\\[2ex]
\large \today
}

\renewcommand*\contentsname{Table of Contents}

\maketitle   


\tableofcontents

\newpage
\section{Preliminaries}
\label{sec:Prelims}

\subsection{Introduction}

Linear matrix inequalities (LMIs) commonly appear in systems, stability, and control applications.  Many analysis and synthesis problems in these areas can be solved as feasibility or optimization problems subject to LMI constraints.  Although most well-known LMI properties and manipulation tricks, such as the Schur complement and the congruence transformation, can be found in standard references~\cite{Boyd1994,Dullerud2000,SchererWeiland2015,Skogestad2005,Duan2013}, many useful LMI properties are scattered throughout the literature.  The purpose of this document is to collect and organize properties, tricks, and applications related to LMIs from a number of references together in a single document.  In this sense, the document can be thought of as an ``LMI encyclopedia'' or ``LMI cookbook.''  
Proofs of the properties presented in this document are not included when they can be found in the cited references in the interest of brevity.  Illustrative examples are included whenever necessary to fully explain a certain property.  Multiple equivalent forms of LMIs are often presented to give the reader a choice of which form may be best suited for a particular problem at hand.  The equivalency of some of the LMIs in this document may be straightforward to more experienced readers, but the authors believe that some readers may benefit from the presentation of multiple equivalent LMIs.

The document is organized as follows.  In the remaining portions of Section~\ref{sec:Prelims}, the notation used throughout the document is presented and some fundamental LMI properties are discussed.  Sections~\ref{sec:BMI_to_LMI} and~\ref{sec:LMIProperties} feature a collection of LMI properties and tricks that are interesting and potentially useful.  
Properties that are primarily aimed at reformulating bilinear matrix inequalities (BMIs) as LMIs are presented in Section~\ref{sec:BMI_to_LMI}, while more general properties and definitions are found in Section~\ref{sec:LMIProperties}.  Applications involving LMIs in systems and stability theory are included in Section~\ref{sec:SystemsStability}.  Section~\ref{sec:OptimalControl} presents a number of LMI-based optimal controller synthesis methods, while Section~\ref{sec:OptimalEstimation} includes LMI-based optimal estimation synthesis methods.

The authors would like to thank the following individuals for alerting us of errors, and providing useful comments and suggestions for improvement: Jad Alsaayed, Logan Anderson, Leila Bridgeman, Jyot Buch, Manash Chakraborty, Steven Dahdah, William Elke, Robyn Fortune, Bruce Lee, and Peter Seiler.

Please note that this document is a work in progress.  If you notice any errors or inaccuracies, or have any suggestions of content that should be included in this document, please email either of the authors at \texttt{rcaverly@umn.edu} or \texttt{james.richard.forbes@mcgill.ca} so that changes to future versions can be made.

\subsection{Notation}

In this document, matrices are denoted by boldface letters (e.g., $\mbf{A} \in \mathbb{R}^{n \times n}$), column matrices are denoted by lowercase boldface letters (e.g., $\mbf{x} \in \mathbb{R}^n$), scalars are denoted by simple letters (e.g., $\gamma \in \mathbb{R}$), and operators are denoted by script letters (e.g., $\mbc{G}: \mathcal{L}_{2e} \to \mathcal{L}_{2e}$).  The set of $n$ by $m$ real matrices is denoted as $\mathbb{R}^{n \times m}$, the set of $n$ by $m$ complex matrices is denoted as $\mathbb{C}^{n \times m}$, and the set of $n$ by $n$ symmetric matrices is denoted as $\mathbb{S}^n$.  The identity matrix \index{identity matrix}is written as $\mbf{1}$ and a matrix filled with zeros is written as $\mbf{0}$.  The dimensions of $\mbf{1}$ and $\mbf{0}$ are specified when necessary.  Repeated blocks within symmetric matrices are replaced by $*$ for brevity and clarity.  The conjugate transpose \index{conjugate transpose}or Hermitian transpose \index{Hermitian transpose}of the matrix $\mbf{V} \in \mathbb{C}^{n \times m}$ is denoted by $\mbf{V}^\herm$.  The notation $\text{He}\{\cdot \}$ is used as a shorthand in situations with limited space, where $\text{He}\{ \cdot \} = \left(\cdot \right) + \left( \cdot \right)^\herm$.  The real and imaginary parts of the complex number $z \in \mathbb{C}$ are denoted as $\text{Re}(z)$ and $\text{Im}(z)$, respectively.  The Kroenecker product\index{Kroenecker product} of two matrices is denoted by $\otimes$.

Consider the square matrix $\mbf{A} \in \mathbb{R}^{n \times n}$.  The eigenvalues \index{eigenvalues}of $\mbf{A}$ are denoted by $\lambda_i(\mbf{A})$, $i=1,2,\ldots,n$.  The matrix $\mbf{A}$ is Hurwitz \index{Hurwitz matrix}if all of its eigenvalues are in the open left-half complex plane (i.e., $\text{Re}\left(\lambda_i(\mbf{A})\right) < 0$, $i=1,\ldots,n$).  A matrix is Schur \index{Schur matrix}if all of its eigenvalues are strictly within a unit disk centered at the origin of the complex plane (i.e., $\abs{\lambda_i(\mbf{A})} < 1$, $i=1,\ldots,n$).  If $\mbf{A} \in \mathbb{S}^n$, then the minimum eigenvalue \index{eigenvalues!minimum eigenvalue}of $\mbf{A}$ is denoted by $\underline{\lambda}(\mbf{A})$ and its maximum eigenvalue \index{eigenvalues!maximum eigenvalue}is denoted by $\bar{\lambda}(\mbf{A})$.  

Consider the matrix $\mbf{B} \in \mathbb{R}^{n \times m}$.  The minimum singular value \index{singular value!minimum singular value}of $\mbf{B}$ is denoted by $\underline{\sigma}(\mbf{B})$ and the maximum singular value \index{singular value!maximum singular value}of $\mbf{B}$ is denoted by $\bar{\sigma}(\mbf{B})$.  The range \index{range}and nullspace \index{nullspace}of $\mbf{B} $ are denoted by $\mathcal{R}(\mbf{B})$ and $\mathcal{N}(\mbf{B})$, respectively.

A state-space realization of the continuous-time linear time-invariant (LTI) system \index{state-space realization!continuous time}
\begin{align*}
\dot{\mbf{x}}(t) &= \mbf{A} \mbf{x}(t) + \mbf{B} \mbf{u}(t), \\
\mbf{y}(t) &= \mbf{C} \mbf{x}(t) + \mbf{D} \mbf{u}(t),
\end{align*}
is often written compactly as $(\mbf{A},\mbf{B},\mbf{C},\mbf{D})$ in this document.  The argument of time is often omitted in continuous-time state-space realizations, unless needed to prevent ambiguity.

A state-space realization of the discrete-time LTI system \index{state-space realization!discrete time}
\begin{align*}
\mbf{x}_{k+1} &= \mbf{A}_\mathrm{d} \mbf{x}_k + \mbf{B}_\mathrm{d} \mbf{u}_k, \\
\mbf{y}_k &= \mbf{C}_\mathrm{d} \mbf{x}_k + \mbf{D}_\mathrm{d} \mbf{u}_k,
\end{align*}
is often written compactly as $(\mbf{A}_\mathrm{d},\mbf{B}_\mathrm{d},\mbf{C}_\mathrm{d},\mbf{D}_\mathrm{d})$.


The inner product spaces $\mathcal{L}_2$ and $\mathcal{L}_{2e}$ for continuous-time signals are defined as

\begin{align*}
	\mathcal{L}_2 &= \left\{
			\mbf{x} : \mathbb{R}_{\geq 0} \to \mathbb{R}^n \; \Big| 
			\norm{\mbf{x}}_2^2 = \int_0^\infty \mbf{x}^\trans(t)\mbf{x}(t) \mathrm{d}t < \infty
		\right\}, \\
	\mathcal{L}_{2e} &= \left\{
			\mbf{x} : \mathbb{R}_{\geq 0} \to \mathbb{R}^n \; \Big| 
			\norm{\mbf{x}}_{2T}^2 =  \int_0^T \mbf{x}^\trans(t)\mbf{x}(t) \mathrm{d}t < \infty,\;\forall T \in \mathbb{R}_{\geq 0}
		\right\}.
\end{align*}

The inner product sequence spaces $\ell_2$ and $\ell_{2e}$ for discrete-time signals are defined as

\begin{align*}
	\ell_2 &= \left\{
			\mbf{x} : \mathbb{Z}_{\geq 0} \to \mathbb{R}^n \; \Big| 
			\norm{\mbf{x}}_2^2 =  \sum_{k=0}^\infty\mbf{x}^\trans_k\mbf{x}_k < \infty
		\right\}, \\
	\ell_{2e} &= \left\{
			\mbf{x} : \mathbb{Z}_{\geq 0} \to \mathbb{R}^n \; \Big| 
			\norm{\mbf{x}}_{2N}^2 =  \sum_{k=0}^{N}\mbf{x}^\trans_{k}\mbf{x}_{k} < \infty,\;\forall N \in \mathbb{Z}_{\geq 0}
		\right\}.
\end{align*}

\subsection{Definitions and Fundamental LMI Properties}

\subsubsection{Definiteness of a Matrix}
\index{definiteness!definition}
\begin{definition}

\cite[pp.~429--430]{Horn2013} Consider the symmetric matrix $\mbf{A} \in \mathbb{S}^{n}$.  The matrix $\mbf{A}$ is

\begin{enumerate}[a)]

\item \emph{positive definite} if $\mbf{x}^\trans\mbf{A}\mbf{x} > 0$, $\forall \mbf{x} \neq \mbf{0} \in \mathbb{R}^n$,

\item \emph{positive semi-definite} if $\mbf{x}^\trans\mbf{A}\mbf{x} \geq 0$, $\forall \mbf{x} \in \mathbb{R}^n$,

\item \emph{negative definite} if $\mbf{x}^\trans\mbf{A}\mbf{x} < 0$, $\forall \mbf{x} \neq \mbf{0} \in \mathbb{R}^n$,

\item \emph{negative semi-definite} if $\mbf{x}^\trans\mbf{A}\mbf{x} \leq 0$, $\forall \mbf{x} \in \mathbb{R}^n$,

\item and indefinite if $\mbf{x}^\trans\mbf{A}\mbf{x}$ is neither positive nor negative.

\end{enumerate}

\end{definition}

\begin{theorem}

\cite[pp.~430--431]{Horn2013},~\cite[p.~703]{BernsteinMatrixBook} Consider the symmetric matrix $\mbf{A}\in \mathbb{S}^{n}$.  The matrix $\mbf{A}$ is

\begin{enumerate}[a)]

\item \emph{positive definite} if and only if $\underline{\lambda}(\mbf{A}) > 0$,

\item \emph{positive semi-definite} if and only if $\underline{\lambda}(\mbf{A}) \geq 0$,

\item \emph{negative definite} if and only if $\bar{\lambda}(\mbf{A}) < 0$,

\item \emph{negative semi-definite} if and only if $\bar{\lambda}(\mbf{A}) \leq 0$,

\item and indefinite if and only if $\underline{\lambda}(\mbf{A}) < 0$ and $\bar{\lambda}(\mbf{A}) > 0$.

\end{enumerate}

\end{theorem}

\proof To see why the sign of $\mbf{x}^\trans\mbf{A}\mbf{x}$ is dictated by the eigenvalues of $\mbf{A}$, let $\mbf{A} = \mbf{V} \mbs{\Lambda}\mbf{V}^{-1}$, where $\mbf{V}^{-1} = \mbf{V}^\trans$ because $\mbf{A}$ is symmetric.  Notice that
\begin{align}
\mbf{x}^\trans\mbf{A}\mbf{x} &= \mbf{x}^\trans\mbf{V}\mbs{\Lambda}\mbf{V}^{-1}\mbf{x} \nonumber \\
&= \left(\mbf{V}^\trans\mbf{x}\right)^\trans\mbs{\Lambda}\mbf{V}^\trans\mbf{x} \nonumber \\
&= \mbf{z}^\trans\mbs{\Lambda}\mbf{z} \nonumber \\
&= \sum_{i = 1}^n \lambda_i(\mbf{A})z_i^2, \nonumber 
\end{align}
where $\mbf{z} = \mbf{V}^\trans\mbf{x} = \bbm z_1 & z_2 & \cdots & z_n \ebm^\trans$.
\endproof

When evaluating the sign of the quadratic form $\mbf{x}^\trans \mbf{A} \mbf{x}$, there is no loss of generality in restricting $\mbf{A}$ to be symmetric.  This is seen through the next two examples.

\begin{example}
  Consider the skew-symmetric matrix $\mbf{A} = -\mbf{A}^\trans \in \mathbb{R}^{n\times n}$.  Evaluating the quadratic form $\mbf{x}^\trans\mbf{A}\mbf{x}$  yields
\begin{align}
\mbf{x}^\trans\mbf{A}\mbf{x} &= \onehalf\mbf{x}^\trans\mbf{A}\mbf{x} +\onehalf \mbf{x}^\trans\mbf{A}\mbf{x} \nonumber \\
&= \onehalf\mbf{x}^\trans\mbf{A}\mbf{x} +\onehalf \left(\mbf{x}^\trans\mbf{A}\mbf{x}\right)^\trans \nonumber \\
&= \onehalf\mbf{x}^\trans\mbf{A}\mbf{x} + \onehalf\mbf{x}^\trans\mbf{A}^\trans\mbf{x} \nonumber \\
&= \onehalf\mbf{x}^\trans\left(\mbf{A}-\mbf{A}\right)\mbf{x} \nonumber \\
&= 0. \nonumber
\end{align}
Therefore, $\mbf{x}^\trans\mbf{A}\mbf{x} = 0$ for all skew-symmetic matrices.

\end{example}

\begin{example}
Consider the matrix $\mbf{A} \in \mathbb{R}^{n\times n}$, which can be decomposed as
\begin{align}
\mbf{A} &= \onehalf \mbf{A} + \onehalf\mbf{A} \nonumber \\
&= \onehalf \mbf{A} + \onehalf\mbf{A} + \onehalf\left(\mbf{A}^\trans-\mbf{A}^\trans\right) \nonumber \\
&= \underbrace{\onehalf\left(\mbf{A}+\mbf{A}^\trans\right)}_{\mbf{A}_{\text{sym}}} + \underbrace{\onehalf\left(\mbf{A}-\mbf{A}^\trans\right)}_{\mbf{A}_{\text{skew}}},  \nonumber 
\end{align}
where $\mbf{A}_{\text{sym}} = \mbf{A}_{\text{sym}}^\trans = \onehalf\left(\mbf{A}+\mbf{A}^\trans\right) $ is the symmetric part of $\mbf{A}$ and $\mbf{A}_{\text{skew}} = -\mbf{A}_{\text{skew}}^\trans = \onehalf\left(\mbf{A}-\mbf{A}^\trans\right) $ is the skew-symmetric part of $\mbf{A}$.  Evaluating the quadratic form $\mbf{x}^\trans\mbf{A}\mbf{x}$ yields
\begin{align}
\mbf{x}^\trans\mbf{A}\mbf{x} &=\mbf{x}^\trans\left(\mbf{A}_{\text{sym}}+\mbf{A}_{\text{skew}}\right)\mbf{x}\nonumber \\
&= \mbf{x}^\trans\mbf{A}_{\text{sym}}\mbf{x} +\cancelto{0}{ \mbf{x}^\trans\mbf{A}_{\text{skew}}\mbf{x}} \nonumber \\
&= \mbf{x}^\trans\mbf{A}_{\text{sym}}\mbf{x}. \nonumber
\end{align}
This confirms that when determining the definiteness of a matrix there is no loss of generality in restricting the matrix to be symmetric.

\end{example}

The positive definiteness and positive semidefiniteness of a matrix are denoted by $>0$ and $\geq 0$, respectively.  That is, $\mbf{A} = \mbf{A}^\trans > 0$ is positive definite and $\mbf{B} = \mbf{B}^\trans \geq 0$ is positive semidefinite.  Similarly, the negative definiteness and negative semidefiniteness of a matrix are denoted by $<0$ and $\leq 0$, respectively.  That is, $\mbf{C} = \mbf{C}^\trans < 0$ is negative definite and $\mbf{D} = \mbf{D}^\trans \leq 0$ is negative semidefinite.  For brevity, the transpose component of a definiteness statement is omitted in this document, for example, $\mbf{A} = \mbf{A}^\trans > 0$ is simply written as $\mbf{A} > 0$.  

\subsubsection{Matrix Inequalities and LMIs}
\label{sec:LMIDefs}

\begin{definition}

A matrix inequality\index{matrix inequality!definition}, $\mbf{G} : \mathbb{R}^m \to \mathbb{S}^{n}$, in the variable $\mbf{x} \in \mathbb{R}^m$ is an expression of the form
\bdis
\mbf{G}(\mbf{x}) = \mbf{G}_0  +  \sum_{i=1}^p f_i(\mbf{x})\mbf{G}_i \leq 0,
\edis
where $\mbf{x}^\trans = \bbm x_1 \cdots x_m \ebm$, $\mbf{G}_0 \in \mathbb{S}^n$, and $\mbf{G}_i  \in \mathbb{R}^{n \times n}$, $i=1,\ldots,p$.

\end{definition}

\begin{definition}

\cite{Boyd1997},~\cite[p.~34]{ElGhaoui2000},~\cite{VanAntwerp2000} A bilinear matrix inequality (BMI)\index{bilinear matrix inequality (BMI)!definition}, $\mbf{H} : \mathbb{R}^m \to \mathbb{S}^{n}$, in the variable $\mbf{x} \in \mathbb{R}^m$ is an expression of the form
\bdis
\mbf{H}(\mbf{x}) = \mbf{H}_0  + \sum_{i=1}^m x_i\mbf{H}_{i} +  \sum_{i=1}^m\sum_{j=1}^m x_ix_j\mbf{H}_{i,j} \leq 0,
\edis
where $\mbf{x}^\trans = \bbm x_1 \cdots x_m \ebm$, and $\mbf{H}_i$,~$\mbf{H}_{i,j}  \in \mathbb{S}^{n}$, $i=0,\ldots,m$, $j=0,\ldots,m$.

\end{definition}

\begin{definition}

\cite[p.~7]{Boyd1994},\cite[p.~17]{SchererWeiland2015} An LMI\index{LMI!definition}, $\mbf{F} : \mathbb{R}^m \to \mathbb{S}^{n}$, in the variable $\mbf{x} \in \mathbb{R}^m$ is an expression of the form
\beq
\label{eq:LMIDef}
\mbf{F}(\mbf{x}) = \mbf{F}_0  +  \sum_{i=1}^m x_i\mbf{F}_i \leq 0,
\eeq
where $\mbf{x}^\trans = \bbm x_1 \cdots x_m \ebm $ and $\mbf{F}_i  \in \mathbb{S}^{n}$, $i=0,\ldots,m$.

\end{definition}

LMIs can alternatively be defined in terms of matrix variables as follows.

\begin{definition}

\cite[p.~125]{Herrmann2007} An LMI\index{LMI!definition}, $\mbf{F} : \mathbb{R}^{p_1 \times q_1} \times \cdots \times \mathbb{R}^{p_r \times q_r} \to \mathbb{S}^{n}$, in the matrix variables $\mbf{X}_i \in \mathbb{R}^{p_i \times q_i}$, $i=1,\ldots,r$, where $m = \sum_{i=1}^rp_iq_i$, is an expression of the form
\beq
\label{eq:LMIDef2}
\mbf{F}(\mbf{X}_1,\ldots,\mbf{X}_r) = \mbf{F}_0  +  \sum_{i=1}^r \left(\mbf{G}_i \mbf{X}_i \mbf{H}_i + \mbf{H}_i^\trans \mbf{X}_i^\trans \mbf{G}_i^\trans \right) \leq 0,
\eeq
where $\mbf{F}_0 \in \mathbb{S}^{n}$, $\mbf{G}_i \in \mathbb{R}^{n \times p_i}$, and $\mbf{H}_i \in \mathbb{R}^{q_i \times n}$, $i = 1,\ldots,r$.

\end{definition}

\begin{example}
\label{Example3}

\cite[pp.~8--9]{Boyd1994} Consider the matrices $\mbf{A} \in \mathbb{R}^{n \times n}$ and $\mbf{Q} \in \mathbb{S}^{n}$, where $\mbf{Q} > 0$.  It is desired to find a symmetric matrix $\mbf{P} \in \mathbb{S}^{n}$ satisfying the matrix inequality
\beq
\label{eq:LMIEx1}
\mbf{P}\mbf{A} + \mbf{A}^\trans\mbf{P} + \mbf{Q} < 0,
\eeq
where $\mbf{P} > 0$.  The matrix $\mbf{P}$ is the design variable in this problem, and this LMI can be directly related to the definition in~\eqref{eq:LMIDef2} by setting $\mbf{F}_0 = \mbf{Q}$, $\mbf{G}_1 = \mbf{1}$, $\mbf{H}_1 = \mbf{A}$, $\mbf{X}_1 = \mbf{P}$, and enforcing the constraint $\mbf{X}_1 = \mbf{X}_1^\trans$.  
This LMI can be reformulated in the form of~\eqref{eq:LMIDef} by defining the scalar entries of the matrix variable $\mbf{P}$ as the design variables.  To illustrate this, consider the case of $n=2$ so that each matrix is of dimension $2 \times 2$, and $\mbf{x} = \bbm p_1 & p_2 & p_3 \ebm^\trans$.  Writing the matrix $\mbf{P}$ in terms of a basis $\mbf{E}_i \in \mathbb{S}^{ 2}$, $i=1,2,3$, yields
\bdis
\mbf{P} = \bbm p_1 & p_2 \\ p_2 & p_3 \ebm = p_1 \underbrace{\bbm 1 & 0 \\ 0 & 0 \ebm}_{\mbf{E}_1} + p_2 \underbrace{\bbm 0 & 1 \\ 1 & 0 \ebm}_{\mbf{E}_2} + p_3 \underbrace{\bbm 0 & 0 \\ 0 & 1\ebm}_{\mbf{E}_3}.
\edis
Note that the matrices $\mbf{E}_i$ are linearly independent and symmetric, thus forming a basis for the symmetric matrix $\mbf{P}$.  The matrix inequality in~\eqref{eq:LMIEx1} can be written as
\bdis
p_1\left(\mbf{E}_1\mbf{A} + \mbf{A}^\trans\mbf{E}_1\right) + p_2\left(\mbf{E}_2\mbf{A} + \mbf{A}^\trans\mbf{E}_2\right)  + p_3\left(\mbf{E}_3\mbf{A} + \mbf{A}^\trans\mbf{E}_3\right) + \mbf{Q} < 0.
\edis
Defining $\mbf{F}_0 = \mbf{Q}$ and $\mbf{F}_i = \mbf{F}_i^\trans = \mbf{E}_i\mbf{A} + \mbf{A}^\trans\mbf{E}_i$, $i=1,2,3$, yields
\bdis
\mbf{F}_0 + \sum_{i=1}^3p_i\mbf{F}_i < 0,
\edis
which now resembles the definition of an LMI in~\eqref{eq:LMIDef}.  Throughout this document, LMIs are typically written in the matrix form of~\eqref{eq:LMIDef2}, rather than the scalar form of~\eqref{eq:LMIDef}.

\end{example}

\subsubsection[Relative Definiteness of a Matrix]{Relative Definiteness of a Matrix}
\index{definiteness!relative definiteness}
The definiteness of a matrix can be found relative to another matrix.  For example, consider the matrices $\mbf{A} \in \mathbb{S}^{n}$ and $\mbf{B} \in \mathbb{S}^{n}$.  The matrix inequality $\mbf{A} < \mbf{B}$ is equivalent to $\mbf{A} - \mbf{B} < 0$ or $\mbf{B} - \mbf{A} > 0$.

Knowing the relative definiteness of matrices can be useful.  For example, if in the previous example we have $\mbf{A} < \mbf{B}$ and also know that $\mbf{A} > 0$, then we know that $\mbf{B} > 0$.  This follows from $0 < \mbf{A} < \mbf{B}$.  For more foundational facts involving the relative definiteness of matrices, see~\cite[pp.~703--704]{BernsteinMatrixBook}.

\subsubsection{Strict and Nonstrict Matrix Inequalities}
\index{LMI!strict LMIs}
\index{LMI!nonstrict LMIs}
A strict matrix inequality can be converted to a nonstrict matrix inequality.  For example, $\mbf{A} > 0$ is implied by $\mbf{A} \geq \epsilon \mbf{1}$, where $\epsilon \in \mathbb{R}_{>0}$.  Similarly, $\mbf{B} < 0$ is implied by $\mbf{B} \leq -\epsilon \mbf{1}$, where $\epsilon \in \mathbb{R}_{>0}$.

Converting a strict matrix inequality into a nonstrict matrix inequality is useful when working with LMI solvers that cannot handle strict constraints.

\subsubsection{Concatenation of LMIs}
\index{LMI!concatenation}
A useful property of LMIs is that multiple LMIs can be concatenated together to form a single LMI.  For example, satisfying the LMIs $\mbf{A} < 0$ and $\mbf{B} < 0$ is equivalent to satisfying the concatenated LMI
\bdis
\bbm \mbf{A} & \mbf{0} \\ \mbf{0} & \mbf{B} \ebm < 0.
\edis
More generally, satisfying the LMIs $\mbf{A}_i < 0$, $i=1,\ldots,n$ is equivalent to satisfying the concatenated LMI $\mathrm{diag}\{\mbf{A}_1,\ldots,\mbf{A}_n\} < 0$.

\subsubsection{Convexity of LMIs}

\begin{definition}
\index{LMI!convexity}
\cite[p.~138]{Lange2013} A set, $\mathcal{S}$, in a real inner product space is convex if for all $\mbf{x},\mbf{y} \in \mathcal{S}$ and $\alpha \in \mathbb{R}$, where $0 \leq \alpha \leq 1$, it holds that $\alpha \mbf{x} + (1-\alpha)\mbf{y} \in \mathcal{S}$.

\end{definition}

\begin{lemma}
\label{lemma:LMI_convex}
\cite{VanAntwerp2000} The set of solutions to an LMI is convex.  That is, the set $\mathcal{S} = \{ \mbf{x} \in \mathbb{R}^{m} \,\, | \,\, \mbf{F}(\mbf{x}) \leq 0 \}$ is a convex set, where $\mbf{F} : \mathbb{R}^m \to\mathbb{S}^{ n}$ is an LMI.

\end{lemma}

\proof 

Consider $\mbf{x},\mbf{y} \in \mathbb{R}^m$ and $\alpha \in [0,1]$, and suppose that $\mbf{x}$ and $\mbf{y}$ satisfy~\eqref{eq:LMIDef}.  The LMI $\mbf{F} : \mathbb{R}^m \to \mathbb{S}^{n}$ is convex, since
\begin{align}
\mbf{F}(\alpha \mbf{x} + (1-\alpha) \mbf{y}) &= \mbf{F}_0 + \sum_{i=1}^m\left(\alpha x_i + (1-\alpha)y_i\right)\mbf{F}_i \nonumber \\
&= \mbf{F}_0 - \alpha \mbf{F}_0 + \alpha\mbf{F}_0 + \alpha \sum_{i=1}^mx_i\mbf{F}_i + (1-\alpha)\sum_{i=1}^my_i\mbf{F}_i \nonumber \\
&= \alpha \mbf{F}_0 + \alpha\sum_{i=1}^mx_i\mbf{F}_i + (1-\alpha)\mbf{F}_0 + (1-\alpha)\sum_{i=1}^my_i\mbf{F}_i \nonumber \\
&= \alpha\mbf{F}(\mbf{x}) + (1-\alpha)\mbf{F}(\mbf{y}). \nonumber 
\end{align}
\endproof

\subsection{Semidefinite Programs (SDPs)}
\index{semidefinite program (SDP)}

A semidefinite program (SDP) is a convex optimization problem of the form~\cite[p.~168]{Boyd2004}
\begin{align}
\min_{\mbf{x} \in \mathbb{R}^m} \quad &\mbf{c}^\trans \mbf{x} \label{eq:SDP1a} \\
\textrm{subject to} \quad & \mbf{F}_0  +  \sum_{i=1}^m x_i\mbf{F}_i \leq 0, \label{eq:SDP1b}
\end{align}
where $\mbf{x}^\trans = \bbm x_1 \cdots x_m \ebm $, $\mbf{c} \in \mathbb{R}^m$, $\mbf{F}_i  \in \mathbb{S}^{n}$, $i=0,\ldots,m$, and~\eqref{eq:SDP1b} is an LMI in the variable $\mbf{x}$.  As shown in Example~\ref{Example3}, the LMI constraint in~\eqref{eq:SDP1b} can be written in matrix form, rather than the standard form.

The dual problem of the SDP described by~\eqref{eq:SDP1a} and~\eqref{eq:SDP1b} is given by~\cite[pp.~168--169]{Boyd2004}
\begin{align*}
\max_{\mbf{Z} \in \mathbb{S}^n}  \quad & \trace \left(\mbf{F}_0 \mbf{Z}\right) \\
\text{subject to} \quad & \trace \left( \mbf{F}_i \mbf{Z} \right) + c_i = 0, \,\, i = 1,\ldots,n,  \\
& \mbf{Z} \geq 0,
\end{align*}
where $\mbf{c}^\trans = \bbm c_1 & \cdots c_m \ebm$.  Within the context of duality, the SDP outlined in~\eqref{eq:SDP1a} and~\eqref{eq:SDP1b} is denoted as the primal problem.  Further details on the use of SDP duality within the context of LTI systems can be found in~\cite{Balakrishnan2003,Balakrishnan2002}.
 
When using matrix variables to describe an SDP's LMI constraints, it may be inconvenient to rewrite the objective function in the form of~\eqref{eq:SDP1a}.  SDP parsers, which will be discussed in Section~\ref{sec:NumTools}, are capable of converting LMIs and linear objective functions in matrix form to the standard form required by most SDP solvers.  An example of a linear objective function in matrix form is
\bdis
\mathcal{J}(\mbf{X}) = \trace \left(\mbf{Q}^\trans \mbf{X} + \mbf{X}^\trans \mbf{R}\right), 
\edis
where $\mbf{X}$,~$\mbf{Q}$,~$\mbf{R} \in \mathbb{R}^{n \times m}$.

More generally, a number of convex objective functions involving matrix variables that are not explicitly written in the standard SDP form can be reformulated as SDPs.  Some SDP parsers are capable of performing this conversion for the user.  Two examples of such objective functions are given, with a brief explanation of how they can be reformulated in the standard SDP form.
 \index{convex objective functions} \index{semidefinite program (SDP)}

\begin{example} \cite[p.~71]{Boyd2004} Consider $\mathcal{J}(\mbf{x}) = \onehalf \mbf{x}^\trans \mbf{P} \mbf{x} + \mbf{q}^\trans \mbf{x} + r$, where $\mbf{x}$,~$\mbf{q} \in \mathbb{R}^{n}$, $\mbf{P} \in \mathbb{S}^n$, $\mbf{P} > 0$, and $r \in \mathbb{R}$.  Two special cases of this objective function are listed next.

\begin{itemize}

\item Special case when $\mbf{q} = \mbf{0}$ and $r = 0$: $\mathcal{J}(\mbf{x}) = \onehalf \mbf{x}^\trans \mbf{P} \mbf{x}$, where $\mbf{x} \in \mathbb{R}^{n}$, $\mbf{P} \in \mathbb{S}^n$, and $\mbf{P} > 0$.

\item Special case when $\mbf{P}=2 \cdot \mbf{1}$, $\mbf{q} = \mbf{0}$, and $r = 0$: $\mathcal{J}(\mbf{x}) =  \mbf{x}^\trans  \mbf{x} = \norm{\mbf{x}}_2^2$, where $\mbf{x} \in \mathbb{R}^{n}$.

\end{itemize}
The optimization problem
\begin{align*}
\min_{\mbf{x} \in \mathbb{R}^m} \quad &\onehalf \mbf{x}^\trans \mbf{P} \mbf{x} + \mbf{q}^\trans \mbf{x} + r \\
\textrm{subject to} \quad & \mbf{F}(\mbf{x})  \leq 0,
\end{align*}
is equivalent to the optimization problem
\begin{align*}
\min_{\mbf{x} \in \mathbb{R}^m, \gamma \in \mathbb{R}} \quad & \gamma \\
\textrm{subject to} \quad & \mbf{F}(\mbf{x})  \leq 0, \\
& \bbm \mbf{q}^\trans \mbf{x} + r - \gamma & \mbf{x}^\trans \\ * & -2 \mbf{P}^{-1} \ebm \leq 0,
\end{align*}
where the Schur complement, presented in Section~\ref{sec:SchurComplement}, is used to reformulate the quadratic objective function into an LMI constraint.

\end{example}

\begin{example}

Consider $\mathcal{J}(\mbf{X}) = \trace \left(\mbf{X}^\trans \mbf{P} \mbf{X} + \mbf{Q}^\trans \mbf{X} + \mbf{X}^\trans \mbf{R} + \mbf{S} \right)$, where $\mbf{X}$,~$\mbf{Q}$,~$\mbf{R} \in \mathbb{R}^{n \times m}$, $\mbf{P} \in \mathbb{S}^n$, $\mbf{S} \in \mathbb{R}^{n \times n}$, and $\mbf{P} \geq 0$.   Four special cases of this objective function are listed next.

\begin{itemize}

\item Special case when $\mbf{Q} = \mbf{R} =  \mbf{0}$ and $\mbf{S} = \mbf{0}$: $\mathcal{J}(\mbf{X}) = \trace \left(\mbf{X}^\trans \mbf{P} \mbf{X} \right)$, where $\mbf{X} \in \mathbb{R}^{n \times m}$, $\mbf{P} \in \mathbb{S}^n$, and $\mbf{P} > 0$.

\item Special case when $\mbf{P} =  \mbf{1}$, $\mbf{Q} = \mbf{R} =  \mbf{0}$, and $\mbf{S} = \mbf{0}$: $\mathcal{J}(\mbf{X}) = \trace \left(\mbf{X}^\trans  \mbf{X} \right) = \norm{\mbf{X}}_\frob^2$,\index{norm!Frobenius norm} where $\mbf{X} \in \mathbb{R}^{n \times m}$.

\itemcite \cite[p.~88]{Boyd1994} Special case when $\mbf{P} = \mbf{0}$, $\mbf{R} =  \mbf{0}$ and $\mbf{S} = \mbf{0}$: $\mathcal{J}(\mbf{X}) = \trace (\mbf{Q}^\trans \mbf{X} )$, where $\mbf{X}$,~$\mbf{Q} \in \mathbb{R}^{n \times m}$.

\itemcite \cite[p.~718]{BernsteinMatrixBook} Special case when $\mbf{P} =  \mbf{1}$, $\mbf{Q} = \mbf{R} =  \mbf{0}$, $\mbf{S} = \mbf{0}$, and $\mbf{X} \in \mathbb{S}^n$: $\mathcal{J}(\mbf{X}) = \trace (\mbf{X}^{2}) $, where $\mbf{X} \in \mathbb{S}^{n}$.

\end{itemize}

The optimization problem
\begin{align*}
\min_{\mbf{X} \in \mathbb{R}^{n \times m}} \quad &\trace \left(\mbf{X}^\trans \mbf{P} \mbf{X} + \mbf{Q}^\trans \mbf{X} + \mbf{X}^\trans \mbf{R} + \mbf{S} \right) \\
\textrm{subject to} \quad & \mbf{F}(\mbf{X})  \leq 0.
\end{align*}
is equivalent to the optimization problem
\begin{align*}
\min_{\mbf{X} \in \mathbb{R}^{n \times m}, \mbf{Z} \in \mathbb{S}^m, \gamma \in \mathbb{R}} \quad & \gamma \\
\textrm{subject to} \quad & \mbf{F}(\mbf{X})  \leq 0, \\
& \bbm \mbf{Q}^\trans \mbf{X} + \mbf{X}^\trans \mbf{R} + \mbf{S}- \mbf{Z} & \mbf{X}^\trans \\ * & -\mbf{P}^{-1} \ebm \leq 0, \\
& \trace(\mbf{Z}) \leq \gamma.
\end{align*}
where a property involving the trace of a symmetric matrix, as discussed in Section~\ref{sec:Trace}, and the Schur complement in Section~\ref{sec:SchurComplement} are used to reformulate the quadratic objective function into an LMI constraint.

\end{example}

Another useful convex objective function is given by $\mathcal{J}(\mbf{X}) = \log \left( \det (\mbf{X}^{-1}) \right) = - \log \left( \det (\mbf{X}) \right)$, where $\mbf{X} \in \mathbb{S}^{n}$ and $\mbf{X} > 0$~\cite[p.~14]{Boyd1994},~\cite{VanAntwerp2000}.  This objective function cannot be readily converted into the standard SDP form, but can be implemented with most SDP solvers and parsers.  In particular, \texttt{SDPT3}~\cite{SDPT3,SDPT3_site} is capable of directly minimizing SDPs with objective functions of the form $- \log \left( \det (\mbf{X}) \right)$.

\subsection{Numerical Tools to Solve SDPs}
\label{sec:NumTools}

There are many semidefinite program \index{semidefinite program (SDP)}solvers that accept LMI constraints.  Most solvers require that LMI constraints be written in the standard form shown in~\eqref{eq:LMIDef}.  This is often not convenient, as it is typical to derive LMI constraints in matrix form, such as the LMI in~\eqref{eq:LMIEx1}.  LMI parsers convert LMIs in matrix form to the standard form in~\eqref{eq:LMIDef}, allowing for a smoother transition from mathematical derivation to numerical implementation.  
A non-exhaustive list of SDP solvers and LMI parsers are included for reference.

\subsubsection{SDP Solvers}
\label{sec:Solvers}
\index{LMI!solvers} \index{semidefinite program (SDP)!solvers}
There are a number of SDP solvers available.  The authors have experience with \texttt{SeDuMi}~\cite{Sedumi,Sedumi_site}, \texttt{SDPT3}~\cite{SDPT3,SDPT3_site}, and \texttt{Mosek}~\cite{Mosek}, though other solvers are available, such as \texttt{CSDP}~\cite{CSDP,CSDP_site}, \texttt{CVXOPT}~\cite{CVXOPT,CVXOPT_site}, \texttt{DDS}~\cite{DDS_2,DDS_site}, \texttt{DSDP}~\cite{DSDP_2005,DSDP_site}, \texttt{LMILab}~\cite{LMILab}, \texttt{PENLAB}~\cite{PENLAB,PENLAB_site}, \texttt{SCS}~\cite{SCS_2016,SCS_site}, \texttt{SDPA}~\cite{SDPA_2003,SDPA_2010,SDPA_site}, \texttt{SMCP}~\cite{SMCP_2010,SMCP_site}, \texttt{SDPNAL}~\cite{SDPNAL_2015,SDPNAL_site}, and \texttt{STRIDE}~\cite{yang2023inexact,STRIDE_site}.  There are advantages and disadvantages to each of these solvers, and sometimes one solver may give a solution to a given problem when others do not.  For this reason, it is useful to have multiple solvers available.  Comparisons of various LMI solvers and benchmark problems are found in~\cite{Mittelmann2002,Arzelier2002,MittelmannSite}.

Many solvers, including \texttt{SeDuMi}, \texttt{SDPT3}, are available for free, while \texttt{Mosek} is a commercial software package.  A free academic license of \texttt{Mosek} can be requested for research in academic institutions or educational purposes.

\subsubsection{LMI Parsers}
\label{sec:Parsers}
\index{LMI!parsers}

LMI parsers allow the user to define the SDP to be solved within standard software environments, and often in a more convenient matrix form.  A number of openly-distributed LMI parsers are available for use within different software environments.  The following is a non-exhaustive list of LMI parsers and the solvers they are known to be compatible with, sorted by software environment.

\begin{itemize}

\item \texttt{Matlab}

\begin{itemize}

\item \texttt{Yalmip}~\cite{Yalmip,Yalmip_site}. Solvers: \texttt{CSDP}, \texttt{DSDP}, \texttt{LMILab}, \texttt{Mosek}, \texttt{PENLAB}, \texttt{SCS}, \texttt{SDPA}, \texttt{SDPT3}, \texttt{SDPNAL}, and \texttt{SeDuMi}.

\item \texttt{CVX}~\cite{CVX_2008,CVX}. Solvers: \texttt{Mosek}, \texttt{SDPT3}, and \texttt{SeDuMi}.

\item \texttt{LMILab}~\cite{LMILab}.  Features an internal solver.

\item \texttt{ROLMIP}~\cite{ROLMIP_2019,ROLMIP_site}.  Parser designed for optimization problems with uncertain polynomial matrices.  Requires \texttt{Yalmip}.  Solvers: \texttt{CSDP}, \texttt{DSDP}, \texttt{LMILab}, \texttt{Mosek}, \texttt{PENLAB}, \texttt{SCS}, \texttt{SDPA}, \texttt{SDPT3}, \texttt{SDPNAL}, and \texttt{SeDuMi}.

\end{itemize}

\item \texttt{Python}

\begin{itemize}

\item \texttt{CVXPY}~\cite{CVXPY_2016,CVXPY_2018,CVXPY_site}.  Solvers: \texttt{SCS}.  Other solvers can be installed separately.

\item \texttt{PICOS}~\cite{PICOS_site}.  Solvers: \texttt{CVXOPT}, \texttt{Mosek}, and \texttt{SMCP}.

\item \texttt{Irene}~\cite{Irene_site}.  Solvers: \texttt{CSDP}, \texttt{CVXOPT}, \texttt{DSDP}, and \texttt{SDPA}.

\item \texttt{PyLMI-SDP}~\cite{PyLMI-SDP_site}.  Solvers: \texttt{CVXOPT} and \texttt{SDPA}.

\end{itemize}

\item \texttt{Julia}

\begin{itemize}

\item\texttt{Convex.jl}~\cite{Convex.jl,Convex.jl_site}.  Solvers: \texttt{Mosek} and \texttt{SCS}.

\item \texttt{JuMP}~\cite{JuMP_2017,JuMP_site}.  Solvers: \texttt{Mosek} and \texttt{SCS}.

\end{itemize}

%
%
%

\item \texttt{NSP}

\begin{itemize}

\item \texttt{NSPYalmip}~\cite{NSPYALMIP,NSPYALMIP_site}.  Solvers: \texttt{CSDP} and \texttt{SeDuMi}.

\end{itemize}

\end{itemize}

%
%
%


\newpage
\section{Properties and Tricks Aimed at Reformulating BMIs as LMIs}
\label{sec:BMI_to_LMI}


\subsection{Introduction}

This section presents a compilation of properties and methods from the literature that are primarily used to reformulate BMI constraints as LMI constraints.  Many of these properties are used in subsequent sections to reformulate LMIs or transform matrix inequalities into LMIs.

The properties discussed in Sections~\ref{sec:ChangeVariables} to~\ref{sec:Dilation} are typically able to reformulate a BMI as an equivalent LMI or LMIs.  Use of these properties is desirable, as they will not introduce any conservatism when reformulating a BMI.  On the other hand, the properties in Sections~\ref{sec:YoungsRelation} to~\ref{sec:CoordDescent} are typically used to obtain an LMI that implies a BMI, generally with conservatism.  This makes the use of these properties a less desirable, yet sometimes unavoidable, option.

A discussion on when and how to use a selection of the properties presented in this section is provided in Section~\ref{sec:Discussion_BMI_LMI}.

\subsection[Change of Variables]{Change of Variables~\cite[pp.~100--101]{Boyd1994},~\cite[Sec.~12.3.1]{Skogestad2005}}
\label{sec:ChangeVariables}
\index{change of variables}
A BMI can sometimes be converted into an LMI using a change of variables.

\begin{example}
\label{ex:ChangeOfVariables}

\cite[Example~12.5, Sec.~12.3.1]{Skogestad2005} Consider $\mbf{A} \in \mathbb{R}^{n \times n}$, $\mbf{B} \in \mathbb{R}^{n \times m}$, $\mbf{K} \in \mathbb{R}^{m \times n}$, and $\mbf{Q} \in \mathbb{S}^n$, where $\mbf{Q} > 0$.  The matrix inequality given by
\bdis
\mbf{Q}\mbf{A}^\trans + \mbf{A}\mbf{Q} -\mbf{Q}\mbf{K}^\trans\mbf{B}^\trans - \mbf{B}\mbf{K}\mbf{Q} < 0,
\edis
is bilinear in the variables $\mbf{Q}$ and $\mbf{K}$.  Define a change of variable as $\mbf{F} = \mbf{K}\mbf{Q}$ to obtain
\bdis
\mbf{Q}\mbf{A}^\trans + \mbf{A}\mbf{Q} - \mbf{F}^\trans\mbf{B}^\trans - \mbf{B}\mbf{F} < 0,
\edis
which is an LMI in the variables $\mbf{Q}$ and $\mbf{F}$.  Once this LMI is solved, the original variable can be recovered by $\mbf{K} = \mbf{F}\mbf{Q}^{-1}$.

\end{example}

It is important that a change of variables is chosen to be a one-to-one mapping in order for the new matrix inequality to be equivalent to the original matrix inequality.  In Example~\ref{ex:ChangeOfVariables} the change of variable $\mbf{F} = \mbf{K} \mbf{Q}$ is a one-to-one mapping since $\mbf{Q}^{-1}$ is invertible due to the constraint $\mbf{Q} > 0$, which gives a unique solution for the reverse change of variable $\mbf{K} = \mbf{F} \mbf{Q}^{-1}$.

\subsection[Congruence Transformation]{Congruence Transformation~\cite[p.~15]{Boyd1994},~\cite[Sec.~12.3.2]{Skogestad2005}}
\index{congruence transformation}Consider $\mbf{Q} \in \mathbb{S}^{n}$ and $\mbf{W} \in \mathbb{R}^{n \times n}$, where $\text{rank}(\mbf{W}) = n$. \index{rank} The matrix inequality $\mbf{Q} < 0$ is satisfied if and only if $\mbf{W} \mbf{Q} \mbf{W}^\trans < 0$ or equivalently $\mbf{W}^\trans \mbf{Q} \mbf{W} < 0$, and is referred to as a congruence transformation.

\begin{example}
\cite[Example~12.6, Sec.~12.3.2]{Skogestad2005}
Consider $\mbf{A} \in \mathbb{R}^{n \times n}$, $\mbf{B} \in \mathbb{R}^{n \times m}$, $\mbf{K} \in \mathbb{R}^{m \times p}$, $\mbf{C}^\trans \in \mathbb{R}^{n \times p}$, $\mbf{P} \in \mathbb{S}^n$, and $\mbf{V} \in \mathbb{S}^p$, where $\mbf{P} > 0$ and $\mbf{V} > 0$.  The matrix inequality given by
\bdis
\mbf{Q} = \bbm \mbf{A}^\trans \mbf{P} + \mbf{P}\mbf{A} & -\mbf{P}\mbf{B}\mbf{K} + \mbf{C}^\trans\mbf{V} \\ * & -2\mbf{V} \ebm < 0,
\edis
is linear in the variable $\mbf{V}$ and bilinear in the variable pair $(\mbf{P},\mbf{K})$.  Choose the matrix $\mbf{W} = \text{diag}\{\mbf{P}^{-1},\mbf{V}^{-1}\}$ to obtain an equivalent BMI given by
\beq
\label{eq:CongTrans1}
\mbf{W}\mbf{Q}\mbf{W}^\trans = \bbm \mbf{P}^{-1}\mbf{A}^\trans + \mbf{A}\mbf{P}^{-1} & -\mbf{B}\mbf{K}\mbf{V}^{-1} + \mbf{P}^{-1}\mbf{C}^\trans \\ * & -2\mbf{V}^{-1} \ebm < 0.
\eeq
Using a change of variable $\mbf{X} = \mbf{P}^{-1}$, $\mbf{U} = \mbf{V}^{-1}$, and $\mbf{F} = \mbf{K}\mbf{V}^{-1}$,~\eqref{eq:CongTrans1} becomes
\beq
\label{eq:CongTrans2}
\mbf{W}\mbf{Q}\mbf{W}^\trans = \bbm \mbf{X}\mbf{A}^\trans + \mbf{A}\mbf{X} & -\mbf{B}\mbf{F} + \mbf{X}\mbf{C}^\trans \\ * & -2\mbf{U} \ebm < 0,
\eeq
which is an LMI in the variables $\mbf{X}$, $\mbf{U}$, and $\mbf{F}$.  Once~\eqref{eq:CongTrans2} is solved in terms of $\mbf{X}$, $\mbf{U}$, and $\mbf{F}$, the original variable $\mbf{K}$ is recovered by the reverse change of variable $\mbf{K} = \mbf{F}\mbf{U}^{-1}$.

\end{example}

A congruence transformation preserves the definiteness of a matrix by ensuring that $\mbf{Q} < 0$ and $\mbf{W} \mbf{Q} \mbf{W}^\trans < 0$ are equivalent.  A congruence transformation is related, but not equivalent to a similarity transformation $\mbf{T} \mbf{Q} \mbf{T}^{-1}$, which preserves not only the definiteness, but also the eigenvalues of a matrix.  A congruence transformation is equivalent to a similarity transformation in the special case when $\mbf{W}^\trans = \mbf{W}^{-1}$.

\subsection{Schur Complement}
\label{sec:SchurComplement}
\index{Schur complement!strict Schur complement lemma}
\subsubsection[Strict Schur Complement]{Strict Schur Complement~\cite[pp.~7--8]{Boyd1994},~\cite[Sec.~12.3.3]{Skogestad2005}}
Consider $\mbf{A} \in \mathbb{S}^{n}$, $\mbf{B} \in \mathbb{R}^{n \times m}$, and $\mbf{C} \in \mathbb{S}^{m}$.  The following statements are equivalent.
\begin{enumerate}[a)]
\item $\bbm \mbf{A} & \mbf{B} \\ \mbf{B}^\trans & \mbf{C} \ebm < 0$.
\item $\mbf{A} - \mbf{B}\mbf{C}^{-1}\mbf{B}^\trans < 0$, $\mbf{C} < 0$.
\item $\mbf{C} - \mbf{B}^\trans\mbf{A}^{-1}\mbf{B} < 0$, $\mbf{A} < 0$.
\end{enumerate}

\subsubsection[Nonstrict Schur Complement]{Nonstrict Schur Complement~\cite[p.~28]{Boyd1994}}
\index{Schur complement!nonstrict Schur complement lemma}
Consider $\mbf{A} \in \mathbb{S}^{n}$, $\mbf{B} \in \mathbb{R}^{n \times m}$, and $\mbf{C} \in \mathbb{S}^{m}$.  The following statements are equivalent.
\begin{enumerate}[a)]
\item $\bbm \mbf{A} & \mbf{B} \\ \mbf{B}^\trans & \mbf{C} \ebm \leq 0$.
\item $\mbf{A} - \mbf{B}\mbf{C}^{+}\mbf{B}^\trans < 0$, $\mbf{C} \leq 0$, $\mbf{B}(\mbf{1} - \mbf{C}\mbf{C}^+) = \mbf{0}$, where $\mbf{C}^+$ is the Moore-Penrose inverse\index{Moore-Penrose inverse} of $\mbf{C}$.
\item $\mbf{C} - \mbf{B}^\trans\mbf{A}^{+}\mbf{B} < 0$, $\mbf{A} \leq 0$, $\mbf{B}^\trans(\mbf{1} - \mbf{A}\mbf{A}^+) = \mbf{0}$, where $\mbf{A}^+$ is the Moore-Penrose inverse of $\mbf{A}$.
\end{enumerate}

\subsubsection{Schur Complement Lemma-Based Properties}
\index{Schur complement!Schur complement-based properties}
\label{sec:SchurProp}
\begin{enumerate}

\itemcite \cite[p.~109]{SchererWeiland2015},\cite[p.~100]{Gu2013} Consider $\mbf{P}_{11} \in \mathbb{S}^n$, $\mbf{P}_{12} \in \mathbb{R}^{n \times m}$, $\mbf{P}_{22}$,~$\mbf{X} \in \mathbb{S}^{m}$, $\mbf{P}_{13} \in \mathbb{R}^{n \times p}$, $\mbf{P}_{23} \in \mathbb{R}^{m \times p}$, and $\mbf{P}_{33} \in \mathbb{S}^p$.  There exists $\mbf{X}$ such that 
\beq
\label{eq:lemma1}
\bbm \mbf{P}_{11} & \mbf{P}_{12} & \mbf{P}_{13} \\ * & \mbf{P}_{22} + \mbf{X} & \mbf{P}_{23} \\ * & * & \mbf{P}_{33}\ebm < 0,
\eeq
if and only if
\bdis
\bbm \mbf{P}_{11} & \mbf{P}_{13} \\ * & \mbf{P}_{33} \ebm < 0.  
\edis
Any matrix $\mbf{X} \in \mathbb{S}^m$ satisfying
\beq
\label{eq:Schur3a}
\mbf{X} < -\mbf{P}_{22} + \bbm \mbf{P}_{12}^\trans & \mbf{P}_{23} \ebm \bbm \mbf{P}_{11} & \mbf{P}_{13} \\ * & \mbf{P}_{33}\ebm^{-1} \bbm \mbf{P}_{12} \\ \mbf{P}_{23}^\trans\ebm
\eeq
is a solution to~\eqref{eq:lemma1}.  That is,~\eqref{eq:Schur3a}$\implies$~\eqref{eq:lemma1}.

\itemcite \label{prop:SchurBased} \cite[pp.~109--110]{SchererWeiland2015},\cite[p.~101]{Gu2013} Consider $\mbf{P}_{11} \in \mathbb{S}^n$, $\mbf{P}_{12}$,~$\mbf{X} \in \mathbb{R}^{n \times m}$, $\mbf{P}_{22} \in \mathbb{S}^{m}$, $\mbf{P}_{13} \in \mathbb{R}^{n \times p}$, $\mbf{P}_{23} \in \mathbb{R}^{m \times p}$, and $\mbf{P}_{33} \in \mathbb{S}^p$.  There exists $\mbf{X}$ such that 
\beq
\label{eq:lemma2}
\bbm \mbf{P}_{11} & \mbf{P}_{12} + \mbf{X}^\trans & \mbf{P}_{13} \\ * & \mbf{P}_{22}  & \mbf{P}_{23} \\ * & * & \mbf{P}_{33}\ebm < 0
\eeq
if and only if
\beq
\bbm \mbf{P}_{11} & \mbf{P}_{13} \\ * & \mbf{P}_{33} \ebm < 0, \hspace{10pt} \text{and} \hspace{10pt} \bbm \mbf{P}_{22} & \mbf{P}_{23} \\ * & \mbf{P}_{33} \ebm < 0.  
\label{eq:lemma2b}
\eeq
If the two matrix inequalities in~\eqref{eq:lemma2b} hold, then a solution to~\eqref{eq:lemma2} is given by
\bdis
\mbf{X} = \mbf{P}_{23}\mbf{P}_{33}^{-1}\mbf{P}_{13}^\trans-\mbf{P}_{12}^\trans.
\edis
\begin{proof}
Necessity (\eqref{eq:lemma2} $\implies$~\eqref{eq:lemma2b}) comes from the requirement that the submatrices corresponding to the principle minors of~\eqref{eq:lemma2} are negative definite.  Sufficiency (\eqref{eq:lemma2b} $\implies$~\eqref{eq:lemma2}) is shown by rewriting the matrix inequalities of~\eqref{eq:lemma2b} in the equivalent form
\beq
\label{eq:lemma2c}
\mbf{P}_{11} - \mbf{P}_{13} \mbf{P}_{33}^{-1} \mbf{P}_{13}^\trans < 0, \hspace{10pt} \text{and} \hspace{10pt} \mbf{P}_{22} - \mbf{P}_{23} \mbf{P}_{33}^{-1} \mbf{P}_{23}^\trans < 0.
\eeq
Concatenating the two matrix inequalities in~\eqref{eq:lemma2c} and choosing $\mbf{X} = \mbf{P}_{23}\mbf{P}_{33}^{-1}\mbf{P}_{13}^\trans-\mbf{P}_{12}^\trans$ gives the equivalent matrix inequality
\bdis
\bbm \mbf{P}_{11} - \mbf{P}_{13} \mbf{P}_{33}^{-1} \mbf{P}_{13}^\trans & \mbf{P}_{12} - \mbf{P}_{13} \mbf{P}_{33}^{-1} \mbf{P}_{23}^\trans + \mbf{X}^\trans \\ * & \mbf{P}_{22} - \mbf{P}_{23} \mbf{P}_{33}^{-1} \mbf{P}_{23}^\trans \ebm < 0,
\edis
or
\bdis
\bbm \mbf{P}_{11} & \mbf{P}_{12} + \mbf{X}^\trans \\ * & \mbf{P}_{22} \ebm - \bbm \mbf{P}_{13} \\ \mbf{P}_{23} \ebm \mbf{P}_{33}^{-1} \bbm \mbf{P}_{13}^\trans & \mbf{P}_{23}^\trans \ebm< 0,
\edis
which is equivalent to~\eqref{eq:lemma2} using the Schur complement lemma.
\end{proof}

Permutation of the columns and rows of~\eqref{eq:lemma2} yields the following equivalent result.

\cite[pp.~41--42]{Duan2013} Consider $\mbf{P}_{11} \in \mathbb{S}^n$, $\mbf{P}_{12}$,~$\mbf{X} \in \mathbb{R}^{n \times m}$, $\mbf{P}_{22} \in \mathbb{S}^{m}$, $\mbf{P}_{13} \in \mathbb{R}^{n \times p}$, $\mbf{P}_{23} \in \mathbb{R}^{m \times p}$, and $\mbf{P}_{33} \in \mathbb{S}^p$.  There exists $\mbf{X}$ such that 
\beq
\label{eq:lemma3}
\bbm \mbf{P}_{11} & \mbf{P}_{12}  & \mbf{P}_{13} \\ * & \mbf{P}_{22}  & \mbf{P}_{23} + \mbf{X}^\trans \\ * & * & \mbf{P}_{33}\ebm < 0
\eeq
if and only if
\beq
\bbm \mbf{P}_{11} & \mbf{P}_{12} \\ * & \mbf{P}_{22} \ebm < 0, \hspace{10pt} \text{and} \hspace{10pt} \bbm \mbf{P}_{11} & \mbf{P}_{13} \\ * & \mbf{P}_{33} \ebm < 0.  
\label{eq:lemma3b}
\eeq
If the matrix inequalities in~\eqref{eq:lemma3b} hold, then a solution to~\eqref{eq:lemma3} is given by
\bdis
\mbf{X} = \mbf{P}_{13}^\trans\mbf{P}_{11}^{-1}\mbf{P}_{12}-\mbf{P}_{23}^\trans.
\edis

\itemcite \label{SchurProp3}\cite[p.~41]{Duan2013} Consider $\mbf{P}_{11}$,~$\mbf{X} \in \mathbb{S}^n$, $\mbf{P}_{12} \in \mathbb{R}^{n \times m}$, and $\mbf{P}_{22} \in \mathbb{S}^{m}$, where $\mbf{X} > 0$.  There exists $\mbf{X}$ such that 
\beq
\label{eq:lemma3_1}
\bbm \mbf{P}_{11} - \mbf{X} & \mbf{P}_{12}  & \mbf{X} \\ * & \mbf{P}_{22}  & \mbf{0} \\ * & * & -\mbf{X} \ebm < 0,
\eeq
if and only if
\beq
\label{eq:lemma3_2}
\bbm \mbf{P}_{11} & \mbf{P}_{12} \\ * & \mbf{P}_{22} \ebm < 0.
\eeq
\begin{proof}
The matrix inequality in~\eqref{eq:lemma3_1} can be rewritten using the Schur complement lemma as
\begin{align*}
\bbm \mbf{P}_{11} - \mbf{X} & \mbf{P}_{12} \\ * & \mbf{P}_{22} \ebm - \bbm \mbf{X} \\ \mbf{0} \ebm \left(-\mbf{X}^{-1}\right) \bbm \mbf{X} & \mbf{0} \ebm &< 0 \\
\bbm \mbf{P}_{11} - \mbf{X} & \mbf{P}_{12} \\ * & \mbf{P}_{22} \ebm + \bbm \mbf{X} & \mbf{0} \\ * & \mbf{0} \ebm &<0 \\
\bbm \mbf{P}_{11} & \mbf{P}_{12} \\ * & \mbf{P}_{22} \ebm &< 0.
\end{align*}
\end{proof}

\itemcite \cite{Gu1999},~\cite[p.~319--320]{Gu2003} Consider $\mbf{P}_{11} \in \mathbb{S}^n$, $\mbf{P}_{12} \in \mathbb{R}^{n \times m}$, $\mbf{P}_{22} \in \mathbb{S}^{m}$, $\mbf{P}_{23} \in \mathbb{R}^{m \times p}$, $\mbf{P}_{33} \in \mathbb{S}^p$, and $\mbf{X} \in \mathbb{R}^{n \times p}$.  There exists $\mbf{X}$ such that 
\bdis
\bbm \mbf{P}_{11}  & \mbf{P}_{12}  & \mbf{X} \\ * & \mbf{P}_{22}  & \mbf{P}_{23} \\ * & * & \mbf{P}_{33} \ebm > 0,
\edis
if and only if
\bdis
\bbm \mbf{P}_{11} & \mbf{P}_{12} \\ * & \mbf{P}_{22} \ebm > 0, \hspace{10pt} \text{and} \hspace{10pt} \bbm \mbf{P}_{22} & \mbf{P}_{23} \\ * & \mbf{P}_{33} \ebm > 0. 
\edis
\begin{proof}
The proof is found in~\cite{Gu2003}.
\end{proof}

\itemcite \cite[p.~320]{Gu2003} Consider $\mbf{P}_{11} \in \mathbb{S}^n$, $\mbf{P}_{12} \in \mathbb{R}^{n \times m}$, $\mbf{P}_{22} \in \mathbb{S}^{m}$, $\mbf{P}_{23} \in \mathbb{R}^{m \times p}$, $\mbf{P}_{33} \in \mathbb{S}^p$, $\mbf{E} \in \mathbb{R}^{p \times n}$, $\mbf{F} \in \mathbb{R}^{p \times m}$, and $\mbf{X} \in \mathbb{R}^{n \times p}$.  There exists $\mbf{X}$ such that 
\bdis
\bbm \mbf{P}_{11} + \mbf{X} \mbf{E} + \mbf{E}^\trans \mbf{X} & \mbf{P}_{12} + \mbf{X} \mbf{F}  & \mbf{X} \\ * & \mbf{P}_{22}  & \mbf{P}_{23} \\ * & * & \mbf{P}_{33} \ebm > 0,
\edis
if and only if
\bdis
\bbm \mbf{P}_{11} + \mbf{E}^\trans \mbf{P}_{33} \mbf{E} & \mbf{P}_{12} - \mbf{E}^\trans \mbf{P}_{23}^\trans + \mbf{E}^\trans \mbf{P}_{33} \mbf{F} \\ * & \mbf{P}_{22} - \mbf{P}_{23} \mbf{F} - \mbf{F}^\trans \mbf{P}_{23}^\trans + \mbf{F}^\trans \mbf{P}_{33} \mbf{F} \ebm > 0, \hspace{10pt} \text{and} \hspace{10pt} \bbm \mbf{P}_{22} & \mbf{P}_{23} \\ * & \mbf{P}_{33} \ebm > 0. 
\edis
\begin{proof}
The proof is found in~\cite{Gu2003}.
\end{proof}

\itemcite \label{sec:SchurProp6} \cite{GeromelNotes} Consider $\mbf{X} \in \mathbb{S}^n$, $\mbf{H} \in \mathbb{R}^{m \times n}$, $\mbf{G} \in \mathbb{R}^{m \times m}$, and $\mbf{P} \in \mathbb{S}^m$, where $\mbf{P} > 0$ and $\mbf{G}$ is invertible.  The matrix inequality given by
\beq
\label{eq:Schur4a}
\bbm \mbf{X} & \mbf{H}^\trans \\ * & \mbf{G}+\mbf{G}^\trans - \mbf{P}\ebm > 0,
\eeq
implies
\beq
\label{eq:Schur4b}
\mbf{X} > \mbf{H}^\trans\mbf{G}^{-1}\mbf{P}\mbf{G}^{-\trans} \mbf{H}.
\eeq
For $\mbf{G} = \mbf{P}$, this relationship becomes the Schur complement lemma.

\begin{proof}
Using the Schur complement lemma on~\eqref{eq:Schur4a} gives
\bdis
\mbf{X} > \mbf{H}^\trans \left( \mbf{G} + \mbf{G}^\trans - \mbf{P} \right)^{-1} \mbf{H}.
\edis
Using the property $\mbf{G} + \mbf{G}^\trans - \mbf{P} \leq \mbf{G}^\trans \mbf{P}^{-1} \mbf{G}$ (see the special case of Young's relation\index{Young's relation} in Section~\ref{sec:YoungsSpecial}), or equivalently $\left(\mbf{G} + \mbf{G}^\trans - \mbf{P}\right)^{-1} \geq \mbf{G}^{-1} \mbf{P} \mbf{G}^{-\trans}$ gives
\bdis
\mbf{X} > \mbf{H}^\trans \left( \mbf{G} + \mbf{G}^\trans - \mbf{P} \right)^{-1} \mbf{H} \geq \mbf{H}^\trans\mbf{G}^{-1}\mbf{P}\mbf{G}^{-\trans} \mbf{H},
\edis
thus implying~\eqref{eq:Schur4b}.
\end{proof}

Variations of this property are listed as follows.

\begin{enumerate}

\itemcite \cite{GeromelNotes} Consider $\mbf{X} \in \mathbb{S}^n$, $\mbf{H} \in \mathbb{R}^{n \times n}$, $\mbf{G} \in \mathbb{R}^{n \times n}$, and $\mbf{P} \in \mathbb{S}^n$, where $\mbf{P} > 0$ and $\mbf{G}$ is invertible.  The matrix inequality given by
\beq
\label{eq:Schur5a}
\bbm \mbf{H}+\mbf{H}^\trans - \mbf{X} & \mbf{G}^\trans \\ * & \mbf{P}\ebm > 0,
\eeq
implies
\beq
\label{eq:Schur5b}
\mbf{X} < \mbf{H}^\trans\mbf{G}^{-1}\mbf{P}\mbf{G}^{-\trans} \mbf{H}.
\eeq

\begin{proof}
Using the Schur complement lemma on~\eqref{eq:Schur5a} gives
\bdis
\mbf{H} + \mbf{H}^\trans - \mbf{X} - \mbf{G}^\trans \mbf{P}^{-1} \mbf{G} > 0.
\edis
Rearranging this expression yields
\beq
\label{eq:Schur5c}
\mbf{X} < \mbf{H} + \mbf{H}^\trans - \mbf{G}^\trans \mbf{P}^{-1} \mbf{G} .
\eeq
The property
\bdis
\mbf{H} + \mbf{H}^\trans \leq \left(\mbf{G}^{-1}\mbf{P} \mbf{G}^{-\trans}\right)^{-1} + \mbf{H}^\trans \left(\mbf{G}^{-1}\mbf{P} \mbf{G}^{-\trans}\right) \mbf{H}
\edis
is a special case of Young's relation\index{Young's relation} in Section~\ref{sec:YoungsSpecial}.  This can be equivalently written 
as 
\bdis
\mbf{H} + \mbf{H}^\trans -\mbf{G}^{\trans}\mbf{P}^{-1} \mbf{G} \leq  \mbf{H}^\trans \mbf{G}^{-1}\mbf{P} \mbf{G}^{-\trans} \mbf{H}.
\edis
Combining this expression with~\eqref{eq:Schur5c} gives
\bdis
\mbf{X} < \mbf{H} + \mbf{H}^\trans - \mbf{G}^\trans \mbf{P}^{-1} \mbf{G} \leq  \mbf{H}^\trans \mbf{G}^{-1}\mbf{P} \mbf{G}^{-\trans} \mbf{H}, 
\edis
thus implying~\eqref{eq:Schur5b}.
\end{proof}

\itemcite \cite{Chang2013} Consider $\mbf{A} \in \mathbb{S}^{n}$, $\mbf{B} \in \mathbb{R}^{n \times m}$, $\mbf{G} \in \mathbb{R}^{m \times m}$, $\mbf{P} \in \mathbb{S}^{m}$, and $\beta \in \mathbb{R}$.  The matrix inequality given by
\bdis
\bbm \mbf{A} & \mbf{B}\mbf{G} \\ * & -\beta \left(\mbf{G} + \mbf{G}^\trans\right) + \beta^2\mbf{P} \ebm < 0,
\edis
implies the matrix inequality $\mbf{A} + \mbf{B}\mbf{P}\mbf{B}^\trans < 0$.

\end{enumerate}

\itemcite \cite{Gu1999,Gu2001},~\cite[p.~321]{Gu2003} Consider $\mbf{P}_1  \in \mathbb{S}^n$, $\mbf{P}_2$,~$\mbf{X} \in \mathbb{S}^q$, $\mbf{Q}_1 \in \mathbb{R}^{n \times m}$, $\mbf{Q}_2 \in \mathbb{R}^{q \times p}$, $\mbf{R}_1 \in \mathbb{S}^m$, and $\mbf{R}_2 \in \mathbb{S}^p$.  The matrix inequalities given by
\beq
\label{eq:Schur9a}
\bbm \mbf{P}_1 - \mbf{L}\mbf{X}\mbf{L}^\trans & \mbf{Q}_1 \\ * & \mbf{R}_1 \ebm > 0, \hspace{10pt} \bbm \mbf{P}_2 + \mbf{X} & \mbf{Q}_2 \\ * & \mbf{R}_2 \ebm   > 0,
\eeq
are satisfied if  and only if
\beq
\label{eq:Schur9b}
\bbm \mbf{P}_1 + \mbf{L}\mbf{P}_2\mbf{L}^\trans & \mbf{Q}_1 & \mbf{L}\mbf{Q}_2 \\ * & \mbf{R}_1 & \mbf{0} \\ * & * & \mbf{R}_2 \ebm > 0.
\eeq
\begin{proof}
The proof is found in~\cite{Gu2001} and is very similar to the proof of Property~\ref{prop:SchurBased}.
\end{proof}

\itemcite \cite{Gu1999,Gu2001} Consider $\mbf{P}  \in \mathbb{S}^n$, $\mbf{R} \in \mathbb{S}^m$, $\mbf{S} \in \mathbb{S}^p$, $\mbf{Q} \in \mathbb{R}^{n \times m}$, $\mbf{X} \in \mathbb{R}^{n \times p}$, $\mbf{V} \in \mathbb{R}^{m \times p}$, and $\mbf{E} \in \mathbb{R}^{p \times m}$.  The matrix inequalities given by
\beq
\label{eq:Schur10a}
\bbm \mbf{P}  & \mbf{Q}  \\ * & \mbf{R}  - \mbf{V} \mbf{E} - \mbf{E}^\trans \mbf{V}^\trans + \mbf{E}^\trans \mbf{S} \mbf{E}\ebm > 0, \hspace{10pt} \bbm \mbf{R} & \mbf{V} \\ * & \mbf{S} \ebm   > 0,
\eeq
are satisfied if  and only if
\beq
\label{eq:Schur10b}
\bbm \mbf{P}& \mbf{Q} + \mbf{X} \mbf{E} & \mbf{X} \\ * & \mbf{R} & \mbf{V} \\ * & * & \mbf{S} \ebm > 0.
\eeq
\begin{proof}
The proof is found in~\cite{Gu2001} and is very similar to the proof of Property~\ref{prop:SchurBased}.
\end{proof}

\itemcite \label{sec:SchurProp7} \cite{Gahinet1994},~\cite[p.~229]{Dullerud2000} Consider $\mbf{P}_1$,~$\mbf{Q} \in \mathbb{S}^n$, $\mbf{P}_2$,~$\mbf{Q}_2 \in \mathbb{R}^{n \times m}$, and $\mbf{P}_3$,~$\mbf{Q}_3 \in \mathbb{S}^m$, where $\mbf{P}_1 > 0$, $\mbf{P}_3 > 0$, $\mbf{Q}_1 > 0$, and $\mbf{Q}_3 > 0$.  There exist $\mbf{P}_2$, $\mbf{P}_3$, $\mbf{Q}_2$, and $\mbf{Q}_3$ such that
\beq
\label{eq:Schur11a}
\bbm \mbf{P}_1 & \mbf{P}_2 \\ * & \mbf{P}_3 \ebm > 0, \hspace{10pt} \bbm \mbf{P}_1 & \mbf{P}_2 \\ * & \mbf{P}_3 \ebm^{-1} = \bbm \mbf{Q}_1 & \mbf{Q}_2 \\ * & \mbf{Q}_3 \ebm,
\eeq
if and only if
\beq
\label{eq:Schur11b}
\bbm \mbf{P}_1 & \mbf{1} \\ * & \mbf{Q}_1 \ebm \geq 0, \hspace{10pt} \mathrm{rank} \left( \bbm \mbf{P}_1 & \mbf{1} \\ * & \mbf{Q}_1 \ebm \right) \leq n + m.
\eeq
\index{rank}Provided $\mbf{P}_1$ and $\mbf{Q}_1$ satisfy~\eqref{eq:Schur11b}, a solution to~\eqref{eq:Schur11a} is given by $\mbf{P}_3 = \mbf{1}$, $\mbf{Q}_2 = - \mbf{Q}_1 \mbf{P}_2$, $\mbf{Q}_3 = \mbf{P}_2^\trans \mbf{Q}_1 \mbf{P}_2 + \mbf{1}$, and $\mbf{P}_2$ satisfies $\mbf{P}_2 \mbf{P}_2^\trans = \mbf{P}_1 - \mbf{Q}_1^{-1}$.

\itemcite \cite[pp.~13--14]{Zhan2002} Consider $\mbf{X}$,~$\mbf{Y} \in \mathbb{R}^{n \times m}$, $\mbf{P}$,~$\mbf{Q} \in \mathbb{S}^{n}$, and $\epsilon \in \mathbb{R}_{>0}$, where $\mbf{P} > 0$, $\mbf{Q} > 0$, and $\epsilon \geq 1$.  The matrix inequality given by
\beq
\label{eq:Schur12}
\epsilon \mbf{X}^\trans \mbf{P}^{-1}\mbf{X} + \epsilon \mbf{Y}^\trans \mbf{Q}^{-1}\mbf{Y} \geq \left(\mbf{X} + \mbf{Y} \right)^\trans\left(\mbf{P} + \mbf{Q} \right)^{-1}\left(\mbf{X} + \mbf{Y} \right)
\eeq
holds.
\begin{proof}
Since $\mbf{P} > 0$, $\mbf{Q} > 0$, and $\epsilon \geq 1$, it is known that $(\epsilon-1)\mbf{X}^\trans\mbf{P}^{-1}\mbf{X} \geq 0$ and $(\epsilon-1)\mbf{Y}^\trans \mbf{Q}^{-1} \mbf{Y} \geq 0$.  These inequalities are rewritten as
\beq
\label{eq:Schur12a}
\epsilon \mbf{X}^\trans \mbf{P}^{-1} \mbf{X} - \mbf{X}^\trans  \mbf{P} ^{-1} \mbf{X} \geq 0, \hspace{20pt} \epsilon \mbf{Y}^\trans \mbf{P}^{-1} \mbf{Y} - \mbf{X}^\trans \ \mbf{Q}^{-1} \mbf{Y} \geq 0.
\eeq
Applying the Schur complement lemma to the expressions in~\eqref{eq:Schur12a} results in
\beq
\label{eq:Schur12b}
\bbm  \mbf{P} & \mbf{X} \\ * & \epsilon \mbf{X}^\trans \mbf{P}^{-1} \mbf{X} \ebm \geq 0, \hspace{20pt} \bbm  \mbf{Q} & \mbf{Y} \\ * & \epsilon \mbf{Y}^\trans \mbf{Q}^{-1} \mbf{Y} \ebm \geq 0.
\eeq
The matrix inequalities in~\eqref{eq:Schur12b} imply
\beq
\label{eq:Schur12c}
\bbm \mbf{P} + \mbf{Q}  & \mbf{X} + \mbf{Y}  \\ *& \epsilon \mbf{X}^\trans \mbf{P}^{-1} \mbf{X} +\epsilon \mbf{Y}^\trans \mbf{Q}^{-1} \mbf{Y} \ebm \geq 0.
\eeq
Applying the Schur complement lemma to~\eqref{eq:Schur12c} yields
\beq
\label{eq:Schur12d}
 \epsilon \mbf{X}^\trans \mbf{P}^{-1} \mbf{X} +\epsilon \mbf{Y}^\trans \mbf{Q}^{-1} \mbf{Y} - \left(\mbf{X} + \mbf{Y} \right)^\trans \left( \mbf{P} + \mbf{Q} \right)^{-1} \left(\mbf{X} + \mbf{Y} \right) \geq 0.
\eeq
Rearranging~\eqref{eq:Schur12d} gives~\eqref{eq:Schur12}.
\end{proof}

\itemcite (\textit{Linearization Lemma}~\cite[pp.~91--92]{SchererWeiland2015})  Consider $\mbf{X} \in \mathbb{R}^{n \times p}$, $\mbf{S} \in \mathbb{R}^{n \times m}$, $\mbf{T} \in \mathbb{R}^{m \times q}$, $\mbf{Y}(v) \in \mathbb{R}^{m \times p}$, $\mbf{Q}(v) \in \mathbb{S}^n$, $\mbf{R}(v) \in \mathbb{S}^m$, and $\mbf{U}(v) \in \mathbb{S}^q$, where $\mbf{Y}(v)$, $\mbf{Q}(v)$, and $\mbf{R}(v)$ depend affinely on the parameter $v$, and $\mbf{R}(v)$ can be decomposed as $\mbf{R}(v) = \mbf{T} \mbf{U}^{-1}(v) \mbf{T}^{-1}$.  The matrix inequalities $\mbf{U}(v) > 0$ and
\bdis
\bbm \mbf{X} \\ \mbf{Y}(v) \ebm^\trans \bbm \mbf{Q}(v) & \mbf{S} \\ \mbf{S}^\trans & \mbf{R}(s) \ebm \bbm \mbf{X} \\ \mbf{Y}(v) \ebm < 0
\edis
are equivalent to
\bdis
\bbm \mbf{X}^\trans \mbf{Q}(v) \mbf{X} + \mbf{X}^\trans \mbf{S} \mbf{Y}(v) + \mbf{Y}^\trans(v) \mbf{S}^\trans \mbf{X} & \mbf{X}^\trans(v) \mbf{T} \\ * & -\mbf{U}(v) \ebm < 0.
\edis

\end{enumerate}

\subsection{Projection Lemma (Matrix Elimination Lemma)}

\subsubsection[Strict Projection Lemma]{Strict Projection Lemma~\cite{Gahinet1994},~\cite[pp.~22--23]{Boyd1994},~\cite[pp.~110--111]{SchererWeiland2015},~\cite[Sec.~12.3.5]{Skogestad2005}}
\index{projection lemma!strict projection lemma}
Consider $\mbs{\Psi} \in \mathbb{S}^n$, $\mbf{G} \in \mathbb{R}^{n \times m}$, $\mbs{\Lambda} \in \mathbb{R}^{m \times p}$, and $\mbf{H} \in \mathbb{R}^{n \times p}$.  There exists $\mbs{\Lambda}$ such that
\beq
\label{eq:proj}
\mbs{\Psi} + \mbf{G}\mbs{\Lambda}\mbf{H}^\trans + \mbf{H}\mbs{\Lambda}^\trans\mbf{G}^\trans < 0,
\eeq
if and only if
\begin{align*}
\mbf{N}_G^\trans\mbs{\Psi}\mbf{N}_G &< 0, \\
\mbf{N}_H^\trans\mbs{\Psi}\mbf{N}_H &< 0,
\end{align*}
where $\mathcal{R}(\mbf{N}_G) = \mathcal{N}(\mbf{G}^\trans)$ and $\mathcal{R}(\mbf{N}_H) = \mathcal{N}(\mbf{H}^\trans)$.

\subsubsection[Nonstrict Projection Lemma]{Nonstrict Projection Lemma~\cite[p.~93]{HelmerssonThesis},~\cite{meijer2024non}}
\label{sec:NonstrictProjectionLemma}
\index{projection lemma!nonstrict projection lemma}
Consider $\mbs{\Psi} \in \mathbb{S}^n$, $\mbf{G} \in \mathbb{R}^{n \times m}$, $\mbs{\Lambda} \in \mathbb{R}^{m \times p}$, and $\mbf{H} \in \mathbb{R}^{n \times p}$, where $\mathcal{R}(\mbf{G})$ and $\mathcal{R}(\mbf{H})$ are linearly independent.  There exists $\mbs{\Lambda}$ such that
\bdis
\mbs{\Psi} + \mbf{G}\mbs{\Lambda}\mbf{H}^\trans + \mbf{H}\mbs{\Lambda}^\trans\mbf{G}^\trans \leq 0,
\edis
if and only if
\begin{align}
\mbf{N}_G^\trans\mbs{\Psi}\mbf{N}_G &\leq 0, \label{eq:NonstricProjection2a}\\
\mbf{N}_H^\trans\mbs{\Psi}\mbf{N}_H &\leq 0, \label{eq:NonstricProjection2b}
\end{align}
where $\mathcal{R}(\mbf{N}_G) = \mathcal{N}(\mbf{G}^\trans)$ and $\mathcal{R}(\mbf{N}_H) = \mathcal{N}(\mbf{H}^\trans)$.

The requirement that $\mathcal{R}(\mbf{G})$ and $\mathcal{R}(\mbf{H})$ are linearly independent can be removed, provided~\eqref{eq:NonstricProjection2a} and~\eqref{eq:NonstricProjection2b} are satisfied with~\cite{meijer2024non} 
\bdis
\mathcal{R}(\mbf{N}_G) \cap \mathcal{R}(\mbf{N}_H) \cap \{\mbs{\xi} \in \mathbb{R}^n \, | \, \mbs{\xi}^\trans \mbs{\Psi} \mbs{\xi} = 0 \} \subset \mathcal{N}(\mbs{\Psi}).
\edis

Note that these results extend to complex matrices.

\subsubsection[Reciprocal Projection Lemma]{Reciprocal Projection Lemma~\cite{Apkarian2001}}
\index{projection lemma!reciprocal projection lemma}
Consider $\mbf{P}$,~$\mbs{\Psi} \in \mathbb{S}^n$ and $\mbf{W}$,~$\mbf{S} \in \mathbb{R}^{n \times n}$.  There exists $\mbf{W}$ such that
\bdis
\bbm \mbs{\Psi} + \mbf{P} - \left(\mbf{W} + \mbf{W}^\trans\right) & \mbf{S}^\trans + \mbf{W}^\trans \\ * & -\mbf{P} \ebm < 0,
\edis
if and only if $\mbs{\Psi} + \mbf{S} + \mbf{S}^\trans < 0$.

\subsubsection{Projection Lemma-Based Properties}

\begin{enumerate}

\itemcite \cite{Chang2011} Consider $\mbf{A} \in \mathbb{S}^{n}$, $\mbf{B}$,~$\mbf{J} \in \mathbb{R}^{n \times m}$, $\mbf{G} \in \mathbb{R}^{m \times m}$, and $\mbf{P} \in \mathbb{S}^{m}$.  The matrix inequality given by
\beq
\label{eq:Schur6a}
\bbm \mbf{A} + \mbf{B}\mbf{J}^\trans + \mbf{J}\mbf{B}^\trans & - \mbf{J} + \mbf{B}\mbf{G} \\ * & -\left(\mbf{G} + \mbf{G}^\trans\right) + \mbf{P} \ebm < 0,
\eeq
implies the matrix inequality
\beq
\label{eq:Schur6b}
\mbf{A} + \mbf{B}\mbf{P}\mbf{B}^\trans < 0.
\eeq
If the matrices $\mbf{J}$ and $\mbf{G}$ are free (i.e., they are design variables), then the matrix inequalities~\eqref{eq:Schur6a} and~\eqref{eq:Schur6b} are equivalent~\cite{Delmotte2007}.

\itemcite \cite{Chang2011a} Consider $\mbf{T} \in \mathbb{S}^{n}$ and $\mbf{A}$,~$\mbf{J}$,~$\mbf{G}$,~$\mbf{P} \in \mathbb{R}^{n \times n}$.  The matrix inequality given by
\beq
\label{eq:Schur7a}
\bbm \mbf{T} + \mbf{A}^\trans\mbf{J}^\trans + \mbf{J}\mbf{A} & \mbf{P} -\mbf{J} + \mbf{A}^\trans\mbf{G} \\ * & -\left(\mbf{G} + \mbf{G}^\trans\right)  \ebm < 0
\eeq
implies the matrix inequality
\beq
\label{eq:Schur7b}
\mbf{T} + \mbf{A}^\trans\mbf{P}^\trans + \mbf{P}\mbf{A} < 0.
\eeq
If the matrices $\mbf{J}$ and $\mbf{G}$ are free (i.e., they are design variables), then the matrix inequalities~\eqref{eq:Schur7a} and~\eqref{eq:Schur7b} are equivalent~\cite{Delmotte2007}.

\itemcite \cite{Delmotte2007} Consider $\mbf{T}_1$,~$\mbf{P} \in \mathbb{S}^{n}$, $\mbf{A}$,~$\mbf{J}_1$,~$\mbf{G} \in \mathbb{R}^{n \times n}$, $\mbf{T}_2 \in \mathbb{R}^{n \times m}$, $\mbf{J}_2 \in \mathbb{R}^{m \times n}$, and $\mbf{T}_3 \in \mathbb{S}^m$, where $\mbf{P} > 0$ and $\mbf{T}_3 < 0$.  The matrix inequality given by
\beq
\label{eq:Schur8a}
\bbm \mbf{T}_1 + \mbf{A}^\trans\mbf{J}_1^\trans + \mbf{J}_1\mbf{A} & \mbf{T}_2 + \mbf{A}^\trans \mbf{J}_2^\trans & \mbf{P} -\mbf{J}_1 + \mbf{A}^\trans\mbf{G} \\ * & \mbf{T}_3 & -\mbf{J}_2 \\ * & * & -\left(\mbf{G} + \mbf{G}^\trans\right)  \ebm < 0
\eeq
implies the matrix inequality
\beq
\label{eq:Schur8b}
\bbm \mbf{T}_1 + \mbf{A}^\trans\mbf{P} + \mbf{P}\mbf{A} & \mbf{T}_2 \\ * & \mbf{T}_3 \ebm < 0.
\eeq
If the matrices $\mbf{J}_1$, $\mbf{J}_2$, and $\mbf{G}$ are free (i.e., they are design variables), then the matrix inequalities~\eqref{eq:Schur8a} and~\eqref{eq:Schur8b} are equivalent.

\itemcite \cite[p.~9]{Chang2014} Consider $\mbf{T} \in \mathbb{S}^{n}$, $\mbf{A} $,~$\mbf{G}$,~$\mbf{P} \in \mathbb{R}^{n \times n}$, and $\beta \in \mathbb{R}$, where $\mbf{T} < 0$.  The matrix inequality given by
\bdis
\bbm \mbf{T} & \beta \mbf{P} + \mbf{A}^\trans\mbf{G} \\ * & -\beta \left(\mbf{G} + \mbf{G}^\trans\right)  \ebm < 0,
\edis
implies the matrix inequality $\mbf{T} + \mbf{A}^\trans\mbf{P}^\trans + \mbf{P}\mbf{A} < 0$.

\end{enumerate}

\subsection{Finsler's Lemma}
\label{sec:FinslersLemma}
\index{Finsler's lemma!lemma}
\subsubsection[Finsler's Lemma]{Finsler's Lemma~\cite[pp.~22--23]{Boyd1994},~\cite[Sec.~12.3.5]{Skogestad2005},~\cite{Finsler1936}}
Consider $\mbs{\Psi} \in \mathbb{S}^n$, $\mbf{G} \in \mathbb{R}^{n \times m}$, $\mbs{\Lambda} \in \mathbb{R}^{m \times p}$, $\mbf{H} \in \mathbb{R}^{n \times p}$, and $\sigma \in \mathbb{R}$.  There exists $\mbs{\Lambda}$ such that
\bdis
\mbs{\Psi} + \mbf{G}\mbs{\Lambda}\mbf{H}^\trans + \mbf{H}\mbs{\Lambda}^\trans\mbf{G}^\trans < 0,
\edis
if and only if there exists $\sigma$ such that
\begin{align*}
\mbs{\Psi} - \sigma \mbf{G}\mbf{G}^\trans &< 0, \\
\mbs{\Psi} - \sigma \mbf{H}\mbf{H}^\trans &< 0.
\end{align*}

\subsubsection[Alternative Form of Finsler's Lemma]{Alternative Form of Finsler's Lemma~\cite{Finsler1936,Petersen1987,deOliveira2001,Gonccalves2025robust},~\cite[pp.~90--97]{Jacobson1977},~\cite[pp.~41--48]{Skelton1998}}
\label{sec:FinslerAlt}
Consider $\mbs{\Psi} \in \mathbb{S}^n$, $\mbf{Z} \in \mathbb{R}^{p \times n}$, and $\mbf{x} \in \mathbb{R}^{n}$, where $\textrm{rank}(\mbf{Z}) < n$.  The following statements are equivalent.

\begin{enumerate}

\item The inequality
\bdis
\mbf{x}^\trans \mbs{\Psi} \mbf{x} < 0
\edis
is satisfied for all $\mbf{x} $ satisfying $\mbf{Z} \mbf{x} = \mbf{0}$, where $\mbf{x} \neq \mbf{0}$.

\item The matrix inequality
\bdis
\mbf{N}_Z^\trans \mbs{\Psi} \mbf{N}_Z < 0
\edis
is satisfied, where $\mathcal{R}(\mbf{N}_Z) = \mathcal{N}(\mbf{Z})$.

\item There exists $\sigma \in \mathbb{R}$ such that
\bdis
\mbs{\Psi} - \sigma \mbf{Z}^\trans\mbf{Z} < 0.
\edis

\item There exists $\mbf{X} \in \mathbb{R}^{n \times p}$ such that
\bdis
\mbs{\Psi} + \mbf{X} \mbf{Z} + \mbf{Z}^\trans \mbf{X}^\trans < 0.
\edis

\item There exist $\mbf{X} \in \mathbb{R}^{n \times p}$, $\mbf{F} \in \mathbb{R}^{n \times p}$, and $\mbf{G} \in \mathbb{R}^{p \times p}$ such that
\bdis
\bbm \mbs{\Psi} & \mbf{X} \\ \mbf{X}^\trans & \mbf{0} \ebm + \bbm \mbf{F} \\ \mbf{G} \ebm \bbm \mbf{Z} & -\mbf{1} \ebm + \bbm \mbf{Z}^\trans \\ -\mbf{1} \ebm \bbm \mbf{F}^\trans & \mbf{G}^\trans \ebm < 0.
\edis

\item There exist $\mbf{F}_1 \in \mathbb{R}^{p \times n}$, $\mbf{F}_2 \in \mathbb{R}^{p \times p}$, $\mbf{G}_1 \in \mathbb{R}^{n \times p}$, and $\mbf{G}_2 \in \mathbb{R}^{p \times p}$ such that
\bdis
\bbm \mbs{\Psi} & \mbf{Z}^\trans \\ \mbf{Z} & \mbf{0} \ebm + \bbm \mbf{F}_1^\trans \\ \mbf{F}_2^\trans \ebm \bbm \mbf{G}_1^\trans & \mbf{G}_1 \ebm + \bbm \mbf{G}_1 \\ \mbf{G}_2 \ebm \bbm \mbf{F}_1 & \mbf{F}_2 \ebm < 0.
\edis

\end{enumerate}

\subsubsection[Non-Strict Finsler's Lemma]{Non-Strict Finsler's Lemma~\cite{Finsler1936,more1993generalizations,zi2010some,anstreicher2000note}}
\label{sec:NonStrictFinsler}
\index{Finsler's lemma!non-strict lemma}
Consider $\mbs{\Psi} \in \mathbb{S}^n$, $\mbf{Z} \in \mathbb{S}^{n \times n}$, and $\mbf{x} \in \mathbb{R}^{n}$, where $\mbf{Z}$ is indefinite.  Then, the inequality
\bdis
\mbf{x}^\trans \mbs{\Psi} \mbf{x} \geq 0
\edis
holds for all $\mbf{x} \in \mathbb{R}^n$ satisfying $\mbf{x}^\trans\mbf{Z} \mbf{x} = \mbf{0}$ if and only if there exists $\sigma \in \mathbb{R}$ such that
\bdis
\mbs{\Psi} - \sigma \mbf{Z} \geq 0.
\edis

An alternative form of this result is presented that does not require the indefiniteness of $\mbf{Z}$~\cite{anstreicher2000note}.  Consider $\mbs{\Psi} \in \mathbb{S}^n$, $\mbf{Z} \in \mathbb{S}^{n \times n}$, and $\mbf{x} \in \mathbb{R}^{n}$, where $\mbf{Z} \geq 0$.  Suppose that the inequality
\bdis
\mbf{x}^\trans \mbs{\Psi} \mbf{x} \geq 0
\edis
holds for all $\mbf{x} \in \mathbb{R}^n$ satisfying $\mbf{x}^\trans\mbf{Z} \mbf{x} = \mbf{0}$.  Then, there exists $\sigma \in \mathbb{R}$ such that
\bdis
\mbs{\Psi} - \sigma \mbf{Z} \geq 0
\edis
if and only if
\bdis
\mathcal{N}(\mbf{N}_Z^\trans \mbs{\Psi} \mbf{N}_Z) = \mathcal{N}(\mbf{N}_Z^\trans \mbs{\Psi}^2 \mbf{N}_Z),
\edis
where $\mathcal{R}(\mbf{N}_Z) = \mathcal{N}(\mbf{Z})$.

\subsubsection[Matrix Finsler's Lemma]{Matrix Finsler's Lemma~\cite{van2021matrix}}
\label{sec:MatrixFinsler}
\index{Finsler's lemma!matrix lemma}
Consider $\mbf{P} \in \mathbb{S}^{n+m}$ and $\mbf{Q} \in \mathbb{S}^{n+m}$ partitioned as
\bdis
\bbm \mbf{P}_{11} & \mbf{P}_{12} \\ \mbf{P}_{12}^\trans & \mbf{P}_{22} \ebm, \quad \bbm \mbf{Q}_{11} & \mbf{Q}_{12} \\ \mbf{Q}_{12}^\trans & \mbf{Q}_{22} \ebm.
\edis
Assume that
\begin{enumerate}

\item $\mbf{P}_{12} = \mbf{0}$ and $\mbf{P}_{22} \geq 0$.

\item $\mbf{Q}_{22} \leq 0$ and $\mbf{Q}_{11} - \mbf{Q}_{12} \mbf{Q}_{22}^\dagger \mbf{Q}_{12}^\trans = 0$, where $\mbf{Q}_{22}^\dagger$ is the pseudoinverse\index{pseudoinverse} of $\mbf{Q}_22$.

\item There exists $\mbf{G} \in \mathbb{R}^{m \times n}$ such that $\mbf{P}_{11} + \mbf{G}^\trans \mbf{P}_{22} \mbf{G} > 0$ and $\mbf{Q}_{22} \mbf{G} = \mbf{Q}_{12}^\trans$.

\end{enumerate}

Then, the matrix inequality
\bdis
\bbm \mbf{1} \\ \mbf{Z} \ebm^\trans \mbf{P} \bbm \mbf{1} \\ \mbf{Z} \ebm \geq 0
\edis
holds for all $\mbf{Z} \in \mathbb{R}^{m \times n}$ satisfying
\bdis
\bbm \mbf{1} \\ \mbf{Z} \ebm^\trans \mbf{Q} \bbm \mbf{1} \\ \mbf{Z} \ebm = \mbf{0}
\edis
if and only if there exists a scalar $\alpha \in \mathbb{R}$ such that $\mbf{P} - \alpha \mbf{Q} \geq 0$.

\subsubsection[Modified Finsler's Lemma]{Modified Finsler's Lemma~\cite[p.~37]{Wu2010},\cite{Xie1992,Xie1996}}
\index{Finsler's lemma!modified lemma}
Consider $\mbs{\Psi} \in \mathbb{S}^n$, $\mbf{G} \in \mathbb{R}^{n \times m}$, $\mbs{\Lambda} \in \mathbb{R}^{m \times p}$, $\mbf{H} \in \mathbb{R}^{n \times p}$, and $\mbf{R} \in \mathbb{S}^p$, where $\mbs{\Lambda}^\trans\mbs{\Lambda} \leq \mbf{R}$ and $\mbf{R} > 0$.  There exists $\mbs{\Lambda}$ such that
\beq
\label{eq:ModFinsler1}
\mbs{\Psi} + \mbf{G}\mbs{\Lambda}\mbf{H}^\trans + \mbf{H}\mbs{\Lambda}^\trans\mbf{G}^\trans < 0,
\eeq
if and only if there exists $\epsilon \in \mathbb{R}_{>0}$ such that
\beq
\label{eq:ModFinsler2}
\mbs{\Psi} +\epsilon^{-1}  \mbf{G}\mbf{G}^\trans + \epsilon \mbf{H}\mbf{R} \mbf{H}^\trans < 0.
\eeq
\begin{proof}
The proof of~\eqref{eq:ModFinsler2} $\implies$~\eqref{eq:ModFinsler1} follows from a completion of the squares argument.  The authors are not aware of a complete proof of~\eqref{eq:ModFinsler1} $\implies$~\eqref{eq:ModFinsler2}, so use this identity with caution.
\end{proof}

\subsubsection[Robust Finsler's Lemma]{Robust Finsler's Lemma~\cite{kussaba2015uniform,ishihara2017existence}}
\label{sec:RobustFinsler}
\index{Finsler's lemma!robust lemma}
Consider $\mathcal{S} \subseteq \mathbb{R}^d$ and the variable $\mbf{s} \in \mathcal{S}$, along with the matrix functions $\mbs{\Psi}(\mbf{s}) \in \mathbb{S}^n$ and $\mbf{Z}(\mbf{s}) \in \mathbb{R}^{p \times n}$.  The following statements are equivalent.

\begin{enumerate}

\item For each value of $\mbf{s} \in \mathcal{S}$, the inequality
\bdis
\mbf{x}^\trans \mbs{\Psi}(\mbf{s}) \mbf{x} < 0
\edis
is satisfied for all $\mbf{x} \in \mathbb{R}^n$ satisfying $\mbf{Z}(\mbf{s}) \mbf{x} = \mbf{0}$, where $\mbf{x} \neq \mbf{0}$.

\item For each value of $\mbf{s} \in \mathcal{S}$, the matrix inequality
\bdis
\mbf{N}_Z^\trans(\mbf{s}) \mbs{\Psi}(\mbf{s}) \mbf{N}_Z(\mbf{s}) < 0
\edis
is satisfied, where $\mathcal{R}(\mbf{N}_Z(\mbf{s})) = \mathcal{N}(\mbf{Z}(\mbf{s}))$.

\item For all $\mbf{s} \in \mathcal{S}$, there exists $\sigma(\mbf{s}) \in \mathbb{R}$ such that
\bdis
\mbs{\Psi}(\mbf{s}) - \sigma(\mbf{s}) \mbf{Z}^\trans(\mbf{s})\mbf{Z}(\mbf{s}) < 0.
\edis

\item For each value of $\mbf{s} \in \mathcal{S}$, there exists $\mbf{X}(\mbf{s}) \in \mathbb{R}^{n \times p}$ such that
\bdis
\mbs{\Psi}(\mbf{s}) + \mbf{X}(\mbf{s}) \mbf{Z}(\mbf{s}) + \mbf{Z}^\trans(\mbf{s}) \mbf{X}^\trans(\mbf{s}) < 0.
\edis
\end{enumerate}

Additional conditions that take advantage of functional structure within $\sigma(\mbf{s})$ and $\mbf{X}(\mbf{s})$ are found in~\cite{kussaba2015uniform,ishihara2017existence}.

\subsubsection[Strict Petersen's Lemma]{Strict Petersen's Lemma~\cite{Petersen1986,Petersen1987,Bisoffi2022}}
\label{sec:StrictPetersen}
\index{Petersen's lemma!strict Petersen's lemma}

Consider $\mbs{\Psi} \in \mathbb{S}^n$, $\mbf{G} \in \mathbb{R}^{n \times m}$, $\mbf{H} \in \mathbb{R}^{n \times p}$, and $\mbf{R} \in \mathbb{S}^p$, where $\mbf{R} \geq 0$.  Also consider the set $\mathcal{F} := \{\mbf{F} \in \mathbb{R}^{m \times p} \,\, | \,\, \mbf{F}^\trans \mbf{F} \leq  \mbf{R}\}$.  The matrix inequality
\bdis
\mbs{\Psi} + \mbf{G}\mbf{F}\mbf{H}^\trans + \mbf{H}\mbf{F}^\trans\mbf{G}^\trans < 0,
\edis
holds for all $\mbf{F} \in \mathcal{F}$ if and only if there exists $\epsilon \in \mathbb{R}_{>0}$ such that
\beq
\label{eq:StrictPetersen1}
\mbs{\Psi} +\epsilon^{-1}  \mbf{G}\mbf{G}^\trans + \epsilon \mbf{H}\mbf{R} \mbf{H}^\trans < 0.
\eeq
The matrix inequality in~\eqref{eq:StrictPetersen1} can be equivalently rewritten using the Schur complement as~\cite{Khlebnikov2018}
\bdis
\bbm \mbs{\Psi} +\epsilon \mbf{H}\mbf{R} \mbf{H}^\trans & \mbf{G} \\ * & -\epsilon \mbf{1} \ebm < 0.
\edis

A modification to the Strict Petersen's lemma found in~\cite{Khlebnikov2018} is stated as follows.  Consider $\mbs{\Psi} \in \mathbb{S}^n$, $\mbf{G} \in \mathbb{R}^{n \times m}$, $\mbf{y} \in \mathbb{R}^{n}$, and $\mbf{R} \in \mathbb{S}^m$, where $\mbf{R} > 0$.  Also consider the set $\mathcal{X} := \{\mbf{x} \in \mathbb{R}^{m } \,\, | \,\, \mbf{x}^\trans \mbf{R} \mbf{x} \leq  1\}$.  The matrix inequality
\bdis
\mbs{\Psi} + \mbf{G}\mbf{x}\mbf{y}^\trans + \mbf{y}\mbf{x}^\trans\mbf{G}^\trans < 0,
\edis
holds for all $\mbf{x} \in \mathcal{X}$ if and only if there exists $\epsilon \in \mathbb{R}_{>0}$ such that
\bdis
\bbm \mbs{\Psi} & \mbf{G} & \mbf{y} \\ * & -\epsilon \mbf{R} & \mbf{0} \\ * & * & -\epsilon^{-1} \mbf{1} \ebm < 0.
\edis

\subsubsection[Nonstrict Petersen's Lemma]{Nonstrict Petersen's Lemma~\cite{Khlebnikov2008,Shcherbakov2008,Bisoffi2022}}
\label{sec:NonstrictPetersen}
\index{Petersen's lemma!nonstrict Petersen's lemma}

Consider $\mbs{\Psi} \in \mathbb{S}^n$, $\mbf{G} \in \mathbb{R}^{n \times m}$, $\mbf{H} \in \mathbb{R}^{n \times p}$, and $\mbf{R} \in \mathbb{S}^p$, where $\mbf{R} > 0$, $\mbf{G} \neq \mbf{0}$, and $\mbf{H} \neq \mbf{0}$.  Also consider the set $\mathcal{F} := \{\mbf{F} \in \mathbb{R}^{m \times p} \,\, | \,\, \mbf{F}^\trans \mbf{F} \leq  \mbf{R}\}$.  The matrix inequality
\beq
\label{eq:NonstrictPetersen1a}
\mbs{\Psi} + \mbf{G}\mbf{F}\mbf{H}^\trans + \mbf{H}\mbf{F}^\trans\mbf{G}^\trans \leq 0,
\eeq
holds for all $\mbf{F} \in \mathcal{F}$ if and only if there exists $\epsilon \in \mathbb{R}_{>0}$ such that
\beq
\label{eq:NonstrictPetersen1}
\mbs{\Psi} +\epsilon^{-1}  \mbf{G}\mbf{G}^\trans + \epsilon \mbf{H}\mbf{R} \mbf{H}^\trans \leq 0.
\eeq
If the assumptions $\mbf{R} > 0$, $\mbf{G} \neq \mbf{0}$, and $\mbf{H} \neq \mbf{0}$ are removed, then the matrix inequality in~\eqref{eq:NonstrictPetersen1a} holds for all if there exists $\epsilon \in \mathbb{R}_{>0}$ satisfying~\eqref{eq:NonstrictPetersen1}.

The matrix inequality in~\eqref{eq:NonstrictPetersen1} can be equivalently rewritten using the Schur complement as~\cite{Khlebnikov2018}
\bdis
\bbm \mbs{\Psi} +\epsilon \mbf{H}\mbf{R} \mbf{H}^\trans & \mbf{G} \\ * & -\epsilon \mbf{1} \ebm \leq 0.
\edis

The following are slight modifications to the original lemma.

\begin{enumerate}

\itemcite \cite{Khlebnikov2014} Consider $\mbs{\Psi} \in \mathbb{S}^n$, $\mbf{G} \in \mathbb{R}^{n \times m}$, and $\mbf{H} \in \mathbb{R}^{n \times p}$, where $\mbf{G} \neq \mbf{0}$ and $\mbf{H} \neq \mbf{0}$.  Also consider the set $\mathcal{F} := \{\mbf{F} \in \mathbb{R}^{m \times p} \,\, | \,\, \norm{\mbf{F}}_\frob \leq  1\}$.  The matrix inequality
\bdis
\mbs{\Psi} + \mbf{G}\mbf{F}\mbf{H}^\trans + \mbf{H}\mbf{F}^\trans\mbf{G}^\trans \leq 0,
\edis
holds for all $\mbf{F} \in \mathcal{F}$ if and only if there exists $\epsilon \in \mathbb{R}_{>0}$ such that
\bdis
\mbs{\Psi} +\epsilon^{-1}  \mbf{G}\mbf{G}^\trans + \epsilon \mbf{H} \mbf{H}^\trans \leq 0.
\edis

\itemcite \cite{Khlebnikov2014} Consider $\mbs{\Psi} \in \mathbb{S}^n$, $\mbf{G} \in \mathbb{R}^{n \times m}$, and $\mbf{H} \in \mathbb{R}^{n \times m}$, where $\mbf{G} \neq \mbf{0}$ and $\mbf{H} \neq \mbf{0}$.  Also consider the set $\mathcal{F} := \{\mbf{F} \in \mathbb{S}^{m \times m} \,\, | \,\, - \mbf{1} \leq \mbf{F} \leq  \mbf{1} \}$.  The matrix inequality
\bdis
\mbs{\Psi} + \mbf{G}\mbf{F}\mbf{H}^\trans + \mbf{H}\mbf{F}^\trans\mbf{G}^\trans \leq 0,
\edis
holds for all $\mbf{F} \in \mathcal{F}$ if and only if there exists $\epsilon \in \mathbb{R}_{>0}$ such that
\bdis
\mbs{\Psi} +\epsilon^{-1}  \mbf{G}\mbf{G}^\trans + \epsilon \mbf{H}\mbf{H}^\trans \leq 0.
\edis

\end{enumerate}

\subsection[Dilation]{Dilation}
\label{sec:Dilation}
\index{dilation}
Matrix inequalities can be dilated to obtain a larger matrix inequality, often with additional design variables.  This can be a useful technique to separate design variables in a BMI.

A common technique to dilate an LMI involves the use the projection lemma in reverse or the reciprocal projection lemma.  For instance, consider the following example taken from~\cite{Apkarian2001} and inspired by the dilated bounded real lemma matrix inequality in~\cite[pp.~153--155]{Duan2013} involving the matrices $\mbf{P}  \in \mathbb{S}^{n}$ and $\mbf{A} \in \mathbb{R}^{n \times n}$, where $\mbf{P} > 0$.  The matrix inequality
\beq
\bbm \mbf{P}\mbf{A} + \mbf{A}^\trans\mbf{P} - \mbf{P} & \mbf{P} \\ * & -\mbf{P} \ebm < 0,
\eeq
can be rewritten as
\beq
\label{eq:Dilation_1}
\bbm \mbf{A}^\trans & \mbf{1} & \mbf{0} \\ \mbf{1} & \mbf{0} & \mbf{1} \ebm \bbm \mbf{0} & \mbf{P} & \mbf{0} \\ * & -\mbf{P} & \mbf{0} \\ * & * & -\mbf{P} \ebm \bbm \mbf{A} & \mbf{1} \\ \mbf{1} & \mbf{0} \\ \mbf{0} & \mbf{1} \ebm < 0.
\eeq
Since $\mbf{P} > 0$, it is also known that
\bdis
\bbm -\mbf{P} & \mbf{0} \\ * & -\mbf{P} \ebm < 0,
\edis
which can be rewritten as
\beq
\label{eq:Dilation_2}
\bbm \mbf{0} & \mbf{1} & \mbf{0} \\ \mbf{0} & \mbf{0} & \mbf{1} \ebm \bbm \mbf{0} & \mbf{P} & \mbf{0} \\ * & -\mbf{P} & \mbf{0} \\ * & * & -\mbf{P} \ebm \bbm \mbf{0} & \mbf{0} \\ \mbf{1} & \mbf{0} \\ \mbf{0} & \mbf{1} \ebm < 0.
\eeq
The matrix inequalities in~\eqref{eq:Dilation_1} and~\eqref{eq:Dilation_2} are in the form of the strict projection lemma.  Specifically,~\eqref{eq:Dilation_1} is in the form of $\mbf{N}_G^\trans(\mbf{A}) \mbs{\Phi}(\mbf{P})\mbf{N}_G(\mbf{A}) < 0$, where
\bdis
\mbs{\Phi}(\mbf{P}) = \bbm \mbf{0} & \mbf{P} & \mbf{0} \\ * & -\mbf{P} & \mbf{0} \\ * & * & -\mbf{P} \ebm, \hspace{10pt} \mbf{N}_G(\mbf{A}) = \bbm \mbf{A} & \mbf{1} \\ \mbf{1} & \mbf{0} \\ \mbf{0} & \mbf{1} \ebm.
\edis
The matrix inequality of~\eqref{eq:Dilation_2} is in the form of $\mbf{N}_H^\trans \mbs{\Phi}(\mbf{P})\mbf{N}_H < 0$, where
\bdis
\mbf{N}_H = \bbm \mbf{0} & \mbf{0} \\ \mbf{1} & \mbf{0} \\ \mbf{0} & \mbf{1} \ebm.
\edis
The projection lemma states that~\eqref{eq:Dilation_1} and~\eqref{eq:Dilation_2} are equivalent to
\beq
\label{eq:Dilation_4}
\mbs{\Phi}(\mbf{P}) + \mbf{G}(\mbf{A}) \mbf{V} \mbf{H}^\trans + \mbf{H} \mbf{V}^\trans \mbf{G}^\trans(\mbf{A}),
\eeq
where $\mathcal{N}(\mbf{G}^\trans(\mbf{A})) = \mathcal{R}(\mbf{N}_G(\mbf{A}))$, $\mathcal{N}(\mbf{H}^\trans) = \mathcal{R}(\mbf{N}_H)$, and $\mbf{V} \in \mathbb{R}^{n \times n}$.  Choosing
\bdis
\mbf{G}(\mbf{A}) = \bbm -\mbf{1} \\ \mbf{A}^\trans \\ \mbf{1} \ebm, \hspace{10pt} \mbf{H} = \bbm \mbf{1} \\ \mbf{0} \\ \mbf{0} \ebm,
\edis
the matrix inequality of~\eqref{eq:Dilation_4} can be rewritten as
\bdis
\bbm \mbf{0} & \mbf{P} & \mbf{0} \\ * & -\mbf{P} & \mbf{0} \\ * & * & -\mbf{P} \ebm + \bbm -\mbf{1} \\ \mbf{A}^\trans \\ \mbf{1} \ebm \mbf{V} \bbm \mbf{1} & \mbf{0} & \mbf{0} \ebm + \bbm \mbf{1} \\ \mbf{0} \\ \mbf{0} \ebm \mbf{V}^\trans \bbm -\mbf{1} & \mbf{A} & \mbf{1} \ebm < 0,
\edis
or equivalently
\beq
\label{eq:Dilation_5}
\bbm -\left(\mbf{V} + \mbf{V}^\trans\right) & \mbf{V}^\trans\mbf{A} + \mbf{P} & \mbf{V}^\trans \\ * & -\mbf{P} & \mbf{0} \\ * & * & -\mbf{P} \ebm < 0.
\eeq
Therefore, the matrix inequality of~\eqref{eq:Dilation_1} with $\mbf{P} > 0$ is equivalent to the dilated matrix inequality of~\eqref{eq:Dilation_5}.

\subsubsection{Examples of Dilated Matrix Inequalities}

Examples of some useful dilated matrix inequalities are presented here, while dilated forms of a number of important matrix inequalities are included as equivalent matrix inequalities in their respective sections.

\begin{enumerate}

\itemcite \cite{Ebihara2004} Consider the matrices $\mbf{A}$,~$\mbf{G} \in \mathbb{R}^{n\times n}$, $\mbs{\Delta} \in \mathbb{R}^{m \times n}$, $\mbf{P} \in \mathbb{S}^{n}$, $\delta_1$,~$\delta_2$,~$a$,~$b \in \mathbb{R}_{> 0}$, where $\mbf{P} > 0$ and $b = a^{-1}$.  The matrix inequality
\beq
\label{eq:Dilation1a}
\mbf{A}\mbf{P} + \mbf{P}\mbf{A}^\trans + \delta_1\mbf{P} + \delta_2\mbf{A}\mbf{P}\mbf{A}^\trans + \mbf{P}\mbs{\Delta}^\trans\mbs{\Delta}\mbf{P} < 0
\eeq
is equivalent to the matrix inequality
\beq
\label{eq:Dilation2a}
\bbm \mbf{0} & -\mbf{P} & \mbf{P} & \mbf{0} & \mbf{P}\mbs{\Delta}^\trans \\ * & \mbf{0} & \mbf{0} & -\mbf{P} & \mbf{0} \\ * & * & -\delta_1^{-1}\mbf{P} & \mbf{0} & \mbf{0} \\ * & * & * & -\delta_2^{-1}\mbf{P} & \mbf{0} \\ * & * & * & * & -\mbf{1} \ebm + \text{He}\left\{ \bbm \mbf{A} \\ \mbf{1} \\ \mbf{0} \\ \mbf{0} \\ \mbf{0} \ebm \mbf{G} \bbm \mbf{1} & -b\mbf{1} & b\mbf{1} & \mbf{1} & b\mbs{\Delta}^\trans \ebm \right\}< 0.
\eeq
Moreover, for every solution $\mbf{P}  > 0$ of~\eqref{eq:Dilation1a}, $\mbf{P}$ and $\mbf{G} = -a\left(\mbf{A} - a\mbf{1}\right)^{-1}\mbf{P}$ will be solutions of~\eqref{eq:Dilation2a}.

\itemcite \cite[pp.~7--8]{Chang2014} Consider the matrices $\mbf{A}$,~$\mbf{V} \in \mathbb{R}^{n \times n}$, $\mbf{P}$,~$\mbf{X} \in \mathbb{S}^{n}$, $\mbf{B} \in \mathbb{R}^{n \times m}$, $\mbf{C} \in \mathbb{R}^{p \times n}$, $\mbf{D} \in \mathbb{R}^{p \times m}$, $\mbf{R} \in \mathbb{S}^{m}$, and $\mbf{S} \in \mathbb{S}^{p}$, where $\mbf{P} > 0$, $\mbf{R} > 0$, $\mbf{S} > 0$, and $\mbf{X} > 0$.  The matrix inequality given by
\bdis
\bbm -\mbf{V} - \mbf{V}^\trans & \mbf{V}\mbf{A} + \mbf{P} & \mbf{V}\mbf{B} & \mbf{0} & \mbf{V} \\ * & -2\mbf{P} + \mbf{X} & \mbf{0} & \mbf{C}^\trans & \mbf{0} \\ * & * & -\mbf{R} & \mbf{D}^\trans & \mbf{0} \\ * & * & * & -\mbf{S} & \mbf{0} \\ * & * & * & * & -\mbf{X} \ebm < 0,
\edis
implies the matrix inequality
\bdis
\bbm \mbf{P} \mbf{A} + \mbf{A}^\trans \mbf{P} & \mbf{P}\mbf{B} & \mbf{C}^\trans \\ * & -\mbf{R} & \mbf{D}^\trans \\ * & * & -\mbf{S} \ebm < 0.
\edis

\itemcite \cite[p.~9]{Chang2014} Consider the matrices $\mbf{A}$,~$\mbf{V} \in \mathbb{R}^{n \times n}$, $\mbf{Q}$,~$\mbf{X} \in \mathbb{S}^{n}$, $\mbf{B} \in \mathbb{R}^{n \times m}$, $\mbf{C} \in \mathbb{R}^{p \times n}$, $\mbf{D} \in \mathbb{R}^{p \times m}$, $\mbf{R} \in \mathbb{S}^{m}$, and $\mbf{S} \in \mathbb{S}^{p}$, where $\mbf{Q} > 0$, $\mbf{R} > 0$, $\mbf{S} > 0$, and $\mbf{X} > 0$.  The matrix inequality given by
\bdis
\bbm -\mbf{V} - \mbf{V}^\trans & \mbf{V}^\trans\mbf{A}^\trans + \mbf{Q} & \mbf{0} & \mbf{V}^\trans\mbf{C} &  \mbf{V}^\trans \\ * & -2\mbf{Q} + \mbf{X} & \mbf{B} & \mbf{0} & \mbf{0} \\ * & * & -\mbf{R} & \mbf{D}^\trans & \mbf{0} \\ * & * & * & -\mbf{S} & \mbf{0} \\ * & * & * & * & -\mbf{X} \ebm < 0
\edis
implies the matrix inequality
\bdis
\bbm  \mbf{A}\mbf{Q} + \mbf{Q}\mbf{A}^\trans & \mbf{B} & \mbf{Q}\mbf{C}^\trans \\ * & -\mbf{R} & \mbf{D}^\trans \\ * & * & -\mbf{S} \ebm < 0.
\edis

\end{enumerate}

\subsection{Young's Relation (Completion of the Squares)}
\label{sec:YoungsRelation}
\index{Young's relation!lemma}
\index{completion of the squares| see {Young's relation}}
\subsubsection[Young's Relation]{Young's Relation~\cite{Zhou1988,Young2016}}
Consider $\mbf{X}$,~$\mbf{Y} \in \mathbb{R}^{n \times m}$ and $\mbf{S} \in \mathbb{S}^{n}$, where $\mbf{S} > 0$.  The matrix inequality given by
\bdis
\mbf{X}^\trans\mbf{Y} + \mbf{Y}^\trans\mbf{X} \leq \mbf{X}^\trans\mbf{S}^{-1}\mbf{X} + \mbf{Y}^\trans\mbf{S}\mbf{Y},
\edis
is known as Young's relation or Young's inequality.

Young's relation can be derived from a completion of the squares as follows.
\begin{align*}
0 &\leq \left( \mbf{X} - \mbf{S} \mbf{Y} \right)^\trans \mbf{S}^{-1} \left(\mbf{X} - \mbf{S} \mbf{Y} \right)  \\
0 &\leq \mbf{X}^\trans \mbf{S}^{-1} \mbf{X} + \mbf{Y}^\trans \mbf{S} \mbf{Y} - \mbf{X}^\trans \mbf{Y} - \mbf{Y}^\trans \mbf{X} \\
\mbf{X}^\trans\mbf{Y} + \mbf{Y}^\trans\mbf{X} &\leq \mbf{X}^\trans\mbf{S}^{-1}\mbf{X} + \mbf{Y}^\trans\mbf{S}\mbf{Y},
\end{align*}
which is Young's relation.

\subsubsection[Reformulation of Young's Relation]{Reformulation of Young's Relation~\cite{Young2016}}
\index{Young's relation!reformulation}
Consider $\mbf{X}$,~$\mbf{Y} \in \mathbb{R}^{n \times m}$ and $\mbf{S} \in \mathbb{S}^{n}$, where $\mbf{S} > 0$.  The matrix inequality given by
\bdis
\mbf{X}^\trans\mbf{Y} + \mbf{Y}^\trans\mbf{X} \leq \onehalf\left(\mbf{X}+\mbf{S}\mbf{Y}\right)^\trans\mbf{S}^{-1}\left(\mbf{X}+\mbf{S}\mbf{Y}\right),
\edis
is a reformulation of Young's relation.

\subsubsection{Special Cases of Young's Relation}
\label{sec:YoungsSpecial}
\index{Young's relation!special cases}
\begin{enumerate}

\item

Consider $\mbf{X}$,$\mbf{Y} \in \mathbb{R}^{n \times m}$.  A special case of Young's relation with $\mbf{S} = \mbf{1}$ is given by
\bdis
\mbf{X}^\trans\mbf{Y} + \mbf{Y}^\trans\mbf{X} \leq \mbf{X}^\trans\mbf{X} + \mbf{Y}^\trans\mbf{Y}.
\edis

\item Consider $\mbfbar{X}$,~$\mbf{Y} \in \mathbb{R}^{n \times m}$ and $\mbf{S} \in \mathbb{S}^{n}$, where $\mbf{S} > 0$.  A special case of Young's relation with $\mbfbar{X} = -\mbf{X}$ is given by
\bdis
-\mbfbar{X}^\trans\mbf{Y} - \mbf{Y}^\trans\mbfbar{X} \leq \mbfbar{X}^\trans\mbf{S}^{-1}\mbfbar{X} + \mbf{Y}^\trans\mbf{S}\mbf{Y}.
\edis

\itemcite \cite{GeromelNotes} Consider $\mbf{G} \in \mathbb{R}^{n \times n}$ and $\mbf{S} \in \mathbb{S}^{n}$, where $\mbf{S}  > 0$.  A special case of Young's relation with $\mbf{X} = \mbf{G}$ and $\mbf{Y} = \mbf{1}$ is given by
\bdis
\mbf{G}^\trans \mbf{S}^{-1}\mbf{G} \geq \mbf{G} + \mbf{G}^\trans - \mbf{S}.
\edis

\itemcite \cite[p.~737]{BernsteinMatrixBook} Consider $\mbf{P}$,~$\mbf{S} \in \mathbb{S}^{n}$, where $\mbf{S}  > 0$.  A special case of Young's relation with $\mbf{X} = \mbf{X}^\trans = \mbf{P}$ and $\mbf{Y} = \mbf{1}$ is given by
\bdis
2 \mbf{P} \leq \mbf{P} \mbf{S}^{-1}\mbf{P} + \mbf{S}.
\edis

\itemcite \cite[p.~732]{BernsteinMatrixBook} Consider $\mbf{G} \in \mathbb{R}^{n \times n}$ and $\alpha \in \mathbb{R}_{>0}$.  A special case of Young's relation with $\mbf{X} = \mbf{G}$, $\mbf{Y} = \mbf{1}$, and $\mbf{S} = \alpha\mbf{1}$ is given by
\bdis
\alpha^{-1}\mbf{G}^\trans \mbf{G} \geq \mbf{G} + \mbf{G}^\trans - \alpha\mbf{1}.
\edis

\itemcite \cite[p.~732]{BernsteinMatrixBook} Consider $\mbf{G} \in \mathbb{R}^{n \times n}$ and $\alpha \in \mathbb{R}_{>0}$.  A special case of Young's relation with $\mbf{X} = \mbf{G}$, $\mbf{Y} = \mbf{G}^\trans$, and $\mbf{S} = \alpha\mbf{1}$ is given by
\bdis
\mbf{G}^2 + \left(\mbf{G}^\trans\right)^2 \leq  \alpha^{-1}\mbf{G}^\trans \mbf{G} + \alpha\mbf{G}\mbf{G}^\trans.
\edis

\itemcite \cite[p.~732]{BernsteinMatrixBook} Consider $\mbf{S} \in \mathbb{S}^{n}$, where $\mbf{S} > 0$.  A special case of Young's relation with $\mbf{X} = \mbf{1}$, $\mbf{Y} = \mbf{1}$ is given by
\bdis
2 \mbf{1} \leq  \mbf{S} + \mbf{S}^{-1}.
\edis

\itemcite \cite{Merco2020} Consider $\mbf{S} \in \mathbb{S}^{n}$ and $\alpha \in \mathbb{R}$, where $\mbf{S} > 0$.  A special case of Young's relation with $\mbf{X} = \mbf{1}$, $\mbf{Y} = \alpha \mbf{1}$ is given by
\bdis
2 \alpha \mbf{1} \leq  \alpha \mbf{S} + \mbf{S}^{-1}.
\edis


\itemcite \label{sec:YoungsSpecial9}\cite[p.~38]{Wu2010} Consider the column matrices $\mbf{x}$,~$\mbf{y} \in \mathbb{R}^n$, and $\mbf{S} \in \mathbb{S}^{n}$, where $\mbf{S} > 0$.  A special case of Young's relation with $\mbf{X} = \mbf{x}$ and $\mbf{Y} = \mbf{y}$ is given by
\beq
\label{eq:SpecialYoung4}
-2 \mbf{x}^\trans\mbf{y} \leq \mbf{x}^\trans\mbf{S}^{-1}\mbf{x} + \mbf{y}^\trans \mbf{S}\mbf{y}.
\eeq

\itemcite \cite{Cao1998} Consider the column matrices $\mbf{x}$,~$\mbf{y} \in \mathbb{R}^n$, and $\mbf{S} \in \mathbb{S}^{n}$, where $\mbf{S} > 0$.  A special case of Young's relation with $\mbf{X} = \mbf{x}$ and $\mbf{Y} = -\mbf{y}$ is given by
\bdis
-2 \mbf{x}^\trans\mbf{y} \leq \mbf{x}^\trans\mbf{S}^{-1}\mbf{x} + \mbf{y}^\trans \mbf{S}\mbf{y}.
\edis

\item Consider $\mbf{X} \in \mathbb{R}^{n \times m}$, $\mbf{F} \in \mathbb{R}^{n \times q}$, $\mbfbar{Y} \in \mathbb{R}^{q \times m}$, and $\mbf{S} \in \mathbb{S}^{n}$, where $\mbf{S} > 0$.  A special case of Young's relation with $\mbf{Y} = \mbf{F}\mbfbar{Y}$ is given by
\beq
\label{eq:SpecialYoung5}
\mbf{X}^\trans\mbf{F}\mbfbar{Y} + \mbfbar{Y}^\trans\mbf{F}^\trans\mbf{X} \leq \mbf{X}^\trans \mbf{S}^{-1}\mbf{X} + \mbfbar{Y}^\trans \mbf{F}^\trans\mbf{S}\mbf{F}\mbfbar{Y}.
\eeq

\itemcite \cite[pp.~29--30]{Duan2013} Consider $\mbf{X} \in \mathbb{R}^{n \times m}$, $\mbfbar{Y} \in \mathbb{R}^{n \times m}$, $\mbf{F} \in \mathbb{S}^{n}$, and $\delta \in \mathbb{R}_{>0}$, where $\mbf{F} > 0$.  A special case of Young's relation with $\mbf{Y} = \mbf{F}\mbfbar{Y}$ and $\mbf{S} = \left(\delta\mbf{F}\right)^{-1}$ is given by
\bdis
\mbf{X}^\trans\mbf{F}\mbfbar{Y} + \mbfbar{Y}^\trans\mbf{F}\mbf{X} \leq \delta \mbf{X}^\trans \mbf{F}\mbf{X} + \delta^{-1}\mbfbar{Y}^\trans\mbf{F}\mbfbar{Y}.
\edis

\itemcite \cite{Petersen1987} Consider $\mbf{X} \in \mathbb{R}^{n \times m}$, $\mbf{F} \in \mathbb{R}^{n \times q}$, $\mbfbar{Y} \in \mathbb{R}^{q \times m}$, and $\epsilon \in \mathbb{R}_{>0}$, where $\mbf{F}^\trans\mbf{F} \leq \mbf{1}$.  A special case of the matrix inequality~\eqref{eq:SpecialYoung5} with $\mbf{S} = \epsilon\mbf{1}$ is given by
\beq
\label{eq:SpecialYoung5a}
\mbf{X}^\trans\mbf{F}\mbfbar{Y} + \mbfbar{Y}^\trans\mbf{F}^\trans\mbf{X} \leq \epsilon^{-1}\mbf{X}^\trans\mbf{X} + \epsilon \mbfbar{Y}^\trans\mbfbar{Y}.
\eeq

\textit{Proof}. Substituting $\mbf{S} = \epsilon\mbf{1}$ into~\eqref{eq:SpecialYoung5} yields
\beq
\label{eq:SpecialYoung5b}
\mbf{X}^\trans\mbf{F}\mbfbar{Y} + \mbfbar{Y}^\trans\mbf{F}^\trans\mbf{X} \leq \epsilon \mbf{X}^\trans \mbf{X} + \epsilon^{-1} \mbfbar{Y}^\trans \mbf{F}^\trans\mbf{F}\mbfbar{Y}.
\eeq
Premultiplying $\mbf{F}^\trans\mbf{F} \leq \mbf{1}$ by $\mbfbar{Y}^\trans$, postmultiplying by $\mbfbar{Y}$, and multiplying both sides by $\epsilon^{-1}$ leads to
\beq
\label{eq:SpecialYoung5c}
 \epsilon^{-1}\mbfbar{Y}^\trans\mbf{F}^\trans\mbf{F}\mbfbar{Y} \leq \epsilon^{-1}\mbfbar{Y}^\trans\mbfbar{Y}.
\eeq
Substituting~\eqref{eq:SpecialYoung5c} into~\eqref{eq:SpecialYoung5b} yields~\eqref{eq:SpecialYoung5a}. \hfill $\Box$

\item Consider $\mbf{X} \in \mathbb{R}^{n \times m}$, $\mbf{F} \in \mathbb{R}^{n \times q}$, $\mbf{Y} \in \mathbb{R}^{q \times m}$, and $\mbf{S} \in \mathbb{S}^{n}$, where $\mbf{S} > 0$.  Applying Young's relation gives the matrix inequality
\beq
\label{eq:SpecialYoung6}
\onehalf\left(\mbf{X}+\mbf{F}\mbf{Y}\right)^\trans\mbf{S}^{-1}\left(\mbf{X}+\mbf{F}\mbf{Y}\right) \leq \mbf{X}^\trans\mbf{S}^{-1}\mbf{X} + \mbf{Y}^\trans\mbf{F}^\trans\mbf{S}^{-1}\mbf{F}\mbf{Y}.
\eeq

\textit{Proof}. Expanding the left-hand side of~\eqref{eq:SpecialYoung6} yields
\beq
\label{eq:SpecialYoung7}
\onehalf\left(\mbf{X}+\mbf{F}\mbf{Y}\right)^\trans\mbf{S}^{-1}\left(\mbf{X}+\mbf{F}\mbf{Y}\right) = \onehalf\left(\mbf{X}^\trans\mbf{S}^{-1}\mbf{X} +  \mbf{X}^\trans\mbf{S}^{-1}\mbf{F}\mbf{Y} + \mbf{Y}^\trans\mbf{F}^{-1}\mbf{S}^{-1}\mbf{X} + \mbf{Y}^\trans\mbf{F}^\trans\mbf{S}^{-1}\mbf{F}\mbf{Y}\right)
\eeq
From Young's relation it can be shown that
\beq
\label{eq:SpecialYoung8}
\mbf{X}^\trans\mbf{S}^{-1}\mbf{F}\mbf{Y} + \mbf{Y}^\trans\mbf{F}^{-1}\mbf{S}^{-1}\mbf{X} \leq \mbf{X}^\trans\mbf{S}^{-1}\mbf{X} + \mbf{Y}^\trans \mbf{F}^\trans \mbf{S}^{-1} \mbf{F} \mbf{Y}.
\eeq
Substituting~\eqref{eq:SpecialYoung8} into~\eqref{eq:SpecialYoung7} gives~\eqref{eq:SpecialYoung6}. \hfill $\Box$

\item Consider $\mbf{X}$,~$\mbf{Y} \in \mathbb{R}^{n \times m}$, and $\mbf{S} \in \mathbb{S}^{n}$, where $\mbf{S} > 0$.  A special case of~\eqref{eq:SpecialYoung6} with $\mbf{F} = \mbf{S}$ is given by
\bdis
\onehalf\left(\mbf{X}+\mbf{S}\mbf{Y}\right)^\trans\mbf{S}^{-1}\left(\mbf{X}+\mbf{S}\mbf{Y}\right) \leq \mbf{X}^\trans\mbf{S}^{-1}\mbf{X} + \mbf{Y}^\trans\mbf{S}\mbf{Y}.
\edis

\itemcite \cite[p.~38]{Wu2010},\cite{Cao1998} Consider $\mbf{X} \in \mathbb{R}^{n \times m}$, $\mbf{D} \in \mathbb{R}^{n \times r}$, $\mbf{F} \in \mathbb{R}^{r \times q}$, $\mbf{E} \in \mathbb{R}^{q \times m}$, $\mbf{P} \in \mathbb{S}^{n}$, and $\epsilon \in \mathbb{R}_{>0}$, where $\mbf{P} > 0$, $\mbf{F}^\trans\mbf{F} \leq \mbf{1}$, and $\mbf{P} - \epsilon \mbf{D}\mbf{D}^\trans > 0$.  Then the matrix inequality given by
\beq
\label{eq:SpecialYoung10}
(\mbf{X} + \mbf{D}\mbf{F}\mbf{E})^\trans\mbf{P}^{-1}(\mbf{X}+\mbf{D}\mbf{F}\mbf{E}) \leq \epsilon^{-1}\mbf{E}^\trans\mbf{E} + \mbf{X}^\trans(\mbf{P} - \epsilon \mbf{D}\mbf{D}^\trans)^{-1}\mbf{X},
\eeq
holds.

\textit{Proof}. Define
\bdis
\mbf{W} = \left(\epsilon^{-1} \mbf{1} - \mbf{D}^\trans\mbf{P}^{-1}\mbf{D}\right)^{-1/2}\mbf{D}^\trans\mbf{P}^{-1}\mbf{X} - \left(\epsilon^{-1} \mbf{1} - \mbf{D}^\trans\mbf{P}^{-1}\mbf{D}\right)^{1/2}\mbf{F}\mbf{E},
\edis
where $\left(\epsilon^{-1} \mbf{1} - \mbf{D}^\trans\mbf{P}^{-1}\mbf{D}\right)^{-1/2}$ exists due to the matrix inversion lemma~\cite[p.~304]{BernsteinMatrixBook} since $\mbf{P} - \epsilon \mbf{D}\mbf{D}^\trans > 0$.  Expanding the terms in $\mbf{W}^\trans\mbf{W} \geq 0$ yields
\begin{multline*}
\mbf{X}^\trans\mbf{P}^{-1}\mbf{D}\left(\epsilon^{-1} \mbf{1} - \mbf{D}^\trans\mbf{P}^{-1}\mbf{D}\right)^{-1}\mbf{D}^\trans\mbf{P}^{-1}\mbf{X} - \mbf{X}^\trans\mbf{P}^{-1}\mbf{D}\mbf{F}\mbf{E} - \mbf{E}^\trans\mbf{F}^\trans\mbf{D}^\trans\mbf{P}^{-1}\mbf{X} \\ + \mbf{E}^\trans\mbf{F}^\trans\left(\epsilon^{-1} \mbf{1} - \mbf{D}^\trans\mbf{P}^{-1}\mbf{D}\right)\mbf{F}\mbf{E} \geq 0.
\end{multline*}
Adding $\mbf{X}^\trans\mbf{P}^{-1}\mbf{X}$ to both sides of the inequality and rearranging gives
\begin{multline}
\mbf{X}^\trans\mbf{P}^{-1}\mbf{X} + \mbf{X}^\trans\mbf{P}^{-1}\mbf{D}\mbf{F}\mbf{E} + \mbf{E}^\trans\mbf{F}^\trans\mbf{D}^\trans\mbf{P}^{-1}\mbf{X} + \mbf{E}^\trans\mbf{F}^\trans\mbf{D}^\trans\mbf{P}^{-1}\mbf{D}\mbf{F}\mbf{E} \\ \leq \epsilon^{-1}\mbf{E}^\trans\mbf{F}^\trans\mbf{F}\mbf{E} + \mbf{X}^\trans\left(\mbf{P}^{-1}\mbf{D}(\epsilon^{-1} \mbf{1} - \mbf{D}^\trans\mbf{P}^{-1}\mbf{D})^{-1}\mbf{D}^\trans\mbf{P}^{-1}+ \mbf{P}^{-1}\right)\mbf{X}. \label{eq:SpecialYoung10a}
\end{multline}
Using the matrix inversion lemma~\cite[p.~304]{BernsteinMatrixBook}, it is known that
\beq
\label{eq:SpecialYoung10b}
(\mbf{P} - \epsilon \mbf{D}\mbf{D}^\trans)^{-1} = \mbf{P}^{-1}\mbf{D}(\epsilon^{-1} \mbf{1} - \mbf{D}^\trans\mbf{P}^{-1}\mbf{D})^{-1}\mbf{D}^\trans\mbf{P}^{-1}+ \mbf{P}^{-1}.
\eeq
Substituting~\eqref{eq:SpecialYoung10b} into~\eqref{eq:SpecialYoung10a}, factoring the left side of the inequality, and knowing $\mbf{F}^\trans \mbf{F} \leq \mbf{1}$ gives~\eqref{eq:SpecialYoung10}. \hfill $\Box$

\itemcite \cite{Wang1992,Cao1998} Consider $\mbf{X} \in \mathbb{R}^{n \times m}$, $\mbf{D} \in \mathbb{R}^{n \times r}$, $\mbf{F} \in \mathbb{R}^{r \times q}$, $\mbf{E} \in \mathbb{R}^{q \times m}$, $\mbf{P} \in \mathbb{S}^{n}$, and $\epsilon \in \mathbb{R}_{>0}$, where $\mbf{P} > 0$, $\mbf{F}^\trans\mbf{F} \leq \mbf{1}$, and $\epsilon \mbf{1} - \mbf{D}^\trans\mbf{P}\mbf{D} > 0$.  Then the matrix inequality given by
\beq
\label{eq:SpecialYoung11}
(\mbf{X} + \mbf{D}\mbf{F}\mbf{E})^\trans\mbf{P}(\mbf{X}+\mbf{D}\mbf{F}\mbf{E}) \leq \epsilon\mbf{E}^\trans\mbf{E} + \mbf{X}^\trans\mbf{P}\mbf{D}(\epsilon \mbf{1} - \mbf{D}^\trans\mbf{P}\mbf{D})^{-1}\mbf{D}^\trans\mbf{P}\mbf{X} + \mbf{X}^\trans \mbf{P}\mbf{X},
\eeq
holds.

\textit{Proof}. Define
\bdis
\mbf{W} = \left(\epsilon \mbf{1} - \mbf{D}^\trans\mbf{P}\mbf{D}\right)^{-1/2}\mbf{D}^\trans\mbf{P}\mbf{X} - \left(\epsilon \mbf{1} - \mbf{D}^\trans\mbf{P}\mbf{D}\right)^{1/2}\mbf{F}\mbf{E},
\edis
where $\left(\epsilon \mbf{1} - \mbf{D}^\trans\mbf{P}\mbf{D}\right)^{-1/2}$ exists since $\epsilon \mbf{1} - \mbf{D}^\trans\mbf{P}\mbf{D} > 0$.  Expanding the terms in $\mbf{W}^\trans\mbf{W} \geq 0$ yields
\bdis
\mbf{X}^\trans\mbf{P}\mbf{D}\left(\epsilon \mbf{1} - \mbf{D}^\trans\mbf{P}\mbf{D}\right)^{-1}\mbf{D}^\trans\mbf{P}\mbf{X} - \mbf{X}^\trans\mbf{P}\mbf{D}\mbf{F}\mbf{E} - \mbf{E}^\trans\mbf{F}^\trans\mbf{D}^\trans\mbf{P}\mbf{X} + \mbf{E}^\trans\mbf{F}^\trans\left(\epsilon \mbf{1} - \mbf{D}^\trans\mbf{P}\mbf{D}\right)\mbf{F}\mbf{E} \geq 0.
\edis
Adding $\mbf{X}^\trans\mbf{P}\mbf{X}$ to both sides of the inequality and rearranging gives
\begin{multline*}
\mbf{X}^\trans\mbf{P}\mbf{X} + \mbf{X}^\trans\mbf{P}\mbf{D}\mbf{F}\mbf{E} + \mbf{E}^\trans\mbf{F}^\trans\mbf{D}^\trans\mbf{P}\mbf{X} + \mbf{E}^\trans\mbf{F}^\trans\mbf{D}^\trans\mbf{P}\mbf{D}\mbf{F}\mbf{E} \\ \leq \epsilon\mbf{E}^\trans\mbf{F}^\trans\mbf{F}\mbf{E} + \mbf{X}^\trans\mbf{P}\mbf{D}(\epsilon \mbf{1} - \mbf{D}^\trans\mbf{P}\mbf{D})^{-1}\mbf{D}^\trans\mbf{P}\mbf{X} + \mbf{X}^\trans \mbf{P}\mbf{X}.
\end{multline*}
Factoring the left side of the inequality and knowing $\mbf{F}^\trans\mbf{F} \geq \mbf{1}$ gives~\eqref{eq:SpecialYoung11}. \hfill $\Box$

\itemcite \cite[p.~11]{Chang2014} Consider $\mbf{N} \in \mathbb{R}^{n \times n}$, $\mbf{E} \in \mathbb{R}^{n \times m}$, $\mbf{H} \in \mathbb{R}^{m \times p}$, $\mbf{F} \in \mathbb{R}^{p \times n}$, $\mbf{J} \in \mathbb{S}^{n}$, and $\epsilon \in \mathbb{R}_{>0}$, where $\mbf{J} > 0$ and $\mbf{F}^\trans\mbf{F} \leq \mbf{1}$.  With some manipulation, a special case of~\eqref{eq:SpecialYoung5} with $\mbf{X} = \mbf{H}^\trans\mbf{E}^\trans\mbf{N}^\trans$ and $\mbfbar{Y} = \mbf{1}$ is given by
\bdis
-\mbf{N}\left(\mbf{1} - \mbf{E}\mbf{H}\mbf{F}\right)\mbf{J}^{-1}\left(\mbf{1} - \mbf{E}\mbf{H}\mbf{F}\right)^\trans \mbf{N}^\trans \leq \mbf{J} - \mbf{N} - \mbf{N}^\trans + \epsilon^{-1} \mbf{N}\mbf{E}\mbf{H}\mbf{H}^\trans\mbf{E}^\trans\mbf{N}^\trans + \epsilon \mbf{1}.
\edis

\itemcite \cite[p.~11]{Chang2014} Consider $\mbf{N} \in \mathbb{R}^{n \times n}$, $\mbf{F} \in \mathbb{R}^{n \times m}$, $\mbf{E} \in \mathbb{R}^{m \times p}$, $\mbf{H} \in \mathbb{R}^{p \times n}$, $\mbf{J} \in \mathbb{S}^{n}$, and $\epsilon \in \mathbb{R}_{>0}$, where $\mbf{J} > 0$ and $\mbf{F}^\trans\mbf{F} \leq \mbf{1}$.  With some manipulation, a special case of~\eqref{eq:SpecialYoung5} with $\mbf{X} = \mbf{N}\mbf{H}\mbf{E}$ and $\mbfbar{Y} = \mbf{1}$ is given by
\bdis
-\mbf{N}^\trans\left(\mbf{1} - \mbf{F}\mbf{E}\mbf{H}\right)^\trans\mbf{J}^{-1}\left(\mbf{1} - \mbf{F}\mbf{E}\mbf{H}\right) \mbf{N} \leq \mbf{J} - \mbf{N} - \mbf{N}^\trans + \epsilon^{-1} \mbf{N}^\trans\mbf{H}^\trans\mbf{E}^\trans\mbf{E}\mbf{H}\mbf{N} + \epsilon \mbf{1}.
\edis

\end{enumerate}

\subsubsection{Young's Relation-Based Properties}
\index{Young's relation!Young's relation-based properties}
\label{sec:YoungProp}
\begin{enumerate}

\itemcite \cite{Tahir2015} Consider $\mbf{X}$,~$\mbf{Y} \in \mathbb{R}^{n \times m}$ and $\mbf{Z} \in \mathbb{S}^{m}$.  The matrix inequality given by
\bdis
 \mbf{Z} + \mbf{X}^\trans\mbf{Y} + \mbf{Y}^\trans\mbf{X} > 0,
\edis
is satisfied if and only if there exist $\mbf{Q} \in \mathbb{S}^m$, $\mbf{P} \in \mathbb{S}^n$, $\mbf{G}_1 \in \mathbb{R}^{n \times n}$, $\mbf{G}_2 \in \mathbb{R}^{n \times m}$, $\mbf{F} \in \mathbb{R}^{m \times n}$, and $\mbf{H} \in \mathbb{R}^{m \times m}$, where $\mbf{Q} > 0$ and $\mbf{P} > 0$, such that
\bdis
\bbm \mbf{P} & \mbf{Y} \\ * & \mbf{Q} \ebm > 0 \hspace{10pt} \text{and} \hspace{10pt} \bbm \mbf{Z} + \mbf{Q} + \mbf{X}^\trans \mbf{P} \mbf{X} & \mbf{F} - \mbf{X}^\trans \mbf{G}_1 & \mbf{H} - \mbf{X}^\trans \mbf{G}_2 \\ * & \mbf{G}_1 + \mbf{G}_1^\trans - \mbf{P} & \mbf{F}^\trans + \mbf{G}_2 - \mbf{Y} \\ * & * & \mbf{H}^\trans + \mbf{H} - \mbf{Q} \ebm > 0.
\edis

\itemcite \cite{Tahir2015} Consider $\mbf{X} \in \mathbb{R}^{n \times n}$ and $\mbf{W} \in \mathbb{S}^{n}$, where $\mbf{X}$ is full rank \index{rank}and $\mbf{W} > 0$.  The matrix inequality given by
\bdis
 \mbf{X}^\trans\mbf{X} - \mbf{W} > 0,
\edis
is satisfied if there exists $\lambda \in \mathbb{R}_{>0}$ such that
\bdis
\bbm \lambda \mbf{1} & \lambda \mbf{1} & \mbf{0} \\ * & \mbf{X} + \mbf{X}^\trans & \mbf{W}^{\onehalf} \\ * & * & \lambda \mbf{1} \ebm > 0.
\edis

\itemcite \cite[p.~737]{BernsteinMatrixBook} Consider $\mbf{P}$,~$\mbf{Q} \in \mathbb{S}^{n}$, where $\mbf{P}  > 0$ and $\mbf{Q} > 0$.  The matrix inequality given by
\bdis
\mbf{P} + \mbf{Q} \leq \mbf{P} \mbf{Q}^{-1} \mbf{P} + \mbf{Q} \mbf{P}^{-1} \mbf{Q}
\edis
holds.

\end{enumerate}

\subsubsection[Convex-Concave Decomposition]{Convex-Concave Decomposition~\cite{dinh2011combining,priuli2022static}}
\label{sec:ConvexConcave}
\index{convex-concave decomposition}
Consider $\mbf{X}$,~$\mbf{Y} \in \mathbb{R}^{n \times m}$ and $\mbf{Q} \in \mathbb{S}^{m}$.  The matrix inequality
\beq
\label{eq:ConvexConcaveDecomp1}
\mbf{Q} + \mbf{X}^\trans \mbf{Y} + \mbf{Y}^\trans \mbf{X} < 0
\eeq
is equivalent to
\beq
\label{eq:ConvexConcaveDecomp2}
\mbf{Q} + \underbrace{\onehalf \left(\mbf{X} + \mbf{Y}\right)^\trans \left(\mbf{X} + \mbf{Y}\right)}_{\mbf{G}(\mbf{X},\mbf{Y})} - \underbrace{\onehalf \left(\mbf{X} - \mbf{Y}\right)^\trans \left(\mbf{X} - \mbf{Y}\right)}_{\mbf{H}(\mbf{X},\mbf{Y})} < 0,
\eeq
where matrix function $\mbf{G}(\mbf{X},\mbf{Y}) = \onehalf \left(\mbf{X} + \mbf{Y}\right)^\trans \left(\mbf{X} + \mbf{Y}\right)$ is convex and the matrix function $-\mbf{H}(\mbf{X},\mbf{Y}) = -\onehalf \left(\mbf{X} - \mbf{Y}\right)^\trans \left(\mbf{X} - \mbf{Y}\right)$ is concave.  Suppose that an initial feasible values of $\mbf{X} = \mbf{X}_0$ and $\mbf{Y} = \mbf{Y}_0$ are known, the matrix inequalities in~\eqref{eq:ConvexConcaveDecomp1} and~\eqref{eq:ConvexConcaveDecomp2} are satisfied with $\mbf{X} = \mbf{X}_0 + \delta \mbf{X}$ and $\mbf{Y} = \mbf{Y}_0 + \delta \mbf{Y}$ if
\beq
\label{eq:ConvexConcaveDecomp3}
\bbm \mbf{Q} - \mbf{H}(\mbf{X}_0,\mbf{Y}_0) - \onehalf \left(\left(\mbf{X}_0 - \mbf{Y}_0\right)^\trans \left( \delta \mbf{X} - \delta \mbf{Y}\right) + \left( \delta \mbf{X} - \delta \mbf{Y}\right)^\trans \left(\mbf{X}_0 - \mbf{Y}_0\right)\right) & \left(\mbf{X} + \mbf{Y}\right)^\trans \\ * & -2 \mbf{1} \ebm < 0.
\eeq
Moreover, the conservatism of~\eqref{eq:ConvexConcaveDecomp3} with respect to the matrix inequalities in~\eqref{eq:ConvexConcaveDecomp1} and~\eqref{eq:ConvexConcaveDecomp2} in the neighborhood of the $\mbf{X}_0$ and $\mbf{Y}_0$ (i.e.,~\eqref{eq:ConvexConcaveDecomp3} becomes equivalent to ~\eqref{eq:ConvexConcaveDecomp1} and~\eqref{eq:ConvexConcaveDecomp2} as $\delta \mbf{X} \to \mbf{0}$ and $\delta \mbf{Y} \to \mbf{0}$).
\begin{proof}
The function $\mbf{H}(\mbf{X},\mbf{Y})$ is rewritten in terms of perturbations from a prior solution $\mbf{X}_0$, $\mbf{Y}_0$ (i.e., $\mbf{X} = \mbf{X}_0 + \delta \mbf{X}$ and $\mbf{Y} = \mbf{Y}_0 + \delta \mbf{Y}$), which yields
\begin{align*}
\mbf{H}(\mbf{X},\mbf{Y}) &= \onehalf \left(\mbf{X}_0 + \delta \mbf{X} - \mbf{Y}_0 - \delta \mbf{Y} \right)^\trans \left(\mbf{X}_0 + \delta \mbf{X} - \mbf{Y}_0 - \delta \mbf{Y} \right) \\
&= \underbrace{\onehalf \left(\mbf{X}_0 - \mbf{Y}_0\right)^\trans \left(\mbf{X}_0 - \mbf{Y}_0\right)}_{\mbf{H}(\mbf{X}_0,\mbf{Y}_0)} + \onehalf \left(\left(\mbf{X}_0 - \mbf{Y}_0\right)^\trans \left( \delta \mbf{X} - \delta \mbf{Y}\right) + \left( \delta \mbf{X} - \delta \mbf{Y}\right)^\trans \left(\mbf{X}_0 - \mbf{Y}_0\right)\right) \\ &\hspace{10pt} + \underbrace{\onehalf \left(\delta \mbf{X} - \delta \mbf{Y}\right)^\trans \left(\delta \mbf{X} - \delta \mbf{Y}\right)}_{\mbf{H}(\delta \mbf{X},\delta \mbf{Y})} \\
&= \mbf{H}(\mbf{X}_0,\mbf{Y}_0) + \onehalf \left(\left(\mbf{X}_0 - \mbf{Y}_0\right)^\trans \left( \delta \mbf{X} - \delta \mbf{Y}\right) + \left( \delta \mbf{X} - \delta \mbf{Y}\right)^\trans \left(\mbf{X}_0 - \mbf{Y}_0\right)\right) + \mbf{H}(\delta \mbf{X},\delta \mbf{Y}).
\end{align*}
Knowing that $\mbf{H}(\delta \mbf{X},\delta \mbf{Y}) \geq 0$ results in
\beq
\label{eq:ConvexConcaveDecomp4}
\mbf{H}(\mbf{X},\mbf{Y}) \geq \mbf{H}(\mbf{X}_0,\mbf{Y}_0) + \onehalf \left(\left(\mbf{X}_0 - \mbf{Y}_0\right)^\trans \left( \delta \mbf{X} - \delta \mbf{Y}\right) + \left( \delta \mbf{X} - \delta \mbf{Y}\right)^\trans \left(\mbf{X}_0 - \mbf{Y}_0\right)\right).
\eeq
Taking the Schur complement of $\mbf{G}(\mbf{X},\mbf{Y}) = \left(\mbf{X} + \mbf{Y}\right)^\trans\left(\onehalf \mbf{1}\right) \left(\mbf{X} + \mbf{Y}\right)$ allows for~\eqref{eq:ConvexConcaveDecomp2} to be equivalently written as
\beq
\label{eq:ConvexConcaveDecomp5}
\bbm \mbf{Q} - \mbf{H}(\mbf{X},\mbf{Y}) & \left(\mbf{X} + \mbf{Y}\right)^\trans \\ * & -2 \mbf{1} \ebm < 0.
\eeq
Making use of~\eqref{eq:ConvexConcaveDecomp4}, results in \eqref{eq:ConvexConcaveDecomp3} implying~\eqref{eq:ConvexConcaveDecomp5}, which is equivalent to~\eqref{eq:ConvexConcaveDecomp1} and~\eqref{eq:ConvexConcaveDecomp2}.
\end{proof}

\subsubsection[Iterative Convex Overbounding]{Iterative Convex Overbounding~\cite{Warner2015,Warner2017}}

Iterative convex overbounding\index{convex overbounding!iterative convex overbounding} is a technique based on Young's relation that is useful when solving an optimization problem with a BMI constraint.

Consider the matrices $\mbf{Q} = \mbf{Q}^\trans \in \mathbb{R}^{n \times n}$, $\mbf{B} \in \mathbb{R}^{n \times m}$, $\mbf{R} \in \mathbb{R}^{m \times p}$, $\mbf{D} \in \mathbb{R}^{p \times q}$, $\mbf{S} \in \mathbb{R}^{q \times r}$, and $\mbf{C} \in \mathbb{R}^{r \times n}$, where $\mbf{S}$ and $\mbf{R}$ are design variables in the BMI given by
\beq
\label{eq:ICO_1}
\mbf{Q} + \mbf{B} \mbf{R} \mbf{D} \mbf{S} \mbf{C} + \mbf{C}^\trans \mbf{S}^\trans \mbf{D}^\trans \mbf{R}^\trans \mbf{B}^\trans < 0.
\eeq
Suppose that $\mbf{S}_0$ and $\mbf{R}_0$ are known to satisfy~\eqref{eq:ICO_1}.  The BMI of~\eqref{eq:ICO_1} is implied by the LMI
\beq
\label{eq:ICO_2}
\bbm \mbf{Q} + \mbs{\phi}(\mbf{R},\mbf{S}) + \mbs{\phi}^\trans(\mbf{R},\mbf{S}) & \mbf{B} \left(\mbf{R} - \mbf{R}_0\right) \mbf{U} & \mbf{C}^\trans \left(\mbf{S} - \mbf{S}_0\right)^\trans \mbf{V}^\trans \\ * & -\mbf{W}^{-1} & \mbf{0} \\ * & * & -\mbf{W} \ebm < 0,
\eeq
where $\mbs{\phi}(\mbf{R},\mbf{S}) = \mbf{B}\left(\mbf{R}\mbf{D}\mbf{S}_0 + \mbf{R}_0\mbf{D}\mbf{S} - \mbf{R}_0\mbf{D}\mbf{S}_0\right)\mbf{C}$, $\mbf{W} > 0$ is an arbitrary matrix, $\mbf{D} = \mbf{U}\mbf{V}$, and the matrices $\mbf{U}$ and $\mbf{V}^\trans$ have full column rank.\index{rank}  The LMI of~\eqref{eq:ICO_2} is equivalent to the BMI of~\eqref{eq:ICO_1} when $\mbf{R} = \mbf{R}_0$ and $\mbf{S} = \mbf{S}_0$, and is therefore non-conservative for values of $\mbf{R}$ and $\mbf{S}$ and are close to the previously known solutions $\mbf{R}_0$ and $\mbf{S}_0$.

Alternatively, the BMI of~\eqref{eq:ICO_1} is implied by the LMI
\beq
\label{eq:ICO_3}
\bbm \mbf{Q} + \mbs{\phi}(\mbf{R},\mbf{S}) + \mbs{\phi}^\trans(\mbf{R},\mbf{S}) & \mbf{Z}^\trans\mbf{U}^\trans\left(\mbf{R} - \mbf{R}_0\right)^\trans\mbf{B}^\trans   + \mbf{V}\left(\mbf{S} - \mbf{S}_0\right)\mbf{C} \\ * & -\mbf{Z} \ebm < 0,
\eeq
where $\mbf{Z} > 0$ is an arbitrary matrix, $\mbf{D} = \mbf{U}\mbf{V}$, and the matrices $\mbf{U}$ and $\mbf{V}^\trans$ have full column rank. \index{rank} Again, the LMI of~\eqref{eq:ICO_3} is equivalent to the BMI of~\eqref{eq:ICO_1} when $\mbf{R} = \mbf{R}_0$ and $\mbf{S} = \mbf{S}_0$, and is therefore non-conservative for values of $\mbf{R}$ and $\mbf{S}$ and are close to the previously known solutions $\mbf{R}_0$ and $\mbf{S}_0$.

A benefit of convex overbounding compared to a linearization approach, is that in addition to ensuring conservatism or error is reduced in the neighborhood of $\mbf{R} = \mbf{R}_0$ and $\mbf{S} = \mbf{S}_0$, the LMIs of~\eqref{eq:ICO_2} and~\eqref{eq:ICO_3} imply~\eqref{eq:ICO_1}.

Iterative convex overbounding is particularly useful when used to solve an optimization problem with BMI constraints.  For example, choose $\mbf{R}_0$ and $\mbf{S}_0$ that are initial feasible solutions to~\eqref{eq:ICO_1}.  Then solve for $\mbf{R}$ and $\mbf{S}$ that minimize a specified objective function and satisfy~\eqref{eq:ICO_2} or~\eqref{eq:ICO_3}, which imply~\eqref{eq:ICO_1} without conservatism when $\mbf{R} = \mbf{R}_0$ and $\mbf{S} = \mbf{S}_0$.  Set $\mbf{R}_0 = \mbf{R}$ and $\mbf{S}_0 = \mbf{S}$, and repeat until the objective function meets a specified stopping criteria.  The benefits of this procedure are that its individual steps are convex optimization problems with very little conservatism in the neighborhood of the solution from the previous iteration, and that it tends to converge quickly to a solution.  However, there is no guarantee that the method will converge to even a local solution.

\begin{example}
Consider a special case of~\eqref{eq:ICO_1} given by
\beq
\label{eq:ICO_4}
\mbf{Q} + \mbf{R}\mbf{S} + \mbf{S}^\trans \mbf{R}^\trans < 0,
\eeq
where $\mbf{Q} \in \mathbb{S}^n$, $\mbf{R} \in \mathbb{R}^{n \times m}$, and $\mbf{S} \in \mathbb{R}^{m \times n}$.  The BMI of~\eqref{eq:ICO_4} is implied by the LMI
\bdis
\bbm \mbf{Q} + \mbf{R}\mbf{S}_0 + \mbf{S}_0^\trans\mbf{R}^\trans + \mbf{R}_0\mbf{S} + \mbf{S}^\trans \mbf{R}_0^\trans - \mbf{R}_0\mbf{S}_0 - \mbf{S}_0^\trans \mbf{R}_0^\trans & \mbf{R}-\mbf{R}_0 & \mbf{S}^\trans - \mbf{S}_0^\trans \\ * & -\mbf{W}^{-1} & \mbf{0} \\ * & * & -\mbf{W} \ebm < 0,
\edis
where $\mbf{W} > 0$ is an arbitrary matrix.  Alternatively, the BMI of~\eqref{eq:ICO_4} is implied by the LMI
\bdis
\bbm \mbf{Q} + \mbf{R}\mbf{S}_0 + \mbf{S}_0^\trans\mbf{R}^\trans + \mbf{R}_0\mbf{S} + \mbf{S}^\trans \mbf{R}_0^\trans - \mbf{R}_0\mbf{S}_0 - \mbf{S}_0^\trans \mbf{R}_0^\trans & \mbf{Z}\left(\mbf{R}-\mbf{R}_0\right)^\trans + \mbf{S} - \mbf{S}_0 \\ * & -\mbf{Z} \ebm < 0,
\edis
where $\mbf{Z} > 0$ is an arbitrary matrix.
\end{example}

\subsection[Penalized Convex Relaxation]{Penalized Convex Relaxation~\cite{kheirandishfard2018convex}}
\label{sec:PenConvexRelax}
\index{penalized convex relaxation}

Consider a BMI constraint in the variable $\mbf{x} \in \mathbb{R}^m$ given by 
\beq
\label{eq:BMI_PenConvexRelax}
\mbf{H}(\mbf{x}) = \mbf{H}_0  + \sum_{i=1}^m x_i\mbf{H}_{i} +  \sum_{i=1}^m\sum_{j=1}^m x_ix_j\mbf{H}_{i,j} \leq 0,
\eeq
where $\mbf{x}^\trans = \bbm x_1 \cdots x_m \ebm$, and $\mbf{H}_i$,~$\mbf{H}_{i,j}  \in \mathbb{S}^{n}$, $i=0,\ldots,m$, $j=0,\ldots,m$.  The BMI in~\eqref{eq:BMI_PenConvexRelax} can be rewritten in terms of a new lifted variable $\mbf{X} \in \mathbb{R}^{m \times m}$ as
\beq
\label{eq:BMI_PenConvexRelax2}
\mbfbar{H}(\mbf{x},\mbf{X}) = \mbf{H}_0  + \sum_{i=1}^m x_i\mbf{H}_{i} +  \sum_{i=1}^m\sum_{j=1}^m X_{ij}\mbf{H}_{i,j} \leq 0,
\eeq
where $X_{ij}$ represents the entry of $\mbf{X}$ in the $i^\textrm{th}$ row and the $j^\textrm{th}$ column and in order to maintain consistency with~\eqref{eq:BMI_PenConvexRelax} the equality constraint $\mbf{X} = \mbf{x}\mbf{x}^\trans$ must be satisfied.

Rather than working with the non-convex constraint $\mbf{X} = \mbf{x}\mbf{x}^\trans$, convex relaxations of this constraint are provided in~\cite{kheirandishfard2018convex}, which include the following options:
\begin{enumerate}
\itemcite \cite{Boyd1997} An LMI relaxation of the form
\bdis
\bbm \mbf{X} & \mbf{x} \\ * & 1 \ebm \geq 0.
\edis
This relaxation can be used to formulate an SDP.

\item A cone relaxation of the form
\begin{align*}
X_{ii} - x_i^2 &\geq 0, \,\, i=1,\ldots,m\\
(X_{ii} - x_i^2)(X_{jj} - x_j^2) &\geq (X_{ij} - x_ix_j)^2, \,\, i=1,\ldots,m, \,\, j=1,\ldots,m.
\end{align*}
This relaxation can be used to formulate a second-order cone program (SOCP).

\item A parabolic relaxation of the form
\begin{align*}
X_{ii} + X_{jj} - 2 X_{ij} &\geq (x_i - x_j)^2, \,\, i=1,\ldots,m, \,\, j=1,\ldots,m, \\
X_{ii} + X_{jj} + 2 X_{ij} &\geq (x_i + x_j)^2, \,\, i=1,\ldots,m, \,\, j=1,\ldots,m.
\end{align*}
This relaxation can be used to formulate an optimization problem with convex quadratic constraints (e.g., a quadratically-constrained quadratic program).

\end{enumerate}

Given that this document focuses on SDPs and LMI constraints, the LMI relaxation is chosen as the focus for the remainder of this section.

\subsubsection[Sequential Penalized Convex Relaxation Optimization]{Sequential Penalized Convex Relaxation Optimization~\cite{kheirandishfard2018convex,kheirandishfard2018convexII}}
\index{penalized convex relaxation!sequential}

The non-convex optimization problem
\begin{align*}
\min_{\mbf{x} \in \mathbb{R}^m} \quad &\mbf{c}^\trans \mbf{x} \\
\text{subject to} \quad & \mbf{H}(\mbf{x}) \leq 0,
\end{align*}
where $\mbf{H}(\mbf{x})$ is a BMI constraint defined in~\eqref{eq:BMI_PenConvexRelax}, can be solved with near-global optimality by formulating a sequential penalized convex relaxation optimization problem of the form
\begin{align*}
\min_{\mbf{x} \in \mathbb{R}^m, \mbf{X} \in \mathbb{R}^{m \times m}} \quad &\mbf{c}^\trans \mbf{x} + \eta \left(\trace(\mbf{X}) - 2 \mbf{x}_0^\trans \mbf{x} + \mbf{x}_0^\trans \mbf{x}_0\right) \\
\text{subject to} \quad & \mbfbar{H}(\mbf{x},\mbf{X}) \leq 0, \\
&\bbm \mbf{X} & \mbf{x} \\ * & 1 \ebm \geq 0,
\end{align*}
where $\mbf{x}_0$ is a prior guess of $\mbf{x}$, $\mbfbar{H}(\mbf{x},\mbf{X})$ is defined in~\eqref{eq:BMI_PenConvexRelax2}, and $\eta > 0$ is a scalar regularization parameter that allows for a tradeoff between the original objective function and the penalty term ensuring tight satisfaction of the constraint $\mbf{X} = \mbf{x}\mbf{x}^\trans$.  As outlined in~\cite{kheirandishfard2018convex}, provided that $\mbf{x}_0$ is a feasible solution to the original optimization problem and a sufficiently large value of $\eta$ is used, the solution to this relaxed optimization problem, denoted as $\mbf{x}^*$ and $\mbf{X}^*$ satisfies $\mbf{X}^* = \mbf{x}^* \mbf{x}^{*^\trans}$ and $\mbf{c}^\trans \mbf{x}^{*} \leq \mbf{c}^\trans \mbf{x}_0$.  Thus, a sequential implementation of this optimization can be formulated, as described in~\cite{kheirandishfard2018convexII} to obtain a near-global solution to the original non-convex optimization problem.

\subsection{Coordinate Descent}
\label{sec:CoordDescent}
\index{coordinate descent}
\index{block coordinate descent}

Coordinate descent, also known as block coordinate descent~\cite{wang2018sequential}, is an iterative technique that can be employed when faced with a BMI that is in fact an LMI when one or more of the design variables are fixed.  For example, consider the design variables $\mbf{P} \in \mbf{S}^n$ and $\mbf{A} \in \mathbb{R}^{n \times n}$ that define the BMI
\beq
\label{eq:BMI_coordinate_descent}
\mbf{P} \mbf{A} + \mbf{A}^\trans \mbf{P} < 0.
\eeq
This BMI in the design variables $\mbf{P}$ and $\mbf{A}$ is an LMI when either $\mbf{P}$ or $\mbf{A}$ is fixed.  When solving an optimization problem with such a BMI constraint, coordinate descent can be used, which involves alternating between fixing one variable and optimizing over the other variable.

To highlight the implementation of a coordinate descent approach, consider the objective function $\phi(\mbf{P},\mbf{A}): (\mbf{S}^n \times \mathbb{R}^{n \times n}) \mapsto \mathbb{R}$, where $\phi(\mbf{P},\mbf{A})$ is convex and a valid SDP objective function for either a fixed value of $\mbf{P}$ or $\mbf{A}$.  The optimization problem of minimizing $\phi(\mbf{P},\mbf{A})$ subject to~\eqref{eq:BMI_coordinate_descent} can be approached using the following iterative process.

\begin{enumerate}

\item Choose an initial value for $\mbf{A}$.

\item Solve for $\mbf{P}$ that minimizes $\phi(\mbf{P},\mbf{A}_0)$ subject to 
\bdis
\mbf{P} \mbf{A}_0 + \mbf{A}_0^\trans \mbf{P} < 0,
\edis
where $\mbf{A}_0$ is the fixed value of $\mbf{A}$ from the previous step.

\item Solve for $\mbf{A}$ that minimizes $\phi(\mbf{P}_0,\mbf{A})$ subject to 
\bdis
\mbf{P}_0 \mbf{A} + \mbf{A}^\trans \mbf{P}_0 < 0,
\edis
where $\mbf{P}_0$ is the fixed value of $\mbf{P}$ from the previous step.

\item Repeat Steps~2 and~3 until the desired convergence or stopping criterion is met.

\end{enumerate}

This type of iterative algorithm is known as coordinate descent.  Coordinate descent is introduced well in~\cite{iwasaki1999dual} and has been used in many applications, including $DK$-iteration~\cite{doyle1985matrix,doyle1985structured} and other control design approaches (e.g.,~\cite{doyle1983synthesis,geromel1993output,rotea1994alternative,iwasaki1995xy,yamada1997lmi,doroudchi2018decentralized,dahdah2022system}).  Although coordinate descent provides a practical approach to iteratively solve a BMI problem, it typically is not capable of guaranteeing convergence to the globally optimal solution.  In general, it will converge to a locally optimal solution that is dependent on the initial guess.  This motivates the need for a good initial guess, ideally in the neighbourhood of the globally optimal solution.

\subsection{Discussion on Reformulating BMIs as LMIs}
\label{sec:Discussion_BMI_LMI}
\index{bilinear matrix inequality (BMI)!discussion}


Properties and tricks were presented Sections~\ref{sec:ChangeVariables} to~\ref{sec:CoordDescent} that can be used to reformulate BMIs into LMIs.  Specifically, the properties in Sections~\ref{sec:ChangeVariables} to~\ref{sec:Dilation} are typically able to reformulate a BMI as an equivalent LMI or LMIs.  The properties in Sections~\ref{sec:YoungsRelation} to~\ref{sec:CoordDescent} are typically used to obtain an LMI that implies a BMI, generally with conservatism.

This section presents examples in which these properties are applied to obtain an LMI that is either equivalent to the original BMI or implies the original BMI.


\subsubsection{Reformulating a BMI as an Equivalent LMI}

\begin{example}
Consider the case of a BMI in the variable $\mbf{Y} \in \mathbb{R}^{m \times n}$ of the form
\beq
\label{eq:SchurYoung1}
\mbf{P} + \mbf{Y}^\trans\mbf{S}\mbf{Y} < 0,
\eeq
where $\mbf{P} \in \mathbb{S}^n$, $\mbf{S} \in \mathbb{S}^m$, and $\mbf{S} > 0$.  The Schur complement is used to obtain an equivalent LMI given by
\bdis
\bbm \mbf{P} & \mbf{Y}^\trans \\ * & -\mbf{S}^{-1} \ebm < 0.
\edis
This LMI can also be written as
\beq
\label{eq:SchurYoung1a}
\bbm \mbf{P} & \mbf{0} \\ * & -\mbf{S}^{-1} \ebm + \bbm \mbf{0} \\ \mbf{1} \ebm \mbf{Y} \bbm \mbf{1} & \mbf{0} \ebm + \bbm \mbf{1} \\ \mbf{0} \ebm \mbf{Y}^\trans \bbm \mbf{0} & \mbf{1} \ebm < 0.
\eeq
Applying the Projection Lemma, it is known that there exists $\mbf{Y}$ satisfying~\eqref{eq:SchurYoung1a} if and only if $\mbf{P} < 0$ and $\mbf{S}^{-1} > 0$, since $\mathcal{N}\left( \bbm \mbf{1} & \mbf{0} \ebm\right) = \mathcal{R}\left(\bbm \mbf{0} \\ \mbf{1} \ebm\right)$, $\mathcal{N}\left( \bbm \mbf{0} & \mbf{1} \ebm\right) = \mathcal{R}\left(\bbm \mbf{1} \\ \mbf{0} \ebm\right)$, and
\bdis
\mbf{P} = \bbm \mbf{1} & \mbf{0} \ebm \bbm \mbf{P} & \mbf{0} \\ * & -\mbf{S}^{-1} \ebm \bbm \mbf{1} \\ \mbf{0} \ebm, \hspace{10pt} -\mbf{S}^{-1} = \bbm \mbf{0} & \mbf{1} \ebm \bbm \mbf{P} & \mbf{0} \\ * & -\mbf{S}^{-1} \ebm \bbm \mbf{0} \\ \mbf{1} \ebm.
\edis
Notice that the Projection Lemma gives two matrix inequalities that do not depend on the variable $\mbf{Y}$.  This is why the Projection Lemma is also known as the Matrix Elimination Lemma.
\end{example}

\subsubsection{Reformulating a BMI as an LMI that Implies the Original BMI}

\begin{example}
As a second example, consider the BMI
\beq
\label{eq:SchurYoung2}
\mbf{P} - \mbf{Y}^\trans\mbf{S}\mbf{Y} < 0,
\eeq
where $\mbf{Y} \in \mathbb{R}^{m \times n}$, $\mbf{P} \in \mathbb{S}^n$, $\mbf{S} \in \mathbb{S}^m$, and $\mbf{S} > 0$.  Young's relation is used to obtain an LMI in $\mbf{Y}$ given by
\beq
\label{eq:SchurYoung3}
\mbf{P} - \mbf{X}^\trans \mbf{Y} - \mbf{X}^\trans \mbf{Y} + \mbf{X}^\trans \mbf{S}^{-1} \mbf{X} < 0,
\eeq
which implies the BMI of~\eqref{eq:SchurYoung2}.  Notice that~\eqref{eq:SchurYoung3} involves a new variable $\mbf{X} \in \mathbb{R}^{m \times n}$.  Using the Schur complement on~\eqref{eq:SchurYoung3} yields
\bdis
\bbm \mbf{P} - \mbf{X}^\trans \mbf{Y} - \mbf{Y}^\trans \mbf{X} & \mbf{X}^\trans \\ * & -\mbf{S} \ebm < 0,
\edis
which is an LMI in $\mbf{Y}$ for a fixed $\mbf{X}$.



It is desirable to use the Schur complement of the Projection Lemma over Young's relation whenever possible, as they provides an LMI or LMIs that are equivalent to the original BMI.  When using Young's relation, the resulting LMI implies the original BMI, but is not equivalent.  This introduces conservatism into an optimization problem.

If a previously-known solution $\mbf{Y}_0$ to~\eqref{eq:SchurYoung2} is available, then the concepts of convex-concave decompositions \index{convex-concave decomposition!discussion}and convex overbounding\index{convex overbounding!discussion} can be used to reduce conservatism in the neighborhood of $\mbf{Y}_0$.  In this particular example, the BMI of~\eqref{eq:SchurYoung2} does not have a convex portion to its decomposition and it can be show that it is equivalent to the BMI
\beq
\label{eq:SchurYoung4}
\mbf{P} - \left(\mbf{Y} - \mbf{Y}_0\right)^\trans \mbf{S} \left(\mbf{Y} - \mbf{Y}_0\right) - \mbf{Y}^\trans\mbf{S}\mbf{Y}_0 - \mbf{Y}_0^\trans\mbf{S}\mbf{Y} + \mbf{Y}_0^\trans \mbf{S} \mbf{Y}_0 < 0.
\eeq
Since the term $\left(\mbf{Y} - \mbf{Y}_0\right)^\trans \mbf{S} \left(\mbf{Y} - \mbf{Y}_0\right)$ is positive definite,~\eqref{eq:SchurYoung4} is implied by the LMI
\beq
\label{eq:SchurYoung5}
\mbf{P} - \mbf{Y}^\trans\mbf{S}\mbf{Y}_0 - \mbf{Y}_0^\trans\mbf{S}\mbf{Y} + \mbf{Y}_0^\trans \mbf{S} \mbf{Y}_0 < 0.
\eeq
The LMI of~\eqref{eq:SchurYoung5} is in general conservative, but this conservatism disappears when $\mbf{Y} = \mbf{Y}_0$ and is reduced when $\mbf{Y}$ is close to $\mbf{Y}_0$.
\end{example}

\newpage
\section{Additional LMI Properties and Tricks}
\label{sec:LMIProperties}


This section presents a compilation of additional LMI properties and tricks from the literature.

\subsection[The S-Lemma (S-Procedure)]{The S-Lemma (S-Procedure)}
\label{sec:SLemma}

\subsubsection[The Non-Strict Scalar S-Lemma]{The Non-Strict Scalar S-Lemma~\cite[pp.~23--24]{Boyd1994},~\cite[Sec.~12.3.4]{Skogestad2005},~\cite{Yakubovich1977,Jonsson2001}}
\index{S-lemma!non-strict scalar S-lemma}
Consider $\mbf{x} \in \mathbb{R}^n$ and the quadratic functions $F_0(\mbf{x}): \mathbb{R}^n \to \mathbb{R}$, $F_i(\mbf{x}): \mathbb{R}^n \to \mathbb{R}$, where $i=1,\ldots,m$.  The inequality $F_0(\mbf{x}) \leq 0$ is satisfied when $F_i(\mbf{x}) \geq 0$, $i=1,\ldots,m$, if there exist $\tau_i \in \mathbb{R}_{\geq 0}$, $i=1,\ldots,m$ such that
\bdis
F_0(\mbf{x}) + \sum_{i=1}^m\tau_iF_i(\mbf{x}) \leq 0.
\edis

If $m = 1$, then this becomes a necessary and sufficient condition, that is, $F_0(\mbf{x}) \leq 0$ is satisfied when $F_1(\mbf{x}) \geq 0$ if and only if there exists $\tau_1 \in \mathbb{R}_{\geq 0}$ such that $F_0(\mbf{x}) + \tau_1F_1(\mbf{x}) \leq 0$.

\begin{example}

\cite[p.~24]{Boyd1994},~\cite[Example~12.8, Sec.~12.3.4]{Skogestad2005} Consider $\mbf{P} \in \mathbb{S}^n$, $\mbf{A} \in \mathbb{R}^{n \times n}$, $\mbf{B} \in \mathbb{R}^{n \times m}$, $\mbf{x} \in \mathbb{R}^n$, $\mbf{u} \in \mathbb{R}^m$, $\gamma \in \mathbb{R}_{>0}$, and $\tau \in \mathbb{R}_{\geq 0}$.  There exists $\mbf{P} > 0$ such that
\bdis
\bbm \mbf{x}^\trans & \mbf{u}^\trans \ebm \bbm \mbf{A}^\trans\mbf{P} + \mbf{P}\mbf{A} & \mbf{P}\mbf{B} \\ * & \mbf{0} \ebm \bbm \mbf{x} \\ \mbf{u} \ebm < 0
\edis
when $\mbf{x} \neq \mbf{0}$ and $\mbf{u}$ satisfy the constraint $\mbf{u}^\trans\mbf{u} \leq \gamma \mbf{x}^\trans\mbf{C}^\trans\mbf{C}\mbf{x}$ if and only if there exist $\mbf{P} > 0$ and $\tau \in \mathbb{R}_{\geq 0}$ such that
\bdis
\bbm \mbf{A}^\trans\mbf{P} + \mbf{P}\mbf{A} + \tau  \mbf{C}^\trans\mbf{C} & \mbf{P}\mbf{B} \\ * & -\tau \gamma^{-1} \mbf{1} \ebm < 0.
\edis

\end{example}

\subsubsection[The Strict Scalar S-Lemma]{The Strict Scalar S-Lemma~\cite{polik2007survey,van2020noisy}}
\index{S-lemma!strict scalar S-lemma}
Consider $\mbf{x} \in \mathbb{R}^n$ and the quadratic functions $F(\mbf{x}): \mathbb{R}^n \to \mathbb{R}$, $G(\mbf{x}): \mathbb{R}^n \to \mathbb{R}$.  Suppose there exists $\mbfbar{x} \in \mathbb{R}^n$ such that $G(\mbfbar{x}) > 0$.  Then, the inequality $F(\mbf{x}) \leq 0$ is satisfied for all $\mbf{x} \in \mathbb{R}^n$ when $G(\mbf{x}) \geq 0$, if and only if there exists $\tau \in \mathbb{R}_{\geq 0}$ such that
\bdis
F(\mbf{x}) + \tau G(\mbf{x}) \leq 0
\edis
for all $\mbf{x} \in \mathbb{R}^n$.

\subsubsection[The Non-Strict Matrix S-Lemma]{The Non-Strict Matrix S-Lemma~\cite{van2020noisy}}
\index{S-lemma!non-strict matrix S-lemma}
Consider $\mbf{P} \in \mathbb{S}^n$, $\mbf{Q} \in \mathbb{S}^n$ and assume that there exists $\mbfbar{X} \in \mathbb{R}^{n \times m}$ such that $\mbfbar{X}^\trans \mbf{Q} \mbfbar{X} > 0$.  The following are equivalent.

\begin{enumerate}

\item $\mbf{X}^\trans \mbf{P} \mbf{X} \geq 0$ for all $\mbf{X} \in \mathbb{R}^{n \times m}$ such that $\mbf{X}^\trans \mbf{Q} \mbf{X} \geq 0$.

\item $\mbf{X}^\trans \mbf{P} \mbf{X} \geq 0$ for all $\mbf{X} \in \mathbb{R}^{n \times m}$ such that $\mbf{X}^\trans \mbf{Q} \mbf{X} > 0$.

\item There exists a scalar $\alpha \in \mathbb{R}_{\geq 0}$ such that $\mbf{P} - \alpha \mbf{Q} \geq 0$. 

\end{enumerate}

An alternate form of the non-strict matrix S-Lemma is stated as follows.  Consider $\mbf{P} \in \mathbb{S}^{n+m}$ and $\mbf{Q} \in \mathbb{S}^{n+m}$, and assume that there exists $\mbfbar{Z} \in \mathbb{R}^{n \times m}$ such that
\bdis
\bbm \mbf{1} \\ \mbfbar{Z} \ebm^\trans \mbf{Q} \bbm \mbf{1} \\ \mbfbar{Z} \ebm >0.
\edis
Then, the following are equivalent.
\begin{enumerate}

\item The matrix inequality
\bdis
\bbm \mbf{1} \\ \mbf{Z} \ebm^\trans \mbf{P} \bbm \mbf{1} \\ \mbf{Z} \ebm \geq 0
\edis
is satisfied for all $\mbf{Z} \in \mathbb{R}^{n \times m}$ satisfying the condition
\bdis
\bbm \mbf{1} \\ \mbf{Z} \ebm^\trans \mbf{Q} \bbm \mbf{1} \\ \mbf{Z} \ebm \geq 0.
\edis

\item The matrix inequality
\bdis
\bbm \mbf{1} \\ \mbf{Z} \ebm^\trans \mbf{P} \bbm \mbf{1} \\ \mbf{Z} \ebm \geq 0
\edis
is satisfied for all $\mbf{Z} \in \mathbb{R}^{n \times m}$ satisfying the condition
\bdis
\bbm \mbf{1} \\ \mbf{Z} \ebm^\trans \mbf{Q} \bbm \mbf{1} \\ \mbf{Z} \ebm > 0.
\edis

\item There exists a scalar $\alpha \in \mathbb{R}_{\geq 0}$ such that $\mbf{P} - \alpha \mbf{Q} \geq 0$. 

\end{enumerate}

\subsubsection[The Strict Matrix S-Lemma]{The Strict Matrix S-Lemma~\cite{van2020noisy}}
\index{S-lemma!strict matrix S-lemma}
Consider $\mbf{P} \in \mathbb{S}^{n+m}$ and $\mbf{Q} \in \mathbb{S}^{n+m}$.  Assume the set
\bdis
\mathcal{S} = \left\{ \mbf{Z} \in \mathbb{R}^{n \times m} \mid \bbm \mbf{1} \\ \mbf{Z} \ebm^\trans \mbf{Q} \bbm \mbf{1} \\ \mbf{Z} \ebm \geq 0\right\}
\edis
is bounded and that there exists $\mbfbar{Z} \in \mathbb{R}^{n \times m}$ such that
\bdis
\bbm \mbf{1} \\ \mbfbar{Z} \ebm^\trans \mbf{Q} \bbm \mbf{1} \\ \mbfbar{Z} \ebm >0.
\edis
Then, the matrix inequality
\bdis
\bbm \mbf{1} \\ \mbf{Z} \ebm^\trans \mbf{P} \bbm \mbf{1} \\ \mbf{Z} \ebm >0
\edis
is satisfied for all $\mbf{Z} \in \mathcal{S}$ if and only if there exists a scalar $\alpha \in \mathbb{R}_{\geq 0}$ such that $\mbf{P} - \alpha \mbf{Q} > 0$. 

%

\subsection[Dualization Lemma]{Dualization Lemma~\cite[pp.~106--107]{SchererWeiland2015}}
\index{dualization lemma}
Consider $\mbf{P} \in \mathbb{S}^n$ and the subspaces $\mathcal{U}$, $\mathcal{V}$, where $\mbf{P}$ is invertible and $\mathcal{U} + \mathcal{V} = \mathbb{R}^n$.  The following are equivalent.

\begin{itemize}

\item $\mbf{x}^\trans \mbf{P} \mbf{x} < 0$ for all $\mbf{x} \in \mathcal{U} \setminus \{0\}$ and $\mbf{x}^\trans \mbf{P} \mbf{x} \geq 0$ for all $\mbf{x} \in \mathcal{V}$.

\item $\mbf{x}^\trans \mbf{P}^{-1} \mbf{x} > 0$ for all $\mbf{x} \in \mathcal{U}^\perp \setminus \{0\}$ and $\mbf{x}^\trans \mbf{P}^{-1} \mbf{x} \leq 0$ for all $\mbf{x} \in \mathcal{V}^\perp$.

\end{itemize}

\begin{example}

\cite[pp.~106--107]{SchererWeiland2015} Consider the matrices $\mbf{Q} \in \mathbb{S}^n$, $\mbf{S} \in \mathbb{R}^{n \times m}$, $\mbf{R} \in \mathbb{S}^m$, $\mbf{M} \in \mathbb{R}^{m \times n}$, where $\mbf{R} \geq 0$, which define the quadratic matrix inequality
\beq
\label{eq:Dualization1}
\bbm \mbf{1} \\ \mbf{M} \ebm^\trans \bbm \mbf{Q} & \mbf{S} \\ \mbf{S}^\trans & \mbf{R} \ebm \bbm \mbf{1} \\ \mbf{M} \ebm < 0.
\eeq
Define $\mbf{P} = \bbm \mbf{Q} & \mbf{S} \\ \mbf{S}^\trans & \mbf{R} \ebm$, $\mathcal{U} = \mathcal{R}\left( \bbm \mbf{1} \\ \mbf{M} \ebm\right)$, and $\mathcal{V} = \mathcal{R}\left( \bbm \mbf{0} \\ \mbf{1} \ebm\right)$, where $\mathcal{U} + \mathcal{V} = \mathbb{R}^{n+m}$.  Notice that~\eqref{eq:Dualization1} is equivalent to $\mbf{x}^\trans \mbf{P} \mbf{x} < 0$ for all $\mbf{x} \in \mathcal{U} \setminus \{0\}$.  Additionally, $\mbf{x}^\trans \mbf{P} \mbf{x} \geq 0$ for all $\mbf{x} \in \mathcal{V}$ is equivalent to
\bdis
\bbm \mbf{0} \\ \mbf{1} \ebm^\trans \bbm \mbf{Q} & \mbf{S} \\ \mbf{S}^\trans & \mbf{R} \ebm \bbm \mbf{0} \\ \mbf{1} \ebm = \mbf{R} \geq 0,
\edis
which is satisfied based on the definition of $\mbf{R}$.  By the dualization lemma,~\eqref{eq:Dualization1} is satisfied with $\mbf{R} \geq 0$ if and only if
\bdis
\bbm -\mbf{M}^\trans \\ \mbf{1} \ebm^\trans \bbm \mbftilde{Q} & \mbftilde{S} \\ \mbftilde{S}^\trans & \mbftilde{R} \ebm \bbm -\mbf{M}^\trans \\ \mbf{1} \ebm > 0, \hspace{20pt} \mbftilde{Q} \leq 0,
\edis
where $ \bbm \mbftilde{Q} & \mbftilde{S} \\ \mbftilde{S}^\trans & \mbftilde{R} \ebm = \bbm \mbf{Q} & \mbf{S} \\ \mbf{S}^\trans & \mbf{R} \ebm^{-1}$, $\mathcal{U}^\perp = \mathcal{N} \left( \bbm \mbf{1} & \mbf{M}^\trans \ebm \right) = \mathcal{R}\left( \bbm -\mbf{M}^\trans \\ \mbf{1} \ebm \right)$, and $\mathcal{V}^\perp = \mathcal{N} \left( \bbm \mbf{0} & \mbf{1} \ebm \right) = \mathcal{R}\left( \bbm \mbf{1} \\ \mbf{0} \ebm \right)$

\end{example}

%

\subsection{Singular Values}

\subsubsection[Maximum Singular Value]{Maximum Singular Value~\cite[p.~8]{Boyd1994},\cite{VanAntwerp2000,Fathi2019}}
\index{singular value!maximum singular value}
Consider $\mbf{A} \in \mathbb{R}^{n \times m}$ and $\gamma \in \mathbb{R}_{>0}$.  The maximum singular value of $\mbf{A}$ is strictly less than $\gamma$ (i.e., $\bar{\sigma}(\mbf{A}) < \gamma$) if and only if $\mbf{A}\mbf{A}^\trans <\gamma^2 \mbf{1}$.  Using the Schur complement, $\mbf{A}\mbf{A}^\trans < \gamma^2 \mbf{1} $ is equivalent to
\bdis
\bbm \gamma \mbf{1} & \mbf{A} \\ * & \gamma \mbf{1} \ebm > 0.
\edis
Equivalently, $\bar{\sigma}(\mbf{A}) < \gamma$ if and only if $\mbf{A}^\trans\mbf{A} <\gamma^2 \mbf{1}$ or
\bdis
\bbm \gamma \mbf{1} & \mbf{A}^\trans \\ * & \gamma \mbf{1} \ebm > 0.
\edis

\subsubsection[Maximum Singular Value of a Complex Matrix]{Maximum Singular Value of a Complex Matrix~\cite{LallNotes3}}
Consider $\mbf{A} \in \mathbb{C}^{n \times m}$ and $\gamma \in \mathbb{R}_{>0}$.  The maximum singular value of $\mbf{A}$ is strictly less than $\gamma$ (i.e., $\bar{\sigma}(\mbf{A}) < \gamma$) if and only if $\mbf{A}\mbf{A}^\herm <\gamma^2 \mbf{1}$.  Using the Schur complement, $\mbf{A}\mbf{A}^\herm < \gamma^2 \mbf{1} $ is equivalent to
\bdis
\bbm \gamma \mbf{1} & \mbf{A} \\ \mbf{A}^\herm & \gamma \mbf{1} \ebm > 0.
\edis
Equivalently, $\bar{\sigma}(\mbf{A}) < \gamma$ if and only if $\mbf{A}^\herm\mbf{A} <\gamma^2 \mbf{1}$ or
\bdis
\bbm \gamma \mbf{1} & \mbf{A}^\herm \\ \mbf{A} & \gamma \mbf{1} \ebm > 0.
\edis

\subsubsection{Minimum Singular Value}
\index{singular value!minimum singular value}
Consider $\mbf{A} \in \mathbb{R}^{n \times m}$ and $\nu \in \mathbb{R}_{\geq 0}$.  If $n \leq m$, the minimum singular value of $\mbf{A}$ is strictly greater than $\nu$ (i.e., $\underline{\sigma}(\mbf{A}) > \nu$) if and only if $\mbf{A}\mbf{A}^\trans > \nu^2 \mbf{1}$.  If $m \leq n$, $\underline{\sigma}(\mbf{A}) > \nu$ if and only if $\mbf{A}^\trans\mbf{A} > \nu^2 \mbf{1}$.

\subsubsection{Minimum Singular Value of a Complex Matrix}
Consider $\mbf{A} \in \mathbb{C}^{n \times m}$ and $\nu \in \mathbb{R}_{\geq 0}$.  If $n \leq m$, the minimum singular value of $\mbf{A}$ is strictly greater than $\nu$ (i.e., $\underline{\sigma}(\mbf{A}) > \nu$) if and only if $\mbf{A}\mbf{A}^\herm > \nu^2 \mbf{1}$.  If $m \leq n$, $\underline{\sigma}(\mbf{A}) > \nu$ if and only if $\mbf{A}^\herm\mbf{A} > \nu^2 \mbf{1}$.

\subsubsection{Frobenius Norm}
\index{singular value!Frobenius norm}
\index{norm!Frobenius norm}
Consider $\mbf{A} \in \mathbb{R}^{n \times m}$ and $\gamma \in \mathbb{R}_{> 0}$.  The Frobenius norm of $\mbf{A}$ is $\norm{\mbf{A}}_\frob = \sqrt{\trace{\left(\mbf{A}^\trans \mbf{A}\right)}} =  \sqrt{\trace{\left(\mbf{A}\mbf{A}^\trans\right)}}$~\cite[pp.~341--342]{Horn2013}.  The Frobenius norm is less than or equal to $\gamma$ if and only if any of the following equivalent conditions are satisfied.

\begin{enumerate}

\item There exists $\mbf{Z} \in \mathbb{S}^n$ such that
\begin{align*}
\bbm \mbf{Z} & \mbf{A}^\trans \\ * & \mbf{1} \ebm \geq 0, \\
\trace(\mbf{Z}) \leq \gamma^2.
\end{align*}

\item There exists $\mbf{Z} \in \mathbb{S}^m$ such that
\begin{align*}
\bbm \mbf{Z} & \mbf{A} \\ * & \mbf{1} \ebm \geq 0, \\
\trace(\mbf{Z}) \leq \gamma^2.
\end{align*}

\end{enumerate}

\subsubsection[Nuclear Norm]{Nuclear Norm~\cite{Fazel2001,Recht2010}}
\index{singular value!nuclear norm}
\index{norm!nuclear norm}
Consider $\mbf{A} \in \mathbb{R}^{n \times m}$ and $\mu \in \mathbb{R}_{> 0}$.  The nuclear norm of $\mbf{A}$ is given by $\norm{\mbf{A}}_* = \sum_{i=1}^p \sigma_i\left(\mbf{A}\right)$, where $p = \min(n,m)$ and $\sigma_i(\mbf{A})$, $i=1,\ldots,p$ are the singular values of $\mbf{A}$~\cite[p.~466]{Horn2013}.  The nuclear norm of $\mbf{A}$ is less than or equal to $\mu$ (i.e., $\norm{\mbf{A}}_* \leq \mu$) if and only if there exist $\mbf{X} \in \mathbb{S}^n$ and $\mbf{Y} \in \mathbb{S}^m$ such that
\begin{align*}
\bbm \mbf{X} & \mbf{A} \\ * & \mbf{Y} \ebm &\geq 0, \\
\onehalf \trace(\mbf{X} + \mbf{Y}) &\leq \mu.
\end{align*}

\subsection{Eigenvalues of Symmetric Matrices}
\index{eigenvalues!maximum eigenvalue}
\subsubsection[Maximum Eigenvalue]{Maximum Eigenvalue~\cite[p.~10]{Boyd1994}}

Consider $\mbf{A} \in \mathbb{S}^{n \times n}$ and $\gamma \in \mathbb{R}$.  The maximum eigenvalue of $\mbf{A}$ is strictly less than $\gamma$ (i.e., $\bar{\lambda}(\mbf{A}) < \gamma$) if and only if $\mbf{A} <\gamma \mbf{1}$.

\subsubsection{Minimum Eigenvalue}
\index{eigenvalues!minimum eigenvalue}
Consider $\mbf{A} \in \mathbb{S}^{n \times n}$ and $\gamma \in \mathbb{R}$.  The minimum eigenvalue of $\mbf{A}$ is strictly greater than $\gamma$ (i.e., $\underline{\lambda}(\mbf{A}) > \gamma$) if and only if $\mbf{A} >\gamma \mbf{1}$.

\subsubsection[Sum of Largest Eigenvalues]{Sum of Largest Eigenvalues~\cite{Alizadeh1995}}
\index{eigenvalues!sum of largest eigenvalues}
Consider $\mbf{A} \in \mathbb{S}^{n \times n}$, $\gamma \in \mathbb{R}$, and $k \in \mathbb{Z}_{>0}$.  The sum of the $k$ largest eigenvalues of $\mbf{A}$, where $k \leq n$, is less than $\gamma$ (i.e., $\sum_{i=1}^k\lambda_i(\mbf{A}) \leq \gamma$) if and only if there exist $\mbf{X} \in \mathbb{S}^n$ and $z \in \mathbb{R}$, where $\mbf{X} \geq 0$, such that
\begin{align*}
z \mbf{1} + \mbf{X} - \mbf{A} & \geq 0, \\
z k + \trace( \mbf{X} ) &\leq \gamma.
\end{align*}

\subsubsection[Sum of Absolute Value of Largest Eigenvalues]{Sum of Absolute Value Largest Eigenvalues~\cite{Alizadeh1995}}
\index{eigenvalues!sum of absolute value of largest eigenvalues}
Consider $\mbf{A} \in \mathbb{S}^{n \times n}$, $\gamma \in \mathbb{R}$, and $k \in \mathbb{Z}_{>0}$.  The sum of the absolute value of the $k$ largest eigenvalues of $\mbf{A}$, where $k \leq n$, is less than $\gamma$ (i.e., $\sum_{i=1}^k\abs{\lambda_i(\mbf{A})} \leq \gamma$) if and only if there exist $\mbf{X}$,~$\mbf{Y} \in \mathbb{S}^n$ and $z \in \mathbb{R}$, where $\mbf{X} \geq 0$ and $\mbf{Y} \geq 0$, such that
\begin{align*}
z \mbf{1} + \mbf{X} - \mbf{A} & \geq 0, \\
z \mbf{1} + \mbf{Y} + \mbf{A} & \geq 0, \\
z k + \trace( \mbf{X} + \mbf{Y}) &\leq \gamma.
\end{align*}

\subsubsection[Weighted Sum of Largest Eigenvalues]{Weighted Sum of Largest Eigenvalues~\cite{Alizadeh1995}}
\label{sec:WeightedSumEigvals}
\index{eigenvalues!weighted sum of largest eigenvalues}
Consider $\mbf{A} \in \mathbb{S}^{n \times n}$, $\gamma \in \mathbb{R}$, $k \in \mathbb{Z}_{>0}$, and $w_i \in \mathbb{R}_{>0}$, $i=1,\ldots,k$, where $0 < w_k \leq w_{k-1} \leq \cdots \leq w_1$.  The weighted sum of the $k$ largest eigenvalues of $\mbf{A}$, where $k \leq n$, is less than $\gamma$ (i.e., $\sum_{i=1}^kw_i\lambda_i(\mbf{A}) \leq \gamma$) if and only if there exist $\mbf{X}_i \in \mathbb{S}^n$ and $z_i \in \mathbb{R}$, $i = 1,\ldots,k$, where $\mbf{X}_i \geq 0$, such that
\begin{align*}
z_i \mbf{1} + \mbf{X}_i - (w_i - w_{i+1}) \mbf{A} & \geq 0, \hspace{10pt} \text{for} \,\, i=1,\ldots,k-1, \\
z_k \mbf{1} + \mbf{X}_k - w_k \mbf{A} & \geq 0, \\
\sum_{i=1}^k \left(i z_i + \trace( \mbf{X} _i)\right) &\leq \gamma.
\end{align*}

\subsubsection[Weighted Sum of Absolute Value of Largest Eigenvalues]{Weighted Sum of Absolute Value of Largest Eigenvalues~\cite{Alizadeh1995}}
\label{sec:WeightedSumAbsEigvals}
\index{eigenvalues!weighted sum of absolute value of largest eigenvalues}
Consider $\mbf{A} \in \mathbb{S}^{n \times n}$, $\gamma \in \mathbb{R}$, $k \in \mathbb{Z}_{>0}$, and $w_i \in \mathbb{R}_{>0}$, $i=1,\ldots,k$, where $0 < w_k \leq w_{k-1} \leq \cdots \leq w_1$.  The weighted sum of the absolute value of the $k$ largest eigenvalues of $\mbf{A}$, where $k \leq n$, is less than $\gamma$ (i.e., $\sum_{i=1}^kw_i\abs{\lambda_i(\mbf{A})} \leq \gamma$) if and only if there exist $\mbf{X}_i$,~$\mbf{Y}_i \in \mathbb{S}^n$ and $z_i \in \mathbb{R}$, $i = 1,\ldots,k$, where $\mbf{X}_i \geq 0$ and $\mbf{Y}_i \geq 0$, such that
\begin{align*}
z_i \mbf{1} + \mbf{X}_i - (w_i - w_{i+1}) \mbf{A} & \geq 0, \hspace{10pt} \text{for} \,\, i=1,\ldots,k-1, \\
z_i \mbf{1} + \mbf{Y}_i + (w_i - w_{i+1}) \mbf{A} & \geq 0, \hspace{10pt} \text{for} \,\, i=1,\ldots,k-1, \\
z_k \mbf{1} + \mbf{X}_k - w_k \mbf{A} & \geq 0, \\
z_k \mbf{1} + \mbf{Y}_k + w_k \mbf{A} & \geq 0, \\
\sum_{i=1}^k \left(i z_i + \trace( \mbf{X} _i + \mbf{Y}_i)\right) &\leq \gamma.
\end{align*}

\subsection{Matrix Condition Number}
\index{condition number}
\subsubsection[Condition Number of a Matrix]{Condition Number of a Matrix~\cite[pp.~37--38]{Boyd1994}}

Consider $\mbf{A} \in \mathbb{R}^{n \times m}$ and $\gamma$,~$\mu \in \mathbb{R}_{>0}$, where the condition number of $\mbf{A}$ is $\kappa(\mbf{A})$.  If $m \leq n$, the inequality $\kappa(\mbf{A}) \leq \gamma$ holds if there exists $\mu$ such that
\bdis
\mu\mbf{1} \leq \mbf{A}^\trans\mbf{A} \leq \gamma^2 \mu\mbf{1}.
\edis
If $n \leq m$, the inequality $\kappa(\mbf{A}) \leq \gamma$ holds if there exists $\mu$ such that
\bdis
\mu\mbf{1} \leq \mbf{A}\mbf{A}^\trans \leq \gamma^2 \mu\mbf{1}.
\edis

\subsubsection[Condition Number of a Positive Definite Matrix]{Condition Number of a Positive Definite Matrix~\cite[p.~38]{Boyd1994}}

Consider $\mbf{A} \in \mathbb{S}^{n}$ and $\gamma$,~$\mu \in \mathbb{R}_{>0}$, where the condition number of $\mbf{A}$ is $\kappa(\mbf{A})$.  The inequality $\kappa(\mbf{A}) \leq \gamma$ holds if there exists $\mu$ such that
\bdis
\mu\mbf{1} \leq \mbf{A} \leq \gamma \mu\mbf{1}.
\edis

\subsection[Spectral Radius]{Spectral Radius~\cite[p.~17]{ElGhaoui2000}}
\index{spectral radius}
Consider $\mbf{A} \in \mathbb{R}^{n \times n}$ and $\delta \in \mathbb{R}_{>0}$.  The spectral radius of $\mbf{A}$ is strictly less than $\delta$ (i.e., $\rho(\mbf{A}) < \delta$) under either of the following necessary and sufficient conditions.

\begin{enumerate}

\item There exists $\mbf{X} \in \mathbb{S}^n$, where $\mbf{X} > 0$, such that
\bdis
\mbf{A}^\trans \mbf{X} \mbf{A} - \delta^2 \mbf{X} < 0.
\edis

\item There exists $\mbf{X} \in \mathbb{S}^n$, where $\mbf{X} > 0$, such that
\bdis
\mbf{A} \mbf{X} \mbf{A}^\trans - \delta^2 \mbf{X} < 0.
\edis

\end{enumerate}

Also see Section~\ref{sec:mu_analysis} for a similar condition related to the structured singular value.

\subsection[Trace of a Symmetric Matrix]{Trace of a Symmetric Matrix}
\label{sec:Trace}
\index{trace}

\subsubsection[Trace of a Matrix with a Slack Variable]{Trace of a Matrix with a Slack Variable}
\index{trace}
\begin{enumerate}

\itemcite \cite[pp.~46--47]{Duan2013}
Consider $\mbf{P}  \in \mathbb{S}^{n}$ and $\gamma \in \mathbb{R}_{>0}$, where $\mbf{P} > 0$ and $\mbf{Z} > 0$.  The inequality given by
\bdis
\trace ( \mbf{P})  < \gamma
\edis
is satisfied if and only if there exists $\mbf{Z} \in \mathbb{S}^{n}$ such that
\bdis
\mbf{P} < \mbf{Z}, \hspace{20pt} \trace ( \mbf{Z} )< \gamma.
\edis

\itemcite \cite[p.~8]{Boyd1994}
Consider $\mbf{P} \in \mathbb{S}^{n}$, $\mbf{X} \in \mathbb{R}^{n \times m}$, and $\gamma \in \mathbb{R}_{>0}$, where $\mbf{P} > 0$ and $\mbf{Z} > 0$.  The matrix inequality given by
\bdis
\trace \left(\mbf{X}^\trans \mbf{P}^{-1} \mbf{X}\right) < \gamma
\edis
is satisfied if and only if there exists $\mbf{Z} \in \mathbb{S}^m$ such that
\bdis
\bbm \mbf{Z} & \mbf{X}^\trans \\ * & \mbf{P} \ebm > 0, \hspace{20pt} \trace (\mbf{Z}) < \gamma.
\edis

\end{enumerate}

\subsubsection[Relative Trace of Matrices]{Relative Trace of Matrices}
\index{trace}
\begin{enumerate}

\itemcite \cite[pp.~46--47]{Duan2013} Consider $\mbf{P}$,~$\mbf{Q} \in \mathbb{S}^{n}$.  The property $\trace(\mbf{P}) < \trace(\mbf{Q})$ holds if the matrix inequality $\mbf{P} < \mbf{Q}$ is satisfied.

\itemcite \cite[p.~768]{BernsteinMatrixBook},~\cite[p.~215]{Zhang2011} Consider $\mbf{P}$,~$\mbf{Q} \in \mathbb{S}^{n}$, where $\mbf{P} \geq 0$, $\mbf{Q} > 0$, and $\mbf{P} \leq \mbf{Q}$.   Then,
\bdis
\frac{\det(\mbf{P})}{\det(\mbf{Q})} \leq \frac{\trace (\mbf{P})}{\trace(\mbf{Q})}.
\edis

\itemcite \cite[p.~771]{BernsteinMatrixBook},~\cite{Bourin1999} Consider $\mbf{P}$,~$\mbf{Q} \in \mathbb{S}^{n}$ and $i$,~$j \in \mathbb{R}_{\geq 0}$, where $\mbf{P} \geq 0$, $\mbf{Q} \geq 0$, and $\mbf{P} \leq \mbf{Q}$.   Then,
\bdis
\trace\left(\mbf{A}^i\mbf{B}^j\right) \leq \trace\left(\mbf{B}^{i+j}\right).
\edis

\itemcite \cite[p.~771]{BernsteinMatrixBook},~\cite{Bourin1999} Consider $\mbf{P}$,~$\mbf{Q} \in \mathbb{S}^{n}$ and $i$,~$j \in \mathbb{R}$, where $\mbf{P} > 0$, $\mbf{Q} > 0$, $\mbf{P} \leq \mbf{Q}$, $j \geq -1$, and $i+j \geq 0$.   Then,
\bdis
\trace\left(\mbf{A}^i\mbf{B}^j\right) \leq \trace\left(\mbf{B}^{i+j}\right).
\edis

\itemcite \cite[p.~771]{BernsteinMatrixBook},~\cite{Travaglia2006} Consider $\mbf{P}$,~$\mbf{Q} \in \mathbb{S}^{n}$ and $i$,~$j \in \mathbb{R}_{>0}$, where $\mbf{P} \geq 0$, $\mbf{Q} \geq 0$, $\mbf{Q} \leq \mbf{1}$, and $i\leq j$.   Then,
\bdis
\trace\left(\mbf{Q} \mbf{P}^i \mbf{Q} \right)^{1/i} \leq \trace\left(\mbf{Q} \mbf{P}^j \mbf{Q} \right)^{1/j}
\edis
and
\bdis
\trace\left(\mbf{Q} \mbf{P}^i \mbf{Q} \right)^{1/i} \leq \trace\left(\mbf{Q}^{i/j} \mbf{P}^i \mbf{Q}^{i/j} \right)^{1/j}.
\edis

\itemcite \cite[p.~773]{BernsteinMatrixBook},~\cite[p.~213]{Zhang2011} Consider $\mbf{P}$,~$\mbf{Q}$,~$\mbf{V} \in \mathbb{S}^{n}$, where $\mbf{P} \geq 0$, $\mbf{Q} \geq 0$, $\mbf{V} \geq 0$, and $\mbf{P} \leq \mbf{Q}$.   Then,
\bdis
\trace\left( \left(\mbf{V} + \mbf{P}\right)^{-1} \mbf{P}\right) \leq \trace\left( \left(\mbf{V} + \mbf{Q}\right)^{-1} \mbf{Q}\right).
\edis

\end{enumerate}

\subsection[Range of a Symmetric Matrix]{Range of a Symmetric Matrix~\cite[p.~714]{BernsteinMatrixBook}}
\index{range}

Consider $\mbf{P} \in \mathbb{S}^{n}$, where $\mbf{P} > 0$.  If $\mbf{P} \leq \mbf{Q}$, then $\mathcal{R}(\mbf{P}) \subseteq \mathcal{R}(\mbf{Q})$.

\subsection[Logarithm of a Positive Definite Matrix]{Logarithm of a Positive Definite Matrix~\cite[p.~715]{BernsteinMatrixBook}}
\index{logarithm}

Consider $\mbf{P}$,~$\mbf{Q}  \in \mathbb{S}^{n}$ and $\alpha \in \mathbb{R}_{>0}$, where $\mbf{P} \geq 0$.  The matrix logarithm of $\mbf{P}$ satisfies the following matrix inequality
\bdis
\mbf{1} - \mbf{P}^{-1} \leq \log (\mbf{P}) \leq \alpha^{-1} \left( \mbf{P}^\alpha - \mbf{1} \right).
\edis

\subsection[Douglas-Fillmore-Williams Lemma]{Douglas-Fillmore-Williams Lemma~\cite[p.~714]{BernsteinMatrixBook}~\cite{Douglas1966,Fillmore1971}}
\label{sec:DouglasFillmore}
\index{Douglas-Fillmore-Williams Lemma}

Consider $\mbf{A} \in \mathbb{R}^{n \times m}$ and $\mbf{B}  \in \mathbb{R}^{n \times p}$.  The following statements are equivalent.

\begin{enumerate}

\item There exists $\mbf{C} \in \mathbb{R}^{p \times m}$ such that $\mbf{A} = \mbf{B} \mbf{C}$.

\item There exists $\alpha \in \mathbb{R}_{>0}$ such that $\mbf{A} \mbf{A}^\trans - \alpha \mbf{B} \mbf{B}^\trans \leq 0$.

\item $\mathcal{R}(\mbf{A}) \subseteq \mathcal{R}(\mbf{B})$.

\end{enumerate}

\subsection[Submatrix Determinants]{Submatrix Determinants~\cite{LallNotes3}}
\index{submatrix}\index{determinant}
Consider $\mbf{A} \in \mathbb{S}^{n}$.  Let $\mbf{A}_k \in \mathbb{S}^{k}$ be a submatrix of $\mbf{A}$ consisting of its first $k$ rows and columns, where $k \leq n$.  The matrix inequality $\mbf{A} > 0$ is satisfied if and only if
\bdis
\text{det}(\mbf{A}_k) > 0, \,\, k=1,\ldots,n.
\edis

\subsection[Imaginary and Real Parts]{Imaginary and Real Parts~\cite[Sec.~12.1.1]{Skogestad2005}}
Consider $\mbf{Q}_R \in \mathbb{S}^{n}$, $\mbf{Q}_I \in \mathbb{R}^{n \times n}$, and $\mbf{Q} = \mbf{Q}^\herm = \mbf{Q}_R + j\mbf{Q}_I \in \mathbb{C}^{n\times n}$\index{Hermitian matrix}.  The matrix inequality $\mbf{Q} > 0$ is equivalent to the matrix inequality given by
\bdis
\bbm \mbf{Q}_R & \mbf{Q}_I \\ -\mbf{Q}_I & \mbf{Q}_R \ebm > 0.
\edis

\subsection{Quadratic Inequalities}

\subsubsection[Weighted Norm]{Weighted Norm~\cite{VanAntwerp2000}}
\index{norm!weighted norm}
Consider $\mbf{W} \in \mathbb{S}^n$, $\mbf{x}$,~$\mbf{y} \in \mathbb{R}^n$, and $\gamma \in \mathbb{R}_{\geq 0}$, where $\mbf{W} > 0$.  The inequality $(\mbf{x} - \mbf{y})^\trans\mbf{W} (\mbf{x}-\mbf{y}) \leq \gamma$ is equivalent to the matrix inequality given by
\bdis
\bbm \gamma & (\mbf{x}-\mbf{y})^\trans \\ * & \mbf{W}^{-1}\ebm \geq 0.
\edis

\subsubsection{Quadratic Inequalities}
\index{quadratic inequality}

\begin{enumerate}

\item Consider $\mbf{W} \in \mathbb{S}^n$, $\mbf{A} \in \mathbb{R}^{n \times m}$, $\mbf{x}$,~$\mbf{c} \in \mathbb{R}^m$, $\mbf{b} \in \mathbb{R}^n$, and $d \in \mathbb{R}$, where $\mbf{W} > 0$.  The quadratic inequality $(\mbf{A}\mbf{x}+\mbf{b})^\trans\mbf{W}(\mbf{A}\mbf{x}+\mbf{b})-\mbf{c}^\trans\mbf{x} - d \leq 0$ with $\mbf{W} > 0$ is equivalent to the matrix inequality given by
\bdis
\bbm \mbf{W}^{-1} & \mbf{A}\mbf{x} + \mbf{b} \\ * & \mbf{c}^\trans\mbf{x} + d \ebm \geq 0.
\edis

\itemcite \cite[p.~731]{BernsteinMatrixBook} Consider $\mbf{x}  \in \mathbb{R}^n$.  The matrix inequality given by
\bdis
\mbf{x} \mbf{x}^\trans - \mbf{x}^\trans \mbf{x} \mbf{1} \leq 0
\edis
holds.

\end{enumerate}

%
%
%
%
%

\subsection{Miscellaneous Properties and Results}

\begin{enumerate}

\itemcite \cite[p.~738]{BernsteinMatrixBook} Consider $\mbf{P}$,~$\mbf{Q} \in \mathbb{S}^{n}$, where $\mbf{P} \geq 0$ and $\mbf{Q} > 0$.  Then, $\mbf{P} \leq \mbf{Q}$ if and only if $\mbf{P} \mbf{Q}^{-1} \mbf{P} \leq \mbf{P}$.

\itemcite \cite[p.~269]{Dym2006},~\cite[p.~738]{BernsteinMatrixBook} Consider $\mbf{P}$,~$\mbf{Q} \in \mathbb{S}^{n}$, where $\mbf{P} \geq 0$, $\mbf{Q} \geq 0$, and $\mbf{P} \leq \mbf{Q}$.  Then, there exists $\mbf{S} \in \mathbb{R}^{n \times n}$ such that $\mbf{P} = \mbf{S}^\trans \mbf{Q} \mbf{S}$ and $\mbf{S}^\trans \mbf{S} \leq \mbf{1}$.

\itemcite \cite{AuYeung1973,Chan1985},~\cite[p.~738]{BernsteinMatrixBook} Consider $\mbf{P}$,~$\mbf{Q}$,~$\mbf{R}$,~$\mbf{S} \in \mathbb{S}^{n}$, where $\mbf{P} \geq 0$, $\mbf{Q} \geq 0$, $\mbf{R} \geq 0$, $\mbf{S} > 0$, $\mbf{S} \leq \mbf{R}$, and $\mbf{Q} \mbf{R} \mbf{Q} \leq \mbf{P} \mbf{S} \mbf{P}$.  Then, $\mbf{Q} \leq \mbf{P}$.

\itemcite \cite[pp.~289--290]{Zhang2011},~\cite[p.~738]{BernsteinMatrixBook} Consider $\mbf{P}$,~$\mbf{Q} \in \mathbb{R}^{n \times m}$.  Then, there exist unitary matrices\index{unitary matrix} $\mbf{S}_1$,~$\mbf{S}_2 \in \mathbb{R}^{m \times m}$ such that
\bdis
\sqrt{\left(\mbf{P} + \mbf{Q}\right)^\trans \left(\mbf{P} + \mbf{Q}\right)} \leq \mbf{S}_1 \sqrt{\mbf{P}^\trans \mbf{P}} \mbf{S}_1^\trans + \mbf{S}_2 \sqrt{\mbf{Q}^\trans \mbf{Q}} \mbf{S}_2^\trans.
\edis
This is a matrix version of the triangle inequality.\index{triangle inequality}

\itemcite \cite{Bhatia1990,Aujla2007},~\cite[p.~739]{BernsteinMatrixBook} Consider $\mbf{P}$,~$\mbf{Q} \in \mathbb{S}^{n}$.  Then, there exists a unitary matrix\index{unitary matrix} $\mbf{S} \in \mathbb{R}^{n \times n}$ such that
\bdis
\sqrt{\mbf{Q}\mbf{P}\mbf{P}\mbf{Q}} \leq \onehalf \mbf{S} \left(\mbf{P}\mbf{P} + \mbf{Q} \mbf{Q} \right) \mbf{S}^\trans.
\edis

\itemcite \cite{Bourin2006},~\cite[p.~739]{BernsteinMatrixBook} Consider $\mbf{P}$,~$\mbf{Q} \in \mathbb{S}^{n}$, where $\mbf{P} \geq 0$, $\mbf{Q} > 0$, $\mbf{P} \leq \mbf{1}$, $\alpha = \lambda_\textrm{min}(\mbf{Q})$, and $\beta = \lambda_\textrm{max}(\mbf{Q})$.  Then, 
\bdis
\mbf{P} \mbf{Q} \mbf{P} \leq \frac{(\alpha + \beta)^2}{4 \alpha \beta} \mbf{Q}.
\edis

\itemcite \cite{Baksalary1991},~\cite[p.~740]{BernsteinMatrixBook} Consider $\mbf{P}$,~$\mbf{Q} \in \mathbb{S}^{n}$, where $\mbf{P} \geq 0$, $\mbf{Q} \geq 0$, $\mbf{P} \leq \mbf{Q}$ and $\mbf{P} \mbf{Q} = \mbf{Q} \mbf{P}$.  Then, $\mbf{P}\mbf{P} \leq \mbf{B} \mbf{B}$.

\itemcite \cite[p.~214]{Zhang2011},~\cite[p.~740]{BernsteinMatrixBook} Consider $\mbf{P}$,~$\mbf{Q} \in \mathbb{S}^{n}$, where $\mbf{P} \geq 0$, $\mbf{Q} \geq 0$, and $\mbf{Q}\mbf{P}\mbf{P} \mbf{Q} \leq \mbf{1}$.  Then, $\sqrt{\mbf{Q}} \mbf{P} \sqrt{\mbf{Q}} \leq \mbf{1}$.

\itemcite \cite[p.~292]{Zhang2011},~\cite[p.~741]{BernsteinMatrixBook} Consider $\mbf{P}$,~$\mbf{Q} \in \mathbb{S}^{n}$.  Then
\bdis
\left(\onehalf\left(\mbf{P} + \mbf{Q}\right)\right)^2 \leq \onehalf \left(\mbf{P}^2 + \mbf{Q}^2\right).
\edis

\itemcite \cite{Kwong1989},~\cite[p.~741]{BernsteinMatrixBook} Consider $\mbf{P}$,~$\mbf{Q} \in \mathbb{S}^{n}$ and $\alpha \in \mathbb{R}$, where $\mbf{P} \geq 0$ and $\mbf{Q} \geq 0$.  If either
\begin{itemize}
\item $\alpha \in [1,2]$, or
\item $\mbf{P} > 0$, $\mbf{Q} > 0$, and $\alpha \in [-1,0] \cup [1,2]$,
\end{itemize}
then,
\bdis
\left(\onehalf\left(\mbf{P} + \mbf{Q}\right)\right)^\alpha \leq \onehalf \left(\mbf{P}^\alpha + \mbf{Q}^\alpha\right).
\edis

\itemcite \cite{Bellman1968,Bhagwat1978},~\cite[p.~741]{BernsteinMatrixBook} Consider $\mbf{P}$,~$\mbf{Q} \in \mathbb{S}^{n}$ and $\alpha$,~$\beta \in \mathbb{R}$, where $\mbf{P} \geq 0$, $\mbf{Q} \geq 0$, and $1 \leq \alpha \leq \beta$.  Then,
\bdis
\left(\onehalf\left(\mbf{P}^\alpha + \mbf{Q}^\alpha\right)\right)^{1/\alpha} \leq \left(\onehalf \left(\mbf{P}^\beta + \mbf{Q}^\beta\right)\right)^{1/\beta}.
\edis
Furthermore,
\bdis
\mu(\mbf{P},\mbf{Q}) \overset{\Delta}{=} \lim_{\gamma \to \infty}  \left(\onehalf \left(\mbf{P}^\gamma + \mbf{Q}^\gamma\right)\right)^{1/\gamma}
\edis
exists and satisfies $\mbf{P} \leq \mu(\mbf{P},\mbf{Q})$ and $\mbf{Q} \leq \mu(\mbf{P},\mbf{Q})$.  Additionally,
\bdis
\lim_{\gamma \to 0}  \left(\onehalf \left(\mbf{P}^\gamma + \mbf{Q}^\gamma\right)\right)^{1/\gamma} = e^{\onehalf\left(\log(\mbf{A}) + \log(\mbf{B})\right)}.
\edis

\itemcite \cite{Petersen1986,Petersen1987} Consider $\mbf{P}$,~$\mbf{Q}$,~$\mbf{Z} \in \mathbb{S}^n$ and $\mbf{x} \in \mathbb{R}^n$, where $\mbf{P} \geq 0$, $\mbf{Q} \geq 0$, and $\mbf{Z} > 0$.  If the inequality
\bdis
\left( \mbf{x}^\trans \mbf{Z} \mbf{x} \right)^2 - 4 \left( \mbf{x}^\trans \mbf{P} \mbf{x} \mbf{x}^\trans \mbf{Q} \mbf{x} \right) > 0
\edis
holds for all $\mbf{x} \neq \mbf{0}$, then there exists $\lambda \in \mathbb{R}_{>0}$ such that
\bdis
\lambda^2 \mbf{P} + \lambda \mbf{Z} + \mbf{Q} < 0.
\edis

\itemcite \cite{Willems1971} Consider $\mbf{A} \in \mathbb{R}^{n \times n}$ and $\mbf{W}$,~$\mbf{Q}  \in \mathbb{S}^n$, where $\mbf{W} > 0$.  If there exists $\mbf{S} \in \mathbb{S}^n$, where $\mbf{S} > 0$, such that
\bdis
\mbf{S} \mbf{W} \mbf{S} = \mbf{S} \mbf{A} + \mbf{A}^\trans \mbf{S} + \mbf{Q},
\edis
then for any $0 < \mbf{W}_1 \leq \mbf{W}$ and $\mbf{Q}_1 \geq \mbf{Q}$ there exists $\mbf{S}_1 \in \mathbb{S}^n$, where $\mbf{S}_1 \geq \mbf{S}$ such that
\bdis
\mbf{S}_1 \mbf{W}_1 \mbf{S}_1 = \mbf{S}_1 \mbf{A} + \mbf{A}^\trans \mbf{S} _1+ \mbf{Q}_1.
\edis

\itemcite \cite{Venkataraman2018} Consider $\mbf{X}$,~$\mbf{Y} \in \mathbb{S}^n$ and $r\in \mathbb{Z}_{>0}$.  There exist $\mbf{X}_2$,~$\mbf{Y}_2 \in \mathbb{R}^{n \times r}$ and $\mbf{X}_3$,~$\mbf{Y}_3 \in \mathbb{S}^{r}$, where $\mbf{X}_3 > 0$ such that
\bdis
\bbm \mbf{X} & \mbf{X}_2 \\ * & \mbf{X}_3 \ebm^{-1} = \bbm \mbf{Y} & \mbf{Y}_2 \\ * & \mbf{Y}_3 \ebm
\edis
if and only if $\mbf{X} - \mbf{Y}^{-1} \geq 0$ and $\text{rank} \left( \mbf{X} - \mbf{Y}^{-1} \right) \leq r$.\index{rank}

\itemcite \cite[p.~19]{Ebihara2015} Consider $\mbf{M}_{11}$,~$\mbf{A} \in \mathbb{S}^{n}$, $\mbf{M}_{12} \in \mathbb{R}^{n \times m}$, $\mbf{M}_{22} \in \mathbb{S}^{m}$, $\mbf{E}$,~$\mbf{F}_1 \in \mathbb{R}^{n \times n}$, and $\mbf{F}_2 \in \mathbb{R}^{m \times n}$, where $\mbf{M}_{11} \geq 0$ and $\mbf{E}$ is invertible.  The matrix inequality
\beq
\label{eq:Misc2a}
\bbm \mbf{E}^{-1}\mbf{A} \\ \mbf{1} \ebm^\trans \bbm \mbf{M}_{11} & \mbf{M}_{12} \\ * & \mbf{M}_{22} \ebm \bbm \mbf{E}^{-1}\mbf{A} \\ \mbf{1} \ebm < 0
\eeq
holds if and only if there exist $\mbf{F}_1$ and $\mbf{F}_2$ such that
\beq
\label{eq:Misc2b}
\bbm \mbf{M}_{11} + \mbf{F}_1\mbf{E} + \mbf{E}^\trans\mbf{F}_1^\trans & \mbf{M}_{12} - \mbf{F}_1\mbf{A} + \mbf{E}^\trans \mbf{F}_2^\trans \\ * & \mbf{M}_{22} - \mbf{F}_2\mbf{A} - \mbf{A}^\trans\mbf{F}_2^\trans \ebm  < 0,
\eeq
Moreover, the following statements hold.
\begin{enumerate}

\item If~\eqref{eq:Misc2a} holds, then~\eqref{eq:Misc2b} holds with $\mbf{F}_1 = -\left(\mbf{M}_{11} + \epsilon\mbf{W}\right)\mbf{E}^{-1}$ and $\mbf{F}_2 = -\mbf{M}_{12}^\trans\mbf{E}^{-1}$, where $\epsilon \in \mathbb{R}_{>0}$ is sufficiently small, $\mbf{W} \in \mathbb{S}^{n}$, and $\mbf{W} > 0$.

\item If~\eqref{eq:Misc2a} holds and $\mbf{M}_{11} > 0$, then~\eqref{eq:Misc2b} holds with $\mbf{F}_1 = \mbf{M}_{11}\mbf{E}^{-1}$ and $\mbf{F}_2 = - \mbf{M}_{12}^\trans \mbf{E}^{-1}$.

\end{enumerate}

\end{enumerate}


\clearpage
\section{LMIs in Systems and Stability Theory}
\label{sec:SystemsStability}

This section presents a compilation of LMIs results that are related to systems and stability theory.

\subsection{Lyapunov Inequalities}
\index{Lyapunov!inequality}
\subsubsection[Lyapunov Stability]{Lyapunov Stability~\cite[pp.~1201--1203]{BernsteinMatrixBook},~\cite[pp.~20--21]{Boyd1994}}
\index{stability!Lyapunov stability}\index{Lyapunov!stability}
Consider the matrices $\mbf{A} \in \mathbb{R}^{n \times n}$ and $\mbf{Q} \in \mathbb{S}^n$, where $\mbf{Q}  \geq 0$.  There exists $\mbf{P} \in \mathbb{S}^n$, where $\mbf{P} > 0$, satisfying the Lyapunov equation\index{Lyapunov!equation}
\bdis
\mbf{A}^\trans\mbf{P}+\mbf{P}\mbf{A} + \mbf{Q} = \mbf{0},
\edis
if and only if there exists $\mbf{P} \in \mathbb{S}^n$, where $\mbf{P} > 0$, such that
\beq
\label{eq:LyapIneq}
\mbf{A}^\trans\mbf{P}+\mbf{P}\mbf{A} \leq 0.
\eeq
If~\eqref{eq:LyapIneq} holds, then $\text{Re}\{\lambda_i(\mbf{A})\} \leq 0$, $i=1,\ldots,n$, and the equilibrium point $\mbfbar{x} = \mbf{0}$ of the system $\dot{\mbf{x}} = \mbf{A} \mbf{x}$ is Lyapunov stable.

The matrix inequality of~\eqref{eq:LyapIneq} is satisfied under any of the following equivalent necessary and sufficient conditions.

\begin{enumerate}

\itemcite There exists $\mbf{X} \in \mathbb{S}^n$, where $\mbf{X} > 0$, such that
\bdis
\mbf{X}\mbf{A}^\trans+\mbf{A}\mbf{X} \leq 0.
\edis



\item There exist $\mbf{X} \in \mathbb{S}^n$ and $\mbf{V} \in \mathbb{R}^n$, where $\mbf{X} > 0$, such that
\bdis
\bbm -\left(\mbf{V} + \mbf{V}^\trans\right) & \mbf{V}^\trans\mbf{A} + \mbf{X} & \mbf{V}^\trans \\ * & -\mbf{X} & \mbf{0} \\ * & * & -\mbf{X} \ebm \leq 0.
\edis
\begin{proof}
Identical to the proof of~\eqref{eq:LyapIneqDil1} in~\cite{Apkarian2001}, except with the use of the Nonstrict Projection Lemma, where $\mbf{G}^\trans = \bbm -\mbf{1} & \mbf{A} & \mbf{1} \ebm$ and $\mbf{H}^\trans = \bbm \mbf{1} & \mbf{0} & \mbf{0} \ebm$, and therefore $\mathcal{R}(\mbf{G})$ and $\mathcal{R}(\mbf{H})$ are linearly independent.
\end{proof}

\item There exist $\mbf{X} \in \mathbb{S}^n$ and $\mbf{V} \in \mathbb{R}^n$, where $\mbf{X} > 0$, such that
\bdis
\bbm -\left(\mbf{V} + \mbf{V}^\trans\right) & \mbf{V}^\trans\mbf{A}^\trans + \mbf{X} & \mbf{V}^\trans \\ * & -\mbf{X} & \mbf{0} \\ * & * & -\mbf{X} \ebm \leq 0.
\edis
\begin{proof}
Identical to the proof of~\eqref{eq:LyapIneqDil2} in~\cite{Apkarian2001}, except with the use of the Nonstrict Projection Lemma, where $\mbf{G}^\trans = \bbm -\mbf{1} & \mbf{A}^\trans & \mbf{1} \ebm$ and $\mbf{H}^\trans = \bbm \mbf{1} & \mbf{0} & \mbf{0} \ebm$, and therefore $\mathcal{R}(\mbf{G})$ and $\mathcal{R}(\mbf{H})$ are linearly independent.
\end{proof}

\itemcite \cite{Balakrishnan2002} There does not exist $\mbf{Z}\in \mathbb{S}^n$, where $\mbf{Z} > 0$, such that
\bdis
\mbf{Z} \mbf{A}^\trans + \mbf{A} \mbf{Z} > 0.
\edis

\end{enumerate}

\subsubsection[Asymptotic Stability]{Asymptotic Stability~\cite[p.~1201--1203]{BernsteinMatrixBook},~\cite[p.~2]{Boyd1994}}
\index{stability!asymptotic stability}
Consider the matrices $\mbf{A} \in \mathbb{R}^{n \times n}$ and $\mbf{Q} \in \mathbb{S}^n$, where $\mbf{Q}  > 0$.  There exists $\mbf{P} \in \mathbb{S}^n$, where $\mbf{P} > 0$, satisfying the Lyapunov equation\index{Lyapunov!equation}
\bdis
\mbf{A}^\trans\mbf{P}+\mbf{P}\mbf{A} + \mbf{Q} = \mbf{0},
\edis
if and only if there exists $\mbf{P} \in \mathbb{S}^n$, where $\mbf{P} > 0$, such that
\beq
\label{eq:LyapStrictIneq}
\mbf{A}^\trans\mbf{P}+\mbf{P}\mbf{A} < 0.
\eeq
If~\eqref{eq:LyapStrictIneq} holds, then $\text{Re}\{\lambda_i(\mbf{A})\} < 0$, $i=1,\ldots,n$, the matrix $\mbf{A}$ is Hurwitz\index{Hurwitz matrix}, and the equilibrium point $\mbfbar{x} = \mbf{0}$ of the system $\dot{\mbf{x}} = \mbf{A} \mbf{x}$ is asymptotically stable.

The matrix inequality of~\eqref{eq:LyapStrictIneq} is satisfied and the matrix $\mbf{A}$ is Hurwitz under any of the following equivalent necessary and sufficient conditions.

\begin{enumerate}

\itemcite There exists $\mbf{X} \in \mathbb{S}^n$, where $\mbf{X} > 0$, such that
\bdis
\mbf{X}\mbf{A}^\trans+\mbf{A}\mbf{X} < 0.
\edis

\item (\textit{The $S$-Variable Approach}~\cite[pp.~2--3]{Ebihara2015},~\cite{Geromel1998}) There exist $\mbf{P} \in \mathbb{S}^{n}$ and $\mbf{F}_1$,~$\mbf{F}_2 \in \mathbb{R}^{n \times n}$, where $\mbf{P} > 0$, such that
\beq
\label{eq:LyapSVariable1}
\bbm \mbf{F}_1\mbf{A} + \mbf{A}^\trans\mbf{F}_1^\trans & \mbf{P} -\mbf{F}_1 + \mbf{A}^\trans\mbf{F}_2^\trans \\ * & -(\mbf{F}_2 + \mbf{F}_2^\trans) \ebm  < 0.
\eeq

\item \cite{Felipe2016} There exist $\mbf{P} \in \mathbb{S}^{n}$ and $\mbf{F}_1$,~$\mbf{F}_2 \in \mathbb{R}^{n \times n}$, where $\mbf{P} > 0$, such that
\beq
\label{eq:LyapSVariable2}
\bbm \mbf{F}_1\mbf{P} + \mbf{P}\mbf{F}_1^\trans & \mbf{A}^\trans -\mbf{F}_1 + \mbf{P}\mbf{F}_2^\trans \\ * & -(\mbf{F}_2 + \mbf{F}_2^\trans) \ebm  < 0.
\eeq

\itemcite \cite{Apkarian2001} There exist $\mbf{Y} \in \mathbb{S}^n$ and $\mbf{W} \in \mathbb{R}^{n \times n}$, where $\mbf{Y} > 0$, such that
\bdis
\bbm \mbf{Y} - \left(\mbf{W} + \mbf{W}^\trans\right) & \mbf{A}\mbf{Y} + \mbf{W}^\trans \\ * & -\mbf{Y} \ebm < 0.
\edis

\itemcite \cite{Apkarian2001} There exist $\mbf{X} \in \mathbb{S}^n$ and $\mbf{V} \in \mathbb{R}^{n \times n}$, where $\mbf{X} > 0$, such that
\beq
\label{eq:LyapIneqDil1}
\bbm -\left(\mbf{V} + \mbf{V}^\trans\right) & \mbf{V}^\trans\mbf{A} + \mbf{X} & \mbf{V}^\trans \\ * & -\mbf{X} & \mbf{0} \\ * & * & -\mbf{X} \ebm < 0.
\eeq

\itemcite \cite{Apkarian2001} There exist $\mbf{X} \in \mathbb{S}^n$ and $\mbf{V} \in \mathbb{R}^{n \times n}$, where $\mbf{X} > 0$, such that
\beq
\label{eq:LyapIneqDil2}
\bbm -\left(\mbf{V} + \mbf{V}^\trans\right) & \mbf{V}^\trans\mbf{A}^\trans + \mbf{X} & \mbf{V}^\trans \\ * & -\mbf{X} & \mbf{0} \\ * & * & -\mbf{X} \ebm < 0.
\eeq

\itemcite \cite{Felipe2016} There exist $\mbf{P} \in \mathbb{S}^{n}$ and $\mbf{F}_1$,~$\mbf{F}_2$,~$\mbf{X}_1$,~$\mbf{X}_2$,~$\mbf{X}_3 \in \mathbb{R}^{n \times n}$, where $\mbf{P} > 0$, such that
\beq
\label{eq:LyapIneqDil3}
\bbm \mbf{X}_1\mbf{F}_1^\trans + \mbf{F}_1\mbf{X}_1^\trans & \mbf{P} + \mbf{X}_1 \mbf{F}_2^\trans + \mbf{F}_1 \mbf{X}_2^\trans & \mbf{A}^\trans -\mbf{X}_1 + \mbf{F}_1\mbf{X}_3^\trans \\ * & \mbf{X}_2 \mbf{F}_2^\trans + \mbf{F}_2 \mbf{X}_2^\trans & - \mbf{1} - \mbf{X}_2 + \mbf{F}_2 \mbf{X}_3^\trans \\ * & * & -(\mbf{X}_3 + \mbf{X}_3^\trans) \ebm  < 0.
\eeq

\item There exist $\mbf{P} \in \mathbb{S}^{n}$ and $\mbf{F}_1$,~$\mbf{F}_2$,~$\mbf{X}_1$,~$\mbf{X}_2$,~$\mbf{X}_3 \in \mathbb{R}^{n \times n}$, where $\mbf{P} > 0$, such that
\bdis
\bbm \mbf{X}_1\mbf{F}_1^\trans + \mbf{F}_1\mbf{X}_1^\trans & \mbf{A}^\trans + \mbf{X}_1 \mbf{F}_2^\trans + \mbf{F}_1 \mbf{X}_2^\trans & \mbf{P} -\mbf{X}_1 + \mbf{F}_1\mbf{X}_3^\trans \\ * & \mbf{X}_2 \mbf{F}_2^\trans + \mbf{F}_2 \mbf{X}_2^\trans & - \mbf{1} - \mbf{X}_2 + \mbf{F}_2 \mbf{X}_3^\trans \\ * & * & -(\mbf{X}_3 + \mbf{X}_3^\trans) \ebm  < 0.
\edis
\begin{proof}
The proof follows the same steps as the proof of~\eqref{eq:LyapIneqDil3} in~\cite{Felipe2016}, beginning with~\eqref{eq:LyapSVariable2} instead of~\eqref{eq:LyapSVariable1}.
\end{proof}

\itemcite \cite{Felipe2020} There exist $\mbf{P} \in \mathbb{S}^{n}$ and $\mbf{Y}_1$,~$\mbf{Y}_2$,~$\mbf{Y}_3$,~$\mbf{X}_1$,~$\mbf{X}_2$,~$\mbf{X}_3 \in \mathbb{R}^{n \times n}$, where $\mbf{P} > 0$, such that
\bdis
\bbm \mbf{X}_1\mbf{Y}_1 + \mbf{Y}_1^\trans\mbf{X}_1^\trans & \mbf{P} + \mbf{X}_1 \mbf{Y}_2 + \mbf{Y}_1^\trans \mbf{X}_2^\trans & \mbf{A}^\trans +\mbf{X}_1\mbf{Y}_3 + \mbf{Y}_1^\trans\mbf{X}_3^\trans \\ * & \mbf{X}_2 \mbf{Y}_2 + \mbf{Y}_2^\trans \mbf{X}_2^\trans & - \mbf{1} + \mbf{X}_2\mbf{Y}_3+ \mbf{Y}_2^\trans \mbf{X}_3^\trans \\ * & * & \mbf{X}_3 \mbf{Y}_3+ \mbf{Y}_3^\trans\mbf{X}_3^\trans \ebm  < 0.
\edis

\itemcite \cite{Balakrishnan2003} There do not exist $\mbf{Z}_1$,~$\mbf{Z}_2 \in \mathbb{S}^n$, where $\mbf{Z}_1 \geq 0$, $\mbf{Z}_2 \geq 0$, $\mbf{Z}_1 \neq \mbf{0}$, and $\mbf{Z}_2 \neq \mbf{0}$, such that
\bdis
\mbf{Z}_1 \mbf{A}^\trans + \mbf{A} \mbf{Z}_1 - \mbf{Z}_2 = \mbf{0}.
\edis

\end{enumerate}

\subsubsection[Discrete-Time Lyapunov Stability]{Discrete-Time Lyapunov Stability~\cite[pp.~1203--1204]{BernsteinMatrixBook}}
\index{stability!Lyapunov stability}\index{Lyapunov!stability}
Consider the matrices $\mbf{A}_\mathrm{d} \in \mathbb{R}^{n \times n}$ and $\mbf{Q} \in \mathbb{S}^n$, where $\mbf{Q}  \geq 0$.  There exists $\mbf{P} \in \mathbb{S}^n$, where $\mbf{P} > 0$, satisfying the discrete-time Lyapunov equation\index{Lyapunov!equation}
\bdis
\mbf{A}_\mathrm{d}^\trans\mbf{P}\mbf{A}_\mathrm{d}-\mbf{P} + \mbf{Q} = \mbf{0}.
\edis
if and only if there exists $\mbf{P} \in \mathbb{S}^n$, where $\mbf{P} > 0$, such that
\beq
\label{eq:DTLyapIneq}
\mbf{A}_\mathrm{d}^\trans\mbf{P}\mbf{A}_\mathrm{d}-\mbf{P} \leq 0.
\eeq
If~\eqref{eq:DTLyapIneq} holds, then $\abs{\lambda_i(\mbf{A}_\mathrm{d})} \leq 1$, $i=1,\ldots,n$, and the equilibrium point $\mbfbar{x} = \mbf{0}$ of the system $\mbf{x}_{k+1} = \mbf{A}_\mathrm{d} \mbf{x}_k$ is Lyapunov stable.

The matrix inequality of~\eqref{eq:DTLyapIneq} is satisfied and the eigenvalues of $\mbf{A}_\mathrm{d}$ satisfy $\abs{\lambda_i(\mbf{A}_\mathrm{d})} \leq 1$, $i=1,\ldots,n$ under any of the following equivalent necessary and sufficient conditions.

\begin{enumerate}

\itemcite There exists $\mbf{X} \in \mathbb{S}^n$, where $\mbf{X} > 0$, such that
\bdis
\mbf{A}_\mathrm{d}\mbf{P}\mbf{A}_\mathrm{d}^\trans-\mbf{P} \leq 0.
\edis

\item There exists $\mbf{P} \in \mathbb{S}^n$, where $\mbf{P} > 0$, such that
\bdis
\bbm \mbf{P} & \mbf{A}_\mathrm{d}\mbf{P}  \\ * & \mbf{P} \ebm \geq 0.
\edis

\item There exists $\mbf{P} \in \mathbb{S}^n$, where $\mbf{P} > 0$, such that
\bdis
\bbm \mbf{P} & \mbf{A}_\mathrm{d}^\trans\mbf{P}  \\ * & \mbf{P} \ebm \geq 0.
\edis

\item There exists $\mbf{P} \in \mathbb{S}^n$, where $\mbf{P} > 0$, such that
\bdis
\bbm \mbf{P} & \mbf{P} \mbf{A}_\mathrm{d}^\trans \\ * & \mbf{P} \ebm \geq 0.
\edis

\item There exists $\mbf{P} \in \mathbb{S}^n$, where $\mbf{P} > 0$, such that
\bdis
\bbm \mbf{P} & \mbf{P} \mbf{A}_\mathrm{d} \\ * & \mbf{P} \ebm \geq 0.
\edis

\itemcite \cite{DeOliveira1999} There exist $\mbf{P} \in \mathbb{S}^n$ and $\mbf{G} \in \mathbb{R}^{n \times n}$, where $\mbf{P} >0$, such that
\bdis
\bbm \mbf{P} & \mbf{A}_\mathrm{d}^\trans \mbf{G}^\trans \\ * & \mbf{G} + \mbf{G}^\trans - \mbf{P} \ebm \geq 0.
\edis

\end{enumerate}

\subsubsection[Discrete-Time Asymptotic Stability]{Discrete-Time Asymptotic Stability~\cite[pp.~1203--1204]{BernsteinMatrixBook},~\cite[pp.~97--98]{Duan2013}}
\index{stability!asymptotic stability}
Consider the matrices $\mbf{A}_\mathrm{d} \in \mathbb{R}^{n \times n}$ and $\mbf{Q} \in \mathbb{S}^n$, where $\mbf{Q}  > 0$.  There exists $\mbf{P} \in \mathbb{S}^n$, where $\mbf{P} > 0$, satisfying the discrete-time Lyapunov equation\index{Lyapunov!equation}
\bdis
\mbf{A}_\mathrm{d}^\trans\mbf{P}\mbf{A}_\mathrm{d}-\mbf{P} + \mbf{Q} = \mbf{0}.
\edis
if and only if there exists $\mbf{P} \in \mathbb{S}^n$, where $\mbf{P} > 0$, such that
\beq
\label{eq:DTLyapStrictIneq}
\mbf{A}_\mathrm{d}^\trans\mbf{P}\mbf{A}_\mathrm{d}-\mbf{P} < 0.
\eeq
If~\eqref{eq:DTLyapStrictIneq} holds, then $\abs{\lambda_i(\mbf{A}_\mathrm{d})} < 1$, $i=1,\ldots,n$, the matrix $\mbf{A}_\mathrm{d}$ is Schur\index{Schur matrix}, and the equilibrium point $\mbfbar{x} = \mbf{0}$ of the system $\mbf{x}_{k+1} = \mbf{A}_\mathrm{d} \mbf{x}_k$ is asymptotically stable.

The matrix inequality of~\eqref{eq:DTLyapStrictIneq} is satisfied and the eigenvalues of $\mbf{A}_\mathrm{d}$ satisfy $\abs{\lambda_i(\mbf{A}_\mathrm{d})} < 1$, $i=1,\ldots,n$ under any of the following equivalent necessary and sufficient conditions.

\begin{enumerate}

\itemcite There exists $\mbf{X} \in \mathbb{S}^n$, where $\mbf{X} > 0$, such that
\bdis
\mbf{A}_\mathrm{d}\mbf{P}\mbf{A}_\mathrm{d}^\trans-\mbf{P} < 0.
\edis

\itemcite \cite[p.~97]{Duan2013} There exists $\mbf{P} \in \mathbb{S}^n$, where $\mbf{P} > 0$, such that
\bdis
\bbm \mbf{P} & \mbf{A}_\mathrm{d}\mbf{P}  \\ * & \mbf{P} \ebm > 0.
\edis

\item There exists $\mbf{P} \in \mathbb{S}^n$, where $\mbf{P} > 0$, such that
\bdis
\bbm \mbf{P} & \mbf{A}_\mathrm{d}^\trans\mbf{P}  \\ * & \mbf{P} \ebm > 0.
\edis

\itemcite \cite[p.~97]{Duan2013} There exists $\mbf{P} \in \mathbb{S}^n$, where $\mbf{P} > 0$, such that
\bdis
\bbm \mbf{P} & \mbf{P} \mbf{A}_\mathrm{d}^\trans \\ * & \mbf{P} \ebm > 0.
\edis

\item There exists $\mbf{P} \in \mathbb{S}^n$, where $\mbf{P} > 0$, such that
\bdis
\bbm \mbf{P} & \mbf{P} \mbf{A}_\mathrm{d} \\ * & \mbf{P} \ebm > 0.
\edis

\itemcite \cite{DeOliveira1999} There exist $\mbf{P} \in \mathbb{S}^n$ and $\mbf{G} \in \mathbb{R}^{n \times n}$, where $\mbf{P} > 0$, such that
\bdis
\bbm \mbf{P} & \mbf{A}_\mathrm{d}^\trans \mbf{G}^\trans \\ * & \mbf{G} + \mbf{G}^\trans - \mbf{P} \ebm > 0.
\edis

\item (\textit{The $S$-Variable Approach}~\cite[p.~3]{Ebihara2015},~\cite{DeOliveira1999b}) There exist $\mbf{P} \in \mathbb{S}^{n}$ and $\mbf{F}_1$,~$\mbf{F}_2 \in \mathbb{R}^{n \times n}$, where $\mbf{P} > 0$, such that
\bdis
\bbm \mbf{F}_1\mbf{A}_\mathrm{d} + \mbf{A}_\mathrm{d}^\trans\mbf{F}_1^\trans - \mbf{P} &  -\mbf{F}_1 + \mbf{A}_\mathrm{d}^\trans\mbf{F}_2^\trans \\ * & \mbf{P} -(\mbf{F}_2 + \mbf{F}_2^\trans) \ebm  < 0.
\edis

\item \cite[pp.~46--47]{Felipe2017},~\cite{Spagolla2019} There exist $\mbf{P} \in \mathbb{S}^{n}$ and $\mbf{F}_1$,~$\mbf{F}_2$,~$\mbf{X}_1$,~$\mbf{X}_2$,~$\mbf{X}_3 \in \mathbb{R}^{n \times n}$, where $\mbf{P} > 0$, such that
\bdis
\bbm \mbf{X}_1\mbf{F}_1 + \mbf{F}_1^\trans\mbf{X}_1^\trans - \mbf{P} &  \mbf{X}_1 \mbf{F}_2 + \mbf{F}_1^\trans \mbf{X}_2^\trans & \mbf{A}_\mathrm{d}^\trans -\mbf{X}_1 + \mbf{F}_1^\trans\mbf{X}_3^\trans \\ * & \mbf{P} + \mbf{X}_2 \mbf{F}_2 + \mbf{F}_2^\trans \mbf{X}_2^\trans & - \mbf{1} - \mbf{X}_2 + \mbf{F}_2^\trans \mbf{X}_3^\trans \\ * & * & -(\mbf{X}_3 + \mbf{X}_3^\trans) \ebm  < 0.
\edis

\item \cite[pp.~46--47]{Felipe2017},~\cite{Spagolla2019,Felipe2020} There exist $\mbf{P} \in \mathbb{S}^{n}$ and $\mbf{Y}_1$,~$\mbf{Y}_2$,~$\mbf{Y}_3$,~$\mbf{X}_1$,~$\mbf{X}_2$,~$\mbf{X}_3 \in \mathbb{R}^{n \times n}$, where $\mbf{P} > 0$, such that
\bdis
\bbm \mbf{X}_1\mbf{Y}_1 + \mbf{Y}_1^\trans\mbf{X}_1^\trans - \mbf{P} &  \mbf{X}_1 \mbf{Y}_2 + \mbf{Y}_1^\trans \mbf{X}_2^\trans & \mbf{A}_\mathrm{d}^\trans + \mbf{X}_1\mbf{Y}_3 + \mbf{Y}_1^\trans\mbf{X}_3^\trans \\ * & \mbf{P} + \mbf{X}_2 \mbf{Y}_2 + \mbf{Y}_2^\trans \mbf{X}_2^\trans & - \mbf{1} + \mbf{X}_2\mbf{Y}_3 + \mbf{Y}_2^\trans \mbf{X}_3^\trans \\ * & * & \mbf{X}_3\mbf{Y}_3 + \mbf{Y}_3^\trans \mbf{X}_3^\trans \ebm  < 0.
\edis

\end{enumerate}

\subsubsection[Descriptor System Admissibility]{Descriptor System Admissibility}
\index{descriptor systems!admissibility}
Consider the descriptor system given by $\mbf{E} \dot{\mbf{x}} = \mbf{A} \mbf{x}$, where $\mbf{E}$,~$\mbf{A} \in \mathbb{R}^{n \times n}$.  The descriptor system is admissible under any of the following equivalent necessary and sufficient conditions.

\begin{enumerate}

\itemcite \cite{Masabuchi1997,Wang1998} There exists $\mbf{X} \in \mathbb{R}^{n\times n}$, satisfying $\mbf{E}^\trans \mbf{X} = \mbf{X}^\trans \mbf{E}\geq 0$ and
\bdis
\mbf{A}^\trans\mbf{X} + \mbf{X}^\trans \mbf{A}< 0.
\edis

\itemcite \cite{Chadli2017} There exists $\mbf{X} \in \mathbb{R}^{n\times n}$, satisfying $\mbf{E} \mbf{X} = \mbf{X}^\trans \mbf{E}^\trans \geq 0$ and
\bdis
\mbf{A}\mbf{X} + \mbf{X}^\trans \mbf{A}^\trans< 0.
\edis

\itemcite \cite{Chadli2017} There exist $\mbf{P} \in \mathbb{S}^n$, $\mbf{X} \in \mathbb{R}^{(n-n_e)\times n}$, and $\mbf{Z} \in \mathbb{R}^{n \times (n-n_e)}$, where $n_e = \text{rank}(\mbf{E})$ \index{rank}and $\mbf{P} > 0$, satisfying $\mbf{E}^\trans \mbf{Z} = \mbf{0}$ and
\bdis
\mbf{A}^\trans \left( \mbf{P} \mbf{E} + \mbf{Z} \mbf{X} \right) + \left( \mbf{P} \mbf{E} + \mbf{Z} \mbf{X} \right)^\trans \mbf{A} < 0.
\edis

\itemcite \cite{Chadli2017} There exist $\mbf{P} \in \mathbb{S}^n$, $\mbf{X} \in \mathbb{R}^{(n-n_e)\times n}$, and $\mbf{Z} \in \mathbb{R}^{n \times (n-n_e)}$, where $n_e = \text{rank}(\mbf{E})$ \index{rank}and $\mbf{P} > 0$, satisfying $\mbf{E} \mbf{Z} = \mbf{0}$ and
\bdis
\mbf{A} \left( \mbf{P} \mbf{E}^\trans + \mbf{Z} \mbf{X} \right) + \left( \mbf{P} \mbf{E}^\trans + \mbf{Z} \mbf{X} \right)^\trans \mbf{A}^\trans < 0.
\edis

\itemcite \cite{Marx2003} There exist $\mbf{P} \in \mathbb{S}^n$, $\mbf{S} \in \mathbb{R}^{(n-n_e)\times (n-n_e}$, and $\mbf{U}$,~$\mbf{V} \in \mathbb{R}^{n \times (n-n_e)}$, where $n_e = \text{rank}(\mbf{E})$, $\mathcal{R}(\mbf{U}) = \mathcal{N}(\mbf{E}^\trans)$, $\mathcal{R}(\mbf{V}) = \mathcal{N}(\mbf{E})$, and $\mbf{P} > 0$, satisfying \index{rank}
\bdis
\mbf{A} \left( \mbf{P} \mbf{E}^\trans + \mbf{V} \mbf{S} \mbf{U}^\trans \right) + \left( \mbf{P} \mbf{E}^\trans + \mbf{V} \mbf{S} \mbf{U}^\trans \right)^\trans \mbf{A}^\trans < 0.
\edis

\itemcite \cite{Chadli2017} There exist $\mbf{P} \in \mathbb{S}^n$, $\mbf{F}$,~$\mbf{G} \in \mathbb{R}^{n\times n}$, $\mbf{X} \in \mathbb{R}^{(n-n_e)\times n}$, and $\mbf{Z} \in \mathbb{R}^{n \times (n-n_e)}$, where $n_e = \text{rank}(\mbf{E})$ and $\mbf{P} > 0$, satisfying $\mbf{E}^\trans \mbf{Z} = \mbf{0}$ and\index{rank}
\bdis
\bbm \mbf{A}^\trans \mbf{G}^\trans + \mbf{G} \mbf{A} & \left( \mbf{P} \mbf{E} + \mbf{Z} \mbf{X} \right)^\trans + \mbf{A}^\trans \mbf{F}^\trans - \mbf{G} \\ * & - \left( \mbf{F} + \mbf{F}^\trans \right) \ebm < 0.
\edis

\itemcite \cite{Chadli2017} There exist $\mbf{P} \in \mathbb{S}^n$, $\mbf{F}$,~$\mbf{G} \in \mathbb{R}^{n\times n}$, $\mbf{X} \in \mathbb{R}^{(n-n_e)\times n}$, and $\mbf{Z} \in \mathbb{R}^{n \times (n-n_e)}$, where $n_e = \text{rank}(\mbf{E})$ and $\mbf{P} > 0$, satisfying $\mbf{E} \mbf{Z} = \mbf{0}$ and\index{rank}
\bdis
\bbm \mbf{A} \mbf{G}+ \mbf{G}^\trans \mbf{A}^\trans & \left( \mbf{P} \mbf{E}^\trans + \mbf{Z} \mbf{X} \right)^\trans + \mbf{A} \mbf{F}- \mbf{G}^\trans \\ * & - \left( \mbf{F} + \mbf{F}^\trans \right) \ebm < 0.
\edis

\end{enumerate}

\subsubsection[Discrete-Time Descriptor System Admissibility]{Discrete-Time Descriptor System Admissibility}
\index{descriptor systems!admissibility}
Consider the discrete-time descriptor system given by $\mbf{E}_\mathrm{d} \mbf{x}_{k+1} = \mbf{A}_\mathrm{d} \mbf{x}_k$, where $\mbf{E}_\mathrm{d}$,~$\mbf{A}_\mathrm{d} \in \mathbb{R}^{n \times n}$.  The discrete-time descriptor system is admissible under any of the following equivalent necessary and sufficient conditions.

\begin{enumerate}

\itemcite \cite{Hsiung1999,Xu1999} There exists $\mbf{P} \in \mathbb{S}^n$, satisfying $\mbf{E}_\mathrm{d}^\trans \mbf{P} \mbf{E}_\mathrm{d} \geq 0$ and
\bdis
\mbf{A}_\mathrm{d}^\trans\mbf{P}\mbf{A}_\mathrm{d}-\mbf{E}_\mathrm{d}^\trans \mbf{P} \mbf{E}_\mathrm{d} < 0.
\edis

\itemcite \cite{Zhang2008,Chadli2012} There exist $\mbf{P}  \in \mathbb{S}^n$, $\mbf{X} \in \mathbb{S}^{(n-n_e)}$, and $\mbf{Z} \in \mathbb{R}^{n \times (n-n_e)}$, where $n_e = \text{rank}(\mbf{E}_\mathrm{d})$ and $\mbf{P} > 0$, satisfying $\mbf{E}_\mathrm{d}^\trans \mbf{Z} = \mbf{0}$ and\index{rank}
\bdis
\mbf{A}_\mathrm{d}^\trans \left( \mbf{P} -  \mbf{Z} \mbf{X} \mbf{Z}^\trans \right) \mbf{A}_\mathrm{d} -  \mbf{E}_\mathrm{d}^\trans \mbf{P} \mbf{E}_\mathrm{d}  < 0.
\edis

\itemcite \cite{Chadli2012} There exist $\mbf{P}  \in \mathbb{S}^n$, $\mbf{X} \in \mathbb{S}^{(n-n_e)}$, and $\mbf{Z} \in \mathbb{R}^{n \times (n-n_e)}$, where $n_e = \text{rank}(\mbf{E}_\mathrm{d})$ \index{rank}and $\mbf{P} > 0$, satisfying $\mbf{E}_\mathrm{d} \mbf{Z} = \mbf{0}$ and
\bdis
\mbf{A}_\mathrm{d} \left( \mbf{P} -  \mbf{Z} \mbf{X} \mbf{Z}^\trans \right) \mbf{A}_\mathrm{d}^\trans -  \mbf{E}_\mathrm{d}^\trans \mbf{P} \mbf{E}_\mathrm{d} < 0.
\edis

\itemcite \cite{Marx2003} There exist $\mbf{P} \in \mathbb{S}^n$, $\mbf{S} \in \mathbb{R}^{(n-n_e)\times (n-n_e}$, and $\mbf{U}$,~$\mbf{V} \in \mathbb{R}^{n \times (n-n_e)}$, where $n_e = \text{rank}(\mbf{E}_\mathrm{d})$, $\mathcal{R}(\mbf{U}) = \mathcal{N}(\mbf{E}_\mathrm{d}^\trans)$, $\mathcal{R}(\mbf{V}) = \mathcal{N}(\mbf{E}_\mathrm{d})$, and $\mbf{P} > 0$, satisfying \index{rank}
\bdis
\bbm -\mbf{E}_\mathrm{d} \mbf{P} \mbf{E}_\mathrm{d}^\trans + \mbf{A}_\mathrm{d} \mbf{V} \mbf{S} \mbf{U}^\trans + \mbf{U} \mbf{S}^\trans \mbf{V}^\trans \mbf{A}_\mathrm{d}^\trans & \mbf{A}_\mathrm{d} \mbf{P} \mbf{E}_\mathrm{d}^\trans + \mbf{A}_\mathrm{d} \mbf{V} \mbf{S} \mbf{U}^\trans + \mbf{U} \mbf{S}^\trans \mbf{V}^\trans \mbf{A}_\mathrm{d}^\trans  \\ * & -\mbf{E}_\mathrm{d} \mbf{P} \mbf{E}_\mathrm{d}^\trans + \mbf{A}_\mathrm{d} \mbf{V} \mbf{S} \mbf{U}^\trans + \mbf{U} \mbf{S}^\trans \mbf{V}^\trans \mbf{A}_\mathrm{d}^\trans  \ebm < 0.
\edis

\itemcite \cite{Xu2004} There exist $\mbf{P}  \in \mathbb{S}^n$, $\mbf{X} \in \mathbb{S}^{(n-n_e)}$, and $\mbf{Z} \in \mathbb{R}^{n \times (n-n_e)}$, where $n_e = \text{rank}(\mbf{E}_\mathrm{d})$ \index{rank}and $\mbf{P} > 0$, satisfying $\mbf{E}_\mathrm{d} \mbf{Z} = \mbf{0}$ and
\bdis
\mbf{A}_\mathrm{d}^\trans \mbf{P}  \mbf{A}_\mathrm{d}  -  \mbf{E}_\mathrm{d}^\trans \mbf{P} \mbf{E}_\mathrm{d} + \mbf{X} \mbf{Z} \mbf{A}_\mathrm{d} + \mbf{A}_\mathrm{d}^\trans \mbf{Z}^\trans \mbf{X}^\trans \mbf< 0.
\edis

\itemcite \cite{Masubuchi2013} There exist $\mbf{X} \in \mathbb{S}^n$ and $\alpha \in \mathbb{R}$, satisfying $\mbf{E}_\mathrm{d}^\trans \mbf{X} = \mbf{X}^\trans  \mbf{E}_\mathrm{d} \geq 0$ and
\bdis
\bbm \mbf{X}^\trans \left( \mbf{E}_\mathrm{d} - \mbf{A}_\mathrm{d} \right) + \left( \mbf{E}_\mathrm{d} - \mbf{A}_\mathrm{d} \right)^\trans \mbf{X} & \left( \mbf{E}_\mathrm{d} - \mbf{A}_\mathrm{d} \right)^\trans \mbf{X} \\ * & \mbf{E}_\mathrm{d}^\trans \mbf{X} + \alpha \left( \mbf{1} - \mbf{E}_\mathrm{d}^\dagger \mbf{E}_\mathrm{d}\right) \ebm> 0,
\edis
where $\mbf{E}_\mathrm{d}^\dagger$ is the pseudoinverse\index{pseudoinverse} of $\mbf{E}_\mathrm{d}$.

\itemcite \cite{Chadli2012} There exist $\mbf{P}  \in \mathbb{S}^n$, $\mbf{X} \in \mathbb{S}^{(n-n_e)}$, $\mbf{F}$,~$\mbf{G} \in \mathbb{R}^{n\times n}$, and $\mbf{Z} \in \mathbb{R}^{n \times (n-n_e)}$, where $n_e = \text{rank}(\mbf{E}_\mathrm{d})$ \index{rank}and $\mbf{P} > 0$, satisfying $\mbf{E}_\mathrm{d}^\trans \mbf{Z} = \mbf{0}$ and
\bdis
\bbm -\mbf{E}_\mathrm{d}^\trans \mbf{P} \mbf{E}_\mathrm{d} + \mbf{A}_\mathrm{d}^\trans \mbf{G}^\trans + \mbf{G} \mbf{A}_\mathrm{d} & - \mbf{G} + \mbf{A}_\mathrm{d}^\trans \mbf{F}^\trans \\ * & \mbf{P} - \mbf{Z} \mbf{X} \mbf{Z}^\trans - \left( \mbf{F} + \mbf{F}^\trans \right) \ebm < 0.
\edis

\itemcite \cite{Chadli2012} There exist $\mbf{P}  \in \mathbb{S}^n$, $\mbf{X} \in \mathbb{S}^{(n-n_e)}$, $\mbf{F}$,~$\mbf{G} \in \mathbb{R}^{n\times n}$, and $\mbf{Z} \in \mathbb{R}^{n \times (n-n_e)}$, where $n_e = \text{rank}(\mbf{E}_\mathrm{d})$ \index{rank}and $\mbf{P} > 0$, satisfying $\mbf{E}_\mathrm{d} \mbf{Z} = \mbf{0}$ and
\bdis
\bbm -\mbf{E}_\mathrm{d} \mbf{P} \mbf{E}_\mathrm{d}^\trans + \mbf{A}_\mathrm{d}\mbf{G}^\trans + \mbf{G} \mbf{A}_\mathrm{d}^\trans & - \mbf{G} + \mbf{A}_\mathrm{d} \mbf{F}^\trans \\ * & \mbf{P} - \mbf{Z}\mbf{X} \mbf{Z}^\trans - \left( \mbf{F} + \mbf{F}^\trans \right) \ebm < 0.
\edis

\end{enumerate}

\subsection{Bounded Real Lemma and the \texorpdfstring{$\mathcal{H}_\infty$}{H-Infinity} Norm}
\index{bounded real lemma!continuous-time}
\index{norm!$\mathcal{H}_\infty$ norm}
\index{norm!induced $\mathcal{L}_2$-$\mathcal{L}_2$ norm}
\subsubsection[Continuous-Time Bounded Real Lemma]{Continuous-Time Bounded Real Lemma~\cite{Gahinet1994},~\cite[pp.~85--86]{Scherer1990}}
\label{sec:BoundedRealLemma}

Consider a continuous-time LTI system, $\mbc{G}: \mathcal{L}_{2e} \to \mathcal{L}_{2e}$, with state-space realization $(\mbf{A},\mbf{B}_,\mbf{C},\mbf{D})$, where $\mbf{A} \in \mathbb{R}^{n \times n}$, $\mbf{B} \in \mathbb{R}^{n \times m}$, $\mbf{C} \in \mathbb{R}^{p \times n}$, and $\mbf{D} \in \mathbb{R}^{p \times m}$.  The $\mathcal{H}_\infty$ norm of $\mbc{G}$ is
\bdis
\norm{\mbc{G}}_\infty = \sup_{\mbf{u} \in \mathcal{L}_2, \mbf{u} \neq \mbf{0}} \frac{\norm{\mbc{G} \mbf{u}}_2}{\norm{\mbf{u}}_2}.
\edis
The inequality $\norm{\mbc{G}}_\infty  < \gamma$ holds under any of the following equivalent necessary and sufficient conditions.

\begin{enumerate}

\item There exist $\mbf{P} \in \mathbb{S}^n $ and $\gamma \in \mathbb{R}_{>0}$, where $\mbf{P} > 0$, such that
\beq
\label{eq:BRL1}
\bbm \mbf{P}\mbf{A} + \mbf{A}^\trans\mbf{P} & \mbf{P}\mbf{B}& \mbf{C}^\trans \\ * & -\gamma \mbf{1} & \mbf{D}^\trans  \\ * & * & -\gamma\mbf{1} \ebm < 0.
\eeq

\item There exist $\mbf{Q} \in \mathbb{S}^n $ and $\gamma \in \mathbb{R}_{>0}$, where $\mbf{Q} > 0$, such that
\beq
\label{eq:BRL2}
\bbm \mbf{A}\mbf{Q} + \mbf{Q}\mbf{A}^\trans & \mbf{B}& \mbf{Q}\mbf{C}^\trans \\ * & -\gamma \mbf{1} & \mbf{D}^\trans  \\ * & * & -\gamma\mbf{1} \ebm < 0.
\eeq

\item There exist $\mbf{P} \in \mathbb{S}^n $ and $\gamma \in \mathbb{R}_{>0}$, where $\mbf{P} > 0$, such that
\bdis
\bbm \mbf{P}\mbf{A} + \mbf{A}^\trans\mbf{P} + \mbf{C}^\trans\mbf{C}& \mbf{P}\mbf{B} + \mbf{C}^\trans\mbf{D}  \\ * & -\gamma^2 \mbf{1} + \mbf{D}^\trans\mbf{D} \ebm < 0.
\edis

\item There exist $\mbf{Q} \in \mathbb{S}^n $ and $\gamma \in \mathbb{R}_{>0}$, where $\mbf{Q} > 0$, such that
\bdis
\bbm \mbf{A}\mbf{Q} + \mbf{Q}\mbf{A}^\trans+ \mbf{B}\mbf{B}^\trans & \mbf{Q}\mbf{C}^\trans + \mbf{B}\mbf{D}^\trans  \\ * & -\gamma^2 \mbf{1} + \mbf{D}\mbf{D}^\trans \ebm < 0.
\edis

\itemcite \cite{Xie2008} There exist $\mbf{P} \in \mathbb{S}^n $, $\mbf{V} \in \mathbb{R}^{n \times n}$, and $r$,~$\gamma \in \mathbb{R}_{>0}$, where $\mbf{P} > 0$, such that
\bdis
\bbm \mbf{A}\mbf{V} + \mbf{V}^\trans \mbf{A}^\trans & \mbf{P} - \mbf{V}^\trans + r \mbf{A} \mbf{V} & \mbf{V}^\trans \mbf{C}^\trans & \mbf{B} \\ * & -r \left( \mbf{V} + \mbf{V}^\trans \right) & r \mbf{V}^\trans \mbf{C}^\trans & \mbf{0} \\ * & * & - \mbf{1} & \mbf{D} \\ * & * & * & - \gamma^2 \mbf{1} \ebm < 0.
\edis

\itemcite \cite{Krokavec2011},~\cite[pp.~46--47]{Lemaire2019} There exist $\mbf{P} \in \mathbb{S}^{n}$, $\mbf{F}_1$,~$\mbf{F}_2 \in \mathbb{R}^{n \times n}$, and $\gamma \in \mathbb{R}_{>0}$, where $\mbf{P} > 0$, such that
\bdis
\bbm \mbf{F}_1\mbf{A} + \mbf{A}^\trans\mbf{F}_1^\trans & \mbf{P} -\mbf{F}_1 + \mbf{A}^\trans\mbf{F}_2^\trans & \mbf{F}_1 \mbf{B} & \mbf{C}^\trans \\ * & -(\mbf{F}_2 + \mbf{F}_2^\trans) & \mbf{F}_2 \mbf{B}  & \mbf{0} \\ * & * & -\gamma \mbf{1} & \mbf{D}^\trans \\ * & * & * & -\gamma \mbf{1} \ebm  < 0.
\edis

\itemcite \cite{Krokavec2011} There exist $\mbf{P} \in \mathbb{S}^{n}$, $\mbf{F}_1$,~$\mbf{F}_2 \in \mathbb{R}^{n \times n}$, and $\gamma \in \mathbb{R}_{>0}$, where $\mbf{P} > 0$, such that
\bdis
\bbm \mbf{P}\mbf{A} + \mbf{A}^\trans\mbf{P} &  \mbf{P} +\mbf{F}_1 + \mbf{A}^\trans\mbf{F}_2  & \mbf{P} \mbf{B} & \mbf{C}^\trans \\ * & \mbf{F}_2 + \mbf{F}_2^\trans & \mbf{F}_2^\trans \mbf{B}  & \mbf{0} \\ * & * & -\gamma^2 \mbf{1} & \mbf{D}^\trans \\ * & * & * & - \mbf{1} \ebm  < 0.
\edis

\itemcite \cite[pp.~46--47]{Lemaire2019} There exist $\mbf{P} \in \mathbb{S}^{n}$, $\mbf{F}_1$,~$\mbf{F}_2$,~$\mbf{X}_1$,~$\mbf{X}_2$,~$\mbf{X}_3 \in \mathbb{R}^{n \times n}$, and $\gamma \in \mathbb{R}_{>0}$, where $\mbf{P} > 0$, such that
\bdis
\bbm \mbf{X}_1\mbf{F}_1 + \mbf{F}_1^\trans\mbf{X}_1^\trans & \mbf{P} + \mbf{X}_1 \mbf{F}_2 + \mbf{F}_1^\trans \mbf{X}_2^\trans & \mbf{A}^\trans - \mbf{X}_1 + \mbf{F}_1^\trans\mbf{X}_3^\trans & \mbf{0} & \mbf{C}^\trans \\ * & \mbf{X}_2 \mbf{F}_2 + \mbf{F}_2^\trans \mbf{X}_2^\trans & - \mbf{1} - \mbf{X}_2 + \mbf{F}_2^\trans \mbf{X}_3^\trans &  \mbf{0} & \mbf{0} \\ * & * & -\left(\mbf{X}_3 + \mbf{X}_3^\trans \right) & \mbf{B} & \mbf{0} \\ * & * & * & -\gamma \mbf{1} & \mbf{D}^\trans \\ * & * & * & * & -\gamma \mbf{1}  \ebm  < 0.
\edis

\itemcite \cite[pp.~46--47]{Lemaire2019} There exist $\mbf{P} \in \mathbb{S}^{n}$, $\mbf{Y}_1$,~$\mbf{Y}_2$,~$\mbf{Y}_3$,~$\mbf{X}_1$,~$\mbf{X}_2$,~$\mbf{X}_3 \in \mathbb{R}^{n \times n}$, and $\gamma \in \mathbb{R}_{>0}$, where $\mbf{P} > 0$, such that
\bdis
\bbm \mbf{X}_1\mbf{Y}_1 + \mbf{Y}_1^\trans\mbf{X}_1^\trans & \mbf{P} + \mbf{X}_1 \mbf{Y}_2 + \mbf{Y}_1^\trans \mbf{X}_2^\trans & \mbf{A}^\trans + \mbf{X}_1\mbf{Y}_3 + \mbf{Y}_1^\trans\mbf{X}_3^\trans & \mbf{0} & \mbf{C}^\trans \\ * & \mbf{X}_2 \mbf{Y}_2 + \mbf{Y}_2^\trans \mbf{X}_2^\trans & - \mbf{1} + \mbf{X}_2\mbf{Y}_3 + \mbf{Y}_2^\trans \mbf{X}_3^\trans &  \mbf{0} & \mbf{0} \\ * & * & \mbf{X}_3\mbf{Y}_3 + \mbf{Y}_3^\trans\mbf{X}_3^\trans & \mbf{B} & \mbf{0} \\ * & * & * & -\gamma \mbf{1} & \mbf{D}^\trans \\ * & * & * & * & -\gamma \mbf{1}  \ebm  < 0.
\edis

\item There exist $\mbf{Q} \in \mathbb{S}^n$, $\mbf{V}_{11} \in \mathbb{R}^{n \times n}$, $\mbf{V}_{12} \in \mathbb{R}^{n \times m}$, $\mbf{V}_{21} \in \mathbb{R}^{m \times n}$, $\mbf{V}_{22} \in \mathbb{R}^{m \times m}$, and $ \gamma \in \mathbb{R}_{>0}$, where $\mbf{Q} > 0$, such that
\bdis
\bbm -(\mbf{V}_{11}+\mbf{V}_{11}^\trans) & \mbf{V}_{11}^\trans \mbf{A}^\trans + \mbf{V}_{21}^\trans\mbf{B}^\trans + \mbf{Q} & \mbf{V}_{11}^\trans\mbf{C}^\trans+\mbf{V}_{21}^\trans\mbf{D}^\trans & \mbf{V}_{11}^\trans & -\mbf{V}_{12}-\mbf{V}_{21}^\trans \\ * & -\mbf{Q} & \mbf{0} & \mbf{0} & \mbf{A}\mbf{V}_{12}+\mbf{B}\mbf{V}_{22} \\ * & * & -\gamma^2 \mbf{1} & \mbf{0} & \mbf{C}\mbf{V}_{12} + \mbf{D}\mbf{V}_{22} \\ * & * & * & -\mbf{Q} & \mbf{V}_{12} \\ * & * & * & * & -\mbf{1} - (\mbf{V}_{22}+\mbf{V}_{22}^\trans) \ebm < 0.
\edis
\begin{proof}
Identical to the proof of~\eqref{eq:HinfDil_Duan} in~\cite[p.~156]{Duan2013}, except with $\mbs{\Omega} = \bbm \mbf{V}_{11} & \mbf{V}_{12} \\ \mbf{V}_{21} & \mbf{V}_{22} \ebm$.
\end{proof}

\item There exist $\mbf{P} \in \mathbb{S}^n$, $\mbf{W}_{11} \in \mathbb{R}^{n \times n}$, $\mbf{W}_{12} \in \mathbb{R}^{n \times p}$, $\mbf{V}_{21} \in \mathbb{R}^{p \times n}$, $\mbf{V}_{22} \in \mathbb{R}^{p \times p}$, and $ \gamma \in \mathbb{R}_{>0}$, where $\mbf{P} > 0$, such that
\bdis
\bbm -(\mbf{W}_{11}+\mbf{W}_{11}^\trans) & \mbf{W}_{11}^\trans \mbf{A} + \mbf{W}_{21}^\trans\mbf{C} + \mbf{P} & \mbf{W}_{11}^\trans\mbf{B}+\mbf{W}_{21}^\trans\mbf{D} & \mbf{W}_{11}^\trans & -(\mbf{W}_{12}+\mbf{W}_{21}^\trans) \\ * & -\mbf{P} & \mbf{0} & \mbf{0} & \mbf{A}^\trans\mbf{W}_{12}+\mbf{C}^\trans\mbf{W}_{22} \\ * & * & -\gamma^2 \mbf{1} & \mbf{0} & \mbf{B}^\trans\mbf{W}_{12} + \mbf{D}^\trans\mbf{W}_{22} \\ * & * & * & -\mbf{P} & \mbf{W}_{12} \\ * & * & * & * &  - (\mbf{1}+\mbf{W}_{22}+\mbf{W}_{22}^\trans) \ebm < 0.
\edis
\begin{proof}
Identical to the proof of~\eqref{eq:HinfDil2}, except with $\mbs{\Omega} = \bbm \mbf{W}_{11} & \mbf{W}_{12} \\ \mbf{W}_{21} & \mbf{W}_{22} \ebm$.
\end{proof}

%

\end{enumerate}

The $\mathcal{H}_\infty$ norm of $\mbc{G}$ is the minimum value of $\gamma \in \mathbb{R}_{>0}$ that satisfies any of the above conditions.  If $(\mbf{A},\mbf{B}_,\mbf{C},\mbf{D})$ is a minimal realization, then the matrix inequalities can be nonstrict~\cite[pp.~26--27]{Boyd1994},~\cite[pp.~308--311]{Anderson1973},~\cite{Rantzer1996}.

The inequality $\norm{\mbc{G}}_\infty < \gamma$ also holds under any of the following equivalent sufficient conditions.

\begin{enumerate}

\itemcite \cite[p.~156]{Duan2013} There exist $\mbf{Q} \in \mathbb{S}^n$, $\mbf{V} \in \mathbb{R}^{n \times n}$, and $\gamma \in \mathbb{R}_{>0}$, where $\mbf{Q} > 0$, such that
\beq
\label{eq:HinfDil_Duan}
\bbm -(\mbf{V}+\mbf{V}^\trans) & \mbf{V}^\trans \mbf{A}^\trans + \mbf{Q} & \mbf{V}^\trans\mbf{C}^\trans & \mbf{V}^\trans & \mbf{0} \\ * & -\mbf{Q} & \mbf{0} & \mbf{0} & \mbf{B} \\ * & * & -\gamma \mbf{1} & \mbf{0} & \mbf{D} \\ * & * & * & -\mbf{Q} & \mbf{0} \\ * & * & * & * & -\gamma \mbf{1} \ebm < 0.
\eeq

\item There exist $\mbf{P} \in \mathbb{S}^n$, $\mbf{W} \in \mathbb{R}^{n \times n}$, and $\gamma \in \mathbb{R}_{>0}$, where $\mbf{P} > 0$, such that
\beq
\label{eq:HinfDil2}
\bbm -(\mbf{W}+\mbf{W}^\trans) & \mbf{W}^\trans \mbf{A} + \mbf{P} & \mbf{W}^\trans\mbf{B} & \mbf{W}^\trans & \mbf{0} \\ * & -\mbf{P} & \mbf{0} & \mbf{0} & \mbf{C}^\trans \\ * & * & -\gamma \mbf{1} & \mbf{0} & \mbf{D}^\trans \\ * & * & * & -\mbf{P} & \mbf{0} \\ * & * & * & * & -\gamma \mbf{1} \ebm < 0.
\eeq
\begin{proof}
Identical to the proof of ~\eqref{eq:HinfDil_Duan} in~\cite[p.~156]{Duan2013}, except starting with the Bounded Real Lemma in the form
\bdis
\bbm \mbf{A} \mbf{Q} + \mbf{Q} \mbf{A}^\trans + \frac{1}{\gamma} \mbf{Q} \mbf{C}^\trans \mbf{C} \mbf{Q} & \mbf{B} + \frac{1}{\gamma} \mbf{Q} \mbf{C}^\trans \mbf{D} \\ * & -\gamma \mbf{1} + \frac{1}{\gamma}\mbf{D}^\trans\mbf{D} \ebm,
\edis
which requires $\mbs{\Phi} = \bbm -\mbf{1} & \mbf{A} & \mbf{B} & \mbf{1} & \mbf{0} \\ \mbf{0} & \mbf{C} & \mbf{D} & \mbf{0} & -\gamma \mbf{1} \ebm$.
\end{proof}

\end{enumerate}

When $\mbf{D} = \mbf{0}$, then the inequality $\norm{\mbc{G}}_\infty > \gamma$ holds if and only if there exist $\mbf{Z}_{11} \in \mathbb{S}^n$, $\mbf{Z}_{12} \in \mathbb{R}^{n \times m}$, and $\mbf{Z}_{22} \in \mathbb{S}^m$ such that~\cite{Balakrishnan2003}
\begin{align*}
\mbf{Z}_{11} \mbf{A}^\trans + \mbf{A} \mbf{Z}_{11} + \mbf{Z}_{12} \mbf{B}^\trans + \mbf{B} \mbf{Z}_{12}^\trans = \mbf{0}, \\
\bbm \mbf{Z}_{11} & \mbf{Z}_{12} \\ * & \mbf{Z}_{22} \ebm &\geq 0, \\
\trace \left( \mbf{Z}_{22} \right) &= 1, \\
\trace \left( \mbf{C} \mbf{Z}_{11} \mbf{C}^\trans \right) &> \gamma.
\end{align*}

\subsubsection[Discrete-Time Bounded Real Lemma]{Discrete-Time Bounded Real Lemma}
\label{sec:DTBoundedRealLemma}
\index{bounded real lemma!discrete-time}
\index{norm!discrete-time $\mathcal{H}_\infty$ norm}
Consider a discrete-time LTI system, $\mbc{G}: \ell_{2e} \to \ell_{2e}$, with state-space realization $(\mbf{A}_\mathrm{d},\mbf{B}_\mathrm{d},\mbf{C}_\mathrm{d},\mbf{D}_\mathrm{d})$, where $\mbf{A}_\mathrm{d} \in \mathbb{R}^{n \times n}$, $\mbf{B}_\mathrm{d} \in \mathbb{R}^{n \times m}$, $\mbf{C}_\mathrm{d} \in \mathbb{R}^{p \times n}$, and $\mbf{D}_\mathrm{d} \in \mathbb{R}^{p \times m}$.  The $\mathcal{H}_\infty$ norm of $\mbc{G}$ is
\bdis
\norm{\mbc{G}}_\infty = \sup_{\mbf{u} \in \ell_2, \mbf{u} \neq \mbf{0}} \frac{\norm{\mbc{G} \mbf{u}}_2}{\norm{\mbf{u}}_2}.
\edis
The inequality $\norm{\mbc{G}}_\infty  < \gamma$ holds under any of the following equivalent necessary and sufficient conditions.

\begin{enumerate}

\itemcite \cite{Gahinet1994} There exist $\mbf{P} \in \mathbb{S}^n $ and $\gamma \in \mathbb{R}_{>0}$, where $\mbf{P} > 0$, such that
\bdis
\bbm \mbf{A}_\mathrm{d}^\trans\mbf{P}\mbf{A}_\mathrm{d} -\mbf{P} & \mbf{A}_\mathrm{d}^\trans\mbf{P}\mbf{B}_\mathrm{d}& \mbf{C}_\mathrm{d}^\trans \\ * & \mbf{B}_\mathrm{d}^\trans\mbf{P}\mbf{B}_\mathrm{d}-\gamma \mbf{1} & \mbf{D}_\mathrm{d}^\trans  \\ * & * & -\gamma\mbf{1} \ebm < 0.
\edis

\itemcite \cite{Xie1996b} There exist $\mbf{Q} \in \mathbb{S}^n$ and $\gamma \in \mathbb{R}_{>0}$, where $\mbf{Q} >0 $, such that
\bdis
\bbm \mbf{A}_\mathrm{d}\mbf{Q}\mbf{A}_\mathrm{d}^\trans -\mbf{Q} & \mbf{B}_\mathrm{d}& \mbf{A}_\mathrm{d}\mbf{Q}\mbf{C}_\mathrm{d}^\trans \\ * &-\gamma \mbf{1} & \mbf{D}_\mathrm{d}^\trans  \\ * & * &  \mbf{C}_\mathrm{d}\mbf{Q}\mbf{C}_\mathrm{d}^\trans -\gamma\mbf{1} \ebm < 0.
\edis

\itemcite \cite{DeOliveira2002} There exist $\mbf{P} \in \mathbb{S}^n$ and $\gamma \in \mathbb{R}_{>0}$, where $\mbf{P} >0 $, such that
\bdis
\bbm \mbf{P} & \mbf{A}_\mathrm{d}\mbf{P} & \mbf{B}_\mathrm{d} & \mbf{0} \\ * & \mbf{P} & \mbf{0} & \mbf{P}\mbf{C}_\mathrm{d}^\trans \\ * & * & \gamma \mbf{1} & \mbf{D}_\mathrm{d}^\trans \\ * & * & * & \gamma \mbf{1} \ebm > 0.
\edis

\itemcite \cite{Masubuchi1995,Masubuchi1998} There exist $\mbf{Q} \in \mathbb{S}^n$ and $\gamma \in \mathbb{R}_{>0}$, where $\mbf{Q} >0 $, such that
\bdis
\bbm \mbf{Q} & \mbf{Q}\mbf{A}_\mathrm{d} & \mbf{Q}\mbf{B}_\mathrm{d} & \mbf{0} \\ * & \mbf{Q} & \mbf{0} & \mbf{C}_\mathrm{d}^\trans \\ * & * & \gamma \mbf{1} & \mbf{D}_\mathrm{d}^\trans \\ * & * & * & \gamma \mbf{1} \ebm > 0.
\edis

\itemcite \cite{Gahinet1994} There exist $\mbf{P} \in \mathbb{S}^n$ and $\gamma \in \mathbb{R}_{>0}$, where $\mbf{P} >0 $, such that
\beq
\label{eq:DT_BRL_4}
\bbm \mbf{P}^{-1} & \mbf{A}_\mathrm{d} & \mbf{B}_\mathrm{d} & \mbf{0} \\ * & \mbf{P} & \mbf{0} & \mbf{C}_\mathrm{d}^\trans \\ * & * & \gamma \mbf{1} & \mbf{D}_\mathrm{d}^\trans \\ * & * & * & \gamma \mbf{1} \ebm > 0.
\eeq

\itemcite \cite{DeOliveira2002} There exist $\mbf{P} \in \mathbb{S}^n$, $\mbf{X} \in \mathbb{R}^{n \times n}$, and $\gamma \in \mathbb{R}_{>0}$, where $\mbf{P} >0 $ and $\mbf{X}$ has full rank, such that\index{rank}
\bdis
\bbm \mbf{P} & \mbf{A}_\mathrm{d}\mbf{X} & \mbf{B}_\mathrm{d} & \mbf{0} \\ * & \mbf{X}^\trans\mbf{P}^{-1}\mbf{X} & \mbf{0} & \mbf{X}\mbf{C}_\mathrm{d}^\trans \\ * & * & \mbf{1} & \mbf{D}_\mathrm{d}^\trans \\ * & * & * & \gamma^2 \mbf{1} \ebm > 0.
\edis

\item  There exist $\mbf{P} \in \mathbb{S}^n$, $\mbf{X} \in \mathbb{R}^{n \times n}$, and $\gamma \in \mathbb{R}_{>0}$, where $\mbf{P} >0 $ and $\mbf{X}$ has full rank, \index{rank}such that
\beq
\label{eq:DT_BRL_6}
\bbm \mbf{X}^\trans\mbf{P}^{-1}\mbf{X} & \mbf{X}\mbf{A}_\mathrm{d}& \mbf{X}\mbf{B}_\mathrm{d} & \mbf{0} \\ * & \mbf{P}& \mbf{0} & \mbf{C}_\mathrm{d}^\trans \\ * & * & \mbf{1} & \mbf{D}_\mathrm{d}^\trans \\ * & * & * & \gamma^2 \mbf{1} \ebm > 0.
\eeq

\begin{proof}
Apply the congruence transformation $\mbf{W} = \mathrm{diag}\{\mbf{X}^\trans,\mbf{1},\mbf{1},\mbf{1}\}$ to~\eqref{eq:DT_BRL_4}, where $\mbf{W}$ has full rank \index{rank}since $\mbf{X}$ has full rank.\index{rank}
\end{proof}

\itemcite \cite{DeOliveira2002,deSouza2006} There exist $\mbf{P} \in \mathbb{S}^n$, $\mbf{X} \in \mathbb{R}^{n \times n}$, and $\gamma \in \mathbb{R}_{>0}$, where $\mbf{P} >0 $, such that
\beq
\label{eq:DT_BRL_7}
\bbm \mbf{P} & \mbf{A}_\mathrm{d}\mbf{X} & \mbf{B}_\mathrm{d} & \mbf{0} \\ * & \mbf{X} + \mbf{X}^\trans - \mbf{P} & \mbf{0} & \mbf{X}\mbf{C}_\mathrm{d}^\trans \\ * & * & \mbf{1} & \mbf{D}_\mathrm{d}^\trans \\ * & * & * & \gamma^2 \mbf{1} \ebm > 0.
\eeq

\item  There exist $\mbf{Q} \in \mathbb{S}^n$, $\mbf{X} \in \mathbb{R}^{n \times n}$, and $\gamma \in \mathbb{R}_{>0}$, where $\mbf{Q} >0 $, such that
\beq
\label{eq:DT_BRL_8}
\bbm \ \mbf{X} + \mbf{X}^\trans - \mbf{Q} & \mbf{X} \mbf{A}_\mathrm{d}& \mbf{X} \mbf{B}_\mathrm{d} & \mbf{0} \\ * &  \mbf{Q} & \mbf{0} &\mbf{C}_\mathrm{d}^\trans \\ * & * & \mbf{1} & \mbf{D}_\mathrm{d}^\trans \\ * & * & * & \gamma^2 \mbf{1} \ebm > 0.
\eeq
\begin{proof}
Same as the proof of~\eqref{eq:DT_BRL_7} in~\cite{DeOliveira2002}, by which it is shown that~\eqref{eq:DT_BRL_8} is equivalent to~\eqref{eq:DT_BRL_6}.
\end{proof}

\itemcite \cite[pp.~48--49]{Spagolla2019b} There exist $\mbf{P} \in \mathbb{S}^{n}$, $\mbf{F}_1$,~$\mbf{F}_2 \in \mathbb{R}^{n \times n}$, and $\gamma \in \mathbb{R}_{>0}$, where $\mbf{P} > 0$, such that
\bdis
\bbm - \mbf{P} + \mbf{A}_\mathrm{d}\mbf{F}_1 + \mbf{F}_1^\trans\mbf{A}_\mathrm{d}^\trans &  \mbf{A}_\mathrm{d} \mbf{F}_2 - \mbf{F}_1^\trans&  \mbf{F}_1^\trans\mbf{C}_\mathrm{d}^\trans  & \mbf{B}_\mathrm{d} \\ * & \mbf{P} - \left(\mbf{F}_2 + \mbf{F}_2^\trans \right) &  \mbf{F}_2^\trans\mbf{C}_\mathrm{d}^\trans  &  \mbf{0} \\ * & * & -\gamma \mbf{1} & \mbf{D}_\mathrm{d} \\  * & * & * & -\gamma \mbf{1}  \ebm  < 0.
\edis

\itemcite \cite[pp.~48--49]{Spagolla2019b} There exist $\mbf{P} \in \mathbb{S}^{n}$, $\mbf{F}_1$,~$\mbf{F}_2$,~$\mbf{X}_1$,~$\mbf{X}_2$,~$\mbf{X}_3 \in \mathbb{R}^{n \times n}$, and $\gamma \in \mathbb{R}_{>0}$, where $\mbf{P} > 0$, such that
\bdis
\bbm - \mbf{P} + \mbf{X}_1\mbf{F}_1 + \mbf{F}_1^\trans\mbf{X}_1^\trans &  \mbf{X}_1 \mbf{F}_2 + \mbf{F}_1^\trans \mbf{X}_2^\trans & \mbf{A}_\mathrm{d} - \mbf{X}_1 + \mbf{F}_1^\trans\mbf{X}_3^\trans & \mbf{B}_\mathrm{d} & \mbf{0} \\ * & \mbf{P} + \mbf{X}_2 \mbf{F}_2 + \mbf{F}_2^\trans \mbf{X}_2^\trans & - \mbf{1} - \mbf{X}_2 + \mbf{F}_2^\trans \mbf{X}_3^\trans &  \mbf{0} & \mbf{0} \\ * & * & -\left( \mbf{X}_3 + \mbf{X}_3^\trans\right) & \mbf{0} & \mbf{C}_\mathrm{d}^\trans \\ * & * & * & -\gamma \mbf{1} & \mbf{D}_\mathrm{d}^\trans \\ * & * & * & * & -\gamma \mbf{1}  \ebm  < 0.
\edis

\itemcite \cite[pp.~48--49]{Spagolla2019b} There exist $\mbf{P} \in \mathbb{S}^{n}$, $\mbf{Y}_1$,~$\mbf{Y}_2$,~$\mbf{Y}_3$,~$\mbf{X}_1$,~$\mbf{X}_2$,~$\mbf{X}_3 \in \mathbb{R}^{n \times n}$, and $\gamma \in \mathbb{R}_{>0}$, where $\mbf{P} > 0$, such that
\bdis
\bbm - \mbf{P} + \mbf{X}_1\mbf{Y}_1 + \mbf{Y}_1^\trans\mbf{X}_1^\trans &  \mbf{X}_1 \mbf{Y}_2 + \mbf{Y}_1^\trans \mbf{X}_2^\trans & \mbf{A}_\mathrm{d} + \mbf{X}_1\mbf{Y}_3 + \mbf{Y}_1^\trans\mbf{X}_3^\trans & \mbf{B}_\mathrm{d} & \mbf{0} \\ * & \mbf{P} + \mbf{X}_2 \mbf{Y}_2 + \mbf{Y}_2^\trans \mbf{X}_2^\trans & - \mbf{1} + \mbf{X}_2\mbf{Y}_3 + \mbf{Y}_2^\trans \mbf{X}_3^\trans &  \mbf{0} & \mbf{0} \\ * & * & \mbf{X}_3\mbf{Y}_3 + \mbf{Y}_3^\trans\mbf{X}_3^\trans & \mbf{0} & \mbf{C}_\mathrm{d}^\trans \\ * & * & * & -\gamma \mbf{1} & \mbf{D}_\mathrm{d}^\trans \\ * & * & * & * & -\gamma \mbf{1}  \ebm  < 0.
\edis

\end{enumerate}

The $\mathcal{H}_\infty$ norm of $\mbc{G}$ is the minimum value of $\gamma \in \mathbb{R}_{>0}$ that satisfies any of the above conditions.  If $(\mbf{A}_\mathrm{d},\mbf{B}_\mathrm{d},\mbf{C}_\mathrm{d},\mbf{D}_\mathrm{d})$ is a minimal realization, then the matrix inequalities can be nonstrict~\cite{Rantzer1996},~\cite{Vaidyanathan1985}.

\subsubsection[Descriptor System Bounded Real Lemma]{Descriptor System Bounded Real Lemma}
\index{descriptor systems!bounded real lemma}

Consider a descriptor system, $\mbc{G}: \mathcal{L}_{2e} \to \mathcal{L}_{2e}$, described by
\begin{align*}
\mbf{E} \dot{\mbf{x}} &= \mbf{A} \mbf{x} + \mbf{B} \mbf{u},\\
\mbf{y} &= \mbf{C} \mbf{x},
\end{align*}
where $\mbf{E}$,~$\mbf{A} \in \mathbb{R}^{n \times n}$, $\mbf{B} \in \mathbb{R}^{n \times m}$, $\mbf{C} \in \mathbb{R}^{p \times n}$, and it is assumed that the system is regular.  The $\mathcal{H}_\infty$ norm of $\mbc{G}$ is
\bdis
\norm{\mbc{G}}_\infty = \sup_{\mbf{u} \in \mathcal{L}_2, \mbf{u} \neq \mbf{0}} \frac{\norm{\mbc{G} \mbf{u}}_2}{\norm{\mbf{u}}_2}.
\edis
The descriptor system is admissible and the inequality $\norm{\mbc{G}}_\infty  < \gamma$ holds under any of the following equivalent necessary and sufficient conditions.

\begin{enumerate}

\itemcite \cite{Masabuchi1997} There exist $\mbf{X} \in \mathbb{R}^{n\times n} $ and $\gamma \in \mathbb{R}_{>0}$, such that $\mbf{E}^\trans \mbf{X} = \mbf{X}^\trans \mbf{E} \geq 0$ and
\bdis
\bbm \mbf{X}^\trans\mbf{A} + \mbf{A}^\trans\mbf{X} + \mbf{C}^\trans \mbf{C} & \mbf{X}^\trans \mbf{B}\\ * & -\gamma^2 \mbf{1}   \ebm < 0.
\edis

\itemcite \cite{Masabuchi1997,Uezato1999} There exist $\mbf{Y} \in \mathbb{R}^{n\times n} $ and $\gamma \in \mathbb{R}_{>0}$, such that $\mbf{Y}\mbf{E}^\trans  =  \mbf{E}\mbf{Y}^\trans \geq 0$ and
\bdis
\bbm \mbf{A}\mbf{Y}^\trans + \mbf{Y}\mbf{A}^\trans + \mbf{B}\mbf{B}^\trans  & \mbf{Y} \mbf{C}^\trans  \\ * & -\gamma \mbf{1} \ebm < 0.
\edis

\itemcite \cite{Masabuchi1997} There exist $\mbf{X} \in \mathbb{R}^{n\times n} $ and $\gamma \in \mathbb{R}_{>0}$, such that $\mbf{E}^\trans \mbf{X} = \mbf{X}^\trans \mbf{E} \geq 0$ and
\bdis
\bbm \mbf{X}^\trans\mbf{A} + \mbf{A}^\trans\mbf{X} & \mbf{X}^\trans \mbf{B}& \mbf{C}^\trans \\ * & -\gamma \mbf{1} & \mbf{0}  \\ * & * & -\gamma\mbf{1} \ebm < 0.
\edis

\itemcite \cite{Masabuchi1997} There exist $\mbf{Y} \in \mathbb{R}^{n\times n} $ and $\gamma \in \mathbb{R}_{>0}$, such that $\mbf{Y}\mbf{E}^\trans  =  \mbf{E}\mbf{Y}^\trans \geq 0$ and
\bdis
\bbm \mbf{A}\mbf{Y}^\trans + \mbf{Y}\mbf{A}^\trans & \mbf{Y} \mbf{C}^\trans& \mbf{B} \\ * & -\gamma \mbf{1} & \mbf{0}  \\ * & * & -\gamma\mbf{1} \ebm < 0.
\edis

\itemcite \cite{Chadli2017} There exist $\mbf{P} \in \mathbb{S}^n$, $\mbf{X} \in \mathbb{R}^{(n-n_e)\times n}$, $\mbf{Z} \in \mathbb{R}^{n \times (n-n_e)}$, and $\gamma \in \mathbb{R}_{>0}$, where $n_e = \text{rank}(\mbf{E})$ \index{rank}and $\mbf{P} > 0$, satisfying $\mbf{E}^\trans \mbf{Z} = \mbf{0}$ and
\bdis
\bbm \mbf{A}^\trans \left( \mbf{P} \mbf{E} + \mbf{Z} \mbf{X} \right) + \left( \mbf{P} \mbf{E} + \mbf{Z} \mbf{X} \right)^\trans \mbf{A} + \mbf{C}^\trans \mbf{C} & \mbf{C} \left( \mbf{P} \mbf{E} + \mbf{Z} \mbf{X} \right)^\trans \mbf{B} \\ * & -\gamma^2 \mbf{1} \ebm \ < 0.
\edis

\itemcite \cite{Uezato1999} There exist $\mbf{P} \in \mathbb{S}^n$, $\mbf{S} \in \mathbb{R}^{(n-n_e)\times (n-n_e}$, $\mbf{U}$,~$\mbf{V} \in \mathbb{R}^{n \times (n-n_e)}$, and $\gamma \in \mathbb{R}_{>0}$, where $n_e = \text{rank}(\mbf{E})$, $\mathcal{R}(\mbf{U}) = \mathcal{N}(\mbf{E}^\trans)$, $\mathcal{R}(\mbf{V}) = \mathcal{N}(\mbf{E})$, and $\mbf{P} > 0$, satisfying \index{rank}
\bdis
\bbm \mbf{A} \left( \mbf{P} \mbf{E}^\trans + \mbf{V} \mbf{S} \mbf{U}^\trans \right) + \left( \mbf{P} \mbf{E}^\trans + \mbf{V} \mbf{S} \mbf{U}^\trans \right)^\trans \mbf{A}^\trans + \mbf{B} \mbf{B}^\trans  & \left( \mbf{P} \mbf{E}^\trans +  \mbf{V} \mbf{S} \mbf{U}^\trans \right)^\trans \mbf{C}^\trans \\ * & -\gamma^2 \mbf{1} \ebm< 0.
\edis

\itemcite \cite{Chadli2017} There exist $\mbf{P} \in \mathbb{S}^n$, $\mbf{X} \in \mathbb{R}^{(n-n_e)\times n}$, $\mbf{Z} \in \mathbb{R}^{n \times (n-n_e)}$, and $\gamma \in \mathbb{R}_{>0}$, where $n_e = \text{rank}(\mbf{E})$ and $\mbf{P} > 0$, satisfying $\mbf{E} \mbf{Z} = \mbf{0}$ and\index{rank}
\bdis
\bbm \mbf{A} \left( \mbf{P} \mbf{E} + \mbf{Z} \mbf{X} \right) + \left( \mbf{P} \mbf{E} + \mbf{Z} \mbf{X} \right)^\trans \mbf{A}^\trans + \mbf{B} \mbf{B}^\trans  & \left( \mbf{P} \mbf{E} + \mbf{Z} \mbf{X} \right)^\trans \mbf{C}^\trans \\ * & -\gamma^2 \mbf{1} \ebm< 0.
\edis

\itemcite \cite{Chadli2017} There exist $\mbf{P} \in \mathbb{S}^n$, $\mbf{X} \in \mathbb{R}^{(n-n_e)\times n}$, $\mbf{Z} \in \mathbb{R}^{(n+m) \times (n-n_e)}$, $\mbf{F}$,~$\mbf{G} \in \mathbb{R}^{(n+m)\times(n+m)}$, and $\gamma \in \mathbb{R}_{>0}$, where $n_e = \text{rank}(\mbf{E})$ and $\mbf{P} > 0$, satisfying $\mbfbar{E}^\trans \mbf{Z} = \mbf{0}$ and\index{rank}
\bdis
\bbm \mbfbar{A}^\trans \mbf{G}^\trans + \mbf{G} \mbfbar{A}  + \mbfbar{C}^\trans \mbfbar{C} & \left( \mbfbar{P} \mbfbar{E} + \mbf{Z} \mbfbar{X} \right)^\trans + \mbfbar{A}^\trans \mbf{F}^\trans - \mbf{G} \\ * & -\left( \mbf{F} + \mbf{F}^\trans \right) \ebm \ < 0,
\edis
where
\bdis
\mbfbar{A} = \bbm \mbf{A} & \mbf{B} \\ \mbf{0} & - \mbf{1} \ebm, \hspace{20pt} \mbfbar{E} = \bbm \mbf{E} & \mbf{0} \\ \mbf{0} & \onehalf \gamma^2 \mbf{1} \ebm, \hspace{20pt} \mbfbar{C} = \bbm \mbf{C} & \mbf{0} \ebm, \hspace{20pt} \mbfbar{P} = \bbm \mbf{P} & \mbf{0} \\ \mbf{0} & \mbf{1} \ebm, \hspace{20pt} \mbfbar{X} = \bbm \mbf{X} & \mbf{0} \ebm. 
\edis

\itemcite \cite{Chadli2017} There exist $\mbf{P} \in \mathbb{S}^n$, $\mbf{X} \in \mathbb{R}^{(n-n_e)\times n}$, $\mbf{Z} \in \mathbb{R}^{(n+p) \times (n-n_e)}$, $\mbf{F}$,~$\mbf{G} \in \mathbb{R}^{(n+p)\times(n+p)}$, and $\gamma \in \mathbb{R}_{>0}$, where $n_e = \text{rank}(\mbf{E})$ and $\mbf{P} > 0$, satisfying $\mbfbar{E} \mbf{Z} = \mbf{0}$ and\index{rank}
\bdis
\bbm \mbfbar{A} \mbf{G}+ \mbf{G}^\trans \mbfbar{A}^\trans  + \mbfbar{B} \mbfbar{B}^\trans & \left( \mbfbar{P} \mbfbar{E}^\trans + \mbf{Z} \mbfbar{X} \right)^\trans + \mbfbar{A} \mbf{F} - \mbf{G}^\trans \\ * & -\left( \mbf{F} + \mbf{F}^\trans \right) \ebm \ < 0,
\edis
where
\bdis
\mbfbar{A} = \bbm \mbf{A} & \mbf{0} \\ \mbf{C} & - \mbf{1} \ebm, \hspace{20pt} \mbfbar{E} = \bbm \mbf{E} & \mbf{0} \\ \mbf{0} & \onehalf \gamma^2 \mbf{1} \ebm, \hspace{20pt} \mbfbar{B} = \bbm \mbf{B} \\ \mbf{0} \ebm, \hspace{20pt} \mbfbar{P} = \bbm \mbf{P} & \mbf{0} \\ \mbf{0} & \mbf{1} \ebm, \hspace{20pt} \mbfbar{X} = \bbm \mbf{X} & \mbf{0} \ebm. 
\edis

\end{enumerate}

\subsubsection[Discrete-Time Descriptor System Bounded Real Lemma]{Discrete-Time Descriptor System Bounded Real Lemma}
\index{descriptor systems!discrete-time bounded real lemma}

Consider a discrete-time descriptor system, $\mbc{G}: \ell_{2e} \to \ell_{2e}$, described by
\begin{align*}
\mbf{E}_\mathrm{d} \mbf{x}_{k+1} &= \mbf{A}_\mathrm{d} \mbf{x}_k + \mbf{B}_\mathrm{d} \mbf{u}_k,\\
\mbf{y}_k &= \mbf{C}_\mathrm{d} \mbf{x}_k + \mbf{D}_\mathrm{d} \mbf{u}_k,
\end{align*}
where $\mbf{E}_\mathrm{d}$,~$\mbf{A}_\mathrm{d} \in \mathbb{R}^{n \times n}$, $\mbf{B}_\mathrm{d} \in \mathbb{R}^{n \times m}$, $\mbf{C}_\mathrm{d} \in \mathbb{R}^{p \times n}$, and $\mbf{D}_\mathrm{d} \in \mathbb{R}^{p \times m}$.  The $\mathcal{H}_\infty$ norm of $\mbc{G}$ is
\bdis
\norm{\mbc{G}}_\infty = \sup_{\mbf{u} \in \ell_2, \mbf{u} \neq \mbf{0}} \frac{\norm{\mbc{G} \mbf{u}}_2}{\norm{\mbf{u}}_2}.
\edis
The descriptor system is admissible and the inequality $\norm{\mbc{G}}_\infty  < \gamma$ holds under any of the following equivalent necessary and sufficient conditions.

\begin{enumerate}

\itemcite \cite{Hsiung1999} There exist $\mbf{P} \in \mathbb{S}^n $ and $\gamma \in \mathbb{R}_{>0}$, where $\mbf{P} > 0$, such that $\mbf{E}_\mathrm{d}^\trans \mbf{P} \mbf{E}_\mathrm{d} \geq 0$ and
\bdis
\bbm \mbf{A}_\mathrm{d}^\trans\mbf{P}\mbf{A}_\mathrm{d} -\mbf{E}_\mathrm{d}^\trans \mbf{P}\mbf{E}_\mathrm{d} & \mbf{A}_\mathrm{d}^\trans\mbf{P}\mbf{B}_\mathrm{d}& \mbf{C}_\mathrm{d}^\trans \\ * & \mbf{B}_\mathrm{d}^\trans\mbf{P}\mbf{B}_\mathrm{d}-\gamma \mbf{1} & \mbf{D}_\mathrm{d}^\trans  \\ * & * & -\gamma\mbf{1} \ebm < 0.
\edis

\itemcite \cite{Rehm2002} There exist $\mbf{Q} \in \mathbb{S}^n $ and $\gamma \in \mathbb{R}_{>0}$, where $\mbf{Q} > 0$, such that $\mbf{E}_\mathrm{d} \mbf{Q} \mbf{E}_\mathrm{d}^\trans \geq 0$ and
\bdis
\bbm \mbf{A}_\mathrm{d}\mbf{Q}\mbf{A}_\mathrm{d}^\trans -\mbf{E}_\mathrm{d} \mbf{Q}\mbf{E}_\mathrm{d}^\trans& \mbf{A}_\mathrm{d}\mbf{Q}\mbf{C}_\mathrm{d}^\trans& \mbf{B}_\mathrm{d} \\ * & \mbf{C}_\mathrm{d}\mbf{Q}\mbf{C}_\mathrm{d}^\trans-\gamma \mbf{1} & \mbf{D}_\mathrm{d}  \\ * & * & -\gamma\mbf{1} \ebm < 0.
\edis

\itemcite \cite{Zhang2008,Chadli2012} There exist $\mbf{P}  \in \mathbb{S}^n$, $\mbf{X} \in \mathbb{S}^{(n-n_e)}$, $\mbf{Z} \in \mathbb{R}^{n \times (n-n_e)}$, and $\gamma \in \mathbb{R}_{>0}$, where $n_e = \text{rank}(\mbf{E}_\mathrm{d})$ and $\mbf{P} > 0$, satisfying $\mbf{E}_\mathrm{d}^\trans \mbf{Z} = \mbf{0}$ and\index{rank}
\bdis
\bbm \mbf{A}_\mathrm{d}^\trans \left( \mbf{P} -  \mbf{Z} \mbf{X} \mbf{Z}^\trans \right) \mbf{A}_\mathrm{d} -  \mbf{E}_\mathrm{d}^\trans \mbf{P} \mbf{E}_\mathrm{d} + \mbf{C}_\mathrm{d}^\trans \mbf{C}_\mathrm{d} &   \mbf{A}_\mathrm{d}^\trans \left( \mbf{P} -  \mbf{Z} \mbf{X} \mbf{Z}^\trans \right) \mbf{B}_\mathrm{d} + \mbf{C}_\mathrm{d}^\trans \mbf{D}_\mathrm{d} \\ * &  \mbf{B}_\mathrm{d}^\trans \left( \mbf{P} -  \mbf{Z} \mbf{X} \mbf{Z}^\trans \right) \mbf{B}_\mathrm{d} - \gamma^2 \mbf{1} + \mbf{D}_\mathrm{d}^\trans \mbf{D}_\mathrm{d} \ebm< 0.
\edis

\itemcite \cite{Chadli2012} There exist $\mbf{P}  \in \mathbb{S}^n$, $\mbf{X} \in \mathbb{S}^{(n-n_e)}$, $\mbf{Z} \in \mathbb{R}^{(n+p) \times (n+p-m-n_e)}$, $\mbf{F} \in \mathbb{R}^{(n + p) \times (n + p)}$, $\mbf{G} \in \mathbb{R}^{(n +m) \times (n+p)}$, and $\gamma \in \mathbb{R}_{>0}$, where $m \leq p$, $n_e = \text{rank}(\mbf{E}_\mathrm{d})$ and $\mbf{P} > 0$, such that $\mbfbar{E}^\trans \mbf{Z} = \mbf{0}$ and\index{rank}
\bdis
\bbm \mbfbar{E}^\trans \mbfbar{P} \mbfbar{E} + \mbf{G} \mbfbar{A} + \mbfbar{A}^\trans \mbf{G}^\trans & - \mbf{G} + \mbfbar{A}^\trans \mbf{F}^\trans \\ * & \mbfbar{P} - \mbf{Z} \mbfbar{X} \mbf{Z}^\trans - \left( \mbf{F} + \mbf{F}^\trans \right) \ebm < 0,
\edis
where
\bdis
\mbfbar{A} = \bbm \mbf{A}_\mathrm{d} & \mbf{B}_\mathrm{d} \\ \mbf{C}_\mathrm{d} & \mbf{D}_\mathrm{d} \ebm, \hspace{20pt} \mbfbar{E} = \bbm \mbf{E}_\mathrm{d} & \mbf{0} \\ \mbf{0} & \gamma \bbm \mbf{1}_{m \times m} \\ \mbf{0}_{p \times m} \ebm \ebm, \hspace{20pt} \mbfbar{P} = \bbm \mbf{P} & \mbf{0} \\ \mbf{0} & \mbf{1} \ebm, \hspace{20pt} \mbfbar{X} = \bbm \mbf{X} & \mbf{0} \\ \mbf{0} & \mbf{0} \ebm.
\edis

\itemcite \cite{Chadli2012} There exist $\mbf{P}  \in \mathbb{S}^n$, $\mbf{X} \in \mathbb{S}^{(n-n_e)}$, $\mbf{Z} \in \mathbb{R}^{(n+m) \times (n-n_e)}$, $\mbf{F} \in \mathbb{R}^{(n + m) \times (n + m)}$, $\mbf{G} \in \mathbb{R}^{(n +p) \times (n+m)}$, and $\gamma \in \mathbb{R}_{>0}$, where $m \leq p$, $n_e = \text{rank}(\mbf{E}_\mathrm{d})$ and $\mbf{P} > 0$, such that $\mbfbar{E} \mbf{Z} = \mbf{0}$ and\index{rank}
\bdis
\bbm \mbfbar{E} \mbfbar{P} \mbfbar{E}^\trans + \mbf{G} \mbfbar{A}^\trans + \mbfbar{A} \mbf{G}^\trans & - \mbf{G} + \mbfbar{A} \mbf{F}^\trans \\ * & \mbfbar{P} - \mbf{Z} \mbf{X} \mbf{Z}^\trans - \left( \mbf{F} + \mbf{F}^\trans \right) \ebm < 0,
\edis
where
\bdis
\mbfbar{A} = \bbm \mbf{A}_\mathrm{d} & \mbf{B}_\mathrm{d} \\ \mbf{C}_\mathrm{d} & \mbf{D}_\mathrm{d} \ebm, \hspace{20pt} \mbfbar{E} = \bbm \mbf{E}_\mathrm{d} & \mbf{0} \\ \mbf{0} & \gamma \mbf{1} \ebm, \hspace{20pt} \mbfbar{P} = \bbm \mbf{P} & \mbf{0} \\ \mbf{0} & \mbf{1} \ebm.
\edis

\end{enumerate}

\subsection{\texorpdfstring{$\mathcal{H}_2$}{H2} Norm}

\subsubsection[Continuous-Time \texorpdfstring{$\mathcal{H}_2$}{H2} Norm]{Continuous-Time $\mathcal{H}_2$ Norm}
\label{sec:H2Norm}
\index{norm!$\mathcal{H}_2$ norm}
Consider a continuous-time LTI system, $\mbc{G}: \mathcal{L}_{2e} \to \mathcal{L}_{2e}$, with state-space realization $(\mbf{A},\mbf{B}_,\mbf{C},\mbf{0})$, where $\mbf{A} \in \mathbb{R}^{n \times n}$, $\mbf{B} \in \mathbb{R}^{n \times m}$, $\mbf{C} \in \mathbb{R}^{p \times n}$, and $\mbf{A}$ is Hurwitz.   The $\mathcal{H}_2$ norm of $\mbc{G}$ is
\bdis
\norm{\mbc{G}}_2 = \sqrt{\trace (\mbf{C} \mbf{W} \mbf{C}^\trans ) } = \sqrt{\trace (\mbf{B}^\trans \mbf{M} \mbf{B} ) }, 
\edis
where $\mbf{W}$,~$\mbf{M} \in \mathbb{S}^n$, $\mbf{W} > 0$, $\mbf{M} > 0$, and
\bdis
\mbf{A} \mbf{W} + \mbf{W} \mbf{A}^\trans + \mbf{B} \mbf{B}^\trans = \mbf{0}, \hspace{10pt} \mbf{M} \mbf{A} + \mbf{A}^\trans \mbf{M} + \mbf{C}^\trans \mbf{C} = \mbf{0}.
\edis
The inequality $\norm{\mbc{G}}_2 < \mu$ holds under any of the following equivalent necessary and sufficient conditions.

\begin{enumerate}

\item \cite[p.~77]{SchererWeiland2015} There exist $\mbf{X} \in \mathbb{S}^n$ and $\mu \in \mathbb{R}_{>0}$, where $\mbf{X} >0 $, such that
\begin{align*}
\mbf{A}\mbf{X} + \mbf{X}\mbf{A}^\trans +  \mbf{B}\mbf{B}^\trans&< 0, \\
\trace\left(\mbf{C}\mbf{X}\mbf{C}^\trans\right) &< \mu^2.
\end{align*}

\itemcite \cite[p.~77]{SchererWeiland2015} There exist $\mbf{Y} \in \mathbb{S}^n$ and $\mu \in \mathbb{R}_{>0}$, where $\mbf{Y} >0 $, such that
\begin{align*}
\mbf{A}^\trans\mbf{Y} + \mbf{Y}\mbf{A} +  \mbf{C}^\trans\mbf{C} &< 0,\\
\trace\left(\mbf{B}^\trans\mbf{Y}\mbf{B}\right) &< \mu^2. 
\end{align*}

%

\itemcite \cite[p.~77]{SchererWeiland2015},\cite{Apkarian2001}  There exist $\mbf{Y} \in \mathbb{S}^n$, $\mbf{Z} \in \mathbb{S}^p$, and $\mu \in \mathbb{R}_{>0}$, where $\mbf{Y} >0 $ and $\mbf{Z} > 0$, such that
\begin{align}
\bbm \mbf{A}^\trans\mbf{Y} + \mbf{Y}\mbf{A}&  \mbf{Y} \mbf{B} \\ * & -\mu\mbf{1} \ebm &< 0, \nonumber \\
\bbm \mbf{Y} & \mbf{C}^\trans \\ * & \mbf{Z} \ebm &> 0, \nonumber  \\
\trace (\mbf{Z} )&< \mu. \nonumber 
\end{align}

\itemcite \cite[p.~77]{SchererWeiland2015} There exist $\mbf{X} \in \mathbb{S}^n$, $\mbf{Z} \in \mathbb{S}^p$, and $\mu \in \mathbb{R}_{>0}$, where $\mbf{X} >0 $ and $\mbf{Z} > 0$, such that
\begin{align}
\bbm \mbf{X}\mbf{A}^\trans + \mbf{A}\mbf{X} & \mbf{X}\mbf{C}^\trans \\ * & -\mu\mbf{1} \ebm &< 0, \nonumber \\
\bbm \mbf{X} & \mbf{B} \\ * & \mbf{Z} \ebm &> 0, \nonumber  \\
\trace (\mbf{Z} ) &< \mu. \nonumber 
\end{align}

\itemcite \cite{Wu2006} There exist $\mbf{Y} \in \mathbb{S}^n$, $\mbf{Z} \in \mathbb{S}^m$, $\mbf{F}$,~$\mbf{G} \in \mathbb{R}^{n \times n}$, and $\mu \in \mathbb{R}_{>0}$, where $\mbf{Y} >0 $ and $\mbf{Z} > 0$, such that
\begin{align*}
\bbm \mbf{F} + \mbf{F}^\trans & \mbf{G} - \mbf{F}^\trans  + \mbf{Y} \mbf{A} & \mbf{0} \\ * & -(\mbf{G} + \mbf{G}^\trans) & \mbf{C}^\trans \\ * & * & -\mbf{1}  \ebm &< 0,  \\
\bbm \mbf{Y} & \mbf{Y}\mbf{B} \\ * & \mbf{Z} \ebm &> 0,  \\
\trace (\mbf{Z}) &< \mu^2.  
\end{align*}

\itemcite \cite{Wu2006} There exist $\mbf{X} \in \mathbb{S}^n$, $\mbf{Z} \in \mathbb{S}^m$, $\mbf{F}$,~$\mbf{G} \in \mathbb{R}^{n \times n}$, and $\mu \in \mathbb{R}_{>0}$, where $\mbf{X} >0 $ and $\mbf{Z} > 0$, such that
\begin{align*}
\bbm \mbf{F} + \mbf{F}^\trans & \mbf{G} - \mbf{F}^\trans  + \mbf{X}\mbf{A}^\trans  & \mbf{0} \\ * & -(\mbf{G} + \mbf{G}^\trans) & \mbf{B} \\ * & * & -\mbf{1}  \ebm &< 0,  \\
\bbm \mbf{X} & \mbf{X}\mbf{C}^\trans \\ * & \mbf{Z} \ebm &> 0,  \\
\trace (\mbf{Z}) &< \mu^2.  
\end{align*}

\itemcite \cite{Wu2006} There exist $\mbf{X} \in \mathbb{S}^n$, $\mbf{Z} \in \mathbb{S}^m$, $\mbf{F}$,~$\mbf{G} \in \mathbb{R}^{n \times n}$, and $\mu \in \mathbb{R}_{>0}$, where $\mbf{X} >0 $ and $\mbf{Z} > 0$, such that
\begin{align*}
\bbm \mbf{F} + \mbf{F}^\trans & \mbf{G} - \mbf{F}^\trans  + \mbf{A} \mbf{X} & \mbf{0} \\ * & -(\mbf{G} + \mbf{G}^\trans) & \mbf{X} \mbf{C}^\trans \\ * & * & -\mbf{1}  \ebm &< 0,  \\
\bbm \mbf{X} & \mbf{B} \\ * & \mbf{Z} \ebm &> 0,  \\
\trace (\mbf{Z}) &< \mu^2.  
\end{align*}

\itemcite \cite{Wu2006} There exist $\mbf{Y} \in \mathbb{S}^n$, $\mbf{Z} \in \mathbb{S}^m$, $\mbf{F}$,~$\mbf{G} \in \mathbb{R}^{n \times n}$, and $\mu \in \mathbb{R}_{>0}$, where $\mbf{Y} >0 $ and $\mbf{Z} > 0$, such that
\begin{align*}
\bbm \mbf{F} + \mbf{F}^\trans & \mbf{G} - \mbf{F}^\trans  +  \mbf{A}^\trans\mbf{Y} & \mbf{0} \\ * & -(\mbf{G} + \mbf{G}^\trans) & \mbf{X} \mbf{B} \\ * & * & -\mbf{1}  \ebm &< 0,  \\
\bbm \mbf{Y} & \mbf{C}^\trans \\ * & \mbf{Z} \ebm &> 0,  \\
\trace (\mbf{Z}) &< \mu^2.  
\end{align*}

\itemcite \cite{Wu2006} There exist $\mbf{X} \in \mathbb{S}^n$, $\mbf{Z} \in \mathbb{S}^m$, $\mbf{F}$,~$\mbf{G} \in \mathbb{R}^{n \times n}$, and $\mu \in \mathbb{R}_{>0}$, where $\mbf{X} >0 $ and $\mbf{Z} > 0$, such that
\begin{align*}
\bbm \mbf{A}\mbf{F} + \mbf{F}^\trans\mbf{A}^\trans & \mbf{X} - \mbf{F}^\trans  + \mbf{A} \mbf{G} & \mbf{F}^\trans \mbf{C}^\trans \\ * & -(\mbf{G} + \mbf{G}^\trans) & \mbf{G}^\trans \mbf{C}^\trans \\ * & * & -\mbf{1}  \ebm &< 0,  \\
\bbm \mbf{X} & \mbf{B} \\ * & \mbf{Z} \ebm &> 0,  \\
\trace (\mbf{Z}) &< \mu^2.  
\end{align*}

\itemcite \cite{Wu2006} There exist $\mbf{Y} \in \mathbb{S}^n$, $\mbf{Z} \in \mathbb{S}^m$, $\mbf{F}$,~$\mbf{G} \in \mathbb{R}^{n \times n}$, and $\mu \in \mathbb{R}_{>0}$, where $\mbf{Y} >0 $ and $\mbf{Z} > 0$, such that
\begin{align*}
\bbm \mbf{A}^\trans\mbf{F} + \mbf{F}^\trans\mbf{A} & \mbf{Y} - \mbf{F}^\trans  + \mbf{A}^\trans \mbf{G} & \mbf{F}^\trans \mbf{B} \\ * & -(\mbf{G} + \mbf{G}^\trans) & \mbf{G}^\trans \mbf{B} \\ * & * & -\mbf{1}  \ebm &< 0,  \\
\bbm \mbf{Y} & \mbf{C}^\trans \\ * & \mbf{Z} \ebm &> 0,  \\
\trace (\mbf{Z}) &< \mu^2.  
\end{align*}

\itemcite \cite{Apkarian2001} There exist $\mbf{X} \in \mathbb{S}^n$, $\mbf{Z} \in \mathbb{S}^p$, $\mbf{V} \in \mathbb{R}^{n \times n}$, and $\mu \in \mathbb{R}_{>0}$, where $\mbf{X} >0 $ and $\mbf{Z} > 0$, such that
\begin{align}
\bbm -\left(\mbf{V} + \mbf{V}^\trans\right) & \mbf{V}^\trans\mbf{A} + \mbf{X} & \mbf{V}^\trans\mbf{B} & \mbf{V}^\trans \\ * & -\mbf{X} & \mbf{0} & \mbf{0} \\ * & * & -\mu^2\mbf{1} & \mbf{0} \\ * & * & * & -\mbf{X} \ebm &< 0, \label{eq:H2_6a} \\
\bbm \mbf{X} & \mbf{C}^\trans \\ * & \mbf{Z} \ebm &> 0, \nonumber  \\
\trace (\mbf{Z}) &< 1. \nonumber 
\end{align}

\itemcite \cite{Apkarian2001} There exist $\mbf{X} \in \mathbb{S}^n$, $\mbf{Z} \in \mathbb{S}^m$, $\mbf{V} \in \mathbb{R}^{n \times n}$, and $\mu \in \mathbb{R}_{>0}$, where $\mbf{X} >0 $ and $\mbf{Z} > 0$, such that
\begin{align}
\bbm -\left(\mbf{V} + \mbf{V}^\trans\right) & \mbf{V}^\trans\mbf{A}^\trans + \mbf{X} & \mbf{V}^\trans\mbf{C}^\trans & \mbf{V}^\trans \\ * & -\mbf{X} & \mbf{0} & \mbf{0} \\ * & * & -\mu^2\mbf{1} & \mbf{0} \\ * & * & * & -\mbf{X} \ebm &< 0, \label{eq:H2_7a} \\
\bbm \mbf{X} & \mbf{B} \\ * & \mbf{Z} \ebm &> 0, \nonumber  \\
\trace (\mbf{Z}) &< 1. \nonumber 
\end{align}

\itemcite \cite{Ebihara2004} There exist $\mbf{X} \in \mathbb{S}^n$, $\mbf{Z} \in \mathbb{S}^m$, $\mbs{\Gamma} \in \mathbb{R}^{n \times n}$, and $\mu \in \mathbb{R}_{>0}$, where $\mbf{X} >0 $ and $\mbf{Z} > 0$, such that
\begin{align}
\bbm \mbf{0} & -\mbf{X} & \mbf{0} \\ * & \mbf{0} & \mbf{0} \\ * & * & -\mbf{1} \ebm + \text{He}\Bigg\{ \bbm \mbf{A} \\ \mbf{1} \\ \mbf{C} \ebm \mbs{\Gamma} \bbm \mbf{1} & -\epsilon \mbf{1} & \mbf{0} \ebm \Bigg\} &< 0, \nonumber \\
\bbm \mbf{Z} & \mbf{B}^\trans \\ * & \mbf{X} \ebm &> 0, \nonumber \\
\trace (\mbf{Z}) &< \mu^2. \nonumber
\end{align}


\end{enumerate}

The $\mathcal{H}_2$ norm of $\mbc{G}$ is the minimum value of $\mu \in \mathbb{R}_{>0}$ that satisfies any of the above conditions.

\subsubsection[Discrete-Time \texorpdfstring{$\mathcal{H}_2$}{H2} Norm Without Feedthrough]{Discrete-Time \texorpdfstring{$\mathcal{H}_2$}{H2} Norm Without Feedthrough}
\index{norm!discrete-time $\mathcal{H}_2$ norm}
Consider a discrete-time LTI system, $\mbc{G}: \ell_{2e} \to \ell_{2e}$, with state-space realization $(\mbf{A}_\mathrm{d},\mbf{B}_\mathrm{d},\mbf{C}_\mathrm{d},\mbf{0})$, where $\mbf{A}_\mathrm{d} \in \mathbb{R}^{n \times n}$, $\mbf{B}_\mathrm{d} \in \mathbb{R}^{n \times m}$, $\mbf{C}_\mathrm{d} \in \mathbb{R}^{p \times n}$, and $\mbf{A}_\mathrm{d}$ is Schur.  The $\mathcal{H}_2$ norm of $\mbc{G}$ is
\bdis
\norm{\mbc{G}}_2 = \sqrt{\trace (\mbf{C}_\mathrm{d} \mbf{W} \mbf{C}_\mathrm{d}^\trans ) } = \sqrt{\trace (\mbf{B}_\mathrm{d}^\trans \mbf{M} \mbf{B}_\mathrm{d} ) },
\edis
where $\mbf{W}$,~$\mbf{M} \in \mathbb{S}^n$, $\mbf{W} > 0$, $\mbf{M} > 0$, and
\bdis
\mbf{A}_\mathrm{d} \mbf{W} \mbf{A}_\mathrm{d}^\trans - \mbf{W} +  \mbf{B}_\mathrm{d} \mbf{B}_\mathrm{d}^\trans = \mbf{0}, \hspace{10pt} \mbf{A}_\mathrm{d}^\trans \mbf{M} \mbf{A}_\mathrm{d} - \mbf{M} +  \mbf{C}_\mathrm{d}^\trans \mbf{C}_\mathrm{d} = \mbf{0}.
\edis
The inequality $\norm{\mbc{G}}_2 < \mu$ holds under any of following equivalent necessary and sufficient conditions.

\begin{enumerate}

\item There exist $\mbf{P} \in \mathbb{S}^n$ and $\mu \in \mathbb{R}_{>0}$, where $\mbf{P} >0 $, such that
\begin{align*}
\mbf{A}_\mathrm{d}\mbf{P}\mbf{A}_\mathrm{d}^\trans - \mbf{P} + \mbf{B}_\mathrm{d}  \mbf{B}_\mathrm{d}^\trans &< 0, \\
\trace \left(\mbf{C}_\mathrm{d} \mbf{P} \mbf{C}_\mathrm{d}^\trans \right)&< \mu^2. 
\end{align*}

\item There exist $\mbf{Q} \in \mathbb{S}^n$ and $\mu \in \mathbb{R}_{>0}$, where $\mbf{Q} >0 $, such that
\begin{align*}
\mbf{A}_\mathrm{d}^\trans\mbf{Q}\mbf{A}_\mathrm{d} - \mbf{Q} + \mbf{C}_\mathrm{d}^\trans  \mbf{C}_\mathrm{d} &< 0, \\
\trace \left(\mbf{B}_\mathrm{d}^\trans \mbf{Q} \mbf{B}_\mathrm{d} \right)&< \mu^2. 
\end{align*}

\itemcite \cite{DeOliveira2002} There exist $\mbf{P} \in \mathbb{S}^n$, $\mbf{Z} \in \mathbb{S}^p$, and $\mu \in \mathbb{R}_{>0}$, where $\mbf{P} >0 $ and $\mbf{Z} > 0$, such that
\begin{align}
\bbm \mbf{P} & \mbf{A}_\mathrm{d}\mbf{P} & \mbf{B}_\mathrm{d} \\ * & \mbf{P} & \mbf{0} \\ * & * & \mbf{1} \ebm &> 0, \label{eq:DT_H2a_0}\\
\bbm \mbf{Z} & \mbf{C}_\mathrm{d}\mbf{P} \\ * & \mbf{P} \ebm &> 0, \label{eq:DT_H2b_0} \\
\trace (\mbf{Z}) &< \mu^2. \nonumber
\end{align}

\item There exist $\mbf{Q} \in \mathbb{S}^n$, $\mbf{Z} \in \mathbb{S}^m$, and $\mu \in \mathbb{R}_{>0}$, where $\mbf{Q} >0 $ and $\mbf{Z} > 0$, such that
\begin{align}
\bbm \mbf{Q} & \mbf{A}_\mathrm{d}^\trans \mbf{Q} & \mbf{C}_\mathrm{d}^\trans \\ * & \mbf{Q} & \mbf{0} \\ * & * & \mbf{1} \ebm &> 0, \label{eq:DT_H2a_01}\\
\bbm \mbf{Z} & \mbf{B}_\mathrm{d}^\trans\mbf{Q} \\ * & \mbf{Q} \ebm &> 0, \label{eq:DT_H2b_01}\\
\trace (\mbf{Z}) &< \mu^2. \nonumber
\end{align}

\item There exist $\mbf{Q} \in \mathbb{S}^n$, $\mbf{Z} \in \mathbb{S}^p$, and $\mu \in \mathbb{R}_{>0}$, where $\mbf{Q} >0 $ and $\mbf{Z} > 0$, such that
\begin{align}
\bbm \mbf{Q} & \mbf{Q}\mbf{A}_\mathrm{d} & \mbf{Q} \mbf{B}_\mathrm{d} \\ * & \mbf{Q} & \mbf{0} \\ * & * & \mbf{1} \ebm &> 0, \label{eq:DT_H2a_1} \\
\bbm \mbf{Z} & \mbf{C}_\mathrm{d} \\ * & \mbf{Q} \ebm &> 0, \nonumber\\
\trace (\mbf{Z}) &< \mu^2. \nonumber
\end{align}
\begin{proof}
Apply the congruence transformation $\mbf{W}_1 = \mathrm{diag}\{\mbf{Q},\mbf{Q},\mbf{1}\}$ to~\eqref{eq:DT_H2a_0} and $\mbf{W}_2 = \mathrm{diag}\{\mbf{1},\mbf{Q}\}$ to~\eqref{eq:DT_H2b_0}, where $\mbf{Q} = \mbf{P}^{-1}$.
\end{proof}

\item There exist $\mbf{P} \in \mathbb{S}^n$, $\mbf{Z} \in \mathbb{S}^m$, and $\mu \in \mathbb{R}_{>0}$, where $\mbf{P} >0 $ and $\mbf{Z} > 0$, such that
\begin{align}
\bbm \mbf{P} & \mbf{P}\mbf{A}_\mathrm{d}^\trans & \mbf{P} \mbf{C}_\mathrm{d}^\trans \\ * & \mbf{P} & \mbf{0} \\ * & * & \mbf{1} \ebm &> 0, \label{eq:DT_H2a_2}\\
\bbm \mbf{Z} & \mbf{B}_\mathrm{d}^\trans \\ * & \mbf{P} \ebm &> 0, \nonumber  \\
\trace (\mbf{Z}) &< \mu^2. \nonumber
\end{align}

\itemcite \cite{DeOliveira2002} There exist $\mbf{P} \in \mathbb{S}^n$, $\mbf{Z} \in \mathbb{S}^p$, $\mbf{X} \in \mathbb{R}^{n \times n}$, and $\mu \in \mathbb{R}_{>0}$, where $\mbf{P} >0 $, $\mbf{Z} > 0$, and $\mbf{X}$ has full rank, such that\index{rank}
\begin{align}
\bbm \mbf{P} & \mbf{A}_\mathrm{d}\mbf{X} & \mbf{B}_\mathrm{d} \\ * & \mbf{X}^\trans\mbf{P}^{-1}\mbf{X} & \mbf{0} \\ * & * & \mbf{1} \ebm &> 0, \nonumber\\
\bbm \mbf{Z} & \mbf{C}_\mathrm{d}\mbf{X} \\ * & \mbf{X}^\trans\mbf{P}^{-1}\mbf{X}  \ebm &> 0, \nonumber\\
\trace (\mbf{Z}) &< \mu^2. \nonumber
\end{align}

\item There exist $\mbf{Q} \in \mathbb{S}^n$, $\mbf{Z} \in \mathbb{S}^m$, $\mbf{X} \in \mathbb{R}^{n \times n}$, and $\mu \in \mathbb{R}_{>0}$, where $\mbf{Q} >0 $ and $\mbf{Z} > 0$, and $\mbf{X}$ has full rank, such that\index{rank}
\begin{align*}
\bbm \mbf{Q} & \mbf{A}_\mathrm{d}^\trans \mbf{X} & \mbf{C}_\mathrm{d}^\trans \\ * & \mbf{X}^\trans \mbf{Q}^{-1}\mbf{X} & \mbf{0} \\ * & * & \mbf{1} \ebm &> 0,\\
\bbm \mbf{Z} & \mbf{B}_\mathrm{d}^\trans\mbf{X} \\ * & \mbf{X}^\trans \mbf{Q}^{-1}\mbf{X} \ebm &> 0, \\
\trace (\mbf{Z}) &< \mu^2. 
\end{align*}
\begin{proof}
Apply the congruence transformation $\mbf{W}_1 = \textrm{diag}\{\mbf{1},\mbf{X}^\trans \mbf{Q}^{-1},\mbf{1}\}$ to~\eqref{eq:DT_H2a_01} and the congruence transformation $\mbf{W}_2 = \textrm{diag}\{\mbf{1},\mbf{X}^\trans \mbf{Q}^{-1}\}$ to~\eqref{eq:DT_H2b_01}, where $\mbf{W}_1$ and $\mbf{W}_2$ have full rank since $\mbf{X}$ has full rank.\index{rank}

\end{proof}

\item There exist $\mbf{Q} \in \mathbb{S}^n$, $\mbf{Z} \in \mathbb{S}^p$, $\mbf{X} \in \mathbb{R}^{n \times n}$, and $\mu \in \mathbb{R}_{>0}$, where $\mbf{Q} >0 $, $\mbf{Z} > 0$, and $\mbf{X}$ has full rank, such that\index{rank}
\begin{align}
\bbm \mbf{X}^\trans\mbf{Q}^{-1}\mbf{X} & \mbf{X}^\trans\mbf{A}_\mathrm{d} & \mbf{X}^\trans \mbf{B}_\mathrm{d} \\ * & \mbf{Q} & \mbf{0} \\ * & * & \mbf{1} \ebm &> 0, \label{eq:DT_H2a_3}\\
\bbm \mbf{Z} & \mbf{C}_\mathrm{d} \\ * & \mbf{Q} \ebm &> 0, \nonumber \\
\trace (\mbf{Z}) &< \mu^2. \nonumber 
\end{align}

\begin{proof}
Apply the congruence transformation $\mbf{W} = \mathrm{diag}\{\mbf{X}^\trans\mbf{Q}^{-1},\mbf{1},\mbf{1}\}$ to~\eqref{eq:DT_H2a_1}, where $\mbf{W}$ has full rank since $\mbf{X}$ has full rank.\index{rank}
\end{proof}

\item There exist $\mbf{P} \in \mathbb{S}^n$, $\mbf{Z} \in \mathbb{S}^m$,  $\mbf{X} \in \mathbb{R}^{n \times n}$, and $\mu \in \mathbb{R}_{>0}$, where $\mbf{P} >0 $ and $\mbf{Z} > 0$, and $\mbf{X}$ has full rank, such that\index{rank}
\begin{align}
\bbm \mbf{X}^\trans  \mbf{P}^{-1} \mbf{X} & \mbf{X}^\trans \mbf{A}_\mathrm{d}^\trans & \mbf{X}^\trans \mbf{C}_\mathrm{d}^\trans \\ * & \mbf{P} & \mbf{0} \\ * & * & \mbf{1} \ebm &> 0, \label{eq:DT_H2a_32}\\
\bbm \mbf{Z} & \mbf{B}_\mathrm{d}^\trans \\ * & \mbf{P} \ebm &> 0, \nonumber \\
\trace (\mbf{Z}) &< \mu^2. \nonumber
\end{align}
\begin{proof}
Apply the congruence transformation $\mbf{W} = \mathrm{diag}\{\mbf{X}^\trans\mbf{P}^{-1},\mbf{1},\mbf{1}\}$ to~\eqref{eq:DT_H2a_2}, where $\mbf{W}$ has full rank since $\mbf{X}$ has full rank.\index{rank}
\end{proof}

\itemcite \cite{DeOliveira2002} There exist $\mbf{P} \in \mathbb{S}^n$, $\mbf{Z} \in \mathbb{S}^p$, $\mbf{X} \in \mathbb{R}^{n \times n}$, and $\mu \in \mathbb{R}_{>0}$, where $\mbf{P} >0 $ and $\mbf{Z} > 0$, such that
\begin{align}
\bbm \mbf{P} & \mbf{A}_\mathrm{d}\mbf{X} & \mbf{B}_\mathrm{d} \\ * & \mbf{X} + \mbf{X}^\trans - \mbf{P} & \mbf{0} \\ * & * & \mbf{1} \ebm &> 0, \label{eq:DT_H2a_4}\\
\bbm \mbf{Z} & \mbf{C}_\mathrm{d}\mbf{X} \\ * & \mbf{X} + \mbf{X}^\trans - \mbf{P} \ebm &> 0, \label{eq:DT_H2b_4} \\
\trace (\mbf{Z}) &< \mu^2. \label{eq:DT_H2c_4}
\end{align}

\item There exist $\mbf{Q} \in \mathbb{S}^n$, $\mbf{Z} \in \mathbb{S}^m$, $\mbf{X} \in \mathbb{R}^{n \times n}$, and $\mu \in \mathbb{R}_{>0}$, where $\mbf{Q} >0 $ and $\mbf{Z} > 0$, and $\mbf{X}$ has full rank, such that\index{rank}
\begin{align*}
\bbm \mbf{Q} & \mbf{A}_\mathrm{d}^\trans \mbf{X} & \mbf{C}_\mathrm{d}^\trans \\ * & \mbf{X} + \mbf{X}^\trans - \mbf{Q} & \mbf{0} \\ * & * & \mbf{1} \ebm &> 0,\\
\bbm \mbf{Z} & \mbf{B}_\mathrm{d}^\trans\mbf{X} \\ * & \mbf{X} + \mbf{X}^\trans - \mbf{Q}  \ebm &> 0, \\
\trace (\mbf{Z}) &< \mu^2. 
\end{align*}
\begin{proof}
Same as the proof of~\eqref{eq:DT_H2a_4},~\eqref{eq:DT_H2b_4},~\eqref{eq:DT_H2c_4} in~\cite{DeOliveira2002}.

\end{proof}

\item There exist $\mbf{Q} \in \mathbb{S}^n$, $\mbf{Z} \in \mathbb{S}^p$, $\mbf{X} \in \mathbb{R}^{n \times n}$, and $\mu \in \mathbb{R}_{>0}$, where $\mbf{Q} >0 $ and $\mbf{Z} > 0$, such that
\begin{align}
\bbm \mbf{X} + \mbf{X}^\trans - \mbf{Q} & \mbf{X}^\trans\mbf{A}_\mathrm{d} & \mbf{X}^\trans \mbf{B}_\mathrm{d} \\ * & \mbf{Q} & \mbf{0} \\ * & * & \mbf{1} \ebm &> 0, \label{eq:DT_H2a_5} \\
\bbm \mbf{Z} & \mbf{C}_\mathrm{d} \\ * & \mbf{Q} \ebm &> 0, \nonumber \\
\trace (\mbf{Z}) &< \mu^2. \nonumber
\end{align}

\begin{proof}
Same as the proof of~\eqref{eq:DT_H2a_4},~\eqref{eq:DT_H2b_4},~\eqref{eq:DT_H2c_4} in~\cite{DeOliveira2002}, by which it is shown that~\eqref{eq:DT_H2a_5} is equivalent to~\eqref{eq:DT_H2a_3}.
\end{proof}

\item There exist $\mbf{P} \in \mathbb{S}^n$, $\mbf{Z} \in \mathbb{S}^m$,  $\mbf{X} \in \mathbb{R}^{n \times n}$, and $\mu \in \mathbb{R}_{>0}$, where $\mbf{P} >0 $ and $\mbf{Z} > 0$, and $\mbf{X}$ has full rank, such that\index{rank}
\begin{align}
\bbm \mbf{X} + \mbf{X}^\trans - \mbf{P} & \mbf{X}^\trans \mbf{A}_\mathrm{d}^\trans & \mbf{X}^\trans \mbf{C}_\mathrm{d}^\trans \\ * & \mbf{P} & \mbf{0} \\ * & * & \mbf{1} \ebm &> 0, \label{eq:DT_H2a_6} \\
\bbm \mbf{Z} & \mbf{B}_\mathrm{d}^\trans \\ * & \mbf{P} \ebm &> 0, \nonumber \\
\trace (\mbf{Z}) &< \mu^2. \nonumber
\end{align}
\begin{proof}
Same as the proof of~\eqref{eq:DT_H2a_4},~\eqref{eq:DT_H2b_4},~\eqref{eq:DT_H2c_4} in~\cite{DeOliveira2002}, by which it is shown that~\eqref{eq:DT_H2a_6} is equivalent to~\eqref{eq:DT_H2a_32}.
\end{proof}

\itemcite \cite[pp.~53--54]{Spagolla2019b} There exist $\mbf{P} \in \mathbb{S}^{n}$, $\mbf{Z} \in \mathbb{S}^p$, $\mbf{F}_1$,~$\mbf{F}_2$,~$\mbf{F}_5 \in \mathbb{R}^{n \times n}$, $\mbf{F}_4 \in \mathbb{R}^{n \times p}$, and $\mu \in \mathbb{R}_{>0}$, where $\mbf{P} > 0$ and $\mbf{Z} > 0$, such that
\begin{align*}
\bbm - \mbf{P} + \mbf{A}_\mathrm{d}\mbf{F}_1 + \mbf{F}_1^\trans\mbf{A}_\mathrm{d}^\trans &  \mbf{A}_\mathrm{d} \mbf{F}_2 - \mbf{F}_1^\trans & \mbf{B}_\mathrm{d} \\ * & \mbf{P} - \left(\mbf{F}_2 + \mbf{F}_2^\trans \right)  &  \mbf{0} \\   * & * & -\gamma \mbf{1}  \ebm  &< 0, \\
\bbm - \mbf{Z} + \mbf{C}_\mathrm{d}\mbf{F}_4 + \mbf{F}_4^\trans\mbf{C}_\mathrm{d}^\trans &  \mbf{C}_\mathrm{d} \mbf{F}_5 - \mbf{F}_4^\trans  \\ * & \mbf{P} - \left(\mbf{F}_5 + \mbf{F}_5^\trans \right)   \ebm  &< 0, \\
\trace (\mbf{Z}) &< \mu^2.
\end{align*}

\itemcite \cite[pp.~53--54]{Spagolla2019b} There exist $\mbf{P} \in \mathbb{S}^{n}$, $\mbf{Z} \in \mathbb{S}^p$, $\mbf{F}_1$,~$\mbf{F}_2$,~$\mbf{F}_5$,~$\mbf{X}_1$,~$\mbf{X}_2$,~$\mbf{X}_3$,~$\mbf{X}_5$,~$\mbf{X}_6 \in \mathbb{R}^{n \times n}$, $\mbf{F}_4 \in \mathbb{R}^{n \times p}$, $\mbf{X}_4 \in \mathbb{R}^{p \times n}$, and $\mu \in \mathbb{R}_{>0}$, where $\mbf{P} > 0$ and $\mbf{Z} > 0$, such that
\begin{align*}
\bbm - \mbf{P} + \mbf{X}_1\mbf{F}_1 + \mbf{F}_1^\trans\mbf{X}_1^\trans &  \mbf{X}_1 \mbf{F}_2 + \mbf{F}_1^\trans \mbf{X}_2^\trans & \mbf{A}_\mathrm{d} - \mbf{X}_1 + \mbf{F}_1^\trans\mbf{X}_3^\trans & \mbf{B}_\mathrm{d} \\ * & \mbf{P} + \mbf{X}_2 \mbf{F}_2 + \mbf{F}_2^\trans \mbf{X}_2^\trans & - \mbf{1}- \mbf{X}_2 + \mbf{F}_2^\trans \mbf{X}_3^\trans &  \mbf{0}  \\ * & * & -\left( \mbf{X}_3+ \mbf{X}_3^\trans\right) & \mbf{0}  \\ * & * & * & - \mbf{1}   \ebm  &< 0, \\
\bbm - \mbf{Z} + \mbf{X}_4 \mbf{F}_4 + \mbf{F}_4^\trans \mbf{X}_4^\trans & \mbf{X}_4 \mbf{F}_5 + \mbf{F}_4^\trans \mbf{X}_5^\trans & \mbf{C}_\mathrm{d} - \mbf{X}_4 + \mbf{F}_4^\trans \mbf{X}_6^\trans \\ * & \mbf{P} + \mbf{X}_5 \mbf{F}_5 + \mbf{F}_5^\trans \mbf{X}_5^\trans & - \mbf{1} - \mbf{X}_5 + \mbf{F}_5^\trans \mbf{X}_6 ^\trans\\ * & * & - \left( \mbf{X}_6  +  \mbf{X}_6^\trans \right) \ebm &< 0, \\
\trace (\mbf{Z}) &< \mu^2.
\end{align*}

\itemcite \cite[pp.~53--54]{Spagolla2019b} There exist $\mbf{P} \in \mathbb{S}^{n}$, $\mbf{Z} \in \mathbb{S}^p$, $\mbf{Y}_1$,~$\mbf{Y}_2$,~$\mbf{Y}_3$,~$\mbf{Y}_5$,~$\mbf{Y}_6$,~$\mbf{X}_1$,~$\mbf{X}_2$,~$\mbf{X}_3$,~$\mbf{X}_5$,~$\mbf{X}_6 \in \mathbb{R}^{n \times n}$, $\mbf{Y}_4 \in \mathbb{R}^{n \times p}$, $\mbf{X}_4 \in \mathbb{R}^{p \times n}$, and $\mu \in \mathbb{R}_{>0}$, where $\mbf{P} > 0$ and $\mbf{Z} > 0$, such that
\begin{align*}
\bbm - \mbf{P} + \mbf{X}_1\mbf{Y}_1 + \mbf{Y}_1^\trans\mbf{X}_1^\trans &  \mbf{X}_1 \mbf{Y}_2 + \mbf{Y}_1^\trans \mbf{X}_2^\trans & \mbf{A}_\mathrm{d} + \mbf{X}_1\mbf{Y}_3 + \mbf{Y}_1^\trans\mbf{X}_3^\trans & \mbf{B}_\mathrm{d} \\ * & \mbf{P} + \mbf{X}_2 \mbf{Y}_2 + \mbf{Y}_2^\trans \mbf{X}_2^\trans & - \mbf{1} + \mbf{X}_2\mbf{Y}_3 + \mbf{Y}_2^\trans \mbf{X}_3^\trans &  \mbf{0}  \\ * & * & \mbf{X}_3\mbf{Y}_3 + \mbf{Y}_3^\trans\mbf{X}_3^\trans & \mbf{0}  \\ * & * & * & - \mbf{1}   \ebm  &< 0, \\
\bbm - \mbf{Z} + \mbf{X}_4 \mbf{Y}_4 + \mbf{Y}_4^\trans \mbf{X}_4^\trans & \mbf{X}_4 \mbf{Y}_5 + \mbf{Y}_4^\trans \mbf{X}_5^\trans & \mbf{C}_\mathrm{d} + \mbf{X}_4\mbf{Y}_6 + \mbf{Y}_4^\trans \mbf{X}_6^\trans \\ * & \mbf{P} + \mbf{X}_5 \mbf{Y}_5 + \mbf{Y}_5^\trans \mbf{X}_5^\trans & - \mbf{1} + \mbf{X}_5 \mbf{Y}_6 + \mbf{Y}_5^\trans \mbf{X}_6 ^\trans\\ * & * & \mbf{X}_6 \mbf{Y}_6 + \mbf{Y}_6^\trans \mbf{X}_6^\trans \ebm &< 0, \\
\trace (\mbf{Z}) &< \mu^2.
\end{align*}

\end{enumerate}

The $\mathcal{H}_2$ norm of $\mbc{G}$ is the minimum value of $\mu \in \mathbb{R}_{>0}$ that satisfies any of the above conditions.

\subsubsection[Discrete-Time \texorpdfstring{$\mathcal{H}_2$}{H2} Norm With Feedthrough]{Discrete-Time \texorpdfstring{$\mathcal{H}_2$}{H2} Norm With Feedthrough}
\index{norm!discrete-time $\mathcal{H}_2$ norm}
Consider a discrete-time LTI system, $\mbc{G}: \ell_{2e} \to \ell_{2e}$, with state-space realization $(\mbf{A}_\mathrm{d},\mbf{B}_\mathrm{d},\mbf{C}_\mathrm{d},\mbf{D}_\mathrm{d})$, where $\mbf{A}_\mathrm{d} \in \mathbb{R}^{n \times n}$, $\mbf{B}_\mathrm{d} \in \mathbb{R}^{n \times m}$, $\mbf{C}_\mathrm{d} \in \mathbb{R}^{p \times n}$, $\mbf{D}_\mathrm{d} \in \mathbb{R}^{p \times m}$, and $\mbf{A}_\mathrm{d}$ is Schur.  The $\mathcal{H}_2$ norm of $\mbc{G}$ is
\bdis
\norm{\mbc{G}}_2 = \sqrt{\trace (\mbf{C}_\mathrm{d} \mbf{W} \mbf{C}_\mathrm{d}^\trans + \mbf{D}_\mathrm{d} \mbf{D}_\mathrm{d}^\trans) } = \sqrt{\trace (\mbf{B}_\mathrm{d}^\trans \mbf{M} \mbf{B}_\mathrm{d} + \mbf{D}_\mathrm{d}^\trans \mbf{D}_\mathrm{d}) },
\edis
where $\mbf{W}$,~$\mbf{M} \in \mathbb{S}^n$, $\mbf{W} > 0$, $\mbf{M} > 0$, and
\bdis
\mbf{A}_\mathrm{d} \mbf{W} \mbf{A}_\mathrm{d}^\trans - \mbf{W} +  \mbf{B}_\mathrm{d} \mbf{B}_\mathrm{d}^\trans = \mbf{0}, \hspace{10pt} \mbf{A}_\mathrm{d}^\trans \mbf{M} \mbf{A}_\mathrm{d} - \mbf{M} +  \mbf{C}_\mathrm{d}^\trans \mbf{C}_\mathrm{d} = \mbf{0}.
\edis
The inequality $\norm{\mbc{G}}_2 < \mu$ holds under any of following equivalent necessary and sufficient conditions.

\begin{enumerate}

\itemcite \cite{Steentjes2020} There exist $\mbf{Q} \in \mathbb{S}^n$ and $\mu \in \mathbb{R}_{>0}$, where $\mbf{Q} >0 $, such that
\begin{align*}
\mbf{A}_\mathrm{d}^\trans \mbf{Q} \mbf{A}_\mathrm{d} - \mbf{Q} +\mbf{C}_\mathrm{d}^\trans \mbf{C}_\mathrm{d} < 0, \\
\trace \left( \mbf{B}_\mathrm{d}^\trans \mbf{Q} \mbf{B}_\mathrm{d} + \mbf{D}_\mathrm{d}^\trans \mbf{D}_\mathrm{d} \right) &< \mu^2. 
\end{align*}

\itemcite \cite{DeCaigny2010} There exist $\mbf{P} \in \mathbb{S}^n$ and $\mu \in \mathbb{R}_{>0}$, where $\mbf{P} >0 $, such that
\begin{align*}
\mbf{A}_\mathrm{d}\mbf{P} \mbf{A}_\mathrm{d}^\trans - \mbf{P} +\mbf{B}_\mathrm{d} \mbf{B}_\mathrm{d}^\trans < 0, \\
\trace \left( \mbf{C}_\mathrm{d} \mbf{P} \mbf{C}_\mathrm{d}^\trans + \mbf{D}_\mathrm{d} \mbf{D}_\mathrm{d}^\trans \right) &< \mu^2. 
\end{align*}

\itemcite \cite{DeCaigny2010} There exist $\mbf{Q} \in \mathbb{S}^n$, $\mbf{Z} \in \mathbb{S}^m$, and $\mu \in \mathbb{R}_{>0}$, where $\mbf{Q} >0 $ and $\mbf{Z} > 0$, such that
\begin{align}
\bbm \mbf{Q}  \mbf{A}_\mathrm{d}^\trans\mbf{Q}\mbf{A}_\mathrm{d} & \mbf{C}_\mathrm{d}^\trans \\  * & \mbf{1} \ebm &> 0, \label{eq:DT_D_H2_5a}\\
\bbm \mbf{Z} - \mbf{D}_\mathrm{d}^\trans\mbf{D}_\mathrm{d} & \mbf{B}_\mathrm{d}^\trans \mbf{Q} \\ * & \mbf{Q}  \ebm &> 0,  \label{eq:DT_D_H2_5b}\\
\trace (\mbf{Z}) &< \mu^2. \nonumber 
\end{align}

\itemcite \cite{DeCaigny2010} There exist $\mbf{P} \in \mathbb{S}^n$, $\mbf{Z} \in \mathbb{S}^p$, and $\mu \in \mathbb{R}_{>0}$, where $\mbf{P} >0 $ and $\mbf{Z} > 0$, such that
\begin{align}
\bbm \mbf{P} - \mbf{A}_\mathrm{d}\mbf{P}\mbf{A}_\mathrm{d}^\trans & \mbf{B}_\mathrm{d} \\  * & \mbf{1} \ebm &> 0,\label{eq:DT_D_H2_6a}\\
\bbm \mbf{Z} -\mbf{D}_\mathrm{d}\mbf{D}_\mathrm{d}^\trans & \mbf{C}_\mathrm{d} \mbf{P}  \\ * & \mbf{P}  \ebm &> 0,  \label{eq:DT_D_H2_6b}\\
\trace (\mbf{Z}) &< \mu^2. \nonumber
\end{align}

\itemcite \cite[p.~25]{Santos2017} There exist $\mbf{Q} \in \mathbb{S}^n$, $\mbf{Z} \in \mathbb{S}^m$, and $\mu \in \mathbb{R}_{>0}$, where $\mbf{Q} >0 $ and $\mbf{Z} > 0$, such that
\begin{align}
\bbm \mbf{Q} & \mbf{A}_\mathrm{d}\mbf{Q} & \mbf{C}_\mathrm{d}^\trans \\ * & \mbf{Q} & \mbf{0} \\ * & * & \mbf{1} \ebm &> 0, \label{eq:DT_D_H2_3a}\\
\bbm \mbf{Z}  & \mbf{B}_\mathrm{d}^\trans \mbf{Q} & \mbf{D}_\mathrm{d}^\trans \\ * & \mbf{Q} & \mbf{0} \\ * & * & \mbf{1} \ebm &> 0,  \label{eq:DT_D_H2_3b}\\
\trace (\mbf{Z}) &< \mu^2. \label{eq:DT_D_H2_3c}
\end{align}
\begin{proof}
Applying the Schur complement to~\eqref{eq:DT_D_H2_5a} and~\eqref{eq:DT_D_H2_5b} yields~\eqref{eq:DT_D_H2_3a} and~\eqref{eq:DT_D_H2_3b}.
\end{proof}

\itemcite \cite[p.~26]{Santos2017} There exist $\mbf{P} \in \mathbb{S}^n$, $\mbf{Z} \in \mathbb{S}^p$, and $\mu \in \mathbb{R}_{>0}$, where $\mbf{P} >0 $ and $\mbf{Z} > 0$, such that
\begin{align}
\bbm \mbf{P} & \mbf{A}_\mathrm{d}^\trans\mbf{P} & \mbf{B}_\mathrm{d} \\ * & \mbf{P} & \mbf{0} \\ * & * & \mbf{1} \ebm &> 0,\label{eq:DT_D_H2_4a}\\
\bbm \mbf{Z}  & \mbf{C}_\mathrm{d} \mbf{P} & \mbf{D}_\mathrm{d} \\ * & \mbf{P} & \mbf{0} \\ * & * & \mbf{1} \ebm &> 0,  \label{eq:DT_D_H2_4b}\\
\trace (\mbf{Z}) &< \mu^2. \label{eq:DT_D_H2_4c}
\end{align}
\begin{proof}
Applying the Schur complement to~\eqref{eq:DT_D_H2_6a} and~\eqref{eq:DT_D_H2_6b} yields~\eqref{eq:DT_D_H2_4a} and~\eqref{eq:DT_D_H2_4b}.
\end{proof}

\item There exist $\mbf{P} \in \mathbb{S}^n$, $\mbf{Z} \in \mathbb{S}^m$, and $\mu \in \mathbb{R}_{>0}$, where $\mbf{P} >0 $ and $\mbf{Z} > 0$, such that
\begin{align*}
\bbm \mbf{P} & \mbf{P} \mbf{A}_\mathrm{d} & \mbf{P}\mbf{C}_\mathrm{d}^\trans \\ * & \mbf{P} & \mbf{0} \\ * & * & \mbf{1} \ebm &> 0, \\
\bbm \mbf{Z}  & \mbf{B}_\mathrm{d}^\trans  & \mbf{D}_\mathrm{d}^\trans \\ * & \mbf{P} & \mbf{0} \\ * & * & \mbf{1} \ebm &> 0,  \\
\trace (\mbf{Z}) &< \mu^2. 
\end{align*}
\begin{proof}
Apply the congruence transformation $\mbf{W}_1 = \mathrm{diag}\{\mbf{P},\mbf{P},\mbf{1}\}$ to~\eqref{eq:DT_D_H2_3a} and $\mbf{W}_2 = \mathrm{diag}\{\mbf{1},\mbf{P},\mbf{1}\}$ to~\eqref{eq:DT_D_H2_3b}, where $\mbf{P} = \mbf{Q}^{-1}$.
\end{proof}

\itemcite \cite{Geromel1993} There exist $\mbf{Q} \in \mathbb{S}^n$, $\mbf{Z} \in \mathbb{S}^p$, and $\mu \in \mathbb{R}_{>0}$, where $\mbf{Q} >0 $ and $\mbf{Z} > 0$, such that
\begin{align*}
\bbm \mbf{Q} & \mbf{Q}\mbf{A}_\mathrm{d}^\trans& \mbf{Q} \mbf{B}_\mathrm{d} \\ * & \mbf{Q} & \mbf{0} \\ * & * & \mbf{1} \ebm &> 0, \\
\bbm \mbf{Z}  & \mbf{C}_\mathrm{d}& \mbf{D}_\mathrm{d} \\ * & \mbf{Q} & \mbf{0} \\ * & * & \mbf{1} \ebm &> 0,  \\
\trace (\mbf{Z}) &< \mu^2. 
\end{align*}

\itemcite \cite{DeCaigny2010} There exist $\mbf{P} \in \mathbb{S}^n$, $\mbf{Z} \in \mathbb{S}^p$, $\mbf{X} \in \mathbb{R}^{n \times n}$, and $\mu \in \mathbb{R}_{>0}$, where $\mbf{P} >0 $ and $\mbf{Z} > 0$, such that
\begin{align*}
\bbm \mbf{P} & \mbf{A}_\mathrm{d}\mbf{X} & \mbf{B}_\mathrm{d} \\ * & \mbf{X} + \mbf{X}^\trans - \mbf{P} & \mbf{0} \\ * & * & \mbf{1} \ebm &> 0, \\
\bbm \mbf{Z} & \mbf{C}_\mathrm{d}\mbf{X} & \mbf{D}_\mathrm{d} \\ * & \mbf{X} + \mbf{X}^\trans - \mbf{P} & \mbf{0} \\ * & * & \mbf{1} \ebm &> 0, \\
\trace (\mbf{Z}) &< \mu^2. 
\end{align*}

\itemcite \cite[pp.~26--27]{Santos2017} There exist $\mbf{Q} \in \mathbb{S}^n$, $\mbf{Z} \in \mathbb{S}^m$, $\mbf{X} \in \mathbb{R}^{n \times n}$, and $\mu \in \mathbb{R}_{>0}$, where $\mbf{Q} >0 $ and $\mbf{Z} > 0$, such that
\begin{align*}
\bbm \mbf{P} & \mbf{A}_\mathrm{d}^\trans \mbf{X} & \mbf{C}_\mathrm{d}^\trans \\ * & \mbf{X} + \mbf{X}^\trans - \mbf{P} & \mbf{0} \\ * & * & \mbf{1} \ebm &> 0, \\
\bbm \mbf{Z} & \mbf{B}_\mathrm{d}^\trans\mbf{X} & \mbf{D}_\mathrm{d}^\trans \\ * & \mbf{X} + \mbf{X}^\trans - \mbf{P} & \mbf{0} \\ * & * & \mbf{1} \ebm &> 0, \\
\trace (\mbf{Z}) &< \mu^2. 
\end{align*}

\itemcite \cite[pp.~53--54]{Spagolla2019b} There exist $\mbf{P} \in \mathbb{S}^{n}$, $\mbf{Z} \in \mathbb{S}^p$, $\mbf{F}_1$,~$\mbf{F}_2$,~$\mbf{F}_5 \in \mathbb{R}^{n \times n}$, $\mbf{F}_4 \in \mathbb{R}^{n \times p}$, and $\mu \in \mathbb{R}_{>0}$, where $\mbf{P} > 0$ and $\mbf{Z} > 0$, such that
\begin{align*}
\bbm - \mbf{P} + \mbf{A}_\mathrm{d}\mbf{F}_1 + \mbf{F}_1^\trans\mbf{A}_\mathrm{d}^\trans &  \mbf{A}_\mathrm{d} \mbf{F}_2 - \mbf{F}_1^\trans & \mbf{B}_\mathrm{d} \\ * & \mbf{P} - \left(\mbf{F}_2 + \mbf{F}_2^\trans \right)  &  \mbf{0} \\   * & * & -\gamma \mbf{1}  \ebm  &< 0, \\
\bbm - \mbf{Z} + \mbf{C}_\mathrm{d}\mbf{F}_4 + \mbf{F}_4^\trans\mbf{C}_\mathrm{d}^\trans &  \mbf{C}_\mathrm{d} \mbf{F}_5 - \mbf{F}_4^\trans  & \mbf{D}_\mathrm{d} \\ * & \mbf{P} - \left(\mbf{F}_5 + \mbf{F}_5^\trans \right) & \mbf{0} \\ * & * & -\mbf{1}  \ebm  &< 0, \\
\trace (\mbf{Z}) &< \mu^2.
\end{align*}

\itemcite \cite[pp.~53--54]{Spagolla2019b} There exist $\mbf{P} \in \mathbb{S}^{n}$, $\mbf{Z} \in \mathbb{S}^p$, $\mbf{F}_1$,~$\mbf{F}_2$,~$\mbf{F}_5$,~$\mbf{X}_1$,~$\mbf{X}_2$,~$\mbf{X}_3$,~$\mbf{X}_5$,~$\mbf{X}_6 \in \mathbb{R}^{n \times n}$, $\mbf{F}_4 \in \mathbb{R}^{n \times p}$, $\mbf{X}_4 \in \mathbb{R}^{p \times n}$, and $\mu \in \mathbb{R}_{>0}$, where $\mbf{P} > 0$ and $\mbf{Z} > 0$, such that
\begin{align*}
\bbm - \mbf{P} + \mbf{X}_1\mbf{F}_1 + \mbf{F}_1^\trans\mbf{X}_1^\trans &  \mbf{X}_1 \mbf{F}_2 + \mbf{F}_1^\trans \mbf{X}_2^\trans & \mbf{A}_\mathrm{d} - \mbf{X}_1 + \mbf{F}_1^\trans\mbf{X}_3^\trans & \mbf{B}_\mathrm{d} \\ * & \mbf{P} + \mbf{X}_2 \mbf{F}_2 + \mbf{F}_2^\trans \mbf{X}_2^\trans & - \mbf{1}- \mbf{X}_2 + \mbf{F}_2^\trans \mbf{X}_3^\trans &  \mbf{0}  \\ * & * & -\left( \mbf{X}_3+ \mbf{X}_3^\trans\right) & \mbf{0}  \\ * & * & * & - \mbf{1}   \ebm  &< 0, \\
\bbm - \mbf{Z} + \mbf{X}_4 \mbf{F}_4 + \mbf{F}_4^\trans \mbf{X}_4^\trans & \mbf{X}_4 \mbf{F}_5 + \mbf{F}_4^\trans \mbf{X}_5^\trans & \mbf{C}_\mathrm{d} - \mbf{X}_4 + \mbf{F}_4^\trans \mbf{X}_6^\trans & \mbf{D}_\mathrm{d} \\ * & \mbf{P} + \mbf{X}_5 \mbf{F}_5 + \mbf{F}_5^\trans \mbf{X}_5^\trans & - \mbf{1} - \mbf{X}_5 + \mbf{F}_5^\trans \mbf{X}_6 ^\trans & \mbf{0} \\ * & * & - \left( \mbf{X}_6  +  \mbf{X}_6^\trans \right) & \mbf{0} \\ * & * & * & -\mbf{1}\ebm &< 0, \\
\trace (\mbf{Z}) &< \mu^2.
\end{align*}

\itemcite \cite[pp.~53--54]{Spagolla2019b} There exist $\mbf{P} \in \mathbb{S}^{n}$, $\mbf{Z} \in \mathbb{S}^p$, $\mbf{Y}_1$,~$\mbf{Y}_2$,~$\mbf{Y}_3$,~$\mbf{Y}_5$,~$\mbf{Y}_6$,~$\mbf{X}_1$,~$\mbf{X}_2$,~$\mbf{X}_3$,~$\mbf{X}_5$,~$\mbf{X}_6 \in \mathbb{R}^{n \times n}$, $\mbf{Y}_4 \in \mathbb{R}^{n \times p}$, $\mbf{X}_4 \in \mathbb{R}^{p \times n}$, and $\mu \in \mathbb{R}_{>0}$, where $\mbf{P} > 0$ and $\mbf{Z} > 0$, such that
\begin{align*}
\bbm - \mbf{P} + \mbf{X}_1\mbf{Y}_1 + \mbf{Y}_1^\trans\mbf{X}_1^\trans &  \mbf{X}_1 \mbf{Y}_2 + \mbf{Y}_1^\trans \mbf{X}_2^\trans & \mbf{A}_\mathrm{d} + \mbf{X}_1\mbf{Y}_3 + \mbf{Y}_1^\trans\mbf{X}_3^\trans & \mbf{B}_\mathrm{d} \\ * & \mbf{P} + \mbf{X}_2 \mbf{Y}_2 + \mbf{Y}_2^\trans \mbf{X}_2^\trans & - \mbf{1} + \mbf{X}_2\mbf{Y}_3 + \mbf{Y}_2^\trans \mbf{X}_3^\trans &  \mbf{0}  \\ * & * & \mbf{X}_3\mbf{Y}_3 + \mbf{Y}_3^\trans\mbf{X}_3^\trans & \mbf{0}  \\ * & * & * & - \mbf{1}   \ebm  &< 0, \\
\bbm - \mbf{Z} + \mbf{X}_4 \mbf{Y}_4 + \mbf{Y}_4^\trans \mbf{X}_4^\trans & \mbf{X}_4 \mbf{Y}_5 + \mbf{Y}_4^\trans \mbf{X}_5^\trans & \mbf{C}_\mathrm{d} + \mbf{X}_4\mbf{Y}_6 + \mbf{Y}_4^\trans \mbf{X}_6^\trans & \mbf{D}_\mathrm{d} \\ * & \mbf{P} + \mbf{X}_5 \mbf{Y}_5 + \mbf{Y}_5^\trans \mbf{X}_5^\trans & - \mbf{1} + \mbf{X}_5 \mbf{Y}_6 + \mbf{Y}_5^\trans \mbf{X}_6 ^\trans & \mbf{0} \\ * & * & \mbf{X}_6 \mbf{Y}_6 + \mbf{Y}_6^\trans \mbf{X}_6^\trans & \mbf{0} \\ * & * & * & -\mbf{1} \ebm &< 0, \\
\trace (\mbf{Z}) &< \mu^2.
\end{align*}

\end{enumerate}

The $\mathcal{H}_2$ norm of $\mbc{G}$ is the minimum value of $\mu \in \mathbb{R}_{>0}$ that satisfies any of the above conditions.

\subsubsection[Descriptor System \texorpdfstring{$\mathcal{H}_2$}{H2} Norm]{Descriptor System \texorpdfstring{$\mathcal{H}_2$}{H2} Norm}
\index{descriptor systems!$\mathcal{H}_2$ norm}

Consider a descriptor system, $\mbc{G}: \mathcal{L}_{2e} \to \mathcal{L}_{2e}$, described by
\begin{align*}
\mbf{E} \dot{\mbf{x}} &= \mbf{A} \mbf{x} + \mbf{B} \mbf{u},\\
\mbf{y} &= \mbf{C} \mbf{x},
\end{align*}
where $\mbf{E}$,~$\mbf{A} \in \mathbb{R}^{n \times n}$, $\mbf{B} \in \mathbb{R}^{n \times m}$, $\mbf{C} \in \mathbb{R}^{p \times n}$, and it is assumed that the system is regular.  The $\mathcal{H}_2$ norm of $\mbc{G}$ is~\cite{Takaba1997,Takaba1998}
\bdis
\norm{\mbc{G}}_2 = \sqrt{\trace \left(\mbfhat{C} \mbf{E} \mbf{W} \mbfhat{C}^\trans \right) } = \sqrt{\trace \left(\mbfhat{B}^\trans \mbf{M} \mbf{E} \mbfhat{B} \right) }, 
\edis
where $\mbfhat{C} \in \mathbb{R}^{p \times n}$, $\mbfhat{B} \in \mathbb{R}^{n \times m}$, $\mbf{W}$,~$\mbf{M} \in \mathbb{R}^{n \times n}$, $\mbf{C} = \mbfhat{C} \mbf{E}$, $\mbf{B} = \mbf{E} \mbfhat{B}$, $ \mbf{W} \mbf{E}^\trans= \mbf{E} \mbf{W}^\trans > 0$, $\mbf{E}^\trans \mbf{M} = \mbf{M}^\trans \mbf{E} > 0$, and
\bdis
\mbf{A} \mbf{W}^\trans + \mbf{W} \mbf{A}^\trans + \mbf{B} \mbf{B}^\trans = \mbf{0}, \hspace{10pt} \mbf{M}^\trans \mbf{A} + \mbf{A}^\trans \mbf{M} + \mbf{C}^\trans \mbf{C} = \mbf{0}.
\edis

The descriptor system is admissible and the inequality $\norm{\mbc{G}}_2  < \mu$ holds under any of the following equivalent necessary and sufficient conditions.

\begin{enumerate}

\itemcite \cite{Ikeda2000} The descriptor state-space matrices satisfy $\mathcal{R}(\mbf{B}) \subseteq \mathcal{R}(\mbf{E})$ and there exist $\mbf{Q} \in \mathbb{S}^n$, $\mbf{S} \in \mathbb{R}^{(n-n_e)\times (n-n_e}$, $\mbf{U}$,~$\mbf{V} \in \mathbb{R}^{n \times (n-n_e)}$, and $\mu \in \mathbb{R}_{>0}$, where $n_e = \text{rank}(\mbf{E})$, $\mathcal{R}(\mbf{U}) = \mathcal{N}(\mbf{E}^\trans)$, $\mathcal{R}(\mbf{V}) = \mathcal{N}(\mbf{E})$, and $\mbf{Q} > 0$, satisfying \index{rank}
\begin{align*}
\mbf{A}^\trans \left( \mbf{Q} \mbf{E}+ \mbf{U} \mbf{S} \mbf{V}^\trans \right) + \left( \mbf{Q} \mbf{E} + \mbf{U} \mbf{S} \mbf{V}^\trans \right)^\trans \mbf{A} + \mbf{C}^\trans \mbf{C} &< 0, \\
\trace \left( \mbf{B}^\trans \mbf{Q} \mbf{B} \right) &< \mu^2.
\end{align*}

\itemcite \cite{Ikeda2000} The descriptor state-space matrices satisfy $\mathcal{N}(\mbf{E}) \subseteq \mathcal{N}(\mbf{C})$ and there exist $\mbf{P} \in \mathbb{S}^n$, $\mbf{S} \in \mathbb{R}^{(n-n_e)\times (n-n_e}$, $\mbf{U}$,~$\mbf{V} \in \mathbb{R}^{n \times (n-n_e)}$, and $\mu \in \mathbb{R}_{>0}$, where $n_e = \text{rank}(\mbf{E})$, $\mathcal{R}(\mbf{U}) = \mathcal{N}(\mbf{E}^\trans)$, $\mathcal{R}(\mbf{V}) = \mathcal{N}(\mbf{E})$, and $\mbf{P} > 0$, satisfying \index{rank}
\begin{align*}
\mbf{A} \left( \mbf{P} \mbf{E}^\trans + \mbf{V} \mbf{S} \mbf{U}^\trans\right) + \left( \mbf{P} \mbf{E}^\trans + \mbf{V} \mbf{S} \mbf{U}^\trans \right)^\trans \mbf{A}^\trans + \mbf{B} \mbf{B}^\trans &< 0, \\
\trace \left( \mbf{C} \mbf{P} \mbf{C}^\trans \right) &< \mu^2.
\end{align*}

\item The descriptor state-space matrices satisfy $\mathcal{R}(\mbf{B}) \subseteq \mathcal{R}(\mbf{E})$ and there exist $\mbf{Q} \in \mathbb{S}^n$, $\mbf{X} \in \mathbb{R}^{(n-n_e)\times n}$, $\mbf{Z} \in \mathbb{R}^{n \times (n-n_e)}$, and $\mu \in \mathbb{R}_{>0}$, where $n_e = \text{rank}(\mbf{E})$ and $\mbf{Q} > 0$, satisfying $\mbf{E}^\trans \mbf{Z} = \mbf{0}$ and\index{rank}
\begin{align*}
\mbf{A}^\trans \left( \mbf{Q} \mbf{E}+ \mbf{Z} \mbf{X} \right) + \left( \mbf{Q} \mbf{E} + \mbf{Z} \mbf{X} \right)^\trans \mbf{A} + \mbf{C}^\trans \mbf{C} &< 0, \\
\trace \left( \mbf{B}^\trans \mbf{Q} \mbf{B} \right) &< \mu^2.
\end{align*}

\itemcite \cite{Yagoubi2009} The descriptor state-space matrices satisfy $\mathcal{N}(\mbf{E}) \subseteq \mathcal{N}(\mbf{C})$ and there exist $\mbf{P} \in \mathbb{S}^n$, $\mbf{X} \in \mathbb{R}^{(n-n_e)\times n}$, $\mbf{Z} \in \mathbb{R}^{n \times (n-n_e)}$, and $\mu \in \mathbb{R}_{>0}$, where $n_e = \text{rank}(\mbf{E})$ and $\mbf{P} > 0$, satisfying $\mbf{E} \mbf{Z} = \mbf{0}$ and\index{rank}
\begin{align*}
\mbf{A} \left( \mbf{P} \mbf{E}^\trans + \mbf{Z} \mbf{X} \right) + \left( \mbf{P} \mbf{E}^\trans + \mbf{Z} \mbf{X} \right)^\trans \mbf{A}^\trans + \mbf{B} \mbf{B}^\trans &< 0, \\
\trace \left( \mbf{C} \mbf{P} \mbf{C}^\trans \right) &< \mu^2.
\end{align*}

\end{enumerate}

\subsubsection[Discrete-Time Descriptor System \texorpdfstring{$\mathcal{H}_2$}{H2} Norm]{Discrete-Time Descriptor System \texorpdfstring{$\mathcal{H}_2$}{H2} Norm}
\index{descriptor systems!discrete-time $\mathcal{H}_2$ norm}

Consider a discrete-time descriptor system, $\mbc{G}: \ell_{2e} \to \ell_{2e}$, described by
\begin{align*}
\mbf{E}_\mathrm{d} \mbf{x}_{k+1} &= \mbf{A}_\mathrm{d} \mbf{x}_k + \mbf{B}_\mathrm{d} \mbf{u}_k,\\
\mbf{y}_k &= \mbf{C}_\mathrm{d} \mbf{x}_k + \mbf{D}_\mathrm{d} \mbf{u}_k,
\end{align*}
where $\mbf{E}_\mathrm{d}$,~$\mbf{A}_\mathrm{d} \in \mathbb{R}^{n \times n}$, $\mbf{B}_\mathrm{d} \in \mathbb{R}^{n \times m}$, $\mbf{C}_\mathrm{d} \in \mathbb{R}^{p \times n}$, and $\mbf{D}_\mathrm{d} \in \mathbb{R}^{p \times m}$.  The $\mathcal{H}_2$ norm of $\mbc{G}$ is~\cite[pp.~87--88]{Belov2018},~\cite{Yang2002}
\bdis
\norm{\mbc{G}}_2 = \sqrt{\trace \left(\mbf{C}_\mathrm{d} \mbf{W} \mbf{C}_\mathrm{d}^\trans + \mbf{D}_\mathrm{d} \mbf{D}_\mathrm{d}^\trans \right) } = \sqrt{\trace \left(\mbf{B}_\mathrm{d}^\trans \mbf{M}  \mbf{B}_\mathrm{d} + \mbf{D}_\mathrm{d}^\trans \mbf{D}_\mathrm{d} \right) }, 
\edis
where $\mbf{W}$,~$\mbf{M} \in \mathbb{R}^{n \times n}$, $ \mbf{W}  > 0$, $ \mbf{M}  > 0$, 
\bdis
\mbf{A}_\mathrm{d} \mbf{W} \mbf{A}_\mathrm{d}^\trans - \mbf{E}_\mathrm{d}\mbf{W} \mbf{E}_\mathrm{d}^\trans  + \mbf{B}_\mathrm{d} \mbf{B}_\mathrm{d}^\trans = \mbf{0}, \hspace{10pt} \mbf{A}_\mathrm{d}^\trans \mbf{M} \mbf{A}_\mathrm{d} - \mbf{E}_\mathrm{d}^\trans \mbf{M} \mbf{E}_\mathrm{d} + \mbf{C}_\mathrm{d}^\trans \mbf{C}_\mathrm{d} = \mbf{0}.
\edis

The descriptor system is admissible and the inequality $\norm{\mbc{G}}_2  < \mu$ holds under any of the following equivalent necessary and sufficient conditions.

\begin{enumerate}

\itemcite \cite{Kang2018} There exist $\mbf{Q} \in \mathbb{S}^n $, $\mbf{Z} \in \mathbb{S}^m$, and $\gamma \in \mathbb{R}_{>0}$, where $\mbf{Q} > 0$, such that $\mbf{E}_\mathrm{d}^\trans \mbf{Q} \mbf{E}_\mathrm{d} \geq 0$, 
\begin{align}
 \mbf{A}_\mathrm{d}^\trans\mbf{Q}\mbf{A}_\mathrm{d} - \mbf{E}_\mathrm{d}^\trans \mbf{Q} \mbf{E}_\mathrm{d} + \mbf{C}_\mathrm{d}^\trans \mbf{C}_\mathrm{d} &< 0, \label{eq:DT_Des_H2_1a}\\
 \mbf{B}_\mathrm{d}^\trans \mbf{Q} \mbf{B}_\mathrm{d} + \mbf{D}_\mathrm{d}^\trans \mbf{D}_\mathrm{d} - \mbf{Z} &< 0, \label{eq:DT_Des_H2_1b}\\
 \trace (\mbf{Z}) & < \mu^2. \nonumber
\end{align}
Note that in~\cite{Kang2018},~\eqref{eq:DT_Des_H2_1a} is missing the $- \mbf{E}_\mathrm{d}^\trans \mbf{P} \mbf{E}_\mathrm{d}$ term.
\begin{proof}
The proof follows from the definition of the $\mathcal{H}_2$ norm using an approach similar to that in~\cite[pp.~201-211, Proposition~6.13]{Dullerud2000}, where $\trace\left( \mbf{B}_\mathrm{d}^\trans \mbf{Q} \mbf{B}_\mathrm{d} + \mbf{D}_\mathrm{d}^\trans \mbf{D}_\mathrm{d}\right) < \mu^2$ is equivalent to $ \mbf{B}_\mathrm{d}^\trans \mbf{Q} \mbf{B}_\mathrm{d} + \mbf{D}_\mathrm{d}^\trans \mbf{D}_\mathrm{d} - \mbf{Z} < 0$ and  $\trace (\mbf{Z})  < \mu^2$.
\end{proof}

\item There exist $\mbf{P} \in \mathbb{S}^n $, $\mbf{Z} \in \mathbb{S}^p$, and $\gamma \in \mathbb{R}_{>0}$, where $\mbf{P} > 0$, such that $\mbf{E}_\mathrm{d}\mbf{P} \mbf{E}_\mathrm{d}^\trans \geq 0$, 
\begin{align}
 \mbf{A}_\mathrm{d}s\mbf{P}\mbf{A}_\mathrm{d}^\trans - \mbf{E}_\mathrm{d} \mbf{P} \mbf{E}_\mathrm{d}^\trans + \mbf{B}_\mathrm{d} \mbf{B}_\mathrm{d}^\trans &< 0, \label{eq:DT_Des_H2_2a}\\
 \mbf{C}_\mathrm{d} \mbf{P} \mbf{C}_\mathrm{d}^\trans + \mbf{D}_\mathrm{d} \mbf{D}_\mathrm{d}^\trans - \mbf{Z} &< 0, \label{eq:DT_Des_H2_2b}\\
 \trace (\mbf{Z}) & < \mu^2. \nonumber
\end{align}
\begin{proof}
The proof follows from the definition of the $\mathcal{H}_2$ norm using an approach similar to that in~\cite[pp.~201-211, Proposition~6.13]{Dullerud2000}, where $\trace\left(  \mbf{C}_\mathrm{d} \mbf{P} \mbf{C}_\mathrm{d}^\trans + \mbf{D}_\mathrm{d} \mbf{D}_\mathrm{d}^\trans \right) < \mu^2$ is equivalent to $ \mbf{C}_\mathrm{d} \mbf{P} \mbf{C}_\mathrm{d}^\trans + \mbf{D}_\mathrm{d} \mbf{D}_\mathrm{d}^\trans - \mbf{Z} < 0$ and  $\trace (\mbf{Z})  < \mu^2$.
\end{proof}

\item There exist $\mbf{Q} \in \mathbb{S}^n $, $\mbf{Z} \in \mathbb{S}^m$, and $\gamma \in \mathbb{R}_{>0}$, where $\mbf{Q} > 0$, such that $\mbf{E}_\mathrm{d}^\trans \mbf{Q} \mbf{E}_\mathrm{d} \geq 0$, 
\begin{align}
\bbm \mbf{E}_\mathrm{d}^\trans \mbf{Q} \mbf{E}_\mathrm{d}  & \mbf{A}_\mathrm{d}\mbf{Q} & \mbf{C}_\mathrm{d}^\trans \\ * & \mbf{Q} & \mbf{0} \\ * & * & \mbf{1} \ebm &> 0, \label{eq:DT_Des_H2_3a}\\
\bbm \mbf{Z}  & \mbf{B}_\mathrm{d}^\trans \mbf{Q} & \mbf{D}_\mathrm{d}^\trans \\ * & \mbf{Q} & \mbf{0} \\ * & * & \mbf{1} \ebm &> 0,  \label{eq:DT_Des_H2_3b}\\
\trace (\mbf{Z}) &< \mu^2. \nonumber
\end{align}
\begin{proof}
Applying the Schur complement to~\eqref{eq:DT_Des_H2_1a} and~\eqref{eq:DT_Des_H2_1b} yields~\eqref{eq:DT_Des_H2_3a} and~\eqref{eq:DT_Des_H2_3b}.  
\end{proof}

\item There exist $\mbf{P} \in \mathbb{S}^n $, $\mbf{Z} \in \mathbb{S}^o$, and $\gamma \in \mathbb{R}_{>0}$, where $\mbf{P} > 0$, such that $\mbf{E}_\mathrm{d}\mbf{P} \mbf{E}_\mathrm{d}^\trans \geq 0$, 
\begin{align}
\bbm \mbf{E}_\mathrm{d} \mbf{P} \mbf{E}_\mathrm{d}^\trans  & \mbf{A}_\mathrm{d}^\trans \mbf{P} & \mbf{B}_\mathrm{d} \\ * & \mbf{P} & \mbf{0} \\ * & * & \mbf{1} \ebm &> 0, \label{eq:DT_Des_H2_4a}\\
\bbm \mbf{Z}  & \mbf{C}_\mathrm{d} \mbf{P} & \mbf{D}_\mathrm{d}\\ * & \mbf{P} & \mbf{0} \\ * & * & \mbf{1} \ebm &> 0,  \label{eq:DT_Des_H2_4b}\\
\trace (\mbf{Z}) &< \mu^2. \nonumber
\end{align}
\begin{proof}
Applying the Schur complement to~\eqref{eq:DT_Des_H2_2a} and~\eqref{eq:DT_Des_H2_2b} yields~\eqref{eq:DT_Des_H2_4a} and~\eqref{eq:DT_Des_H2_4b}.  
\end{proof}

\end{enumerate}

\subsection[Generalized \texorpdfstring{$\mathcal{H}_2$}{H2} Norm (Induced \texorpdfstring{$\mathcal{L}_2$}{L2}-\texorpdfstring{$\mathcal{L}_\infty$}{L-Infinity} Norm)]{Generalized \texorpdfstring{$\mathcal{H}_2$}{H2} Norm (Induced \texorpdfstring{$\mathcal{L}_2$}{L2}-\texorpdfstring{$\mathcal{L}_\infty$}{L-Infinity} Norm)}
\index{norm!generalized $\mathcal{H}_2$ norm}
\index{norm!induced $\mathcal{L}_2$-$\mathcal{L}_\infty$ norm}

Consider a continuous-time LTI system, $\mbc{G}: \mathcal{L}_{2e} \to \mathcal{L}_{2e}$, with state-space realization $(\mbf{A},\mbf{B}_,\mbf{C},\mbf{0})$, where $\mbf{A} \in \mathbb{R}^{n \times n}$, $\mbf{B} \in \mathbb{R}^{n \times m}$, $\mbf{C} \in \mathbb{R}^{p \times n}$, and $\mbf{A}$ is Hurwitz.  The generalized $\mathcal{H}_2$ norm of $\mbc{G}$ is
\bdis
\norm{\mbc{G}}_{2,\infty} = \sup_{\mbf{u} \in \mathcal{L}_2, \mbf{u} \neq \mbf{0}} \frac{\norm{\mbc{G} \mbf{u}}_\infty}{\norm{\mbf{u}}_2}.
\edis
The inequality $\norm{\mbc{G}}_{2,\infty}  < \mu$ holds under any of following equivalent necessary and sufficient conditions.

\begin{enumerate}

\itemcite \cite[p.~79]{SchererWeiland2015},~\cite{Scherer1997} There exist $\mbf{P} \in \mathbb{S}^n$ and $\mu \in \mathbb{R}_{>0}$, where $\mbf{P} >0 $, such that
\begin{align*}
\bbm \mbf{A}^\trans \mbf{P} + \mbf{P}\mbf{A} & \mbf{P}\mbf{B} \\ * & -\mu \mbf{1} \ebm &< 0,  \\
\bbm \mbf{P} & \mbf{C}^\trans \\ * & \mu \mbf{1} \ebm &> 0.
\end{align*}

\itemcite \cite{Rotea1993} There exist $\mbf{Q} \in \mathbb{S}^n$ and $\mu \in \mathbb{R}_{>0}$, where $\mbf{Q} >0 $, such that
\begin{align*}
\bbm \mbf{Q}\mbf{A}^\trans  + \mbf{A}\mbf{Q} & \mbf{B} \\ * & -\mu \mbf{1} \ebm &< 0, \\
\bbm \mbf{Q} & \mbf{Q}\mbf{C}^\trans \\ * & \mu \mbf{1} \ebm &> 0.
\end{align*}

\item  There exist $\mbf{P} \in \mathbb{S}^n$, $\mbf{V} \in \mathbb{R}^{n \times n}$, and $\mu \in \mathbb{R}_{>0}$, where $\mbf{P} >0 $, such that
\begin{align}
\bbm -\left(\mbf{V} + \mbf{V}^\trans\right) & \mbf{V}^\trans\mbf{A} + \mbf{P} & \mbf{V}^\trans\mbf{B} & \mbf{V}^\trans \\ * & -\mbf{P} & \mbf{0} & \mbf{0} \\ * & * & -\mu \mbf{1} & \mbf{0} \\ * & * & * & -\mbf{P} \ebm &< 0, \nonumber \\
\bbm \mbf{P} & \mbf{C}^\trans \\ * & \mu \mbf{1} \ebm &> 0. \nonumber
\end{align}

\begin{proof}
Identical to the proof in~\cite{Apkarian2001} used to obtain the dilated matrix inequality in~\eqref{eq:H2_6a}.
\end{proof}

%

\end{enumerate}

The generalized $\mathcal{H}_2$ norm of $\mbc{G}$ is the minimum value of $\mu \in \mathbb{R}_{>0}$ that satisfies any of the above conditions.

\subsection[Peak-to-Peak Norm (Induced \texorpdfstring{$\mathcal{L}_\infty$}{L2}-\texorpdfstring{$\mathcal{L}_\infty$}{L-Infinity} Norm)]{Peak-to-Peak Norm (Induced \texorpdfstring{$\mathcal{L}_\infty$}{L2}-\texorpdfstring{$\mathcal{L}_\infty$}{L-Infinity} Norm)~\cite[p.~80]{SchererWeiland2015},~\cite{Scherer1997}}
\index{norm!peak-to-peak norm}
\index{norm!induced $\mathcal{L}_\infty$-$\mathcal{L}_\infty$ norm}

Consider a continuous-time LTI system, $\mbc{G}: \mathcal{L}_{2e} \to \mathcal{L}_{2e}$, with state-space realization $(\mbf{A},\mbf{B}_,\mbf{C},\mbf{D})$, where $\mbf{A} \in \mathbb{R}^{n \times n}$, $\mbf{B} \in \mathbb{R}^{n \times m}$, $\mbf{C} \in \mathbb{R}^{p \times n}$, $\mbf{D} \in \mathbb{R}^{p \times m}$, and $\mbf{A}$ is Hurwitz.  The peak-to-peak norm of $\mbc{G}$ is
\bdis
\norm{\mbc{G}}_{\infty,\infty} = \sup_{\mbf{u} \in \mathcal{L}_\infty, \mbf{u} \neq \mbf{0}} \frac{\norm{\mbc{G} \mbf{u}}_\infty}{\norm{\mbf{u}}_\infty}.
\edis
The inequality $\norm{\mbc{G}}_{\infty,\infty}  < \mu$ holds under any of the following equivalent sufficient conditions.
\begin{enumerate}

\item There exist $\mbf{P} \in \mathbb{S}^n$ and $\lambda$,~$\epsilon$,~$\mu \in \mathbb{R}_{>0}$, where $\mbf{P} >0 $, such that
\begin{align*}
\bbm \mbf{A}^\trans \mbf{P} + \mbf{P}\mbf{A} + \lambda \mbf{P} & \mbf{P}\mbf{B} \\ * & -\epsilon \mbf{1} \ebm &< 0,  \\
\bbm \lambda \mbf{P} & \mbf{0} & \mbf{C}^\trans \\ * & (\mu - \epsilon)\mbf{1} & \mbf{D}^\trans \\ * & * & \mu \mbf{1} \ebm &> 0.
\end{align*}

\item There exist $\mbf{Q} \in \mathbb{S}^n$ and $\lambda$,~$\epsilon$,~$\mu \in \mathbb{R}_{>0}$, where $\mbf{Q} >0 $, such that
\begin{align*}
\bbm \mbf{Q}\mbf{A}^\trans + \mbf{A} \mbf{Q} + \lambda \mbf{Q} & \mbf{B} \\ * & -\epsilon \mbf{1} \ebm &< 0, \\
\bbm \lambda \mbf{Q} & \mbf{0} & \mbf{Q}\mbf{C}^\trans \\ * & (\mu - \epsilon)\mbf{1} & \mbf{D}^\trans \\ * & * & \mu \mbf{1} \ebm &> 0. 
\end{align*}

\end{enumerate}

The peak-to-peak norm of $\mbc{G}$ is smaller than any $\mu \in \mathbb{R}_{>0}$ that satisfies either of the above conditions.


\subsection{Kalman-Yakubovich-Popov (KYP) Lemma}
\index{Kalman-Yakubovich-Popov (KYP) lemma}

\subsubsection[KYP Lemma for QSR Dissipative Systems]{KYP Lemma for QSR Dissipative Systems~\cite{Willems1971,Rantzer1996,Kottenstette2014}}
Consider a continuous-time LTI system, $\mbc{G}: \mathcal{L}_{2e} \to \mathcal{L}_{2e}$, with minimal state-space realization $(\mbf{A},\mbf{B}_,\mbf{C},\mbf{D})$, where $\mbf{A} \in \mathbb{R}^{n \times n}$, $\mbf{B} \in \mathbb{R}^{n \times m}$, $\mbf{C} \in \mathbb{R}^{p \times n}$, and $\mbf{D} \in \mathbb{R}^{p \times m}$.  The system $\mbc{G}$ is QSR dissipative~\cite{Willems1972,Hill1976} \index{QSR dissipative} if
\bdis
\int_0^T \left(\mbf{y}^\trans(t) \mbf{Q} \mbf{y}(t) + 2 \mbf{y}^\trans(t) \mbf{S} \mbf{u}(t) + \mbf{u}^\trans(t) \mbf{R} \mbf{u}(t) \right) \mathrm{d} t \geq 0, \quad \forall \mbf{u} \in \mathcal{L}_{2e}, \quad \forall T \in \mathbb{R}_{\geq 0},
\edis
where $\mbf{u}(t)$ is the input to $\mbc{G}$, $\mbf{y}(t)$ is the output of $\mbc{G}$, $\mbf{Q} \in \mathbb{S}^p$, $\mbf{S} \in \mathbb{R}^{p \times m}$, and $\mbf{R} \in \mathbb{S}^m$.  The system $\mbc{G}$ is also QSR dissipative if and only if there exists $\mbf{P} \in \mathbb{S}^{n}$, where $\mbf{P}  > 0$, such that
\bdis
\bbm \mbf{P}\mbf{A} + \mbf{A}^\trans\mbf{P} - \mbf{C}^\trans \mbf{Q} \mbf{C} &  \mbf{P}\mbf{B} - \mbf{C}^\trans \mbf{S} - \mbf{C}^\trans \mbf{Q} \mbf{D} \\ * &  -\mbf{D}^\trans\mbf{Q} \mbf{D} - \left(\mbf{D}^\trans \mbf{S} + \mbf{S}^\trans \mbf{D} \right) - \mbf{R} \ebm \leq 0.
\edis

Note that the Bounded Real Lemma (Section~\ref{sec:BoundedRealLemma}) is a special case of the KYP Lemma for QSR dissipative systems with $\mbf{Q} = - \mbf{1}$, $\mbf{S} = \mbf{0}$, and $\mbf{R} = \gamma^2 \mbf{1}$.

\subsubsection[Discrete-Time KYP Lemma for QSR Dissipative Systems]{Discrete-Time KYP Lemma for QSR Dissipative Systems~\cite{Kottenstette2014},~\cite[p.~495]{Goodwin1984}}
Consider a discrete-time LTI system, $\mbc{G}: \ell_{2e} \to \ell_{2e}$, with minimal state-space realization $(\mbf{A}_\mathrm{d},\mbf{B}_\mathrm{d},\mbf{C}_\mathrm{d},\mbf{D}_\mathrm{d})$, where $\mbf{A}_\mathrm{d} \in \mathbb{R}^{n \times n}$, $\mbf{B}_\mathrm{d} \in \mathbb{R}^{n \times m}$, $\mbf{C}_\mathrm{d} \in \mathbb{R}^{p \times n}$, and $\mbf{D}_\mathrm{d} \in \mathbb{R}^{p \times m}$.  The system $\mbc{G}$ is QSR dissipative~\cite{Willems1972,Hill1976} \index{QSR dissipative} if
\bdis
\sum_{i=0}^k \left(\mbf{y}^\trans_i \mbf{Q} \mbf{y}_i + 2 \mbf{y}^\trans_i \mbf{S} \mbf{u}_i + \mbf{u}^\trans_i \mbf{R} \mbf{u}_i \right)  \geq 0, \quad \forall \mbf{u} \in \ell_{2e}, \quad \forall k \in \mathbb{Z}_{\geq 0},
\edis
where $\mbf{u}_k$ is the input to $\mbc{G}$, $\mbf{y}_k$ is the output of $\mbc{G}$, $\mbf{Q} \in \mathbb{S}^p$, $\mbf{S} \in \mathbb{R}^{p \times m}$, and $\mbf{R} \in \mathbb{S}^m$.  The system $\mbc{G}$ is also QSR dissipative if and only if there exists $\mbf{P} \in \mathbb{S}^{n}$, where $\mbf{P}  > 0$, such that
\bdis
\bbm \mbf{A}_\mathrm{d}^\trans\mbf{P}\mbf{A}_\mathrm{d} - \mbf{P} - \mbf{C}_\mathrm{d}^\trans \mbf{Q} \mbf{C}_\mathrm{d} &  \mbf{A}_\mathrm{d}^\trans\mbf{P}\mbf{B}_\mathrm{d} - \mbf{C}_\mathrm{d}^\trans\mbf{S} - \mbf{C}_\mathrm{d}^\trans \mbf{Q} \mbf{D}_\mathrm{d} \\ * & \mbf{B}_\mathrm{d}^\trans\mbf{P}\mbf{B}_\mathrm{d} - \mbf{D}_\mathrm{d}^\trans \mbf{Q} \mbf{D}_\mathrm{d} -\left(\mbf{D}_\mathrm{d}^\trans\mbf{S} + \mbf{S}^\trans\mbf{D}_\mathrm{d} \right) - \mbf{R}  \ebm \leq 0.
\edis

Note that the Discrete-Time Bounded Real Lemma (Section~\ref{sec:DTBoundedRealLemma}) is a special case of the Discrete-Time KYP Lemma for QSR dissipative systems with $\mbf{Q} = - \mbf{1}$, $\mbf{S} = \mbf{0}$, and $\mbf{R} = \gamma^2 \mbf{1}$.

\subsubsection[KYP (Positive Real) Lemma Without Feedthrough]{KYP (Positive Real) Lemma Without Feedthrough~\cite[p.~219]{Marquez2003},~\cite{Anderson1967},~\cite[p.~14]{Bao2007}}
Consider a square, continuous-time LTI system, $\mbc{G}: \mathcal{L}_{2e} \to \mathcal{L}_{2e}$, with minimal state-space realization $(\mbf{A},\mbf{B}_,\mbf{C},\mbf{0})$, where $\mbf{A} \in \mathbb{R}^{n \times n}$, $\mbf{B} \in \mathbb{R}^{n \times m}$, and $\mbf{C} \in \mathbb{R}^{m \times n}$.   The system $\mbc{G}$ is positive real (PR)\index{positive real (PR)} under either of the following equivalent necessary and sufficient conditions.

\begin{enumerate}

\item There exists $\mbf{P} \in \mathbb{S}^{n}$, where $\mbf{P}  > 0$, such that
\begin{align}
\mbf{P}\mbf{A} + \mbf{A}^\trans\mbf{P} &\leq 0, \nonumber \\
\mbf{P}\mbf{B} &= \mbf{C}^\trans. \nonumber
\end{align}
\item There exists $\mbf{Q} \in \mathbb{S}^{n}$, where $\mbf{Q}  > 0$, such that
\begin{align}
\mbf{A}\mbf{Q} + \mbf{Q}\mbf{A}^\trans &\leq 0, \nonumber \\
\mbf{B} &= \mbf{Q}\mbf{C}^\trans. \nonumber
\end{align}
\end{enumerate}
This is a special case of the KYP Lemma for QSR dissipative systems with $\mbf{Q} = \mbf{0}$, $\mbf{S} = \onehalf \cdot \mbf{1}$, and $\mbf{R} = \mbf{0}$.

The system $\mbc{G}$ is strictly positive real (SPR)\index{strictly positive real (SPR)} under either of the following necessary and sufficient conditions.

\begin{enumerate}

\item There exists $\mbf{P} \in \mathbb{S}^{n}$, where $\mbf{P}  > 0$, such that
\begin{align}
\mbf{P}\mbf{A} + \mbf{A}^\trans\mbf{P} &< 0, \nonumber \\
\mbf{P}\mbf{B} &= \mbf{C}^\trans. \nonumber
\end{align}

\item There exists $\mbf{Q} \in \mathbb{S}^{n}$, where $\mbf{Q}  > 0$, such that
\begin{align}
\mbf{A}\mbf{Q} + \mbf{Q}\mbf{A}^\trans &< 0, \nonumber \\
\mbf{B} &= \mbf{Q}\mbf{C}^\trans. \nonumber
\end{align}

\end{enumerate}
This is a special case of the KYP Lemma for QSR dissipative systems with $\mbf{Q} = \epsilon \cdot \mbf{1}$, $\mbf{S} = \onehalf \cdot \mbf{1}$, and $\mbf{R} = \mbf{0}$, where $\epsilon \in \mathbb{R}_{>0}$.

\subsubsection[KYP (Positive Real) Lemma With Feedthrough]{KYP (Positive Real) Lemma With Feedthrough~\cite[p.~25]{Boyd1994},~\cite[p.~218]{Marquez2003},~\cite{Anderson1967},~\cite[pp.~79--80]{Brogliato2007}}
Consider a square, continuous-time LTI system, $\mbc{G}: \mathcal{L}_{2e} \to \mathcal{L}_{2e}$, with minimal state-space realization $(\mbf{A},\mbf{B}_,\mbf{C},\mbf{D})$, where $\mbf{A} \in \mathbb{R}^{n \times n}$, $\mbf{B} \in \mathbb{R}^{n \times m}$, $\mbf{C} \in \mathbb{R}^{m \times n}$, and $\mbf{D} \in \mathbb{R}^{m \times m}$.  The system $\mbc{G}$ is positive real (PR)\index{positive real (PR)} under either of the following equivalent necessary and sufficient conditions.

\begin{enumerate}

\item There exists $\mbf{P} \in \mathbb{S}^{n}$, where $\mbf{P}  > 0$, such that
\bdis
\bbm \mbf{P}\mbf{A} + \mbf{A}^\trans\mbf{P} & \mbf{P}\mbf{B} - \mbf{C}^\trans  \\ * & -\left(\mbf{D} + \mbf{D}^\trans\right) \ebm \leq 0.
\edis

\item There exists $\mbf{Q} \in \mathbb{S}^n$, where $\mbf{Q}  > 0$, such that
\bdis
\bbm \mbf{A}\mbf{Q} +\mbf{Q} \mbf{A}^\trans & \mbf{B} - \mbf{Q}\mbf{C}^\trans  \\ * & -\left(\mbf{D} + \mbf{D}^\trans\right) \ebm \leq 0.
\edis
\end{enumerate}
This is a special case of the KYP Lemma for QSR dissipative systems with $\mbf{Q} = \mbf{0}$, $\mbf{S} = \onehalf \cdot \mbf{1}$, and $\mbf{R} = \mbf{0}$.

The system $\mbc{G}$ is strictly positive real (SPR)\index{strictly positive real (SPR)} under either of the following equivalent necessary and sufficient conditions.

\begin{enumerate}

\item There exists $\mbf{P} \in \mathbb{S}^{n}$, where $\mbf{P}  > 0$, such that
\bdis
\bbm \mbf{P}\mbf{A} + \mbf{A}^\trans\mbf{P} & \mbf{P}\mbf{B} - \mbf{C}^\trans  \\ * & -\left(\mbf{D} + \mbf{D}^\trans\right) \ebm < 0.
\edis

\item There exists $\mbf{Q} \in \mathbb{S}^n$, where $\mbf{Q}  > 0$, such that
\bdis
\bbm \mbf{A}\mbf{Q} +\mbf{Q} \mbf{A}^\trans & \mbf{B} - \mbf{Q}\mbf{C}^\trans  \\ * & -\left(\mbf{D} + \mbf{D}^\trans\right) \ebm < 0.
\edis

\end{enumerate}
This is a special case of the KYP Lemma for QSR dissipative systems with $\mbf{Q} = \epsilon  \mbf{1}$, $\mbf{S} = \onehalf \cdot \mbf{1}$, and $\mbf{R} = \mbf{0}$, where $\epsilon \in \mathbb{R}_{>0}$.

\subsubsection[Discrete-Time KYP (Positive Real) Lemma With Feedthrough]{Discrete-Time KYP (Positive Real) Lemma With Feedthrough~\cite[pp.~171--172]{Brogliato2007},~\cite{Hitz1969},~\cite{Haddad1994}}
Consider a square, discrete-time LTI system, $\mbc{G}: \ell_{2e} \to \ell_{2e}$, with minimal state-space realization $(\mbf{A}_\mathrm{d},\mbf{B}_\mathrm{d},\mbf{C}_\mathrm{d},\mbf{D}_\mathrm{d})$, where $\mbf{A}_\mathrm{d} \in \mathbb{R}^{n \times n}$, $\mbf{B}_\mathrm{d} \in \mathbb{R}^{n \times m}$, $\mbf{C}_\mathrm{d} \in \mathbb{R}^{m \times n}$, and $\mbf{D}_\mathrm{d} \in \mathbb{R}^{m \times m}$.  The system $\mbc{G}$ is positive real (PR)\index{positive real (PR)} under any of the following equivalent necessary and sufficient conditions.

\begin{enumerate}

\itemcite \cite{Wu1996} There exists $\mbf{P} \in \mathbb{S}^{n}$, where $\mbf{P}  > 0$, such that
\bdis
\bbm \mbf{A}_\mathrm{d}^\trans\mbf{P}\mbf{A}_\mathrm{d} - \mbf{P} &  \mbf{A}_\mathrm{d}^\trans\mbf{P}\mbf{B}_\mathrm{d} - \mbf{C}_\mathrm{d}^\trans \\ * & \mbf{B}_\mathrm{d}^\trans\mbf{P}\mbf{B}_\mathrm{d} -\left(\mbf{D}_\mathrm{d} + \mbf{D}_\mathrm{d}^\trans\right)   \ebm \leq 0.
\edis

\item There exists $\mbf{Q} \in \mathbb{S}^{n}$, where $\mbf{Q}  > 0$, such that
\bdis
\bbm \mbf{A}_\mathrm{d}\mbf{Q}\mbf{A}_\mathrm{d}^\trans - \mbf{Q} &  \mbf{A}_\mathrm{d}\mbf{Q}\mbf{C}_\mathrm{d}^\trans - \mbf{B}_\mathrm{d} \\ * & \mbf{C}_\mathrm{d} \mbf{Q}\mbf{C}_\mathrm{d}^\trans -\left(\mbf{D}_\mathrm{d} + \mbf{D}_\mathrm{d}^\trans\right)   \ebm \leq 0.
\edis

\itemcite \cite{Masubuchi1998} There exists $\mbf{P} \in \mathbb{S}^{n}$, where $\mbf{P}  > 0$, such that
\bdis
\bbm   \mbf{P} & \mbf{P}\mbf{A}_\mathrm{d}  &  \mbf{P}\mbf{B}_\mathrm{d}  \\ * & \mbf{P} & \mbf{C}_\mathrm{d}^\trans \\ * & * & \mbf{D}_\mathrm{d} + \mbf{D}_\mathrm{d}^\trans   \ebm \geq 0.
\edis

\item There exists $\mbf{Q} \in \mathbb{S}^{n}$, where $\mbf{Q}  > 0$, such that
\bdis
\bbm  \mbf{Q} &  \mbf{A}_\mathrm{d} \mbf{Q} &\mbf{B}_\mathrm{d} \\ * & \mbf{Q} & \mbf{Q}\mbf{C}_\mathrm{d}^\trans \\ * & * & \mbf{D}_\mathrm{d} + \mbf{D}_\mathrm{d}^\trans   \ebm \geq 0.
\edis

\end{enumerate}
This is a special case of the Discrete-Time KYP Lemma for QSR dissipative systems with $\mbf{Q} = \mbf{0}$, $\mbf{S} = \onehalf \cdot \mbf{1}$, and $\mbf{R} = \mbf{0}$.

The system $\mbc{G}$ is strictly positive real (SPR)\index{strictly positive real (SPR)} under any of the following equivalent necessary and sufficient conditions.
\begin{enumerate}

\item There exists $\mbf{P} \in \mathbb{S}^{n}$, where $\mbf{P}  > 0$, such that
\bdis
\bbm \mbf{A}_\mathrm{d}^\trans\mbf{P}\mbf{A}_\mathrm{d} - \mbf{P} &  \mbf{A}_\mathrm{d}^\trans\mbf{P}\mbf{B}_\mathrm{d} - \mbf{C}_\mathrm{d}^\trans \\ * & \mbf{B}_\mathrm{d}^\trans\mbf{P}\mbf{B}_\mathrm{d} -\left(\mbf{D}_\mathrm{d} + \mbf{D}_\mathrm{d}^\trans\right)   \ebm < 0.
\edis

\item There exists $\mbf{Q} \in \mathbb{S}^{n}$, where $\mbf{Q}  > 0$, such that
\bdis
\bbm \mbf{A}_\mathrm{d}\mbf{Q}\mbf{A}_\mathrm{d}^\trans - \mbf{Q} &  \mbf{A}_\mathrm{d}\mbf{Q}\mbf{C}_\mathrm{d}^\trans - \mbf{B}_\mathrm{d} \\ * & \mbf{C}_\mathrm{d} \mbf{Q}\mbf{C}_\mathrm{d}^\trans -\left(\mbf{D}_\mathrm{d} + \mbf{D}_\mathrm{d}^\trans\right)   \ebm < 0.
\edis

\item There exists $\mbf{P} \in \mathbb{S}^{n}$, where $\mbf{P}  > 0$, such that
\bdis
\bbm   \mbf{P} & \mbf{P}\mbf{A}_\mathrm{d}  &  \mbf{P}\mbf{B}_\mathrm{d}  \\ * & \mbf{P} & \mbf{C}_\mathrm{d}^\trans \\ * & * & \mbf{D}_\mathrm{d} + \mbf{D}_\mathrm{d}^\trans   \ebm > 0.
\edis

\item There exists $\mbf{Q} \in \mathbb{S}^{n}$, where $\mbf{Q}  > 0$, such that
\bdis
\bbm  \mbf{Q} &  \mbf{A}_\mathrm{d} \mbf{Q} &\mbf{B}_\mathrm{d} \\ * & \mbf{Q} & \mbf{Q}\mbf{C}_\mathrm{d}^\trans \\ * & * & \mbf{D}_\mathrm{d} + \mbf{D}_\mathrm{d}^\trans   \ebm > 0.
\edis

\end{enumerate}
This is a special case of the Discrete-Time KYP Lemma for QSR dissipative systems with $\mbf{Q} = \epsilon \mbf{1}$, $\mbf{S} = \onehalf \cdot \mbf{1}$, and $\mbf{R} = \mbf{0}$, where $\epsilon \in \mathbb{R}_{>0}$.

\subsubsection[KYP Lemma for Descriptor Systems]{KYP Lemma for Descriptor Systems~\cite[pp.~91--93]{Brogliato2007},~\cite{Masubuchi2006}}
\index{descriptor systems!KYP lemma}
Consider a square, LTI descriptor system\index{descriptor systems} given by
\begin{align}
\mbf{E}\dot{\mbf{x}} &= \mbf{A}\mbf{x} + \mbf{B}\mbf{u}, \nonumber \\
\mbf{y} &= \mbf{C}\mbf{x}+ \mbf{D} \mbf{u}, \nonumber
\end{align}
where $\mbf{E}$,~$\mbf{A} \in \mathbb{R}^{n \times n}$, $\mbf{B} \in \mathbb{R}^{n \times m}$, $\mbf{C} \in \mathbb{R}^{m \times n}$, and $\mbf{D} \in \mathbb{R}^{m \times m}$.  The system is extended strictly positive real (ESPR)\index{extended strictly positive real (ESPR)} if and only if there exist $\mbf{X} \in \mathbb{R}^{n \times n}$ and $\mbf{W} \in \mathbb{R}^{n \times m}$ such that $\mbf{E}^\trans\mbf{X} = \mbf{X}^\trans\mbf{E} \geq 0$, $\mbf{E}^\trans\mbf{W} = \mbf{0}$, and
\bdis
\bbm \mbf{X}^\trans\mbf{A} + \mbf{A}^\trans\mbf{X} & \mbf{A}^\trans\mbf{W} + \mbf{X}^\trans\mbf{B} - \mbf{C}^\trans \\ * & \mbf{W}^\trans\mbf{B} + \mbf{B}^\trans\mbf{W} -\left(\mbf{D} + \mbf{D}^\trans\right) \ebm < 0.
\edis
The system is also ESPR if there exists $\mbf{X} \in \mathbb{R}^{n \times n}$ such that $\mbf{E}^\trans\mbf{X} = \mbf{X}^\trans\mbf{E} \geq 0$ and~\cite{Freund2004} 
\bdis
\bbm \mbf{X}^\trans\mbf{A} + \mbf{A}^\trans\mbf{X} &  \mbf{X}^\trans\mbf{B} - \mbf{C}^\trans \\ * & -\left(\mbf{D} + \mbf{D}^\trans\right) \ebm < 0.
\edis

\subsubsection[Discrete-Time KYP Lemma for Descriptor Systems]{Discrete-Time KYP Lemma for Descriptor Systems~\cite{Zhang2002,Lee2003}}
Consider a square, discrete-time LTI descriptor system\index{descriptor systems} given by
\begin{align}
\mbf{E}_\mathrm{d}\mbf{x}_{k+1} &= \mbf{A}_\mathrm{d}\mbf{x}_k + \mbf{B}_\mathrm{d}\mbf{u}_k, \nonumber \\
\mbf{y}_k &= \mbf{C}_\mathrm{d}\mbf{x}_k+ \mbf{D}_\mathrm{d} \mbf{u}_k, \nonumber
\end{align}
where $\mbf{E}_\mathrm{d}$,~$\mbf{A}_\mathrm{d} \in \mathbb{R}^{n \times n}$, $\mbf{B}_\mathrm{d} \in \mathbb{R}^{n \times m}$, $\mbf{C}_\mathrm{d} \in \mathbb{R}^{m \times n}$, and $\mbf{D}_\mathrm{d} \in \mathbb{R}^{m \times m}$.  The system is extended strictly positive real (ESPR)\index{extended strictly positive real (ESPR)} if and only if there exists $\mbf{X} \in \mathbb{S}^{n}$ such that $\mbf{E}^\trans\mbf{X}\mbf{E} \geq 0$ and
\bdis
\bbm \mbf{A}_\mathrm{d}^\trans\mbf{X}\mbf{A}_\mathrm{d} - \mbf{E}_\mathrm{d}^\trans \mbf{X} \mbf{E}_\mathrm{d} &  \mbf{A}_\mathrm{d}^\trans\mbf{X}\mbf{B}_\mathrm{d} - \mbf{C}_\mathrm{d}^\trans \\ * & -\left(\mbf{D}_\mathrm{d} + \mbf{D}_\mathrm{d}^\trans - \mbf{B}_\mathrm{d}^\trans \mbf{X} \mbf{B}_\mathrm{d}\right) \ebm < 0.
\edis

\subsubsection{QSR Dissipativity-Related Properties}

\begin{enumerate}

\itemcite \cite{Tang2019} Consider a QSR-dissipative continuous-time LTI system, $\mbc{G}: \mathcal{L}_{2e} \to \mathcal{L}_{2e}$, with minimal state-space realization $(\mbf{A},\mbf{B}_,\mbf{C},\mbf{D})$, where $\mbf{A} \in \mathbb{R}^{n \times n}$, $\mbf{B} \in \mathbb{R}^{n \times m}$, $\mbf{C} \in \mathbb{R}^{p \times n}$, and $\mbf{D} \in \mathbb{R}^{p \times m}$.  The $\mathcal{H}_\infty$ norm of $\mbc{G}$ is less than $\gamma$ (i.e., $\norm{\mbc{G}}_\infty  < \gamma$) if there exist $\alpha$,~$\gamma \in \mathbb{R}_{>0}$ such that $\mbf{1} + \alpha \mbf{Q} < 0$ and
\bdis
\bbm \mbf{1} + \alpha \mbf{Q} & \alpha \mbf{S} \\ * & \alpha \mbf{R} - \gamma^2 \mbf{1} \ebm \leq 0.
\edis

\end{enumerate}

\subsection{Conic Sectors}

\subsubsection[Conic Sector Lemma]{Conic Sector Lemma}
\index{conic sectors!conic sector lemma}
Consider a square, continuous-time LTI system, $\mbc{G}: \mathcal{L}_{2e} \to \mathcal{L}_{2e}$, with minimal state-space realization $(\mbf{A},\mbf{B}_,\mbf{C},\mbf{D})$, where $\mbf{A} \in \mathbb{R}^{n \times n}$, $\mbf{B} \in \mathbb{R}^{n \times m}$, $\mbf{C} \in \mathbb{R}^{m \times n}$, and $\mbf{D} \in \mathbb{R}^{m \times m}$.

The system $\mbc{G}$ is inside the cone $[a,b]$, where $a$,~$b \in \mathbb{R}$, and $a < b$, under any of the following equivalent necessary and sufficient conditions.

\begin{enumerate}

\itemcite \cite{Gupta1994} There exists $\mbf{P} \in \mathbb{S}^n$, where $\mbf{P}  > 0$, such that
\beq
\label{eq:CSL_1a}
 \bbm \mbf{P}\mbf{A}+\mbf{A}^\trans\mbf{P}+\mbf{C}^\trans\mbf{C} & \mbf{P}\mbf{B}-\frac{a+b}{2}\mbf{C}^\trans+\mbf{C}^\trans\mbf{D} \\ * & \mbf{D}^\trans\mbf{D}-\frac{a+b}{2}\left(\mbf{D}+\mbf{D}^\trans\right) + ab\mbf{1} \ebm \leq 0.
\eeq
Note that the matrix inequality of~\eqref{eq:CSL_1a} does not allow for the case where the upper bound $b$ is infinite.

\itemcite \cite[p.~28]{Forbes2011} There exists $\mbf{P} \in \mathbb{S}^n$, where $\mbf{P}  > 0$, such that
\bdis
\bbm \mbf{P}\mbf{A}+\mbf{A}^\trans\mbf{P}+\frac{1}{b}\mbf{C}^\trans\mbf{C} & \mbf{P}\mbf{B}-\onehalf\left(\frac{a}{b}+1\right)\mbf{C}^\trans+\frac{1}{b}\mbf{C}^\trans\mbf{D} \\ * & \frac{1}{b}\mbf{D}^\trans\mbf{D}-\onehalf\left(\frac{a}{b}+1\right)\left(\mbf{D}+\mbf{D}^\trans\right) + a\mbf{1} \ebm \leq 0.
\edis


\itemcite \cite{Bridgeman2014b} There exists $\mbf{P} \in \mathbb{S}^n$, where $\mbf{P}  > 0$, such that
\bdis
\bbm \mbf{P} \mbf{A} + \mbf{A}^\trans\mbf{P} & \mbf{P}\mbf{B} & \mbf{C}^\trans \\ * & -\frac{(a-b)^2}{4b} \mbf{1} & \mbf{D}^\trans - \frac{a+b}{2}\mbf{1} \\ * & * & -b\mbf{1} \ebm \leq 0.
\edis

\item There exists $\mbf{Q} \in \mathbb{S}^n$, where $\mbf{Q}  > 0$, such that
\bdis
\bbm \mbf{A}\mbf{Q} + \mbf{Q}\mbf{A}^\trans & \mbf{B} & \mbf{Q}\mbf{C}^\trans \\ * & -\frac{(a-b)^2}{4b} \mbf{1} & \mbf{D}^\trans - \frac{a+b}{2}\mbf{1} \\ * & * & -b\mbf{1} \ebm \leq 0.
\edis

\end{enumerate}

The system $\mbc{G}$ is inside the cone of radius $r$ centered at $c$, where $r \in \mathbb{R}_{>0}$ and $b \in \mathbb{R}$, under any of the following equivalent necessary and sufficient conditions.

\begin{enumerate}

\itemcite \cite{Joshi2002},~\cite[pp.~23--24]{BridgemanThesis} There exists $\mbf{P} \in \mathbb{S}^n$, where $\mbf{P}  > 0$, such that
\beq
\label{eq:CSL_1}
 \bbm \mbf{P}\mbf{A}+\mbf{A}^\trans\mbf{P}+\mbf{C}^\trans\mbf{C} & \mbf{P}\mbf{B}-c\mbf{C}^\trans+\mbf{C}^\trans\mbf{D} \\ * & \mbf{D}^\trans\mbf{D}-c\left(\mbf{D}+\mbf{D}^\trans\right) + \left(c^2-r^2 \right)\mbf{1} \ebm \leq 0.
\eeq
Note that the matrix inequality of~\eqref{eq:CSL_1} does not allow for the case where the upper bound $b$ is infinite.

\end{enumerate}

The Conic Sector Lemma is a special case of the KYP Lemma for QSR dissipative systems with $\mbf{Q} = - \mbf{1}$, $\mbf{S} = \frac{a+b}{2} \mbf{1} = c \mbf{1}$, and $\mbf{R} = -ab \mbf{1} = \left(r^2-c^2\right) \mbf{1}$.

\subsubsection[Exterior Conic Sector Lemma]{Exterior Conic Sector Lemma}
\index{conic sectors!exterior conic sector lemma}
Consider a square, continuous-time LTI system, $\mbc{G}: \mathcal{L}_{2e} \to \mathcal{L}_{2e}$, with state-space realization $(\mbf{A},\mbf{B}_,\mbf{C},\mbf{D})$, where $\mbf{A} \in \mathbb{R}^{n \times n}$, $\mbf{B} \in \mathbb{R}^{n \times m}$, $\mbf{C} \in \mathbb{R}^{m \times n}$, and $\mbf{D} \in \mathbb{R}^{m \times m}$.  The system $\mbc{G}$ is in the exterior cone of radius $r$ centered at $c$ (i.e., $\mbc{G} \in \text{excone}_r(c)$), where $r \in \mathbb{R}_{> 0}$ and $c \in \mathbb{R}$, under either of the following equivalent necessary and sufficient conditions.

\begin{enumerate}

\itemcite \cite{Bridgeman2015} There exists $\mbf{P} \in \mathbb{S}^n$, where $\mbf{P}  \geq 0$, such that
\beq
\label{eq:ECSL_1}
\bbm \mbf{P}\mbf{A}+\mbf{A}^\trans\mbf{P}-\mbf{C}^\trans\mbf{C} & \mbf{P}\mbf{B}-\mbf{C}^\trans\left(\mbf{D}-c\mbf{1}\right) \\ * & r^2\mbf{1}-\left(\mbf{D}-c\mbf{1}\right)^\trans\left(\mbf{D}-c\mbf{1}\right) \ebm \leq 0.
\eeq

\item There exists $\mbf{P} \in \mathbb{S}^n$, where $\mbf{P}  \geq 0$, such that
\beq
\label{eq:ECSL_2}
\bbm \mbf{P}\mbf{A}+\mbf{A}^\trans\mbf{P}-\mbf{C}^\trans\mbf{C} & \mbf{P}\mbf{B}-\mbf{C}^\trans\left(\mbf{D}-c\mbf{1}\right) & \mbf{0} \\ * & -\left(\mbf{D}-c\mbf{1}\right)^\trans\left(\mbf{D}-c\mbf{1}\right) & r\mbf{1} \\ * & * & -\mbf{1} \ebm \leq 0.
\eeq
\begin{proof}
Applying the Schur complement lemma to the $r^2 \mbf{1}$ term in~\eqref{eq:ECSL_1} gives~\eqref{eq:ECSL_2}.
\end{proof}

\end{enumerate}

\subsubsection{Modified Exterior Conic Sector Lemma}
\index{conic sectors!modified exterior conic sector lemma}
Consider a square, continuous-time LTI system, $\mbc{G}: \mathcal{L}_{2e} \to \mathcal{L}_{2e}$, with state-space realization $(\mbf{A},\mbf{B}_,\mbf{C},\mbf{D})$, where $\mbf{A} \in \mathbb{R}^{n \times n}$, $\mbf{B} \in \mathbb{R}^{n \times m}$, $\mbf{C} \in \mathbb{R}^{m \times n}$, and $\mbf{D} \in \mathbb{R}^{m \times m}$.  The system $\mbc{G}$ is in the exterior cone of radius $r$ centered at $c$ (i.e., $\mbc{G} \in \text{excone}_r(c)$), where $r \in \mathbb{R}_{> 0}$ and $c \in \mathbb{R}$, under either of the following equivalent sufficient conditions.

\begin{enumerate}

\item There exists $\mbf{P} \in \mathbb{S}^n$, where $\mbf{P}  \geq 0$, such that
\beq
\label{eq:MECSL_1}
\bbm \mbf{P}\mbf{A}+\mbf{A}^\trans\mbf{P} & \mbf{P}\mbf{B}-\mbf{C}^\trans\left(\mbf{D}-c\mbf{1}\right) \\ * & r^2\mbf{1}-\left(\mbf{D}-c\mbf{1}\right)^\trans\left(\mbf{D}-c\mbf{1}\right) \ebm \leq 0.
\eeq
\begin{proof}
The term $-\mbf{C}^\trans \mbf{C}$ in~\eqref{eq:ECSL_1} makes the matrix inequality ``more'' negative definite.  Therefore,
\bdis
\bbm \mbf{P}\mbf{A}+\mbf{A}^\trans\mbf{P}-\mbf{C}^\trans\mbf{C} & \mbf{P}\mbf{B}-\mbf{C}^\trans\left(\mbf{D}-c\mbf{1}\right) \\ * & r^2\mbf{1}-\left(\mbf{D}-c\mbf{1}\right)^\trans\left(\mbf{D}-c\mbf{1}\right) \ebm \leq \bbm \mbf{P}\mbf{A}+\mbf{A}^\trans\mbf{P} & \mbf{P}\mbf{B}-\mbf{C}^\trans\left(\mbf{D}-c\mbf{1}\right) \\ * & r^2\mbf{1}-\left(\mbf{D}-c\mbf{1}\right)^\trans\left(\mbf{D}-c\mbf{1}\right) \ebm,
\edis
and~\eqref{eq:MECSL_1} implies~\eqref{eq:ECSL_1}.
\end{proof}

\item There exists $\mbf{P} \in \mathbb{S}^n$, where $\mbf{P}  \geq 0$, such that
\beq
\label{eq:MECSL_2}
\bbm \mbf{P}\mbf{A}+\mbf{A}^\trans\mbf{P} & \mbf{P}\mbf{B}-\mbf{C}^\trans\left(\mbf{D}-c\mbf{1}\right) & \mbf{0} \\ * & -\left(\mbf{D}-c\mbf{1}\right)^\trans\left(\mbf{D}-c\mbf{1}\right) & r\mbf{1} \\ * & * & -\mbf{1} \ebm \leq 0.
\eeq
\begin{proof}
Applying the Schur complement lemma to the $r^2 \mbf{1}$ term in~\eqref{eq:MECSL_1} gives~\eqref{eq:MECSL_2}.
\end{proof}
\end{enumerate}
A system satisfying the Modified Exterior Conic Sector Lemma is Lyapunov stable if the additional restriction $\mbf{P} >0$ is made, which is not necessarily true for a system satisfying the Exterior Conic Sector Lemma.

The system $\mbc{G}$ is also in the exterior cone of radius $r$ centered at $c$, where $r \in \mathbb{R}_{> 0}$ and $c \in \mathbb{R}$, under either of the following equivalent sufficient conditions.

\begin{enumerate}

\item There exists $\mbf{Q} \in \mathbb{S}^n$, where $\mbf{Q} > 0$, such that
\bdis
\bbm \mbf{A}\mbf{Q}+\mbf{Q}\mbf{A}^\trans & \mbf{B}-\mbf{Q}\mbf{C}^\trans\left(\mbf{D}-c\mbf{1}\right) \\ * & r^2\mbf{1}-\left(\mbf{D}-c\mbf{1}\right)^\trans\left(\mbf{D}-c\mbf{1}\right) \ebm \leq 0.
\edis

\item There exists $\mbf{Q} \in \mathbb{S}^n$, where $\mbf{Q} > 0$, such that
\bdis
\bbm \mbf{A}\mbf{Q}+\mbf{Q}\mbf{A}^\trans & \mbf{B}-\mbf{Q}\mbf{C}^\trans\left(\mbf{D}-c\mbf{1}\right) & \mbf{0} \\ * & -\left(\mbf{D}-c\mbf{1}\right)^\trans\left(\mbf{D}-c\mbf{1}\right) & r\mbf{1} \\ * & * & -\mbf{1} \ebm \leq 0.
\edis

\end{enumerate}

\subsubsection[Generalized KYP (GKYP) Lemma for Conic Sectors]{Generalized KYP (GKYP) Lemma for Conic Sectors}
\label{sec:GKYP_Lemma}
\index{generalized KYP Lemma (GKYP)}

Consider a square, continuous-time LTI system, $\mbc{G}: \mathcal{L}_{2e} \to \mathcal{L}_{2e}$, with state-space realization $(\mbf{A},\mbf{B}_,\mbf{C},\mbf{D})$, where $\mbf{A} \in \mathbb{R}^{n \times n}$, $\mbf{B} \in \mathbb{R}^{n \times m}$, $\mbf{C} \in \mathbb{R}^{m \times n}$, and $\mbf{D} \in \mathbb{R}^{m \times m}$.  Also consider $\mbs{\Pi}_c(a,b) \in \mathbb{S}^{m}$, which is defined as
\bdis
\mbs{\Pi}_c(a,b) = \bbm \frac{1}{b} \mbf{1} & -\onehalf \left(1 + \frac{a}{b}\right) \mbf{1} \\ * & a \mbf{1} \ebm,
\edis
where $a \in \mathbb{R}$, $b \in \mathbb{R}_{>0}$, and $a < b$.  The following generalized KYP Lemmas give conditions for $\mbc{G}$ to be inside the cone $[a,b]$ within finite frequency bandwidths.

\begin{enumerate}

\item (\textit{Low Frequency Range}~\cite{Iwasaki2003}) The system $\mbc{G}$ is inside the cone $[a,b]$ for all $\omega \in \{\omega \in \mathbb{R} \,\, | \,\, \abs{\omega} < \omega_1, \,\, \det (j\omega \mbf{1} - \mbf{A}) \neq 0 \}$, where $\omega_1 \in \mathbb{R}_{>0}$, $a \in \mathbb{R}$, $b \in \mathbb{R}_{>0}$, and $a < b$, if there exist $\mbf{P}$,~$\mbf{Q} \in \mathbb{S}^n$ and $\bar{\omega}_1 \in \mathbb{R}_{>0}$, where $\mbf{Q}  \geq 0$, such that
\beq
\label{eq:GKYP_ConicLow}
\bbm \mbf{A} & \mbf{B} \\ \mbf{1} & \mbf{0} \ebm^\trans \bbm -\mbf{Q} & \mbf{P} \\ * & (\omega_1 - \bar{\omega}_1)^2 \mbf{Q} \ebm \bbm \mbf{A} & \mbf{B} \\ \mbf{1} & \mbf{0} \ebm + \bbm \mbf{C} & \mbf{D} \\ \mbf{0} & \mbf{1} \ebm^\trans \mbs{\Pi}_c(a,b) \bbm \mbf{C} & \mbf{D} \\ \mbf{0} & \mbf{1} \ebm < 0.
\eeq
If $\omega_1 \to \infty$, $\mbf{P} > 0$, and $\mbf{Q} = \mbf{0}$, then the traditional Conic Sector Lemma is recovered~\cite{Iwasaki2005b}.

The parameter $\bar{\omega}_1$ is included in~\eqref{eq:GKYP_ConicLow} to effectively transform $\abs{\omega} \leq (\omega_1 - \bar{\omega}_1)$ into the strict inequality $\abs{\omega} < \omega_1$.

\item (\textit{Intermediate Frequency Range}~\cite{Hara2003,Iwasaki2005b,Iwasaki2005}) The system $\mbc{G}$ is inside the cone $[a,b]$ for all $\omega \in \{\omega \in \mathbb{R} \,\, | \,\, \omega_1 \leq \abs{\omega} < \omega_2, \,\, \det (j\omega \mbf{1} - \mbf{A}) \neq 0 \}$, where $\omega_1$,~$\omega_2 \in \mathbb{R}_{>0}$, $a \in \mathbb{R}$, $b \in \mathbb{R}_{>0}$, and $a < b$, if there exist $\mbf{P}$,~$\mbf{Q} \in \mathbb{C}^n$, $\bar{\omega}_2 \in \mathbb{R}_{>0}$, and $\hat{\omega}_2 = \left(\omega_1 + (\omega_2 - \bar{\omega}_2)\right)/2$, where $\mbf{P}^\herm = \mbf{P}$, $\mbf{Q}^\herm = \mbf{Q}$, and $\mbf{Q}  \geq 0$, such that
\beq
\label{eq:GKYP_ConicInter}
\bbm \mbf{A} & \mbf{B} \\ \mbf{1} & \mbf{0} \ebm^\trans \bbm -\mbf{Q} & \mbf{P} + j \hat{\omega}_2 \mbf{Q} \\ \mbf{P} - j\hat{\omega}_2 \mbf{Q} & -\omega_1(\omega_2 - \bar{\omega}-2) \mbf{Q} \ebm \bbm \mbf{A} & \mbf{B} \\ \mbf{1} & \mbf{0} \ebm + \bbm \mbf{C} & \mbf{D} \\ \mbf{0} & \mbf{1} \ebm^\trans \mbs{\Pi}_c(a,b) \bbm \mbf{C} & \mbf{D} \\ \mbf{0} & \mbf{1} \ebm < 0.
\eeq
The parameter $\bar{\omega}_2$ is included in~\eqref{eq:GKYP_ConicInter} to effectively transform $ \omega_1 \leq \abs{\omega} \leq (\omega_2 - \bar{\omega}_2)$ into the strict inequality $\omega_1 \leq \abs{\omega} < \omega_2$.

\item (\textit{High Frequency Range}~\cite{Hara2003}) The system $\mbc{G}$ is inside the cone $[a,b]$ for all $\omega \in \{\omega \in \mathbb{R} \,\, | \,\, \omega_2 \leq  \abs{\omega}, \,\, \det (j\omega \mbf{1} - \mbf{A}) \neq 0 \}$, where $\omega_2 \in \mathbb{R}_{>0}$, $a \in \mathbb{R}$, $b \in \mathbb{R}_{>0}$, and $a < b$, if there exist $\mbf{P}$,~$\mbf{Q} \in \mathbb{S}^n$, where $\mbf{Q}  \geq 0$, such that
\beq
\label{eq:GKYP_ConicHigh}
\bbm \mbf{A} & \mbf{B} \\ \mbf{1} & \mbf{0} \ebm^\trans \bbm \mbf{Q} & \mbf{P} \\ * & -\omega_2^2\mbf{Q} \ebm \bbm \mbf{A} & \mbf{B} \\ \mbf{1} & \mbf{0} \ebm + \bbm \mbf{C} & \mbf{D} \\ \mbf{0} & \mbf{1} \ebm^\trans \mbs{\Pi}_c(a,b) \bbm \mbf{C} & \mbf{D} \\ \mbf{0} & \mbf{1} \ebm < 0.
\eeq

\end{enumerate}
If $(\mbf{A},\mbf{B}_,\mbf{C},\mbf{D})$ is a minimal realization, then the matrix inequalities in~\eqref{eq:GKYP_ConicLow},~\eqref{eq:GKYP_ConicInter}, and~\eqref{eq:GKYP_ConicHigh} can be nonstrict~\cite{Iwasaki2003}.

\subsection{Minimum Gain}
\index{minimum gain!minimum gain lemma}
\subsubsection[Minimum Gain Lemma]{Minimum Gain Lemma} 
\label{sec:MinimumGainLemma}

Consider a continuous-time LTI system, $\mbc{G}: \mathcal{L}_{2e} \to \mathcal{L}_{2e}$, with state-space realization $(\mbf{A},\mbf{B}_,\mbf{C},\mbf{D})$, where $\mbf{A} \in \mathbb{R}^{n \times n}$, $\mbf{B} \in \mathbb{R}^{n \times m}$, $\mbf{C} \in \mathbb{R}^{p \times n}$, and $\mbf{D} \in \mathbb{R}^{p \times m}$.  The system $\mbc{G}$ has minimum gain $\nu $ under any of the following equivalent sufficient conditions.

\begin{enumerate}

\itemcite \cite{Bridgeman2014} There exist $\mbf{P} \in \mathbb{S}^n$ and $\nu \in \mathbb{R}_{\geq0}$, where $\mbf{P}  \geq 0$, such that
\bdis
\bbm \mbf{P} \mbf{A} + \mbf{A}^\trans \mbf{P} - \mbf{C}^\trans\mbf{C} & \mbf{P} \mbf{B} - \mbf{C}^\trans\mbf{D} \\ * & \nu^2 \mbf{1} - \mbf{D}^\trans\mbf{D} \ebm \leq 0.
\edis

\itemcite \cite{Caverly2018} There exist $\mbf{P} \in \mathbb{S}^n$ and $\nu \in \mathbb{R}_{\geq0}$, where $\mbf{P}  \geq 0$, such that

\bdis
\label{eq:MinGain1}
\bbm \mbf{P} \mbf{A} + \mbf{A}^\trans \mbf{P} - \mbf{C}^\trans\mbf{C} & \mbf{P} \mbf{B} - \mbf{C}^\trans\mbf{D} & \mbf{0} \\ * &  - \mbf{D}^\trans\mbf{D} & \nu \mbf{1} \\ * & * & -\mbf{1} \ebm \leq 0.
\edis

\end{enumerate}

If $\mbc{G}$ is a square system (i.e., $m=p$) or $\text{span}(\mbf{C}) \subseteq \text{span}(\mbf{D})$, then the preceding conditions are necessary and sufficient for $\mbc{G}$ to have minimum gain $\nu \in \mathbb{R}_{\geq 0}$~\cite{Bridgeman2014}.  The minimum gain lemma is a special case of the exterior conic sector lemma with $a = -\nu$ and $b = \nu$.

The system $\mbc{G}$ also has minimum gain $\nu $ under any of the following sufficient conditions.

\begin{enumerate}

\item There exist $\mbf{P} \in \mathbb{S}^n$, $\mbf{V}_{11} \in \mathbb{R}^{n \times n}$, $\mbf{V}_{12} \in \mathbb{R}^{n \times m}$, $\mbf{V}_{21} \in \mathbb{R}^{p \times n}$, $\mbf{V}_{22} \in \mathbb{R}^{p \times m}$, and $\nu \in \mathbb{R}_{\geq 0}$, where $\mbf{P}  \geq 0$, such that
\beq
\label{eq:MinGainDil1}
\bbm -(\mbf{V}_{11}+\mbf{V}_{11}^\trans) & \mbf{V}_{11}^\trans\mbf{A} + \mbf{V}_{21}^\trans\mbf{C} + \mbf{P} & \mbf{V}_{11}^\trans\mbf{B} + \mbf{V}_{21}^\trans\mbf{D} - \mbf{V}_{12} & \mbf{V}_{11}^\trans & \nu \mbf{V}_{21}^\trans \\ * & -\mbf{P} & \mbf{C}^\trans\mbf{V}_{22} + \mbf{A}^\trans\mbf{V}_{12} & \mbf{0} & \mbf{0} \\ * & * & \nu \mbf{1} +\mbf{V}_{22}\mbf{D}+\mbf{D}^\trans\mbf{V}_{22}^\trans + \mbf{V}_{12}^\trans\mbf{B} + \mbf{B}^\trans\mbf{V}_{12} & \mbf{V}_{12}^\trans & \nu \mbf{V}_{22}^\trans \\ * & * & * & -\mbf{P} & \mbf{0} \\ * & * & * & * & - \nu\mbf{1} \ebm \leq 0.
\eeq
\begin{proof}
Applying the congruence transformation $\mbf{W} = \text{diag}\{\nu^{-1/2}\mbf{1},\nu^{-1/2}\mbf{1}\}$ and defining $\mbfbar{P} = \nu^{-1}\mbf{P}$, the matrix inequality of~\eqref{eq:MinGain1} can be rewritten as
\beq
\label{eq:MinGainDil1a}
\bbm \mbfbar{P} \mbf{A} + \mbf{A}^\trans \mbfbar{P} - \nu^{-1}\mbf{C}^\trans\mbf{C} & \mbfbar{P} \mbf{B} - \nu^{-1}\mbf{C}^\trans\mbf{D} \\ * & \nu \mbf{1} - \nu^{-1}\mbf{D}^\trans\mbf{D} \ebm \leq 0.
\eeq
Using Property~\ref{SchurProp3} from Section~\ref{sec:SchurProp} and making the assumption that $\mbfbar{P}$ is invertible,~\eqref{eq:MinGainDil1a} is equivalent to
\bdis
\bbm \mbfbar{P} \mbf{A} + \mbf{A}^\trans \mbfbar{P} - \mbfbar{P} - \nu^{-1}\mbf{C}^\trans\mbf{C} & \mbfbar{P} \mbf{B} - \nu^{-1}\mbf{C}^\trans\mbf{D} & \mbfbar{P} \\ * & \nu \mbf{1} - \nu^{-1}\mbf{D}^\trans\mbf{D} & \mbf{0} \\ * & * & - \mbfbar{P} \ebm \leq 0.
\edis
which is rewritten as
\beq
\label{eq:Dilation_1_MinGain}
\bbm \mbf{A}^\trans & \mbf{1} & \mbf{0} & \mbf{0} & -\nu^{-1} \mbf{C}^\trans \\ \mbf{B}^\trans & \mbf{0} & \mbf{1} & \mbf{0} & -\nu^{-1} \mbf{D}^\trans\\ \mbf{1} & \mbf{0} & \mbf{0} & \mbf{1} & \mbf{0} \ebm \bbm \mbf{0} & \mbfbar{P} & \mbf{0} & \mbf{0} & \mbf{0}  \\ * & -\mbfbar{P} & \mbf{0} & \mbf{0} & \mbf{0}  \\ * & * & \nu \mbf{1} & \mbf{0} & \mbf{0} \\ * & * & * & -\mbfbar{P} & \mbf{0} \\ * & * & * & * & -\nu \mbf{1} \ebm \bbm \mbf{A} & \mbf{B} & \mbf{1} \\ \mbf{1} & \mbf{0} & \mbf{0}\\ \mbf{0} & \mbf{1} & \mbf{0} \\ \mbf{0} & \mbf{0} & \mbf{1} \\ -\nu^{-1}\mbf{C} & -\nu^{-1} \mbf{D} & \mbf{0} \ebm \leq 0.
\eeq
Since $\mbfbar{P} > 0$ and $\nu \in \mathbb{R}_{\geq 0}$, it is also known that
\bdis
\bbm -\mbfbar{P} & \mbf{0} & \mbf{0} \\ * & -\mbfbar{P} & \mbf{0} \\ *  & * & -\nu \mbf{1} \ebm \leq 0,
\edis
which can be rewritten as
\beq
\label{eq:Dilation_2_MinGain}
\bbm \mbf{0} & \mbf{1} & \mbf{0} & \mbf{0} & \mbf{0}  \\ \mbf{0} & \mbf{0} & \mbf{0} & \mbf{1} & \mbf{0} \\ \mbf{0} & \mbf{0} & \mbf{0} & \mbf{0} & \mbf{1}  \ebm \bbm \mbf{0} & \mbfbar{P} & \mbf{0} & \mbf{0} & \mbf{0}  \\ * & -\mbfbar{P} & \mbf{0} & \mbf{0} & \mbf{0}  \\ * & * & \nu \mbf{1} & \mbf{0} & \mbf{0} \\ * & * & * & -\mbfbar{P} & \mbf{0} \\ * & * & * & * & -\nu \mbf{1} \ebm \bbm \mbf{0} & \mbf{0} & \mbf{0} \\ \mbf{1} & \mbf{0} & \mbf{0} \\ \mbf{0} & \mbf{0} & \mbf{0} \\ \mbf{0} & \mbf{1} & \mbf{0} \\ \mbf{0} & \mbf{0} & \mbf{1} \ebm \leq 0.
\eeq
The matrix inequalities in~\eqref{eq:Dilation_1_MinGain} and~\eqref{eq:Dilation_2_MinGain} are in the form of the nonstrict projection lemma.  Specifically,~\eqref{eq:Dilation_1_MinGain} is in the form of $\mbf{N}_G^\trans \mbs{\Phi}\mbf{N}_G \leq 0$, where
\bdis
\mbs{\Phi} = \bbm \mbf{0} & \mbfbar{P} & \mbf{0} & \mbf{0} & \mbf{0}  \\ * & -\mbfbar{P} & \mbf{0} & \mbf{0} & \mbf{0}  \\ * & * & \nu \mbf{1} & \mbf{0} & \mbf{0} \\ * & * & * & -\mbfbar{P} & \mbf{0} \\ * & * & * & * & -\nu \mbf{1} \ebm, \hspace{10pt} \mbf{N}_G = \bbm \mbf{A} & \mbf{B} & \mbf{1} \\ \mbf{1} & \mbf{0} & \mbf{0}\\ \mbf{0} & \mbf{1} & \mbf{0} \\ \mbf{0} & \mbf{0} & \mbf{1} \\ -\nu^{-1}\mbf{C} & -\nu^{-1} \mbf{D} & \mbf{0} \ebm.
\edis
The matrix inequality of~\eqref{eq:Dilation_2_MinGain} is in the form of $\mbf{N}_H^\trans \mbs{\Phi}\mbf{N}_H < 0$, where
\bdis
\mbf{N}_H = \bbm \mbf{0} & \mbf{0} & \mbf{0} \\ \mbf{1} & \mbf{0} & \mbf{0} \\ \mbf{0} & \mbf{0} & \mbf{0} \\ \mbf{0} & \mbf{1} & \mbf{0} \\ \mbf{0} & \mbf{0} & \mbf{1} \ebm.
\edis
The nonstrict projection lemma states that~\eqref{eq:Dilation_1_MinGain} and~\eqref{eq:Dilation_2_MinGain} are equivalent to
\beq
\label{eq:Dilation_4_MinGain}
\mbs{\Phi} + \mbf{G} \mbf{V} \mbf{H}^\trans + \mbf{H} \mbf{V}^\trans \mbf{G}^\trans,
\eeq
where $\mathcal{N}(\mbf{G}^\trans) = \mathcal{R}(\mbf{N}_G)$, $\mathcal{N}(\mbf{H}^\trans) = \mathcal{R}(\mbf{N}_H)$, $\mbf{V} \in \mathbb{R}^{n \times n}$, and $\mathcal{R}(\mbf{G})$, $\mathcal{R}(\mbf{H})$ are linearly independent.  Choosing
\bdis
\mbf{G}^\trans = \bbm -\mbf{1} & \mbf{A} & \mbf{B} & \mbf{1} & \mbf{0} \\ \mbf{0} & \mbf{C} & \mbf{D} & \mbf{0} & \nu \mbf{1} \ebm, \hspace{10pt} \mbf{H}^\trans = \bbm \mbf{1} & \mbf{0} & \mbf{0} & \mbf{0} & \mbf{0} \\ \mbf{0} & \mbf{0} & \mbf{1} & \mbf{0} & \mbf{0}  \ebm, \hspace{10pt} \mbf{V} = \bbm \mbf{V}_{11} & \mbf{V}_{12} \\ \mbf{V}_{21} & \mbf{V}_{22} \ebm,
\edis
where $\mathcal{R}(\mbf{G})$ and $\mathcal{R}(\mbf{H})$ are in fact linearly independent, the matrix inequality of~\eqref{eq:Dilation_4_MinGain} can be rewritten as
\begin{multline*}
\bbm \mbf{0} & \mbfbar{P} & \mbf{0} & \mbf{0} & \mbf{0}  \\ * & -\mbfbar{P} & \mbf{0} & \mbf{0} & \mbf{0}  \\ * & * & \nu \mbf{1} & \mbf{0} & \mbf{0} \\ * & * & * & -\mbfbar{P} & \mbf{0} \\ * & * & * & * & -\nu \mbf{1} \ebm + \bbm -\mbf{1} & \mbf{0} \\ \mbf{A}^\trans & \mbf{C}^\trans \\ \mbf{B}^\trans & \mbf{D}^\trans  \\ \mbf{1} & \mbf{0} \\ \mbf{0} & \nu \mbf{1}  \ebm \bbm \mbf{V}_{11} & \mbf{V}_{12} \\ \mbf{V}_{21} & \mbf{V}_{22} \ebm \bbm \mbf{1} & \mbf{0} & \mbf{0} & \mbf{0} & \mbf{0} \\ \mbf{0} & \mbf{0} & \mbf{1} & \mbf{0} & \mbf{0}  \ebm \\ + \bbm \mbf{1} & \mbf{0} \\ \mbf{0} & \mbf{0}  \\ \mbf{0} & \mbf{1} \\ \mbf{0} & \mbf{0} \\ \mbf{0} & \mbf{0} \ebm \bbm \mbf{V}_{11}^\trans & \mbf{V}_{21}^\trans \\ \mbf{V}_{12}^\trans & \mbf{V}_{22}^\trans \ebm \bbm -\mbf{1} & \mbf{A} & \mbf{B} & \mbf{1} & \mbf{0} \\ \mbf{0} & \mbf{C} & \mbf{D} & \mbf{0} & \nu \mbf{1} \ebm < 0,
\end{multline*}
or equivalently
\beq
\label{eq:Dilation_5_MinGain}
\bbm -(\mbf{V}_{11}+\mbf{V}_{11}^\trans) & \mbf{V}_{11}^\trans\mbf{A} + \mbf{V}_{21}^\trans\mbf{C} + \mbfbar{P} & \mbf{V}_{11}^\trans\mbf{B} + \mbf{V}_{21}^\trans\mbf{D} - \mbf{V}_{12} & \mbf{V}_{11}^\trans & \nu \mbf{V}_{21}^\trans \\ * & -\mbfbar{P} & \mbf{C}^\trans\mbf{V}_{22} + \mbf{A}^\trans\mbf{V}_{12} & \mbf{0} & \mbf{0} \\ * & * & \nu \mbf{1} +\mbf{V}_{22}\mbf{D}+\mbf{D}^\trans\mbf{V}_{22}^\trans + \mbf{V}_{12}^\trans\mbf{B} + \mbf{B}^\trans\mbf{V}_{12} & \mbf{V}_{12}^\trans & \nu\mbf{V}_{22}^\trans \\ * & * & * & -\mbfbar{P} & \mbf{0} \\ * & * & * & * & - \nu\mbf{1} \ebm \leq 0.
\eeq
Redefining $\mbf{P} = \mbfbar{P}$,~\eqref{eq:Dilation_5_MinGain} is identical to~\eqref{eq:MinGainDil1}.
\end{proof}


\item There exist $\mbf{P} \in \mathbb{S}^n$, $\mbf{V}_{11} \in \mathbb{R}^{n \times n}$, and $\nu \in \mathbb{R}_{\geq 0}$, where $\mbf{P}  \geq 0$, such that
\beq
\label{eq:MinGainDil2}
\bbm -(\mbf{V}+\mbf{V}^\trans) & \mbf{V}^\trans\mbf{A} + \mbf{P} & \mbf{V}^\trans\mbf{B} & \mbf{V}^\trans  \\ * & -\mbf{P} & -\mbf{C}^\trans & \mbf{0}  \\ * & * & 2\nu \mbf{1} - (\mbf{D}+\mbf{D}^\trans) & \mbf{0} \\ * & * & * & -\mbf{P}   \ebm  < 0.
\eeq
\begin{proof}
The matrix inequality of ~\eqref{eq:MinGainDil2} is derived from~\eqref{eq:MinGainDil1} with $\mbf{V}_{11} = \mbf{V}$, $\mbf{V}_{12} = \mbf{0}$, $\mbf{V}_{21} = \mbf{0}$, and $\mbf{V}_{22} = -\mbf{1}$.  The dilation in~\eqref{eq:MinGainDil1} relies on the projection lemma and becomes only a sufficient condition in this case due to the structure imposed on $\mbf{V}_{11}$, $\mbf{V}_{12}$, $\mbf{V}_{21}$, and $\mbf{V}_{22}$.
\end{proof}

\end{enumerate}

\subsubsection[Modified Minimum Gain Lemma]{Modified Minimum Gain Lemma}
\index{minimum gain!modified minimum gain lemma}
\label{sec:ModMinimumGainLemma}
Consider a continuous-time LTI system, $\mbc{G}: \mathcal{L}_{2e} \to \mathcal{L}_{2e}$, with state-space realization $(\mbf{A},\mbf{B}_,\mbf{C},\mbf{D})$, where $\mbf{A} \in \mathbb{R}^{n \times n}$, $\mbf{B} \in \mathbb{R}^{n \times m}$, $\mbf{C} \in \mathbb{R}^{p \times n}$, and $\mbf{D} \in \mathbb{R}^{p \times m}$.  The system $\mbc{G}$ has minimum gain $\nu $ under any of the following equivalent sufficient conditions.

\begin{enumerate}

\itemcite \cite{Caverly2016} There exist $\mbf{P} \in \mathbb{S}^n$ and $\nu \in \mathbb{R}_{\geq0}$, where $\mbf{P}  \geq 0$, such that
\beq
\label{eq:MMGL_1}
\bbm \mbf{P} \mbf{A} + \mbf{A}^\trans \mbf{P}  & \mbf{P} \mbf{B} - \mbf{C}^\trans\mbf{D} \\ * & \nu^2 \mbf{1} - \mbf{D}^\trans\mbf{D} \ebm \leq 0.
\eeq

\item There exist $\mbf{P} \in \mathbb{S}^n$ and $\nu \in \mathbb{R}_{\geq0}$, where $\mbf{P}  \geq 0$, such that

\beq
\label{eq:MMGL_2}
\bbm \mbf{P} \mbf{A} + \mbf{A}^\trans \mbf{P}  & \mbf{P} \mbf{B} - \mbf{C}^\trans\mbf{D} & \mbf{0} \\ * &  - \mbf{D}^\trans\mbf{D} & \nu \mbf{1} \\ * & * & -\mbf{1} \ebm \leq 0.
\eeq
\begin{proof}
Applying the Schur complement lemma to the $\nu^2 \mbf{1}$ term in~\eqref{eq:MMGL_1} gives~\eqref{eq:MMGL_2}.
\end{proof}

\end{enumerate}

A system satisfying the Modified Minimum Gain Lemma is Lyapunov stable if the additional restriction $\mbf{P} >0$ is made, which is not necessarily true for a system satisfying the Minimum Gain Lemma.  %

The system $\mbc{G}$ also has minimum gain $\nu$ under any of the following equivalent sufficient conditions.
\begin{enumerate}

\item There exist $\mbf{Q} \in \mathbb{S}^n$ and $\nu \in \mathbb{R}_{\geq0}$, where $\mbf{Q}  > 0$, such that
\bdis
\label{eq:MMGT3}
\bbm \mbf{A} \mbf{Q} + \mbf{Q}\mbf{A}^\trans & \mbf{B} - \mbf{Q}\mbf{C}^\trans\mbf{D} \\ * & \nu^2 \mbf{1} - \mbf{D}^\trans\mbf{D} \ebm \leq 0.
\edis

\item There exist $\mbf{Q} \in \mathbb{S}^n$ and $\nu \in \mathbb{R}_{\geq0}$, where $\mbf{Q}  > 0$, such that
\bdis
\bbm \mbf{A} \mbf{Q} + \mbf{Q}\mbf{A}^\trans & \mbf{B} - \mbf{Q}\mbf{C}^\trans\mbf{D} & \mbf{0}  \\ * &  - \mbf{D}^\trans\mbf{D} & \nu \mbf{1} \\ * & * & -\mbf{1} \ebm \leq 0.
\edis

\end{enumerate}

\subsubsection[Discrete-Time Minimum Gain Lemma]{Discrete-Time Minimum Gain Lemma}
\index{minimum gain!discrete-time minimum gain lemma}
Consider a discrete-time LTI system, $\mbc{G}: \ell_{2e} \to \ell_{2e}$, with state-space realization $(\mbf{A}_\mathrm{d},\mbf{B}_\mathrm{d},\mbf{C}_\mathrm{d},\mbf{D}_\mathrm{d})$, where $\mbf{A}_\mathrm{d} \in \mathbb{R}^{n \times n}$, $\mbf{B}_\mathrm{d} \in \mathbb{R}^{n \times m}$, $\mbf{C}_\mathrm{d} \in \mathbb{R}^{p \times n}$, and $\mbf{D}_\mathrm{d} \in \mathbb{R}^{p \times m}$.  The system $\mbc{G}$ has minimum gain $\nu $ under any of the following equivalent sufficient conditions.

\begin{enumerate}

\itemcite \cite[p.~30]{CaverlyPhDThesis} There exist $\mbf{P} \in \mathbb{S}^n$ and $\nu \in \mathbb{R}_{\geq0}$, where $\mbf{P}  \geq 0$, such that
\beq
\label{eq:DT_MGL1}
\bbm \mbf{A}_\mathrm{d}^\trans\mbf{P} \mbf{A}_\mathrm{d} - \mbf{P} - \mbf{C}_\mathrm{d}^\trans\mbf{C}_\mathrm{d} & \mbf{A}_\mathrm{d}^\trans\mbf{P} \mbf{B}_\mathrm{d} - \mbf{C}_\mathrm{d}^\trans\mbf{D}_\mathrm{d} \\ * & \mbf{B}_\mathrm{d}^\trans\mbf{P}\mbf{B}_\mathrm{d} + \nu^2 \mbf{1} - \mbf{D}_\mathrm{d}^\trans\mbf{D}_\mathrm{d}  \ebm \leq 0.
\eeq

\item There exist $\mbf{P} \in \mathbb{S}^n$ and $\nu \in \mathbb{R}_{\geq0}$, where $\mbf{P} \geq 0$, such that
\beq
\label{eq:DT_MGL1a}
\bbm \mbf{A}_\mathrm{d}^\trans\mbf{P} \mbf{A}_\mathrm{d} - \mbf{P} - \mbf{C}_\mathrm{d}^\trans\mbf{C}_\mathrm{d} & \mbf{A}_\mathrm{d}^\trans\mbf{P} \mbf{B}_\mathrm{d} - \mbf{C}_\mathrm{d}^\trans\mbf{D}_\mathrm{d} & \mbf{0} \\ * & \mbf{B}_\mathrm{d}^\trans\mbf{P}\mbf{B}_\mathrm{d}  - \mbf{D}_\mathrm{d}^\trans\mbf{D}_\mathrm{d} & \nu \mbf{1} \\ * & * & \mbf{1} \ebm \leq 0.
\eeq
\begin{proof}
Applying the Schur complement lemma to the $\nu^2 \mbf{1}$ term in~\eqref{eq:DT_MGL1} gives~\eqref{eq:DT_MGL1a}.
\end{proof}

\end{enumerate}

The system $\mbc{G}$ also has minimum gain $\nu $ under any of the following equivalent sufficient conditions.

\begin{enumerate}

\item There exist $\mbf{P} \in \mathbb{S}^n$ and $\nu \in \mathbb{R}_{\geq0}$, where $\mbf{P} > 0$, such that
\beq
\label{eq:DT_MGL2}
\bbm \mbf{P} & \mbf{P}\mbf{A}_\mathrm{d} & \mbf{P}\mbf{B}_\mathrm{d} \\ * &  \mbf{P} + \mbf{C}_\mathrm{d}^\trans\mbf{C}_\mathrm{d} &   \mbf{C}_\mathrm{d}^\trans\mbf{D}_\mathrm{d} \\ * & * & \mbf{D}_\mathrm{d}^\trans\mbf{D}_\mathrm{d} - \nu^2 \mbf{1}  \ebm \geq 0.
\eeq
\begin{proof}
Under the assumption that $\mbf{P} >0$, the nonstrict Schur complement lemma is applied to~\eqref{eq:DT_MGL1} to yield~\eqref{eq:DT_MGL2}.
\end{proof}

\item There exist $\mbf{P} \in \mathbb{S}^n$ and $\nu \in \mathbb{R}_{\geq0}$, where $\mbf{P} > 0$, such that
\beq
\label{eq:DT_MGL2a}
\bbm \mbf{P} & \mbf{P}\mbf{A}_\mathrm{d} & \mbf{P}\mbf{B}_\mathrm{d} & \mbf{0} \\ * &  \mbf{P} + \mbf{C}_\mathrm{d}^\trans\mbf{C}_\mathrm{d} &   \mbf{C}_\mathrm{d}^\trans\mbf{D}_\mathrm{d}  & \mbf{0}\\ * & * & \mbf{D}_\mathrm{d}^\trans\mbf{D}_\mathrm{d} & \nu \mbf{1} \\ * & * & * & \mbf{1}  \ebm \geq 0.
\eeq
\begin{proof}
Applying the Schur complement lemma to the $\nu^2 \mbf{1}$ term in~\eqref{eq:DT_MGL2} gives~\eqref{eq:DT_MGL2a}.
\end{proof}

\end{enumerate}

\subsubsection{Discrete-Time Modified Minimum Gain Lemma}
\index{minimum gain!discrete-time modified minimum gain lemma}
Consider a discrete-time LTI system, $\mbc{G}: \ell_{2e} \to \ell_{2e}$, with state-space realization $(\mbf{A}_\mathrm{d},\mbf{B}_\mathrm{d},\mbf{C}_\mathrm{d},\mbf{D}_\mathrm{d})$, where $\mbf{A}_\mathrm{d} \in \mathbb{R}^{n \times n}$, $\mbf{B}_\mathrm{d} \in \mathbb{R}^{n \times m}$, $\mbf{C}_\mathrm{d} \in \mathbb{R}^{p \times n}$, and $\mbf{D}_\mathrm{d} \in \mathbb{R}^{p \times m}$.  The system $\mbc{G}$ has minimum gain $\nu $ under any of the following equivalent sufficient conditions.
\begin{enumerate}

\item There exist $\mbf{P} \in \mathbb{S}^n$ and $\nu \in \mathbb{R}_{\geq0}$, where $\mbf{P}  \geq 0$, such that
\beq
\label{eq:DT_MMGL1}
\bbm \mbf{A}_\mathrm{d}^\trans\mbf{P} \mbf{A}_\mathrm{d} - \mbf{P}  & \mbf{A}_\mathrm{d}^\trans\mbf{P} \mbf{B}_\mathrm{d} - \mbf{C}_\mathrm{d}^\trans\mbf{D}_\mathrm{d} \\ * & \mbf{B}_\mathrm{d}^\trans\mbf{P}\mbf{B}_\mathrm{d} + \nu^2 \mbf{1} - \mbf{D}_\mathrm{d}^\trans\mbf{D}_\mathrm{d}  \ebm \leq 0.
\eeq
\begin{proof}
The term $-\mbf{C}_\mathrm{d}^\trans \mbf{C}_\mathrm{d}$ in~\eqref{eq:DT_MGL1} makes the matrix inequality ``more'' negative definite.  Therefore,
\bdis
\bbm \mbf{A}_\mathrm{d}^\trans\mbf{P} \mbf{A}_\mathrm{d} - \mbf{P} - \mbf{C}_\mathrm{d}^\trans\mbf{C}_\mathrm{d} & \mbf{A}_\mathrm{d}^\trans\mbf{P} \mbf{B}_\mathrm{d} - \mbf{C}_\mathrm{d}^\trans\mbf{D}_\mathrm{d} \\ * & \mbf{B}_\mathrm{d}^\trans\mbf{P}\mbf{B}_\mathrm{d} + \nu^2 \mbf{1} - \mbf{D}_\mathrm{d}^\trans\mbf{D}_\mathrm{d}  \ebm \leq \bbm \mbf{A}_\mathrm{d}^\trans\mbf{P} \mbf{A}_\mathrm{d} - \mbf{P}  & \mbf{A}_\mathrm{d}^\trans\mbf{P} \mbf{B}_\mathrm{d} - \mbf{C}_\mathrm{d}^\trans\mbf{D}_\mathrm{d} \\ * & \mbf{B}_\mathrm{d}^\trans\mbf{P}\mbf{B}_\mathrm{d} + \nu^2 \mbf{1} - \mbf{D}_\mathrm{d}^\trans\mbf{D}_\mathrm{d}  \ebm,
\edis
and~\eqref{eq:DT_MMGL1} implies~\eqref{eq:DT_MGL1}.
\end{proof}

\item There exist $\mbf{P} \in \mathbb{S}^n$ and $\nu \in \mathbb{R}_{\geq0}$, where $\mbf{P} > 0$, such that
\beq
\label{eq:DT_MMGL1a}
\bbm \mbf{A}_\mathrm{d}^\trans\mbf{P} \mbf{A}_\mathrm{d} - \mbf{P}  & \mbf{A}_\mathrm{d}^\trans\mbf{P} \mbf{B}_\mathrm{d} - \mbf{C}_\mathrm{d}^\trans\mbf{D}_\mathrm{d} & \mbf{0} \\ * & \mbf{B}_\mathrm{d}^\trans\mbf{P}\mbf{B}_\mathrm{d}  - \mbf{D}_\mathrm{d}^\trans\mbf{D}_\mathrm{d} & \nu \mbf{1} \\ * & * & \mbf{1} \ebm \leq 0.
\eeq
\begin{proof}
Applying the Schur complement lemma to the $\nu^2 \mbf{1}$ term in~\eqref{eq:DT_MMGL1} gives~\eqref{eq:DT_MMGL1a}.
\end{proof}

\end{enumerate}

A system satisfying the Discrete-Time Modified Minimum Gain Lemma is Lyapunov stable if the additional restriction $\mbf{P} >0$ is made, which is not necessarily true for a system satisfying the Discrete-Time Minimum Gain Lemma.  

The system $\mbc{G}$ also has minimum gain $\nu$ under any of the following sufficient conditions.

\begin{enumerate}

\item There exist $\mbf{P} \in \mathbb{S}^n$ and $\nu \in \mathbb{R}_{\geq0}$, where $\mbf{P} > 0$, such that
\beq
\label{eq:DT_MMGL2}
\bbm \mbf{P} & \mbf{P}\mbf{A}_\mathrm{d} & \mbf{P}\mbf{B}_\mathrm{d} \\ * &  \mbf{P}  &   \mbf{C}_\mathrm{d}^\trans\mbf{D}_\mathrm{d} \\ * & * & \mbf{D}_\mathrm{d}^\trans\mbf{D}_\mathrm{d} - \nu^2 \mbf{1}  \ebm \geq 0.
\eeq
\begin{proof}
Under the assumption that $\mbf{P} >0$, the nonstrict Schur complement lemma is applied to~\eqref{eq:DT_MMGL1} to yield~\eqref{eq:DT_MMGL2}.
\end{proof}

\item There exist $\mbf{P} \in \mathbb{S}^n$ and $\nu \in \mathbb{R}_{\geq0}$, where $\mbf{P} > 0$, such that
\beq
\label{eq:DT_MMGL3}
\bbm \mbf{P} & \mbf{P}\mbf{A}_\mathrm{d} & \mbf{P}\mbf{B}_\mathrm{d} & \mbf{0} \\ * &  \mbf{P}  &   \mbf{C}_\mathrm{d}^\trans\mbf{D}_\mathrm{d} & \mbf{0} \\ * & * & \mbf{D}_\mathrm{d}^\trans\mbf{D}_\mathrm{d} & \nu \mbf{1} \\ * & * & * & \mbf{1}  \ebm \geq 0.
\eeq
\begin{proof}
Applying the Schur complement lemma to the $\nu^2 \mbf{1}$ term in~\eqref{eq:DT_MMGL2} gives~\eqref{eq:DT_MMGL3}.
\end{proof}

\item There exist $\mbf{Q} \in \mathbb{S}^n$ and $\nu \in \mathbb{R}_{\geq0}$, where $\mbf{Q} > 0$, such that
\bdis
\bbm \mbf{Q} & \mbf{A}_\mathrm{d}\mbf{Q} & \mbf{B}_\mathrm{d} \\ * &  \mbf{Q}  &   \mbf{Q}\mbf{C}_\mathrm{d}^\trans\mbf{D}_\mathrm{d} \\ * & * & \mbf{D}_\mathrm{d}^\trans\mbf{D}_\mathrm{d} - \nu^2 \mbf{1}  \ebm \geq 0.
\edis

\item There exist $\mbf{Q} \in \mathbb{S}^n$ and $\nu \in \mathbb{R}_{\geq0}$, where $\mbf{Q} > 0$, such that
\bdis
\bbm \mbf{Q} & \mbf{A}_\mathrm{d}\mbf{Q} & \mbf{B}_\mathrm{d} & \mbf{0} \\ * &  \mbf{Q}  &   \mbf{Q}\mbf{C}_\mathrm{d}^\trans\mbf{D}_\mathrm{d} & \mbf{0} \\ * & * & \mbf{D}_\mathrm{d}^\trans\mbf{D}_\mathrm{d} & \nu \mbf{1} \\ * & * & * & \mbf{1} \ebm \geq 0.
\edis

\end{enumerate}

\subsection[Negative Imaginary Systems]{Negative Imaginary Systems}
\index{negative imaginary systems}

\subsubsection[Negative Imaginary Lemma]{Negative Imaginary Lemma~\cite{Lanzon2008,Song2012}}
\index{negative imaginary systems!negative imaginary lemma}
Consider a square, continuous-time LTI system, $\mbc{G}: \mathcal{L}_{2e} \to \mathcal{L}_{2e}$, with state-space realization $(\mbf{A},\mbf{B}_,\mbf{C},\mbf{D})$, where $\mbf{A} \in \mathbb{R}^{n \times n}$, $\mbf{B} \in \mathbb{R}^{n \times m}$, $\mbf{C} \in \mathbb{R}^{m \times n}$, and $\mbf{D} \in \mathbb{S}^{m}$.  The system $\mbc{G}$ is negative imaginary under either of the following equivalent necessary and sufficient conditions.

\begin{enumerate}

\item There exists $\mbf{P} \in \mathbb{S}^n$, where $\mbf{P}  \geq 0$, such that
\beq
\label{eq:NI_Lemma1}
 \bbm \mbf{P}\mbf{A}+\mbf{A}^\trans\mbf{P} & \mbf{P}\mbf{B}-\mbf{A}^\trans\mbf{C}^\trans \\ * & -\left(\mbf{C}\mbf{B}+\mbf{B}^\trans\mbf{C}^\trans\right) \ebm \leq 0.
\eeq
\item There exists $\mbf{Q} \in \mathbb{S}^n$, where $\mbf{Q} \geq 0$, such that
\beq
\label{eq:NI_Lemma2}
 \bbm \mbf{A}\mbf{Q}+\mbf{Q}\mbf{A}^\trans & \mbf{B}-\mbf{Q}\mbf{A}^\trans\mbf{C}^\trans \\ * & -\left(\mbf{C}\mbf{B}+\mbf{B}^\trans\mbf{C}^\trans\right) \ebm \leq 0.
\eeq
\end{enumerate}
%
The system $\mbc{G}$ is strictly negative imaginary if $\det(\mbf{A}) \neq 0$ and either~\eqref{eq:NI_Lemma1} is satisfied with $\mbf{P} > 0$ or~\eqref{eq:NI_Lemma2} is satisfied with $\mbf{Q} > 0$.
%
%
%
%

\subsubsection[Discrete-Time Negative Imaginary Lemma]{Discrete-Time Negative Imaginary Lemma}
\index{negative imaginary systems!discrete-time negative imaginary lemma}

Consider a square, discrete-time LTI system, $\mbc{G}: \ell_{2e} \to \ell_{2e}$, with state-space realization $(\mbf{A}_\mathrm{d},\mbf{B}_\mathrm{d},\mbf{C}_\mathrm{d},\mbf{D}_\mathrm{d})$, where $\mbf{A}_\mathrm{d} \in \mathbb{R}^{n \times n}$, $\mbf{B}_\mathrm{d} \in \mathbb{R}^{n \times m}$, $\mbf{C}_\mathrm{d} \in \mathbb{R}^{m \times n}$, $\mbf{D}_\mathrm{d} \in \mathbb{R}^{m \times m}$, $\mbf{C}_\mathrm{d} \left(z \mbf{1} - \mbf{A}_\mathrm{d} \right)^{-1} \mbf{B}_\mathrm{d} + \mbf{D}_\mathrm{d} = \mbf{B}_\mathrm{d}^\trans \left(z \mbf{1} - \mbf{A}_\mathrm{d}^\trans \right)^{-1} \mbf{C}_\mathrm{d}^\trans + \mbf{D}_\mathrm{d}^\trans$, $\text{det} \left( \mbf{1} + \mbf{A} \right) \neq 0$, and $\text{det} \left( \mbf{1} - \mbf{A} \right) \neq 0$.  The system $\mbc{G}$ is negative imaginary under either of the following equivalent necessary and sufficient conditions.

\begin{enumerate}

\itemcite \cite{Ferrante2017,Liu2017} There exists $\mbf{P} \in \mathbb{S}^n$, where $\mbf{P}  > 0$, such that
\begin{align*}
\mbf{A}_\mathrm{d}^\trans \mbf{P} \mbf{A}_\mathrm{d} - \mbf{P} & \leq 0, \\
\mbf{C}_\mathrm{d} + \mbf{B}_\mathrm{d}^\trans \left( \mbf{A}_\mathrm{d}^\trans - \mbf{1} \right)^{-1} \mbf{P}\left(\mbf{A}_\mathrm{d} + \mbf{1} \right) &= \mbf{0}.
\end{align*}

\itemcite \cite{Ferrante2017} There exists $\mbf{Q} \in \mathbb{S}^n$, where $\mbf{Q}  > 0$, such that
\begin{align*}
\mbf{A}_\mathrm{d} \mbf{Q} \mbf{A}_\mathrm{d}^\trans - \mbf{Q} & \leq 0, \\
\mbf{B}_\mathrm{d} +  \left( \mbf{A}_\mathrm{d}- \mbf{1} \right)^{-1} \mbf{Q}\left(\mbf{A}_\mathrm{d}^\trans  + \mbf{1} \right)\mbf{C}_\mathrm{d}^\trans &= \mbf{0}.
\end{align*}

\end{enumerate}

\subsubsection[Generalized Negative Imaginary Lemma]{Generalized Negative Imaginary Lemma}
\index{negative imaginary systems!generalized negative imaginary lemma}

Consider a square, continuous-time LTI system, $\mbc{G}: \mathcal{L}_{2e} \to \mathcal{L}_{2e}$, with state-space realization $(\mbf{A},\mbf{B}_,\mbf{C},\mbf{D})$, where $\mbf{A} \in \mathbb{R}^{n \times n}$, $\mbf{B} \in \mathbb{R}^{n \times m}$, $\mbf{C} \in \mathbb{R}^{m \times n}$, and $\mbf{D} \in \mathbb{S}^{m}$.  Also consider $\mbs{\Pi}_p \in \mathbb{S}^{m}$, which is defined as
\bdis
\mbs{\Pi}_p = \bbm \mbf{0} & \mbf{1} \\ \mbf{1} & \mbf{0} \ebm,
\edis
The following generalized KYP Lemmas give conditions for $\mbc{G}$ to be negative imaginary within finite frequency bandwidths.

\begin{enumerate}

\item (\textit{Low Frequency Range}~\cite{Xiong2012}) The system $\mbc{G}$ is negative imaginary for all $\omega \in \{\omega \in \mathbb{R} \,\, | \,\, \abs{\omega} < \omega_1, \,\, \det (j\omega \mbf{1} - \mbf{A}) \neq 0 \}$, where $\omega_1 \in \mathbb{R}_{>0}$, if $\mbf{D} = \mbf{D}^\trans$ and there exist $\mbf{P}$,~$\mbf{Q} \in \mathbb{S}^n$ and $\bar{\omega}_1 \in \mathbb{R}_{>0}$, where $\mbf{Q}  \geq 0$, such that
\beq
\label{eq:Gen_NegImagLow}
\bbm \mbf{A} & \mbf{B} \\ \mbf{1} & \mbf{0} \ebm^\trans \bbm -\mbf{Q} & \mbf{P} \\ * & (\omega_1 - \bar{\omega}_1)^2 \mbf{Q} \ebm \bbm \mbf{A} & \mbf{B} \\ \mbf{1} & \mbf{0} \ebm - \bbm \mbf{C}\mbf{A} & \mbf{C}\mbf{B} \\ \mbf{0} & \mbf{1} \ebm^\trans \mbs{\Pi}_p \bbm \mbf{C}\mbf{A} & \mbf{C}\mbf{B} \\ \mbf{0} & \mbf{1} \ebm < 0.
\eeq
If $\omega_1 \to \infty$, $\mbf{P} > 0$, and $\mbf{Q} = \mbf{0}$, then the traditional Negative Imaginary Lemma is recovered~\cite{Xiong2012}.

The parameter $\bar{\omega}_1$ is included in~\eqref{eq:Gen_NegImagLow} to effectively transform $\abs{\omega} \leq (\omega_1 - \bar{\omega}_1)$ into the strict inequality $\abs{\omega} < \omega_1$.

\item (\textit{Intermediate Frequency Range}) The system $\mbc{G}$ is negative imaginary for all $\omega \in \{\omega \in \mathbb{R} \,\, | \,\, \omega_1 \leq \abs{\omega} < \omega_2, \,\, \det (j\omega \mbf{1} - \mbf{A}) \neq 0 \}$, where $\omega_1$,~$\omega_2 \in \mathbb{R}_{>0}$, if $\mbf{D} = \mbf{D}^\trans$ and there exist $\mbf{P}$,~$\mbf{Q} \in \mathbb{C}^n$, $\bar{\omega}_2 \in \mathbb{R}_{>0}$, and $\hat{\omega}_2 = \left(\omega_1 + (\omega_2 - \bar{\omega}_2)\right)/2$, where $\mbf{P}^\herm = \mbf{P}$, $\mbf{Q}^\herm = \mbf{Q}$, and $\mbf{Q}  \geq 0$, such that
\beq
\label{eq:Gen_NegImagInter}
\bbm \mbf{A} & \mbf{B} \\ \mbf{1} & \mbf{0} \ebm^\trans \bbm -\mbf{Q} & \mbf{P} + j \hat{\omega}_2 \mbf{Q} \\ \mbf{P} - j\hat{\omega}_2 \mbf{Q} & -\omega_1(\omega_2 - \bar{\omega}-2) \mbf{Q} \ebm \bbm \mbf{A} & \mbf{B} \\ \mbf{1} & \mbf{0} \ebm -\bbm \mbf{C}\mbf{A} & \mbf{C}\mbf{B} \\ \mbf{0} & \mbf{1} \ebm^\trans \mbs{\Pi}_p \bbm \mbf{C}\mbf{A} & \mbf{C}\mbf{B} \\ \mbf{0} & \mbf{1} \ebm< 0.
\eeq
The parameter $\bar{\omega}_2$ is included in~\eqref{eq:Gen_NegImagInter} to effectively transform $ \omega_1 \leq \abs{\omega} \leq (\omega_2 - \bar{\omega}_2)$ into the strict inequality $\omega_1 \leq \abs{\omega} < \omega_2$.

\item (\textit{High Frequency Range}) The system $\mbc{G}$ is negative imaginary for all $\omega \in \{\omega \in \mathbb{R} \,\, | \,\, \omega_2 \leq  \abs{\omega}, \,\, \det (j\omega \mbf{1} - \mbf{A}) \neq 0 \}$, where $\omega_2 \in \mathbb{R}_{>0}$, if $\mbf{D} = \mbf{D}^\trans$ and there exist $\mbf{P}$,~$\mbf{Q} \in \mathbb{S}^n$, where $\mbf{Q}  \geq 0$, such that
\beq
\label{eq:Gen_NegImagHigh}
\bbm \mbf{A} & \mbf{B} \\ \mbf{1} & \mbf{0} \ebm^\trans \bbm \mbf{Q} & \mbf{P} \\ * & -\omega_2^2\mbf{Q} \ebm \bbm \mbf{A} & \mbf{B} \\ \mbf{1} & \mbf{0} \ebm - \bbm \mbf{C}\mbf{A} & \mbf{C}\mbf{B} \\ \mbf{0} & \mbf{1} \ebm^\trans \mbs{\Pi}_p \bbm \mbf{C}\mbf{A} & \mbf{C}\mbf{B} \\ \mbf{0} & \mbf{1} \ebm < 0.
\eeq

\end{enumerate}

\subsubsection[Negative Imaginary System DC Constraint]{Negative Imaginary System DC Constraint~\cite{Caverly2019,Lee2019},~\cite[pp.~32--34]{Lee_MS_thesis}}

Consider an NI transfer matrix $\mbf{G}_{1}(s)$ and an SNI transfer matrix $\mbf{G}_{2}(s)  = \mbf{C}_2 \left(s\mbf{1} - \mbf{A}_{2}\right)^{-1}\mbf{B}_{2} + \mbf{D}_{2}$.  The condition $\bar{\lambda}(\mbf{G}_{1}(0)\mbf{G}_{2}(0)) < 1$ is satisfied if and only if
\bdis
	\mbf{S}^{\trans}(-\mbf{C}_{2}\mbf{A}_{2}^{-1}\mbf{B}_{2} + \mbf{D}_{2})\mbf{S} < \mbf{1},
\edis
where $\mbf{S}\mbf{S}^{\trans} = \mbf{G}_{1}(0)$.

\subsection{Algebraic Riccati Inequalities}

\subsubsection[Algebraic Riccati Inequality]{Algebraic Riccati Inequality~\cite{Willems1971}}
Consider $\mbf{A} \in \mathbb{R}^{n \times n}$, $\mbf{B} \in \mathbb{R}^{n \times m}$, $\mbf{P}$,~$\mbf{Q} \in \mathbb{S}^n$, $\mbf{N} \in \mathbb{R}^{n \times m}$, and $\mbf{R} \in \mathbb{S}^m$, where $\mbf{P} >0$, $\mbf{Q} \geq 0$, and $\mbf{R} > 0$.  The algebraic Riccati inequality given by
\bdis
\mbf{A}^\trans \mbf{P} + \mbf{P}\mbf{A} - \left(\mbf{P}\mbf{B} + \mbf{N}^\trans\right)\mbf{R}^{-1}\left(\mbf{B}^\trans\mbf{P} + \mbf{N}\right) + \mbf{Q} \geq 0,
\edis
can be rewritten using the Schur complement lemma as
\bdis
\bbm \mbf{A}^\trans\mbf{P}+\mbf{P}\mbf{A}+\mbf{Q} & \mbf{P}\mbf{B} + \mbf{N}^\trans \\ * & \mbf{R} \ebm \geq 0.
\edis

\subsubsection[Discrete-Time Algebraic Riccati Inequality]{Discrete-Time Algebraic Riccati Inequality~\cite{Hung1998}}
Consider $\mbf{A}_\mathrm{d} \in \mathbb{R}^{n \times n}$, $\mbf{B}_\mathrm{d} \in \mathbb{R}^{n \times m}$, $\mbf{P}$,~$\mbf{Q} \in \mathbb{S}^n$, and $\mbf{R} \in \mathbb{S}^m$, where $\mbf{P} >0$, $\mbf{Q} \geq 0$, and $\mbf{R} > 0$.  The discrete-time algebraic Riccati inequality given by
\bdis
\mbf{A}_\mathrm{d}^\trans \mbf{P}\mbf{A}_\mathrm{d} - \mbf{A}_\mathrm{d}^\trans\mbf{P}\mbf{B}_\mathrm{d}\left(\mbf{R} + \mbf{B}_\mathrm{d}^\trans\mbf{P}\mbf{B}_\mathrm{d}\right)^{-1}\mbf{B}_\mathrm{d}^\trans\mbf{P}\mbf{A}_\mathrm{d} + \mbf{Q} - \mbf{P} \geq 0,
\edis
can be rewritten using the Schur complement lemma as
\bdis
\bbm \mbf{A}_\mathrm{d}^\trans \mbf{P}\mbf{A}_\mathrm{d} - \mbf{P} + \mbf{Q} & \mbf{A}_\mathrm{d}^\trans\mbf{P}\mbf{B}_\mathrm{d} \\ * & \mbf{R} + \mbf{B}_\mathrm{d}^\trans\mbf{P}\mbf{B}_\mathrm{d} \ebm \geq 0.
\edis
Equivalently, this discrete-time algebraic Riccati inequality is satisfied under any of the following necessary and sufficient conditions.
\begin{enumerate}

\item There exist $\mbf{P}$,~$\mbf{Q} \in \mathbb{S}^n$, and $\mbf{R} \in \mathbb{S}^m$, where $\mbf{P} >0$, $\mbf{Q} \geq 0$, and $\mbf{R} > 0$, such that
\bdis
\bbm  \mbf{Q}  & \mbf{0} & \mbf{A}_\mathrm{d}^\trans\mbf{P} & \mbf{P}  \\ * & \mbf{R} &  \mbf{B}_\mathrm{d}^\trans\mbf{P} & \mbf{0} \\ * & * & \mbf{P} & \mbf{0} \\ * & * & * & \mbf{P} \ebm \geq 0.
\edis

\item There exist $\mbf{X}$,~$\mbf{Q} \in \mathbb{S}^n$, and $\mbf{R} \in \mathbb{S}^m$, where $\mbf{X} >0$, $\mbf{Q} \geq 0$, and $\mbf{R} > 0$, such that
\bdis
\bbm  \mbf{Q}  & \mbf{0} & \mbf{A}_\mathrm{d}^\trans & \mbf{1}  \\ * & \mbf{R} &  \mbf{B}_\mathrm{d}^\trans & \mbf{0} \\ * & * & \mbf{X} & \mbf{0} \\ * & * & * & \mbf{X} \ebm \geq 0.
\edis

\end{enumerate}

\subsection[Stabilizability]{Stabilizability}
\index{stabilizability}
\subsubsection[Continuous-Time Stabilizability]{Continuous-Time Stabilizability~\cite[pp.~166--168]{Duan2013}}

Consider a continuous-time LTI system, $\mbc{G}: \mathcal{L}_{2e} \to \mathcal{L}_{2e}$, with state-space realization $(\mbf{A},\mbf{B}_,\mbf{C},\mbf{D})$, where $\mbf{A} \in \mathbb{R}^{n \times n}$, $\mbf{B} \in \mathbb{R}^{n \times m}$, $\mbf{C} \in \mathbb{R}^{p \times n}$, and $\mbf{D} \in \mathbb{R}^{p \times m}$.  The system $\mbc{G}$ is stabilizable if and only if there exists $\mbf{P} \in \mathbb{S}^n$, where $\mbf{P} > 0$, such that
\bdis
\mbf{A}\mbf{P} + \mbf{P}\mbf{A}^\trans - \mbf{B}\mbf{B}^\trans < 0.
\edis
The matrix $\mbf{A} + \mbf{B} \mbf{K}$ is Hurwitz\index{Hurwitz matrix} with $\mbf{K} = -\onehalf \mbf{B}^\trans \mbf{P}^{-1}$.  Equivalently, $\mbc{G}$ is stabilizable if and only if there exist $\mbf{P} \in \mathbb{S}^n$ and $\mbf{W} \in \mathbb{R}^{m \times n}$, where $\mbf{P} > 0$, such that
\bdis
\mbf{A}\mbf{P} + \mbf{P}\mbf{A}^\trans + \mbf{B} \mbf{W} + \mbf{W}^\trans \mbf{B}^\trans < 0.
\edis
The matrix $\mbf{A} + \mbf{B} \mbf{K}$ is Hurwitz\index{Hurwitz matrix} with $\mbf{K} = \mbf{W} \mbf{P}^{-1}$.

\subsubsection[Discrete-Time Stabilizability]{Discrete-Time Stabilizability~\cite[pp.~172--176]{Duan2013}}

Consider a discrete-time LTI system, $\mbc{G}: \ell_{2e} \to \ell_{2e}$, with state-space realization $(\mbf{A}_\mathrm{d},\mbf{B}_\mathrm{d},\mbf{C}_\mathrm{d},\mbf{D}_\mathrm{d})$, where $\mbf{A}_\mathrm{d} \in \mathbb{R}^{n \times n}$, $\mbf{B}_\mathrm{d} \in \mathbb{R}^{n \times m}$, $\mbf{C}_\mathrm{d} \in \mathbb{R}^{p \times n}$, and $\mbf{D}_\mathrm{d} \in \mathbb{R}^{p \times m}$.  The system $\mbc{G}$ is stabilizable if and only if there exists $\mbf{P} \in \mathbb{S}^n$, where $\mbf{P} > 0$, such that
\bdis
\bbm \mbf{P} & \mbf{P}\mbf{A}_\mathrm{d}^\trans \\ * & \mbf{P}  + \mbf{B}_\mathrm{d} \mbf{B}_\mathrm{d}^\trans \ebm > 0.
\edis
The matrix $\mbf{A}_\mathrm{d} + \mbf{B}_\mathrm{d} \mbf{K}_\mathrm{d}$ is Schur\index{Schur matrix} with $\mbf{K}_\mathrm{d} = -\left(2\mbf{1} + \mbf{B}_\mathrm{d}^\trans\mbf{P}^{-1}\mbf{B}_\mathrm{d}\right)^{-1}\mbf{B}_\mathrm{d}^\trans \mbf{P}^{-1}\mbf{A}_\mathrm{d}$.  Equivalently, $\mbc{G}$ is stabilizable if and only if there exist $\mbf{P} \in \mathbb{S}^n$ and $\mbf{W} \in \mathbb{R}^{m \times n}$, where $\mbf{P} > 0$, such that
\bdis
\bbm \mbf{P} & \mbf{A}_\mathrm{d}\mbf{P} + \mbf{B}_\mathrm{d} \mbf{W} \\ * & \mbf{P} \ebm > 0.
\edis
The matrix $\mbf{A}_\mathrm{d} + \mbf{B}_\mathrm{d} \mbf{K}_\mathrm{d}$ is Schur\index{Schur matrix} with $\mbf{K}_\mathrm{d} = \mbf{W} \mbf{P}^{-1}$.

\subsection[Detectability]{Detectability}
\index{detectability}
\subsubsection[Continuous-Time Detectability]{Continuous-Time Detectability~\cite[pp.~170--171]{Duan2013}}

Consider a continuous-time LTI system, $\mbc{G}: \mathcal{L}_{2e} \to \mathcal{L}_{2e}$, with state-space realization $(\mbf{A},\mbf{B}_,\mbf{C},\mbf{D})$, where $\mbf{A} \in \mathbb{R}^{n \times n}$, $\mbf{B} \in \mathbb{R}^{n \times m}$, $\mbf{C} \in \mathbb{R}^{p \times n}$, and $\mbf{D} \in \mathbb{R}^{p \times m}$.  The system $\mbc{G}$ is detectable if and only if there exists $\mbf{P} \in \mathbb{S}^n$, where $\mbf{P} > 0$, such that
\bdis
\mbf{P}\mbf{A} + \mbf{A}^\trans\mbf{P}  - \mbf{C}^\trans\mbf{C} < 0.
\edis
The matrix $\mbf{A} + \mbf{L} \mbf{C}$ is Hurwitz\index{Hurwitz matrix} with $\mbf{L} = -\onehalf \mbf{P}^{-1} \mbf{C}^\trans $.  Equivalently, $\mbc{G}$ is detectable if and only if there exist $\mbf{P} \in \mathbb{S}^n$ and $\mbf{W} \in \mathbb{R}^{p \times n}$, where $\mbf{P} > 0$, such that
\bdis
\mbf{P}\mbf{A} + \mbf{A}^\trans\mbf{P}  + \mbf{W}^\trans \mbf{C} + \mbf{C}^\trans \mbf{W}< 0.
\edis
The matrix $\mbf{A} + \mbf{L} \mbf{C}$ is Hurwitz\index{Hurwitz matrix} with $\mbf{L} = -\onehalf \mbf{P}^{-1} \mbf{W}^\trans $.

\subsubsection[Discrete-Time Detectability]{Discrete-Time Detectability~\cite[pp.~177--178]{Duan2013}}

Consider a discrete-time LTI system, $\mbc{G}: \ell_{2e} \to \ell_{2e}$, with state-space realization $(\mbf{A}_\mathrm{d},\mbf{B}_\mathrm{d},\mbf{C}_\mathrm{d},\mbf{D}_\mathrm{d})$, where $\mbf{A}_\mathrm{d} \in \mathbb{R}^{n \times n}$, $\mbf{B}_\mathrm{d} \in \mathbb{R}^{n \times m}$, $\mbf{C}_\mathrm{d} \in \mathbb{R}^{p \times n}$, and $\mbf{D}_\mathrm{d} \in \mathbb{R}^{p \times m}$.  The system $\mbc{G}$ is detectable if and only if there exists $\mbf{P} \in \mathbb{S}^n$, where $\mbf{P} > 0$, such that
\bdis
\bbm \mbf{P} & \mbf{P}\mbf{A}_\mathrm{d} \\ * & \mbf{P}  +\mbf{C}_\mathrm{d}^\trans \mbf{C}_\mathrm{d} \ebm > 0.
\edis
The matrix $\mbf{A}_\mathrm{d} + \mbf{L} \mbf{C}_\mathrm{d}$ is Schur\index{Schur matrix} with $\mbf{L} = -\mbf{A}_\mathrm{d}\mbf{P}^{-1}\mbf{C}_\mathrm{d}^\trans\left(2\mbf{1} + \mbf{C}_\mathrm{d}\mbf{P}^{-1}\mbf{C}_\mathrm{d}^\trans\right)^{-1} $.  Equivalently, $\mbc{G}$ is detectable if and only if there exist $\mbf{P} \in \mathbb{S}^n$ and $\mbf{W} \in \mathbb{R}^{m \times n}$, where $\mbf{P} > 0$, such that
\bdis
\bbm \mbf{P} & \mbf{A}_\mathrm{d}^\trans\mbf{P} + \mbf{C}_\mathrm{d}^\trans \mbf{W} \\ * & \mbf{P} \ebm > 0.
\edis
The matrix $\mbf{A}_\mathrm{d} + \mbf{L} \mbf{C}_\mathrm{d}$ is Schur\index{Schur matrix} with $\mbf{L} =  \mbf{P}^{-1}\mbf{W}$.

\subsection{Static Output Feedback Stabilizability}
\index{static output feedback!stabilizability}
\index{stabilizability!static output feedback stabilizability}
\subsubsection[Continuous-Time Static Output Feedback Stabilizability]{Continuous-Time Static Output Feedback Stabilizability~\cite{Cao1998b,Kucera1995},~\cite[p.~120]{Fathi2019}}

Consider a continuous-time LTI system, $\mbc{G}: \mathcal{L}_{2e} \to \mathcal{L}_{2e}$, with state-space realization $(\mbf{A},\mbf{B}_,\mbf{C},\mbf{0})$, where $\mbf{A} \in \mathbb{R}^{n \times n}$, $\mbf{B} \in \mathbb{R}^{n \times m}$, and $\mbf{C} \in \mathbb{R}^{p \times n}$.  The system $\mbc{G}$ is static output feedback stabilizable under any of the following equivalent necessary and sufficient conditions.
\begin{enumerate}

\item There exist $\mbf{K} \in \mathbb{R}^{m \times p}$ and $\mbf{P} \in \mathbb{S}^{n}$, where $\mbf{P} >0$, such that
\bdis
\bbm \mbf{A}^\trans\mbf{P} + \mbf{P}\mbf{A} - \mbf{P}\mbf{B}\mbf{B}^\trans\mbf{P}& \mbf{P}\mbf{B} + \mbf{C}^\trans\mbf{K}^\trans \\ * & -\mbf{1}  \ebm <0.
\edis

\item There exist $\mbf{K} \in \mathbb{R}^{m \times p}$ and $\mbf{Q} \in \mathbb{S}^{n}$, where $\mbf{Q} >0$, such that
\bdis
\bbm \mbf{Q}\mbf{A}^\trans + \mbf{A}\mbf{Q} -\mbf{Q}\mbf{C}^\trans\mbf{C} \mbf{Q} & \mbf{B}\mbf{K} + \mbf{Q}\mbf{C}^\trans \\ * & -\mbf{1}  \ebm <0.
\edis

\item There exist $\mbf{K} \in \mathbb{R}^{m \times p}$ and $\mbf{Q} \in \mathbb{S}^{n}$, where $\mbf{Q} >0$, such that
\bdis
\bbm \mbf{Q}\mbf{A}^\trans + \mbf{A}\mbf{Q} -\mbf{B}\mbf{B}^\trans& \mbf{B} + \mbf{Q}\mbf{C}^\trans\mbf{K}^\trans \\ * & -\mbf{1}  \ebm <0.
\edis

\item There exist $\mbf{K} \in \mathbb{R}^{m \times p}$ and $\mbf{P} \in \mathbb{S}^{n}$, where $\mbf{P} >0$, such that
\bdis
\bbm \mbf{A}^\trans\mbf{P} + \mbf{P}\mbf{A} -\mbf{C}^\trans\mbf{C} & \mbf{P} \mbf{B}\mbf{K} + \mbf{C}^\trans \\ * & -\mbf{1}  \ebm <0.
\edis

\item There exist $\mbf{K} \in \mathbb{R}^{m \times p}$, $\mbf{P}$,~$\mbf{X} \in \mathbb{S}^{n}$, where $\mbf{P} >0$ and $\mbf{X} > 0$, such that
\bdis
\bbm \mbf{A}^\trans\mbf{X} + \mbf{X}\mbf{A} - \mbf{P} \mbf{B} \mbf{B}^\trans \mbf{X} - \mbf{X} \mbf{B}\mbf{B}^\trans \mbf{P} + \mbf{X}\mbf{B}\mbf{B}^\trans\mbf{X}& \mbf{P}\mbf{B} + \mbf{C}^\trans\mbf{K}^\trans \\ * & -\mbf{1}  \ebm <0.
\edis

\item There exist $\mbf{K} \in \mathbb{R}^{m \times p}$ and $\mbf{Q}$,~$\mbf{X} \in \mathbb{S}^{n}$, where $\mbf{Q} >0$ and $\mbf{X} > 0$, such that
\bdis
\bbm \mbf{Q}\mbf{A}^\trans + \mbf{A}\mbf{Q} - \mbf{Q}\mbf{C}^\trans\mbf{C} \mbf{X} - \mbf{X}\mbf{C}^\trans\mbf{C} \mbf{Q} + \mbf{X}\mbf{C}^\trans\mbf{C} \mbf{X}& \mbf{B}\mbf{K} + \mbf{Q}\mbf{C}^\trans \\ * & -\mbf{1}  \ebm <0.
\edis

\end{enumerate}

\subsubsection[Discrete-Time Static Output Feedback Stabilizability]{Discrete-Time Static Output Feedback Stabilizability}
Consider a discrete-time LTI system, $\mbc{G}: \ell_{2e} \to \ell_{2e}$, with state-space realization $(\mbf{A}_\mathrm{d},\mbf{B}_\mathrm{d},\mbf{C}_\mathrm{d},\mbf{0})$, where $\mbf{A}_\mathrm{d} \in \mathbb{R}^{n \times n}$, $\mbf{B}_\mathrm{d} \in \mathbb{R}^{n \times m}$, and $\mbf{C}_\mathrm{d} \in \mathbb{R}^{p \times n}$.  The system $\mbc{G}$ is static output feedback stabilizable under any of the following equivalent necessary and sufficient conditions.
\begin{enumerate}

\item There exist $\mbf{K}_\mathrm{d} \in \mathbb{R}^{m \times p}$ and $\mbf{P} \in \mathbb{S}^{n}$, where $\mbf{P} >0$, such that
\beq
\label{eq:DT_SOF_Stab}
\bbm -\mbf{P} &\left(\mbf{A}_\mathrm{d} + \mbf{B}_\mathrm{d} \mbf{K}_\mathrm{d} \mbf{C}_\mathrm{d} \right)\mbf{P} \\ * & -\mbf{P}  \ebm <0.
\eeq

\item There exist $\mbf{K}_\mathrm{d} \in \mathbb{R}^{m \times p}$ and $\mbf{P} \in \mathbb{S}^{n}$, where $\mbf{P} >0$, such that
\beq
\label{eq:DT_SOF_Stab2}
\bbm -\mbf{A}_\mathrm{d} \mbf{P} \mbf{P} \mbf{A}_\mathrm{d}^\trans & \mbf{A}_\mathrm{d}\mbf{P} + \mbf{B}_\mathrm{d} \mbf{K}_\mathrm{d} \mbf{C}_\mathrm{d} & \mbf{A}_\mathrm{d} \mbf{P} \\ * & -\mbf{1} & \mbf{0} \\ * & * & -\mbf{P}  \ebm <0.
\eeq
\begin{proof}
Applying the reverse Schur complement lemma to~\eqref{eq:DT_SOF_Stab} yields
\bdis
\left(\mbf{A}_\mathrm{d} + \mbf{B}_\mathrm{d} \mbf{K}_\mathrm{d} \mbf{C}_\mathrm{d} \right)\mbf{P}\left(\mbf{A}_\mathrm{d} + \mbf{B}_\mathrm{d} \mbf{K}_\mathrm{d} \mbf{C}_\mathrm{d} \right)^\trans - \mbf{P} < 0.
\edis
Multiplying out this matrix inequality and adding $ \mbf{0} = \mbf{A}_\mathrm{d} \mbf{P} \mbf{P} \mbf{A}_\mathrm{d} - \mbf{A}_\mathrm{d} \mbf{P} \mbf{P} \mbf{A}_\mathrm{d}$ to the left-hand side gives
\bdis
\mbf{A}_\mathrm{d} \mbf{P} \mbf{A}_\mathrm{d}^\trans - \mbf{A}_\mathrm{d} \mbf{P} \mbf{P} \mbf{A}_\mathrm{d}^\trans + \left(\mbf{A}_\mathrm{d}\mbf{P} + \mbf{B}_\mathrm{d} \mbf{K}_\mathrm{d} \mbf{C}_\mathrm{d}\right) \left(\mbf{A}_\mathrm{d}\mbf{P} + \mbf{B}_\mathrm{d} \mbf{K}_\mathrm{d} \mbf{C}_\mathrm{d}\right)^\trans < 0.
\edis 
Applying the Schur complement lemma twice gives~\eqref{eq:DT_SOF_Stab2}.
\end{proof}

\end{enumerate}

The system $\mbc{G}$ is also static output feedback stabilizable if there exist $\mbf{K}_\mathrm{d} \in \mathbb{R}^{m \times p}$ and $\mbf{P}$,~$\mbf{X} \in \mathbb{S}^{n}$, where $\mbf{P} >0$ and $\mbf{X} > 0$, such that
\beq
\label{eq:DT_SOF_Stab3}
\bbm -\mbf{A}_\mathrm{d} \left(\mbf{X} \mbf{P} + \mbf{P} \mbf{X} \right)\mbf{A}_\mathrm{d}^\trans & \mbf{A}_\mathrm{d}\mbf{P} + \mbf{B}_\mathrm{d} \mbf{K}_\mathrm{d} \mbf{C}_\mathrm{d} & \mbf{A}_\mathrm{d} \mbf{P} & \mbf{A}_\mathrm{d} \mbf{X} \\ * & -\mbf{1} & \mbf{0} & \mbf{0} \\ * & * & -\mbf{P} & \mbf{0} \\ * & * & * & -\mbf{1}  \ebm <0.
\eeq
\begin{proof}
Using completion of the squares, it can be shown that
\beq
\label{eq:DT_SOF_Stab4}
-\mbf{A}_\mathrm{d} \mbf{P} \mbf{P} \mbf{A}_\mathrm{d}^\trans \leq -\mbf{A}_\mathrm{d} \left(\mbf{X} \mbf{P} + \mbf{P} \mbf{X} \right)\mbf{A}_\mathrm{d}^\trans + \mbf{A}_\mathrm{d} \mbf{X} \mbf{X} \mbf{A}_\mathrm{d}^\trans.
\eeq
Substituting~\eqref{eq:DT_SOF_Stab4} into~\eqref{eq:DT_SOF_Stab2} and using the Schur complement lemma yields~\eqref{eq:DT_SOF_Stab3}.  The matrix inequality in~\eqref{eq:DT_SOF_Stab3} is only a sufficient condition for static output feedback stabilizability since~\eqref{eq:DT_SOF_Stab4} is an inequality.
\end{proof}

\subsection[Strong Stabilizability]{Strong Stabilizability}
\index{stabilizability!strong stabilizability}

\subsubsection[Continuous-Time Strong Stabilizability]{Continuous-Time Strong Stabilizability~\cite{Gumussoy2005}}

Consider a continuous-time LTI system, $\mbc{G}: \mathcal{L}_{2e} \to \mathcal{L}_{2e}$, with state-space realization $(\mbf{A},\mbf{B}_,\mbf{C},\mbf{0})$, where $\mbf{A} \in \mathbb{R}^{n \times n}$, $\mbf{B} \in \mathbb{R}^{n \times m}$, and $\mbf{C} \in \mathbb{R}^{p \times n}$, and it is assumed that $(\mbf{A},\mbf{B})$ is stabilizable, $(\mbf{A},\mbf{C})$ is detectable, and the transfer matrix $\mbf{G}(s) = \mbf{C} \left(s \mbf{1} - \mbf{A} \right)^{-1} \mbf{B} $ has no poles on the imaginary axis.  The system $\mbc{G}$ is strongly stabilizable if there exist $\mbf{P} \in \mathbb{S}^n$, $\mbf{Z} \in \mathbb{R}^{n \times p}$, and $\gamma \in \mathbb{R}_{>0}$, where $\mbf{P} > 0$, such that
\begin{align*}
\mbf{P}\mbf{A} + \mbf{A}^\trans \mbf{P}  + \mbf{Z} \mbf{C} + \mbf{C}^\trans \mbf{Z}^\trans &< 0, \\
\bbm \mbf{P}\left(\mbf{A} + \mbf{B}\mbf{F}\right) + \left(\mbf{A} + \mbf{B}\mbf{F}\right)^\trans \mbf{P} + \mbf{Z} \mbf{C} + \mbf{C}^\trans \mbf{Z}^\trans & - \mbf{Z} & - \mbf{X} \mbf{B} \\ * & - \gamma \mbf{1} & \mbf{0} \\ * & * & - \gamma \mbf{1} \ebm &< 0,
\end{align*}
where $\mbf{F} = - \mbf{B}^\trans \mbf{X}$ and $\mbf{X} \in \mathbb{S}_n$, $\mbf{X} \geq 0$ is the solution to the Lyapunov equation given by
\bdis
\mbf{X} \mbf{A} + \mbf{A}^\trans \mbf{X} - \mbf{X} \mbf{B} \mbf{B}^\trans \mbf{X} = \mbf{0}.
\edis
Moreover, a controller that strongly stabilizes $\mbc{G}$ is given by the state-space realization
\begin{align*}
\dot{\mbf{x}}_c &= \left(\mbf{A} + \mbf{B}\mbf{F} + \mbf{P}^{-1} \mbf{Z} \mbf{C}\right) \mbf{x} - \mbf{P}^{-1} \mbf{Z} \mbf{u}, \\
\mbf{y}_c &= -\mbf{B}^\trans \mbf{X} \mbf{x}.
\end{align*}

\subsubsection[Discrete-Time Strong Stabilizability]{Discrete-Time Strong Stabilizability}

Consider a discrete-time LTI system, $\mbc{G}: \ell_{2e} \to \ell_{2e}$, with state-space realization $(\mbf{A}_\mathrm{d},\mbf{B}_\mathrm{d},\mbf{C}_\mathrm{d},\mbf{0})$, where $\mbf{A}_\mathrm{d} \in \mathbb{R}^{n \times n}$, $\mbf{B}_\mathrm{d} \in \mathbb{R}^{n \times m}$, and $\mbf{C}_\mathrm{d} \in \mathbb{R}^{p \times n}$, and it is assumed that $(\mbf{A}_\mathrm{d},\mbf{B}_\mathrm{d})$ is stabilizable, $(\mbf{A}_\mathrm{d},\mbf{C}_\mathrm{d})$ is detectable, and the transfer matrix $\mbf{G}(z) = \mbf{C}_\mathrm{d} \left(z \mbf{1} - \mbf{A}_\mathrm{d} \right)^{-1} \mbf{B}_\mathrm{d} $ has no poles on the unit circle.  The system $\mbc{G}$ is strongly stabilizable if there exist $\mbf{P} \in \mathbb{S}^n$, $\mbf{Z} \in \mathbb{R}^{n \times p}$, and $\gamma \in \mathbb{R}_{>0}$, where $\mbf{P} > 0$, such that
\begin{align}
\bbm \mbf{A}_\mathrm{d}^\trans \mbf{P} \mbf{A}_\mathrm{d} - \mbf{P} - \mbf{A}_\mathrm{d}^\trans \mbf{Z} \mbf{C}_\mathrm{d} - \mbf{C}_\mathrm{d}^\trans \mbf{Z}^\trans \mbf{A}_\mathrm{d} & \mbf{C}_\mathrm{d}^\trans \mbf{Z}^\trans \\ * & -\mbf{P} \ebm &< 0, \label{eq:DT_StrongStab1}\\
\bbm \mbf{N}_{11} & \left(\mbf{A}_\mathrm{d} + \mbf{B}_\mathrm{d} \mbf{F} \right)^\trans\mbf{Z} & \mbf{X}\mbf{B}_\mathrm{d} & \mbf{C}_\mathrm{d}^\trans \mbf{Z}^\trans \\ * & -\gamma \mbf{1} & \mbf{0} & \mbf{Z}^\trans  \\ * & * & -\gamma\mbf{1} & \mbf{0} \\ * & * & * & -\mbf{P} \ebm &< 0, \label{eq:DT_StrongStab2}
\end{align}
where $\mbf{N}_{11} = \left(\mbf{A}_\mathrm{d} + \mbf{B}_\mathrm{d} \mbf{F} \right)^\trans\mbf{P}\left(\mbf{A}_\mathrm{d} + \mbf{B}_\mathrm{d} \mbf{F} \right) -\mbf{P} + \left(\mbf{A}_\mathrm{d} + \mbf{B}_\mathrm{d} \mbf{F} \right)^\trans \mbf{Z} \mbf{C}_\mathrm{d} + \mbf{C}_\mathrm{d}^\trans \mbf{Z}^\trans\left(\mbf{A}_\mathrm{d} + \mbf{B}_\mathrm{d} \mbf{F} \right)$, $\mbf{F} = - \mbf{B}_\mathrm{d}^\trans \mbf{X}$, $\mbf{X} = \mbf{Y}$, and $\mbf{Y} \in \mathbb{S}_n$, $\mbf{Y} \geq 0$ is the solution to the discrete-time Lyapunov equation given by
\bdis
\mbf{A}_\mathrm{d}\mbf{Y}\mbf{A}_\mathrm{d}^\trans - \mbf{Y} -\mbf{B}_\mathrm{d} \mbf{B}_\mathrm{d}^\trans = \mbf{0}.
\edis
Moreover, a discrete-time controller that strongly stabilizes $\mbc{G}$ is given by the state-space realization
\begin{align}
\mbf{x}_{c,k+1} &= \left(\mbf{A}_\mathrm{d} + \mbf{B}_\mathrm{d}\mbf{F} + \mbf{P}^{-1} \mbf{Z} \mbf{C}_\mathrm{d}\right) \mbf{x}_k - \mbf{P}^{-1} \mbf{Z} \mbf{u}_k, \label{eq:DT_StrongStab3}\\
\mbf{y}_{c,k} &= -\mbf{B}_\mathrm{d}^\trans \mbf{X} \mbf{x}_k. \label{eq:DT_StrongStab4}
\end{align}

\begin{proof}
The proof follows the same procedure as in~\cite{Gumussoy2005} for the continuous-time case, where~\eqref{eq:DT_StrongStab1} ensures that the feedback controller defined by~\eqref{eq:DT_StrongStab3} and~\eqref{eq:DT_StrongStab4} renders the closed-loop system asymptotically stable and~\eqref{eq:DT_StrongStab2} ensures that the feedback controller defined by~\eqref{eq:DT_StrongStab3} and~\eqref{eq:DT_StrongStab4} has a finite $\mathcal{H}_\infty$ norm, and thus is asymptotically stable.
\end{proof}

\subsection[System Zeros]{System Zeros}

\subsubsection[System Zeros without Feedthrough]{System Zeros without Feedthrough~\cite{Kouvaritakis1976a}}

Consider a continuous-time LTI system, $\mbc{G}: \mathcal{L}_{2e} \to \mathcal{L}_{2e}$, with minimal state-space realization $(\mbf{A},\mbf{B},\mbf{C},\mbf{0})$, where $\mbf{A} \in \mathbb{R}^{n \times n}$, $\mbf{B} \in \mathbb{R}^{n \times m}$, and $\mbf{C} \in \mathbb{R}^{p \times n}$.  The transmission zeros\index{transmission zeros} of $\mbf{G}(s) = \mbf{C}\left(s\mbf{1} - \mbf{A}\right)^{-1}\mbf{B}$ are the eigenvalues\index{eigenvalues} of $\mbf{N}\mbf{A}\mbf{M}$, where $\mbf{N} \in \mathbb{R}^{q \times n}$, $\mbf{M} \in \mathbb{R}^{n \times q}$, $\mbf{C}\mbf{M} = \mbf{0}$, $\mbf{N}\mbf{B} = \mbf{0}$, and $\mbf{N}\mbf{M} = \mbf{1}$.  Therefore, $\mbf{G}(s)$ is minimum phase\index{minimum phase} if and only if there exists $\mbf{P} \in \mathbb{S}^{q}$, where $\mbf{P} > 0$, such that
\bdis
\mbf{P}\mbf{N}\mbf{A}\mbf{M} + \mbf{M}^\trans\mbf{A}^\trans\mbf{N}^\trans\mbf{P} < 0.
\edis

\subsubsection{System Zeros with Feedthrough}
\label{sec:Zeros_Feedthrough}

Consider a continuous-time LTI system, $\mbc{G}: \mathcal{L}_{2e} \to \mathcal{L}_{2e}$, with minimal state-space realization $(\mbf{A},\mbf{B},\mbf{C},\mbf{D})$, where $\mbf{A} \in \mathbb{R}^{n \times n}$, $\mbf{B} \in \mathbb{R}^{n \times m}$, $\mbf{C} \in \mathbb{R}^{p \times n}$, $\mbf{D} \in \mathbb{R}^{p \times m}$, $m \leq p$, and $\mbf{D}$ is full rank.  The transmission zeros\index{transmission zeros} of $\mbf{G}(s) = \mbf{C}\left(s\mbf{1} - \mbf{A}\right)^{-1}\mbf{B} + \mbf{D}$ are the eigenvalues\index{eigenvalues} of $\mbf{A} - \mbf{B}\left(\mbf{D}^\trans \mbf{D}\right)^{-1}\mbf{D}^\trans \mbf{C}$.\index{rank}

\begin{enumerate}

\item $\mbf{G}(s)$ is minimum phase\index{minimum phase} if and only if there exists $\mbf{P} \in \mathbb{S}^{n}$, where $\mbf{P} > 0$, such that
\beq
\label{eq:Zero_D}
\mbf{P}\left(\mbf{A} - \mbf{B}\left(\mbf{D}^\trans \mbf{D}\right)^{-1}\mbf{D}^\trans \mbf{C}\right)+ \left(\mbf{A} - \mbf{B}\left(\mbf{D}^\trans \mbf{D}\right)^{-1}\mbf{D}^\trans \mbf{C}\right)^\trans\mbf{P} < 0.
\eeq
If the system is square ($m = p$), then $\mbf{D}$ full rank implies $\mbf{D}^{-1}$ exists and~\eqref{eq:Zero_D} simplifies to\index{rank}
\beq
\label{eq:Zero_D1}
\mbf{P}\left(\mbf{A} - \mbf{B}\mbf{D}^{-1} \mbf{C}\right)+ \left(\mbf{A} - \mbf{B}\mbf{D}^{-1} \mbf{C}\right)^\trans\mbf{P} < 0.
\eeq
\begin{proof}
The system $\mbc{G}$ can be written in state-space form as
\begin{align}
\dot{\mbf{x}} &= \mbf{A} \mbf{x} + \mbf{B} \mbf{u}, \label{eq:Zero_proof1a1}\\
\mbf{y} &= \mbf{C} \mbf{x} + \mbf{D} \mbf{u}. \label{eq:Zero_proof1a}
\end{align}
Left-multiplying~\eqref{eq:Zero_proof1a} by $\mbf{D}^\trans$ and rearranging yields
\beq
\label{eq:Zero_proof1b}
\mbf{D}^\trans \mbf{D} \mbf{u} = -\mbf{D}^\trans \mbf{C} \mbf{x} + \mbf{D}^\trans \mbf{y}.
\eeq
Since $\mbf{D}$ is full rank, $\left(\mbf{D}^\trans \mbf{D}\right)^{-1}$ exists.  Therefore, left-multiplying~\eqref{eq:Zero_proof1b} by $\left(\mbf{D}^\trans \mbf{D}\right)^{-1}$ gives\index{rank}
\beq
\label{eq:Zero_proof1c}
\mbf{u} = - \left(\mbf{D}^\trans \mbf{D}\right)^{-1} \mbf{D}^\trans \mbf{C} \mbf{x} + \left(\mbf{D}^\trans \mbf{D}\right)^{-1}\mbf{D}^\trans \mbf{y}.
\eeq
Substituting~\eqref{eq:Zero_proof1c} into~\eqref{eq:Zero_proof1a1} gives the following state-space representation of the inverted transfer matrix from $\mbf{y}$ to $\mbf{u}$.
\begin{align}
\dot{\mbf{x}} &=\left(\mbf{A} - \mbf{B}\left(\mbf{D}^\trans \mbf{D}\right)^{-1}\mbf{D}^\trans \mbf{C}\right)\mbf{x} + \mbf{B} \left(\mbf{D}^\trans \mbf{D}\right)^{-1}\mbf{D}^\trans \mbf{y}, \label{eq:Zero_proof1d}\\
\mbf{u} &= - \left(\mbf{D}^\trans \mbf{D}\right)^{-1} \mbf{D}^\trans \mbf{C} \mbf{x} + \left(\mbf{D}^\trans \mbf{D}\right)^{-1}\mbf{D}^\trans \mbf{y}. \label{eq:Zero_proof1e}
\end{align}
The transmission zeros of $\mbf{G}(s)$ are the poles of the inverted transfer matrix from $\mbf{y}$ to $\mbf{u}$, which are the eigenvalues of $\left(\mbf{A} - \mbf{B}\left(\mbf{D}^\trans \mbf{D}\right)^{-1}\mbf{D}^\trans \mbf{C}\right)$.  Substituting this matrix into a Lyapunov inequality gives the desired inequality in~\eqref{eq:Zero_D}.

If the system is square and $\mbf{D}^{-1}$ exists, then $\left(\mbf{D}^\trans \mbf{D}\right)^{-1} \mbf{D}^\trans = \mbf{D}^{-1}$ and~\eqref{eq:Zero_D} simplifies to~\eqref{eq:Zero_D1}.
\end{proof}

\item The transfer matrix $\mbf{G}(s)$ is also minimum phase\index{minimum phase} if and only if there exist $\mbf{P} \in \mathbb{S}^{n}$ and $\mbf{Q} \in \mathbb{S}^{n}$, where $\mbf{P} > 0$ and $\mbf{Q} = \mbf{P}^{-1}$, such that
\begin{align}
\mbf{M}^\trans\left(\mbf{P}\mbf{A} + \mbf{A}^\trans \mbf{P}\right)\mbf{M} &< 0, \label{eq:Zero_D2} \\
\mbf{N}\left( \mbf{A}\mbf{Q} + \mbf{Q} \mbf{A}^\trans\right) \mbf{N}^\trans &< 0, \label{eq:Zero_D3}
\end{align}
where $\mbf{N} \in \mathbb{R}^{q \times n}$, $\mbf{M} \in \mathbb{R}^{n \times q}$, $\mathcal{R}(\mbf{N}^\trans) = \mathcal{N}(\mbf{B}^\trans)$, and $\mathcal{R}(\mbf{M}) = \mathcal{N}(\mbf{C})$.
\begin{proof}
Applying the Strict Projection Lemma to~\eqref{eq:Zero_D} yields~\eqref{eq:Zero_D2} and~\eqref{eq:Zero_D3}.
\end{proof}

\end{enumerate}

%

\subsubsection[Discrete-Time System Zeros with Feedthrough]{Discrete-Time System Zeros with Feedthrough}

Consider a discrete-time LTI system, $\mbc{G}: \mathcal{\ell}_{2e} \to \mathcal{\ell}_{2e}$, with minimal state-space realization $(\mbf{A}_\mathrm{d},\mbf{B}_\mathrm{d},\mbf{C}_\mathrm{d},\mbf{D}_\mathrm{d})$, where $\mbf{A}_\mathrm{d} \in \mathbb{R}^{n \times n}$, $\mbf{B}_\mathrm{d} \in \mathbb{R}^{n \times m}$, $\mbf{C}_\mathrm{d} \in \mathbb{R}^{p \times n}$, $\mbf{D}_\mathrm{d} \in \mathbb{R}^{p \times m}$, $m \leq p$, and $\mbf{D}_\mathrm{d}$ is full rank\index{rank}.  The transmission zeros\index{transmission zeros} of $\mbf{G}(z) = \mbf{C}_\mathrm{d}\left(z\mbf{1} - \mbf{A}_\mathrm{d}\right)^{-1}\mbf{B}_\mathrm{d} + \mbf{D}_\mathrm{d}$ are the eigenvalues\index{eigenvalues} of $\mbf{A}_\mathrm{d} - \mbf{B}_\mathrm{d}\left(\mbf{D}_\mathrm{d}^\trans \mbf{D}_\mathrm{d}\right)^{-1}\mbf{D}_\mathrm{d}^\trans \mbf{C}_\mathrm{d}$.  Therefore, $\mbf{G}(z)$ is minimum phase\index{minimum phase} if and only if there exists $\mbf{P} \in \mathbb{S}^{n}$, where $\mbf{P} > 0$, such that
\beq
\label{eq:DTZero_D}
\bbm \mbf{P} & \left(\mbf{A}_\mathrm{d} - \mbf{B}_\mathrm{d}\left(\mbf{D}_\mathrm{d}^\trans \mbf{D}_\mathrm{d}\right)^{-1}\mbf{D}_\mathrm{d}^\trans \mbf{C}_\mathrm{d}\right)\mbf{P} \\ * & \mbf{P} \ebm > 0.
\eeq
If the system is square ($m = p$), then $\mbf{D}_\mathrm{d}$ full rank\index{rank} implies $\mbf{D}_\mathrm{d}^{-1}$ exists and~\eqref{eq:DTZero_D} simplifies to
\bdis
\bbm \mbf{P} & \left(\mbf{A}_\mathrm{d} - \mbf{B}_\mathrm{d}\mbf{D}_\mathrm{d}^{-1} \mbf{C}_\mathrm{d}\right)\mbf{P} \\ * & \mbf{P} \ebm > 0.
\edis
\begin{proof}
The proof follows the same procedure used in the proof of the continuous-time result in Section~\ref{sec:Zeros_Feedthrough}.
\end{proof}


\subsection[\texorpdfstring{$\mathcal{D}$}{D}-Stability]{\texorpdfstring{$\mathcal{D}$}{D}-Stability}
\index{stability!$\mathcal{D}$-stability}

\subsubsection[General LMI Region \texorpdfstring{$\mathcal{D}$}{D}-Stability]{General LMI Region \texorpdfstring{$\mathcal{D}$}{D}-Stability~\cite[pp.~107--108]{Duan2013},~\cite{ChilaliGahinet1996}}
Consider $\mbf{A} \in \mathbb{R}^{n \times n}$.  The eigenvalues\index{eigenvalues} of a $\mathcal{D}$-stable matrix lie within the LMI region\index{LMI!region} $\mathcal{D}$ of the complex plane, which is defined as $\mathcal{D} = \{z \in \mathbb{C} : f_\mathcal{D}(z) < 0\}$, where
\bdis
f_\mathcal{D}(z) := \mbs{\Lambda} + z\mbs{\Phi} + \overline{z}\mbs{\Phi}^\trans = [\lambda_{kl} + \phi_{kl}z+\phi_{lk}\overline{z}]_{1\leq k,l\leq m},
\edis
$\mbs{\Lambda}  \in \mathbb{S}^{m}$, $\mbs{\Phi} \in \mathbb{R}^{m \times m}$, and $\overline{z}$ is the complex conjugate\index{complex conjugate} of $z$.

The matrix $\mbf{A}$ is $\mathcal{D}$-stable if and only if any of the following equivalent conditions are satisfied.

\begin{enumerate}

\item There exists $\mbf{P} \in \mathbb{S}^n$, where $\mbf{P}  > 0$, such that
\bdis
[\lambda_{kl}\mbf{P} + \phi_{kl}\mbf{A}\mbf{P} + \phi_{lk}\mbf{P}\mbf{A}^\trans]_{1\leq k,l\leq m} < 0,
\edis
\item There exists $\mbf{P} \in \mathbb{S}^n$, where $\mbf{P}  > 0$, such that
\beq
\label{eq:DStability2}
\mbs{\Lambda} \otimes \mbf{P} + \mbs{\Phi} \otimes \left(\mbf{A}\mbf{P}\right) + \mbs{\Phi}^\trans \otimes \left(\mbf{P} \mbf{A}^\trans\right)< 0,
\eeq
where $\otimes$ is the Kroenecker product\index{Kroenecker product}.

\end{enumerate}

Alternatively, consider the LMI region\index{LMI!region} $\mathcal{D}$ of the complex plane defined by~\cite[p.~70]{SchererWeiland2015}
\bdis
\mathcal{D} = \{z \in \mathbb{C} : \bbm \mbf{1} \\ z \mbf{1} \ebm^\herm \bbm \mbf{Q} & \mbf{S} \\ \mbf{S}^\trans & \mbf{Q} \ebm \bbm \mbf{1} \\ z \mbf{1} \ebm < 0\},
\edis
where $\mbf{Q}$,~$\mbf{R} \in \mathbb{S}^m$ and $\mbf{S} \in \mathbb{R}^{m \times m}$.  The matrix $\mbf{A}$ is $\mathcal{D}$-stable if and only if there exists $\mbf{P}$ such that
\bdis
\bbm \mbf{1} \\ \mbf{A} \otimes \mbf{1} \ebm^\trans \bbm \mbf{P} \otimes \mbf{Q} & \mbf{P} \otimes \mbf{S} \\ \mbf{P} \otimes \mbf{S}^\trans & \mbf{P} \otimes \mbf{R} \ebm \bbm \mbf{1} \\ \mbf{A} \otimes \mbf{1} \ebm < 0.
\edis

\subsubsection[\texorpdfstring{$\alpha$}{Alpha}-Stability Region]{\texorpdfstring{$\alpha$}{Alpha}-Stability Region}
\index{stability!$\alpha$-stability}
\index{stability!exponential stability}
Consider $\mbf{A} \in \mathbb{R}^{n \times n}$ and $\alpha \in \mathbb{R}_{>0}$. The matrix $\mbf{A}$ satisfies $\lambda(\mbf{A}) \subset \mathcal{D}(\alpha)$, where $\mathcal{D}(\alpha) := \{ z \in \mathbb{C} : \text{Re}(z) < - \alpha\}$ if and only if any of the following equivalent conditions are satisfied.

\begin{enumerate}
\itemcite \cite[pp.~66-67]{Boyd1994},~\cite[p.~99]{Duan2013},~\cite{Ohara1991,Yedavalli1993,ChilaliGahinet1996} There exists $\mbf{P} \in \mathbb{S}^n$, where $\mbf{P} > 0$, such that
\beq
\label{eq:Dilation3}
\mbf{A}\mbf{P} + \mbf{P}\mbf{A}^\trans + 2\alpha \mbf{P} < 0.
\eeq
\item There exists $\mbf{P} \in \mathbb{S}^n$, where $\mbf{P} > 0$, such that
\beq
\label{eq:Dilation3a}
\bbm \mbf{A}\mbf{P} + \mbf{P}\mbf{A}^\trans & \alpha \mbf{P} \\ * & -\onehalf  \mbf{P}\ebm < 0.
\eeq
\begin{proof}
Equation~\eqref{eq:Dilation3} is rewritten as
\bdis
\mbf{A}\mbf{P} + \mbf{P}\mbf{A}^\trans - \left(\alpha \mbf{P}\right) \left(-\onehalf \alpha \mbf{P}\right)^{-1}\left(\alpha \mbf{P} \right) < 0,
\edis
which is equivalent to ~\eqref{eq:Dilation3a} using the Schur complement.
\end{proof}
\itemcite \cite{Ebihara2004} There exist $\mbf{X} \in \mathbb{S}^n$, $\epsilon \in \mathbb{R}_{>0}$, and $\mbf{F} \in \mathbb{R}^{n \times n}$, where $\mbf{X} > 0$, such that
\beq
\label{eq:Dilation4}
\bbm \mbf{0} & -\mbf{X} & \mbf{X} \\ * & \mbf{0} & \mbf{0} \\ * & * & -\onehalf\alpha^{-1} \mbf{X} \ebm + \text{He}\left\{ \bbm \mbf{A} \\ \mbf{1} \\ \mbf{0} \ebm \mbf{F} \bbm \mbf{1} & -\epsilon\mbf{1} & \epsilon\mbf{1} \ebm \right\} < 0.
\eeq
Moreover, for every $\mbf{X}$ that satisfies~\eqref{eq:Dilation3}, $\mbf{X}$ and $\mbf{F} = -\epsilon^{-1}\left(\mbf{A}-\epsilon^{-1}\mbf{1}\right)^{-1}\mbf{X}$ are solutions to~\eqref{eq:Dilation4}.

\itemcite \cite{Felipe2020} There exist $\mbf{P} \in \mathbb{S}^{n}$, $\mbf{Y}_1$,~$\mbf{Y}_2$,~$\mbf{Y}_3$,~$\mbf{X}_1$,~$\mbf{X}_2$,~$\mbf{X}_3 \in \mathbb{R}^{n \times n}$, and $\gamma \in \mathbb{R}_{>0}$, where $\mbf{P} > 0$, such that
\bdis
\bbm \mbf{X}_1\mbf{Y}_1 + \mbf{Y}_1^\trans\mbf{X}_1^\trans & \mbf{P} + \mbf{X}_1 \mbf{Y}_2 + \mbf{Y}_1^\trans \mbf{X}_2^\trans & \mbf{A}^\trans - \alpha \mbf{1} +\mbf{X}_1\mbf{Y}_3 + \mbf{Y}_1^\trans\mbf{X}_3^\trans \\ * & \mbf{X}_2 \mbf{Y}_2 + \mbf{Y}_2^\trans \mbf{X}_2^\trans & - \gamma \mbf{1} + \mbf{X}_2\mbf{Y}_3+ \mbf{Y}_2^\trans \mbf{X}_3^\trans \\ * & * & \mbf{X}_3 \mbf{Y}_3+ \mbf{Y}_3^\trans\mbf{X}_3^\trans \ebm  < 0.
\edis

\end{enumerate}

If $\lambda(\mbf{A}) \subset \mathcal{D}(\alpha)$, then the solution to $\dot{\mbf{x}} = \mbf{A} \mbf{x}$, $\mbf{x}(0) = \mbf{x}_0$ satisfies $\norm{\mbf{x}(t)}_2 \leq \sqrt{\kappa(\mbf{P})} \norm{\mbf{x}_0}_2 e^{-\alpha t}$, where $\kappa(\mbf{P})$ is the condition number \index{condition number}of $\mbf{P}$.  This system is exponentially stable \index{stability!exponential stability}with exponential decay rate $\alpha$.

\subsubsection[Vertical Band]{Vertical Band~\cite[p.~99]{Duan2013},~\cite{Ohara1991,Yedavalli1993,ChilaliGahinet1996}}

Consider $\mbf{A} \in \mathbb{R}^{n \times n}$ and $\alpha$,~$\beta \in \mathbb{R}_{>0}$. The matrix $\mbf{A}$ satisfies $\lambda(\mbf{A}) \subset \mathcal{D}(\alpha,\beta)$, where $\mathcal{D}(\alpha,\beta) := \{ z \in \mathbb{C} : -\beta < \text{Re}(z) < - \alpha\}$ if and only if there exists $\mbf{P} \in \mathbb{S}^n$, where $\mbf{P} > 0$, such that
\begin{align*}
\mbf{A}\mbf{P} + \mbf{P}\mbf{A}^\trans + 2\alpha \mbf{P} &< 0, \\
\mbf{A}\mbf{P} + \mbf{P}\mbf{A}^\trans + 2\beta \mbf{P} &> 0.
\end{align*}

If $\lambda(\mbf{A}) \subset \mathcal{D}(\alpha,\beta)$, then the solution to $\dot{\mbf{x}} = \mbf{A} \mbf{x}$, $\mbf{x}(0) = \mbf{x}_0$ satisfies $\norm{\mbf{x}(t)}_2 \leq \sqrt{\kappa(\mbf{P})} \norm{\mbf{x}_0}_2 e^{-\alpha t}$, where $\kappa(\mbf{P})$ is the condition number \index{condition number}of $\mbf{P}$.  This system is exponentially stable \index{stability!exponential stability}with exponential decay rate $\alpha$.

\subsubsection[Conic Sector Region]{Conic Sector Region}
Consider $\mbf{A} \in \mathbb{R}^{n \times n}$ and $\theta \in \mathbb{R}_{>0}$.  The matrix $\mbf{A}$ satisfies $\lambda(\mbf{A}) \subset\mathcal{D}(k)$, where $\mathcal{D}(k) := \{ z \in \mathbb{C} : \abs{\text{Im}(z)} < -\tan(\theta)\text{Re}(z), \, 0 < \theta < \pi/2 \}$, if and only if any of the following equivalent conditions are satisfied.

\begin{enumerate}
\itemcite \cite[pp.~105--106]{Duan2013},~\cite{ChilaliGahinet1996}  There exists $\mbf{P} \in \mathbb{S}^n$, where $\mbf{P}  > 0$, such that
\bdis
\bbm \sin(\theta) \left(\mbf{A}\mbf{P} + \mbf{P}\mbf{A}^\trans\right) & \cos(\theta) \left(\mbf{A}\mbf{P} - \mbf{P}\mbf{A}^\trans\right) \\ * &  \sin(\theta) \left(\mbf{A}\mbf{P} + \mbf{P}\mbf{A}^\trans\right)  \ebm < 0.
\edis

\itemcite \cite{Ebihara2004}  There exists $\mbf{P} \in \mathbb{S}^n$, where $\mbf{P}  > 0$, such that
\beq
\label{eq:Dilation7}
\bbm k\left(\mbf{A}\mbf{P} + \mbf{P}\mbf{A}^\trans\right) & \mbf{A}\mbf{P} - \mbf{P}\mbf{A}^\trans \\ * &  k\left(\mbf{A}\mbf{P} + \mbf{P}\mbf{A}^\trans\right)  \ebm < 0,
\eeq
where $k = \tan(\theta)$.
\itemcite \cite{Ebihara2004} There exist $\mbf{X} \in \mathbb{S}^n$, $\epsilon \in \mathbb{R}_{>0}$, and $\mbf{F} \in \mathbb{R}^{n \times n}$, where $\mbf{X} > 0$, such that
\beq
\label{eq:Dilation8}
\bbm \mbf{0} & -k\mbf{X} & \mbf{X} & \mbf{0} \\ * & \mbf{0} & \mbf{0} & -\mbf{X} \\ * & * & \mbf{0} & -k\mbf{X} \\ * & * & * & \mbf{0} \ebm + \text{He}\left\{ \bbm \mbf{A} & \mbf{0} \\ \mbf{1} & \mbf{0} \\ \mbf{0} & \mbf{1} \\ \mbf{0} & \mbf{A} \ebm \bbm \mbf{F} & \mbf{0} \\ \mbf{0} & \mbf{F} \ebm \bbm k\mbf{1} & -\epsilon k\mbf{1} & \epsilon\mbf{1} & \mbf{1} \\ -\mbf{1} & -\epsilon \mbf{1} & \epsilon k\mbf{1} & k\mbf{1}\ebm \right\} < 0,
\eeq
where $k = \tan(\theta)$.  Moreover, for every $\mbf{X}$ that satisfies~\eqref{eq:Dilation7}, $\mbf{X}$ and $\mbf{F} = -\epsilon^{-1}\left(\mbf{A}-\epsilon^{-1}\mbf{1}\right)^{-1}\mbf{X}$ are solutions to~\eqref{eq:Dilation8}.
\end{enumerate}

\subsubsection[Circular Region]{Circular Region}

Consider $\mbf{A} \in \mathbb{R}^{n \times n}$, $r \in \mathbb{R}_{>0}$, and $c \in \mathbb{R}_{<0}$, where $c < -r$.  The matrix $\mbf{A}$ satisfies $\lambda(\mbf{A}) \subset\mathcal{D}(c,r)$, where $\mathcal{D}(c,r) := \{ z \in \mathbb{C} : \left(\text{Re}(z) - c\right)^2 + \left(\text{Im}(z)\right)^2 < r^2\}$, if and only if any of the following equivalent conditions are satisfied.

\begin{enumerate}
\itemcite \cite[p.~101]{Duan2013},~\cite{Yedavalli1993,ChilaliGahinet1996}  There exists $\mbf{P} \in \mathbb{S}^n$, where $\mbf{P} > 0$, such that
\bdis
\bbm -r \mbf{P} & -c \mbf{P} + \mbf{A} \mbf{P} \\ * & -r \mbf{P} \ebm < 0.
\edis
\itemcite \cite{Ebihara2004}   There exists $\mbf{P} \in \mathbb{S}^n$, where $\mbf{P} > 0$, such that
\beq
\label{eq:Dilation5}
\mbf{A}\mbf{P} + \mbf{P}\mbf{A}^\trans -\frac{c^2-r^2}{c}\mbf{P} - \frac{1}{c}\mbf{A}\mbf{P}\mbf{A}^\trans < 0.
\eeq
\itemcite \cite{Ebihara2004} There exist $\mbf{X} \in \mathbb{S}^n$, $\epsilon \in \mathbb{R}_{>0}$, and $\mbf{F} \in \mathbb{R}^{n \times n}$, where $\mbf{X} > 0$, such that
\beq
\label{eq:Dilation6}
\bbm \mbf{0} & -\mbf{X} & \mbf{X} & \mbf{0} \\ * & \mbf{0} & \mbf{0} & -\mbf{X} \\ * & * & \frac{c}{c^2-r^2} \mbf{X} & \mbf{0} \\ * & * & * & c\mbf{X} \ebm + \text{He}\left\{ \bbm \mbf{A} \\ \mbf{1} \\ \mbf{0} \\ \mbf{0} \ebm \mbf{F} \bbm \mbf{1} & -\epsilon\mbf{1} & \epsilon\mbf{1} & \mbf{1} \ebm \right\} < 0.
\eeq
Moreover, for every $\mbf{X}$ that satisfies~\eqref{eq:Dilation5}, $\mbf{X}$ and $\mbf{F} = -\epsilon^{-1}\left(\mbf{A}-\epsilon^{-1}\mbf{1}\right)^{-1}\mbf{X}$ are solutions to~\eqref{eq:Dilation6}.
\end{enumerate}

\subsubsection[Horizontal Band]{Horizontal Band~\cite{Ohara1991},~\cite[p.~164]{Chadli2013},~\cite[p.~48]{Xue2019}}

Consider $\mbf{A} \in \mathbb{R}^{n \times n}$ and $\gamma \in \mathbb{R}_{>0}$. The matrix $\mbf{A}$ satisfies $\lambda(\mbf{A}) \subset \mathcal{D}(\gamma)$, where $\mathcal{D}(\gamma) := \{ z \in \mathbb{C} :  \abs{\text{Im}(z)} <  \gamma \}$, if and only if there exists $\mbf{P} \in \mathbb{S}^n$, where $\mbf{P} > 0$, such that
\bdis
\bbm -2 \gamma \mbf{P} & \mbf{A} \mbf{P} - \mbf{P}\mbf{A}^\trans \\ * & -2 \gamma \mbf{P} \ebm < 0.
\edis

\subsubsection[Elliptic Region]{Elliptic Region~\cite[p.~31]{Iwasaki2007}}
\label{sec:DStability_Elliptic}

Consider $\mbf{A} \in \mathbb{R}^{n \times n}$, $a$,~$b \in \mathbb{R}_{>0}$, and $c \in \mathbb{R}$. The matrix $\mbf{A}$ satisfies $\lambda(\mbf{A}) \subset \mathcal{D}(\gamma)$, where $\mathcal{D}(a,b,c) := \{ z \in \mathbb{C} : \left(\frac{\text{Re}(z) - c}{a}\right)^2 + \left(\frac{\text{Im}(z)}{b}\right)^2 < 1 \}$, if and only if there exists $\mbf{P} \in \mathbb{S}^n$, where $\mbf{P} > 0$, such that
\bdis
\bbm 2ab  \mbf{P} & (a+b) \mbf{P}\mbf{A} + (b-a) \mbf{A}^\trans\mbf{P} \\ * & 2 ab \mbf{P} \ebm > 0.
\edis
The parameter $c$ is the center of the ellipse region on the real axis, $a$ is the semi-major axis, and $b$ is the semi-minor axis.

\subsubsection[Hyperbolic Region]{Hyperbolic Region~\cite[p.~32]{Iwasaki2007}}
\label{sec:DStability_Hyperbolic}

Consider $\mbf{A} \in \mathbb{R}^{n \times n}$, $a$,~$b \in \mathbb{R}_{>0}$, and $c \in \mathbb{R}$. The matrix $\mbf{A}$ satisfies $\lambda(\mbf{A}) \subset \mathcal{D}(\gamma)$, where $\mathcal{D}(a,b,c) := \{ z \in \mathbb{C} : \left(\frac{\text{Re}(z) - c}{a}\right)^2 - \left(\frac{\text{Im}(z)}{b}\right)^2 > 1 \}$, if and only if there exists $\mbf{P} \in \mathbb{S}^n$, where $\mbf{P} > 0$, such that
\bdis
\bbm 2bc  \mbf{P} - b\left(\mbf{P} \mbf{A} + \mbf{A}^\trans \mbf{P}\right)& 2ab\mbf{P} + a \left(\mbf{P} \mbf{A} - \mbf{A}^\trans \mbf{P}\right) \\ * & 2 bc \mbf{P} - b\left(\mbf{P} \mbf{A} + \mbf{A}^\trans \mbf{P}\right) \ebm > 0.
\edis
The parameter $c$ is the center of the hyperbolic region on the real axis, $a$ is the semi-major axis, and $b$ is the semi-minor axis.

\subsection[\texorpdfstring{$\mathcal{D}$}{D}-Admissibility]{\texorpdfstring{$\mathcal{D}$}{D}-Admissibility}
\index{descriptor systems!$\mathcal{D}$-admissibility}

\subsubsection[General LMI Region \texorpdfstring{$\mathcal{D}$}{D}-Admissibility]{General LMI Region \texorpdfstring{$\mathcal{D}$}{D}-Admissibility}

Consider $\mbf{A}$,~$\mbf{E} \in \mathbb{R}^{n \times n}$.  The pair $(\mbf{E},\mbf{A})$ is $\mathcal{D}$-admissible if it is regular and causal, and the eigenvalues\index{eigenvalues} of $(\mbf{E},\mbf{A})$ lie within the LMI region\index{LMI!region} $\mathcal{D}$ of the complex plane, which is defined as $\mathcal{D} = \{z \in \mathbb{C} : f_\mathcal{D}(z) < 0\}$, where
\bdis
f_\mathcal{D}(z) := \mbs{\Lambda} + z\mbs{\Phi} + \overline{z}\mbs{\Phi}^\trans = [\lambda_{kl} + \phi_{kl}z+\phi_{lk}\overline{z}]_{1\leq k,l\leq m},
\edis
$\mbs{\Lambda}  \in \mathbb{S}^{m}$, $\mbs{\Phi} \in \mathbb{R}^{m \times m}$, and $\overline{z}$ is the complex conjugate\index{complex conjugate} of $z$.

The pair $(\mbf{E},\mbf{A})$ is $\mathcal{D}$-admissible if and only if any of the following equivalent conditions are satisfied.

\begin{enumerate}

\itemcite \cite{Marx2003} There exist $\mbf{P} \in \mathbb{S}^n$, $\mbf{S} \in \mathbb{R}^{(n-n_e)\times (n-n_e}$, and $\mbf{U}$,~$\mbf{V} \in \mathbb{R}^{n \times (n-n_e)}$, where $n_e = \text{rank}(\mbf{E})$, $\mathcal{R}(\mbf{U}) = \mathcal{N}(\mbf{E}^\trans)$, $\mathcal{R}(\mbf{V}) = \mathcal{N}(\mbf{E})$, and $\mbf{P} > 0$, satisfying \index{rank}
\bdis
[\lambda_{kl} \mbf{E} \mbf{P} \mbf{E}^\trans  + \phi_{kl}\mbf{A}\mbf{P}\mbf{E} + \phi_{lk}\mbf{E}^\trans\mbf{P}\mbf{A}^\trans + \mbf{A} \mbf{V}\mbf{S} \mbf{U}^\trans + \mbf{U} \mbf{S}^\trans \mbf{V}^\trans \mbf{A}^\trans]_{1\leq k,l\leq m} < 0,
\edis

\itemcite \cite{Kuo2004} There exist $\mbf{P}$,~$\mbf{Q} \in \mathbb{S}^n$, where $\mbf{P} > 0$, satisfying $\mbf{E}^\trans \mbf{Q} \mbf{E} \geq 0$ and
\bdis
[\lambda_{kl} \mbf{E} \mbf{P} \mbf{E}^\trans  + \phi_{kl}\mbf{A}\mbf{P}\mbf{E} + \phi_{lk}\mbf{E}^\trans\mbf{P}\mbf{A}^\trans + \mbf{A}^\trans \mbf{Q} \mbf{A}]_{1\leq k,l\leq m} < 0,
\edis

\itemcite \cite{Kuo2004} There exist $\mbf{P}  \in \mathbb{S}^n$, $\mbf{S} \in \mathbb{R}^{(n-n_e)\times (n-n_e}$, $\mbf{U} \in \mathbb{R}^{n \times (n-n_e)}$, where $n_e = \text{rank}(\mbf{E})$, $\mbf{U} \mbf{E} = \mbf{0}$, and $\mbf{P} > 0$, satisfying \index{rank}
\bdis
[\lambda_{kl} \mbf{E} \mbf{P} \mbf{E}^\trans  + \phi_{kl}\mbf{A}\mbf{P}\mbf{E} + \phi_{lk}\mbf{E}^\trans\mbf{P}\mbf{A}^\trans + \mbf{A}^\trans \mbf{U}^\trans \mbf{S} \mbf{U} \mbf{A}]_{1\leq k,l\leq m} < 0,
\edis

\itemcite \cite{Marx2003} There exist $\mbf{P} \in \mathbb{S}^n$, $\mbf{S} \in \mathbb{R}^{(n-n_e)\times (n-n_e}$, and $\mbf{U}$,~$\mbf{V} \in \mathbb{R}^{n \times (n-n_e)}$, where $n_e = \text{rank}(\mbf{E})$, $\mathcal{R}(\mbf{U}) = \mathcal{N}(\mbf{E}^\trans)$, $\mathcal{R}(\mbf{V}) = \mathcal{N}(\mbf{E})$, and $\mbf{P} > 0$, satisfying \index{rank}
\bdis
\mbs{\Lambda} \otimes \mbf{E} \mbf{P} \mbf{E}^\trans + \mbs{\Phi} \otimes \left(\mbf{A}\mbf{P} \mbf{E} \right) + \mbs{\Phi}^\trans \otimes \left(\mbf{E} \mbf{P} \mbf{A}^\trans\right) + \mbf{1}_{mm} \otimes \left( \mbf{A} \mbf{V}\mbf{S} \mbf{U}^\trans + \mbf{U} \mbf{S}^\trans \mbf{V}^\trans \mbf{A}^\trans \right) < 0,
\edis
where $\otimes$ is the Kroenecker product\index{Kroenecker product} and $\mbf{1}_{mm}$ is an $m \times m$ matrix filled with ones.

\itemcite \cite{Kuo2004} There exist $\mbf{P}$,~$\mbf{Q} \in \mathbb{S}^n$, where $\mbf{P} > 0$, satisfying $\mbf{E}^\trans \mbf{Q} \mbf{E} \geq 0$ and
\bdis
\mbs{\Lambda} \otimes \mbf{E} \mbf{P} \mbf{E}^\trans + \mbs{\Phi} \otimes \left(\mbf{A}\mbf{P} \mbf{E} \right) + \mbs{\Phi}^\trans \otimes \left(\mbf{E} \mbf{P} \mbf{A}^\trans\right) + \mbf{1}_{mm} \otimes \left(\mbf{A}^\trans \mbf{Q} \mbf{A}  \right) < 0,
\edis
where $\otimes$ is the Kroenecker product\index{Kroenecker product} and $\mbf{1}_{mm}$ is an $m \times m$ matrix filled with ones.

\itemcite \cite{Kuo2004} There exist $\mbf{P}  \in \mathbb{S}^n$, $\mbf{S} \in \mathbb{R}^{(n-n_e)\times (n-n_e}$, $\mbf{U} \in \mathbb{R}^{n \times (n-n_e)}$, where $n_e = \text{rank}(\mbf{E})$, $\mbf{U} \mbf{E} = \mbf{0}$, and $\mbf{P} > 0$, satisfying \index{rank}
\bdis
\mbs{\Lambda} \otimes \mbf{E} \mbf{P} \mbf{E}^\trans + \mbs{\Phi} \otimes \left(\mbf{A}\mbf{P} \mbf{E} \right) + \mbs{\Phi}^\trans \otimes \left(\mbf{E} \mbf{P} \mbf{A}^\trans\right) + \mbf{1}_{mm} \otimes \left(  \mbf{A}^\trans \mbf{U}^\trans \mbf{S} \mbf{U} \mbf{A} \right) < 0,
\edis
where $\otimes$ is the Kroenecker product\index{Kroenecker product} and $\mbf{1}_{mm}$ is an $m \times m$ matrix filled with ones.


\end{enumerate}

\subsubsection[Circular Region]{Circular Region~\cite{Masubuchi2013}}

Consider $\mbf{A}$,~$\mbf{E} \in \mathbb{R}^{n \times n}$, $a$,~$b \in \mathbb{R}$, and $d \in \mathbb{R}_{>0}$, where $b \neq 0$. The pair $(\mbf{E},\mbf{A})$ is $\mathcal{D}$-admissible with $\mathcal{D} = \{ z \in \mathbb{C} : a  + 2 b \text{Re}(z) + d \abs{z}^2 < 0\}$ if and only if there exist $\mbf{X} \in \mathbb{R}^{n \times n}$ and $\alpha \in \mathbb{R}$ such that $\mbf{E}^\trans \mbf{X} = \mbf{X}^\trans \mbf{E} \geq 0$ and
\bdis
\bbm -a \mbf{E}^\trans \mbf{X} - b\left( \mbf{X}^\trans \mbf{A} + \mbf{A}^\trans \mbf{X}\right) & \mbf{A}^\trans \mbf{X} \\ * & d^{-1} \mbf{E}^\trans \mbf{X} + \alpha \left( \mbf{1} - \mbf{E}^\dagger \mbf{E} \right) \ebm > 0,
\edis
where $\mbf{E}^\dagger$ is the pseudoinverse\index{pseudoinverse} of $\mbf{E}$.  The region $\mathcal{D}$ describes a circular region of the complex plane with radius $r = \sqrt{-a/d+ b^2/d^2}$ centered at $(c,0)$, where $c = -b/d$.

\subsection[DC Gain of a Transfer Matrix]{DC Gain of a Transfer Matrix}
\index{DC gain}

Consider $\gamma \in \mathbb{R}_{>0}$ and a continuous-time LTI system, $\mbc{G}: \mathcal{L}_{2e} \to \mathcal{L}_{2e}$, with transfer matrix $\mbf{G}(s) = \mbf{C} \left(s\mbf{1} - \mbf{A}\right)^{-1}\mbf{B} + \mbf{D}$, where $\mbf{A} \in \mathbb{R}^{n \times n}$, $\mbf{B} \in \mathbb{R}^{n \times m}$, $\mbf{C} \in \mathbb{R}^{p \times n}$, and $\mbf{D} \in \mathbb{R}^{p \times m}$.  The DC gain of $\mbc{G}$ is strictly less than $\gamma$ (i.e., $\bar{\sigma}(\mbf{G}(0)) < \gamma$) if and only if
\beq
\label{eq:DC_Gain1}
\bbm \gamma \mbf{1} & -\mbf{C} \mbf{A}^{-1} \mbf{B} + \mbf{D} \\ * & \gamma \mbf{1} \ebm > 0,
\eeq
or
\beq
\label{eq:DC_Gain2}
\bbm \gamma \mbf{1} & -\mbf{B}^\trans \mbf{A}^{-\trans} \mbf{C}^\trans + \mbf{D}^\trans \\ * & \gamma \mbf{1} \ebm > 0.
\eeq
\begin{proof}
$\bar{\sigma}(\mbf{G}(0)) < \gamma$ if and only if $\bar{\lambda}\left(\mbf{G}(0)\mbf{G}^\trans(0)\right) < \gamma^2$, or equivalently
\begin{align}
\mbf{G}(0)\mbf{G}^\trans(0) - \gamma^2 \mbf{1} &< 0 \nonumber \\
\mbf{G}(0)(-\gamma^{-1}\mbf{1})\mbf{G}^\trans(0) - \gamma \mbf{1} &< 0 \nonumber \\
\gamma \mbf{1} - \mbf{G}(0)(\gamma^{-1}\mbf{1})\mbf{G}^\trans(0) &> 0 \nonumber \\
\bbm \gamma \mbf{1} & \mbf{G}(0) \\ * & \gamma \mbf{1} \ebm &> 0. \label{eq:DC_Gain3}
\end{align}
Substituting $\mbf{G}(0) = - \mbf{C} \mbf{A}^{-1}\mbf{B} + \mbf{D}$ into~\eqref{eq:DC_Gain3} gives~\eqref{eq:DC_Gain1}.  Starting with $\bar{\sigma}(\mbf{G}(0)) < \gamma \iff \bar{\lambda}\left(\mbf{G}^\trans(0)\mbf{G}(0)\right) < \gamma^2$ in the first step of the proof and following the same steps yields~\eqref{eq:DC_Gain2}.
\end{proof}

\subsection[Transient Bounds]{Transient Bounds}

\subsubsection[Transient State Bound for Autonomous LTI Systems]{Transient State Bound for Autonomous LTI Systems~\cite[p.~88]{Boyd1994},~\cite{Whidborne2007,Polyak2015}}
\index{transient!state bound}
\label{sec:TransStateBound}

Consider the continuous-time LTI system with state-space realization
\bdis
\dot{\mbf{x}} = \mbf{A} \mbf{x},
\edis
where $\mbf{A}  \in \mathbb{R}^{n \times n}$ and $\mbf{x}(0) = \mbf{x}_0$.  The Euclidean norm \index{norm!Euclidean norm}of the state satisfies
\bdis
\norm{\mbf{x}(T)}_2 \leq \gamma \norm{\mbf{x}_0}_2, \,\,\forall T \in \mathbb{R}_{\geq 0}
\edis
if there exist $\mbf{P} \in \mathbb{S}^n$ and $\gamma \in \mathbb{R}_{>0}$, where $\mbf{P} > 0$, such that
\begin{align}
\mbf{P} - \gamma \mbf{1}&\leq 0, \label{eq:State_Bound_1} \\
\bbm \mbf{P} & \mbf{1} \\ * & \gamma \mbf{1} \ebm &\geq 0, \label{eq:State_Bound_2} \\
\mbf{P} \mbf{A} + \mbf{A}^\trans \mbf{P} &\leq 0.  \label{eq:State_Bound_3}
\end{align}

\begin{proof}
Define $V = \mbf{x}^\trans \mbf{P} \mbf{x}$.  Evaluating $\dot{V}$ and substituting in the matrix inequality from~\eqref{eq:State_Bound_3} results in $\dot{V} \leq 0$.  Integrating both sides of this inequality from $t = 0$ to $t = T$, where $T \in \mathbb{R}_{\geq 0}$ gives
\begin{align}
V(T) &\leq V(0) \nonumber \\
\mbf{x}^\trans(T) \mbf{P} \mbf{x}(T) &\leq \mbf{x}_0^\trans \mbf{P} \mbf{x}_0. \label{eq:State_Bound_4}
\end{align}
Using the non-strict Schur complement,~\eqref{eq:State_Bound_2} can be rewritten as $\gamma^{-1} \mbf{1} \leq \mbf{P}$.  Substituting this and~\eqref{eq:State_Bound_1} into~\eqref{eq:State_Bound_4} yields
\begin{align*}
\gamma^{-1}\mbf{x}^\trans(T) \mbf{x} (T) &\leq \gamma \mbf{x}_0^\trans \mbf{x}_0 \\
\norm{\mbf{x}(T)}_2 &\leq \gamma \norm{\mbf{x}_0}_2.
\end{align*}
\end{proof}

\subsubsection[Transient State Bound for Discrete-Time Autonomous LTI Systems]{Transient State Bound for Discrete-Time Autonomous LTI Systems}
\index{transient!discrete-time state bound}
\label{sec:DTTransStateBound}

Consider the discrete-time LTI system with state-space realization
\bdis
\mbf{x}_{k+1}= \mbf{A}_\mathrm{d} \mbf{x}_k,
\edis
where $\mbf{A}_\mathrm{d}  \in \mathbb{R}^{n \times n}$.  The Euclidean norm \index{norm!Euclidean norm}of the state satisfies \bdis
\norm{\mbf{x}_k}_2 \leq \gamma \norm{\mbf{x}_0}_2, \,\, \forall k \in \mathbb{Z}_{\geq 0}
\edis
if there exist $\mbf{P} \in \mathbb{S}^n$ and $\gamma \in \mathbb{R}_{>0}$, where $\mbf{P} > 0$, such that 
\begin{align}
\mbf{P} - \gamma \mbf{1}&\leq 0, \label{eq:DT_State_Bound_1} \\
\bbm \mbf{P} & \mbf{1} \\ * & \gamma \mbf{1} \ebm &\geq 0, \label{eq:DT_State_Bound_2} \\
\mbf{A}_\mathrm{d}^\trans \mbf{P} \mbf{A}_\mathrm{d} - \mbf{P} &\leq 0. \label{eq:DT_State_Bound_3}
\end{align}

\begin{proof}
Define $V(k) = \mbf{x}_k^\trans \mbf{P} \mbf{x}_k$.  Evaluating $V(k+1) - V(k)$ and substituting in the matrix inequality from~\eqref{eq:DT_State_Bound_3} results in
\begin{align*}
V(k+1) &\leq V(k) \\
\mbf{x}_{k+1}^\trans \mbf{P} \mbf{x}_{k+1} &\leq \mbf{x}_k^\trans \mbf{P} \mbf{x}_k.
\end{align*}
Using induction, this inequality implies
\beq
\label{eq:DT_State_Bound_4}
\mbf{x}_k^\trans \mbf{P} \mbf{x}_k \leq \mbf{x}_0^\trans \mbf{P} \mbf{x}_0.
\eeq
Using the non-strict Schur complement,~\eqref{eq:DT_State_Bound_2} can be rewritten as $\gamma^{-1} \mbf{1} \leq \mbf{P}$.  Substituting this and~\eqref{eq:DT_State_Bound_1} into~\eqref{eq:DT_State_Bound_4} yields
\begin{align*}
\gamma^{-1} \mbf{x}_k^\trans \mbf{x}_k &\leq \gamma \mbf{x}_0^\trans \mbf{x}_0 \\
\norm{\mbf{x}_k}_2 &\leq \gamma \norm{\mbf{x}_0}_2.
\end{align*}
\end{proof}

\subsubsection[Transient State Bound for Non-Autonomous LTI Systems]{Transient State Bound for Non-Autonomous LTI Systems~\cite[p.~77--78]{Boyd1994}}
\index{transient!state bound}
\label{sec:TransStateBoundNonAuton}

Consider the continuous-time LTI system with state-space realization
\bdis
\dot{\mbf{x}} = \mbf{A} \mbf{x} + \mbf{B} \mbf{u},
\edis
where $\mbf{A}  \in \mathbb{R}^{n \times n}$, $\mbf{B}  \in \mathbb{R}^{n \times m}$ and $\mbf{x}(0) = \mbf{x}_0$.  The Euclidean norm \index{norm!Euclidean norm}of the state satisfies
\bdis
\norm{\mbf{x}(T)}_2^2 \leq \gamma^2 \left(\norm{\mbf{x}_0}_2^2 + \norm{\mbf{u}}_{2T}^2\right), \,\,\forall T \in \mathbb{R}_{\geq 0}
\edis
if there exist $\mbf{P} \in \mathbb{S}^n$ and $\gamma \in \mathbb{R}_{>0}$, where $\mbf{P} > 0$, such that
\begin{align}
\mbf{P} - \gamma \mbf{1}&\leq 0, \label{eq:State_Bound_NonAuton_1} \\
\bbm \mbf{P} & \mbf{1} \\ * & \gamma \mbf{1} \ebm &\geq 0, \label{eq:State_Bound_NonAuton_2} \\
\bbm \mbf{P} \mbf{A} + \mbf{A}^\trans \mbf{P} & \mbf{P} \mbf{B} \\ * & -\gamma \mbf{1} \ebm &\leq 0.  \label{eq:State_Bound_NonAuton_3}
\end{align}

If $\mbf{x}_0 = \mbf{0}$ and $\mbf{u}$ is a unit-energy input (i.e., $\norm{\mbf{u}}_{2T} \leq 1$, $\forall T \in \mathbb{R}_{\geq 0}$), then the preceding conditions ensure that $\norm{\mbf{x}(T)}_2 \leq \gamma$, $\forall T \in \mathbb{R}_{\geq 0}$.

\begin{proof}
Define $V = \mbf{x}^\trans \mbf{P} \mbf{x}$.  Evaluating $\dot{V}$ results in
\begin{align}
\dot{V} &= \bbm \mbf{x}^\trans & \mbf{u}^\trans \ebm \bbm \mbf{P} \mbf{A} + \mbf{A}^\trans \mbf{P} & \mbf{P} \mbf{B} \\ * & \mbf{0} \ebm \bbm \mbf{x} \\ \mbf{u} \ebm \nonumber \\
&=  \bbm \mbf{x}^\trans & \mbf{u}^\trans \ebm \bbm \mbf{P} \mbf{A} + \mbf{A}^\trans \mbf{P} & \mbf{P} \mbf{B} \\ * & -\gamma \mbf{1} \ebm \bbm \mbf{x} \\ \mbf{u} \ebm + \gamma \mbf{u}^\trans \mbf{u}. \label{eq:State_Bound_NonAuton_4}
\end{align}
Substituting~\eqref{eq:State_Bound_NonAuton_3} into~\eqref{eq:State_Bound_NonAuton_4} gives $\dot{V} \leq \gamma \mbf{u}^\trans \mbf{u}$.  Integrating both sides of this inequality from $t = 0$ to $t = T$, where $T \in \mathbb{R}_{\geq 0}$ yields
\beq
\mbf{x}^\trans(T) \mbf{P} \mbf{x}(T) \leq \mbf{x}_0^\trans \mbf{P} \mbf{x}_0 + \gamma \norm{\mbf{u}}_{2T}^2. \label{eq:State_Bound_NonAuton_5}
\eeq
Substituting ~\eqref{eq:State_Bound_NonAuton_1} and~\eqref{eq:State_Bound_NonAuton_2} into~\eqref{eq:State_Bound_NonAuton_5} results in
\begin{align*}
\gamma^{-1}\mbf{x}^\trans(T) \mbf{x} (T) &\leq \gamma \mbf{x}_0^\trans \mbf{x}_0 + \gamma \norm{\mbf{u}}_{2T}^2\\
\norm{\mbf{x}(T)}_2^2 &\leq \gamma^2 \left(\norm{\mbf{x}_0}_2^2 + \norm{\mbf{u}}_{2T}^2\right).
\end{align*}
\end{proof}

\subsubsection[Transient State Bound for Discrete-Time Non-Autonomous LTI Systems]{Transient State Bound for Discrete-Time Non-Autonomous LTI Systems}
\index{transient!discrete-time state bound}
\label{sec:DTTransStateBoundNonAuton}

Consider the discrete-time LTI system with state-space realization
\bdis
\mbf{x}_{k+1}= \mbf{A}_\mathrm{d} \mbf{x}_k + \mbf{B}_\mathrm{d} \mbf{u}_k,
\edis
where $\mbf{A}_\mathrm{d}  \in \mathbb{R}^{n \times n}$ and $\mbf{B}_\mathrm{d}  \in \mathbb{R}^{n \times m}$.  The Euclidean norm \index{norm!Euclidean norm}of the state satisfies \bdis
\norm{\mbf{x}_k}_2^2 \leq \gamma^2\left( \norm{\mbf{x}_0}_2^2 + \norm{\mbf{u}}_{2k}^2\right), \,\, \forall k \in \mathbb{Z}_{\geq 0}
\edis
if there exist $\mbf{P} \in \mathbb{S}^n$ and $\gamma \in \mathbb{R}_{>0}$, where $\mbf{P} > 0$, such that 
\begin{align}
\mbf{P} - \gamma \mbf{1}&\leq 0, \label{eq:DT_State_Bound_NonAuton_1} \\
\bbm \mbf{P} & \mbf{1} \\ * & \gamma \mbf{1} \ebm &\geq 0, \label{eq:DT_State_Bound_NonAuton_2} \\
\bbm \mbf{A}_\mathrm{d}^\trans \mbf{P} \mbf{A}_\mathrm{d} - \mbf{P} & \mbf{A}_\mathrm{d}^\trans \mbf{P} \mbf{B}_\mathrm{d} \\ * & \mbf{B}_\mathrm{d}^\trans \mbf{B}_\mathrm{d} - \gamma \mbf{1} \ebm &\leq 0. \label{eq:DT_State_Bound_NonAuton_3}
\end{align}

If $\mbf{x}_0 = \mbf{0}$ and $\mbf{u}$ is a unit-energy input (i.e., $\norm{\mbf{u}}_{2k} \leq 1$, $\forall k \in \mathbb{Z}_{\geq 0}$), then the preceding conditions ensure that $\norm{\mbf{x}_k}_2 \leq \gamma$, $\forall k \in \mathbb{Z}_{\geq 0}$.

\begin{proof}
Define $V(k) = \mbf{x}_k^\trans \mbf{P} \mbf{x}_k$.  Evaluating $V(k+1) - V(k)$ results in
\begin{align}
V(k+1) - V(k) &= \bbm \mbf{x}_k^\trans & \mbf{u}_k^\trans \ebm \bbm \mbf{A}_\mathrm{d}^\trans \mbf{P} \mbf{A}_\mathrm{d} - \mbf{P} & \mbf{A}_\mathrm{d}^\trans \mbf{P} \mbf{B}_\mathrm{d} \\ * & \mbf{B}_\mathrm{d}^\trans \mbf{B}_\mathrm{d}  \ebm \bbm \mbf{x}_k \\ \mbf{u}_k \ebm \nonumber \\
&=  \bbm \mbf{x}_k^\trans & \mbf{u}_k^\trans \ebm \bbm \mbf{A}_\mathrm{d}^\trans \mbf{P} \mbf{A}_\mathrm{d} - \mbf{P} & \mbf{A}_\mathrm{d}^\trans \mbf{P} \mbf{B}_\mathrm{d} \\ * & \mbf{B}_\mathrm{d}^\trans \mbf{B}_\mathrm{d} - \gamma \mbf{1} \ebm \bbm \mbf{x}_k \\ \mbf{u}_k \ebm + \gamma \mbf{u}_k^\trans \mbf{u}_k. \label{eq:DT_State_Bound_NonAuton_4}
\end{align}
Substituting in~\eqref{eq:DT_State_Bound_NonAuton_3} and using induction gives
\beq
\label{eq:DT_State_Bound_NonAuton_5}
\mbf{x}_k^\trans \mbf{P} \mbf{x}_k \leq \mbf{x}_0^\trans \mbf{P} \mbf{x}_0 + \gamma \sum_{i=0}^k\mbf{u}_i^\trans \mbf{u}_i.
\eeq
Substituting~\eqref{eq:DT_State_Bound_NonAuton_1} and~\eqref{eq:DT_State_Bound_NonAuton_2} into~\eqref{eq:DT_State_Bound_NonAuton_5} yields
\begin{align*}
\gamma^{-1} \mbf{x}_k^\trans \mbf{x}_k &\leq \gamma \mbf{x}_0^\trans \mbf{x}_0 + \gamma \norm{\mbf{u}}_{2k}^2 \\
\norm{\mbf{x}_k}_2^2 &\leq \gamma^2\left( \norm{\mbf{x}_0}_2^2 + \norm{\mbf{u}}_{2k}^2\right).
\end{align*}
\end{proof}

\subsubsection[Transient Output Bound for Autonomous LTI Systems]{Transient Output Bound for Autonomous LTI Systems~\cite[p.~88]{Boyd1994},~\cite{Hayes2020}}
\index{transient!output bound}
\label{sec:TransOutputBound}

Consider the continuous-time LTI system with state-space realization
\begin{align*}
\dot{\mbf{x}} &= \mbf{A} \mbf{x}, \\
\mbf{y} &= \mbf{C} \mbf{x},
\end{align*}
where $\mbf{A}  \in \mathbb{R}^{n \times n}$, $\mbf{C}  \in \mathbb{R}^{p \times n}$ and $\mbf{x}(0) = \mbf{x}_0$.  The Euclidean norm \index{norm!Euclidean norm}of the output satisfies
\bdis
\norm{\mbf{y}(T)}_2 \leq \gamma \norm{\mbf{x}_0}_2, \,\, \forall T \in \mathbb{R}_{\geq 0}
\edis
if there exist $\mbf{P} \in \mathbb{S}^p$ and $\gamma \in \mathbb{R}_{>0}$, where $\mbf{P} > 0$, such that
\begin{align}
\mbf{P} - \gamma \mbf{1} &\leq 0, \label{eq:Output_Bound_1}\\
\bbm \mbf{P} & \mbf{C}^\trans \\ * & \gamma \mbf{1} \ebm &\geq 0, \label{eq:Output_Bound_2} \\
 \mbf{P} \mbf{A}+   \mbf{A}^\trans \mbf{P}  &\leq 0. \nonumber
\end{align}

\begin{proof}
The proof follows the same procedure as the proof in Section~\ref{sec:TransStateBound}, except the inequalities in~\eqref{eq:Output_Bound_1} and~\eqref{eq:Output_Bound_2} are substituted in to the inequality of~\eqref{eq:State_Bound_4}.
\end{proof}

\subsubsection[Transient Output Bound for Discrete-Time Autonomous LTI Systems]{Transient Output Bound for Discrete-Time Autonomous LTI Systems}
\index{transient!discrete-time output bound}
\label{sec:DTTransOutputBound}

Consider the discrete-time LTI system with state-space realization
\begin{align*}
\mbf{x}_{k+1} &= \mbf{A}_\mathrm{d} \mbf{x}_k, \\
\mbf{y}_k &= \mbf{C}_\mathrm{d} \mbf{x}_k,
\end{align*}
where $\mbf{A}_\mathrm{d}  \in \mathbb{R}^{n \times n}$ and $\mbf{C}_\mathrm{d}  \in \mathbb{R}^{p \times n}$.  The Euclidean norm \index{norm!Euclidean norm}of the output satisfies
\bdis
\norm{\mbf{y}_k}_2 \leq \gamma \norm{ \mbf{x}_0}_2, \,\, \forall k \in \mathbb{Z}_{\geq 0}
\edis
if there exist $\mbf{P} \in \mathbb{S}^n$ and $\gamma \in \mathbb{R}_{>0}$, where $\mbf{P} > 0$, such that
\begin{align}
\mbf{P} - \gamma \mbf{1} &\leq 0, \label{eq:DT_Output_Bound_1} \\
\bbm \mbf{P} & \mbf{C}_\mathrm{d}^\trans \\ * & \gamma \mbf{1} \ebm &\geq 0, \label{eq:DT_Output_Bound_2} \\
\mbf{A}_\mathrm{d}^\trans \mbf{P} \mbf{A}_\mathrm{d} - \mbf{P}  &\leq 0. \nonumber
\end{align}
\begin{proof}
The proof follows the same procedure as the proof in Section~\ref{sec:DTTransStateBound}, except the inequalities in~\eqref{eq:DT_Output_Bound_1} and~\eqref{eq:DT_Output_Bound_2} are substituted in to the inequality of~\eqref{eq:DT_State_Bound_4}.
\end{proof}

\subsubsection[Transient Output Bound for Non-Autonomous LTI Systems]{Transient Output Bound for Non-Autonomous LTI Systems}
\index{transient!output bound}
\label{sec:TransOutputBoundNonAuton}

Consider the continuous-time LTI system with state-space realization
\begin{align*}
\dot{\mbf{x}} &= \mbf{A} \mbf{x} + \mbf{B} \mbf{u}, \\
\mbf{y} &= \mbf{C} \mbf{x},
\end{align*}
where $\mbf{A}  \in \mathbb{R}^{n \times n}$, $\mbf{B}  \in \mathbb{R}^{n \times m}$, $\mbf{C}  \in \mathbb{R}^{p \times n}$, and $\mbf{x}(0) = \mbf{x}_0$.  The Euclidean norm \index{norm!Euclidean norm}of the output satisfies
\bdis
\norm{\mbf{y}(T)}_2^2 \leq \gamma^2\left( \norm{\mbf{x}_0}_2^2 + \norm{\mbf{u}}_{2T}^2\right), \,\, \forall T \in \mathbb{R}_{\geq 0}
\edis
if there exist $\mbf{P} \in \mathbb{S}^p$ and $\gamma \in \mathbb{R}_{>0}$, where $\mbf{P} > 0$, such that
\begin{align}
\mbf{P} - \gamma \mbf{1} &\leq 0, \label{eq:Output_Bound_NonAuton_1}\\
\bbm \mbf{P} & \mbf{C}^\trans \\ * & \gamma \mbf{1} \ebm &\geq 0, \label{eq:Output_Bound_NonAuton_2} \\
\bbm \mbf{P} \mbf{A} + \mbf{A}^\trans \mbf{P} & \mbf{P} \mbf{B} \\ * & -\gamma \mbf{1} \ebm &\leq 0. \nonumber
\end{align}

If $\mbf{x}_0 = \mbf{0}$ and $\mbf{u}$ is a unit-energy input (i.e., $\norm{\mbf{u}}_{2T} \leq 1$, $\forall T \in \mathbb{R}_{\geq 0}$), then the preceding conditions ensure that $\norm{\mbf{y}(T)}_2 \leq \gamma$, $\forall T \in \mathbb{R}_{\geq 0}$.
\begin{proof}
The proof follows the same procedure as the proof in Section~\ref{sec:TransStateBoundNonAuton}, except the inequalities in~\eqref{eq:Output_Bound_NonAuton_1} and~\eqref{eq:Output_Bound_NonAuton_2} are substituted in to the inequality of~\eqref{eq:State_Bound_NonAuton_5}.
\end{proof}

\subsubsection[Transient Output Bound for Discrete-Time Non-Autonomous LTI Systems]{Transient Output Bound for Discrete-Time Non-Autonomous LTI Systems}
\index{transient!discrete-time output bound}
\label{sec:DTTransOutputBoundNonAuton}

Consider the discrete-time LTI system with state-space realization
\begin{align*}
\mbf{x}_{k+1} &= \mbf{A}_\mathrm{d} \mbf{x}_k + \mbf{C}_\mathrm{d} \mbf{u}_k, \\
\mbf{y}_k &= \mbf{C}_\mathrm{d} \mbf{x}_k,
\end{align*}
where $\mbf{A}_\mathrm{d}  \in \mathbb{R}^{n \times n}$, $\mbf{B}_\mathrm{d}  \in \mathbb{R}^{n \times m}$, and $\mbf{C}_\mathrm{d}  \in \mathbb{R}^{p \times n}$.  The Euclidean norm \index{norm!Euclidean norm}of the output satisfies
\bdis
\norm{\mbf{y}_k}_2^2 \leq \gamma^2\left( \norm{\mbf{x}_0}_2^2 + \norm{\mbf{u}}_{2k}^2\right), \,\, \forall k \in \mathbb{Z}_{\geq 0}
\edis
if there exist $\mbf{P} \in \mathbb{S}^n$ and $\gamma \in \mathbb{R}_{>0}$, where $\mbf{P} > 0$, such that
\begin{align}
\mbf{P} - \gamma \mbf{1} &\leq 0, \label{eq:DT_Output_Bound_NonAuton_1} \\
\bbm \mbf{P} & \mbf{C}_\mathrm{d}^\trans \\ * & \gamma \mbf{1} \ebm &\geq 0, \label{eq:DT_Output_Bound_NonAuton_2} \\
\bbm \mbf{A}_\mathrm{d}^\trans \mbf{P} \mbf{A}_\mathrm{d} - \mbf{P} & \mbf{A}_\mathrm{d}^\trans \mbf{P} \mbf{B}_\mathrm{d} \\ * & \mbf{B}_\mathrm{d}^\trans \mbf{B}_\mathrm{d} - \gamma \mbf{1} \ebm &\leq 0. \nonumber
\end{align}

If $\mbf{x}_0 = \mbf{0}$ and $\mbf{u}$ is a unit-energy input (i.e., $\norm{\mbf{u}}_{2k} \leq 1$, $\forall k \in \mathbb{Z}_{\geq 0}$), then the preceding conditions ensure that $\norm{\mbf{y}_k}_2 \leq \gamma$, $\forall k \in \mathbb{Z}_{\geq 0}$.

\begin{proof}
The proof follows the same procedure as the proof in Section~\ref{sec:DTTransStateBoundNonAuton}, except the inequalities in~\eqref{eq:DT_Output_Bound_NonAuton_1} and~\eqref{eq:DT_Output_Bound_NonAuton_2} are substituted in to the inequality of~\eqref{eq:DT_State_Bound_NonAuton_5}.
\end{proof}

\subsubsection[Transient Impulse Response Bound]{Transient Impulse Response Bound~\cite{Scherer1997}}
\index{transient!impulse response bound}
\index{impulse response} 

Consider the single-input multi-output continuous-time LTI system with state-space realization
\begin{align*}
\dot{\mbf{x}} &= \mbf{A} \mbf{x} + \mbf{B} u, \\
\mbf{y} &= \mbf{C} \mbf{x},
\end{align*}
where $\mbf{A}  \in \mathbb{R}^{n \times n}$, $\mbf{B}  \in \mathbb{R}^{n \times 1}$, and $\mbf{C}  \in \mathbb{R}^{p \times n}$.  Let $\mbf{z}(t) = \mbf{C} e^{\mbf{A}t} \mbf{B}$ be the unit impulse response of the system.  The Euclidean norm \index{norm!Euclidean norm}of the impulse response satisfies
\bdis
\norm{\mbf{z}(T)}_2 \leq \gamma , \,\, \forall T \in \mathbb{R}_{\geq 0}
\edis
if there exist $\mbf{P} \in \mathbb{S}^p$ and $\gamma \in \mathbb{R}_{>0}$, where $\mbf{P} > 0$, such that
\begin{align}
\bbm \mbf{P} & \mbf{P} \mbf{B} \\ * & \gamma \ebm &\geq 0, \label{eq:Impulse_Bound_1}\\
\bbm \mbf{P} & \mbf{C}^\trans \\ * & \gamma \mbf{1} \ebm &\geq 0, \label{eq:Impulse_Bound_2} \\
 \mbf{P} \mbf{A}+   \mbf{A}^\trans \mbf{P}  &\leq 0. \nonumber
\end{align}

\begin{proof}
The proof follows the same procedure as the proof in Section~\ref{sec:TransOutputBound}, where the initial condition is chosen as $\mbf{x}_0 = \mbf{B}$.  This yields the result
\beq
\label{eq:Impulse_Bound_4}
\mbf{x}^\trans(T) \mbf{P} \mbf{x}(T) \leq \mbf{B}^\trans \mbf{P} \mbf{B}.
\eeq
Using the non-strict Schur complement, the matrix inequality in~\eqref{eq:Impulse_Bound_1} is equivalent to $\mbf{B}^\trans \mbf{P} \mbf{B} \leq \gamma$.  Substituting this and~\eqref{eq:Impulse_Bound_2} into~\eqref{eq:Impulse_Bound_4} gives the desired result.
\end{proof}

\subsubsection[Discrete-Time Transient Impulse Response Bound]{Discrete-Time Transient Impulse Response Bound}
\index{transient!discrete-time impulse response bound}
\index{impulse response} 

Consider the single-input multi-output discrete-time LTI system with state-space realization
\begin{align*}
\mbf{x}_{k+1} &= \mbf{A}_\mathrm{d} \mbf{x}_k + \mbf{B}_\mathrm{d} u_k, \\
\mbf{y}_k &= \mbf{C}_\mathrm{d} \mbf{x}_k,
\end{align*}
where $\mbf{A}_\mathrm{d}  \in \mathbb{R}^{n \times n}$, $\mbf{B}_\mathrm{d}  \in \mathbb{R}^{n \times 1}$, $\mbf{C}_\mathrm{d}  \in \mathbb{R}^{p \times n}$, and it is assumed that $\mbf{A}_\mathrm{d}$ is invertible.  Let $\mbf{z}_k = \mbf{C}_\mathrm{d} \mbf{A}_\mathrm{d}^{k-1} \mbf{B}_\mathrm{d}$ be the unit impulse response of the system.  The Euclidean norm \index{norm!Euclidean norm}of the impulse response satisfies
\bdis
\norm{\mbf{z}_k}_2 \leq \gamma , \,\, \forall k \in \mathbb{Z}_{\geq 0}
\edis
if there exist $\mbf{P} \in \mathbb{S}^n$ and $\gamma \in \mathbb{R}_{>0}$, where $\mbf{P} > 0$, such that 
\begin{align}
\bbm \mbf{P} & \mbf{P} \mbf{A}_\mathrm{d}^{-1} \mbf{B}_\mathrm{d} \\ * & \gamma \ebm &\geq 0, \label{eq:DT_Impulse_Bound_1} \\
\bbm \mbf{P} & \mbf{C}_\mathrm{d}^\trans \\ * & \gamma \mbf{1} \ebm &\geq 0, \label{eq:DT_Impulse_Bound_2} \\
\mbf{A}_\mathrm{d}^\trans \mbf{P} \mbf{A}_\mathrm{d} - \mbf{P}  &\leq 0. \nonumber
\end{align}
\begin{proof}
The proof follows the same procedure as the proof in Section~\ref{sec:DTTransOutputBound}, where the initial condition is chosen as $\mbf{x}_0 = \mbf{A}_\mathrm{d}^{-1}\mbf{B}_\mathrm{d}$ so that the unit impulse response matching the free response $\mbf{z}_k = \mbf{C}_\mathrm{d} \mbf{A}_\mathrm{d}^k \mbf{x}_0$.  This yields the result
\beq
\label{eq:DT_Impulse_Bound_4}
\mbf{x}_k^\trans \mbf{P} \mbf{x}_k \leq \mbf{B}_\mathrm{d}^\trans \mbf{A}_\mathrm{d}^{-\trans} \mbf{P} \mbf{A}_\mathrm{d}^{-1} \mbf{B}_\mathrm{d}.
\eeq
Using the non-strict Schur complement, the matrix inequality in~\eqref{eq:DT_Impulse_Bound_1} is equivalent to the inequality $ \mbf{B}_\mathrm{d}^\trans \mbf{A}_\mathrm{d}^{-\trans} \mbf{P} \mbf{A}_\mathrm{d}^{-1} \mbf{B}_\mathrm{d} \leq \gamma$.  Substituting this and~\eqref{eq:DT_Impulse_Bound_2} into~\eqref{eq:DT_Impulse_Bound_4} gives the desired result.
\end{proof}

\subsection[Output Energy Bounds]{Output Energy Bounds}

\subsubsection[Output Energy Bound for Autonomous LTI Systems]{Output Energy Bound for Autonomous LTI Systems~\cite[pp.~85--86]{Boyd1994}}
\index{energy bound!discrete-time output energy bound}
\label{sec:OutputEnergyBound}

Consider the continuous-time LTI system with state-space realization
\begin{align*}
\dot{\mbf{x}} &= \mbf{A} \mbf{x}, \\
\mbf{y} &= \mbf{C} \mbf{x},
\end{align*}
where $\mbf{A}  \in \mathbb{R}^{n \times n}$, $\mbf{C}  \in \mathbb{R}^{p \times n}$ and $\mbf{x}(0) = \mbf{x}_0$.  The output satisfies
\bdis
\sqrt{\int_0^T \mbf{y}^\trans \mbf{y} \mathrm{d}t} = \norm{\mbf{y}}_{2T} \leq \gamma \norm{\mbf{x}_0}_2, \,\, \forall T \in \mathbb{R}_{\geq 0}
\edis
if there exist $\mbf{P} \in \mathbb{S}^p$ and $\gamma \in \mathbb{R}_{>0}$, where $\mbf{P} > 0$, such that
\begin{align}
\mbf{P} - \gamma \mbf{1} &\leq 0, \label{eq:Output_Energy_Bound_1}\\
\bbm  \mbf{P} \mbf{A}+   \mbf{A}^\trans \mbf{P}  & \mbf{C}^\trans \\ * & -\gamma \mbf{1} \ebm &\leq 0. \label{eq:Output_Energy_Bound_2}
\end{align}

\begin{proof}
Define $V = \mbf{x}^\trans \mbf{P} \mbf{x}$.  Evaluating $\dot{V}$ results in
\begin{align}
\dot{V} &= \mbf{x}^\trans \left(\mbf{P} \mbf{A} + \mbf{A}^\trans \mbf{P} \right) \mbf{x} \nonumber \\
&=  \mbf{x}^\trans \left(\mbf{P} \mbf{A} + \mbf{A}^\trans \mbf{P} + \gamma^{-1} \mbf{C}^\trans \mbf{C} \right) \mbf{x}  - \gamma^{-1} \mbf{y}^\trans \mbf{y}. \label{eq:Output_Energy_Bound_4}
\end{align}
Using the Schur complement lemma and substituting~\eqref{eq:Output_Energy_Bound_2} into~\eqref{eq:Output_Energy_Bound_4} gives $\dot{V} \leq - \gamma^{-1} \mbf{y}^\trans \mbf{y}$.  Integrating both sides of this inequality from $t = 0$ to $t = T$, where $T \in \mathbb{R}_{\geq 0}$ yields
\begin{align}
\gamma^{-1} \norm{\mbf{y}}_{2T}^2 &\leq -\mbf{x}^\trans(T) \mbf{P} \mbf{x}(T) + \mbf{x}_0^\trans \mbf{P} \mbf{x}_0 \nonumber \\
&\leq  \mbf{x}_0^\trans \mbf{P} \mbf{x}_0 \label{eq:Output_Energy_Bound_5}
\end{align}
Substituting ~\eqref{eq:Output_Energy_Bound_1} into~\eqref{eq:Output_Energy_Bound_5} results in
\begin{align*}
\gamma^{-1} \norm{\mbf{y}}_{2T}^2 &\leq \gamma \mbf{x}_0^\trans \mbf{x}_0 \\
\norm{\mbf{y}}_{2T} &\leq \gamma\norm{\mbf{x}_0}_2.
\end{align*}
\end{proof}

\subsubsection[Output Energy Bound for Discrete-Time Autonomous LTI Systems]{Output Energy Bound for Discrete-Time Autonomous LTI Systems}
\index{energy bound!discrete-time output energy bound}
\label{sec:DTOutputEnergyBound}

Consider the discrete-time LTI system with state-space realization
\begin{align*}
\mbf{x}_{k+1} &= \mbf{A}_\mathrm{d} \mbf{x}_k, \\
\mbf{y}_k &= \mbf{C}_\mathrm{d} \mbf{x}_k,
\end{align*}
where $\mbf{A}_\mathrm{d}  \in \mathbb{R}^{n \times n}$ and $\mbf{C}_\mathrm{d}  \in \mathbb{R}^{p \times n}$.  The output satisfies
\bdis
\norm{\mbf{y}}_{2k} \leq \gamma \norm{ \mbf{x}_0}_2, \,\, \forall k \in \mathbb{Z}_{\geq 0}
\edis
if there exist $\mbf{P} \in \mathbb{S}^n$ and $\gamma \in \mathbb{R}_{>0}$, where $\mbf{P} > 0$, such that
\begin{align}
\mbf{P} - \gamma \mbf{1} &\leq 0, \label{eq:DT_Output_Energy_Bound_1} \\
\bbm \mbf{A}_\mathrm{d}^\trans \mbf{P} \mbf{A}_\mathrm{d} - \mbf{P} & \mbf{C}_\mathrm{d}^\trans \\ * & -\gamma \mbf{1} \ebm &\leq 0. \label{eq:DT_Output_Energy_Bound_2} 
\end{align}

\begin{proof}
Define $V(k) = \mbf{x}_k^\trans \mbf{P} \mbf{x}_k$.  Evaluating $V(k+1) - V(k)$ results in
\begin{align}
V(k+1) - V(k) &= \mbf{x}_k^\trans \left(\mbf{A}_\mathrm{d}^\trans \mbf{P} \mbf{A}_\mathrm{d} - \mbf{P} \right) \mbf{x}_k \nonumber \\
&=  \mbf{x}_k^\trans \left(\mbf{A}_\mathrm{d}^\trans \mbf{P} \mbf{A}_\mathrm{d} - \mbf{P} + \gamma^{-1} \mbf{C}_\mathrm{d}^\trans \mbf{C}_\mathrm{d} \right) \mbf{x}_k - \gamma^{-1} \mbf{y}_k^\trans \mbf{y}_k. \label{eq:DT_Output_Energy_Bound_4}
\end{align}
Using the Schur complement lemma, substituting~\eqref{eq:DT_Output_Energy_Bound_2} into~\eqref{eq:DT_Output_Energy_Bound_4}, and using induction gives
\begin{align}
 \gamma^{-1} \sum_{i=0}^k\mbf{y}_i^\trans \mbf{y}_i   &\leq - \mbf{x}_k^\trans \mbf{P} \mbf{x}_k + \mbf{x}_0^\trans \mbf{P} \mbf{x}_0 \nonumber \\
\gamma^{-1}\norm{\mbf{y}}_{2k}^2  &\leq - \mbf{x}_k^\trans \mbf{P} \mbf{x}_k + \mbf{x}_0^\trans \mbf{P} \mbf{x}_0 \nonumber \\
&\leq \mbf{x}_0^\trans \mbf{P} \mbf{x}_0 \label{eq:DT_Output_Energy_Bound_5}
\end{align}
Substituting~\eqref{eq:DT_Output_Energy_Bound_1} into~\eqref{eq:DT_Output_Energy_Bound_5} yields
\begin{align*}
\gamma^{-1}\norm{\mbf{y}}_{2k}^2 &\leq \gamma \mbf{x}_0^\trans \mbf{x}_0  \\
\norm{\mbf{y}}_{2k} &\leq \gamma\norm{\mbf{x}_0}_2.
\end{align*}
\end{proof}

\subsubsection[Output Energy Bound for Non-Autonomous LTI Systems]{Output Energy Bound for Non-Autonomous LTI Systems}
\index{energy bound!output energy bound}
\label{sec:OutputEnergyBoundNonAuton}

Consider the continuous-time LTI system with state-space realization
\begin{align*}
\dot{\mbf{x}} &= \mbf{A} \mbf{x} + \mbf{B} \mbf{u}, \\
\mbf{y} &= \mbf{C} \mbf{x} + \mbf{D} \mbf{u},
\end{align*}
where $\mbf{A}  \in \mathbb{R}^{n \times n}$, $\mbf{B}  \in \mathbb{R}^{n \times m}$, $\mbf{C}  \in \mathbb{R}^{p \times n}$, $\mbf{D}  \in \mathbb{R}^{p \times m}$, and $\mbf{x}(0) = \mbf{x}_0$.  The output satisfies
\bdis
\int_0^T \mbf{y}^\trans \mbf{y} \mathrm{d}t = \norm{\mbf{y}}_{2T}^2 \leq \gamma^2\left( \norm{\mbf{x}_0}_2^2 + \norm{\mbf{u}}_{2T}^2\right), \,\, \forall T \in \mathbb{R}_{\geq 0}
\edis
if there exist $\mbf{P} \in \mathbb{S}^p$ and $\gamma \in \mathbb{R}_{>0}$, where $\mbf{P} > 0$, such that
\begin{align}
\mbf{P} - \gamma \mbf{1} &\leq 0, \label{eq:Output_Energy_Bound_NonAuton_1}\\
\bbm \mbf{P} \mbf{A} + \mbf{A}^\trans \mbf{P} & \mbf{P} \mbf{B} & \mbf{C}^\trans \\ * & -\gamma \mbf{1} & \mbf{D}^\trans \\ * & * & -\gamma \mbf{1} \ebm &\leq 0. \label{eq:Output_Energy_Bound_NonAuton_2}
\end{align}

If $\mbf{x}_0 = \mbf{0}$, then the preceding conditions match the Bounded Real Lemma and ensure that $\norm{\mbf{y}}_{2T} \leq \gamma \norm{\mbf{u}}_{2T}$, $\forall T \in \mathbb{R}_{\geq 0}$.

\begin{proof}
Define $V = \mbf{x}^\trans \mbf{P} \mbf{x}$.  Evaluating $\dot{V}$ results in
\begin{align}
\dot{V} &= \bbm \mbf{x}^\trans & \mbf{u}^\trans \ebm \bbm \mbf{P} \mbf{A} + \mbf{A}^\trans \mbf{P} & \mbf{P} \mbf{B} \\ * & \mbf{0} \ebm \bbm \mbf{x} \\ \mbf{u} \ebm \nonumber \\
&=  \bbm \mbf{x}^\trans & \mbf{u}^\trans \ebm \bbm \mbf{P} \mbf{A} + \mbf{A}^\trans \mbf{P} + \gamma^{-1} \mbf{C}^\trans \mbf{C} & \mbf{P} \mbf{B} + \gamma^{-1}\mbf{C}^\trans \mbf{D} \\ * & -\gamma \mbf{1} + \gamma^{-1} \mbf{D}^\trans \mbf{D}\ebm \bbm \mbf{x} \\ \mbf{u} \ebm + \gamma \mbf{u}^\trans \mbf{u} - \gamma^{-1} \mbf{y}^\trans \mbf{y}. \label{eq:Output_Energy_Bound_NonAuton_4}
\end{align}
Using the Schur complement lemma and substituting~\eqref{eq:Output_Energy_Bound_NonAuton_2} into~\eqref{eq:Output_Energy_Bound_NonAuton_4} gives $\dot{V} \leq \gamma \mbf{u}^\trans \mbf{u} - \gamma^{-1} \mbf{y}^\trans \mbf{y}$.  Integrating both sides of this inequality from $t = 0$ to $t = T$, where $T \in \mathbb{R}_{\geq 0}$ yields
\begin{align}
\gamma^{-1} \norm{\mbf{y}}_{2T}^2 &\leq -\mbf{x}^\trans(T) \mbf{P} \mbf{x}(T) + \mbf{x}_0^\trans \mbf{P} \mbf{x}_0 + \gamma \norm{\mbf{u}}_{2T}^2 \nonumber \\
&\leq  \mbf{x}_0^\trans \mbf{P} \mbf{x}_0 + \gamma \norm{\mbf{u}}_{2T}^2 \label{eq:Output_Energy_Bound_NonAuton_5}
\end{align}
Substituting ~\eqref{eq:Output_Energy_Bound_NonAuton_1} into~\eqref{eq:Output_Energy_Bound_NonAuton_5} results in
\begin{align*}
\gamma^{-1} \norm{\mbf{y}}_{2T}^2 &\leq \gamma \mbf{x}_0^\trans \mbf{x}_0 + \gamma \norm{\mbf{u}}_{2T}^2\\
\norm{\mbf{y}}_{2T}^2 &\leq \gamma^2 \left(\norm{\mbf{x}_0}_2^2 + \norm{\mbf{u}}_{2T}^2\right).
\end{align*}
\end{proof}

\subsubsection[Output Energy Bound for Discrete-Time Non-Autonomous LTI Systems]{Output Energy Bound for Discrete-Time Non-Autonomous LTI Systems}
\index{energy bound!discrete-time output energy bound}
\label{sec:DTOutputEnergyBoundNonAuton}

Consider the discrete-time LTI system with state-space realization
\begin{align*}
\mbf{x}_{k+1} &= \mbf{A}_\mathrm{d} \mbf{x}_k + \mbf{B}_\mathrm{d} \mbf{u}_k, \\
\mbf{y}_k &= \mbf{C}_\mathrm{d} \mbf{x}_k + \mbf{D}_\mathrm{d} \mbf{u}_k,
\end{align*}
where $\mbf{A}_\mathrm{d}  \in \mathbb{R}^{n \times n}$, $\mbf{B}_\mathrm{d}  \in \mathbb{R}^{n \times m}$, $\mbf{C}_\mathrm{d}  \in \mathbb{R}^{p \times n}$, and $\mbf{D}_\mathrm{d}  \in \mathbb{R}^{p \times m}$.  The output satisfies
\bdis
\sum_{i=0}^k\mbf{y}_i^\trans \mbf{y}_i = \norm{\mbf{y}}_{2k}^2 \leq \gamma^2\left( \norm{\mbf{x}_0}_2^2 + \norm{\mbf{u}}_{2k}^2\right), \,\, \forall k \in \mathbb{Z}_{\geq 0}
\edis
if there exist $\mbf{P} \in \mathbb{S}^n$ and $\gamma \in \mathbb{R}_{>0}$, where $\mbf{P} > 0$, such that
\begin{align}
\mbf{P} - \gamma \mbf{1} &\leq 0, \label{eq:DT_Output_Energy_Bound_NonAuton_1} \\
\bbm \mbf{A}_\mathrm{d}^\trans \mbf{P} \mbf{A}_\mathrm{d} - \mbf{P} & \mbf{A}_\mathrm{d}^\trans \mbf{P} \mbf{B}_\mathrm{d} & \mbf{C}_\mathrm{d}^\trans \\ * & \mbf{B}_\mathrm{d}^\trans \mbf{B}_\mathrm{d} - \gamma \mbf{1} & \mbf{D}_\mathrm{d}^\trans \\ * & * & -\gamma \mbf{1} \ebm &\leq 0. \label{eq:DT_Output_Energy_Bound_NonAuton_2}
\end{align}

If $\mbf{x}_0 = \mbf{0}$, then the preceding conditions match the Bounded Real Lemma and ensure that $\norm{\mbf{y}}_{2k} \leq \gamma \norm{\mbf{u}}_{2k}$, $\forall k \in \mathbb{Z}_{\geq 0}$.

\begin{proof}
Define $V(k) = \mbf{x}_k^\trans \mbf{P} \mbf{x}_k$.  Evaluating $V(k+1) - V(k)$ results in
\begin{align}
V(k+1) - V(k) &= \bbm \mbf{x}_k^\trans & \mbf{u}_k^\trans \ebm \bbm \mbf{A}_\mathrm{d}^\trans \mbf{P} \mbf{A}_\mathrm{d} - \mbf{P} & \mbf{A}_\mathrm{d}^\trans \mbf{P} \mbf{B}_\mathrm{d} \\ * & \mbf{B}_\mathrm{d}^\trans \mbf{B}_\mathrm{d}  \ebm \bbm \mbf{x}_k \\ \mbf{u}_k \ebm \nonumber \\
&= \bbm \mbf{x}_k^\trans & \mbf{u}_k^\trans \ebm \bbm \mbf{A}_\mathrm{d}^\trans \mbf{P} \mbf{A}_\mathrm{d} - \mbf{P} + \gamma^{-1} \mbf{C}_\mathrm{d}^\trans \mbf{C}_\mathrm{d} & \mbf{A}_\mathrm{d}^\trans \mbf{P} \mbf{B}_\mathrm{d} + \gamma^{-1} \mbf{C}_\mathrm{d}^\trans \mbf{D}_\mathrm{d}\\ * & \mbf{B}_\mathrm{d}^\trans \mbf{B}_\mathrm{d} - \gamma \mbf{1}  + \gamma^{-1} \mbf{D}_\mathrm{d}^\trans \mbf{D}_\mathrm{d}\ebm \bbm \mbf{x}_k \\ \mbf{u}_k \ebm \nonumber \\ & \hspace{20pt} + \gamma \mbf{u}_k^\trans \mbf{u}_k - \gamma^{-1} \mbf{y}_k^\trans \mbf{y}_k. \label{eq:DT_Output_Energy_Bound_NonAuton_4}
\end{align}
Using the Schur complement lemma, substituting~\eqref{eq:DT_Output_Energy_Bound_NonAuton_2} into~\eqref{eq:DT_Output_Energy_Bound_NonAuton_4}, and using induction gives
\begin{align}
\gamma^{-1}\sum_{i=0}^k\mbf{y}_i^\trans \mbf{y}_i  &\leq - \mbf{x}_k^\trans \mbf{P} \mbf{x}_k + \mbf{x}_0^\trans \mbf{P} \mbf{x}_0 + \gamma \sum_{i=0}^k\mbf{u}_i^\trans \mbf{u}_i \nonumber \\
\gamma^{-1}\norm{\mbf{y}}_{2k}^2  &\leq - \mbf{x}_k^\trans \mbf{P} \mbf{x}_k + \mbf{x}_0^\trans \mbf{P} \mbf{x}_0 + \gamma \norm{\mbf{u}}_{2k}^2 \nonumber \\
&\leq \mbf{x}_0^\trans \mbf{P} \mbf{x}_0 + \gamma \norm{\mbf{u}}_{2k}^2 \label{eq:DT_Output_Energy_Bound_NonAuton_5}
\end{align}
Substituting~\eqref{eq:DT_Output_Energy_Bound_NonAuton_1} into~\eqref{eq:DT_Output_Energy_Bound_NonAuton_5} yields
\begin{align*}
\gamma^{-1}\norm{\mbf{y}}_{2k}^2 &\leq \gamma \mbf{x}_0^\trans \mbf{x}_0 + \gamma \norm{\mbf{u}}_{2k}^2  \\
\norm{\mbf{y}}_{2k}^2 &\leq \gamma^2 \left( \norm{\mbf{x}_0}_2^2 + \gamma \norm{\mbf{u}}_{2k}^2\right).
\end{align*}
\end{proof}

\subsection[Kharitonov-Bernstein-Haddad (KBH) Theorem]{Kharitonov-Bernstein-Haddad (KBH) Theorem~\cite{Bernstein1992}}
\index{Kharitonov-Bernstein-Haddad (KBH) theorem}
Consider the set of matrices
\beq
\label{eq:KBH1}
\mbc{A} =  \Bigg\{ \mbf{A} = \bbm  \mbf{0}_{(n-1)\times 1} & & \mbf{1}_{(n-1)\times (n-1)}   \\ - a_0 & \cdots & -a_{n-1} \ebm \,\, | \,\, \munderbar{a}_j \leq a_j \leq \bar{a}_j, \quad j=0,1,2,\ldots,n-1 \Bigg\}.
\eeq
Every matrix in the set $\mbc{A}$ is Hurwitz\index{Hurwitz matrix} if and only if there exist $\mbf{P}_i \in \mathbb{S}^n$, $i = 1,2,3,4$, where $\mbf{P}_i  > 0$, $i = 1,2,3,4$, such that
\bdis
\mbf{P}_i \mbf{A}_i + \mbf{A}_i ^\trans \mbf{P}_i < 0, \quad i = 1,2,3,4,
\edis
where
\begin{align}
\mbf{A}_i &= \bbm \bbm \mbf{0}_{(n-1)\times 1} & \mbf{1}_{(n-1)\times (n-1)} \ebm \\ \mbf{a}_i \ebm, \quad i = 1,2,3,4,\nonumber \\
\mbf{a}_1 &= -\bbm \munderbar{a}_0 & \munderbar{a}_1 & \bar{a}_2 & \bar{a}_3 & \cdots & \munderbar{a}_{n-4} & \munderbar{a}_{n-3} & \bar{a}_{n-2} & \bar{a}_{n-1} \ebm, \nonumber \\
 \mbf{a}_2 &= -\bbm \munderbar{a}_0 & \bar{a}_1 & \bar{a}_2 & \munderbar{a}_3 & \cdots & \munderbar{a}_{n-4} & \bar{a}_{n-3} & \bar{a}_{n-2} & \munderbar{a}_{n-1} \ebm, \nonumber \\
\mbf{a}_3 &= -\bbm \bar{a}_0 & \munderbar{a}_1 & \munderbar{a}_2 & \bar{a}_3 & \cdots & \bar{a}_{n-4} & \munderbar{a}_{n-3} & \munderbar{a}_{n-2} & \bar{a}_{n-1} \ebm, \nonumber \\
\mbf{a}_4 &= -\bbm \bar{a}_0 & \bar{a}_1 & \munderbar{a}_2 & \munderbar{a}_3 & \cdots & \bar{a}_{n-4} & \bar{a}_{n-3} & \munderbar{a}_{n-2} & \munderbar{a}_{n-1} \ebm. \nonumber 
\end{align}
Equivalently, every matrix in the set $\mbc{A}$ is Hurwitz if and only if there exist $\mbf{Q}_i \in \mathbb{S}^n$, $i = 1,2,3,4$, where $\mbf{Q}_i  > 0$, $i = 1,2,3,4$, such that

\bdis
\mbf{A}_i\mbf{Q}_i + \mbf{Q}_i\mbf{A}_i ^\trans < 0, \quad i = 1,2,3,4.
\edis

\subsection{Stability of Discrete-Time System with Polytopic Uncertainty}
\index{polytopic uncertainty}
\subsubsection[Open-Loop Robust Stability]{Open-Loop Robust Stability~\cite{DeOliveira1999}}

Consider the set of matrices
\bdis
\mbc{A} =  \Bigg\{ \mbf{A}_\mathrm{d}(\alpha) \in \mathbb{R}^{n \times n} \,\, | \,\, \mbf{A}_\mathrm{d}(\alpha) = \sum_{i=1}^n \alpha_i\mbf{A}_{\mathrm{d},i}, \,\, \mbf{A}_{\mathrm{d},i} \in \mathbb{R}^{n \times n}, \,\, \alpha_i \in \mathbb{R}_{\geq 0}, \,\, \sum_{i=1}^n\alpha_i = 1  \Bigg\}.
\edis
The discrete-time LTI system $\mbf{x}_{k+1} = \mbf{A}_\mathrm{d}(\alpha)\mbf{x}_k $ is asymptotically stable for all $\mbf{A}_\mathrm{d}(\alpha) \in \mbc{A}$ if there exist $\mbf{P}_i \in \mathbb{S}^n$, $i = 1,\ldots,n$, and $\mbf{G} \in \mathbb{R}^{n \times n}$, where $\mbf{P}_i  > 0$, $i = 1,\ldots,n$, such that
\bdis
\bbm \mbf{P}_i & \mbf{A}_{\mathrm{d},i}^\trans\mbf{G}^\trans \\ * & \mbf{G} + \mbf{G}^\trans - \mbf{P}_i \ebm< 0, \quad i = 1,\ldots,n.
\edis

\subsubsection[Closed-Loop Robust Stability]{Closed-Loop Robust Stability~\cite{DeOliveira1999}}

Consider the set of matrices
\bdis
\mbc{A} =  \Bigg\{ \mbf{A}_\mathrm{d}(\alpha) \in \mathbb{R}^{n \times n} \,\, | \,\, \mbf{A}_\mathrm{d}(\alpha) = \sum_{i=1}^n \alpha_i\mbf{A}_{\mathrm{d},i}, \,\, \mbf{A}_{\mathrm{d},i} \in \mathbb{R}^{n \times n}, \,\, \alpha_i \in \mathbb{R}_{\geq 0}, \,\, \sum_{i=1}^n\alpha_i = 1  \Bigg\}.
\edis
and
\bdis
\mbc{B} =  \Bigg\{ \mbf{B}_\mathrm{d}(\beta) \in \mathbb{R}^{n \times m} \,\, | \,\, \mbf{B}_\mathrm{d}(\beta) = \sum_{i=1}^p \beta_i\mbf{B}_{\mathrm{d},i}, \mbf{B}_{\mathrm{d},i} \in \mathbb{R}^{n \times m}, \,\, \beta_i \in \mathbb{R}_{\geq 0}, \,\, \sum_{i=1}^m\beta_i = 1  \Bigg\}.
\edis
The discrete-time LTI system $\mbf{x}_{k+1} = \mbf{A}_\mathrm{d}(\alpha)\mbf{x}_k + \mbf{B}_\mathrm{d}(\beta)\mbf{u}_k$ is asymptotically stabilized by the state feedback control law $\mbf{u}_k = -\mbf{L}\mbf{G}^{-1}\mbf{u}_k$ for all $\mbf{A}_\mathrm{d}(\alpha) \in \mbc{A}$ and $\mbf{B}_\mathrm{d}(\alpha) \in \mbc{B}$ if there exist $\mbf{P}_{ij}  \in \mathbb{S}^n$, $i = 1,\ldots,n$, $j = 1,\ldots,p$, $\mbf{G} \in \mathbb{R}^{n \times n}$, and $\mbf{L} \in \mathbb{R}^{m \times n}$, where $\mbf{P}_{ij}  > 0$, $i = 1,\ldots,n$, $j = 1,\ldots,p$ and $\mbf{G}$ is invertible, such that
\bdis
\bbm \mbf{P}_{ij} & \mbf{A}_{\mathrm{d},i}\mbf{G} - \mbf{B}_{\mathrm{d},j}\mbf{L} \\ * & \mbf{G} + \mbf{G}^\trans - \mbf{P}_{ij} \ebm< 0, \quad i = 1,\ldots,n, \quad j = 1,\ldots,p.
\edis

\subsection{Quadratic Stability}
\index{stability!quadratic stability}
\subsubsection[Continuous-Time Quadratic Stability]{Continuous-Time Quadratic Stability~\cite[pp.~112--115]{Duan2013}}

Consider the uncertain continuous-time linear system with state-space representation
\beq
\label{eq:CTQS1}
\dot{\mbf{x}} = \left(\mbf{A}_0 + \Delta \mbf{A}(\mbs{\delta}(t))\right)\mbf{x},
\eeq
where $\mbf{A}_0 \in \mathbb{R}^{n \times n}$, $\Delta \mbf{A}(\mbs{\delta}(t)) = \sum_{i=1}^k \delta_i(t)\mbf{A}_i \in \mathbb{R}^{n \times n}$, $\delta_i \in \mathbb{R}$, $i=1, \ldots, k$, $\mbf{A}_i \in \mathbb{R}^{n \times n}$, $i=1,\ldots,k$, $\mbs{\delta}^\trans(t) = \bbm \delta_1(t) & \delta_2(t) & \cdots & \delta_k(t) \ebm \in \mbs{\Delta}$, and $\mbs{\Delta}$ is the set of perturbation parameters.  The uncertain system in~\eqref{eq:CTQS1} is quadratically stable if there exists $\mbf{P} \in \mathbb{S}^n$, where $\mbf{P} > 0$, such that
\bdis
\left(\mbf{A}_0 + \Delta \mbf{A}(\mbs{\delta}(t))\right)^\trans \mbf{P} + \mbf{P} \left(\mbf{A}_0 + \Delta \mbf{A}(\mbs{\delta}(t))\right) < 0, \hspace{10pt} \forall \mbs{\delta}(t) \in \mbs{\Delta}.
\edis
The following statements can be made for particular sets of perturbations.
\begin{enumerate}

\item Consider the case where the set of perturbation parameters is defined by a regular polyhedron as
\bdis
\mbs{\Delta} = \{ \mbs{\delta}(t) = \bbm \delta_1(t) & \delta_2(t) & \cdots & \delta_k(t) \ebm \in \mathbb{R}^k \,\, | \,\, \delta_i(t), \, \munderbar{\delta}_i, \, \bar{\delta}_i \in \mathbb{R}, \,\, \munderbar{\delta}_i \leq \delta_i(t) \leq \bar{\delta}_i ] \}.
\edis
The uncertain system in~\eqref{eq:CTQS1} is quadratically stable if and only if there exists $\mbf{P} \in \mathbb{S}^n$, where $\mbf{P} > 0$, such that
\bdis
\left(\mbf{A}_0 + \Delta \mbf{A}(\mbs{\delta}(t))\right)^\trans \mbf{P} + \mbf{P} \left(\mbf{A}_0 + \Delta \mbf{A}(\mbs{\delta}(t))\right) < 0, \hspace{10pt} \forall \delta_i(t) \in \{ \munderbar{\delta}_i, \bar{\delta}_i\}, \,\, i=1,\ldots,k.
\edis

\item Consider the case where the set of perturbation parameters is defined by a polytope as
\bdis
\mbs{\Delta} = \{ \mbs{\delta}(t) = \bbm \delta_1(t) & \delta_2(t) & \cdots & \delta_k(t) \ebm \in \mathbb{R}^k \,\, | \,\, \delta_i(t) \in \mathbb{R}_{\geq 0}, \,\, \sum_{i=1}^k \delta_i(t) = 1 \}.
\edis
The uncertain system in~\eqref{eq:CTQS1} is quadratically stable if and only if there exists $\mbf{P} \in \mathbb{S}^n$, where $\mbf{P} > 0$, such that
\bdis
\left(\mbf{A}_0 + \mbf{A}_i\right)^\trans \mbf{P} + \mbf{P} \left(\mbf{A}_0 + \mbf{A}_i\right) < 0, \hspace{10pt} i=1,\ldots,k.
\edis

\end{enumerate}

\subsubsection[Discrete-Time Quadratic Stability]{Discrete-Time Quadratic Stability~\cite[pp.~116--118]{Duan2013}}

Consider the uncertain discrete-time linear system with state-space representation
\beq
\label{eq:DTQS1}
\mbf{x}_{k+1} = \left(\mbf{A}_{\mathrm{d},0} + \Delta \mbf{A}_\mathrm{d}(\mbs{\delta}(t))\right)\mbf{x}_k,
\eeq
where $\mbf{A}_{\mathrm{d},0} \in \mathbb{R}^{n \times n}$, $\Delta \mbf{A}_\mathrm{d}(\mbs{\delta}(t)) = \sum_{i=1}^k \delta_i(t)\mbf{A}_{\mathrm{d},i} \in \mathbb{R}^{n \times n}$, $\delta_i \in \mathbb{R}$, $i=1, \ldots, k$, $\mbf{A}_{\mathrm{d},i} \in \mathbb{R}^{n \times n}$, $i=1,\ldots,k$, $\mbs{\delta}^\trans(t) = \bbm \delta_1(t) & \delta_2(t) & \cdots & \delta_k(t) \ebm \in \mbs{\Delta}$, and $\mbs{\Delta}$ is the set of perturbation parameters.  The uncertain system in~\eqref{eq:CTQS1} is quadratically stable if there exists $\mbf{P} \in \mathbb{S}^n$, where $\mbf{P} > 0$, such that
\bdis
\left(\mbf{A}_{\mathrm{d},0} + \Delta \mbf{A}_\mathrm{d}(\mbs{\delta}(t))\right)^\trans \mbf{P} \left(\mbf{A}_{\mathrm{d},0} + \Delta \mbf{A}_\mathrm{d}(\mbs{\delta}(t))\right) - \mbf{P} < 0, \hspace{10pt} \forall \mbs{\delta}(t) \in \mbs{\Delta}.
\edis
The following statements can be made for particular sets of perturbations.
\begin{enumerate}

\item Consider the case where the set of perturbation parameters is defined by a regular polyhedron as
\bdis
\mbs{\Delta} = \{ \mbs{\delta}(t) = \bbm \delta_1(t) & \delta_2(t) & \cdots & \delta_k(t) \ebm \in \mathbb{R}^k \,\, | \,\, \delta_i(t), \, \munderbar{\delta}_i, \, \bar{\delta}_i \in \mathbb{R}, \,\, \munderbar{\delta}_i \leq \delta_i(t) \leq \bar{\delta}_i ] \}.
\edis
The uncertain system in~\eqref{eq:CTQS1} is quadratically stable if and only if there exists $\mbf{P} \in \mathbb{S}^n$, where $\mbf{P} > 0$, such that
\bdis
 \left(\mbf{A}_{\mathrm{d},0} + \Delta \mbf{A}_\mathrm{d}(\mbs{\delta}(t))\right)^\trans \mbf{P} \left(\mbf{A}_{\mathrm{d},0} + \Delta \mbf{A}_\mathrm{d}(\mbs{\delta}(t))\right)  - \mbf{P}  < 0, \hspace{10pt} \forall \delta_i(t) \in \{ \munderbar{\delta}_i, \bar{\delta}_i\}, \,\, i=1,2,\ldots,k.
\edis

\item Consider the case where the set of perturbation parameters is defined by a polytope as
\bdis
\mbs{\Delta} = \{ \mbs{\delta}(t) = \bbm \delta_1(t) & \delta_2(t) & \cdots & \delta_k(t) \ebm \in \mathbb{R}^k \,\, | \,\, \delta_i(t) \in \mathbb{R}_{\geq 0}, \,\, \sum_{i=1}^k \delta_i(t) = 1 \}.
\edis
The uncertain system in~\eqref{eq:CTQS1} is quadratically stable if and only if there exists $\mbf{P} \in \mathbb{S}^n$, where $\mbf{P} > 0$, such that
\bdis
\left(\mbf{A}_{\mathrm{d},0} + \mbf{A}_{\mathrm{d},i}\right)^\trans \mbf{P}\left(\mbf{A}_{\mathrm{d},0} + \mbf{A}_{\mathrm{d},i}\right) - \mbf{P}  < 0, \hspace{10pt} i=1,2,\ldots,k.
\edis

\end{enumerate}

\subsection[Stability of Time-Delay Systems]{Stability of Time-Delay Systems}

Consider the continuous-time linear time-delay\index{time delay} system with state-space representation
\beq
\label{eq:TD1}
\dot{\mbf{x}}(t) = \mbf{A} \mbf{x}(t) + \mbf{A}_\textrm{d} \mbf{x}(t - d),
\eeq
where $\mbf{A}$,~$\mbf{A}_\mathrm{d} \in \mathbb{R}^{n \times n}$, $d$,~$\bar{d} \in \mathbb{R}_{>0}$, and the initial condition is given by $\mbf{x}(t) = \mbs{\phi}(t)$, $t \in [-d,0]$, where $\bar{d}$ is a known upper-bound on the time-delay (i.e., $0 < d \leq \bar{d}$).

\subsubsection[Delay-Independent Condition]{Delay-Independent Condition~\cite[p.~126]{Duan2013},~\cite[pp.~18--19]{Dey2018}}

The time-delay system in~\eqref{eq:TD1} is asymptotically stable if there exist $\mbf{P}$,~$\mbf{S} \in \mathbb{S}^n$, where $\mbf{P} > 0$ and $\mbf{S} > 0$, such that
\bdis
\bbm \mbf{A}^\trans \mbf{P} + \mbf{P} \mbf{A} + \mbf{S} & \mbf{P} \mbf{A}_\textrm{d} \\ * & -\mbf{S} \ebm < 0.
\edis

\subsubsection[Delay-Dependent Condition]{Delay-Dependent Condition}
\label{sec:TimeDelayDependent}
The time-delay system in~\eqref{eq:TD1} is uniformly asymptotically stable under either of the following sufficient conditions.

\begin{enumerate}

\itemcite \cite[pp.~128--129]{Duan2013} There exist $\mbf{X} \in \mathbb{S}^n$ and $\beta \in \mathbb{R}_{>0}$, where $\mbf{X} > 0$ and $\beta < 1$, such that
\bdis
\bbm \mbf{X} \left(\mbf{A} + \mbf{A}_\textrm{d}\right)^\trans + \left(\mbf{A} + \mbf{A}_\textrm{d}\right) \mbf{X} + \bar{d} \mbf{A}_\textrm{d} \mbf{A}_\textrm{d}^\trans & \bar{d} \mbf{X} \mbf{A}^\trans & \bar{d} \mbf{X} \mbf{A}_\textrm{d}^\trans \\ * & -\bar{d} \beta \mbf{1} & \mbf{0} \\ * & * & -\bar{d} (1-\beta) \mbf{1} \ebm < 0.
\edis

\itemcite \cite[pp.~19--21]{Dey2018} There exist $\mbf{X}$,~$\mbf{Q}_1$,~$\mbf{Q}_2 \in \mathbb{S}^n$, where $\mbf{X} > 0$, $\mbf{Q}_1 > 0$, and $\mbf{Q}_2 > 0$, such that
\bdis
\bbm \mbf{X} \left(\mbf{A} + \mbf{A}_\textrm{d}\right)^\trans + \left(\mbf{A} + \mbf{A}_\textrm{d}\right) \mbf{X} + \bar{d} \left(\mbf{Q}_1 + \mbf{Q}_2\right) & \bar{d} \mbf{X} \mbf{A}_\textrm{d} & \bar{d} \mbf{X} \mbf{A}_\textrm{d} \\ * & -\mbf{Q}_1 & \mbf{0} \\ * & * & -\mbf{Q}_2 \ebm < 0.
\edis

\end{enumerate}

\subsection[\texorpdfstring{$\mu$}{Mu}-Analysis]{\texorpdfstring{$\mu$}{Mu}-Analysis~\cite[p.~38--39]{Boyd1994},~\cite{Doyle1991}}
\label{sec:mu_analysis}
\index{structured singular value}
Consider the matrix $\mbf{A} \in \mathbb{C}^{n \times n}$ and the invertible matrix $\mbf{D} \in \mathbb{C}^{n \times n}$.  The inequality $\bar{\sigma}\left(\mbf{D}\mbf{A}\mbf{D}^{-1}\right) < \gamma$\index{singular value!maximum singular value} holds if and only if there exist $\mbf{X} \in \mathbb{C}^{n \times n}$ and $\gamma \in \mathbb{R}_{>0}$, where $\mbf{X} = \mbf{X}^\herm  >0$, satisfying
\beq
\label{eq:MuAnalysis}
\mbf{A}^\trans\mbf{X}\mbf{A} - \gamma^2\mbf{X} < 0.
\eeq
The inequality $\bar{\sigma}\left(\mbf{D}\mbf{A}\mbf{D}^{-1}\right) < \gamma$ holds for $\mbf{D}$ satisfying $\mbf{X} = \mbf{D}^\herm \mbf{D}$ and $\mbf{X}$ satisfying~\eqref{eq:MuAnalysis}.

\subsection[Static Output Feedback Algebraic Loop]{Static Output Feedback Algebraic Loop\cite[p.~1284]{BernsteinMatrixBook},~\cite[pp.~39--40]{CaverlyPhDThesis}}
\index{static output feedback}
\index{algebraic loop}
Consider a continuous-time LTI system, $\mbc{G}: \mathcal{L}_{2e} \to \mathcal{L}_{2e}$, with state-space realization 
\begin{align}
\dot{\mbf{x}} &= \mbf{A} \mbf{x} + \mbf{B}_1 \mbf{w} + \mbf{B}_2 \mbf{u}, \label{eq:Loop1a} \\
\mbf{z} &= \mbf{C}_1 \mbf{x} + \mbf{D}_{11} \mbf{w} + \mbf{D}_{12} \mbf{u}, \label{eq:Loop1b}\\
\mbf{y} &= \mbf{C}_2 \mbf{x} + \mbf{D}_{21} \mbf{w} + \mbf{D}_{22} \mbf{u}, \nonumber
\end{align}
where $\mbf{x}(t) \in \mathbb{R}^{n_x}$ is the system state, $\mbf{z}(t) \in \mathbb{R}^{n_z}$ is the performance signal, $\mbf{y}(t) \in \mathbb{R}^{n_y}$ is the measurement signal, $\mbf{w}(t) \in \mathbb{R}^{n_w}$ is the exogenous signal, $\mbf{u}(t) \in \mathbb{R}^{n_u}$ is the control input signal, and the state-space matrices are real matrices with appropriate dimensions.  Additionally, consider a static output feedback controller of the form $\mbf{u} = \mbf{K} \mbf{y}$, where $\mbf{K} \in \mathbb{R}^{n_u \times n_y}$ and it is assumed that the feedback interconnection is well-posed, that is, $\det(\mbf{1} - \mbf{K} \mbf{D}_{22}) \neq 0$.  The closed-loop system can be described by the following state-space realization.
\begin{align}
\dot{\mbf{x}} &= \left(\mbf{A} + \mbf{B}_2 \mbfbar{K} \mbf{C}_2 \right)\mbf{x} + \left(\mbf{B}_1 + \mbf{B}_2 \mbfbar{K} \mbf{D}_{21} \right) \mbf{w}, \label{eq:Loop2a} \\
\mbf{z} &= \left(\mbf{C}_1 + \mbf{D}_{12} \mbfbar{K} \mbf{C}_2 \right) \mbf{x} + \left(\mbf{D}_{11} + \mbf{D}_{12} \mbfbar{K} \mbf{D}_{21} \right)\mbf{w} , \label{eq:Loop2b}
\end{align}
where $\mbfbar{K} = \left(\mbf{1} - \mbf{K} \mbf{D}_{22} \right)^{-1} \mbf{K}$.

The change of variable $\mbfbar{K} = \left(\mbf{1} - \mbf{K} \mbf{D}_{22} \right)^{-1} \mbf{K}$ allows for the simplification of matrix inequalities involving the closed-loop system.
\begin{proof}
Substituting the expression for $\mbf{y}$ into $\mbf{u} = \mbf{K} \mbf{y}$ gives
\bdis
\mbf{u} = \mbf{K} \left(\mbf{C}_2 \mbf{x} + \mbf{D}_{21} \mbf{w} + \mbf{D}_{22} \mbf{u} \right).
\edis
Bringing the terms with $\mbf{u}$ to the left-hand-side of the equation, left-multiplying by $\left(\mbf{1} - \mbf{K} \mbf{D}_{22} \right)^{-1}$, and defining $\mbfbar{K} = \left(\mbf{1} - \mbf{K} \mbf{D}_{22} \right)^{-1} \mbf{K}$ yields
\begin{align}
\left(\mbf{1} - \mbf{K} \mbf{D}_{22} \right) \mbf{u} &= \mbf{K} \mbf{C}_2 \mbf{x} + \mbf{K}\mbf{D}_{21} \mbf{w} \nonumber \\
\mbf{u} &= \left(\mbf{1} - \mbf{K} \mbf{D}_{22} \right)^{-1} \mbf{K} \mbf{C}_2 \mbf{x} + \left(\mbf{1} - \mbf{K} \mbf{D}_{22} \right)^{-1}\mbf{K}\mbf{D}_{21} \mbf{w}  \nonumber \\
\mbf{u} &= \mbfbar{K} \mbf{C}_2 \mbf{x} +\mbfbar{K}\mbf{D}_{21} \mbf{w}. \label{eq:Loop3}
\end{align}
Substituting~\eqref{eq:Loop3} into~\eqref{eq:Loop1a} and~\eqref{eq:Loop1b} gives~\eqref{eq:Loop2a} and~\eqref{eq:Loop2b}.
\end{proof}

\clearpage
\section{LMIs in Optimal Control}
\label{sec:OptimalControl}

This section presents controller synthesis methods using LMIs for a number of well-known optimal control problems.  The derivation of the LMIs used for controller synthesis is provided in some cases, while longer derivations can be found in the cited references.

\subsection{The Generalized Plant}
\index{generalized plant}
\subsubsection{The Continuous-Time Generalized Plant}

\begin{figure}[H]
\centering
\includegraphics[width=0.5\textwidth]{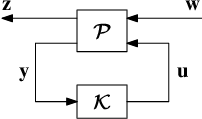}
\caption{Block diagram of the generalized plant $\mbc{P}$ with the controller $\mbc{K}$.} \label{fig:GP}
\end{figure}

Consider the generalized LTI plant $\mbc{P}: \mathcal{L}_{2e} \to \mathcal{L}_{2e}$, shown in Figure~\ref{fig:GP}, with a minimal state-space realization~\cite[pp.~1291--1292]{BernsteinMatrixBook},~\cite[Section~3.8]{Skogestad2005},~\cite[p.~141]{Green2012},~\cite[pp.~14--16]{Francis1987},~\cite[pp.~809--817]{Ogata2010}
\begin{align*}
\dot{\mbf{x}} &= \mbf{A} \mbf{x} + \mbf{B}_1 \mbf{w} + \mbf{B}_2\mbf{u}, \\
\mbf{z} &= \mbf{C}_1 \mbf{x} + \mbf{D}_{11} \mbf{w}+ \mbf{D}_{12} \mbf{u}, \\
\mbf{y} &= \mbf{C}_2 \mbf{x} + \mbf{D}_{21} \mbf{w} + \mbf{D}_{22} \mbf{u}, 
\end{align*}
where $\mbf{x}(t) \in \mathbb{R}^{n_x}$ is the system state, $\mbf{z}(t) \in \mathbb{R}^{n_z}$ is the performance signal, $\mbf{y}(t) \in \mathbb{R}^{n_y}$ is the measurement signal, $\mbf{w}(t) \in \mathbb{R}^{n_w}$ is the exogenous signal, $\mbf{u}(t) \in \mathbb{R}^{n_u}$ is the control input signal, and the state-space matrices are real matrices with appropriate dimensions. 
The generalized LTI plant can also be written in transfer matrix form as
\bdis
\bbm \mbf{z}(s) \\ \mbf{y}(s) \ebm = \mbf{P}(s) \bbm \mbf{w}(s) \\ \mbf{u}(s) \ebm,
\edis
where the transfer matrix $\mbf{P}(s) \in \mathbb{C}^{(n_z + n_y) \times (n_w+n_u)}$ is partitioned as
\bdis
\mbf{P}(s) = \bbm \mbf{P}_{zw}(s) & \mbf{P}_{zu}(s) \\ \mbf{P}_{yw}(s) & \mbf{P}_{yu}(s) \ebm = \bbm \mbf{C}_1 \left(s \mbf{1} - \mbf{A} \right)^{-1} \mbf{B}_1 + \mbf{D}_{11} &  \mbf{C}_1 \left(s \mbf{1} - \mbf{A} \right)^{-1} \mbf{B}_2 + \mbf{D}_{12}  \\   \mbf{C}_2 \left(s \mbf{1} - \mbf{A} \right)^{-1} \mbf{B}_1 + \mbf{D}_{21} &   \mbf{C}_2 \left(s \mbf{1} - \mbf{A} \right)^{-1} \mbf{B}_2 + \mbf{D}_{22} \ebm .
\edis
The generalized plant, also known as the standard control problem in~\cite[pp.~1291--1292]{BernsteinMatrixBook},~\cite[pp.~14--16]{Francis1987},~\cite{DSB580Notes}, is useful, as it is possible to represent a number of LTI systems in this form, as shown in the following example.

\begin{figure}[t]
\centering
\includegraphics[width=0.9\textwidth]{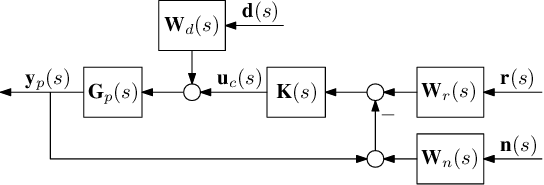}
\caption{Block diagram of the basic servo loop with plant $\mbf{G}_p(s)$, controller $\mbf{K}(s)$, and weighting transfer matrices $\mbf{W}_r(s)$, $\mbf{W}_d(s)$, and $\mbf{W}_n(s)$.} \label{fig:BSL_1}
\end{figure}

\begin{example}[Basic Servo Loop Tracking~{\cite[p.~18]{CaverlyPhDThesis},~\cite[p.~18]{Francis1987},~\cite{DSB580Notes}}]
\label{ex:ServoLoop}

Consider the basic servo loop\index{basic servo loop} shown in Figure~\ref{fig:BSL_1} involving the LTI controller $\mbf{K}(s) \in \mathbb{C}^{n_{y_c} \times n_{u_c}}$ and the plant $\mbf{G}_p(s) \in \mathbb{C}^{n_{y_p} \times n_{u_p}}$, where the weighting transfer matrices are simply chosen as $\mbf{W}_r(s) = \mbf{1}$, $\mbf{W}_d(s) = \mbf{1}$, and $\mbf{W}_n(s) = \mbf{1}$.  The plant $\mbf{G}_p(s)$ has a minimal state-space realization $(\mbf{A}_p,\mbf{B}_p,\mbf{C}_p,\mbf{D}_p)$ and the state $\mbf{x}_p(t)$.  The performance variables are the true tracking error $\mbf{z}_1(t) = \mbf{e}(t) = \mbf{r}(t) - \mbf{y}_p(t)$ and the control effort $\mbf{z}_2(t) = \mbf{u}_c(t)$, where $\mbf{z}^\trans(t) = \bbm \mbf{z}_1^\trans(t) & \mbf{z}_2^\trans(t) \ebm$.  The generalized plant can be formulated with minimal state-space representation
\begin{align*}
\dot{\mbf{x}} &= \mbf{A}_p \mbf{x} + \bbm \mbf{0} & \mbf{B}_p & \mbf{0} \ebm \mbf{w} + \mbf{B}_p\mbf{u}, \\
\mbf{z} &= \bbm -\mbf{C}_p \\ \mbf{0} \ebm \mbf{x} +\bbm \mbf{1} &  -\mbf{D}_p & \mbf{0} \\ \mbf{0} & \mbf{0} & \mbf{0} \ebm \mbf{w} + \bbm - \mbf{D}_p \\ \mbf{1} \ebm\mbf{u}, \\
\mbf{y} &=-\mbf{C}_p  \mbf{x} + \bbm \mbf{1} & -\mbf{D}_p & -\mbf{1}  \ebm  \mbf{w}  - \mbf{D}_p  \mbf{u}, 
\end{align*}
where $\mbf{x}(t) = \mbf{x}_p(t)$, $\mbf{w}^\trans(t) = \bbm \mbf{r}^\trans(t) & \mbf{d}^\trans(t) & \mbf{n}^\trans(t) \ebm$, $\mbf{u}(t) = \mbf{u}_c(t)$, and $\mbf{y}(t) = \mbf{r}(t) - \mbf{y}_p(t) - \mbf{n}(t)$.
\end{example}

\begin{example}[Basic Servo Loop Tracking with Weights~{\cite[Section~9.3.6]{Skogestad2005},~\cite[p.~19]{CaverlyPhDThesis},~\cite[pp.~169--170]{Zhou1998}}]
\label{ex:ServoLoopWeights}

Consider the same basic servo loop\index{basic servo loop} shown in Figure~\ref{fig:BSL_1} involving the LTI controller $\mbf{K}(s) \in \mathbb{C}^{n_{y_c} \times n_{u_c}}$, the plant $\mbf{G}_p(s) \in \mathbb{C}^{n_{y_p} \times n_{u_p}}$, and the weighting transfer matrices $\mbf{W}_r(s) \in \mathbb{C}^{n_r \times n_r}$, $\mbf{W}_d(s) \in \mathbb{C}^{n_d \times n_d}$, and $\mbf{W}_n(s) \in \mathbb{C}^{n_n \times n_n}$.   The plant $\mbf{G}_p(s)$ has a minimal state-space realization $(\mbf{A}_p,\mbf{B}_p,\mbf{C}_p,\mbf{D}_p)$ and the weighting transfer matrices $\mbf{W}_r(s)$, $\mbf{W}_d(s)$, and $\mbf{W}_n(s)$ have minimal state-space realizations $(\mbf{A}_r,\mbf{B}_r,\mbf{C}_r,\mbf{D}_r)$, $(\mbf{A}_d,\mbf{B}_d,\mbf{C}_d,\mbf{D}_d)$, and $(\mbf{A}_n,\mbf{B}_n,\mbf{C}_n,\mbf{D}_n)$, respectively.  The performance variable is defined as the weighted true tracking error $\mbf{z}_1(s) = \mbf{W}_e(s) \mbf{e}(s) = \mbf{W}_e(s) \left(\mbf{W}_r(s)\mbf{r}(s) - \mbf{y}_p(s)\right)$ and the weighted control effort $\mbf{z}_2(s) = \mbf{W}_u(s)\mbf{u}_c(s)$, where $\mbf{z}^\trans(s) = \bbm \mbf{z}_1^\trans(s) & \mbf{z}_2^\trans(s) \ebm$ and $\mbf{W}_e(s) \in \mathbb{C}^{n_e \times n_e}$, $\mbf{W}_u(s) \in \mathbb{C}^{n_u \times n_u}$ are weighting transfer matrices with minimal state-space realizations $(\mbf{A}_e,\mbf{B}_e,\mbf{C}_e,\mbf{D}_e)$ and $(\mbf{A}_u,\mbf{B}_u,\mbf{C}_u,\mbf{D}_u)$, respectively.  The generalized plant can be formulated with minimal state-space representation
\begin{align*}
\dot{\mbf{x}} &= \bbm \mbf{A}_p & \mbf{0} & \mbf{B}_p\mbf{C}_d & \mbf{0} & \mbf{0} & \mbf{0} \\ \mbf{0} & \mbf{A}_r & \mbf{0} & \mbf{0} & \mbf{0} & \mbf{0} \\ \mbf{0} & \mbf{0} & \mbf{A}_d & \mbf{0} & \mbf{0} & \mbf{0} \\ \mbf{0} & \mbf{0} & \mbf{0} & \mbf{A}_n & \mbf{0} & \mbf{0} \\ -\mbf{B}_e \mbf{C}_p & \mbf{B}_e \mbf{C}_r & - \mbf{B}_e \mbf{D}_p\mbf{C}_d & \mbf{0} & \mbf{A}_e & \mbf{0} \\ \mbf{0} & \mbf{0} & \mbf{0} & \mbf{0} & \mbf{0} & \mbf{A}_u \ebm \mbf{x}  + \bbm \mbf{0} & \mbf{B}_p\mbf{D}_d & \mbf{0} \\ \mbf{B}_r & \mbf{0} & \mbf{0} \\ \mbf{0} & \mbf{B}_d & \mbf{0} \\ \mbf{0} & \mbf{0} & \mbf{B}_n \\ \mbf{B}_e \mbf{D}_r & -\mbf{B}_e \mbf{D}_p\mbf{D}_d & \mbf{0} \\ \mbf{0} & \mbf{0} & \mbf{0} \ebm \mbf{w} + \bbm \mbf{B}_p \\ \mbf{0} \\ \mbf{0} \\ \mbf{0} \\ -\mbf{B}_e \mbf{D}_p \\ \mbf{B}_u \ebm \mbf{u}, \\
\mbf{z} &= \bbm -\mbf{D}_e \mbf{C}_p & \mbf{D}_e \mbf{C}_r & - \mbf{D}_e \mbf{D}_p \mbf{C}_d & \mbf{0} & \mbf{C}_e & \mbf{0} \\ \mbf{0} & \mbf{0} & \mbf{0} & \mbf{0} & \mbf{0} & \mbf{C}_u \ebm  \mbf{x} +\bbm \mbf{D}_e \mbf{D}_r &  -\mbf{D}_e\mbf{D}_p\mbf{D}_d & \mbf{0} \\ \mbf{0} & \mbf{0} & \mbf{0} \ebm \mbf{w} + \bbm -\mbf{D}_e\mbf{D}_p \\ \mbf{D}_u \ebm \mbf{u}, \\
\mbf{y} &= \bbm -\mbf{C}_p & \mbf{C}_r & - \mbf{D}_p\mbf{C}_d & -\mbf{C}_n & \mbf{0} & \mbf{0} \ebm \mbf{x} + \bbm \mbf{D}_r & -\mbf{D}_p\mbf{D}_d & -\mbf{D}_n \ebm  \mbf{w} - \mbf{D}_p \mbf{u}, 
\end{align*}
where $\mbf{x}^\trans(t) = \bbm \mbf{x}_p^\trans(t) & \mbf{x}_r^\trans(t) & \mbf{x}_d^\trans(t) & \mbf{x}_n^\trans(t) & \mbf{x}_e^\trans(t) & \mbf{x}_u^\trans(t) \ebm$, $\mbf{w}^\trans(t) = \bbm \mbf{r}^\trans(t) & \mbf{d}^\trans(t) & \mbf{n}^\trans(t) \ebm$, $\mbf{u}(t) = \mbf{u}_c(t)$, $\mbf{y}(s) = \mbf{W}_r(s)\mbf{r}(s) - \mbf{y}_p(s) - \mbf{W}_n(s)\mbf{n}(s)$, and $\mbf{x}_r(t)$, $\mbf{x}_d(t)$, $\mbf{x}_n(t)$, $\mbf{x}_e(t)$, and $\mbf{x}_u(t)$ are the states associated with the state-space realizations of the weighting transfer matrices $\mbf{W}_r(s)$, $\mbf{W}_d(s)$, $\mbf{W}_n(s)$, $\mbf{W}_e(s)$, and $\mbf{W}_u(s)$, respectively.
\end{example}
\subsubsection{The Discrete-Time Generalized Plant}

The discrete-time generalized LTI plant $\mbc{P}: \ell_{2e} \to \ell_{2e}$, shown in Figure~\ref{fig:GP}, is described by the state-space realization
\begin{align*}
\mbf{x}_{k+1} &= \mbf{A}_\mathrm{d} \mbf{x}_k + \mbf{B}_{\mathrm{d}1} \mbf{w}_k + \mbf{B}_{\mathrm{d}2}\mbf{u}_k, \\
\mbf{z}_k &= \mbf{C}_{\mathrm{d}1} \mbf{x}_k + \mbf{D}_{\mathrm{d}11} \mbf{w}_k+ \mbf{D}_{\mathrm{d}12} \mbf{u}_k, \\
\mbf{y}_k &= \mbf{C}_{\mathrm{d}2} \mbf{x}_k + \mbf{D}_{\mathrm{d}21} \mbf{w}_k + \mbf{D}_{\mathrm{d}22} \mbf{u}_k, 
\end{align*}
where $\mbf{x}_k \in \mathbb{R}^{n_x}$ is the system state at time step $k$, $\mbf{z}_k \in \mathbb{R}^{n_z}$ is the performance signal at time step $k$, $\mbf{y}_k \in \mathbb{R}^{n_y}$ is the measurement signal at time step $k$, $\mbf{w}_k \in \mathbb{R}^{n_w}$ is the exogenous signal at time step $k$, $\mbf{u}_k \in \mathbb{R}^{n_u}$ is the control input signal at time step $k$, and the state-space matrices have appropriate dimensions. 
The generalized LTI plant can also be written in discrete-time transfer matrix form as
\bdis
\bbm \mbf{z}(z) \\ \mbf{y}(z) \ebm = \mbf{P}(z) \bbm \mbf{w}(z) \\ \mbf{u}(z) \ebm,
\edis
where the transfer matrix $\mbf{P}(z) \in \mathbb{C}^{(n_z + n_y) \times (n_w+n_u)}$ is partitioned as
\bdis
\mbf{P}(z) = \bbm \mbf{P}_{zw}(z) & \mbf{P}_{zu}(z) \\ \mbf{P}_{yw}(z) & \mbf{P}_{yu}(z) \ebm = \bbm \mbf{C}_{\mathrm{d}1} \left(z \mbf{1} - \mbf{A}_\mathrm{d} \right)^{-1} \mbf{B}_{\mathrm{d}1} + \mbf{D}_{\mathrm{d}11} &  \mbf{C}_{\mathrm{d}1} \left(z \mbf{1} - \mbf{A}_\mathrm{d} \right)^{-1} \mbf{B}_{\mathrm{d}2} + \mbf{D}_{\mathrm{d}12}  \\   \mbf{C}_{\mathrm{d}2} \left(z \mbf{1} - \mbf{A}_\mathrm{d} \right)^{-1} \mbf{B}_{\mathrm{d}1} + \mbf{D}_{\mathrm{d}21} &   \mbf{C}_{\mathrm{d}2} \left(z \mbf{1} - \mbf{A}_\mathrm{d} \right)^{-1} \mbf{B}_{\mathrm{d}2} + \mbf{D}_{\mathrm{d}22} \ebm .
\edis


\subsection{\texorpdfstring{$\mathcal{H}_2$}{H2}-Optimal Control}

The goal of $\mathcal{H}_2$-optimal control is to design a controller that minimizes the $\mathcal{H}_2$ norm of the closed-loop transfer matrix from $\mbf{w}$ to $\mbf{z}$.

\subsubsection[\texorpdfstring{$\mathcal{H}_2$}{H2}-Optimal Full-State Feedback Control]{\texorpdfstring{$\mathcal{H}_2$}{H2}-Optimal Full-State Feedback Control~\cite[pp.~257--258]{Duan2013}}
\index{full-state feedback!$\mathcal{H}_2$-optimal}
Consider the continuous-time generalized LTI plant $\mbc{P}$ with state-space realization
\begin{align}
\dot{\mbf{x}} &= \mbf{A} \mbf{x} + \mbf{B}_1 \mbf{w} + \mbf{B}_2\mbf{u}, \label{eq:LQR1a} \\
\mbf{z} &= \mbf{C}_1 \mbf{x} + \mbf{D}_{12} \mbf{u}, \label{eq:LQR1b} \\
\mbf{y} &= \mbf{x} , \nonumber
\end{align}
where it is assumed that ($\mbf{A}$,$\mbf{B}_2$) is stabilizable.  A full-state feedback controller $\mbc{K} = \mbf{K} \in \mathbb{R}^{n_u \times n_x}$ (i.e., $\mbf{u} = \mbf{K} \mbf{x}$) is to be designed to minimize the $\mathcal{H}_2$ norm of the closed loop transfer matrix from the exogenous input $\mbf{w}$ to the performance output $\mbf{z}$.  Substituting the full-state feedback controller into~\eqref{eq:LQR1a} and~\eqref{eq:LQR1b} yields
 \begin{align*}
\dot{\mbf{x}} &= \left(\mbf{A}+\mbf{B}_2\mbf{K}\right) \mbf{x} + \mbf{B}_1 \mbf{w}, \\
\mbf{z} &= \left(\mbf{C}_1 + \mbf{D}_{12}\mbf{K}\right) \mbf{x}, 
\end{align*}
and a closed-loop transfer matrix
\bdis
\mbf{T}(s) = \left(\mbf{C}_1 + \mbf{D}_{12}\mbf{K}\right)\left(s\mbf{1} - \left(\mbf{A}+\mbf{B}_2\mbf{K}\right)\right)^{-1}\mbf{B}_1.
\edis
Minimizing the $\mathcal{H}_2$ norm of the transfer matrix $\mbf{T}(s)$ is equivalent to minimizing $\mathcal{J}(\mu) = \mu^2$ subject to 
\begin{align}
\bbm \left(\mbf{A}+\mbf{B}_2\mbf{K}\right)\mbf{P} + \mbf{P}\left(\mbf{A}+\mbf{B}_2\mbf{K}\right)^\trans & \mbf{P}\left(\mbf{C}_1 + \mbf{D}_{12}\mbf{K}\right)^\trans \\ * & -\mbf{1} \ebm &< 0, \label{eq:LQR3a}\\
\bbm \mbf{Z} & \mbf{B}_1^\trans \\ * & \mbf{P} \ebm &> 0, \label{eq:LQR3b} \\
\trace (\mbf{Z}) &< \mu^2, \label{eq:LQR3c}
\end{align}
where $\mbf{P} \in \mathbb{S}^{n_x}$, $\mbf{Z} \in \mathbb{S}^{n_w}$, $\mu \in \mathbb{R}_{>0}$, $\mbf{P} > 0$, and $\mbf{Z} > 0$.  A change of variables is performed with $\mbf{F} = \mbf{K}\mbf{P}$ and $\nu = \mu^2$, which transforms~\eqref{eq:LQR3a} and~\eqref{eq:LQR3c} into LMIs in the variables $\mbf{P}$, $\mbf{F}$, $\mbf{Z}$, and $\nu$ given by
\begin{align}
\bbm \mbf{A}\mbf{P} + \mbf{P}\mbf{A}^\trans+\mbf{B}_2\mbf{F}+\mbf{F}^\trans\mbf{B}_2^\trans & \mbf{P}\mbf{C}_1^\trans + \mbf{F}^\trans\mbf{D}_{12}^\trans \\ * & -\mbf{1} \ebm &< 0, \label{eq:LQR4a}\\
\trace (\mbf{Z}) &< \nu. \label{eq:LQR4c}
\end{align}
\begin{SynthMeth}
The $\mathcal{H}_2$-optimal full-state feedback controller is synthesized by solving for $\mbf{P} \in \mathbb{S}^{n_x}$, $\mbf{Z} \in \mathbb{S}^{n_w}$, $\mbf{F} \in \mathbb{R}^{n_u \times n_x}$, and $\nu \in \mathbb{R}_{>0}$ that minimize $\mathcal{J}(\nu) = \nu$ subject to $\mbf{P} >0 $, $\mbf{Z} > 0$,~\eqref{eq:LQR3b},~\eqref{eq:LQR4a}, and~\eqref{eq:LQR4c}.  The $\mathcal{H}_2$-optimal full-state feedback gain is recovered by $\mbf{K} = \mbf{F}\mbf{P}^{-1}$ and the $\mathcal{H}_2$ norm of $\mbf{T}(s)$ is $\mu = \sqrt{\nu}$.
\end{SynthMeth}

\subsubsection{Discrete-Time \texorpdfstring{$\mathcal{H}_2$}{H2}-Optimal Full-State Feedback Control}
\index{full-state feedback!discrete-time $\mathcal{H}_2$-optimal}
Consider the discrete-time generalized LTI plant $\mbc{P}$ with state-space realization
\begin{align*}
\mbf{x}_{k+1} &= \mbf{A}_\mathrm{d} \mbf{x}_k + \mbf{B}_{\mathrm{d}1} \mbf{w}_k + \mbf{B}_{\mathrm{d}2}\mbf{u}_k,  \\
\mbf{z}_k &= \mbf{C}_{\mathrm{d}1} \mbf{x}_k + \mbf{D}_{\mathrm{d}12} \mbf{u}_k,  \\
\mbf{y}_k &= \mbf{x}_k , \nonumber
\end{align*}
where it is assumed that ($\mbf{A}_\mathrm{d}$,$\mbf{B}_{\mathrm{d}2}$) is stabilizable.  A full-state feedback controller $\mbc{K} = \mbf{K}_\mathrm{d} \in \mathbb{R}^{n_u \times n_x}$ (i.e., $\mbf{u}_k = \mbf{K}_\mathrm{d}\mbf{x}_k$) is to be designed to minimize the $\mathcal{H}_2$ norm of the closed loop transfer matrix from the exogenous input $\mbf{w}_k$ to the performance output $\mbf{z}_k$, given by
\bdis
\mbf{T}(z) = \left(\mbf{C}_{\mathrm{d}1} + \mbf{D}_{\mathrm{d}12}\mbf{K}_\mathrm{d}\right)\left(z\mbf{1} - \left(\mbf{A}_\mathrm{d}+\mbf{B}_{\mathrm{d}2}\mbf{K}_\mathrm{d}\right)\right)^{-1}\mbf{B}_{\mathrm{d}1}.
\edis

\begin{SynthMeth}
The discrete-time $\mathcal{H}_2$-optimal full-state feedback controller is synthesized by solving for $\mbf{P} \in \mathbb{S}^{n_x} $, $\mbf{Z} \in \mathbb{S}^{n_z}$, $\mbf{F}_\mathrm{d} \in \mathbb{R}^{n_u \times n_x}$, and $\nu \in \mathbb{R}_{>0}$ that minimize $\mathcal{J}(\nu) = \nu$ subject to $\mbf{P} >0 $, $\mbf{Z} > 0$,
\begin{align*}
\bbm \mbf{P} & \mbf{A}_\mathrm{d}\mbf{P}+ \mbf{B}_{\mathrm{d}2}\mbf{F}_\mathrm{d} & \mbf{B}_{\text{d}1} \\ * & \mbf{P} & \mbf{0} \\ * & * & \mbf{1} \ebm &> 0,\\
\bbm \mbf{Z} & \mbf{C}_{\text{d}1}\mbf{P} + \mbf{D}_{\mathrm{d}12}\mbf{F}_\mathrm{d} \\ * & \mbf{P} \ebm &> 0. \\
\trace (\mbf{Z}) &< \nu. 
\end{align*}
The $\mathcal{H}_2$-optimal full-state feedback gain is recovered by $\mbf{K}_\mathrm{d} = \mbf{F}_\mathrm{d}\mbf{P}^{-1}$ and the $\mathcal{H}_2$ norm of $\mbf{T}(z)$ is $\mu = \sqrt{\nu}$.
\end{SynthMeth}

\subsubsection[\texorpdfstring{$\mathcal{H}_2$}{H2}-Optimal Dynamic Output Feedback Control]{\texorpdfstring{$\mathcal{H}_2$}{H2}-Optimal Dynamic Output Feedback Control~\cite{Scherer1997,PeetOptNotes22}}
\index{dynamic output feedback!$\mathcal{H}_2$-optimal}
\label{sec:H2DynOutputCTSynth}
Consider the continuous-time generalized LTI plant $\mbc{P}$ with minimal state-space realization
\begin{align*}
\dot{\mbf{x}} &= \mbf{A} \mbf{x} + \mbf{B}_1 \mbf{w} + \mbf{B}_2\mbf{u}, \\ 
\mbf{z} &= \mbf{C}_1 \mbf{x} + \mbf{D}_{11} \mbf{w}+ \mbf{D}_{12} \mbf{u}, \\ 
\mbf{y} &= \mbf{C}_2 \mbf{x} + \mbf{D}_{21} \mbf{w} + \mbf{D}_{22} \mbf{u}.  
\end{align*}
A continuous-time dynamic output feedback LTI controller with state-space realization $(\mbf{A}_c,\mbf{B}_c,\mbf{C}_c,\mbf{D}_c)$ is to be designed to minimize the $\mathcal{H}_2$ norm of the closed-loop system transfer matrix from $\mbf{w}$ to $\mbf{z}$, given by
\bdis
\mbf{T}(s) = \mbf{C}_\du{\mathrm{CL}}\left(s\mbf{1} - \mbf{A}_\du{\mathrm{CL}}\right)^{-1}\mbf{B}_\du{\mathrm{CL}} + \mbf{D}_\du{\mathrm{CL}},
\edis
where
\begin{align*}
\mbf{A}_\du{\mathrm{CL}} &= \bbm \mbf{A} + \mbf{B}_2 \mbf{D}_c\mbftilde{D}^{-1} \mbf{C}_2 & \mbf{B}_2 \left(\mbf{1} + \mbf{D}_c\mbftilde{D}^{-1}\mbf{D}_{22} \right)\mbf{C}_c \\ \mbf{B}_c  \mbftilde{D}^{-1}\mbf{C}_2 & \mbf{A}_c + \mbf{B}_c \mbftilde{D}^{-1}\mbf{D}_{22}\mbf{C}_c \ebm, \\
\mbf{B}_\du{\mathrm{CL}} &= \bbm  \mbf{B}_1 + \mbf{B}_2\mbf{D}_c\mbftilde{D}^{-1}\mbf{D}_{21}  \\ \mbf{B}_c \mbftilde{D}^{-1}\mbf{D}_{21} \ebm, \\
\mbf{C}_\du{\mathrm{CL}} &= \bbm \mbf{C}_1 + \mbf{D}_{12}\mbf{D}_c\mbftilde{D}^{-1} \mbf{C}_2 & \mbf{D}_{12} \left(\mbf{1} + \mbf{D}_c\mbftilde{D}^{-1}\mbf{D}_{22} \right)\mbf{C}_c \ebm, \\
\mbf{D}_\du{\mathrm{CL}} &= \mbf{D}_{11} + \mbf{D}_{12} \mbf{D}_c  \mbftilde{D}^{-1}\mbf{D}_{21}, 
\end{align*}
and $\mbftilde{D} = \mbf{1} - \mbf{D}_{22}\mbf{D}_c$.
\begin{SynthMeth}
Solve for $\mbf{A}_n \in \mathbb{R}^{n_x \times n_x}$, $\mbf{B}_n \in \mathbb{R}^{n_x \times n_y}$, $\mbf{C}_n \in \mathbb{R}^{n_u \times n_x}$, $\mbf{D}_n \in \mathbb{R}^{n_u \times n_y}$, $\mbf{X}_1$,~$\mbf{Y}_1 \in \mathbb{S}^{n_x}$, $\mbf{Z} \in \mathbb{S}^{n_z}$, and $\nu \in \mathbb{R}_{>0}$ that minimize $\mathcal{J}(\nu) = \nu$ subject to $\mbf{X}_1 > 0$, $\mbf{Y}_1 > 0$, $\mbf{Z} > 0$,
\begin{align}
\bbm \mbf{A} \mbf{Y}_1 + \mbf{Y}_1\mbf{A}^\trans + \mbf{B}_2\mbf{C}_n + \mbf{C}_n^\trans\mbf{B}_2^\trans & \mbf{A} + \mbf{A}_n^\trans + \mbf{B}_2\mbf{D}_n\mbf{C}_2 & \mbf{B}_1 + \mbf{B}_2\mbf{D}_n\mbf{D}_{21} \\ * & \mbf{X}_1\mbf{A} + \mbf{A}^\trans\mbf{X}_1 + \mbf{B}_n \mbf{C}_2 + \mbf{C}_2^\trans \mbf{B}_n^\trans & \mbf{X}_1\mbf{B}_1 + \mbf{B}_n\mbf{D}_{21} \\ * & * & -\mbf{1} \ebm &< 0, \nonumber \\
\bbm \mbf{Y}_1 & \mbf{1} & \mbf{X}_1\mbf{C}_1^\trans + \mbf{C}_n^\trans \mbf{D}_{12}^\trans \\ * & \mbf{Y}_1 & \mbf{C}_1^\trans + \mbf{C}_2^\trans \mbf{D}_n^\trans \mbf{D}_{12}^\trans \\ * & * & \mbf{Z} \ebm &>0, \nonumber\\
\mbf{D}_{11} + \mbf{D}_{12}\mbf{D}_n\mbf{D}_{21} &= \mbf{0}, \label{eq:H2_D}\\
\bbm \mbf{X}_1 & \mbf{1} \\ * & \mbf{Y}_1 \ebm &> 0, \nonumber\\
\trace (\mbf{Z}) &< \nu. \nonumber
\end{align}
The controller is recovered by
\begin{align*}
\mbf{A}_c &= \mbf{A}_\du{K} - \mbf{B}_c\left(\mbf{1} - \mbf{D}_{22}\mbf{D}_c\right)^{-1}\mbf{D}_{22}\mbf{C}_c, \\
\mbf{B}_c &= \mbf{B}_\du{K}\left(\mbf{1} - \mbf{D}_{22}\mbf{D}_{c}\right), \\
\mbf{C}_c &= \left(\mbf{1} - \mbf{D}_c\mbf{D}_{22}\right)\mbf{C}_\du{K}, \\
\mbf{D}_c &= \left(\mbf{1} + \mbf{D}_\du{K}\mbf{D}_{22}\right)^{-1}\mbf{D}_\du{K},
\end{align*}
where
\bdis
\bbm \mbf{A}_\du{K} & \mbf{B}_\du{K} \\ \mbf{C}_\du{K} & \mbf{D}_\du{K} \ebm = \bbm \mbf{X}_2 & \mbf{X}_1\mbf{B}_2 \\ \mbf{0} & \mbf{1} \ebm^{-1} \left( \bbm \mbf{A}_n & \mbf{B}_n \\ \mbf{C}_n & \mbf{D}_n \ebm - \bbm \mbf{X}_1\mbf{A}\mbf{Y}_1 & \mbf{0} \\ \mbf{0} & \mbf{0}\ebm \right)\bbm \mbf{Y}_2^\trans & \mbf{0} \\ \mbf{C}_2 \mbf{Y}_1 & \mbf{1} \ebm^{-1},
\edis
and the matrices $\mbf{X}_2$ and $\mbf{Y}_2$ satisfy $\mbf{X}_2\mbf{Y}_2^\trans = \mbf{1} - \mbf{X}_1\mbf{Y}_1$.  If $\mbf{D}_{22} = \mbf{0}$, then $\mbf{A}_c = \mbf{A}_\du{K}$, $\mbf{B}_c = \mbf{B}_\du{K}$, $\mbf{C}_c = \mbf{C}_\du{K}$, and $\mbf{D}_c = \mbf{D}_\du{K}$.

Given $\mbf{X}_1$ and $\mbf{Y}_1$, the matrices $\mbf{X}_2$ and $\mbf{Y}_2$ can be found using a matrix decomposition, such as a LU decomposition or a Cholesky decomposition.

If $\mbf{D}_{11} = \mbf{0}$, $\mbf{D}_{12} \neq \mbf{0}$, and $\mbf{D}_{21} \neq \mbf{0}$, then it is often simplest to choose $\mbf{D}_n = \mbf{0}$ in order to satisfy the equality constraint of~\eqref{eq:H2_D}.
\end{SynthMeth}

\subsubsection[Discrete-Time \texorpdfstring{$\mathcal{H}_2$}{H2}-Optimal Dynamic Output Feedback Control]{Discrete-Time \texorpdfstring{$\mathcal{H}_2$}{H2}-Optimal Dynamic Output Feedback Control}
\label{sec:DT_H2_DynSynth}
\index{dynamic output feedback!discrete-time $\mathcal{H}_2$-optimal}

Consider the discrete-time generalized LTI plant $\mbc{P}$ with state-space realization
\begin{align*}
\mbf{x}_{k+1} &= \mbf{A}_\mathrm{d} \mbf{x}_k + \mbf{B}_{\mathrm{d}1} \mbf{w}_k + \mbf{B}_{\mathrm{d}2}\mbf{u}_k,  \\
\mbf{z}_k &= \mbf{C}_{\mathrm{d}1} \mbf{x}_k + \mbf{D}_{\mathrm{d}11}\mbf{w}_k + \mbf{D}_{\mathrm{d}12} \mbf{u}_k,  \\
\mbf{y}_k &= \mbf{C}_{\mathrm{d}2} \mbf{x}_k + \mbf{D}_{\mathrm{d}21}\mbf{w}_k + \mbf{D}_{\mathrm{d}22} \mbf{u}_k, \nonumber
\end{align*}
A discrete-time dynamic output feedback LTI controller with state-space realization $(\mbf{A}_{\mathrm{d}c},\mbf{B}_{\mathrm{d}c},\mbf{C}_{\mathrm{d}c},\mbf{D}_{\mathrm{d}c})$ is to be designed to minimize the $\mathcal{H}_2$ norm of the closed-loop system transfer matrix from $\mbf{w}_k$ to $\mbf{z}_k$, given by
\bdis
\mbf{T}(z) = \mbf{C}_{\mathrm{d}_\mathrm{CL}}\left(z\mbf{1} - \mbf{A}_{\mathrm{d}_\mathrm{CL}}\right)^{-1}\mbf{B}_{\mathrm{d}_\mathrm{CL}} + \mbf{D}_{\mathrm{d}_\mathrm{CL}},
\edis
where
\begin{align*}
\mbf{A}_{\mathrm{d}_\mathrm{CL}} &= \bbm \mbf{A}_\mathrm{d} + \mbf{B}_{\mathrm{d}2} \mbf{D}_{\mathrm{d}c}\mbftilde{D}_\mathrm{d}^{-1} \mbf{C}_{\mathrm{d}2} & \mbf{B}_{\mathrm{d}2} \left(\mbf{1} + \mbf{D}_{\mathrm{d}c}\mbftilde{D}_\mathrm{d}^{-1}\mbf{D}_{\mathrm{d}22} \right)\mbf{C}_{\mathrm{d}c} \\ \mbf{B}_{\mathrm{d}c}  \mbftilde{D}_\mathrm{d}^{-1}\mbf{C}_{\mathrm{d}2} & \mbf{A}_{\mathrm{d}c} + \mbf{B}_{\mathrm{d}c} \mbftilde{D}_\mathrm{d}^{-1}\mbf{D}_{\mathrm{d}22}\mbf{C}_{\mathrm{d}c} \ebm, \\
\mbf{B}_{\mathrm{d}_\mathrm{CL}} &= \bbm  \mbf{B}_{\mathrm{d}1} + \mbf{B}_{\mathrm{d}2}\mbf{D}_{\mathrm{d}c}\mbftilde{D}_\mathrm{d}^{-1}\mbf{D}_{\mathrm{d}21}  \\ \mbf{B}_{\mathrm{d}c} \mbftilde{D}_\mathrm{d}^{-1}\mbf{D}_{\mathrm{d}21} \ebm, \\
\mbf{C}_{\mathrm{d}_\mathrm{CL}} &= \bbm \mbf{C}_{\mathrm{d}1} + \mbf{D}_{\mathrm{d}12}\mbf{D}_{\mathrm{d}c}\mbftilde{D}_\mathrm{d}^{-1} \mbf{C}_{\mathrm{d}2} & \mbf{D}_{\mathrm{d}12} \left(\mbf{1} + \mbf{D}_{\mathrm{d}c}\mbftilde{D}_\mathrm{d}^{-1}\mbf{D}_{\mathrm{d}22} \right)\mbf{C}_{\mathrm{d}c} \ebm, \\
\mbf{D}_{\mathrm{d}_\mathrm{CL}} &= \mbf{D}_{\mathrm{d}11} + \mbf{D}_{\mathrm{d}12} \mbf{D}_{\mathrm{d}c}  \mbftilde{D}_\mathrm{d}^{-1}\mbf{D}_{\mathrm{d}21}, 
\end{align*}
and $\mbftilde{D}_\mathrm{d} = \mbf{1} - \mbf{D}_{\mathrm{d}22}\mbf{D}_{\mathrm{d}c}$.

\begin{SynthMeth}
\cite{DeOliveira2002} Solve for $\mbf{A}_{\mathrm{d}n} \in \mathbb{R}^{n_x \times n_x}$, $\mbf{B}_{\mathrm{d}n} \in \mathbb{R}^{n_x \times n_y}$, $\mbf{C}_{\mathrm{d}n} \in \mathbb{R}^{n_u \times n_x}$, $\mbf{D}_{\mathrm{d}n} \in \mathbb{R}^{n_u \times n_y}$, $\mbf{X}_1$,~$\mbf{Y}_1 \in \mathbb{S}^{n_x}$, $\mbf{Z} \in \mathbb{S}^{n_z}$, $\mbf{G}$,~$\mbf{H}$,~$\mbf{J}$,~$\mbf{S} \in \mathbb{R}^{n_x \times n_x}$, and $\nu \in \mathbb{R}_{>0}$ that minimize $\mathcal{J}(\nu) = \nu$ subject to $\mbf{X}_1 > 0$, $\mbf{Y}_1 > 0$, $\mbf{Z} > 0$,
\begin{align}
\bbm \mbf{X}_1 & \mbf{J}^\trans & \mbf{H} \mbf{A}_\mathrm{d} + \mbf{B}_{\mathrm{d}n}\mbf{C}_{\mathrm{d}2} & \mbf{A}_{\mathrm{d}n} & \mbf{H} \mbf{B}_{\mathrm{d}1} + \mbf{B}_{\mathrm{d}n} \mbf{D}_{\mathrm{d}21}   \\ *  & \mbf{Y}_1 & \mbf{A}_\mathrm{d} + \mbf{B}_{\mathrm{d}2} \mbf{D}_{\mathrm{d}n} \mbf{C}_{\mathrm{d}2} & \mbf{A}_\mathrm{d} \mbf{G} + \mbf{B}_{\mathrm{d}2} \mbf{C}_{\mathrm{d}n} & \mbf{B}_{\mathrm{d}1} + \mbf{B}_{\mathrm{d}2} \mbf{D}_{\mathrm{d}n} \mbf{D}_{\mathrm{d}21}  \\ * & * & \mbf{H} + \mbf{H}^\trans - \mbf{X}_1 & \mbf{1} + \mbf{S} - \mbf{J}^\trans & \mbf{0} \\ * & * & * & \mbf{G} + \mbf{G}^\trans -\mbf{Y}_1 & \mbf{0}  \\ * & * & * & * &  \mbf{1} \ebm &> 0, \label{eq:H2_DT_Output1a} \\
\bbm \mbf{Z} & \mbf{C}_{\mathrm{d}1} + \mbf{D}_{\mathrm{d}12} \mbf{D}_{\mathrm{d}n}\mbf{C}_{\mathrm{d}2} & \mbf{C}_{\mathrm{d}1}\mbf{G} + \mbf{D}_{\mathrm{d}12}\mbf{C}_{\mathrm{d}n} \\ * & \mbf{H} + \mbf{H}^\trans - \mbf{X}_1 & \mbf{1} + \mbf{S} - \mbf{J}^\trans \\ * & * & \mbf{G} + \mbf{G}^\trans - \mbf{Y}_1 \ebm &>0, \label{eq:H2_DT_Output1b} \\
\mbf{D}_{\mathrm{d}11} + \mbf{D}_{\mathrm{d}12} \mbf{D}_{\mathrm{d}n}  \mbf{D}_{\mathrm{d}21} &= \mbf{0}, \label{eq:H2_DT_Output1c}\\
\trace (\mbf{Z}) &< \nu. \nonumber
\end{align}
The controller is recovered by
\begin{align*}
\mbf{A}_{\mathrm{d}c} &= \mbf{A}_{\mathrm{d}_K} - \mbf{B}_{\mathrm{d}c}\left(\mbf{1} - \mbf{D}_{\mathrm{d}22}\mbf{D}_{\mathrm{d}c}\right)^{-1}\mbf{D}_{\mathrm{d}22}\mbf{C}_{\mathrm{d}c}, \\
\mbf{B}_{\mathrm{d}c} &= \mbf{B}_{\mathrm{d}_K}\left(\mbf{1} - \mbf{D}_{\mathrm{d}22}\mbf{D}_{\mathrm{d}c}\right), \\
\mbf{C}_{\mathrm{d}c} &= \left(\mbf{1} - \mbf{D}_{\mathrm{d}c}\mbf{D}_{\mathrm{d}22}\right)\mbf{C}_{\mathrm{d}_K}, \\
\mbf{D}_{\mathrm{d}c} &= \left(\mbf{1} + \mbf{D}_{\mathrm{d}_K}\mbf{D}_{\mathrm{d}22}\right)^{-1}\mbf{D}_{\mathrm{d}_K},
\end{align*}
where
\bdis
\bbm \mbf{A}_{\mathrm{d}_K} & \mbf{B}_{\mathrm{d}_K} \\ \mbf{C}_{\mathrm{d}_K} & \mbf{D}_{\mathrm{d}_K} \ebm = \bbm \mbf{Y}_2^{-\trans} & \mbf{Y}_2^{-\trans}\mbf{H}\mbf{B}_{\mathrm{d}2} \\ \mbf{0} & \mbf{1} \ebm \left( \bbm \mbf{A}_{\mathrm{d}n}& \mbf{B}_{\mathrm{d}n} \\ \mbf{C}_{\mathrm{d}n} & \mbf{D}_{\mathrm{d}n} \ebm - \bbm \mbf{H}\mbf{A}_\mathrm{d}\mbf{G} & \mbf{0} \\ \mbf{0} & \mbf{0}\ebm \right)\bbm \mbf{X}_2^{-1} & \mbf{0} \\ -\mbf{C}_{\mathrm{d}2} \mbf{G} \mbf{X}_2^{-1} & \mbf{1} \ebm,
\edis
and the matrices $\mbf{X}_2$ and $\mbf{Y}_2$ satisfy $\mbf{X}_2\mbf{Y}_2^\trans = \mbf{1} - \mbf{H}\mbf{G}$.  If $\mbf{D}_{\mathrm{d}22} = \mbf{0}$, then $\mbf{A}_{\mathrm{d}c} = \mbf{A}_{\mathrm{d}_K}$, $\mbf{B}_{\mathrm{d}c} = \mbf{B}_{\mathrm{d}_K}$, $\mbf{C}_{\mathrm{d}c} = \mbf{C}_{\mathrm{d}_K}$, and $\mbf{D}_{\mathrm{d}c} = \mbf{D}_{\mathrm{d}_K}$.

Given $\mbf{G}$ and $\mbf{H}$, the matrices $\mbf{X}_2$ and $\mbf{Y}_2$ can be found using a matrix decomposition, such as a LU decomposition or a Cholesky decomposition.

If $\mbf{D}_{\mathrm{d}11} = \mbf{0}$, $\mbf{D}_{\mathrm{d}12} \neq \mbf{0}$, and $\mbf{D}_{\mathrm{d}21} \neq \mbf{0}$, then it is often simplest to choose $\mbf{D}_{\mathrm{d}n} = \mbf{0}$ in order to satisfy the equality constraint of~\eqref{eq:H2_DT_Output1c}.

The LMI in~\eqref{eq:H2_DT_Output1a} is derived from the LMI in Theorem~7 of~\cite{DeOliveira2002} by performing a congruence transformation involving a multiplication on the left and right by the symmetric matrix
\bdis
\mbf{W}_1 = \text{diag}\Big\{\bbm \mbf{0} & \mbf{1} \\ \mbf{1} & \mbf{0} \ebm,\bbm \mbf{0} &   \mbf{1} \\ \mbf{1} & \mbf{0} \ebm,  \mbf{1}\Big\}.
\edis
Similarly, the LMI in~\eqref{eq:H2_DT_Output1b} is derived from the LMI in Theorem~7 of~\cite{DeOliveira2002} by performing a congruence transformation involving a multiplication on the left and right by the symmetric matrix
\bdis
\mbf{W}_2 = \text{diag}\Big\{\mbf{1}, \bbm \mbf{0} & \mbf{1} \\ \mbf{1} & \mbf{0} \ebm \Big\}.
\edis

\end{SynthMeth}

\begin{SynthMeth}
\label{SynthMeth:H2_Dyn_Output_2}
Solve for $\mbf{A}_{\mathrm{d}n} \in \mathbb{R}^{n_x \times n_x}$, $\mbf{B}_{\mathrm{d}n} \in \mathbb{R}^{n_x \times n_y}$, $\mbf{C}_{\mathrm{d}n} \in \mathbb{R}^{n_u \times n_x}$, $\mbf{D}_{\mathrm{d}n} \in \mathbb{R}^{n_u \times n_y}$, $\mbf{X}_1$,~$\mbf{Y}_1 \in \mathbb{S}^{n_x}$, $\mbf{Z} \in \mathbb{S}^{n_z}$, and $\nu \in \mathbb{R}_{>0}$ that minimize $\mathcal{J}(\nu) = \nu$ subject to $\mbf{X}_1 > 0$, $\mbf{Y}_1 > 0$, $\mbf{Z} >0$,
\begin{align}
\bbm \mbf{X}_1 & \mbf{1} & \mbf{X}_1 \mbf{A}_\mathrm{d} + \mbf{B}_{\mathrm{d}n}\mbf{C}_{\mathrm{d}2} & \mbf{A}_{\mathrm{d}n} & \mbf{X}_1 \mbf{B}_{\mathrm{d}1} + \mbf{B}_{\mathrm{d}n} \mbf{D}_{\mathrm{d}21}   \\ *  & \mbf{Y}_1 & \mbf{A}_\mathrm{d} + \mbf{B}_{\mathrm{d}2} \mbf{D}_{\mathrm{d}n} \mbf{C}_{\mathrm{d}2} & \mbf{A}_\mathrm{d} \mbf{Y}_1 + \mbf{B}_{\mathrm{d}2} \mbf{C}_{\mathrm{d}n} & \mbf{B}_{\mathrm{d}1} + \mbf{B}_{\mathrm{d}2} \mbf{D}_{\mathrm{d}n} \mbf{D}_{\mathrm{d}21}  \\ * & * & \mbf{X}_1 & \mbf{1} & \mbf{0}  \\ * & * & * & \mbf{Y}_1 & \mbf{0}  \\ * & * & * & * &  \mbf{1}  \ebm &> 0, \nonumber\\
\bbm \mbf{Z} & \mbf{C}_{\mathrm{d}1} + \mbf{D}_{\mathrm{d}12} \mbf{D}_{\mathrm{d}n}\mbf{C}_{\mathrm{d}2} & \mbf{C}_{\mathrm{d}1}\mbf{Y}_1 + \mbf{D}_{\mathrm{d}12}\mbf{C}_{\mathrm{d}n} \\ * & \mbf{X}_1 & \mbf{1} \\ * & * & \mbf{Y}_1 \ebm &>0, \label{eq:H2_D_DTa}\\
\mbf{D}_{\mathrm{d}11} + \mbf{D}_{\mathrm{d}12} \mbf{D}_{\mathrm{d}n}  \mbf{D}_{\mathrm{d}21} &= \mbf{0}, \label{eq:H2_D_DT}\\
\bbm \mbf{X}_1 & \mbf{1} \\ * & \mbf{Y}_1 \ebm &> 0,  \label{eq:H2_D_DTc}\\
\trace (\mbf{Z}) &< \nu. \nonumber
\end{align}
The controller is recovered by
\begin{align*}
\mbf{A}_{\mathrm{d}c} &= \mbf{A}_{\mathrm{d}_K} - \mbf{B}_{\mathrm{d}c}\left(\mbf{1} - \mbf{D}_{\mathrm{d}22}\mbf{D}_{\mathrm{d}c}\right)^{-1}\mbf{D}_{\mathrm{d}22}\mbf{C}_{\mathrm{d}c}, \\
\mbf{B}_{\mathrm{d}c} &= \mbf{B}_{\mathrm{d}_K}\left(\mbf{1} - \mbf{D}_{\mathrm{d}22}\mbf{D}_{\mathrm{d}c}\right), \\
\mbf{C}_{\mathrm{d}c} &= \left(\mbf{1} - \mbf{D}_{\mathrm{d}c}\mbf{D}_{\mathrm{d}22}\right)\mbf{C}_{\mathrm{d}_K}, \\
\mbf{D}_{\mathrm{d}c} &= \left(\mbf{1} + \mbf{D}_{\mathrm{d}_K}\mbf{D}_{\mathrm{d}22}\right)^{-1}\mbf{D}_{\mathrm{d}_K},
\end{align*}
where
\bdis
\bbm \mbf{A}_{\mathrm{d}_K} & \mbf{B}_{\mathrm{d}_K} \\ \mbf{C}_{\mathrm{d}_K} & \mbf{D}_{\mathrm{d}_K} \ebm = \bbm \mbf{X}_2 & \mbf{X}_1\mbf{B}_{\mathrm{d}2} \\ \mbf{0} & \mbf{1} \ebm^{-1} \left( \bbm \mbf{A}_{\mathrm{d}n}& \mbf{B}_{\mathrm{d}n} \\ \mbf{C}_{\mathrm{d}n} & \mbf{D}_{\mathrm{d}n} \ebm - \bbm \mbf{X}_1\mbf{A}_\mathrm{d}\mbf{Y}_1 & \mbf{0} \\ \mbf{0} & \mbf{0}\ebm \right)\bbm \mbf{Y}_2^\trans & \mbf{0} \\ \mbf{C}_{\mathrm{d}2} \mbf{Y}_1 & \mbf{1} \ebm^{-1},
\edis
and the matrices $\mbf{X}_2$ and $\mbf{Y}_2$ satisfy $\mbf{X}_2\mbf{Y}_2^\trans = \mbf{1} - \mbf{X}_1\mbf{Y}_1$.  If $\mbf{D}_{\mathrm{d}22} = \mbf{0}$, then $\mbf{A}_{\mathrm{d}c} = \mbf{A}_{\mathrm{d}_K}$, $\mbf{B}_{\mathrm{d}c} = \mbf{B}_{\mathrm{d}_K}$, $\mbf{C}_{\mathrm{d}c} = \mbf{C}_{\mathrm{d}_K}$, and $\mbf{D}_{\mathrm{d}c} = \mbf{D}_{\mathrm{d}_K}$.

Given $\mbf{X}_1$ and $\mbf{Y}_1$, the matrices $\mbf{X}_2$ and $\mbf{Y}_2$ can be found using a matrix decomposition, such as a LU decomposition or a Cholesky decomposition.

If $\mbf{D}_{\mathrm{d}11} = \mbf{0}$, $\mbf{D}_{\mathrm{d}12} \neq \mbf{0}$, and $\mbf{D}_{\mathrm{d}21} \neq \mbf{0}$, then it is often simplest to choose $\mbf{D}_{\mathrm{d}n} = \mbf{0}$ in order to satisfy the equality constraint of~\eqref{eq:H2_D_DT}.

The LMIs in~\eqref{eq:H2_D_DTa} and~\eqref{eq:H2_D_DT} are derived from~\eqref{eq:H2_DT_Output1a} and~\eqref{eq:H2_DT_Output1b} using the change of variables $\mbf{S} = \mbf{J} = \mbf{1}$, $\mbf{H} = \mbf{X}_1$, $\mbf{G} = \mbf{Y}_1$.  The LMI in~\eqref{eq:H2_D_DTc} is added to ensure that $\mbf{1} - \mbf{X}_1\mbf{Y}_1 \geq 0$ in a similar fashion to the approach used in~\cite{Scherer1997}.

An alternate formulation of this synthesis method involves replacing~\eqref{eq:H2_D_DTa} and~\eqref{eq:H2_D_DT} with
\beq
\label{eq:H2_D_DTd}
\bbm \mbf{Z} & \mbf{C}_{\mathrm{d}1} + \mbf{D}_{\mathrm{d}12} \mbf{D}_{\mathrm{d}n}\mbf{C}_{\mathrm{d}2} & \mbf{C}_{\mathrm{d}1}\mbf{Y}_1 + \mbf{D}_{\mathrm{d}12}\mbf{C}_{\mathrm{d}n} & \mbf{D}_{\mathrm{d}11} + \mbf{D}_{\mathrm{d}12} \mbf{D}_{\mathrm{d}n}  \mbf{D}_{\mathrm{d}21} \\ * & \mbf{X}_1 & \mbf{1} & \mbf{0} \\ * & * & \mbf{Y}_1 & \mbf{0} \\ * & * & * & \mbf{1} \ebm >0.
\eeq
The matrix inequality in~\eqref{eq:H2_D_DTd} is derived by performing the same procedure used in~\cite{DeOliveira2002} with the change of variables $\mbf{S} = \mbf{J} = \mbf{1}$, $\mbf{H} = \mbf{X}_1$, $\mbf{G} = \mbf{Y}_1$, but instead starting with the matrix inequality formulation of the $\mathcal{H}_2$ that allows for a non-zero feedthrough term in~\cite[p.~25]{Santos2017} (summarized by~\eqref{eq:DT_D_H2_4a},~\eqref{eq:DT_D_H2_4b}, and~\eqref{eq:DT_D_H2_4c}).  In general, the matrix inequality in~\eqref{eq:H2_D_DTd} is less conservative than~\eqref{eq:H2_D_DTa} and~\eqref{eq:H2_D_DT}, as it allows for the resulting closed-loop system to have non-zero feedthrough, which, for a discrete-time system, is possible while maintaining a finite $\mathcal{H}_2$ norm.

\end{SynthMeth}

\subsection{\texorpdfstring{$\mathcal{H}_\infty$}{H-Infinity}-Optimal Control}
\label{sec:HinfSynth}
The goal of $\mathcal{H}_\infty$-optimal control is to design a controller that minimizes the $\mathcal{H}_\infty$ norm of the closed-loop transfer matrix from $\mbf{w}$ to $\mbf{z}$.

\subsubsection[\texorpdfstring{$\mathcal{H}_\infty$}{H-Infinity}-Optimal Full-State Feedback Control]{\texorpdfstring{$\mathcal{H}_\infty$}{H-Infinity}-Optimal Full-State Feedback Control~\cite[pp.~251--252]{Duan2013}}
\index{full-state feedback!$\mathcal{H}_\infty$-optimal}
Consider the continuous-time generalized LTI plant $\mbc{P}$ with state-space realization
\begin{align}
\dot{\mbf{x}} &= \mbf{A} \mbf{x} + \mbf{B}_1 \mbf{w} + \mbf{B}_2\mbf{u}, \label{eq:Hinf1a} \\
\mbf{z} &= \mbf{C}_1 \mbf{x} + \mbf{D}_{11} \mbf{w} + \mbf{D}_{12} \mbf{u}, \label{eq:Hinf1b} \\
\mbf{y} &= \mbf{x} , \nonumber
\end{align}
where it is assumed that ($\mbf{A}$,$\mbf{B}_2$) is stabilizable.  A full-state feedback controller $\mbc{K} = \mbf{K} \in \mathbb{R}^{n_u \times n_x}$ (i.e., $\mbf{u} = \mbf{K} \mbf{x}$) is to be designed to minimize $\mathcal{H}_\infty$ norm of the closed loop transfer matrix from the exogenous input $\mbf{w}$ to the performance output $\mbf{z}$.  Substituting the full-state feedback controller into~\eqref{eq:Hinf1a} and~\eqref{eq:Hinf1b} yields
\begin{align}
\dot{\mbf{x}} &= \left(\mbf{A} + \mbf{B}_2\mbf{K}\right) \mbf{x} + \mbf{B}_{1}\mbf{w} , \nonumber \\
\mbf{z} &= \left(\mbf{C}_1 + \mbf{D}_{12}\mbf{K}\right) \mbf{x} + \mbf{D}_{11}\mbf{w},  \nonumber
\end{align}
and a closed-loop transfer matrix
\bdis
\mbf{T}(s) = \left(\mbf{C}_1 + \mbf{D}_{12}\mbf{K}\right)\left(s\mbf{1} - \left(\mbf{A}+\mbf{B}_2\mbf{K}\right)\right)^{-1}\mbf{B}_1 + \mbf{D}_{11}.
\edis
From the Bounded Real Lemma in Section~\ref{sec:BoundedRealLemma}, the $\mathcal{H}_\infty$ of the closed-loop system is the minimum value of $\gamma \in \mathbb{R}_{>0}$ that satisfies
\beq
\label{eq:HinfFSF3}
\bbm  \mbf{P}\left(\mbf{A} + \mbf{B}_2\mbf{K}\right) + \left(\mbf{A} + \mbf{B}_2\mbf{K}\right)^\trans\mbf{P}  & \mbf{P}\mbf{B}_1 & \left(\mbf{C}_1 + \mbf{D}_{12}\mbf{K}\right)^\trans \\ *& -\gamma\mbf{1} & \mbf{D}_{11}^\trans \\ * & * & -\gamma\mbf{1} \ebm < 0,
\eeq
where $\mbf{P} \in \mathbb{S}^{n_x}$ and $\mbf{P} > 0$.  A congruence transformation is performed on~\eqref{eq:HinfFSF3} with $\mbf{W} = \text{diag}\{\mbf{P}^{-1},\mbf{1},\mbf{1}\}$ and a change of variables is made with $\mbf{Q} = \mbf{P}^{-1}$ and $\mbf{F} = \mbf{K}\mbf{Q}$.  This yields an LMI in the design variables $\mbf{Q}$, $\mbf{F}$, and $\gamma$, given by 
\beq
\label{eq:HinfFSF4}
\bbm  \mbf{A}\mbf{Q} + \mbf{Q}\mbf{A}^\trans + \mbf{B}_2\mbf{F} + \mbf{F}^\trans\mbf{B}_2^\trans & \mbf{B}_1 & \mbf{Q}\mbf{C}_1^\trans + \mbf{F}^\trans\mbf{D}_{12}^\trans \\ *& -\gamma \mbf{1} & \mbf{D}_{11}^\trans \\ * & * & -\gamma \mbf{1} \ebm < 0.
\eeq
\begin{SynthMeth}
The $\mathcal{H}_\infty$-optimal full-state feedback controller is synthesized by solving for $\mbf{Q} \in \mathbb{S}^{n_x}$ and $\mbf{F} \in \mathbb{R}^{n_u \times n_x}$ that minimize $\mathcal{J}(\gamma) = \gamma$ subject to $\mbf{Q} >0$ and~\eqref{eq:HinfFSF4}.  The $\mathcal{H}_\infty$-optimal full-state feedback controller gain is recovered by $\mbf{K} = \mbf{F}\mbf{Q}^{-1}$ and the $\mathcal{H}_\infty$ norm of $\mbf{T}(s)$ is $\gamma$.
\end{SynthMeth}

\subsubsection{Discrete-Time \texorpdfstring{$\mathcal{H}_\infty$}{H-Infinity}-Optimal Full-State Feedback Control}
\index{full-state feedback!discrete-time $\mathcal{H}_\infty$-optimal}

Consider the discrete-time generalized LTI plant $\mbc{P}$ with state-space realization
\begin{align*}
\mbf{x}_{k+1} &= \mbf{A}_\mathrm{d} \mbf{x}_k + \mbf{B}_{\mathrm{d}1} \mbf{w}_k + \mbf{B}_{\mathrm{d}2}\mbf{u}_k,  \\
\mbf{z}_k &= \mbf{C}_{\mathrm{d}1} \mbf{x}_k + \mbf{D}_{\mathrm{d}12} \mbf{u}_k,  \\
\mbf{y}_k &= \mbf{x}_k , \nonumber
\end{align*}
where it is assumed that ($\mbf{A}_\mathrm{d}$,$\mbf{B}_{\mathrm{d}2}$) is stabilizable.  A full-state feedback controller $\mbc{K} = \mbf{K}_\mathrm{d} \in \mathbb{R}^{n_u \times n_x}$ (i.e., $\mbf{u}_k = \mbf{K}_\mathrm{d}\mbf{x}_k$) is to be designed to minimize the $\mathcal{H}_\infty$ norm of the closed loop transfer matrix from the exogenous input $\mbf{w}_k$ to the performance output $\mbf{z}_k$, given by
\bdis
\mbf{T}(z) = \left(\mbf{C}_{\mathrm{d}1} + \mbf{D}_{\mathrm{d}12}\mbf{K}_\mathrm{d}\right)\left(z\mbf{1} - \left(\mbf{A}_\mathrm{d}+\mbf{B}_{\mathrm{d}2}\mbf{K}_\mathrm{d}\right)\right)^{-1}\mbf{B}_{\mathrm{d}1}.
\edis

\begin{SynthMeth}

The discrete-time $\mathcal{H}_\infty$-optimal full-state feedback controller is synthesized by solving for $\mbf{P} \in \mathbb{S}^{n_x} $, $\mbf{F}_\mathrm{d} \in \mathbb{R}^{n_u \times n_x}$, and $\gamma \in \mathbb{R}_{>0}$ that minimize $\mathcal{J}(\gamma) = \gamma$ subject to $\mbf{P} >0 $,
\bdis
\bbm \mbf{P}_\mathrm{d} & \mbf{A}_\mathrm{d}\mbf{P}_\mathrm{d} + \mbf{B}_{\mathrm{d}2}\mbf{F}_\mathrm{d} & \mbf{B}_{\text{d}1} & \mbf{0} \\ * & \mbf{P}_\mathrm{d} & \mbf{0} & \mbf{P}_\mathrm{d}\mbf{C}_{\text{d}1}^\trans + \mbf{F}_\mathrm{d}^\trans\mbf{D}_{\mathrm{d}12}^\trans \\ * & * & \gamma \mbf{1} & \mbf{D}_{\text{d}11}^\trans \\ * & * & * & \gamma \mbf{1} \ebm > 0.
\edis
The $\mathcal{H}_\infty$-optimal full-state feedback gain is recovered by $\mbf{K}_\mathrm{d} = \mbf{F}_\mathrm{d}\mbf{P}^{-1}$ and the $\mathcal{H}_\infty$ norm of $\mbf{T}(z)$ is $\gamma$.
\end{SynthMeth}


\subsubsection[\texorpdfstring{$\mathcal{H}_\infty$}{H-Infinity}-Optimal Dynamic Output Feedback Control]{\texorpdfstring{$\mathcal{H}_\infty$}{H-Infinity}-Optimal Dynamic Output Feedback Control}
\index{dynamic output feedback!$\mathcal{H}_\infty$-optimal}
Consider the continuous-time generalized LTI plant $\mbc{P}$ with minimal state-space realization
\begin{align*}
\dot{\mbf{x}} &= \mbf{A} \mbf{x} + \mbf{B}_1 \mbf{w} + \mbf{B}_2\mbf{u}, \\ 
\mbf{z} &= \mbf{C}_1 \mbf{x} + \mbf{D}_{11} \mbf{w}+ \mbf{D}_{12} \mbf{u}, \\ 
\mbf{y} &= \mbf{C}_2 \mbf{x} + \mbf{D}_{21} \mbf{w} + \mbf{D}_{22} \mbf{u}.  
\end{align*}
A continuous-time dynamic output feedback LTI controller with state-space realization $(\mbf{A}_c,\mbf{B}_c,\mbf{C}_c,\mbf{D}_c)$ is to be designed to minimize the $\mathcal{H}_\infty$ norm of the closed-loop system transfer matrix from $\mbf{w}$ to $\mbf{z}$, given by
\bdis
\mbf{T}(s) = \mbf{C}_\du{\mathrm{CL}}\left(s\mbf{1} - \mbf{A}_\du{\mathrm{CL}}\right)^{-1}\mbf{B}_\du{\mathrm{CL}} + \mbf{D}_\du{\mathrm{CL}},
\edis
where
\begin{align*}
\mbf{A}_\du{\mathrm{CL}} &= \bbm \mbf{A} + \mbf{B}_2 \mbf{D}_c\mbftilde{D}^{-1} \mbf{C}_2 & \mbf{B}_2 \left(\mbf{1} + \mbf{D}_c\mbftilde{D}^{-1}\mbf{D}_{22} \right)\mbf{C}_c \\ \mbf{B}_c  \mbftilde{D}^{-1}\mbf{C}_2 & \mbf{A}_c + \mbf{B}_c \mbftilde{D}^{-1}\mbf{D}_{22}\mbf{C}_c \ebm, \\
\mbf{B}_\du{\mathrm{CL}} &= \bbm  \mbf{B}_1 + \mbf{B}_2\mbf{D}_c\mbftilde{D}^{-1}\mbf{D}_{21}  \\ \mbf{B}_c \mbftilde{D}^{-1}\mbf{D}_{21} \ebm, \\
\mbf{C}_\du{\mathrm{CL}} &= \bbm \mbf{C}_1 + \mbf{D}_{12}\mbf{D}_c\mbftilde{D}^{-1} \mbf{C}_2 & \mbf{D}_{12} \left(\mbf{1} + \mbf{D}_c\mbftilde{D}^{-1}\mbf{D}_{22} \right)\mbf{C}_c \ebm, \\
\mbf{D}_\du{\mathrm{CL}} &= \mbf{D}_{11} + \mbf{D}_{12} \mbf{D}_c  \mbftilde{D}^{-1}\mbf{D}_{21}, 
\end{align*}
and $\mbftilde{D} = \mbf{1} - \mbf{D}_{22}\mbf{D}_c$.

Two different synthesis methods for the $\mathcal{H}_\infty$-optimal dynamic output feedback control problem are presented as follows.
\begin{SynthMeth}\cite{Scherer1997,PeetOptNotes21,LallNotes16}
Solve for $\mbf{A}_n \in \mathbb{R}^{n_x \times n_x}$, $\mbf{B}_n \in \mathbb{R}^{n_x \times n_y}$, $\mbf{C}_n \in \mathbb{R}^{n_u \times n_x}$, $\mbf{D}_n \in \mathbb{R}^{n_u \times n_y}$, $\mbf{X}_1$,~$\mbf{Y}_1 \in \mathbb{S}^{n_x}$, and $\gamma \in \mathbb{R}_{>0}$ that minimize $\mathcal{J}(\gamma) = \gamma$ subject to $\mbf{X}_1 > 0$, $\mbf{Y}_1 > 0$,
\begin{align*}
\bbm \mbf{N}_{11} & \mbf{A} + \mbf{A}_n^\trans + \mbf{B}_2\mbf{D}_n\mbf{C}_2 & \mbf{B}_1 + \mbf{B}_2\mbf{D}_n\mbf{D}_{21} & \mbf{Y}_1^\trans\mbf{C}_1^\trans + \mbf{C}_n^\trans\mbf{D}_{12}^\trans \\ * & \mbf{X}_1\mbf{A} + \mbf{A}^\trans\mbf{X}_1 + \mbf{B}_n \mbf{C}_2 + \mbf{C}_2^\trans \mbf{B}_n^\trans & \mbf{X}_1\mbf{B}_1 + \mbf{B}_n\mbf{D}_{21} & \mbf{C}_1^\trans + \mbf{C}_2^\trans\mbf{D}_n^\trans\mbf{D}_{12}^\trans \\ * & * & -\gamma \mbf{1} & \mbf{D}_{11}^\trans + \mbf{D}_{21}^\trans\mbf{D}_n^\trans\mbf{D}_{12}^\trans \\ * & * & * & -\gamma \mbf{1} \ebm &< 0, \\
\bbm \mbf{X}_1 & \mbf{1} \\ * & \mbf{Y}_1 \ebm &> 0,
\end{align*}
where $\mbf{N}_{11} = \mbf{A} \mbf{Y}_1 + \mbf{Y}_1\mbf{A}^\trans + \mbf{B}_2\mbf{C}_n + \mbf{C}_n^\trans\mbf{B}_2^\trans$.  The controller is recovered by
\begin{align*}
\mbf{A}_c &= \mbf{A}_\du{K} - \mbf{B}_c\left(\mbf{1} - \mbf{D}_{22}\mbf{D}_c\right)^{-1}\mbf{D}_{22}\mbf{C}_c, \\
\mbf{B}_c &= \mbf{B}_\du{K}\left(\mbf{1} - \mbf{D}_{22}\mbf{D}_c\right), \\
\mbf{C}_c &= \left(\mbf{1} - \mbf{D}_c\mbf{D}_{22}\right)\mbf{C}_\du{K}, \\
\mbf{D}_c &= \left(\mbf{1} + \mbf{D}_\du{K}\mbf{D}_{22}\right)^{-1}\mbf{D}_\du{K},
\end{align*}
where
\bdis
\bbm \mbf{A}_\du{K} & \mbf{B}_\du{K} \\ \mbf{C}_\du{K} & \mbf{D}_\du{K} \ebm = \bbm \mbf{X}_2 & \mbf{X}_1\mbf{B}_2 \\ \mbf{0} & \mbf{1} \ebm^{-1} \left( \bbm \mbf{A}_n & \mbf{B}_n \\ \mbf{C}_n & \mbf{D}_n \ebm - \bbm \mbf{X}_1\mbf{A}\mbf{Y}_1 & \mbf{0} \\ \mbf{0} & \mbf{0}\ebm \right)\bbm \mbf{Y}_2^\trans & \mbf{0} \\ \mbf{C}_2 \mbf{Y}_1 & \mbf{1} \ebm^{-1},
\edis
and the matrices $\mbf{X}_2$ and $\mbf{Y}_2$ satisfy $\mbf{X}_2\mbf{Y}_2^\trans = \mbf{1} - \mbf{X}_1\mbf{Y}_1$.  If $\mbf{D}_{22} = \mbf{0}$, then $\mbf{A}_c = \mbf{A}_\du{K}$, $\mbf{B}_c = \mbf{B}_\du{K}$, $\mbf{C}_c = \mbf{C}_\du{K}$, and $\mbf{D}_c = \mbf{D}_\du{K}$.

Given $\mbf{X}_1$ and $\mbf{Y}_1$, the matrices $\mbf{X}_2$ and $\mbf{Y}_2$ can be found using a matrix decomposition, such as a LU decomposition or a Cholesky decomposition.
\end{SynthMeth}


\begin{SynthMeth} \cite{Gahinet1994},\cite[pp.~224--232]{Dullerud2000}
The controller is solved for in the following two steps.
\begin{enumerate}
\item Solve for $\mbf{P}$,~$\mbf{Q} \in \mathbb{S}^{n_x}$ and $\gamma \in \mathbb{R}_{>0}$, where $\mbf{P} > 0$ and $\mbf{Q} > 0$, that minimize $\mathcal{J}(\gamma) = \gamma$ subject to
\begin{align}
\bbm \mbf{N_o} & \mbf{0} \\ \mbf{0} & \mbf{1} \ebm^\trans \bbm \mbf{P}\mbf{A} + \mbf{A}^\trans\mbf{P} & \mbf{P}\mbf{B}_1 & \mbf{C}_1^\trans \\ * & -\gamma \mbf{1} & \mbf{D}_{11}^\trans \\ * & * & -\gamma \mbf{1} \ebm \bbm \mbf{N_o} & \mbf{0} \\ \mbf{0} & \mbf{1} \ebm &< 0, \nonumber \\
\bbm \mbf{N_c} & \mbf{0} \\ \mbf{0} & \mbf{1} \ebm^\trans \bbm \mbf{A}\mbf{Q} + \mbf{Q}\mbf{A}^\trans & \mbf{Q}\mbf{C}_1^\trans & \mbf{B}_1 \\ * & -\gamma \mbf{1} & \mbf{D}_{11} \\ * & * & -\gamma \mbf{1} \ebm \bbm \mbf{N_c} & \mbf{0} \\ \mbf{0} & \mbf{1} \ebm &< 0, \nonumber \\
\bbm \mbf{P} & \mbf{1} \\ * & \mbf{Q} \ebm &\geq 0, \label{eq:HinfPQ}
\end{align}
where $\mathcal{R}\left(\mbf{N}_o\right) = \mathcal{N}\left(\bbm \mbf{C}_2 & \mbf{D}_{21} \ebm\right)$ and $\mathcal{R}\left(\mbf{N}_c\right) = \mathcal{N}\left(\bbm \mbf{B}_2^\trans & \mbf{D}_{12}^\trans \ebm\right)$.  Define $\mbf{P}_\du{\rm CL} = \bbm \mbf{P} & \mbf{P}_2^\trans \\ * & \mbf{1} \ebm$, where $\mbf{P}_2\mbf{P}_2^\trans = \mbf{P} - \mbf{Q}^{-1}$.

\item Fix $\mbf{P}_\du{\rm CL}$ and solve for $\mbf{A}_n \in \mathbb{R}^{n_x \times n_x}$, $\mbf{B}_n \in \mathbb{R}^{n_x \times n_y}$, $\mbf{C}_n \in \mathbb{R}^{n_u \times n_x}$, $\mbf{D}_n \in \mathbb{R}^{n_u \times n_y}$, and $\gamma \in \mathbb{R}_{>0}$ that minimize $\mathcal{J}(\gamma) = \gamma$ subject to
\begin{multline*}
\bbm \mbf{P}_{\du{\rm CL}}\mbfbar{A} + \mbfbar{A}^\trans \mbf{P}_{\du{\rm CL}} & \mbf{P}_{\du{\rm CL}}\mbfbar{B} & \mbfbar{C}^\trans \\ * & -\gamma \mbf{1} & \mbf{D}_{11}^\trans \\ * & * & -\gamma \mbf{1} \ebm + \bbm \mbf{P}_{\du{\rm CL}}\underline{\mbf{B}} \\ \mbf{0} \\ \underline{\mbf{D}}_{12} \ebm \bbm \mbf{A}_{n} & \mbf{B}_{n} \\ \mbf{C}_{n} & \mbf{D}_{n} \ebm \bbm \underline{\mbf{C}} & \underline{\mbf{D}}_{21} & \mbf{0} \ebm \\ + \bbm \underline{\mbf{C}}^\trans \\ \underline{\mbf{D}}_{21}^\trans \\ \mbf{0} \ebm \bbm \mbf{A}_{n} & \mbf{B}_{n} \\ \mbf{C}_{n} & \mbf{D}_{n} \ebm^\trans \bbm \underline{\mbf{B}}^\trans \mbf{P}_{\du{\rm CL}} & \mbf{0} & \underline{\mbf{D}}_{12}^\trans \ebm < 0, 
\end{multline*}
where
\begin{alignat*}{3}
\mbfbar{A}&= \bbm \mbf{A}   & \mbf{0} \\ \mbf{0} & \mbf{0} \ebm, &\mbfbar{B} &= \bbm \mbf{B}_1 - \mbf{B}_2\mbfbar{D}_c\mbf{D}_{21} \\ \mbf{0}  \ebm, \nonumber \\
\mbfbar{C} &= \bbm \mbf{C}_1 & \mbf{0} \ebm, &\mbfubar{C} &= \bbm \mbf{0} & \mbf{1} \\ \mbf{C}_2 & \mbf{0}  \ebm, \nonumber \\
\mbfubar{B} &= \bbm \mbf{0} & -\mbf{B}_2\\ \mbf{1} & \mbf{0} \ebm, \hspace{20pt} &\mbfubar{D}_{12} &= \bbm \mbf{0} & -\mbf{D}_{12} \ebm, \nonumber \\
 \mbfubar{D}_{21} &= \bbm \mbf{0} \\ \mbf{D}_{21} \ebm. && \nonumber 
\end{alignat*}

\end{enumerate}
The controller is recovered by
\begin{align*}
\mbf{A}_c &= \mbf{A}_{n} - \mbf{B}_c\left(\mbf{1} - \mbf{D}_{22}\mbf{D}_c\right)^{-1}\mbf{D}_{22}\mbf{C}_c, \\
\mbf{B}_c &= \mbf{B}_{n}\left(\mbf{1} - \mbf{D}_{22}\mbf{D}_c\right), \\
\mbf{C}_c &= \left(\mbf{1} - \mbf{D}_c\mbf{D}_{22}\right)\mbf{C}_{n}, \\
\mbf{D}_c &= \left(\mbf{1} + \mbf{D}_{n}\mbf{D}_{22}\right)^{-1}\mbf{D}_{n}.
\end{align*}
If $\mbf{D}_{22} = \mbf{0}$, then $\mbf{A}_c = \mbf{A}_n$, $\mbf{B}_c = \mbf{B}_n$, $\mbf{C}_c = \mbf{C}_n$, and $\mbf{D}_c = \mbf{D}_n$.

\end{SynthMeth}

Note that the purpose of the matrix inequality $\bbm \mbf{P} & \mbf{1} \\ * & \mbf{Q} \ebm \geq 0$ in~\eqref{eq:HinfPQ} is to ensure that there exists $\mbf{P}_\du{\rm CL} = \bbm \mbf{P} & \mbf{P}_2^\trans \\ * & \mbf{1} \ebm > 0$ and $\mbf{P}_\du{\rm CL}^{-1} = \bbm \mbf{Q} & - \mbf{Q} \mbf{P}_2 \\ * & \mbf{P}_2^\trans \mbf{Q} \mbf{P}_2 + \mbf{1} \ebm$.  This follows from Property~\ref{sec:SchurProp7} in Section~\ref{sec:SchurProp}.

\subsubsection[Discrete-Time \texorpdfstring{$\mathcal{H}_\infty$}{H-Infinity}-Optimal Dynamic Output Feedback Control]{Discrete-Time \texorpdfstring{$\mathcal{H}_\infty$}{H-Infinity}-Optimal Dynamic Output Feedback Control}
\index{dynamic output feedback!discrete-time $\mathcal{H}_\infty$-optimal}
Consider the discrete-time generalized LTI plant $\mbc{P}$ with minimal state-space realization
\begin{align*}
\mbf{x}_{k+1} &= \mbf{A}_\mathrm{d} \mbf{x}_k + \mbf{B}_{\mathrm{d}1} \mbf{w}_k + \mbf{B}_{\mathrm{d}2}\mbf{u}_k,  \\
\mbf{z}_k &= \mbf{C}_{\mathrm{d}1} \mbf{x}_k + \mbf{D}_{\mathrm{d}11}\mbf{w}_k + \mbf{D}_{\mathrm{d}12} \mbf{u}_k,  \\
\mbf{y}_k &= \mbf{C}_{\mathrm{d}2} \mbf{x}_k + \mbf{D}_{\mathrm{d}21}\mbf{w}_k + \mbf{D}_{\mathrm{d}22} \mbf{u}_k, \nonumber
\end{align*}
A discrete-time dynamic output feedback LTI controller with state-space realization $(\mbf{A}_{\mathrm{d}c},\mbf{B}_{\mathrm{d}c},\mbf{C}_{\mathrm{d}c},\mbf{D}_{\mathrm{d}c})$ is to be designed to minimize the $\mathcal{H}_\infty$ norm of the closed-loop system transfer matrix from $\mbf{w}$ to $\mbf{z}$, given by
\bdis
\mbf{T}(z) = \mbf{C}_{\mathrm{d}_\mathrm{CL}}\left(z\mbf{1} - \mbf{A}_{\mathrm{d}_\mathrm{CL}}\right)^{-1}\mbf{B}_{\mathrm{d}_\mathrm{CL}} + \mbf{D}_{\mathrm{d}_\mathrm{CL}},
\edis
where
\begin{align*}
\mbf{A}_{\mathrm{d}_\mathrm{CL}} &= \bbm \mbf{A}_\mathrm{d} + \mbf{B}_{\mathrm{d}2} \mbf{D}_{\mathrm{d}c}\mbftilde{D}_\mathrm{d}^{-1} \mbf{C}_{\mathrm{d}2} & \mbf{B}_{\mathrm{d}2} \left(\mbf{1} + \mbf{D}_{\mathrm{d}c}\mbftilde{D}_\mathrm{d}^{-1}\mbf{D}_{\mathrm{d}22} \right)\mbf{C}_{\mathrm{d}c} \\ \mbf{B}_{\mathrm{d}c}  \mbftilde{D}_\mathrm{d}^{-1}\mbf{C}_{\mathrm{d}2} & \mbf{A}_{\mathrm{d}c} + \mbf{B}_{\mathrm{d}c} \mbftilde{D}_\mathrm{d}^{-1}\mbf{D}_{\mathrm{d}22}\mbf{C}_{\mathrm{d}c} \ebm, \\
\mbf{B}_{\mathrm{d}_\mathrm{CL}} &= \bbm  \mbf{B}_{\mathrm{d}1} + \mbf{B}_{\mathrm{d}2}\mbf{D}_{\mathrm{d}c}\mbftilde{D}_\mathrm{d}^{-1}\mbf{D}_{\mathrm{d}21}  \\ \mbf{B}_{\mathrm{d}c} \mbftilde{D}_\mathrm{d}^{-1}\mbf{D}_{\mathrm{d}21} \ebm, \\
\mbf{C}_{\mathrm{d}_\mathrm{CL}} &= \bbm \mbf{C}_{\mathrm{d}1} + \mbf{D}_{\mathrm{d}12}\mbf{D}_{\mathrm{d}c}\mbftilde{D}_\mathrm{d}^{-1} \mbf{C}_{\mathrm{d}2} & \mbf{D}_{\mathrm{d}12} \left(\mbf{1} + \mbf{D}_{\mathrm{d}c}\mbftilde{D}_\mathrm{d}^{-1}\mbf{D}_{\mathrm{d}22} \right)\mbf{C}_{\mathrm{d}c} \ebm, \\
\mbf{D}_{\mathrm{d}_\mathrm{CL}} &= \mbf{D}_{\mathrm{d}11} + \mbf{D}_{\mathrm{d}12} \mbf{D}_{\mathrm{d}c}  \mbftilde{D}_\mathrm{d}^{-1}\mbf{D}_{\mathrm{d}21}, 
\end{align*}
and $\mbftilde{D}_\mathrm{d} = \mbf{1} - \mbf{D}_{\mathrm{d}22}\mbf{D}_{\mathrm{d}c}$.

\begin{SynthMeth}
\cite{DeOliveira2002} Solve for $\mbf{A}_{\mathrm{d}n} \in \mathbb{R}^{n_x \times n_x}$, $\mbf{B}_{\mathrm{d}n} \in \mathbb{R}^{n_x \times n_y}$, $\mbf{C}_{\mathrm{d}n} \in \mathbb{R}^{n_u \times n_x}$, $\mbf{D}_{\mathrm{d}n} \in \mathbb{R}^{n_u \times n_y}$, $\mbf{X}_1$,~$\mbf{Y}_1 \in \mathbb{S}^{n_x}$, $\mbf{G}$,~$\mbf{H}$,~$\mbf{J}$,~$\mbf{S} \in \mathbb{R}^{n_x \times n_x}$, and $\gamma \in \mathbb{R}_{>0}$ that minimize $\mathcal{J}(\gamma) = \gamma$ subject to $\mbf{X}_1 > 0$, $\mbf{Y}_1 > 0$,
\beq
\bbm \mbf{X}_1 & \mbf{J}^\trans & \mbf{H} \mbf{A}_\mathrm{d} + \mbf{B}_{\mathrm{d}n}\mbf{C}_{\mathrm{d}2} & \mbf{A}_{\mathrm{d}n} & \mbf{H} \mbf{B}_{\mathrm{d}1} + \mbf{B}_{\mathrm{d}n} \mbf{D}_{\mathrm{d}21} & \mbf{0}  \\ *  & \mbf{Y}_1 & \mbf{A}_\mathrm{d} + \mbf{B}_{\mathrm{d}2} \mbf{D}_{\mathrm{d}n} \mbf{C}_{\mathrm{d}2} & \mbf{A}_\mathrm{d} \mbf{G} + \mbf{B}_{\mathrm{d}2} \mbf{C}_{\mathrm{d}n} & \mbf{B}_{\mathrm{d}1} + \mbf{B}_{\mathrm{d}2} \mbf{D}_{\mathrm{d}n} \mbf{D}_{\mathrm{d}21} & \mbf{0} \\ * & * & \mbf{H} + \mbf{H}^\trans - \mbf{X}_1 & \mbf{1} + \mbf{S} - \mbf{J}^\trans & \mbf{0} & \mbf{C}_{\mathrm{d}1}^\trans + \mbf{C}_{\mathrm{d}2}^\trans \mbf{D}_{\mathrm{d}n}^\trans \mbf{D}_{\mathrm{d}12}^\trans \\ * & * & * & \mbf{G} + \mbf{G}^\trans -\mbf{Y}_1 & \mbf{0} & \mbf{G}^\trans \mbf{C}_{\mathrm{d}1}^\trans + \mbf{C}_{\mathrm{d}n}^\trans \mbf{D}_{\mathrm{d}12}^\trans \\ * & * & * & * & \gamma \mbf{1} & \mbf{D}_{\mathrm{d}11}^\trans + \mbf{D}_{\mathrm{d}21}^\trans\mbf{D}_{\mathrm{d}n}^\trans\mbf{D}_{\mathrm{d}12}^\trans \\ * & * & * & * & * & \gamma \mbf{1} \ebm > 0. \label{eq:Hinf_DT_Output1a}
\eeq
The controller is recovered by
\begin{align*}
\mbf{A}_{\mathrm{d}c} &= \mbf{A}_{\mathrm{d}_K} - \mbf{B}_{\mathrm{d}c}\left(\mbf{1} - \mbf{D}_{\mathrm{d}22}\mbf{D}_{\mathrm{d}c}\right)^{-1}\mbf{D}_{\mathrm{d}22}\mbf{C}_{\mathrm{d}c}, \\
\mbf{B}_{\mathrm{d}c} &= \mbf{B}_{\mathrm{d}_K}\left(\mbf{1} - \mbf{D}_{\mathrm{d}22}\mbf{D}_{\mathrm{d}c}\right), \\
\mbf{C}_{\mathrm{d}c} &= \left(\mbf{1} - \mbf{D}_{\mathrm{d}c}\mbf{D}_{\mathrm{d}22}\right)\mbf{C}_{\mathrm{d}_K}, \\
\mbf{D}_{\mathrm{d}c} &= \left(\mbf{1} + \mbf{D}_{\mathrm{d}_K}\mbf{D}_{\mathrm{d}22}\right)^{-1}\mbf{D}_{\mathrm{d}_K},
\end{align*}
where
\bdis
\bbm \mbf{A}_{\mathrm{d}_K} & \mbf{B}_{\mathrm{d}_K} \\ \mbf{C}_{\mathrm{d}_K} & \mbf{D}_{\mathrm{d}_K} \ebm = \bbm \mbf{Y}_2^{-\trans} & \mbf{Y}_2^{-\trans}\mbf{H}\mbf{B}_{\mathrm{d}2} \\ \mbf{0} & \mbf{1} \ebm \left( \bbm \mbf{A}_{\mathrm{d}n}& \mbf{B}_{\mathrm{d}n} \\ \mbf{C}_{\mathrm{d}n} & \mbf{D}_{\mathrm{d}n} \ebm - \bbm \mbf{H}\mbf{A}_\mathrm{d}\mbf{G} & \mbf{0} \\ \mbf{0} & \mbf{0}\ebm \right)\bbm \mbf{X}_2^{-1} & \mbf{0} \\ -\mbf{C}_{\mathrm{d}2} \mbf{G} \mbf{X}_2^{-1} & \mbf{1} \ebm,
\edis
and the matrices $\mbf{X}_2$ and $\mbf{Y}_2$ satisfy $\mbf{X}_2\mbf{Y}_2^\trans = \mbf{1} - \mbf{H}\mbf{G}$.  If $\mbf{D}_{\mathrm{d}22} = \mbf{0}$, then $\mbf{A}_{\mathrm{d}c} = \mbf{A}_{\mathrm{d}_K}$, $\mbf{B}_{\mathrm{d}c} = \mbf{B}_{\mathrm{d}_K}$, $\mbf{C}_{\mathrm{d}c} = \mbf{C}_{\mathrm{d}_K}$, and $\mbf{D}_{\mathrm{d}c} = \mbf{D}_{\mathrm{d}_K}$.

Given $\mbf{G}$ and $\mbf{H}$, the matrices $\mbf{X}_2$ and $\mbf{Y}_2$ can be found using a matrix decomposition, such as a LU decomposition or a Cholesky decomposition.

The LMI in~\eqref{eq:Hinf_DT_Output1a} is derived from the LMI in Theorem~8 of~\cite{DeOliveira2002} by performing a congruence transformation involving a multiplication on the left and right by the symmetric matrix
\bdis
\mbf{W} = \text{diag}\Big\{\bbm \mbf{0} & \sqrt{\gamma} \mbf{1} \\ \frac{1}{ \sqrt{\gamma} } \mbf{1} & \mbf{0} \ebm,\bbm \mbf{0} &  \sqrt{\gamma}  \mbf{1} \\ \frac{1}{ \sqrt{\gamma} } \mbf{1} & \mbf{0} \ebm, \sqrt{\gamma}  \mbf{1}, \frac{1}{ \sqrt{\gamma} } \mbf{1} \Big\},
\edis
followed by the change of variables $\gamma = \mu^2$, $\mbf{X}_1 = \gamma \mbf{H}$, $\mbf{Y}_1 = \gamma^{-1} \mbf{P}$.

\end{SynthMeth}

\begin{SynthMeth}
\label{SynthMeth:Hinf_Dyn_Output_2}
Solve for $\mbf{A}_{\mathrm{d}n} \in \mathbb{R}^{n_x \times n_x}$, $\mbf{B}_{\mathrm{d}n} \in \mathbb{R}^{n_x \times n_y}$, $\mbf{C}_{\mathrm{d}n} \in \mathbb{R}^{n_u \times n_x}$, $\mbf{D}_{\mathrm{d}n} \in \mathbb{R}^{n_u \times n_y}$, $\mbf{X}_1$,~$\mbf{Y}_1 \in \mathbb{S}^{n_x}$, and $\gamma \in \mathbb{R}_{>0}$ that minimize $\mathcal{J}(\gamma) = \gamma$ subject to $\mbf{X}_1 > 0$, $\mbf{Y}_1 > 0$,
\begin{align}
\bbm \mbf{X}_1 & \mbf{1} & \mbf{X}_1 \mbf{A}_\mathrm{d} + \mbf{B}_{\mathrm{d}n}\mbf{C}_{\mathrm{d}2} & \mbf{A}_{\mathrm{d}n} & \mbf{X}_1 \mbf{B}_{\mathrm{d}1} + \mbf{B}_{\mathrm{d}n} \mbf{D}_{\mathrm{d}21} & \mbf{0}  \\ *  & \mbf{Y}_1 & \mbf{A}_\mathrm{d} + \mbf{B}_{\mathrm{d}2} \mbf{D}_{\mathrm{d}n} \mbf{C}_{\mathrm{d}2} & \mbf{A}_\mathrm{d} \mbf{Y}_1 + \mbf{B}_{\mathrm{d}2} \mbf{C}_{\mathrm{d}n} & \mbf{B}_{\mathrm{d}1} + \mbf{B}_{\mathrm{d}2} \mbf{D}_{\mathrm{d}n} \mbf{D}_{\mathrm{d}21} & \mbf{0} \\ * & * & \mbf{X}_1 & \mbf{1} & \mbf{0} & \mbf{C}_{\mathrm{d}1}^\trans + \mbf{C}_{\mathrm{d}2}^\trans \mbf{D}_{\mathrm{d}n}^\trans \mbf{D}_{\mathrm{d}12}^\trans \\ * & * & * & \mbf{Y}_1 & \mbf{0} & \mbf{Y}_1 \mbf{C}_{\mathrm{d}1}^\trans + \mbf{C}_{\mathrm{d}n}^\trans \mbf{D}_{\mathrm{d}12}^\trans \\ * & * & * & * & \gamma \mbf{1} & \mbf{D}_{\mathrm{d}11}^\trans + \mbf{D}_{\mathrm{d}21}^\trans\mbf{D}_{\mathrm{d}n}^\trans\mbf{D}_{\mathrm{d}12}^\trans \\ * & * & * & * & * & \gamma \mbf{1} \ebm &> 0, \label{eq:Hinf_DT_Output2a}\\
\bbm \mbf{X}_1 & \mbf{1} \\ * & \mbf{Y}_1 \ebm &> 0. \label{eq:Hinf_DT_Output2b}
\end{align}
The controller is recovered by
\begin{align*}
\mbf{A}_{\mathrm{d}c} &= \mbf{A}_{\mathrm{d}_K} - \mbf{B}_{\mathrm{d}c}\left(\mbf{1} - \mbf{D}_{\mathrm{d}22}\mbf{D}_{\mathrm{d}c}\right)^{-1}\mbf{D}_{\mathrm{d}22}\mbf{C}_{\mathrm{d}c}, \\
\mbf{B}_{\mathrm{d}c} &= \mbf{B}_{\mathrm{d}_K}\left(\mbf{1} - \mbf{D}_{\mathrm{d}22}\mbf{D}_{\mathrm{d}c}\right), \\
\mbf{C}_{\mathrm{d}c} &= \left(\mbf{1} - \mbf{D}_{\mathrm{d}c}\mbf{D}_{\mathrm{d}22}\right)\mbf{C}_{\mathrm{d}_K}, \\
\mbf{D}_{\mathrm{d}c} &= \left(\mbf{1} + \mbf{D}_{\mathrm{d}_K}\mbf{D}_{\mathrm{d}22}\right)^{-1}\mbf{D}_{\mathrm{d}_K},
\end{align*}
where
\bdis
\bbm \mbf{A}_{\mathrm{d}_K} & \mbf{B}_{\mathrm{d}_K} \\ \mbf{C}_{\mathrm{d}_K} & \mbf{D}_{\mathrm{d}_K} \ebm = \bbm \mbf{X}_2 & \mbf{X}_1\mbf{B}_{\mathrm{d}2} \\ \mbf{0} & \mbf{1} \ebm^{-1} \left( \bbm \mbf{A}_{\mathrm{d}n}& \mbf{B}_{\mathrm{d}n} \\ \mbf{C}_{\mathrm{d}n} & \mbf{D}_{\mathrm{d}n} \ebm - \bbm \mbf{X}_1\mbf{A}_\mathrm{d}\mbf{Y}_1 & \mbf{0} \\ \mbf{0} & \mbf{0}\ebm \right)\bbm \mbf{Y}_2^\trans & \mbf{0} \\ \mbf{C}_{\mathrm{d}2} \mbf{Y}_1 & \mbf{1} \ebm^{-1},
\edis
and the matrices $\mbf{X}_2$ and $\mbf{Y}_2$ satisfy $\mbf{X}_2\mbf{Y}_2^\trans = \mbf{1} - \mbf{X}_1\mbf{Y}_1$.  If $\mbf{D}_{\mathrm{d}22} = \mbf{0}$, then $\mbf{A}_{\mathrm{d}c} = \mbf{A}_{\mathrm{d}_K}$, $\mbf{B}_{\mathrm{d}c} = \mbf{B}_{\mathrm{d}_K}$, $\mbf{C}_{\mathrm{d}c} = \mbf{C}_{\mathrm{d}_K}$, and $\mbf{D}_{\mathrm{d}c} = \mbf{D}_{\mathrm{d}_K}$.

Given $\mbf{X}_1$ and $\mbf{Y}_1$, the matrices $\mbf{X}_2$ and $\mbf{Y}_2$ can be found using a matrix decomposition, such as a LU decomposition or a Cholesky decomposition.

The LMI in~\eqref{eq:Hinf_DT_Output2a} is derived from~\eqref{eq:Hinf_DT_Output1a} using the change of variables $\mbf{S} = \mbf{J} = \mbf{1}$, $\mbf{H} = \mbf{X}_1$, $\mbf{G} = \mbf{Y}_1$.  The LMI in~\eqref{eq:Hinf_DT_Output2b} is added to ensure that $\mbf{1} - \mbf{X}_1\mbf{Y}_1 \geq 0$ in a similar fashion to the approach used in~\cite{Scherer1997}.

\end{SynthMeth}

\subsection{Mixed \texorpdfstring{$\mathcal{H}_2$-$\mathcal{H}_\infty$}{H2 H-Infinity}-Optimal Control}
\label{sec:MixedSynth}
The goal of mixed $\mathcal{H}_2$-$\mathcal{H}_\infty$-optimal control is to design a controller that minimizes the $\mathcal{H}_2$ norm of the closed-loop transfer matrix from $\mbf{w}_1$ to $\mbf{z}_1$, while ensuring that the $\mathcal{H}_\infty$ norm of the closed-loop transfer function from $\mbf{w}_2$ to $\mbf{z}_2$ is below a specified bound.

\subsubsection[Mixed \texorpdfstring{$\mathcal{H}_2$-$\mathcal{H}_\infty$}{H2 H-Infinity}-Optimal Full-State Feedback Control]{Mixed \texorpdfstring{$\mathcal{H}_2$-$\mathcal{H}_\infty$}{H2 H-Infinity}-Optimal Full-State Feedback Control~\cite[pp.~329--330]{Duan2013}}
\label{sec:MixedFullStateSynth}
\index{full-state feedback!mixed $\mathcal{H}_2$-$\mathcal{H}_\infty$-optimal}

Consider the continuous-time generalized LTI plant $\mbc{P}$ with state-space realization
\begin{align*}
\dot{\mbf{x}} &= \mbf{A} \mbf{x} + \bbm \mbf{B}_{1,1} & \mbf{B}_{1,2} \ebm \bbm \mbf{w}_1 \\ \mbf{w}_2 \ebm + \mbf{B}_2\mbf{u},  \\ 
\bbm \mbf{z}_1 \\ \mbf{z}_2 \ebm  &= \bbm \mbf{C}_{1,1} \\ \mbf{C}_{1,2} \ebm \mbf{x} + \bbm \mbf{0} & \mbf{D}_{11,12} \\ \mbf{D}_{11,21} & \mbf{D}_{11,22} \ebm \bbm \mbf{w}_1 \\ \mbf{w}_2 \ebm + \bbm \mbf{D}_{12,1} \\ \mbf{D}_{12,2} \ebm \mbf{u},  \\ 
\mbf{y} &=  \mbf{x},   
\end{align*}
where it is assumed that $(\mbf{A},\mbf{B}_2)$ is stabilizable.  A full-state feedback controller $\mbc{K} = \mbf{K} \in \mathbb{R}^{n_u \times n_x}$ (i.e., $\mbf{u} = \mbf{K} \mbf{x}$) is to be designed to minimize the $\mathcal{H}_2$ norm of the closed-loop transfer matrix $\mbf{T}_{11}(s)$ from the exogenous input $\mbf{w}_1$ to the performance output $\mbf{z}_1$ while ensuring the $\mathcal{H}_\infty$ norm of the closed-loop transfer matrix $\mbf{T}_{22}(s)$ from the exogenous input $\mbf{w}_2$ to the performance output $\mbf{z}_2$ is less than $\gamma_d$, where
 \begin{align*}
\mbf{T}_{11}(s) &= \left(\mbf{C}_{1,1} + \mbf{D}_{12,1}\mbf{K}\right)\left(s\mbf{1} - \left(\mbf{A}+\mbf{B}_2\mbf{K}\right)\right)^{-1}\mbf{B}_{1,1}, \\
\mbf{T}_{22}(s) &= \left(\mbf{C}_{1,2} + \mbf{D}_{12,2}\mbf{K}\right)\left(s\mbf{1} - \left(\mbf{A}+\mbf{B}_2\mbf{K}\right)\right)^{-1}\mbf{B}_{1,2} + \mbf{D}_{11,22}. 
\end{align*}

\begin{SynthMeth}
The mixed $\mathcal{H}_2$-$\mathcal{H}_\infty$-optimal full-state feedback controller is synthesized by solving for $\mbf{P} \in \mathbb{S}^{n_x}$, $\mbf{Z} \in \mathbb{S}^{n_w}$, $\mbf{F} \in \mathbb{R}^{n_u \times n_x}$, and $\nu \in \mathbb{R}_{>0}$ that minimize $\mathcal{J}(\nu) = \nu$ subject to $\mbf{P} >0 $, $\mbf{Z} > 0$,
\begin{align*}
\bbm \mbf{A}\mbf{P} + \mbf{P}\mbf{A}^\trans+\mbf{B}_2\mbf{F}+\mbf{F}^\trans\mbf{B}_2^\trans & \mbf{P}\mbf{C}_{1,1}^\trans + \mbf{F}^\trans\mbf{D}_{12,1}^\trans \\ * & -\mbf{1} \ebm &< 0, \\
\bbm  \mbf{A}\mbf{P} + \mbf{P}\mbf{A}^\trans+\mbf{B}_2\mbf{F} + \mbf{F}^\trans\mbf{B}_2^\trans & \mbf{B}_{1,2} & \mbf{P}\mbf{C}_{1,2}^\trans + \mbf{F}^\trans\mbf{D}_{12,2}^\trans \\ *& -\gamma_d\mbf{1} & \mbf{D}_{11,22}^\trans \\ * & * & -\gamma_d\mbf{1} \ebm &< 0, \\
\bbm \mbf{Z} & \mbf{B}_{1,1}^\trans \\ * & \mbf{P} \ebm &> 0, \\
\trace (\mbf{Z}) &< \nu. 
\end{align*}
The $\mathcal{H}_2$-optimal full-state feedback gain is recovered by $\mbf{K} = \mbf{F}\mbf{P}^{-1}$, the $\mathcal{H}_2$ norm of $\mbf{T}_{11}(s)$ is less than $\mu = \sqrt{\nu}$, and the $\mathcal{H}_\infty$ norm of $\mbf{T}_{22}(s)$ is less than $\gamma_d$.
\end{SynthMeth}

\subsubsection{Discrete-Time Mixed \texorpdfstring{$\mathcal{H}_2$-$\mathcal{H}_\infty$}{H2 H-Infinity}-Optimal Full-State Feedback Control}
\index{full-state feedback!discrete-time mixed $\mathcal{H}_2$-$\mathcal{H}_\infty$-optimal}

Consider the discrete-time generalized LTI plant $\mbc{P}$ with state-space realization
\begin{align*}
\mbf{x}_{k+1} &= \mbf{A}_\mathrm{d} \mbf{x}_k + \bbm \mbf{B}_{\mathrm{d}1,1} & \mbf{B}_{\mathrm{d}1,2} \ebm \bbm \mbf{w}_{1,k} \\ \mbf{w}_{2,k} \ebm + \mbf{B}_{\mathrm{d}2}\mbf{u}_k,  \\ 
\bbm \mbf{z}_{1,k} \\ \mbf{z}_{2,k} \ebm  &= \bbm \mbf{C}_{\mathrm{d}1,1} \\ \mbf{C}_{\mathrm{d}1,2} \ebm \mbf{x}_k + \bbm \mbf{0} & \mbf{D}_{\mathrm{d}11,12} \\ \mbf{D}_{\mathrm{d}11,21} & \mbf{D}_{\mathrm{d}11,22} \ebm \bbm \mbf{w}_{1,k} \\ \mbf{w}_{2,k} \ebm + \bbm \mbf{D}_{\mathrm{d}12,1} \\ \mbf{D}_{\mathrm{d}12,2} \ebm \mbf{u}_k,  \\ 
\mbf{y}_k &=  \mbf{x}_k,   
\end{align*}
where it is assumed that ($\mbf{A}_\mathrm{d}$,$\mbf{B}_{\mathrm{d}2}$) is stabilizable.  A full-state feedback controller $\mbc{K} = \mbf{K}_\mathrm{d} \in \mathbb{R}^{n_u \times n_x}$ (i.e., $\mbf{u}_k = \mbf{K}_\mathrm{d}\mbf{x}_k$) is to be designed to minimize the $\mathcal{H}_2$ norm of the closed loop transfer matrix $\mbf{T}_{11}(z)$ from the exogenous input $\mbf{w}_{1,k}$ to the performance output $\mbf{z}_{1,k}$ while ensuring the $\mathcal{H}_\infty$ norm of the closed-loop transfer matrix $\mbf{T}_{22}(z)$ from the exogenous input $\mbf{w}_{2,k}$ to the performance output $\mbf{z}_{2,k}$ is less than $\gamma_d$, where
 \begin{align*}
\mbf{T}_{11}(z) &= \left(\mbf{C}_{\mathrm{d}1,1} + \mbf{D}_{\mathrm{d}12,1}\mbf{K}_\mathrm{d}\right)\left(z\mbf{1} - \left(\mbf{A}_\mathrm{d}+\mbf{B}_{\mathrm{d}2}\mbf{K}_\mathrm{d}\right)\right)^{-1}\mbf{B}_{\mathrm{d}1,1}, \\
\mbf{T}_{22}(z) &= \left(\mbf{C}_{\mathrm{d}1,2} + \mbf{D}_{\mathrm{d}12,2}\mbf{K}_\mathrm{d}\right)\left(z\mbf{1} - \left(\mbf{A}_\mathrm{d}+\mbf{B}_{\mathrm{d}2}\mbf{K}_\mathrm{d}\right)\right)^{-1}\mbf{B}_{\mathrm{d}1,2} + \mbf{D}_{\mathrm{d}11,22}. 
\end{align*}
\begin{SynthMeth}
The discrete-time mixed $\mathcal{H}_2$-$\mathcal{H}_\infty$-optimal full-state feedback controller is synthesized by solving for $\mbf{P} \in \mathbb{S}^{n_x}$, $\mbf{Z} \in \mathbb{S}^{n_w}$, $\mbf{F}_\mathrm{d} \in \mathbb{R}^{n_u \times n_x}$, and $\nu \in \mathbb{R}_{>0}$ that minimize $\mathcal{J}(\nu) = \nu$ subject to $\mbf{P} >0 $, $\mbf{Z} > 0$, 
\begin{align*}
\bbm \mbf{P} & \mbf{A}_\mathrm{d}\mbf{P} + \mbf{B}_{\mathrm{d}2}\mbf{F}_\mathrm{d} & \mbf{B}_{\text{d}1,1} \\ * & \mbf{P} & \mbf{0} \\ * & * & \mbf{1} \ebm &> 0,\\
\bbm \mbf{P} & \mbf{A}_\mathrm{d}\mbf{P} + \mbf{B}_{\mathrm{d}2}\mbf{F}_\mathrm{d} & \mbf{B}_{\text{d}1,2} & \mbf{0} \\ * & \mbf{P} & \mbf{0} & \mbf{P}\mbf{C}_{\text{d}1,2}^\trans + \mbf{F}_\mathrm{d}^\trans\mbf{D}_{\mathrm{d}12,2}^\trans \\ * & * & \gamma_d \mbf{1} & \mbf{D}_{\text{d}11,22}^\trans \\ * & * & * & \gamma_d \mbf{1} \ebm &>0, \\
\bbm \mbf{Z} & \mbf{C}_{\text{d}1,1}\mbf{P} + \mbf{D}_{\mathrm{d}12,1}\mbf{F}_\mathrm{d} \\ * & \mbf{P} \ebm &> 0. \\
\trace (\mbf{Z}) &< \nu. 
\end{align*}
The $\mathcal{H}_2$-optimal full-state feedback gain is recovered by $\mbf{K}_\mathrm{d} = \mbf{F}_\mathrm{d}\mbf{P}^{-1}$, the $\mathcal{H}_2$ norm of $\mbf{T}_{11}(z)$ is less than $\mu = \sqrt{\nu}$, and the $\mathcal{H}_\infty$ norm of $\mbf{T}_{22}(z)$ is less than $\gamma_d$.
\end{SynthMeth}

\subsubsection[Mixed \texorpdfstring{$\mathcal{H}_2$-$\mathcal{H}_\infty$}{H2 H-Infinity}-Optimal Dynamic Output Feedback Control]{Mixed \texorpdfstring{$\mathcal{H}_2$-$\mathcal{H}_\infty$}{H2 H-Infinity}-Optimal Dynamic Output Feedback Control~\cite{Scherer1997,PeetOptRobNotes11}}
\label{sec:MixedDynSynth}
\index{dynamic output feedback!mixed $\mathcal{H}_2$-$\mathcal{H}_\infty$-optimal}

Consider the continuous-time generalized LTI plant $\mbc{P}$ with minimal state-space realization
\begin{align*}
\dot{\mbf{x}} &= \mbf{A} \mbf{x} + \bbm \mbf{B}_{1,1} & \mbf{B}_{1,2} \ebm \bbm \mbf{w}_1 \\ \mbf{w}_2 \ebm + \mbf{B}_2\mbf{u},  \\ 
\bbm \mbf{z}_1 \\ \mbf{z}_2 \ebm  &= \bbm \mbf{C}_{1,1} \\ \mbf{C}_{1,2} \ebm \mbf{x} + \bbm \mbf{D}_{11,11} & \mbf{D}_{11,12} \\ \mbf{D}_{11,21} & \mbf{D}_{11,22} \ebm \bbm \mbf{w}_1 \\ \mbf{w}_2 \ebm + \bbm \mbf{D}_{12,1} \\ \mbf{D}_{12,2} \ebm \mbf{u},  \\ 
\mbf{y} &= \mbf{C}_2 \mbf{x} + \bbm \mbf{D}_{21,1} & \mbf{D}_{21,2} \ebm \bbm \mbf{w}_1 \\ \mbf{w}_2 \ebm + \mbf{D}_{22} \mbf{u}.   
\end{align*}
A continuous-time dynamic output feedback LTI controller with state-space realization $(\mbf{A}_c,\mbf{B}_c,\mbf{C}_c,\mbf{D}_c)$ is to be designed to minimize the $\mathcal{H}_2$ norm of the closed-loop transfer matrix $\mbf{T}_{11}(s)$ from the exogenous input $\mbf{w}_1$ to the performance output $\mbf{z}_1$ while ensuring the $\mathcal{H}_\infty$ norm of the closed-loop transfer matrix $\mbf{T}_{22}(s)$ from the exogenous input $\mbf{w}_2$ to the performance output $\mbf{z}_2$ is less than $\gamma_d$, where
 \begin{align*}
\mbf{T}_{11}(s) &= \mbf{C}_\dup{\mathrm{CL}}{1,1}\left(s\mbf{1} - \mbf{A}_\du{\mathrm{CL}}\right)^{-1}\mbf{B}_\dup{\mathrm{CL}}{1,1}, \\
\mbf{T}_{22}(s) &= \mbf{C}_\dup{\mathrm{CL}}{1,2} \left(s\mbf{1} - \mbf{A}_\du{\mathrm{CL}}\right)^{-1}\mbf{B}_\dup{\mathrm{CL}}{1,2} + \mbf{D}_\dup{\mathrm{CL}}{11,22}, 
\end{align*}
\begin{align*}
\mbf{A}_\du{\mathrm{CL}} &= \bbm \mbf{A} + \mbf{B}_2 \mbf{D}_c\mbftilde{D}^{-1} \mbf{C}_2 & \mbf{B}_2 \left(\mbf{1} + \mbf{D}_c\mbftilde{D}^{-1}\mbf{D}_{22} \right)\mbf{C}_c \\ \mbf{B}_c  \mbftilde{D}^{-1}\mbf{C}_2 & \mbf{A}_c + \mbf{B}_c \mbftilde{D}^{-1}\mbf{D}_{22}\mbf{C}_c \ebm, \\
\mbf{B}_\dup{\mathrm{CL}}{1,1} &= \bbm  \mbf{B}_{1,1} + \mbf{B}_2\mbf{D}_c\mbftilde{D}^{-1}\mbf{D}_{21,1}  \\ \mbf{B}_c \mbftilde{D}^{-1}\mbf{D}_{21,1} \ebm, \\
\mbf{B}_\dup{\mathrm{CL}}{1,2} &= \bbm  \mbf{B}_{1,2} + \mbf{B}_2\mbf{D}_c\mbftilde{D}^{-1}\mbf{D}_{21,2}  \\ \mbf{B}_c \mbftilde{D}^{-1}\mbf{D}_{21,2} \ebm, \\
\mbf{C}_\dup{\mathrm{CL}}{1,1} &= \bbm \mbf{C}_{1,1} + \mbf{D}_{12,1}\mbf{D}_c\mbftilde{D}^{-1} \mbf{C}_{2,1} & \mbf{D}_{12,1} \left(\mbf{1} + \mbf{D}_c\mbftilde{D}^{-1}\mbf{D}_{22} \right)\mbf{C}_c \ebm, \\
\mbf{C}_\dup{\mathrm{CL}}{1,2} &= \bbm \mbf{C}_{1,2} + \mbf{D}_{12,2}\mbf{D}_c\mbftilde{D}^{-1} \mbf{C}_{2,2} & \mbf{D}_{12,2} \left(\mbf{1} + \mbf{D}_c\mbftilde{D}^{-1}\mbf{D}_{22} \right)\mbf{C}_c \ebm, \\
\mbf{D}_\dup{\mathrm{CL}}{11,22} &= \mbf{D}_{11,22} + \mbf{D}_{12,2} \mbf{D}_c  \mbftilde{D}^{-1}\mbf{D}_{21,2}, 
\end{align*}
and $\mbftilde{D} = \mbf{1} - \mbf{D}_{22}\mbf{D}_c$.

\begin{SynthMeth}
Solve for $\mbf{A}_n \in \mathbb{R}^{n_x \times n_x}$, $\mbf{B}_n \in \mathbb{R}^{n_x \times n_y}$, $\mbf{C}_n \in \mathbb{R}^{n_u \times n_x}$, $\mbf{D}_n \in \mathbb{R}^{n_u \times n_x}$, $\mbf{X}_1$,~$\mbf{Y}_1 \in \mathbb{S}^{n_x}$, $\mbf{Z} \in \mathbb{S}^{n_{z_1}}$, and $\nu \in \mathbb{R}_{>0}$ that minimize $\mathcal{J}(\nu) = \nu$ subject to $\mbf{X}_1 > 0$, $\mbf{Y}_1 > 0$, $\mbf{Z} > 0$,
\begin{align}
\bbm \mbf{N}_{11} & \mbf{A} + \mbf{A}_n^\trans + \mbf{B}_2\mbf{D}_n\mbf{C}_2 & \mbf{B}_{1,1} + \mbf{B}_2\mbf{D}_n\mbf{D}_{21,1}  \\ * & \mbf{X}_1\mbf{A} + \mbf{A}^\trans\mbf{X}_1 + \mbf{B}_n \mbf{C}_2 + \mbf{C}_2^\trans \mbf{B}_n^\trans & \mbf{X}_1\mbf{B}_{1,1} + \mbf{B}_n\mbf{D}_{21,1}  \\ * & * & - \mbf{1} \ebm &< 0, \nonumber\\
\bbm \mbf{N}_{11} & \mbf{A} + \mbf{A}_n^\trans + \mbf{B}_2\mbf{D}_n\mbf{C}_2 & \mbf{B}_{1,2} + \mbf{B}_2\mbf{D}_n\mbf{D}_{21,2} & \mbf{Y}_1\mbf{C}_{1,2}^\trans + \mbf{C}_n^\trans\mbf{D}_{12,2}^\trans \\ * & \mbf{X}_1\mbf{A} + \mbf{A}^\trans\mbf{X}_1 + \mbf{B}_n \mbf{C}_2 + \mbf{C}_2^\trans \mbf{B}_n^\trans & \mbf{X}_1\mbf{B}_{1,2} + \mbf{B}_n\mbf{D}_{21,2} & \mbf{C}_{1,2}^\trans + \mbf{C}_2^\trans\mbf{D}_n^\trans\mbf{D}_{12,2}^\trans \\ * & * & -\gamma_d \mbf{1} & \mbf{D}_{11,22}^\trans + \mbf{D}_{21,2}^\trans\mbf{D}_n^\trans\mbf{D}_{12,2}^\trans \\ * & * & * & -\gamma_d \mbf{1} \ebm &< 0, \nonumber\\
\bbm \mbf{Y}_1 & \mbf{1} & \mbf{Y}_1\mbf{C}_{1,1}^\trans + \mbf{C}_n^\trans\mbf{D}_{12,1}^\trans \\ * & \mbf{X}_1 & \mbf{C}_{1,1}^\trans + \mbf{C}_2^\trans \mbf{D}_n^\trans\mbf{D}_{12,1}^\trans \\ * & * & \mbf{Z} \ebm &> 0, \nonumber\\
\bbm \mbf{X}_1 & \mbf{1} \\ * & \mbf{Y}_1 \ebm &> 0, \nonumber\\
\mbf{D}_{11,11} + \mbf{D}_{12,1}\mbf{D}_n\mbf{D}_{21,1} &= \mbf{0}, \label{eq:H2Hinf_D}\\
\trace (\mbf{Z}) &< \nu, \nonumber
\end{align}
where $\mbf{N}_{11} = \mbf{A} \mbf{Y}_1 + \mbf{Y}_1\mbf{A}^\trans + \mbf{B}_2\mbf{C}_n + \mbf{C}_n^\trans\mbf{B}_2^\trans$.  The controller is recovered by
\begin{align*}
\mbf{A}_c &= \mbf{A}_\du{K} - \mbf{B}_c\left(\mbf{1} - \mbf{D}_{22}\mbf{D}_c\right)^{-1}\mbf{D}_{22}\mbf{C}_c, \\
\mbf{B}_c &= \mbf{B}_\du{K}\left(\mbf{1} - \mbf{D}_{22}\mbf{D}_c\right), \\
\mbf{C}_c &= \left(\mbf{1} - \mbf{D}_c\mbf{D}_{22}\right)\mbf{C}_\du{K}, \\
\mbf{D}_c &= \left(\mbf{1} + \mbf{D}_\du{K}\mbf{D}_{22}\right)^{-1}\mbf{D}_\du{K},
\end{align*}
where
\bdis
\bbm \mbf{A}_\du{K} & \mbf{B}_\du{K} \\ \mbf{C}_\du{K} & \mbf{D}_\du{K} \ebm = \bbm \mbf{X}_2 & \mbf{X}_1\mbf{B}_2 \\ \mbf{0} & \mbf{1} \ebm^{-1} \left( \bbm \mbf{A}_n & \mbf{B}_n \\ \mbf{C}_n & \mbf{D}_n \ebm - \bbm \mbf{X}_1\mbf{A}\mbf{Y}_1 & \mbf{0} \\ \mbf{0} & \mbf{0}\ebm \right)\bbm \mbf{Y}_2^\trans & \mbf{0} \\ \mbf{C}_2 \mbf{Y}_1 & \mbf{1} \ebm^{-1},
\edis
and the matrices $\mbf{X}_2$ and $\mbf{Y}_2$ satisfy $\mbf{X}_2\mbf{Y}_2^\trans = \mbf{1} - \mbf{X}_1\mbf{Y}_1$.  If $\mbf{D}_{22} = \mbf{0}$, then $\mbf{A}_c = \mbf{A}_\du{K}$, $\mbf{B}_c = \mbf{B}_\du{K}$, $\mbf{C}_c = \mbf{C}_\du{K}$, and $\mbf{D}_c = \mbf{D}_\du{K}$.

Given $\mbf{X}_1$ and $\mbf{Y}_1$, the matrices $\mbf{X}_2$ and $\mbf{Y}_2$ can be found using a matrix decomposition, such as a LU decomposition or a Cholesky decomposition.

If $\mbf{D}_{11,11} = \mbf{0}$, $\mbf{D}_{12,1} \neq \mbf{0}$, and $\mbf{D}_{21,1} \neq \mbf{0}$, then it is often simplest to choose $\mbf{D}_n = \mbf{0}$ in order to satisfy the equality constraint of~\eqref{eq:H2Hinf_D}.
\end{SynthMeth}

\subsubsection[Discrete-Time Mixed \texorpdfstring{$\mathcal{H}_2$-$\mathcal{H}_\infty$}{H2 H-Infinity}-Optimal Dynamic Output Feedback Control]{Discrete-Time Mixed \texorpdfstring{$\mathcal{H}_2$-$\mathcal{H}_\infty$}{H2 H-Infinity}-Optimal Dynamic Output Feedback Control}
\label{sec:DT_MixedDynSynth}
\index{dynamic output feedback!discrete-time mixed $\mathcal{H}_2$-$\mathcal{H}_\infty$-optimal}

Consider the discrete-time generalized LTI plant $\mbc{P}$ with minimal state-space realization
\begin{align*}
\mbf{x}_{k+1} &= \mbf{A}_\mathrm{d} \mbf{x}_k + \bbm \mbf{B}_{\mathrm{d}1,1} & \mbf{B}_{\mathrm{d}1,2} \ebm \bbm \mbf{w}_{1,k} \\ \mbf{w}_{2,k} \ebm + \mbf{B}_{\mathrm{d}2}\mbf{u}_k,  \\ 
\bbm \mbf{z}_{1,k} \\ \mbf{z}_{2,k} \ebm  &= \bbm \mbf{C}_{\mathrm{d}1,1} \\ \mbf{C}_{\mathrm{d}1,2} \ebm \mbf{x}_k + \bbm \mbf{D}_{\mathrm{d}11,11} & \mbf{D}_{\mathrm{d}11,12} \\ \mbf{D}_{\mathrm{d}11,21} & \mbf{D}_{\mathrm{d}11,22} \ebm \bbm \mbf{w}_{1,k} \\ \mbf{w}_{2,k} \ebm + \bbm \mbf{D}_{\mathrm{d}12,1} \\ \mbf{D}_{\mathrm{d}12,2} \ebm \mbf{u}_k,  \\ 
\mbf{y}_k &= \mbf{C}_{\mathrm{d}2} \mbf{x}_k + \bbm \mbf{D}_{\mathrm{d}21,1} & \mbf{D}_{\mathrm{d}21,2} \ebm \bbm \mbf{w}_{1,k} \\ \mbf{w}_{2,k} \ebm + \mbf{D}_{\mathrm{d}22} \mbf{u}_k.   
\end{align*}
A discrete-time dynamic output feedback LTI controller with state-space realization $(\mbf{A}_{\mathrm{d}c},\mbf{B}_{\mathrm{d}c},\mbf{C}_{\mathrm{d}c},\mbf{D}_{\mathrm{d}c})$ is to be designed to minimize the $\mathcal{H}_2$ norm of the closed loop transfer matrix $\mbf{T}_{11}(z)$ from the exogenous input $\mbf{w}_{1,k}$ to the performance output $\mbf{z}_{1,k}$ while ensuring the $\mathcal{H}_\infty$ norm of the closed-loop transfer matrix $\mbf{T}_{22}(z)$ from the exogenous input $\mbf{w}_{2,k}$ to the performance output $\mbf{z}_{2,k}$ is less than $\gamma_d$, where
 \begin{align*}
\mbf{T}_{11}(z) &= \mbf{C}_{\mathrm{d}_\mathrm{CL}1,1}\left(z\mbf{1} - \mbf{A}_{\mathrm{d}_\mathrm{CL}}\right)^{-1}\mbf{B}_{\mathrm{d}_\mathrm{CL}1,1}, \\
\mbf{T}_{22}(z) &= \mbf{C}_{\mathrm{d}_\mathrm{CL}1,2} \left(z\mbf{1} - \mbf{A}_{\mathrm{d}_\mathrm{CL}}\right)^{-1}\mbf{B}_{\mathrm{d}_\mathrm{CL}1,2} + \mbf{D}_{\mathrm{d}_\mathrm{CL}11,22}, 
\end{align*}
\begin{align*}
\mbf{A}_{\mathrm{d}_\mathrm{CL}} &= \bbm \mbf{A}_\mathrm{d} + \mbf{B}_{\mathrm{d}2} \mbf{D}_{\mathrm{d}c}\mbftilde{D}_\mathrm{d}^{-1} \mbf{C}_{\mathrm{d}2} & \mbf{B}_{\mathrm{d}2} \left(\mbf{1} + \mbf{D}_{\mathrm{d}c}\mbftilde{D}_\mathrm{d}^{-1}\mbf{D}_{\mathrm{d}22} \right)\mbf{C}_{\mathrm{d}c} \\ \mbf{B}_{\mathrm{d}c}  \mbftilde{D}_\mathrm{d}^{-1}\mbf{C}_{\mathrm{d}2} & \mbf{A}_{\mathrm{d}c} + \mbf{B}_{\mathrm{d}c} \mbftilde{D}_\mathrm{d}^{-1}\mbf{D}_{\mathrm{d}22}\mbf{C}_{\mathrm{d}c} \ebm, \\
\mbf{B}_{\mathrm{d}_\mathrm{CL}1,1} &= \bbm  \mbf{B}_{\mathrm{d}1,1} + \mbf{B}_{\mathrm{d}2}\mbf{D}_{\mathrm{d}c}\mbftilde{D}_\mathrm{d}^{-1}\mbf{D}_{\mathrm{d}21,1}  \\ \mbf{B}_{\mathrm{d}c} \mbftilde{D}_\mathrm{d}^{-1}\mbf{D}_{\mathrm{d}21,1} \ebm, \\
\mbf{B}_{\mathrm{d}_\mathrm{CL}1,2} &= \bbm  \mbf{B}_{\mathrm{d}1,2} + \mbf{B}_{\mathrm{d}2}\mbf{D}_{\mathrm{d}c}\mbftilde{D}_\mathrm{d}^{-1}\mbf{D}_{\mathrm{d}21,2}  \\ \mbf{B}_{\mathrm{d}c} \mbftilde{D}_\mathrm{d}^{-1}\mbf{D}_{\mathrm{d}21,2} \ebm, \\
\mbf{C}_{\mathrm{d}_\mathrm{CL}1,1} &= \bbm \mbf{C}_{\mathrm{d}1,1} + \mbf{D}_{\mathrm{d}12,1}\mbf{D}_{\mathrm{d}c}\mbftilde{D}_\mathrm{d}^{-1} \mbf{C}_{\mathrm{d}2,1} & \mbf{D}_{\mathrm{d}12,1} \left(\mbf{1} + \mbf{D}_{\mathrm{d}c}\mbftilde{D}_\mathrm{d}^{-1}\mbf{D}_{\mathrm{d}22} \right)\mbf{C}_{\mathrm{d}c} \ebm, \\
\mbf{C}_{\mathrm{d}_\mathrm{CL}1,2} &= \bbm \mbf{C}_{\mathrm{d}1,2} + \mbf{D}_{\mathrm{d}12,2}\mbf{D}_{\mathrm{d}c}\mbftilde{D}_\mathrm{d}^{-1} \mbf{C}_{\mathrm{d}2,2} & \mbf{D}_{\mathrm{d}12,2} \left(\mbf{1} + \mbf{D}_{\mathrm{d}c}\mbftilde{D}_\mathrm{d}^{-1}\mbf{D}_{\mathrm{d}22} \right)\mbf{C}_{\mathrm{d}c} \ebm, \\
\mbf{D}_{\mathrm{d}_\mathrm{CL}11,22} &= \mbf{D}_{\mathrm{d}11,22} + \mbf{D}_{\mathrm{d}12,2} \mbf{D}_{\mathrm{d}c}  \mbftilde{D}_\mathrm{d}^{-1}\mbf{D}_{\mathrm{d}21,2}, 
\end{align*}
and $\mbftilde{D}_\mathrm{d} = \mbf{1} - \mbf{D}_{\mathrm{d}22}\mbf{D}_{\mathrm{d}c}$.

\begin{SynthMeth}
Solve for $\mbf{A}_{\mathrm{d}n} \in \mathbb{R}^{n_x \times n_x}$, $\mbf{B}_{\mathrm{d}n} \in \mathbb{R}^{n_x \times n_y}$, $\mbf{C}_{\mathrm{d}n} \in \mathbb{R}^{n_u \times n_x}$, $\mbf{D}_{\mathrm{d}n} \in \mathbb{R}^{n_u \times n_y}$, $\mbf{X}_1$,~$\mbf{Y}_1 \in \mathbb{S}^{n_x}$, $\mbf{Z} \in \mathbb{S}^{n_{z_1}}$, and $\nu \in \mathbb{R}_{>0}$ that minimize $\mathcal{J}(\nu) = \nu$ subject to $\mbf{X}_1 > 0$, $\mbf{Y}_1 > 0$, $\mbf{Z} >0$,
\begin{align}
\bbm \mbf{X}_1 & \mbf{1} & \mbf{X}_1 \mbf{A}_\mathrm{d} + \mbf{B}_{\mathrm{d}n}\mbf{C}_{\mathrm{d}2} & \mbf{A}_{\mathrm{d}n} & \mbf{X}_1 \mbf{B}_{\mathrm{d}1,1} + \mbf{B}_{\mathrm{d}n} \mbf{D}_{\mathrm{d}21,1}   \\ *  & \mbf{Y}_1 & \mbf{A}_\mathrm{d} + \mbf{B}_{\mathrm{d}2} \mbf{D}_{\mathrm{d}n} \mbf{C}_{\mathrm{d}2} & \mbf{A}_\mathrm{d} \mbf{Y}_1 + \mbf{B}_{\mathrm{d}2} \mbf{C}_{\mathrm{d}n} & \mbf{B}_{\mathrm{d}1,1} + \mbf{B}_{\mathrm{d}2} \mbf{D}_{\mathrm{d}n} \mbf{D}_{\mathrm{d}21,1}  \\ * & * & \mbf{X}_1 & \mbf{1} & \mbf{0}  \\ * & * & * & \mbf{Y}_1 & \mbf{0}  \\ * & * & * & * &  \mbf{1}  \ebm &> 0, \nonumber\\
\bbm \mbf{X}_1 & \mbf{1} & \mbf{X}_1 \mbf{A}_\mathrm{d} + \mbf{B}_{\mathrm{d}n}\mbf{C}_{\mathrm{d}2} & \mbf{A}_{\mathrm{d}n} & \mbf{X}_1 \mbf{B}_{\mathrm{d}1,2} + \mbf{B}_{\mathrm{d}n} \mbf{D}_{\mathrm{d}21,2} & \mbf{0}  \\ *  & \mbf{Y}_1 & \mbf{A}_\mathrm{d} + \mbf{B}_{\mathrm{d}2} \mbf{D}_{\mathrm{d}n} \mbf{C}_{\mathrm{d}2} & \mbf{A}_\mathrm{d} \mbf{Y}_1 + \mbf{B}_{\mathrm{d}2} \mbf{C}_{\mathrm{d}n} & \mbf{B}_{\mathrm{d}1,2} + \mbf{B}_{\mathrm{d}2} \mbf{D}_{\mathrm{d}n} \mbf{D}_{\mathrm{d}21,2} & \mbf{0} \\ * & * & \mbf{X}_1 & \mbf{1} & \mbf{0} & \mbf{C}_{\mathrm{d}1,2}^\trans + \mbf{C}_{\mathrm{d}2}^\trans \mbf{D}_{\mathrm{d}n}^\trans \mbf{D}_{\mathrm{d}12,2}^\trans \\ * & * & * & \mbf{Y}_1 & \mbf{0} & \mbf{Y}_1 \mbf{C}_{\mathrm{d}1,2}^\trans + \mbf{C}_{\mathrm{d}n}^\trans \mbf{D}_{\mathrm{d}12,2}^\trans \\ * & * & * & * & \gamma_d \mbf{1} & \mbf{D}_{\mathrm{d}11,22}^\trans + \mbf{D}_{\mathrm{d}21,2}^\trans\mbf{D}_{\mathrm{d}n}^\trans\mbf{D}_{\mathrm{d}12,2}^\trans \\ * & * & * & * & * & \gamma_d \mbf{1} \ebm &> 0, \nonumber \\
\bbm \mbf{Z} & \mbf{C}_{\mathrm{d}1,1} + \mbf{D}_{\mathrm{d}12,1} \mbf{D}_{\mathrm{d}n}\mbf{C}_{\mathrm{d}2} & \mbf{C}_{\mathrm{d}1,1}\mbf{Y}_1 + \mbf{D}_{\mathrm{d}12,1}\mbf{C}_{\mathrm{d}n} \\ * & \mbf{X}_1 & \mbf{1} \\ * & * & \mbf{Y}_1 \ebm &>0, \label{eq:H2Hinf_D_DTa}\\
\mbf{D}_{\mathrm{d}11,11} + \mbf{D}_{\mathrm{d}12,1} \mbf{D}_{\mathrm{d}n}  \mbf{D}_{\mathrm{d}21,1} &= \mbf{0}, \label{eq:H2Hinf_D_DT}\\
\bbm \mbf{X}_1 & \mbf{1} \\ * & \mbf{Y}_1 \ebm &> 0, \nonumber\\
\trace (\mbf{Z}) &< \nu. \nonumber
\end{align}
The controller is recovered by
\begin{align*}
\mbf{A}_{\mathrm{d}c} &= \mbf{A}_{\mathrm{d}_K} - \mbf{B}_{\mathrm{d}c}\left(\mbf{1} - \mbf{D}_{\mathrm{d}22}\mbf{D}_{\mathrm{d}c}\right)^{-1}\mbf{D}_{\mathrm{d}22}\mbf{C}_{\mathrm{d}c}, \\
\mbf{B}_{\mathrm{d}c} &= \mbf{B}_{\mathrm{d}_K}\left(\mbf{1} - \mbf{D}_{\mathrm{d}22}\mbf{D}_{\mathrm{d}c}\right), \\
\mbf{C}_{\mathrm{d}c} &= \left(\mbf{1} - \mbf{D}_{\mathrm{d}c}\mbf{D}_{\mathrm{d}22}\right)\mbf{C}_{\mathrm{d}_K}, \\
\mbf{D}_{\mathrm{d}c} &= \left(\mbf{1} + \mbf{D}_{\mathrm{d}_K}\mbf{D}_{\mathrm{d}22}\right)^{-1}\mbf{D}_{\mathrm{d}_K},
\end{align*}
where
\bdis
\bbm \mbf{A}_{\mathrm{d}_K} & \mbf{B}_{\mathrm{d}_K} \\ \mbf{C}_{\mathrm{d}_K} & \mbf{D}_{\mathrm{d}_K} \ebm = \bbm \mbf{X}_2 & \mbf{X}_1\mbf{B}_{\mathrm{d}2} \\ \mbf{0} & \mbf{1} \ebm^{-1} \left( \bbm \mbf{A}_{\mathrm{d}n}& \mbf{B}_{\mathrm{d}n} \\ \mbf{C}_{\mathrm{d}n} & \mbf{D}_{\mathrm{d}n} \ebm - \bbm \mbf{X}_1\mbf{A}_\mathrm{d}\mbf{Y}_1 & \mbf{0} \\ \mbf{0} & \mbf{0}\ebm \right)\bbm \mbf{Y}_2^\trans & \mbf{0} \\ \mbf{C}_{\mathrm{d}2} \mbf{Y}_1 & \mbf{1} \ebm^{-1},
\edis
and the matrices $\mbf{X}_2$ and $\mbf{Y}_2$ satisfy $\mbf{X}_2\mbf{Y}_2^\trans = \mbf{1} - \mbf{X}_1\mbf{Y}_1$.  If $\mbf{D}_{\mathrm{d}22} = \mbf{0}$, then $\mbf{A}_{\mathrm{d}c} = \mbf{A}_{\mathrm{d}_K}$, $\mbf{B}_{\mathrm{d}c} = \mbf{B}_{\mathrm{d}_K}$, $\mbf{C}_{\mathrm{d}c} = \mbf{C}_{\mathrm{d}_K}$, and $\mbf{D}_{\mathrm{d}c} = \mbf{D}_{\mathrm{d}_K}$.

Given $\mbf{X}_1$ and $\mbf{Y}_1$, the matrices $\mbf{X}_2$ and $\mbf{Y}_2$ can be found using a matrix decomposition, such as a LU decomposition or a Cholesky decomposition.

If $\mbf{D}_{\mathrm{d}11,11} = \mbf{0}$, $\mbf{D}_{\mathrm{d}12,1} \neq \mbf{0}$, and $\mbf{D}_{\mathrm{d}21,1} \neq \mbf{0}$, then it is often simplest to choose $\mbf{D}_{\mathrm{d}n} = \mbf{0}$ in order to satisfy the equality constraint of~\eqref{eq:H2Hinf_D_DT}.

An alternate formulation of this synthesis method involves replacing~\eqref{eq:H2Hinf_D_DTa} and~\eqref{eq:H2Hinf_D_DT} with
\beq
\label{eq:H2Hinf_D_DTd}
\bbm \mbf{Z} & \mbf{C}_{\mathrm{d}1,1} + \mbf{D}_{\mathrm{d}12,1} \mbf{D}_{\mathrm{d}n}\mbf{C}_{\mathrm{d}2} & \mbf{C}_{\mathrm{d}1,1}\mbf{Y}_1 + \mbf{D}_{\mathrm{d}12,1}\mbf{C}_{\mathrm{d}n} & \mbf{D}_{\mathrm{d}11,11} + \mbf{D}_{\mathrm{d}12,1} \mbf{D}_{\mathrm{d}n}  \mbf{D}_{\mathrm{d}21,1}\\ * & \mbf{X}_1 & \mbf{1} & \mbf{0} \\ * & * & \mbf{Y}_1 & \mbf{0} \\ * & * & * & \mbf{1} \ebm >0.
\eeq
The matrix inequality in~\eqref{eq:H2Hinf_D_DTd} is derived by performing the same procedure used in~\cite{DeOliveira2002} with the change of variables $\mbf{S} = \mbf{J} = \mbf{1}$, $\mbf{H} = \mbf{X}_1$, $\mbf{G} = \mbf{Y}_1$, but instead starting with the matrix inequality formulation of the $\mathcal{H}_2$ that allows for a non-zero feedthrough term in~\cite[p.~25]{Santos2017} (summarized by~\eqref{eq:DT_D_H2_4a},~\eqref{eq:DT_D_H2_4b}, and~\eqref{eq:DT_D_H2_4c}).  In general, the matrix inequality in~\eqref{eq:H2Hinf_D_DTd} is less conservative than~\eqref{eq:H2Hinf_D_DTa} and~\eqref{eq:H2Hinf_D_DT}, as it allows for the resulting closed-loop system to have non-zero feedthrough, which, for a discrete-time system, is possible while maintaining a finite $\mathcal{H}_2$ norm.
\end{SynthMeth}

\clearpage
\section{LMIs in Optimal Estimation and Filtering}
\label{sec:OptimalEstimation}

This section presents controller synthesis methods using LMIs for a number of well-known optimal state-estimation and filtering problems.  The derivation of the LMIs used for synthesis is provided in some cases, while longer derivations can be found in the cited references.

\subsection{\texorpdfstring{$\mathcal{H}_2$}{H2}-Optimal State Estimation}
\index{estimation!$\mathcal{H}_2$-optimal}

The goal of $\mathcal{H}_2$-optimal state estimation is to design an observer that minimizes the $\mathcal{H}_2$ norm of the closed-loop transfer matrix from $\mbf{w}$ to $\mbf{z}$.

\subsubsection[\texorpdfstring{$\mathcal{H}_2$}{H2}-Optimal Observer]{\texorpdfstring{$\mathcal{H}_2$}{H2}-Optimal Observer~\cite[p.~296]{Duan2013}}
\index{observer!$\mathcal{H}_2$-optimal}
Consider the continuous-time generalized plant $\mbc{P}$ with state-space realization
\begin{align*}
\dot{\mbf{x}} &= \mbf{A} \mbf{x} + \mbf{B}_1\mbf{w}, \\
\mbf{y} &= \mbf{C}_2\mbf{x} + \mbf{D}_{21}\mbf{w} ,
\end{align*}
where it is assumed that ($\mbf{A}$,$\mbf{C}_2$) is detectable.  An observer of the form
\begin{align*}
\dot{\mbfhat{x}} &= \mbf{A} \mbfhat{x} + \mbf{L}\left(\mbf{y} - \mbfhat{y}\right), \\
\mbfhat{y} &= \mbf{C}_2\mbfhat{x} ,
\end{align*}
is to be designed, where $\mbf{L} \in \mathbb{R}^{n_x \times n_y}$ is the observer gain.  Defining the error state $\mbf{e} = \mbf{x} - \mbfhat{x}$, the error dynamics are found to be
\bdis
\dot{\mbf{e}} = \left(\mbf{A}-\mbf{L}\mbf{C}_2\right)\mbf{e} + \left(\mbf{B}_1 - \mbf{L}\mbf{D}_{21}\right)\mbf{w}, 
\edis
and the performance output is defined as
\bdis
\mbf{z} = \mbf{C}_1 \mbf{e}.
\edis
The observer gain $\mbf{L}$ is to be designed such that the $\mathcal{H}_2$ norm of the transfer matrix from $\mbf{w}$ to $\mbf{z}$, given by
\bdis
\mbf{T}(s) = \mbf{C}_1\left(s\mbf{1} - \left(\mbf{A}-\mbf{L}\mbf{C}_2\right)\right)^{-1}\left(\mbf{B}_1 - \mbf{L}\mbf{D}_{21}\right),
\edis
is minimized.  Minimizing the $\mathcal{H}_2$ norm of the transfer matrix $\mbf{T}(s)$ is equivalent to minimizing $\mathcal{J}(\mu) = \mu^2$ subject to 
\begin{align}
\bbm \mbf{P}\left(\mbf{A}-\mbf{L}\mbf{C}_2\right) + \left(\mbf{A}-\mbf{L}\mbf{C}_2\right)^\trans\mbf{P} & \mbf{P}\left(\mbf{B}_1 - \mbf{L}\mbf{D}_{21}\right) \\ * & -\mbf{1} \ebm &< 0, \label{eq:LQE3a}\\
\bbm \mbf{P} & \mbf{C}_1^\trans \\ * & \mbf{Z} \ebm &> 0, \label{eq:LQE3b} \\
\trace (\mbf{Z}) &< \mu^2, \label{eq:LQE3c}
\end{align}
where $\mbf{P} \in \mathbb{S}^{n_x}$, $\mbf{Z} \in \mathbb{S}^{n_z}$, $\mu \in \mathbb{R}_{>0}$, $\mbf{P} > 0$, and $\mbf{Z} > 0$.  A change of variables is performed with $\mbf{G} = \mbf{P}\mbf{L}$ and $\nu = \mu^2$, which transforms~\eqref{eq:LQE3a} and~\eqref{eq:LQE3c} into LMIs in the variables $\mbf{P}$, $\mbf{G}$, $\mbf{Z}$, and $\nu$ given by
\begin{align}
\bbm  \mbf{P}\mbf{A} + \mbf{A}^\trans\mbf{P} - \mbf{G}\mbf{C}_2 - \mbf{C}_2^\trans\mbf{G}^\trans & \mbf{P}\mbf{B}_1 - \mbf{G}\mbf{D}_{21} \\ * & -\mbf{1} \ebm &< 0, \label{eq:LQE4a}\\
\trace (\mbf{Z}) &< \nu. \label{eq:LQE4c}
\end{align}
\begin{SynthMeth}
The $\mathcal{H}_2$-optimal observer gain is synthesized by solving for $\mbf{P} \in \mathbb{S}^{n_x}$, $\mbf{Z} \in \mathbb{S}^{n_z}$, $\mbf{G} \in \mathbb{R}^{n_x \times n_y}$, and $\nu \in \mathbb{R}_{>0}$ that minimize $\mathcal{J}(\nu) = \nu$ subject to $\mbf{P} >0 $, $\mbf{Z} > 0$,~\eqref{eq:LQE3b},~\eqref{eq:LQE4a}, and~\eqref{eq:LQE4c}.  The $\mathcal{H}_2$-optimal observer gain is recovered by $\mbf{L} = \mbf{P}^{-1}\mbf{G}$ and the $\mathcal{H}_2$ norm of $\mbf{T}(s)$ is $\mu = \sqrt{\nu}$.
\end{SynthMeth}

\subsubsection{Discrete-Time \texorpdfstring{$\mathcal{H}_2$}{H2}-Optimal Observer}
\label{sec:DT_H2_Observer}
\index{observer!discrete-time $\mathcal{H}_2$-optimal}
Consider the discrete-time generalized LTI plant $\mbc{P}$ with state-space realization
\begin{align*}
\mbf{x}_{k+1} &= \mbf{A}_\mathrm{d} \mbf{x}_k + \mbf{B}_{\mathrm{d}1}\mbf{w}_k,  \\
\mbf{y}_k &= \mbf{C}_{\mathrm{d}2}\mbf{x}_k + \mbf{D}_{\mathrm{d}21}\mbf{w}_k , 
\end{align*}
where it is assumed that ($\mbf{A}_\mathrm{d}$,$\mbf{C}_{\mathrm{d}2}$) is detectable.  An observer of the form
\begin{align*}
\mbfhat{x}_{k+1} &= \mbf{A}_\mathrm{d} \mbfhat{x}_k + \mbf{L}_\mathrm{d}\left(\mbf{y}_k - \mbfhat{y}_k\right), \\
\mbfhat{y}_k &= \mbf{C}_{\mathrm{d}2}\mbfhat{x}_k , 
\end{align*}
is to be designed, where $\mbf{L}_\mathrm{d} \in \mathbb{R}^{n_x \times n_y}$ is the observer gain.  Defining the error state $\mbf{e}_k = \mbf{x}_k - \mbfhat{x}_k$, the error dynamics are found to be
\bdis
\mbf{e}_{k+1} = \left(\mbf{A}_\mathrm{d}-\mbf{L}_\mathrm{d}\mbf{C}_{\mathrm{d}2}\right)\mbf{e}_k + \left(\mbf{B}_{\mathrm{d}1} - \mbf{L}_\mathrm{d}\mbf{D}_{\mathrm{d}21}\right)\mbf{w}_k, \\
\edis
and the performance output is defined as
\bdis
\mbf{z}_k = \mbf{C}_{\mathrm{d}1} \mbf{e}_k.
\edis
The observer gain $\mbf{L}_\mathrm{d}$ is to be designed such that the $\mathcal{H}_2$ of the transfer matrix from $\mbf{w}_k$ to $\mbf{z}_k$, given by
\bdis
\mbf{T}(z) = \mbf{C}_{\mathrm{d}1}\left(z\mbf{1} - \left(\mbf{A}_\mathrm{d}-\mbf{L}_\mathrm{d}\mbf{C}_{\mathrm{d}2}\right)\right)^{-1}\left(\mbf{B}_{\mathrm{d}1} - \mbf{L}_\mathrm{d}\mbf{D}_{\mathrm{d}21}\right),
\edis
is minimized.

\begin{SynthMeth}
The discrete-time $\mathcal{H}_2$-optimal observer gain is synthesized by solving for $\mbf{P} \in \mathbb{S}^{n_x}$, $\mbf{Z} \in \mathbb{S}^{n_z}$, $\mbf{G}_\mathrm{d} \in \mathbb{R}^{n_x \times n_y}$, and $\nu \in \mathbb{R}_{>0}$ that minimize $\mathcal{J}(\nu) = \nu$ subject to $\mbf{P} >0 $, $\mbf{Z} > 0$,
\begin{align*}
\bbm \mbf{P} & \mbf{P}\mbf{A}_\mathrm{d} - \mbf{G}_\mathrm{d}\mbf{C}_{\mathrm{d}2} & \mbf{P}\mbf{B}_{\text{d}1} - \mbf{G}_\mathrm{d}\mbf{D}_{\mathrm{d}21} \\ * & \mbf{P} & \mbf{0} \\ * & * & \mbf{1} \ebm &> 0,\\
\bbm \mbf{Z} & \mbf{C}_{\text{d}1} \\ * & \mbf{P} \ebm &> 0. \\
\trace (\mbf{Z}) &< \nu. 
\end{align*}
The $\mathcal{H}_2$-optimal observer gain is recovered by $\mbf{L}_\mathrm{d} = \mbf{P}^{-1}\mbf{G}_\mathrm{d}$ and the $\mathcal{H}_2$ norm of $\mbf{T}(z)$ is $\mu = \sqrt{\nu}$.
\end{SynthMeth}

\subsection{\texorpdfstring{$\mathcal{H}_\infty$}{H-inf}-Optimal State Estimation}
\index{estimation!$\mathcal{H}_\infty$-optimal}

The goal of $\mathcal{H}_\infty$-optimal state estimation is to design an observer that minimizes the $\mathcal{H}_\infty$ norm of the closed-loop transfer matrix from $\mbf{w}$ to $\mbf{z}$.

\subsubsection[\texorpdfstring{$\mathcal{H}_\infty$}{H-inf}--Optimal Observer]{\texorpdfstring{$\mathcal{H}_\infty$}{H-inf}--Optimal Observer~\cite[p.~295]{Duan2013}}
\index{observer!$\mathcal{H}_\infty$-optimal}
Consider the continuous-time generalized plant $\mbc{P}$ with state-space realization
\begin{align*}
\dot{\mbf{x}} &= \mbf{A} \mbf{x} + \mbf{B}_1\mbf{w}, \\
\mbf{y} &= \mbf{C}_2\mbf{x} + \mbf{D}_{21}\mbf{w} ,
\end{align*}
where it is assumed that ($\mbf{A}$,$\mbf{C}_2$) is detectable.  An observer of the form
\begin{align*}
\dot{\mbfhat{x}} &= \mbf{A} \mbfhat{x} + \mbf{L}\left(\mbf{y} - \mbfhat{y}\right), \\
\mbfhat{y} &= \mbf{C}_2\mbfhat{x} ,
\end{align*}
is to be designed, where $\mbf{L} \in \mathbb{R}^{n_x \times n_y}$ is the observer gain.  Defining the error state $\mbf{e} = \mbf{x} - \mbfhat{x}$, the error dynamics are found to be
\bdis
\dot{\mbf{e}} = \left(\mbf{A}-\mbf{L}\mbf{C}_2\right)\mbf{e} + \left(\mbf{B}_1 - \mbf{L}\mbf{D}_{21}\right)\mbf{w},
\edis
and the performance output is defined as
\bdis
\mbf{z} = \mbf{C}_1 \mbf{e} + \mbf{D}_{11} \mbf{w}.
\edis
The observer gain $\mbf{L}$ is to be designed such that the $\mathcal{H}_\infty$ of the transfer matrix from $\mbf{w}$ to $\mbf{z}$, given by
\bdis
\mbf{T}(s) = \mbf{C}_1\left(s\mbf{1} - \left(\mbf{A}-\mbf{L}\mbf{C}_2\right)\right)^{-1}\left(\mbf{B}_1 - \mbf{L}\mbf{D}_{21}\right) + \mbf{D}_{11},
\edis
is minimized.
\begin{SynthMeth}
The $\mathcal{H}_\infty$-optimal observer gain is synthesized by solving for $\mbf{P} \in \mathbb{S}^{n_x}$, $\mbf{G} \in \mathbb{R}^{n_x \times n_y}$, and $\gamma \in \mathbb{R}_{>0}$ that minimize $\mathcal{J}(\gamma) = \gamma$ subject to $\mbf{P} >0 $ and
\bdis
\bbm  \mbf{P}\mbf{A} + \mbf{A}^\trans\mbf{P} - \mbf{G}\mbf{C}_2 - \mbf{C}_2^\trans\mbf{G}^\trans  & \mbf{P}\mbf{B}_1 - \mbf{G}\mbf{D}_{21} & \mbf{C}_1^\trans \\ *& -\gamma \mbf{1} & \mbf{D}_{11}^\trans \\ * & * & -\gamma \mbf{1} \ebm < 0.
\edis
The $\mathcal{H}_\infty$-optimal observer gain is recovered by $\mbf{L} = \mbf{P}^{-1}\mbf{G}$ and the $\mathcal{H}_\infty$ norm of $\mbf{T}(s)$ is $\gamma$.
\end{SynthMeth}

\subsubsection{Discrete-Time \texorpdfstring{$\mathcal{H}_\infty$}{H-inf}--Optimal Observer}
\index{observer!discrete-time$\mathcal{H}_\infty$-optimal}
Consider the discrete-time LTI plant $\mbc{G}$ with state-space realization
\begin{align*}
\mbf{x}_{k+1} &= \mbf{A}_\mathrm{d} \mbf{x}_k + \mbf{B}_{\mathrm{d}1}\mbf{w}_k,  \\
\mbf{y}_k &= \mbf{C}_{\mathrm{d}2}\mbf{x}_k + \mbf{D}_{\mathrm{d}21}\mbf{w}_k , \nonumber
\end{align*}
where it is assumed that ($\mbf{A}_\mathrm{d}$,$\mbf{C}_{\mathrm{d}2}$) is detectable.  An observer of the form
\begin{align*}
\mbfhat{x}_{k+1} &= \mbf{A}_\mathrm{d} \mbfhat{x}_k + \mbf{L}_\mathrm{d}\left(\mbf{y}_k - \mbfhat{y}_k\right), \\
\mbfhat{y}_k &= \mbf{C}_{\mathrm{d}2}\mbfhat{x}_k , 
\end{align*}
is to be designed, where $\mbf{L}_\mathrm{d} \in \mathbb{R}^{n_x \times n_y}$ is the observer gain.  Defining the error state $\mbf{e}_k = \mbf{x}_k - \mbfhat{x}_k$, the error dynamics are found to be
\bdis
\mbf{e}_{k+1}= \left(\mbf{A}_\mathrm{d}-\mbf{L}_\mathrm{d}\mbf{C}_{\mathrm{d}2}\right)\mbf{e}_k + \left(\mbf{B}_{\mathrm{d}1} - \mbf{L}_\mathrm{d}\mbf{D}_{\mathrm{d}21}\right)\mbf{w}_k,
\edis
and the performance output is defined as
\bdis
\mbf{z}_k = \mbf{C}_{\mathrm{d}1} \mbf{e}_k + \mbf{D}_{\mathrm{d}11}\mbf{w}_k.
\edis
The observer gain $\mbf{L}_\mathrm{d}$ is to be designed such that the $\mathcal{H}_\infty$ of the transfer matrix from $\mbf{w}_k$ to $\mbf{z}_k$, given by
\bdis
\mbf{T}(z) = \mbf{C}_{\mathrm{d}1}\left(z\mbf{1} - \left(\mbf{A}_\mathrm{d}-\mbf{L}_\mathrm{d}\mbf{C}_{\mathrm{d}2}\right)\right)^{-1}\left(\mbf{B}_{\mathrm{d}1} - \mbf{L}_\mathrm{d}\mbf{D}_{\mathrm{d}21}\right) + \mbf{D}_{\mathrm{d}11},
\edis
is minimized.

\begin{SynthMeth}
The $\mathcal{H}_\infty$-optimal observer gain is synthesized by solving for $\mbf{P} \in \mathbb{S}^{n_x}$, $\mbf{G}_\mathrm{d} \in \mathbb{R}^{n_x \times n_y}$, and $\gamma \in \mathbb{R}_{>0}$ that minimize $\mathcal{J}(\gamma) = \gamma$ subject to $\mbf{P} >0 $ and
\bdis
\bbm \mbf{P}& \mbf{P}\mbf{A}_\mathrm{d} - \mbf{G}_\mathrm{d}\mbf{C}_{\mathrm{d}2} & \mbf{P} \mbf{B}_{\text{d}1} - \mbf{G}_\mathrm{d} \mbf{D}_{\mathrm{d}21} & \mbf{0} \\ * & \mbf{P} & \mbf{0} & \mbf{C}_{\text{d}1}^\trans \\ * & * & \gamma \mbf{1} & \mbf{D}_{\text{d}11}^\trans \\ * & * & * & \gamma \mbf{1} \ebm > 0.
\edis
The $\mathcal{H}_\infty$-optimal observer gain is recovered by $\mbf{L}_\mathrm{d} = \mbf{P}^{-1}\mbf{G}_\mathrm{d}$ and the $\mathcal{H}_\infty$ norm of $\mbf{T}(z)$ is $\gamma$.
\end{SynthMeth}

\subsection{Mixed \texorpdfstring{$\mathcal{H}_2$-$\mathcal{H}_\infty$}{H2 H-Infinity}-Optimal State Estimation}
\index{estimation!mixed $\mathcal{H}_2$-$\mathcal{H}_\infty$-optimal}

The goal of mixed $\mathcal{H}_2$-$\mathcal{H}_\infty$-optimal state estimation is to design an observer that minimizes the $\mathcal{H}_2$ norm of the closed-loop transfer matrix from $\mbf{w}_1$ to $\mbf{z}_1$, while ensuring that the $\mathcal{H}_\infty$ norm of the closed-loop transfer matrix from $\mbf{w}_2$ to $\mbf{z}_2$ is below a specified bound.

\subsubsection{Mixed \texorpdfstring{$\mathcal{H}_2$-$\mathcal{H}_\infty$}{H2 H-Infinity}-Optimal Observer}
\index{observer!mixed $\mathcal{H}_2$-$\mathcal{H}_\infty$-optimal}

Consider the continuous-time generalized plant $\mbc{P}$ with state-space realization
\begin{align*}
\dot{\mbf{x}} &= \mbf{A} \mbf{x} + \mbf{B}_{1,1}\mbf{w}_1 + \mbf{B}_{1,2} \mbf{w}_2, \\
\mbf{y} &= \mbf{C}_2\mbf{x} + \mbf{D}_{21,1}\mbf{w}_1 + \mbf{D}_{21,1} \mbf{w}_2 ,
\end{align*}
where it is assumed that ($\mbf{A}$,$\mbf{C}_2$) is detectable.  An observer of the form
\begin{align*}
\dot{\mbfhat{x}} &= \mbf{A} \mbfhat{x} + \mbf{L}\left(\mbf{y} - \mbfhat{y}\right), \\
\mbfhat{y} &= \mbf{C}_2\mbfhat{x} ,
\end{align*}
is to be designed, where $\mbf{L} \in \mathbb{R}^{n_x \times n_y}$ is the observer gain.  Defining the error state $\mbf{e} = \mbf{x} - \mbfhat{x}$, the error dynamics are found to be
\bdis
\dot{\mbf{e}} = \left(\mbf{A}-\mbf{L}\mbf{C}_2\right)\mbf{e} + \left(\mbf{B}_{1,1} - \mbf{L}\mbf{D}_{21,1}\right)\mbf{w}_1 + \left(\mbf{B}_{1,2} - \mbf{L}\mbf{D}_{21,2}\right)\mbf{w}_2,
\edis
and the performance output is defined as
\bdis
\bbm \mbf{z}_1 \\ \mbf{z}_2 \ebm  = \bbm \mbf{C}_{1,1} \\ \mbf{C}_{1,2} \ebm \mbf{e} + \bbm \mbf{0} & \mbf{D}_{11,12} \\ \mbf{D}_{11,21} & \mbf{D}_{11,22} \ebm \bbm \mbf{w}_1 \\ \mbf{w}_2 \ebm.
\edis
The observer gain $\mbf{L}$ is to be designed to minimize the $\mathcal{H}_2$ norm of the closed-loop transfer matrix $\mbf{T}_{11}(s)$ from the exogenous input $\mbf{w}_1$ to the performance output $\mbf{z}_1$ while ensuring the $\mathcal{H}_\infty$ norm of the closed-loop transfer matrix $\mbf{T}_{22}(s)$ from the exogenous input $\mbf{w}_2$ to the performance output $\mbf{z}_2$ is less than $\gamma_d$, where
 \begin{align*}
\mbf{T}_{11}(s) &= \mbf{C}_{1,1} \left(s\mbf{1} - \left(\mbf{A}-\mbf{L}\mbf{C}_2\right)\right)^{-1}\left(\mbf{B}_{1,1} - \mbf{L} \mbf{D}_{21,1}\right), \\
\mbf{T}_{22}(s) &= \mbf{C}_{1,2} \left(s\mbf{1} - \left(\mbf{A}-\mbf{L}\mbf{C}_2\right)\right)^{-1}\left(\mbf{B}_{1,2} - \mbf{L} \mbf{D}_{21,2}\right) + \mbf{D}_{11,22}. 
\end{align*}
\begin{SynthMeth}
The mixed $\mathcal{H}_2$-$\mathcal{H}_\infty$-optimal observer gain is synthesized by solving for $\mbf{P} \in \mathbb{S}^{n_x}$, $\mbf{Z} \in \mathbb{S}^{n_z}$, $\mbf{G} \in \mathbb{R}^{n_x \times n_y}$, and $\nu \in \mathbb{R}_{>0}$ that minimize $\mathcal{J}(\nu) = \nu$ subject to $\mbf{P} >0 $, $\mbf{Z} > 0$,
\begin{align*}
\bbm  \mbf{P}\mbf{A} + \mbf{A}^\trans\mbf{P} - \mbf{G}\mbf{C}_2 - \mbf{C}_2^\trans\mbf{G}^\trans & \mbf{P}\mbf{B}_{1,1} - \mbf{G}\mbf{D}_{21,1} \\ * & -\mbf{1} \ebm &< 0, \\
\bbm  \mbf{P}\mbf{A} + \mbf{A}^\trans\mbf{P} - \mbf{G}\mbf{C}_2 - \mbf{C}_2^\trans\mbf{G}^\trans  & \mbf{P}\mbf{B}_{1,2} - \mbf{G}\mbf{D}_{21,2} & \mbf{C}_{1,2}^\trans \\ *& -\gamma_d \mbf{1} & \mbf{D}_{11,22}^\trans \\ * & * & -\gamma_d \mbf{1} \ebm &< 0, \\
\bbm \mbf{P} & \mbf{C}_{1,1}^\trans \\ * & \mbf{Z} \ebm &> 0, \\
\trace (\mbf{Z}) &< \nu. 
\end{align*}
The mixed-$\mathcal{H}_2$-$\mathcal{H}_\infty$-optimal observer gain is recovered by $\mbf{L} = \mbf{P}^{-1}\mbf{G}$, the $\mathcal{H}_2$ norm of $\mbf{T}_{11}(s)$ is less than $\mu = \sqrt{\nu}$, and the $\mathcal{H}_\infty$ norm of $\mbf{T}_{22}(s)$ is less than $\gamma_d$.
\end{SynthMeth}

\subsubsection{Discrete-Time Mixed \texorpdfstring{$\mathcal{H}_2$-$\mathcal{H}_\infty$}{H2 H-Infinity}-Optimal Observer}
\index{observer!discrete-time mixed $\mathcal{H}_2$-$\mathcal{H}_\infty$-optimal}

Consider the discrete-time generalized LTI plant $\mbc{P}$ with state-space realization
\begin{align*}
\mbf{x}_{k+1} &= \mbf{A}_\mathrm{d} \mbf{x}_k + \mbf{B}_{\mathrm{d}1,1}\mbf{w}_{1,k} + \mbf{B}_{\mathrm{d}1,1}\mbf{w}_{1,k},  \\
\mbf{y}_k &= \mbf{C}_{\mathrm{d}2}\mbf{x}_k + \mbf{D}_{\mathrm{d}21,1}\mbf{w}_{1,k} + \mbf{D}_{\mathrm{d}21,2}\mbf{w}_{2,k}, 
\end{align*}
where it is assumed that ($\mbf{A}_\mathrm{d}$,$\mbf{C}_{\mathrm{d}2}$) is detectable.  An observer of the form
\begin{align*}
\mbfhat{x}_{k+1} &= \mbf{A}_\mathrm{d} \mbfhat{x}_k + \mbf{L}_\mathrm{d}\left(\mbf{y}_k - \mbfhat{y}_k\right), \\
\mbfhat{y}_k &= \mbf{C}_{\mathrm{d}2}\mbfhat{x}_k , 
\end{align*}
is to be designed, where $\mbf{L}_\mathrm{d} \in \mathbb{R}^{n_x \times n_y}$ is the observer gain.  Defining the error state $\mbf{e}_k = \mbf{x}_k - \mbfhat{x}_k$, the error dynamics are found to be
\bdis
\mbf{e}_{k+1} = \left(\mbf{A}_\mathrm{d}-\mbf{L}_\mathrm{d}\mbf{C}_{\mathrm{d}2}\right)\mbf{e}_k + \left(\mbf{B}_{\mathrm{d}1,1} - \mbf{L}_\mathrm{d}\mbf{D}_{\mathrm{d}21,1}\right)\mbf{w}_{1,k} + \left(\mbf{B}_{\mathrm{d}1,2} - \mbf{L}_\mathrm{d}\mbf{D}_{\mathrm{d}21,2}\right)\mbf{w}_{2,k}, \\
\edis
and the performance output is defined as
\bdis
\bbm \mbf{z}_{1,k} \\ \mbf{z}_{2,k} \ebm  = \bbm \mbf{C}_{\mathrm{d}1,1} \\ \mbf{C}_{\mathrm{d}1,2} \ebm \mbf{e}_k + \bbm \mbf{0} & \mbf{D}_{\mathrm{d}11,12} \\ \mbf{D}_{\mathrm{d}11,21} & \mbf{D}_{\mathrm{d}11,22} \ebm \bbm \mbf{w}_{1,k} \\ \mbf{w}_{2,k} \ebm .
\edis
The observer gain $\mbf{L}_\mathrm{d}$ is to be designed to minimize the $\mathcal{H}_2$ norm of the closed loop transfer matrix $\mbf{T}_{11}(z)$ from the exogenous input $\mbf{w}_{1,k}$ to the performance output $\mbf{z}_{1,k}$ while ensuring the $\mathcal{H}_\infty$ norm of the closed-loop transfer matrix $\mbf{T}_{22}(z)$ from the exogenous input $\mbf{w}_{2,k}$ to the performance output $\mbf{z}_{2,k}$ is less than $\gamma_d$, where
 \begin{align*}
\mbf{T}_{11}(z) &= \mbf{C}_{\mathrm{d}1,1}\left(z\mbf{1} - \left(\mbf{A}_\mathrm{d}-\mbf{L}_{\mathrm{d}}\mbf{C}_{\mathrm{d}2}\right)\right)^{-1}\left(\mbf{B}_{\mathrm{d}1,1} - \mbf{L}_\mathrm{d} \mbf{D}_{\mathrm{d}21,1}\right), \\
\mbf{T}_{22}(z) &= \mbf{C}_{\mathrm{d}1,2} \left(z\mbf{1} - \left(\mbf{A}_\mathrm{d}-\mbf{L}_{\mathrm{d}}\mbf{C}_{\mathrm{d}2}\right)\right)^{-1}\left(\mbf{B}_{\mathrm{d}1,2} - \mbf{L}_\mathrm{d} \mbf{D}_{\mathrm{d}21,2}\right) + \mbf{D}_{\mathrm{d}11,22}. 
\end{align*}

\begin{SynthMeth}
The discrete-time mixed-$\mathcal{H}_2$-$\mathcal{H}_\infty$-optimal observer gain is synthesized by solving for $\mbf{P} \in \mathbb{S}^{n_x}$, $\mbf{Z} \in \mathbb{S}^{n_z}$, $\mbf{G}_\mathrm{d} \in \mathbb{R}^{n_x \times n_y}$, and $\nu \in \mathbb{R}_{>0}$ that minimize $\mathcal{J}(\nu) = \nu$ subject to $\mbf{P} >0 $, $\mbf{Z} > 0$,
\begin{align*}
\bbm \mbf{P} & \mbf{P}\mbf{A}_\mathrm{d} - \mbf{G}_\mathrm{d}\mbf{C}_{\mathrm{d}2} & \mbf{P}\mbf{B}_{\text{d}1,1} - \mbf{G}_\mathrm{d}\mbf{D}_{\mathrm{d}21,1} \\ * & \mbf{P} & \mbf{0} \\ * & * & \mbf{1} \ebm &> 0,\\
\bbm \mbf{P} & \mbf{P}\mbf{A}_\mathrm{d} - \mbf{G}_\mathrm{d}\mbf{C}_{\mathrm{d}2} & \mbf{P} \mbf{B}_{\text{d}1,2} - \mbf{G}_\mathrm{d} \mbf{D}_{\mathrm{d}21,2} & \mbf{0} \\ * & \mbf{P} & \mbf{0} & \mbf{C}_{\text{d}1,2}^\trans \\ * & * & \gamma_d \mbf{1} & \mbf{D}_{\text{d}11,22}^\trans \\ * & * & * & \gamma_d \mbf{1} \ebm &> 0, \\
\bbm \mbf{Z} & \mbf{C}_{\text{d}1,1} \\ * & \mbf{P} \ebm &> 0. \\
\trace (\mbf{Z}) &< \nu. 
\end{align*}
The mixed-$\mathcal{H}_2$-$\mathcal{H}_\infty$-optimal observer gain is recovered by $\mbf{L}_\mathrm{d} = \mbf{P}^{-1}\mbf{G}_\mathrm{d}$, the $\mathcal{H}_2$ norm of $\mbf{T}_{11}(z)$ is less than $\mu = \sqrt{\nu}$, and the $\mathcal{H}_\infty$ norm of $\mbf{T}_{22}(z)$ is less than $\gamma_d$.
\end{SynthMeth}

\subsection{Continuous-Time and Discrete-Time Optimal Filtering}

The goal of optimal filtering is to design a filter that acts on the output $\mbf{z}$ of the generalized plant and optimizes the transfer matrix from $\mbf{w}$ to the filtered output.  

\textbf{Continuous-Time Filtering}: Consider the continuous-time generalized LTI plant with minimal states-space realization
\begin{align*}
\dot{\mbf{x}} &= \mbf{A} \mbf{x} + \mbf{B}_1 \mbf{w}, \\ 
\mbf{z} &= \mbf{C}_1 \mbf{x} + \mbf{D}_{11} \mbf{w}, \\ 
\mbf{y} &= \mbf{C}_2 \mbf{x} + \mbf{D}_{21} \mbf{w},  
\end{align*}
where it is assumed that $\mbf{A}$ is Hurwitz.  A continuous-time dynamic LTI filter with state-space realization
\begin{align*}
\dot{\mbf{x}}_f &= \mbf{A}_f \mbf{x}_f + \mbf{B}_f \mbf{y}, \\
\mbfhat{z} &= \mbf{C}_f \mbf{x}_f + \mbf{D}_f \mbf{y},
\end{align*}
is to be designed to optimize the transfer function from $\mbf{w}$ to $\mbftilde{z} = \mbf{z} - \mbfhat{z}$, given by
\beq
\label{eq:TM_Filt}
\mbftilde{P}(s) = \mbftilde{C}_1 \left(s\mbf{1} - \mbftilde{A}\right)^{-1} \mbftilde{B}_1 + \mbftilde{D}_{11},
\eeq
where
\bdis
\mbftilde{A} = \bbm \mbf{A} & \mbf{0} \\ \mbf{B}_f \mbf{C}_2 & \mbf{A}_f \ebm, \hspace{20pt} \mbftilde{B}_1 = \bbm \mbf{B}_1 \\ \mbf{B}_f \mbf{D}_{21} \ebm, \hspace{20pt} \mbftilde{C}_1 = \bbm \mbf{C}_1 - \mbf{D}_f \mbf{C}_2 & - \mbf{C}_f \ebm, \hspace{20pt} \mbftilde{D}_{11} = \mbf{D}_{11}-\mbf{D}_f \mbf{D}_{21}.
\edis
This can alternatively be formulated as a special case of synthesizing a dynamic output ``feedback'' controller for the generalized plant given by
\begin{align*}
\dot{\mbf{x}} &= \mbf{A} \mbf{x} + \mbf{B}_1 \mbf{w}, \\ 
\mbf{z} &= \mbf{C}_1 \mbf{x} + \mbf{D}_{11} \mbf{w} - \mbf{u}, \\ 
\mbf{y} &= \mbf{C}_2 \mbf{x} + \mbf{D}_{21} \mbf{w}.  
\end{align*}
The controller in this case is not truly a feedback controller, as it only appears as a feedthrough term in the performance channel.  The synthesis methods presented in this subsection take advantage of this fact, resulting in a simpler formulation than applying the controller synthesis methods in Section~\ref{sec:OptimalControl}.

\textbf{Discrete-Time Filtering}: Consider the discrete-time generalized LTI plant with minimal states-space realization
\begin{align*}
\mbf{x}_{k+1} &= \mbf{A}_\mathrm{d} \mbf{x}_k + \mbf{B}_{\mathrm{d} 1} \mbf{w}_k, \\ 
\mbf{z}_k &= \mbf{C}_{\mathrm{d}1} \mbf{x}_k + \mbf{D}_{\mathrm{d}11} \mbf{w}_k, \\ 
\mbf{y}_k &= \mbf{C}_{\mathrm{d}2} \mbf{x}_k + \mbf{D}_{\mathrm{d}21} \mbf{w}_k,  
\end{align*}
where it is assumed that $\mbf{A}_\mathrm{d}$ is Schur.  A discrete-time dynamic LTI filter with state-space realization
\begin{align*}
\mbf{x}_{f,k+1} &= \mbf{A}_{f} \mbf{x}_{f,k} + \mbf{B}_f \mbf{y}_k, \\
\mbfhat{z}_k &= \mbf{C}_f \mbf{x}_{f,k} + \mbf{D}_f \mbf{y}_k,
\end{align*}
is to be designed to optimize the transfer function from $\mbf{w}_k$ to $\mbftilde{z}_k = \mbf{z}_k - \mbfhat{z}_k$, given by
\beq
\label{eq:TM_Filt2}
\mbftilde{P}(z) = \mbftilde{C}_{\mathrm{d}1} \left(z\mbf{1} - \mbftilde{A}_\mathrm{d}\right)^{-1} \mbftilde{B}_{\mathrm{d}1} + \mbftilde{D}_{\mathrm{d}11},
\eeq
where
\bdis
\mbftilde{A}_\mathrm{d} = \bbm \mbf{A}_\mathrm{d} & \mbf{0} \\ \mbf{B}_f \mbf{C}_{\mathrm{d}2} & \mbf{A}_f \ebm, \hspace{5pt} \mbftilde{B}_{\mathrm{d}1} = \bbm \mbf{B}_{\mathrm{d}1} \\ \mbf{B}_f \mbf{D}_{\mathrm{d}21} \ebm, \hspace{5pt} \mbftilde{C}_{\mathrm{d}1} = \bbm \mbf{C}_{\mathrm{d}1} - \mbf{D}_f \mbf{C}_{\mathrm{d}2} & - \mbf{C}_f \ebm, \hspace{5pt} \mbftilde{D}_{\mathrm{d}11} = \mbf{D}_{\mathrm{d}11}-\mbf{D}_f \mbf{D}_{\mathrm{d}21}.
\edis
This can alternatively be formulated as a special case of synthesizing a dynamic output ``feedback'' controller for the generalized plant given by
\begin{align*}
\mbf{x}_{k+1} &= \mbf{A}_\mathrm{d} \mbf{x}_k + \mbf{B}_{\mathrm{d} 1} \mbf{w}_k, \\ 
\mbf{z}_k &= \mbf{C}_{\mathrm{d}1} \mbf{x}_k + \mbf{D}_{\mathrm{d}11} \mbf{w}_k - \mbf{u}_k, \\ 
\mbf{y}_k &= \mbf{C}_{\mathrm{d}2} \mbf{x}_k + \mbf{D}_{\mathrm{d}21} \mbf{w}_k.  
\end{align*}

\subsubsection[\texorpdfstring{$\mathcal{H}_2$}{H2}-Optimal Filter]{\texorpdfstring{$\mathcal{H}_2$}{H2}-Optimal Filter}
\index{filtering!$\mathcal{H}_2$-optimal}

An $\mathcal{H}_2$-optimal filter is designed to minimize the $\mathcal{H}_2$ norm of $\mbftilde{P}(s)$ in~\eqref{eq:TM_Filt}.  
\begin{SynthMeth}
\cite[pp.~309--310]{Duan2013} Solve for $\mbf{A}_n \in \mathbb{R}^{n_x \times n_x}$, $\mbf{B}_n \in \mathbb{R}^{n_x \times n_y}$, $\mbf{C}_f \in \mathbb{R}^{n_z \times n_x}$, $\mbf{D}_f \in \mathbb{R}^{n_z \times n_y}$, $\mbf{X}$,~$\mbf{Y} \in \mathbb{S}^{n_x}$, $\mbf{Z} \in \mathbb{S}^{n_z}$, and $\nu \in \mathbb{R}_{>0}$ that minimize $\mathcal{J}(\nu) = \nu$ subject to $\mbf{X} > 0$, $\mbf{Y} > 0$, $\mbf{Z} > 0$,
\begin{align}
\bbm \mbf{Y} \mbf{A}  + \mbf{A}^\trans \mbf{Y} + \mbf{B}_n \mbf{C}_2 + \mbf{C}_2^\trans \mbf{B}_n^\trans & \mbf{A}_n + \mbf{C}_2^\trans \mbf{B}_n^\trans + \mbf{A}^\trans \mbf{X} & \mbf{Y} \mbf{B}_1 + \mbf{B}_n \mbf{D}_{21} \\ * &\mbf{A}_n + \mbf{A}_n^\trans & \mbf{X} \mbf{B}_1 + \mbf{B}_n \mbf{D}_{21} \\ * & * & -\mbf{1} \ebm &< 0, \nonumber \\
\bbm -\mbf{Z} & \mbf{C}_1- \mbf{D}_f \mbf{C}_2 & - \mbf{C}_f \\ * & - \mbf{Y} & -\mbf{X} \\ * & * & -\mbf{X} \ebm &<0, \nonumber\\
\mbf{D}_{11} - \mbf{D}_{f}  \mbf{D}_{21} &= \mbf{0}, \label{eq:H2filt_1c}\\
\mbf{Y} - \mbf{X} &> 0, \nonumber\\
\trace (\mbf{Z}) &< \nu. \nonumber
\end{align}
The filter is recovered by the state-space matrices $\mbf{A}_f= \mbf{X}^{-1} \mbf{A}_n$, $\mbf{B}_f = \mbf{X}^{-1} \mbf{B}_n$, $\mbf{C}_f$, and $\mbf{D}_f$.

If $\mbf{D}_{11} = \mbf{0}$ and $\mbf{D}_{21} \neq \mbf{0}$, then it is often simplest to choose $\mbf{D}_f = \mbf{0}$ in order to satisfy the equality constraint of~\eqref{eq:H2filt_1c}.
\end{SynthMeth}

\begin{SynthMeth}
\cite[pp.~309--310]{Duan2013} Solve for $\mbf{A}_n \in \mathbb{R}^{n_x \times n_x}$, $\mbf{B}_n \in \mathbb{R}^{n_x \times n_y}$, $\mbf{C}_f \in \mathbb{R}^{n_z \times n_x}$, $\mbf{D}_f \in \mathbb{R}^{n_z \times n_y}$, $\mbf{X}$,~$\mbf{Y} \in \mathbb{S}^{n_x}$, $\mbf{Z} \in \mathbb{S}^{n_z}$, and $\nu \in \mathbb{R}_{>0}$ that minimize $\mathcal{J}(\nu) = \nu$ subject to $\mbf{X} > 0$, $\mbf{Y} > 0$, $\mbf{Z} > 0$,
\begin{align}
\bbm \mbf{Y} \mbf{A}  + \mbf{A}^\trans \mbf{Y} + \mbf{B}_n \mbf{C}_2 + \mbf{C}_2^\trans \mbf{B}_n^\trans & \mbf{A}_n + \mbf{C}_2^\trans \mbf{B}_n^\trans + \mbf{A}^\trans \mbf{X} & \mbf{C}_1^\trans-  \mbf{C}_2^\trans \mbf{D}_f^\trans  \\ * &\mbf{A}_n + \mbf{A}_n^\trans & -\mbf{C}_f^\trans \\ * & * & -\mbf{1} \ebm &< 0, \nonumber \\
\bbm -\mbf{Z} &\mbf{B}_1^\trans \mbf{Y}^\trans + \mbf{D}_{21} ^\trans\mbf{B}_n^\trans  &  \mbf{B}_1 ^\trans \mbf{X}^\trans +  \mbf{D}_{21}^\trans \mbf{B}_n^\trans \\ * & - \mbf{Y} & -\mbf{X} \\ * & * & -\mbf{X} \ebm &<0, \nonumber\\
\mbf{D}_{11} - \mbf{D}_{f}  \mbf{D}_{21} &= \mbf{0}, \label{eq:H2filt_2c}\\
\mbf{Y} - \mbf{X} &> 0, \nonumber\\
\trace (\mbf{Z}) &< \nu. \nonumber
\end{align}
The filter is recovered by the state-space matrices $\mbf{A}_f= \mbf{X}^{-1} \mbf{A}_n$, $\mbf{B}_f = \mbf{X}^{-1} \mbf{B}_n$, $\mbf{C}_f$, and $\mbf{D}_f$.

If $\mbf{D}_{11} = \mbf{0}$ and $\mbf{D}_{21} \neq \mbf{0}$, then it is often simplest to choose $\mbf{D}_f = \mbf{0}$ in order to satisfy the equality constraint of~\eqref{eq:H2filt_2c}.
\end{SynthMeth}

\subsubsection[Discrete-Time \texorpdfstring{$\mathcal{H}_2$}{H2}-Optimal Filter]{Discrete-Time \texorpdfstring{$\mathcal{H}_2$}{H2}-Optimal Filter}
\index{filtering!discrete-time $\mathcal{H}_2$-optimal}

\begin{SynthMeth} 
\cite{Geromel2000} Consider the case where $\mbf{D}_{\mathrm{d}11} = \mbf{0}$ and $\mbf{D}_f = \mbf{0}$.  Solve for $\mbf{A}_{\mathrm{d}n} \in \mathbb{R}^{n_x \times n_x}$, $\mbf{B}_{\mathrm{d}n} \in \mathbb{R}^{n_x \times n_y}$, $\mbf{C}_{\mathrm{d}n} \in \mathbb{R}^{n_u \times n_x}$, $\mbf{X}$,~$\mbf{Y} \in \mathbb{S}^{n_x}$, $\mbf{Z} \in \mathbb{S}^{n_z}$, and $\nu \in \mathbb{R}_{>0}$ that minimize $\mathcal{J}(\nu) = \nu$ subject to $\mbf{X} > 0$, $\mbf{Y} > 0$, $\mbf{Z} >0$,
\begin{align*}
\bbm \mbf{X} & \mbf{X} & \mbf{X} \mbf{A}_\mathrm{d}  & \mbf{X}\mbf{A}_{\mathrm{d}} & \mbf{X} \mbf{B}_{\mathrm{d}1} \\ *  & \mbf{Y} & \mbf{Y} \mbf{A}_\mathrm{d} + \mbf{B}_{\mathrm{d}n} \mbf{C}_{\mathrm{d}1} + \mbf{A}_{\mathrm{d}n} &\mbf{Y} \mbf{A}_\mathrm{d} + \mbf{B}_{\mathrm{d}n} \mbf{C}_{\mathrm{d}1}  & \mbf{Y} \mbf{B}_{\mathrm{d}1} + \mbf{B}_{\mathrm{d}n} \mbf{D}_{\mathrm{d}21}    \\ * & * & \mbf{X} & \mbf{X} & \mbf{0}  \\ * & * & * & \mbf{Y} & \mbf{0}  \\ * & * & * & * &  \mbf{1}  \ebm &> 0, \nonumber\\
\bbm \mbf{Z} & \mbf{C}_{\mathrm{d}1}  & \mbf{C}_{\mathrm{d}1} - \mbf{C}_{\mathrm{d}n}  \\ * & \mbf{Y} & \mbf{X} \\ * & * & \mbf{X} \ebm &>0, \nonumber\\
\bbm \mbf{Y} & \mbf{X} \\ * & \mbf{X} \ebm &> 0, \nonumber\\
\trace (\mbf{Z})&< \nu. \nonumber
\end{align*}
The filter is recovered by $\mbf{A}_f= -\mbf{Y}^{-1} \mbf{A}_{\mathrm{d}n}\left(\mbf{1} - \mbf{Y}^{-1} \mbf{X}\right)^{-1}$, $\mbf{B}_f = -\mbf{Y}^{-1} \mbf{B}_{\mathrm{d}n}$, and $\mbf{C}_f = \mbf{C}_{\mathrm{d}n} \left(\mbf{1} - \mbf{Y}^{-1} \mbf{X}\right)^{-1}$.
\end{SynthMeth}

\begin{SynthMeth} 
Solve for $\mbf{A}_{\mathrm{d}n} \in \mathbb{R}^{n_x \times n_x}$, $\mbf{B}_{\mathrm{d}n} \in \mathbb{R}^{n_x \times n_y}$, $\mbf{C}_{\mathrm{d}n} \in \mathbb{R}^{n_u \times n_x}$, $\mbf{D}_{f} \in \mathbb{R}^{n_u \times n_y}$, $\mbf{X}_1$,~$\mbf{Y}_1 \in \mathbb{S}^{n_x}$, $\mbf{Z} \in \mathbb{S}^{n_z}$, and $\nu \in \mathbb{R}_{>0}$ that minimize $\mathcal{J}(\nu) = \nu$ subject to $\mbf{X}_1 > 0$, $\mbf{Y}_1 > 0$, $\mbf{Z} >0$,
\begin{align}
\bbm \mbf{X}_1 & \mbf{1} & \mbf{X}_1 \mbf{A}_\mathrm{d} + \mbf{B}_{\mathrm{d}n} \mbf{C}_{\mathrm{d}2} & \mbf{A}_{\mathrm{d}n} & \mbf{X}_1 \mbf{B}_{\mathrm{d}1} + \mbf{B}_{\mathrm{d}n} \mbf{D}_{\mathrm{d}21}   \\ *  & \mbf{Y}_1 & \mbf{A}_\mathrm{d} & \mbf{A}_\mathrm{d} \mbf{Y}_1 & \mbf{B}_{\mathrm{d}1}   \\ * & * & \mbf{X}_1 & \mbf{1} & \mbf{0}  \\ * & * & * & \mbf{Y}_1 & \mbf{0}  \\ * & * & * & * &  \mbf{1}  \ebm &> 0, \nonumber\\
\bbm \mbf{Z} & \mbf{C}_{\mathrm{d}1} - \mbf{D}_{f} \mbf{C}_{\mathrm{d}2} & \mbf{C}_{\mathrm{d}1}\mbf{Y}_1 - \mbf{C}_{\mathrm{d}n} \\ * & \mbf{X}_1 & \mbf{1} \\ * & * & \mbf{Y}_1 \ebm &>0, \nonumber\\
\mbf{D}_{\mathrm{d}11} - \mbf{D}_{f}  \mbf{D}_{\mathrm{d}21} &= \mbf{0}, \label{eq:H2filt_DT_Output1c}\\
\bbm \mbf{X}_1 & \mbf{1} \\ * & \mbf{Y}_1 \ebm &> 0, \nonumber\\
\trace (\mbf{Z})&< \nu. \nonumber
\end{align}
The filter state-space matrices are recovered by $\mbf{A}_f = \mbf{X}_2^{-1} \left( \mbf{A}_{\mathrm{d}n} - \mbf{X}_1 \mbf{A}_\mathrm{d} \mbf{Y}_1 \right) \mbf{Y}_2^{-\trans}$, $\mbf{B}_f = \mbf{X}_2^{-1} \mbf{B}_{\mathrm{d}n}$, $\mbf{C}_f = \mbf{C}_{\mathrm{d}n} \mbf{Y}_2^{-\trans}$, and $\mbf{D}_f$, where the matrices $\mbf{X}_2$ and $\mbf{Y}_2$ satisfy $\mbf{X}_2\mbf{Y}_2^\trans = \mbf{1} - \mbf{X}_1\mbf{Y}_1$.  Given $\mbf{X}_1$ and $\mbf{Y}_1$, the matrices $\mbf{X}_2$ and $\mbf{Y}_2$ can be found using a matrix decomposition, such as a LU decomposition or a Cholesky decomposition.

If $\mbf{D}_{\mathrm{d}11} = \mbf{0}$ and $\mbf{D}_{\mathrm{d}21} \neq \mbf{0}$, then it is often simplest to choose $\mbf{D}_{f} = \mbf{0}$ in order to satisfy the equality constraint of~\eqref{eq:H2filt_DT_Output1c}.

This synthesis method is derived from the discrete-time $\mathcal{H}_2$-optimal dynamic output feedback controller synthesis method in Synthesis Method~\ref{SynthMeth:H2_Dyn_Output_2} using the fact that $\mathcal{H}_2$-optimal filter synthesis is a special case of this problem.

\end{SynthMeth}

\subsubsection[\texorpdfstring{$\mathcal{H}_\infty$}{H-inf}-Optimal Filter]{\texorpdfstring{$\mathcal{H}_\infty$}{H-inf}-Optimal Filter}
\index{filtering!$\mathcal{H}_\infty$-optimal}

An $\mathcal{H}_\infty$-optimal filter is designed to minimize the $\mathcal{H}_\infty$ norm of $\mbftilde{P}(s)$ in~\eqref{eq:TM_Filt}.

\begin{SynthMeth}
\cite[pp.~303--304]{Duan2013} Solve for $\mbf{A}_n \in \mathbb{R}^{n_x \times n_x}$, $\mbf{B}_n \in \mathbb{R}^{n_x \times n_y}$, $\mbf{C}_f \in \mathbb{R}^{n_z \times n_x}$, $\mbf{D}_f \in \mathbb{R}^{n_z \times n_y}$, $\mbf{X}$,~$\mbf{Y} \in \mathbb{S}^{n_x}$, and $\gamma \in \mathbb{R}_{>0}$ that minimize $\mathcal{J}(\gamma) = \gamma$ subject to $\mbf{X} > 0$, $\mbf{Y} > 0$,
\begin{align}
\bbm \mbf{Y} \mbf{A}  + \mbf{A}^\trans \mbf{Y} + \mbf{B}_n \mbf{C}_2 + \mbf{C}_2^\trans \mbf{B}_n^\trans & \mbf{A}_n + \mbf{C}_2^\trans \mbf{B}_n^\trans + \mbf{A}^\trans \mbf{X} & \mbf{Y} \mbf{B}_1 + \mbf{B}_n \mbf{D}_{21} & \mbf{C}_1^\trans - \mbf{C}_2^\trans \mbf{D}_f^\trans \\ * &\mbf{A}_n + \mbf{A}_n^\trans & \mbf{X} \mbf{B}_1 + \mbf{B}_n \mbf{D}_{21} & - \mbf{C}_f^\trans \\ * & * & -\gamma \mbf{1} & \mbf{D}_{11}^\trans-\mbf{D}_{21}^\trans \mbf{D}_f^\trans \\ * & * & * & -\gamma \mbf{1}\ebm &< 0, \nonumber \\
\mbf{Y} - \mbf{X} &> 0. \nonumber
\end{align}
The filter is recovered by $\mbf{A}_f= \mbf{X}^{-1} \mbf{A}_n$ and $\mbf{B}_f = \mbf{X}^{-1} \mbf{B}_n$.
\end{SynthMeth}

\subsubsection[Discrete-Time \texorpdfstring{$\mathcal{H}_\infty$}{H-inf}-Optimal Filter]{Discrete-Time \texorpdfstring{$\mathcal{H}_\infty$}{H-inf}-Optimal Filter}
\index{filtering!discrete-time $\mathcal{H}_\infty$-optimal}

\begin{SynthMeth}
\cite{Geromel2000} Consider the case where $\mbf{D}_{\mathrm{d}11} = \mbf{0}$ and $\mbf{D}_f = \mbf{0}$.  Solve for $\mbf{A}_{\mathrm{d}n} \in \mathbb{R}^{n_x \times n_x}$, $\mbf{B}_{\mathrm{d}n} \in \mathbb{R}^{n_x \times n_y}$, $\mbf{C}_{\mathrm{d}n} \in \mathbb{R}^{n_u \times n_x}$, $\mbf{X}$,~$\mbf{Y} \in \mathbb{S}^{n_x}$, and $\gamma \in \mathbb{R}_{>0}$ that minimize $\mathcal{J}(\gamma) = \gamma$ subject to $\mbf{X} > 0$, $\mbf{Y} > 0$,
\begin{align*}
\bbm \mbf{X} & \mbf{X} & \mbf{X} \mbf{A}_\mathrm{d}  & \mbf{X} \mbf{A}_{\mathrm{d}} & \mbf{X} \mbf{B}_{\mathrm{d}1} & \mbf{0}  \\ *  & \mbf{Y} & \mbf{Y}\mbf{A}_\mathrm{d} + \mbf{B}_{\mathrm{d}n} \mbf{C}_{\mathrm{d}1} + \mbf{A}_{\mathrm{d}n} & \mbf{Y}\mbf{A}_\mathrm{d} + \mbf{B}_{\mathrm{d}n} \mbf{C}_{\mathrm{d}1}  & \mbf{Y}\mbf{B}_{\mathrm{d}1} + \mbf{B}_{\mathrm{d}n} \mbf{D}_{\mathrm{d}21} & \mbf{0} \\ * & * & \mbf{X} & \mbf{X} & \mbf{0} & \mbf{C}_{\mathrm{d}1}^\trans - \mbf{C}_{\mathrm{d}n}^\trans   \\ * & * & * & \mbf{Y} & \mbf{0} &  \mbf{C}_{\mathrm{d}1}^\trans  \\ * & * & * & * &  \mbf{1} & \mbf{0} \\ * & * & * & * & * & \gamma \mbf{1} \ebm &> 0, \\
\bbm \mbf{Y} & \mbf{X} \\ * & \mbf{X} \ebm &> 0. 
\end{align*}
The filter is recovered by $\mbf{A}_f= -\mbf{Y}^{-1} \mbf{A}_{\mathrm{d}n}\left(\mbf{1} - \mbf{Y}^{-1} \mbf{X}\right)^{-1}$, $\mbf{B}_f = -\mbf{Y}^{-1} \mbf{B}_{\mathrm{d}n}$, and $\mbf{C}_f = \mbf{C}_{\mathrm{d}n} \left(\mbf{1} - \mbf{Y}^{-1} \mbf{X}\right)^{-1}$.
\end{SynthMeth}

\begin{SynthMeth}
Solve for $\mbf{A}_{\mathrm{d}n} \in \mathbb{R}^{n_x \times n_x}$, $\mbf{B}_{\mathrm{d}n} \in \mathbb{R}^{n_x \times n_y}$, $\mbf{C}_{\mathrm{d}n} \in \mathbb{R}^{n_u \times n_x}$, $\mbf{D}_{\mathrm{d}n} \in \mathbb{R}^{n_u \times n_y}$, $\mbf{X}_1$,~$\mbf{Y}_1 \in \mathbb{S}^{n_x}$, and $\gamma \in \mathbb{R}_{>0}$ that minimize $\mathcal{J}(\gamma) = \gamma$ subject to $\mbf{X}_1 > 0$, $\mbf{Y}_1 > 0$,
\begin{align}
\bbm \mbf{X}_1 & \mbf{1} & \mbf{X}_1 \mbf{A}_\mathrm{d} + \mbf{B}_{\mathrm{d}n}\mbf{C}_{\mathrm{d}2} & \mbf{A}_{\mathrm{d}n} & \mbf{X}_1 \mbf{B}_{\mathrm{d}1} + \mbf{B}_{\mathrm{d}n} \mbf{D}_{\mathrm{d}21} & \mbf{0}  \\ *  & \mbf{Y}_1 & \mbf{A}_\mathrm{d} & \mbf{A}_\mathrm{d} \mbf{Y}_1 & \mbf{B}_{\mathrm{d}1}  & \mbf{0} \\ * & * & \mbf{X}_1 & \mbf{1} & \mbf{0} & \mbf{C}_{\mathrm{d}1}^\trans - \mbf{C}_{\mathrm{d}2}^\trans \mbf{D}_{\mathrm{d}n}^\trans  \\ * & * & * & \mbf{Y}_1 & \mbf{0} & \mbf{Y}_1 \mbf{C}_{\mathrm{d}1}^\trans - \mbf{C}_{\mathrm{d}n}^\trans  \\ * & * & * & * & \gamma \mbf{1} & \mbf{D}_{\mathrm{d}11}^\trans - \mbf{D}_{\mathrm{d}21}^\trans\mbf{D}_{\mathrm{d}n}^\trans \\ * & * & * & * & * & \gamma \mbf{1} \ebm &> 0, \nonumber \\
\bbm \mbf{X}_1 & \mbf{1} \\ * & \mbf{Y}_1 \ebm &> 0. \nonumber  
\end{align}
The filter state-space matrices are recovered by $\mbf{A}_f = \mbf{X}_2^{-1} \left( \mbf{A}_{\mathrm{d}n} - \mbf{X}_1 \mbf{A}_\mathrm{d} \mbf{Y}_1 \right) \mbf{Y}_2^{-\trans}$, $\mbf{B}_f = \mbf{X}_2^{-1} \mbf{B}_{\mathrm{d}n}$, $\mbf{C}_f = \mbf{C}_{\mathrm{d}n} \mbf{Y}_2^{-\trans}$, and $\mbf{D}_f$, where the matrices $\mbf{X}_2$ and $\mbf{Y}_2$ satisfy $\mbf{X}_2\mbf{Y}_2^\trans = \mbf{1} - \mbf{X}_1\mbf{Y}_1$.  Given $\mbf{X}_1$ and $\mbf{Y}_1$, the matrices $\mbf{X}_2$ and $\mbf{Y}_2$ can be found using a matrix decomposition, such as a LU decomposition or a Cholesky decomposition.

This synthesis method is derived from the discrete-time $\mathcal{H}_\infty$-optimal dynamic output feedback controller synthesis method in Synthesis Method~\ref{SynthMeth:Hinf_Dyn_Output_2} using the fact that $\mathcal{H}_\infty$-optimal filter synthesis is a special case of this problem.

\end{SynthMeth}

\newpage
\appendix

\section{Version History}

\subsection{Updates in Version 5 (\today)}

\subsubsection*{Section 2}
\noindent
Sec. \ref{sec:SchurProp}.\ref{sec:SchurProp6}: Updated references by removing a broken link and added proofs of the properties. \\
Sec. \ref{sec:NonstrictProjectionLemma}: Updated results of nonstrict projection lemma with less restrictive assumptions. \\
Sec. \ref{sec:FinslersLemma}: The Finsler's lemma section was reorganized and expanded to include additional results regarding an alternative form, a non-strict version, a matrix version, and a robust version. \\
Sec. \ref{sec:NonstrictPetersen}: Added a sufficient condition to the non-strict Petersen lemma that is satisfied when some assumptions are removed.

\subsubsection*{Section 3}
\noindent
Sec. \ref{sec:SLemma}: The S-lemma section was reorganized and expanded to account for both strict and non-strict results, as well as matrix S-lemma results.

\subsubsection*{Section 5}
\noindent
Sec. \ref{sec:H2DynOutputCTSynth}: Fixed type involving swapped $\mbf{X}_1$ and $\mbf{Y}_1$ in second LMI of synthesis method.

\hspace{50pt}
\subsection{Updates in Version 4 (May 22, 2024)}

The major structural update in Version 4 is that Section 2 from Version 3 has been split up into two sections (Section 2 and Section 3).  This change helps separate the LMI properties and tricks that are primarily focused on reformulating BMIs as LMIs (Section 2) from other LMI properties (Section 3).  Also, the subsections of Section~2 have been re-ordered slightly, so that methods that typically reformulate BMIs as equivalent LMIs are presented first, followed by methods that are typically used to derive LMI conditions that imply the original BMI conditions.  The section headings referred to in this update section correspond to those in Version~4.

\subsubsection*{Section 1}
\noindent
Sec. \ref{sec:LMIDefs}: Added missing ``$\mbf{Q} < 0$'' to LMI in Example~\ref{Example3}. \\
Sec. \ref{sec:Solvers}: Added new solver (STRIDE). \\
Sec. \ref{sec:Parsers}: Added new parser (ROLMIP) and removed Scilab parser due to inaccessible code.

\subsubsection*{Section 2}
\noindent
Sec. \ref{sec:SchurProp}.\ref{prop:SchurBased}: Fixed typo. Transposes were swapped on some terms in the proof.\\
Sec. \ref{sec:YoungsSpecial}.\ref{sec:YoungsSpecial9}: Fixed typo.\\
Sec. \ref{sec:StrictPetersen}: Added Strict Petersen's lemma. \\
Sec. \ref{sec:NonstrictPetersen}: Added Nonstrict Petersen's lemma. \\
Sec. \ref{sec:ConvexConcave}: Added convex-concave decomposition. \\
Sec. \ref{sec:PenConvexRelax}: Added penalized convex relaxation conditions. \\
Sec. \ref{sec:CoordDescent}: Added section on coordinate descent. \\
Sec. \ref{sec:Discussion_BMI_LMI}: The discussion on how to reformulate BMIs as LMIs was extended and rewritten.

\subsubsection*{Section 3}
\noindent
Secs. \ref{sec:WeightedSumEigvals}, \ref{sec:WeightedSumAbsEigvals}: Fixed wrong ordering of weights.\\
Sec. \ref{sec:DouglasFillmore}: Fixed typo in Douglas-Fillmore-Williams Lemma.

%
%

\subsubsection*{Section 4}
\noindent
Sec. \ref{sec:BoundedRealLemma}: Added new results.\\
Sec. \ref{sec:H2Norm}: Added new results.\\
Sec. \ref{sec:GKYP_Lemma}: Slight adjustment in the presentation of the results.\\
Sec. \ref{sec:DStability_Elliptic}: Added elliptic region to $\mathcal{D}$-stability results. \\
Sec. \ref{sec:DStability_Hyperbolic}: Added hyperbolic region to $\mathcal{D}$-stability results. \\
Sec. \ref{sec:TimeDelayDependent}: Added additional delay-dependent condition. \\
Sec. \ref{sec:mu_analysis}: Fixed typo in definition of variables.

\subsubsection*{Section 5}
\noindent
Sec. \ref{sec:DT_H2_DynSynth}: Added alternative formulation that allows for the closed-loop system to have non-zero feedthrough.\\
Secs. \ref{sec:HinfSynth}-\ref{sec:MixedSynth}: Fixed typos regarding positive vs negative feedback.\\
Sec. \ref{sec:MixedFullStateSynth}: Fixed typo in synthesis method.\\
Sec. \ref{sec:MixedDynSynth}: Fixed typo in synthesis method.\\
Sec. \ref{sec:DT_MixedDynSynth}: Added alternative formulation that allows for the closed-loop system to have non-zero feedthrough.

\subsubsection*{Section 6}
\noindent
Sec. \ref{sec:DT_H2_Observer}: Fixed typo in synthesis method.

\hspace{50pt}
\subsection{Updates in Version 3 (April 2, 2021)}

\subsubsection*{Section 1}
\noindent
Sec. 1.3.2: Added matrix variable form of LMI definition.\\
Sec. 1.4: New section with improved discussion on SDPs.\\
Sec. 1.5: Edited LMI Solvers section to include more details on solvers/parsers.

\subsubsection*{Section 2}
\noindent
Sec. 2.3.3: Added Linearization Lemma.\\
Sec. 2.6: Updated Finsler's Lemma.\\
Sec. 2.7: Fixed a typo and swapped variables to be consistent with Young's Relation.\\
Sec. 2.9: Slight adjustment to S-Procedure.\\
Sec. 2.10: Added Dualization Lemma.\\
Sec. 2.11: Added Frobenius norm and nuclear norm.\\
Sec. 2.12: Added additional eigenvalue properties (sum, sum of absolute values, weighted sum, weighted sum of absolute values).\\
Sec. 2.14: Added spectral radius.\\
Sec. 2.15: Added more details on the trace of a symmetric matrix. Fixed typos. Added new fact from Duan and Yu~\cite{Duan2013}.\\
Sec. 2.16: Added fact on the range of a symmetric matrix.\\
Sec. 2.17: Added the Douglas-Fillmore-Williams Lemma.

\subsubsection*{Section 3}
\noindent
Sec. 3.1.2, 3.1.4: Added dilated results.\\
Sec. 3.1.5, 3.1.6: Added descriptor system admissibility.\\
Sec. 3.2.1, 3.2.2: Added dilated results.\\
Sec. 3.2.3, 3.2.4: Added descriptor system Bounded Real Lemma.\\
Sec. 3.3.1-3.3.3: Added dilated results.\\
Secs. 3.3.2, 3.3.3: Added new results for $\mathcal{H}_2$ norm of discrete-time systems.\\
Secs. 3.3.4, 3.3.5: Added new results for $\mathcal{H}_2$ norm of descriptor systems.\\
Sec. 3.4: Updated reference for Generalized $\mathcal{H}_2$ Norm and added ``(Induced $\mathcal{L}_2$-$\mathcal{L}_\infty$ Norm)'' to Title.\\
Sec. 3.5: Updated reference for Peak-to-Peak Norm and added ``(Induced $\mathcal{L}_\infty$-$\mathcal{L}_\infty$ Norm)'' to Title.\\
Sec. 3.6.7: Added the Discrete-Time KYP Lemma for descriptor systems.\\
Sec. 3.6.8: Added a QSR dissipativity-related property.\\
Sec. 3.9.2: Added the discrete-time NI Lemma.\\
Sec. 3.9.3: Added the negative imaginary system DC constraint.\\
Sec. 3.15: Added discrete-time zeros condition.\\
Sec. 3.16: Improved organization of D-Stability section.\\
Sec. 3.17: Added D-Admissibility section.\\
Sec. 3.19: Added transient bounds on state and output for autonomous and non-autonomous LTI systems. Also added transient bounds on unit impulse response.\\
Sec. 3.20: Added output energy bounds for autonomous and non-autonomous LTI systems

\subsubsection*{Section 4}
\noindent
Sec. 4.1.1: Fixed typo in generalized plant of Example 4.2.\\
Secs. 4.2.3, 4.2.4, 4.3.3, 4.3.4, 4.4.3, 4.4.4: Fixed typos in reformulation of $\mbf{B}_c$.\\
Sec. 4.2.4: Added a second synthesis method for discrete-time $\mathcal{H}_\infty$-optimal dynamic output feedback control.\\
Sec. 4.3.4: Added a second synthesis method for discrete-time $\mathcal{H}_2$-optimal dynamic output feedback control.

\subsubsection*{Section 5}
\noindent
Sec. 5.4: Added discrete-time optimal filtering results.\\
Secs. 5.2.1, 5.3.1: Fixed missing transpose on matrices $\mbf{C}_1$ and $\mbf{C}_{1,2}$

\newpage
\addcontentsline{toc}{section}{References}
\bibliographystyle{IEEEtran}
\bibliography{Bibliography}

\begin{thebibliography}{100}
\providecommand{\url}[1]{#1}
\csname url@samestyle\endcsname
\providecommand{\newblock}{\relax}
\providecommand{\bibinfo}[2]{#2}
\providecommand{\BIBentrySTDinterwordspacing}{\spaceskip=0pt\relax}
\providecommand{\BIBentryALTinterwordstretchfactor}{4}
\providecommand{\BIBentryALTinterwordspacing}{\spaceskip=\fontdimen2\font plus
\BIBentryALTinterwordstretchfactor\fontdimen3\font minus
  \fontdimen4\font\relax}
\providecommand{\BIBforeignlanguage}[2]{{%
\expandafter\ifx\csname l@#1\endcsname\relax
\typeout{** WARNING: IEEEtran.bst: No hyphenation pattern has been}%
\typeout{** loaded for the language `#1'. Using the pattern for}%
\typeout{** the default language instead.}%
\else
\language=\csname l@#1\endcsname
\fi
#2}}
\providecommand{\BIBdecl}{\relax}
\BIBdecl

\bibitem{Boyd1994}
S.~Boyd, L.~{El Ghaoui}, E.~Feron, and V.~Balakrishnan, \emph{Linear Matrix
  Inequalities in System and Control Theory}.\hskip 1em plus 0.5em minus
  0.4em\relax Philadelphia, PA: Society for Industrial and Applied Mathematics,
  1994.

\bibitem{Dullerud2000}
G.~E. Dullerud and F.~Paganini, \emph{A Course in Robust Control Theory: A
  Convex Approach}, ser. Texts in Applied Mathematics.\hskip 1em plus 0.5em
  minus 0.4em\relax New York, NY: Springer, 2000, no.~36.

\bibitem{SchererWeiland2015}
\BIBentryALTinterwordspacing
C.~Scherer and S.~Weiland, ``Linear matrix inequalities in control,'' January
  2015. [Online]. Available:
  \url{https://www.imng.uni-stuttgart.de/mst/files/LectureNotes.pdf}
\BIBentrySTDinterwordspacing

\bibitem{Skogestad2005}
S.~Skogestad and I.~Postlethwaite, \emph{Multivariable Feedback Control:
  Analysis and Design}, 2nd~ed.\hskip 1em plus 0.5em minus 0.4em\relax Hoboken,
  NJ: Wiley, 2005.

\bibitem{Duan2013}
G.-R. Duan and H.-H. Yu, \emph{LMIs in Control Systems: Analysis, Design and
  Applications}.\hskip 1em plus 0.5em minus 0.4em\relax Boca Raton, FL: CRC
  Press, 2013.

\bibitem{Horn2013}
R.~A. Horn and C.~R. Johnson, \emph{Matrix Analysis}, 2nd~ed.\hskip 1em plus
  0.5em minus 0.4em\relax New York, NY: Cambridge University Press, 2013.

\bibitem{BernsteinMatrixBook}
D.~S. Bernstein, \emph{Scalar, Vector, and Matrix Mathematics: Theory, Facts,
  and Formulas}.\hskip 1em plus 0.5em minus 0.4em\relax Princeton, NJ:
  Princeton University Press, 2018.

\bibitem{Boyd1997}
S.~Boyd and L.~Vandenberghe, \emph{Semidefinite Programming Relaxations of
  Non-Convex Problems in Control and Combinatorial Optimization}.\hskip 1em
  plus 0.5em minus 0.4em\relax Boston, MA: Springer US, 1997, pp. 279--287.

\bibitem{ElGhaoui2000}
L.~{El Ghaoui} and S.-I. Niculescu, \emph{Advances in Linear Matrix Inequality
  Methods in Control}, ser. Advances in Design and Control.\hskip 1em plus
  0.5em minus 0.4em\relax Philadelphia, PA: Society for Industrial and Applied
  Mathematics, 2000, ch. Robust Decision Problems in Engineering: A Linear
  Matrix Inequality Approach.

\bibitem{VanAntwerp2000}
J.~G. VanAntwerp and R.~D. Braatz, ``A tutorial on linear and bilinear matrix
  inequalities,'' \emph{Journal of Process Control}, vol.~10, pp. 363--385,
  2000.

\bibitem{Herrmann2007}
G.~Herrmann, M.~C. Turner, and I.~Postlethwaite, ``Linear matrix inequalities
  in control,'' in \emph{Mathematical Methods for Robust and Nonlinear Control:
  {EPSRC} Summer School}, ser. Lecture Notes in Control and Information
  Sciences, M.~C. Turner and D.~G. Bates, Eds.\hskip 1em plus 0.5em minus
  0.4em\relax Berlin, Germany: Springer-Verlag, 2007, vol. 367, pp. 123--142.

\bibitem{Lange2013}
K.~Lange, \emph{Optimization}.\hskip 1em plus 0.5em minus 0.4em\relax New York,
  NY: Springer, 2013.

\bibitem{Boyd2004}
S.~Boyd and L.~Vandenberghe, \emph{Convex Optimization}.\hskip 1em plus 0.5em
  minus 0.4em\relax Cambridge, UK: Cambridge University Press, 2004.

\bibitem{Balakrishnan2003}
V.~Balakrishnan and L.~Vandenberghe, ``Semidefinite programming duality and
  linear time-invariant systems,'' \emph{IEEE Transactions on Automatic
  Control}, vol.~48, no.~1, pp. 30--41, 2003.

\bibitem{Balakrishnan2002}
------, ``Semidefinite programming duality and linear time-invariant systems,''
  Department of Electrical and Computer Engineering, Purdue University, West
  Lafayette, IN, Tech. Rep. TR-ECE-02-02, 2002.

\bibitem{SDPT3}
K.~C. Toh, M.~J. Todd, and R.~H. T{\"u}t{\"u}nc{\"u}, ``\texttt{SDPT3} - a
  \texttt{MATLAB} software package for semidefinite programming,''
  \emph{Optimization Methods and Software}, vol.~11, no. 1--4, pp. 545--581,
  1999.

\bibitem{SDPT3_site}
\BIBentryALTinterwordspacing
K.~C. Toh, R.~H. T{\"u}t{\"u}nc{\"u}, and M.~J. Todd, ``{SDPT}$^3$ a matlab
  software package for semidefinite-quadratic-linear pgrogramming.'' [Online].
  Available: \url{http://www.math.cmu.edu/~reha/sdpt3.html}
\BIBentrySTDinterwordspacing

\bibitem{Sedumi}
J.~Strum, ``Using \texttt{SeDuMi} 1.02, a \texttt{MATLAB} toolbox for
  optimization over symmetric cones,'' \emph{Optimization Methods and Software:
  Special Issue on Interior Point Methods}, vol.~11, no. 1--4, pp. 625--653,
  1999.

\bibitem{Sedumi_site}
\BIBentryALTinterwordspacing
``{SeDuMi}.'' [Online]. Available: \url{http://sedumi.ie.lehigh.edu/}
\BIBentrySTDinterwordspacing

\bibitem{Mosek}
{MOSEK ApS}, ``The mosek optimization software,'' \emph{Online at
  http://www.mosek.com}, 2018.

\bibitem{CSDP}
B.~Borchers, ``{CSDP}, a {C} library for semidefinite programming,''
  \emph{Optimization Methods and Software}, vol.~11, no.~1, pp. 613--623, 1999.

\bibitem{CSDP_site}
\BIBentryALTinterwordspacing
------, ``{CSDP},'' 2018. [Online]. Available:
  \url{https://github.com/coin-or/Csdp}
\BIBentrySTDinterwordspacing

\bibitem{CVXOPT}
M.~Andersen, J.~Dahl, Z.~Liu, L.~Vandenberghe, S.~Sra, S.~Nowozin, and
  S.~Wright, ``Interior-point methods for large-scale cone programming,'' in
  \emph{Optimization for Machine Learning}, S.~Sra, S.~Nowozin, and S.~J.
  Wright, Eds.\hskip 1em plus 0.5em minus 0.4em\relax Cambridge, MA: MIT Press,
  2012, vol. 5583, ch.~3, pp. 55--83.

\bibitem{CVXOPT_site}
\BIBentryALTinterwordspacing
M.~Andersen, J.~Dahl, and L.~Vandenberghe, ``{CVXOPT}: Python software for
  convex optimization,'' 2020. [Online]. Available:
  \url{http://cvxopt.org/index.html}
\BIBentrySTDinterwordspacing

\bibitem{DDS_2}
M.~Karimi and L.~Tun{\c{c}}el, ``{Domain-Driven Solver (DDS) Version 2.1}: a
  {MATLAB}-based software package for convex optimization problems in
  domain-driven form,'' \emph{Mathematical Programming Computation}, vol.~16,
  no.~1, pp. 37--92, 2024.

\bibitem{DDS_site}
\BIBentryALTinterwordspacing
M.~Karimi and L.~Tun\c{c}el, ``{DDS} users' guide.'' [Online]. Available:
  \url{http://www.math.uwaterloo.ca/~m7karimi/DDS.html}
\BIBentrySTDinterwordspacing

\bibitem{DSDP_2005}
S.~J. Benson and Y.~Ye, ``{DSDP5}: Software for semidefinite programming,''
  \emph{ACM Transactions on Mathematical Software}, vol.~34, no.~3, pp.
  16:1--20, 2005.

\bibitem{DSDP_site}
\BIBentryALTinterwordspacing
``{DSDP}: Software for semidefinite programming,'' 2006. [Online]. Available:
  \url{https://www.mcs.anl.gov/hs/software/DSDP/}
\BIBentrySTDinterwordspacing

\bibitem{LMILab}
P.~Gahinet, A.~Nemirovskii, A.~J. Laub, and M.~Chilali, ``The {LMI} control
  toolbox,'' in \emph{Proc. IEEE Conference on Decision and Control}, Lake
  Buena Vista, FL, 1994, pp. 2038--2041.

\bibitem{PENLAB}
\BIBentryALTinterwordspacing
J.~Fiala, M.~Ko{\v{c}}vara, and M.~Stingl, ``{PENLAB}: A {MATLAB} solver for
  nonlinear semidefinite optimization,'' \emph{arXiv}, 2013. [Online].
  Available: \url{https://arxiv.org/abs/1311.5240}
\BIBentrySTDinterwordspacing

\bibitem{PENLAB_site}
\BIBentryALTinterwordspacing
M.~Ko{\v{c}}vara, ``{PENLAB},'' 2017. [Online]. Available:
  \url{http://web.mat.bham.ac.uk/kocvara/penlab/}
\BIBentrySTDinterwordspacing

\bibitem{SCS_2016}
B.~O'Donoghue, E.~Chu, N.~Parikh, and S.~Boyd, ``Conic optimization via
  operator splitting and homogeneous self-dual embedding,'' \emph{Journal of
  Optimization Theory and Applications}, vol. 169, no.~3, pp. 1042--1068, 2016.

\bibitem{SCS_site}
\BIBentryALTinterwordspacing
------, ``{SCS}: Splitting conic solver, version 2.1.2,'' 2019. [Online].
  Available: \url{https://github.com/cvxgrp/scs}
\BIBentrySTDinterwordspacing

\bibitem{SDPA_2003}
M.~Yamashita, K.~Fujisawa, and M.~Kojima, ``Implementation and evaluation of
  {SDPA} 6.0 ({SemiDefinite Programming Algorithm} 6.0),'' \emph{Optimization
  Methods and Software}, vol.~18, no.~4, pp. 491--505, 2003.

\bibitem{SDPA_2010}
M.~Yamashita, K.~Fujisawa, K.~Nakata, M.~Nakata, M.~Fukuda, K.~Kobayashi, and
  K.~Goto, ``A high-performance software package for semidefinite programs:
  {SDPA} 7,'' Dept. of Mathematical and Computing Science, Tokyo Institute of
  Technology, Tokyo, Japan, Tech. Rep. B-460, 2010.

\bibitem{SDPA_site}
\BIBentryALTinterwordspacing
K.~Fujisawa, M.~Fukuda, Y.~Futakata, K.~Kobayashi, M.~Kojima, K.~Nakata,
  M.~Nakata, and M.~Yamashita, ``{SDPA} official page,'' 2020. [Online].
  Available: \url{http://sdpa.sourceforge.net/index.html}
\BIBentrySTDinterwordspacing

\bibitem{SMCP_2010}
M.~S. Andersen, J.~Dahl, and L.~Vandenberghe, ``Implementation of nonsymmetric
  interior-point methods for linear optimization over sparse matrix cones,''
  \emph{Mathematical Programming Computation}, vol.~2, no. 3--4, pp. 167--201,
  2010.

\bibitem{SMCP_site}
\BIBentryALTinterwordspacing
M.~S. Andersen and L.~Vandenberghe, ``{SMCP - Python} extension for sparse
  matrix cone programs,'' 2018. [Online]. Available:
  \url{https://smcp.readthedocs.io/en/latest/}
\BIBentrySTDinterwordspacing

\bibitem{SDPNAL_2015}
L.~Q. Yang, D.~F. Sun, and K.~C. Toh, ``{SDPNAL}+: A majorized semismooth
  {Newton-CG} augmented {Lagrangian} method for semidefinite programming with
  nonnegative constraints,'' \emph{Mathematical Programming Computation},
  vol.~7, pp. 331--366, 2015.

\bibitem{SDPNAL_site}
\BIBentryALTinterwordspacing
D.~F. Sun and K.~C. Toh, ``{SDPNALplus}.'' [Online]. Available:
  \url{https://blog.nus.edu.sg/mattohkc/softwares/sdpnalplus/}
\BIBentrySTDinterwordspacing

\bibitem{yang2023inexact}
H.~Yang, L.~Liang, L.~Carlone, and K.-C. Toh, ``An inexact projected gradient
  method with rounding and lifting by nonlinear programming for solving
  rank-one semidefinite relaxation of polynomial optimization,''
  \emph{Mathematical Programming}, vol. 201, no. 1-2, pp. 409--472, 2023.

\bibitem{STRIDE_site}
\BIBentryALTinterwordspacing
H.~Yang and L.~Liang, ``{STRIDE: SpecTrahedRal Inexact projected gradient
  Descent along vErtices},'' 2022. [Online]. Available:
  \url{https://github.com/MIT-SPARK/STRIDE}
\BIBentrySTDinterwordspacing

\bibitem{Mittelmann2002}
H.~D. Mittelmann, ``An independent benchmarking of {SDP} and {SOCP} solvers,''
  \emph{Mathematical Programming}, vol.~95, no.~2, pp. 407--430, 2002.

\bibitem{Arzelier2002}
\BIBentryALTinterwordspacing
D.~Arzelier, D.~Peaucelle, and D.~Henrion, ``Some notes on standard {LMI}
  solvers,'' 2002. [Online]. Available:
  \url{http://homepages.laas.fr/arzelier/publis/2002/prague102.pdf}
\BIBentrySTDinterwordspacing

\bibitem{MittelmannSite}
\BIBentryALTinterwordspacing
H.~D. Mittelmann, ``Decision tree for optimization software,'' 2018. [Online].
  Available: \url{http://plato.la.asu.edu/bench.html}
\BIBentrySTDinterwordspacing

\bibitem{Yalmip}
J.~L\"{o}ftberg, ``\texttt{YALMIP}: A toolbox for modeling and optimization in
  \texttt{MATLAB},'' in \emph{IEEE International Symposium on Computer Aided
  Control Systems Design}, 2004.

\bibitem{Yalmip_site}
\BIBentryALTinterwordspacing
------, ``Yalmip,'' 2020. [Online]. Available: \url{https://yalmip.github.io/}
\BIBentrySTDinterwordspacing

\bibitem{CVX_2008}
M.~Grant and S.~Boyd, ``Graph implementations for nonsmooth convex programs,''
  in \emph{Recent Advances in Learning and Control}, ser. Lecture Notes in
  Control and Information Sciences, V.~Blondel, S.~Boyd, and H.~Kimura,
  Eds.\hskip 1em plus 0.5em minus 0.4em\relax Springer-Verlag Limited, 2008,
  pp. 95--110.

\bibitem{CVX}
\BIBentryALTinterwordspacing
------, ``{CVX}: Matlab software for disciplined convex programming, version
  2.1,'' 2014. [Online]. Available: \url{http://cvxr.com/cvx}
\BIBentrySTDinterwordspacing

\bibitem{ROLMIP_2019}
C.~M. Agulhari, A.~Felipe, R.~C.~L.~F. Oliveira, and P.~L.~D. Peres,
  ``Algorithm 998: {The Robust LMI Parser} - a toolbox to construct {LMI}
  conditions for uncertain systems,'' \emph{ACM Transactions on Mathematical
  Software}, vol.~45, no.~3, p.~36, 2019.

\bibitem{ROLMIP_site}
\BIBentryALTinterwordspacing
------, ``Robust {LMI} parser,'' October 2020. [Online]. Available:
  \url{https://rolmip.github.io/}
\BIBentrySTDinterwordspacing

\bibitem{CVXPY_2016}
S.~Diamond and S.~Boyd, ``{CVXPY}: A {P}ython-embedded modeling language for
  convex optimization,'' \emph{Journal of Machine Learning Research}, vol.~17,
  no.~83, pp. 1--5, 2016.

\bibitem{CVXPY_2018}
A.~Agrawal, R.~Verschueren, S.~Diamond, and S.~Boyd, ``A rewriting system for
  convex optimization problems,'' \emph{Journal of Control and Decision},
  vol.~5, no.~1, pp. 42--60, 2018.

\bibitem{CVXPY_site}
\BIBentryALTinterwordspacing
S.~Diamond and A.~Agrawal, ``{Welcome to CVXPY} 1.0,'' 2019. [Online].
  Available: \url{https://www.cvxpy.org/index.html}
\BIBentrySTDinterwordspacing

\bibitem{PICOS_site}
\BIBentryALTinterwordspacing
G.~Sagnol and M.~Stahlberg, ``A {Python} interface to conic optimization
  solvers,'' 2020. [Online]. Available:
  \url{https://picos-api.gitlab.io/picos/introduction.html}
\BIBentrySTDinterwordspacing

\bibitem{Irene_site}
\BIBentryALTinterwordspacing
M.~Ghasemi, ``{Irene 1.2.3} documentation,'' 2017. [Online]. Available:
  \url{https://irene.readthedocs.io/en/latest/index.html}
\BIBentrySTDinterwordspacing

\bibitem{PyLMI-SDP_site}
\BIBentryALTinterwordspacing
C.~D. Sousa, ``{PyLMI-SDP 0.2},'' 2013. [Online]. Available:
  \url{https://pypi.org/project/PyLMI-SDP}
\BIBentrySTDinterwordspacing

\bibitem{Convex.jl}
M.~Udell, K.~Mohan, D.~Zeng, J.~Hong, S.~Diamond, and S.~Boyd, ``Convex
  optimization in {Julia},'' in \emph{First Workshop for High Performance
  Technical Computing in Dynamic Languages}, New Orleans, LA, 2014, pp. 18--28.

\bibitem{Convex.jl_site}
\BIBentryALTinterwordspacing
J.~Hong, K.~Mohan, M.~Udell, and D.~Zeng, ``Convex.jl - convex optimization in
  {Julia},'' 2019. [Online]. Available:
  \url{https://www.juliaopt.org/Convex.jl/stable/}
\BIBentrySTDinterwordspacing

\bibitem{JuMP_2017}
I.~Dunning, J.~Huchette, and M.~Lubin, ``{JuMP: A Modeling Language for
  Mathematical Optimization},'' \emph{SIAM Review}, vol.~59, no.~2, pp.
  295--320, 2017.

\bibitem{JuMP_site}
\BIBentryALTinterwordspacing
------, ``{JuMP}.'' [Online]. Available:
  \url{https://www.juliaopt.org/JuMP.jl/stable/}
\BIBentrySTDinterwordspacing

\bibitem{NSPYALMIP}
J.~P. Chancelier, P.~V. Pakshin, and S.~G. Soloviev, ``{LMI} parse for {NSP}
  software package,'' \emph{IFAC Proceedings Volumes: 18th IFAC World
  Congress}, vol.~44, no.~1, pp. 14\,253--14\,258, 2011.

\bibitem{NSPYALMIP_site}
\BIBentryALTinterwordspacing
J.~P. Chancelier, ``Nsp toolboxes,'' 2016. [Online]. Available:
  \url{https://cermics.enpc.fr/~jpc/nsp-tiddly/}
\BIBentrySTDinterwordspacing

\bibitem{Gu2013}
D.~W. Gu, P.~H. Petkov, and M.~M. Konstantinov, \emph{Robust Control Design
  with {MATLAB}}, 2nd~ed.\hskip 1em plus 0.5em minus 0.4em\relax London, UK:
  Springer, 2013.

\bibitem{Gu1999}
K.~Gu, ``Partial solution of {LMI} in stability problem of time-delay
  systems,'' in \emph{Proc. IEEE Conference on Decision and Control}, Phoenix,
  AZ, 1999, pp. 227--232.

\bibitem{Gu2003}
K.~Gu, V.~L. Kharitonov, and J.~Chen, \emph{Stability of Time-Delay
  Systems}.\hskip 1em plus 0.5em minus 0.4em\relax Boston, MA: Birkhauser
  Boston, 2003.

\bibitem{GeromelNotes}
J.~C. Geromel, ``Robustness of linear dynamic systems,'' August 2005, {Lecture
  Notes}.

\bibitem{Chang2013}
X.~H. Chang and G.~H. Yang, ``New results on output feedback control for linear
  discrete-time systems,'' \emph{IEEE Transactions on Automatic Control},
  vol.~59, no.~5, pp. 1355--1359, 2013.

\bibitem{Gu2001}
K.~Gu, ``A further refinement of discretized lyapunov functional method for the
  stability of time-delay systems,'' \emph{International Journal of Control},
  vol.~74, no.~10, pp. 967--976, 2001.

\bibitem{Gahinet1994}
P.~Gahinet and P.~Apkarian, ``A linear matrix inequality approach to
  $\mathcal{H}_\infty$ control,'' \emph{International Journal of Robust and
  Nonlinear Control}, vol.~4, no.~4, pp. 421--448, 1994.

\bibitem{Zhan2002}
X.~Zhan, \emph{Matrix Inequalities}, ser. Lecture Notes in Mathematics.\hskip
  1em plus 0.5em minus 0.4em\relax Berlin, Germany: Springer-Verlag, 2002, vol.
  1790.

\bibitem{HelmerssonThesis}
A.~Helmersson, ``Methods for robust gain scheduling,'' Ph.D. dissertation,
  Link\"{o}ping University, Link\"{o}ping, Sweden, Nov. 1995.

\bibitem{meijer2024non}
T.~J. Meijer, T.~Holicki, S.~Van Den~Eijnden, C.~W. Scherer, and W.~M. Heemels,
  ``The non-strict projection lemma,'' \emph{IEEE Transactions on Automatic
  Control}, vol.~69, no.~8, pp. 5584--5590, 2024.

\bibitem{Apkarian2001}
P.~Apkarian, H.~D. Tuan, and J.~Bernussou, ``Continuous-time analysis,
  eigenstructure assignment, and $\mathcal{H}_2$ synthesis with enhanced linear
  matrix inequalities ({LMI}) characterizations,'' \emph{IEEE Transactions on
  Automatic Control}, vol.~46, no.~12, pp. 1941--1946, 2001.

\bibitem{Chang2011}
X.~H. Chang and G.~H. Yang, ``A descriptor representation approach to
  observer-based $\mathcal{H}_\infty$ control synthesis for discrete-time fuzzy
  systems,'' \emph{Fuzzy Sets and Systems}, vol. 185, no.~1, pp. 38--51, 2011.

\bibitem{Delmotte2007}
F.~Delmotte, T.~M. Guerra, and M.~Ksantini, ``Continuous {Takagi-Sugeno}'s
  models: Reduction of the number of {LMI} conditions in various fuzzy control
  design technics,'' \emph{IEEE Transactions on Fuzzy Systems}, vol.~15, no.~3,
  pp. 426--438, 2007.

\bibitem{Chang2011a}
X.~H. Chang and G.~H. Yang, ``Nonfragile $\mathcal{H}_\infty$ filtering of
  continuous-time fuzzy systems,'' \emph{IEEE Transactions on Signal
  Processing}, vol.~59, no.~4, pp. 1528--1538, 2011.

\bibitem{Chang2014}
X.-H. Chang, \emph{Robust Output Feedback $\mathcal{H}_\infty$ Control and
  Filtering for Uncertain Linear Systems}.\hskip 1em plus 0.5em minus
  0.4em\relax Berlin, Germany: Springer, 2014.

\bibitem{Finsler1936}
P.~Finsler, ``{\"U}ber das vorkommen definiter und semidefiniter formen in
  scharen quadratischer formen,'' \emph{Commentarii Mathematici Helvetici},
  vol.~9, no.~1, pp. 188--192, 1936.

\bibitem{Petersen1987}
I.~R. Petersen, ``A stabilization algorithm for a class of uncertain linear
  systems,'' \emph{Systems \& Control Letters}, vol.~8, no.~4, pp. 351--357,
  1987.

\bibitem{deOliveira2001}
M.~C. {de Oliveira} and R.~E. Skelton, ``Stability tests for constrained linear
  systems,'' in \emph{Perspectives in Robust Control}, ser. Lecture Notes in
  Control and Information Sciences, S.~P. Moheimani, Ed.\hskip 1em plus 0.5em
  minus 0.4em\relax London, UK: Springer, 2001, vol. 268.

\bibitem{Gonccalves2025robust}
E.~N. Gon{\c{c}}alves, D.~J. Pereira, and F.~B. Dos~Santos, ``Robust
  $\mathcal{H}_\infty$ control of discrete-time uncertain systems based on the
  extended {Finsler}'s lemma,'' \emph{IEEE Access}, vol.~13, pp.
  216\,428--216\,437, 2025.

\bibitem{Jacobson1977}
D.~H. Jacobson, \emph{Extensions of Linear-Quadratic Control, Optimization and
  Matrix Theory}, ser. Mathematics in Science and Engineering.\hskip 1em plus
  0.5em minus 0.4em\relax New York, NY: Academic Press, 1977, vol. 133.

\bibitem{Skelton1998}
R.~E. Skelton, T.~Iwasaki, and K.~Grigoriadis, \emph{A Unified Algebraic
  Approach to Linear Control Design}.\hskip 1em plus 0.5em minus 0.4em\relax
  London, UK: Taylor \& Francis, 1998.

\bibitem{more1993generalizations}
J.~J. Mor{\'e}, ``Generalizations of the trust region problem,''
  \emph{Optimization Methods and Software}, vol.~2, no. 3-4, pp. 189--209,
  1993.

\bibitem{zi2010some}
Y.~Zi-Zong and G.~Jin-Hai, ``Some equivalent results with {Yakubovich's
  S-lemma},'' \emph{SIAM Journal on Control and Optimization}, vol.~48, no.~7,
  pp. 4474--4480, 2010.

\bibitem{anstreicher2000note}
K.~M. Anstreicher and M.~H. Wright, ``A note on the augmented {Hessian} when
  the reduced {Hessian} is semidefinite,'' \emph{SIAM Journal on Optimization},
  vol.~11, no.~1, pp. 243--253, 2000.

\bibitem{van2021matrix}
H.~J. Van~Waarde and M.~K. Camlibel, ``A matrix {Finsler}'s lemma with
  applications to data-driven control,'' in \emph{Conference on Decision and
  Control}, Austin, TX, 2021, pp. 5777--5782.

\bibitem{Wu2010}
M.~Wu, Y.~He, and J.~H. She, \emph{Stability Analysis and Robust Control of
  Time-Delay Systems}.\hskip 1em plus 0.5em minus 0.4em\relax Berlin,
  Heidelberg: Springer, 2010.

\bibitem{Xie1992}
L.~Xie, M.~Fu, and C.~{de Souza}, ``$\mathcal{H}_\infty$ control and quadratic
  stabilization of systems with parameter uncertainty via output feedback,''
  \emph{IEEE Transactions on Automatic Control}, vol.~37, no.~8, pp.
  1253--1256, 1992.

\bibitem{Xie1996}
L.~Xie, ``Output feedback $\mathcal{H}_\infty$ control of systems with
  parameter uncertainty,'' \emph{International Journal of Control}, vol.~63,
  no.~4, pp. 741--750, 1996.

\bibitem{kussaba2015uniform}
H.~Kussaba, J.~Y. Ishihara, and R.~A. Borges, ``Uniform versions of {Finsler}'s
  lemma,'' in \emph{IEEE Conference on Decision and Control}.\hskip 1em plus
  0.5em minus 0.4em\relax Osaka, Japan: IEEE, 2015, pp. 7292--7297.

\bibitem{ishihara2017existence}
J.~Y. Ishihara, H.~T.~M. Kussaba, and R.~A. Borges, ``Existence of continuous
  or constant {Finsler}'s variables for parameter-dependent systems,''
  \emph{IEEE Transactions on Automatic Control}, vol.~62, no.~8, pp.
  4187--4193, 2017.

\bibitem{Petersen1986}
I.~R. Petersen and C.~V. Hollot, ``A {Riccati} equation approach to the
  stabilization of uncertain linear systems,'' \emph{Automatica}, vol.~22,
  no.~4, pp. 397--411, 1986.

\bibitem{Bisoffi2022}
A.~Bisoffi, C.~De~Persis, and P.~Tesi, ``Data-driven control via {Petersen's}
  lemma,'' \emph{Automatica}, vol. 145, p. 110537, November 2022.

\bibitem{Khlebnikov2018}
M.~V. Khlebnikov, ``Quadratic stabilization of discrete-time bilinear
  systems,'' \emph{Automation and Remote Control}, vol.~79, no.~7, pp.
  1222--1239, 2018.

\bibitem{Khlebnikov2008}
M.~V. Khlebnikov and P.~S. Shcherbakov, ``{Petersen's} lemma on matrix
  uncertainty and its generalizations,'' \emph{Automation and Remote Control},
  vol.~69, no.~11, pp. 1932--1945, 2008.

\bibitem{Shcherbakov2008}
P.~Shcherbakov and M.~Topunov, ``Extensions of {Petersen's} lemma on matrix
  uncertainty,'' \emph{IFAC Proceedings Volumes}, vol.~41, no.~2, pp.
  11\,385--11\,390, 2008.

\bibitem{Khlebnikov2014}
M.~V. Khlebnikov, ``New generalizations of the {Petersen} lemma,''
  \emph{Automation and Remote Control}, vol.~75, no.~5, pp. 917--921, 2014.

\bibitem{Ebihara2004}
Y.~Ebihara and T.~Hagiwara, ``New dilated {LMI} characterizations for
  continuous-time multiobjective controller synthesis,'' \emph{Automatica},
  vol.~10, pp. 2003--2009, 2004.

\bibitem{Zhou1988}
K.~Zhou and P.~P. Khargonekar, ``Robust stabilization of linear systems with
  norm-bounded time-varying uncertainty,'' \emph{Systems \& Control Letters},
  vol.~10, no.~1, pp. 17--20, 1988.

\bibitem{Young2016}
A.~Zemouche, R.~Rajamani, B.~Boulkroune, H.~Rafaralahy, and M.~Zasadzinski,
  ``$\mathcal{H}_\infty$ circle criterion observer design for {Lipschitz}
  nonlinear systems with enhanced {LMI} conditions,'' in \emph{Proc. American
  Control Conference}, Boston, MA, 2016, pp. 131--136.

\bibitem{Merco2020}
R.~Merco, F.~Ferrante, R.~G. Sanfelice, and P.~Pisu, ``{LMI}-based output
  feedback control design in the presence of sporadic measurements,'' in
  \emph{Proc. American Control Conference}, Denver, CO, 2020, pp. 3331--3336.

\bibitem{Cao1998}
Y.~Y. Cao, Y.~X. Sun, and C.~Cheng, ``Delay-dependent robust stabilization of
  uncertain systems with multiple state delays,'' \emph{IEEE Transactions on
  Automatic Control}, vol.~43, no.~11, pp. 1608--1612, 1998.

\bibitem{Wang1992}
Y.~Wang, L.~Xie, and C.~E. {de Souza}, ``Robust control of a class of
  uncertaint nonlinear systems,'' \emph{Systems \& Control Letters}, vol.~19,
  no.~2, pp. 139--149, 1992.

\bibitem{Tahir2015}
F.~Tahir and I.~M. Jaimoukha, ``Low-complexity polytopic invariant sets for
  linear systems subject to norm-bounded uncertainty,'' \emph{IEEE Transactions
  on Automatic Control}, vol.~60, no.~5, pp. 1416--1421, 2015.

\bibitem{dinh2011combining}
Q.~T. Dinh, S.~Gumussoy, W.~Michiels, and M.~Diehl, ``Combining convex--concave
  decompositions and linearization approaches for solving {BMIs}, with
  application to static output feedback,'' \emph{IEEE Transactions on Automatic
  Control}, vol.~57, no.~6, pp. 1377--1390, 2011.

\bibitem{priuli2022static}
A.~Priuli, S.~Tarbouriech, and L.~Zaccarian, ``Static linear anti-windup design
  with sign-indefinite quadratic forms,'' \emph{IEEE Control Systems Letters},
  vol.~6, pp. 3158--3163, 2022.

\bibitem{Warner2015}
E.~C. Warner and J.~T. Scruggs, ``Control of vibratory networks with passive
  and regenerative systems,'' in \emph{Proc. American Control Conference},
  Chicago, IL, 2015, pp. 5502--5508.

\bibitem{Warner2017}
------, ``Iterative convex overbounding algorithms for {BMI} optimization
  problems,'' \emph{IFAC PapersOnline}, vol.~50, no.~1, pp. 10\,449--10\,455,
  2017.

\bibitem{kheirandishfard2018convex}
M.~Kheirandishfard, F.~Zohrizadeh, and R.~Madani, ``Convex relaxation of
  bilinear matrix inequalities part i: Theoretical results,'' in \emph{IEEE
  Conference on Decision and Control}, Miami, FL, 2018, pp. 67--74.

\bibitem{kheirandishfard2018convexII}
M.~Kheirandishfard, F.~Zohrizadeh, M.~Adil, and R.~Madani, ``Convex relaxation
  of bilinear matrix inequalities part ii: Applications to optimal control
  synthesis,'' in \emph{IEEE Conference on Decision and Control}, Miami, FL,
  2018, pp. 75--82.

\bibitem{wang2018sequential}
Y.~Wang, A.~Zemouche, and R.~Rajamani, ``A sequential {LMI} approach to design
  a {BMI}-based multi-objective nonlinear observer,'' \emph{European Journal of
  Control}, vol.~44, pp. 50--57, 2018.

\bibitem{iwasaki1999dual}
T.~Iwasaki, ``The dual iteration for fixed-order control,'' \emph{IEEE
  Transactions on Automatic Control}, vol.~44, no.~4, pp. 783--788, 1999.

\bibitem{doyle1985matrix}
J.~C. Doyle and C.-C. Chu, ``Matrix interpolation and $\mathcal{H}_\infty$
  performance bounds,'' in \emph{American Control Conference}, Boston, MA,
  1985, pp. 129--134.

\bibitem{doyle1985structured}
J.~C. Doyle, ``Structured uncertainty in control system design,'' in \emph{IEEE
  Conference on Decision and Control}, Fort Lauderdale, FL, 1985, pp. 260--265.

\bibitem{doyle1983synthesis}
------, ``Synthesis of robust controllers and filters,'' in \emph{IEEE
  Conference on Decision and Control}, San Antonio, TX, 1983, pp. 109--114.

\bibitem{geromel1993output}
J.~Geromel, P.~Peres, and S.~Souza, ``Output feedback stabilization of
  uncertain systems through a min/max problem,'' \emph{IFAC Proceedings
  Volumes}, vol.~26, no.~2, pp. 215--218, 1993.

\bibitem{rotea1994alternative}
M.~A. Rotea and T.~Iwasaki, ``An alternative to the {DK} iteration?'' in
  \emph{American Control Conference}, vol.~1, Baltimore, MD, 1994, pp. 53--57.

\bibitem{iwasaki1995xy}
T.~Iwasaki and R.~Skelton, ``The {XY}-centring algorithm for the dual {LMI}
  problem: a new approach to fixed-order control design,'' \emph{International
  Journal of Control}, vol.~62, no.~6, pp. 1257--1272, 1995.

\bibitem{yamada1997lmi}
Y.~Yamada and S.~Hara, ``An {LMI} approach to local optimization for constantly
  scaled $\mathcal{H}_\infty$ control problems,'' \emph{International Journal
  of Control}, vol.~67, no.~2, pp. 233--250, 1997.

\bibitem{doroudchi2018decentralized}
A.~Doroudchi, S.~Shivakumar, R.~E. Fisher, H.~Marvi, D.~Aukes, X.~He,
  S.~Berman, and M.~M. Peet, ``Decentralized control of distributed actuation
  in a segmented soft robot arm,'' in \emph{IEEE Conference on Decision and
  Control}, Miami, FL, 2018, pp. 7002--7009.

\bibitem{dahdah2022system}
S.~Dahdah and J.~R. Forbes, ``System norm regularization methods for {Koopman}
  operator approximation,'' \emph{Proceedings of the Royal Society A}, vol.
  478, no. 2265, p. 20220162, 2022.

\bibitem{Yakubovich1977}
V.~A. Yakubovich, ``The {S}-procedure in non-linear control theory,''
  \emph{Vestnik Leningrad University, Mathematics}, vol.~4, pp. 73--93, 1977.

\bibitem{Jonsson2001}
\BIBentryALTinterwordspacing
U.~T. J{\"o}nsson, ``A lecture on the {S}-procedure,'' \emph{Lecture Notes at
  the Royal Institute of Technology}, 2001. [Online]. Available:
  \url{https://people.kth.se/~uj/5B5746/Lecture.ps}
\BIBentrySTDinterwordspacing

\bibitem{polik2007survey}
I.~P{\'o}lik and T.~Terlaky, ``A survey of the {S}-lemma,'' \emph{SIAM Review},
  vol.~49, no.~3, pp. 371--418, 2007.

\bibitem{van2020noisy}
H.~J. Van~Waarde, M.~K. Camlibel, and M.~Mesbahi, ``From noisy data to feedback
  controllers: Nonconservative design via a matrix {S}-lemma,'' \emph{IEEE
  Transactions on Automatic Control}, vol.~67, no.~1, pp. 162--175, 2020.

\bibitem{Fathi2019}
M.~Fathi and H.~Bevrani, \emph{Optimization in Electrical Engineering}.\hskip
  1em plus 0.5em minus 0.4em\relax Cham, Switzerland: Springer, 2019.

\bibitem{LallNotes3}
\BIBentryALTinterwordspacing
S.~Lall, ``Engr210a lecture 3: Singular values and {LMI}s,'' August 2001.
  [Online]. Available:
  \url{https://lall.stanford.edu/engr210a/lectures/lecture3_2001_10_08_01.pdf}
\BIBentrySTDinterwordspacing

\bibitem{Fazel2001}
M.~Fazel, H.~Hindi, and S.~P. Boyd, ``A rank minimization heuristic with
  application to minimum order system approximation,'' in \emph{Proc. American
  Control Conference}, Arlington, VA, 2001, pp. 4734--4739.

\bibitem{Recht2010}
B.~Recht, M.~Fazel, and P.~A. Parrilo, ``Guaranteed minimum-rank solutions of
  linear matrix equations via nuclear norm minimization,'' \emph{SIAM Review},
  vol.~52, no.~3, pp. 471--501, 2010.

\bibitem{Alizadeh1995}
F.~Alizadeh, ``Interior point methods in semidefinite programming with
  applications to combinatorial optimization,'' \emph{SIAM Journal on
  Optimization}, vol.~5, no.~1, pp. 13--51, 1995.

\bibitem{Zhang2011}
F.~Zhang, \emph{Matrix Theory: Basic Resuls and Techniques}, 2nd~ed.\hskip 1em
  plus 0.5em minus 0.4em\relax New York, NY: Springer, 2011.

\bibitem{Bourin1999}
J.-C. Bourin, ``Some inequalities for norms on matrices and operators,''
  \emph{Linear Algebra and its Applications}, vol. 292, no. 1--3, pp. 139--154,
  1999.

\bibitem{Travaglia2006}
M.~V. Travaglia, ``On an inequality involving power and contraction matrices
  with and without trace,'' \emph{Journal of Inequalities in Pure and Applied
  Mathematics}, vol.~7, no.~2, p.~65, 2006.

\bibitem{Douglas1966}
R.~G. Douglas, ``On majorization, factorication, and range inclusion of
  operators on {Hilbert} space,'' \emph{Proc. American Mathematics Society},
  vol.~17, no.~2, pp. 413--415, 1966.

\bibitem{Fillmore1971}
P.~A. Fillmore and J.~P. Williams, ``On operator ranges,'' \emph{Advances in
  Mathematics}, vol.~7, no.~3, pp. 254--281, 1971.

\bibitem{Dym2006}
H.~Dym, \emph{Linear Algebra in Action}.\hskip 1em plus 0.5em minus 0.4em\relax
  Providence, RI: American Mathematical Society, 2006.

\bibitem{AuYeung1973}
Y.-H. Au-Yeung, ``Some inequalities for the rational power of a nonnegative
  definite matrix,'' \emph{Linear Algebra and its Applications}, vol.~7, no.~4,
  pp. 347--350, 1973.

\bibitem{Chan1985}
N.~N. Chan and M.~K. Kwong, ``Hermitian matrix inequalities and a conjecture,''
  \emph{American Mathematical Monthly}, vol.~92, no.~8, pp. 533--541, 1985.

\bibitem{Bhatia1990}
R.~Bhatia and F.~Kittaneh, ``On the singular values of a product of
  operators,'' \emph{SIAM Journal on Matrix Analysis and Applications},
  vol.~11, no.~2, pp. 272--277, 1990.

\bibitem{Aujla2007}
J.~S. Aujla and J.-C. Bourin, ``Eigenvalue inequalities for convex and
  log-convex functions,'' \emph{Linear Algebra and its Applications}, vol. 424,
  no.~1, pp. 25--35, 2007.

\bibitem{Bourin2006}
J.-C. Bourin, ``Reverse rearrangement inequalities via matrix technics,''
  \emph{Journal of Inequalities in Pure and Applied Mathematics}, vol.~7,
  no.~2, p.~43, 2006.

\bibitem{Baksalary1991}
J.~K. Baksalary and F.~Pukelsheim, ``On the {L\"{o}wner}, minus, and star
  partial orderings of nonnegative definite matrices,'' \emph{Linear Algebra
  and its Applications}, vol. 151, pp. 135--141, June 1991.

\bibitem{Kwong1989}
M.~K. Kwong, ``Some results on matrix monotone functions,'' \emph{Linear
  Algebra and its Applications}, vol. 118, pp. 129--153, June 1989.

\bibitem{Bellman1968}
R.~Bellman, ``Some inequalities for the square root of a positive definite
  matrix,'' \emph{Linear Algebra and its Applications}, vol.~1, no.~3, pp.
  321--324, 1968.

\bibitem{Bhagwat1978}
K.~V. Bhagwat and R.~Subramanian, ``Inequalities between means of positive
  operators,'' \emph{Mathematical Proceedings of the Campbridge Philosophical
  Society}, vol.~83, no.~3, pp. 393--401, 1978.

\bibitem{Willems1971}
J.~C. Willems, ``Least squares stationary optimal control and the algebraic
  {Riccati} equation,'' \emph{IEEE Transactions on Automatic Control}, vol.~16,
  no.~6, pp. 621--634, 1971.

\bibitem{Venkataraman2018}
R.~Venkataraman and P.~Seiler, ``Convex {LPV} synthesis of estimators and
  feedforwards using dualuty and integral quadratic constraints,''
  \emph{International Journal of Robust and Nonlinear Control}, vol.~28, no.~3,
  pp. 953--975, 2018.

\bibitem{Ebihara2015}
Y.~Ebihara, D.~Peaucelle, and D.~Arzelier, \emph{$S$-Variable Approach to
  LMI-Based Robust Control}.\hskip 1em plus 0.5em minus 0.4em\relax London, UK:
  Springer, 2015.

\bibitem{Geromel1998}
J.~C. Geromel, M.~C. {de Oliveira}, and L.~Hsu, ``{LMI} characterization of
  structural and robust stability,'' \emph{Linear Algebra and its
  Applications}, vol. 285, no. 1--3, pp. 69--80, 1998.

\bibitem{Felipe2016}
A.~Felipe, R.~C.~L.~F. Oliveira, and P.~L.~D. Peres, ``An iterative {LMI} based
  procedure for robust stabilization of continuous-time polytopic systems,'' in
  \emph{Proc. American Control Conference}, Boston, MA, 2016, pp. 3826--3831.

\bibitem{Felipe2020}
A.~Felipe and R.~C.~L.~F. Oliveira, ``An {LMI}-based algorithm to compute
  robust stabilizing feedback gains directly as optimization variables,''
  \emph{IEEE Transactions on Automatic Control}, 2020, in press.

\bibitem{DeOliveira1999}
M.~C. {De Oliveira}, J.~Bernussou, and J.~C. Geromel, ``A new discrete-time
  robust stability conditions,'' \emph{Systems \& Control Letters}, vol.~37,
  no.~4, pp. 261--265, 1999.

\bibitem{DeOliveira1999b}
M.~C. {De Oliveira}, J.~C. Geromel, and L.~Hsu, ``{LMI} characterization of
  structural and robust stability: The discrete-time case,'' \emph{Linear
  Algebra and its Applications}, vol. 296, no. 1--3, pp. 27--38, 1999.

\bibitem{Felipe2017}
A.~Felipe, ``Um algoritmo de busca local baseado em {LMIs} para computar ganhos
  de realimenta\c{c}\~{a}o estabilizantes diretamente como vari\'{a}veis de
  otimiza\c{c}\~{a}o,'' Master's thesis, Universidade Estuadual de Campinas,
  Campinas, Brazil, 2017.

\bibitem{Spagolla2019}
A.~Spagolla, C.~F. Morais, R.~C.~L.~F. Oliveira, and P.~L.~D. Peres,
  ``Realimenta\c{c}\~{a}o est\'{a}tica de sa\'{ı}da de sistemas {LPV}
  positivos a tempo discreto,'' in \emph{Simp\'{o}sio Brasileiro de
  Automa\c{c}\~{a}o Inteligente}, Ouro Preto, Brazil, 2019, pp. 774--779.

\bibitem{Masabuchi1997}
I.~Masubuchi, Y.~Kamitane, A.~Ohara, and N.~Suda, ``$\mathcal{H}_\infty$
  control for descriptor systems: A matrix inequalities approach,''
  \emph{Automatica}, vol.~33, no.~4, pp. 669--673, 1997.

\bibitem{Wang1998}
H.-S. Wang, C.-F. Yung, and F.-R. Chang, ``Bounded real lemma and
  $\mathcal{H}_\infty$ control for descriptor systems,'' \emph{IEE Proceedings
  - Control Theory and Applications}, vol. 145, no.~3, pp. 316--322, 1998.

\bibitem{Chadli2017}
M.~Chadli, P.~Shi, Z.~Feng, and J.~Lam, ``New bounded real lemma formulation
  and $\mathcal{H}_\infty$ control for continuous-time descriptor systems,''
  \emph{Asian Journal of Control}, vol.~19, no.~6, pp. 2192--2198, 2017.

\bibitem{Marx2003}
B.~Marx, D.~Koenig, and D.~Georges, ``Robust pole-clustering for descriptor
  systems a strict {LMI} characterization,'' in \emph{Proc. European Control
  Conference}, Cambridge, UK, 2003, pp. 1117--1122.

\bibitem{Hsiung1999}
K.-L. Hsiung and L.~Lee, ``Lyapunov inequality and bounded real lemma for
  discrete-time descriptor systems,'' \emph{IEE Proceedings - Control Theory
  and Applications}, vol. 146, no.~4, pp. 327--331, 1999.

\bibitem{Xu1999}
S.~Xu and C.~Yang, ``Stabilization of discrete-time singular systems: A matrix
  inequalities approach,'' \emph{Automatica}, vol.~35, no.~9, pp. 1613--1617,
  1999.

\bibitem{Zhang2008}
G.~Zhang, Y.~Xia, and P.~Shi, ``New bounded real lemma for discrete-time
  singular systems,'' \emph{Automatica}, vol.~44, no.~3, pp. 886--890, 2008.

\bibitem{Chadli2012}
M.~Chadli and M.~Darouach, ``Novel bounded real lemma for discrete-time
  descriptor systems: Application to $\mathcal{H}_\infty$ control design,''
  \emph{Automatica}, vol.~48, no.~2, pp. 449--453, 2012.

\bibitem{Xu2004}
S.~Xu and J.~Lam, ``Robust stability and stabilization of discrete singular
  systems: An equivalent characterization,'' \emph{IEEE Transactions on
  Automatic Control}, vol.~49, no.~4, pp. 568--574, 2004.

\bibitem{Masubuchi2013}
I.~Masubuchi and Y.~Ohta, ``Stability and stabilization of discrete-time
  descriptor systems with several extensions,'' in \emph{Proc. European Control
  Conference}, Z\"{u}rich, Switzerland, 2013, pp. 3378--3383.

\bibitem{Scherer1990}
C.~Scherer, ``The {Riccati} inequality and state-space
  $\mathcal{H}_\infty$-optimal control,'' Ph.D. dissertation, Julius
  Maximilians University W{\"u}rzburg, W{\"u}rzburg, Germany, 1990.

\bibitem{Xie2008}
W.~Xie, ``An equivalent {LMI} representation of bounded real lemma for
  continuous-time systems,'' \emph{Journal of Inequalities and Applications},
  vol. 2008, p. 672905, 2008.

\bibitem{Krokavec2011}
D.~Krokavec and A.~Filasov\'{a}, ``Equivalent representations of bounded real
  lemma,'' in \emph{18th International Conference on Process Control},
  Tatransk\'{a} Lomniva, Slovakia, 2011, pp. 106--110.

\bibitem{Lemaire2019}
A.~A. Lemaire, ``M\'{e}todos iterativos baseados em desigualdades matriciais
  lineares para controle de sistemas lineares incertos positivos cont\'{i}nuos
  no tempo,'' Master's thesis, Universidade Estuadual de Campinas, Campinas,
  Brazil, 2019.

\bibitem{Anderson1973}
B.~D.~O. Anderson and S.~Vongpanitlerd, \emph{Network Analysis and Synthesis: A
  Modern Systems Theory Approach}, ser. Networks Series, R.~W. Newcomb,
  Ed.\hskip 1em plus 0.5em minus 0.4em\relax Englewood Cliffs, NJ:
  Prentice-Hall, 1973.

\bibitem{Rantzer1996}
A.~Rantzer, ``On the {Kalman}-{Yakubovich}-{Popov} lemma,'' \emph{Systems \&
  Control Letters}, vol.~28, no.~1, pp. 7--10, 1996.

\bibitem{Xie1996b}
L.~Xie, C.~E. {de Souza}, and Y.~Wang, ``Robust filtering for a class of
  discrete-time uncertain nonlinear systems: An $\mathcal{H}_\infty$
  approach,'' \emph{International Journal of Robust and Nonlinear Control},
  vol.~6, no.~4, pp. 297--312, 1996.

\bibitem{DeOliveira2002}
M.~C. {De Oliveira}, J.~C. Geromel, and J.~Bernussou, ``Extended
  $\mathcal{H}_2$ and $\mathcal{H}_\infty$ norm characterization and controller
  parameterization for discrete-time systems,'' \emph{International Journal of
  Control}, vol.~75, no.~9, pp. 666--679, 2002.

\bibitem{Masubuchi1995}
I.~Masubuchi, A.~Ohara, and N.~Suda, ``{LMI}-based output feedback controller
  design,'' in \emph{Proc. American Control Conference}, Seattle, WA, 1995, pp.
  3473--3477.

\bibitem{Masubuchi1998}
------, ``{LMI}-based controller synthesis: A unified formulation and
  solution,'' \emph{International Journal of Robust and Nonlinear Control},
  vol.~8, no.~8, pp. 669--686, 1998.

\bibitem{deSouza2006}
C.~E. {de Souza}, K.~A. Barbosa, and A.~T. Neto, ``Robust $\mathcal{H}_\infty$
  filtering for discrete-time linear systems with uncertain time-varying
  parameters,'' \emph{IEEE Transactions on Signal Processing}, vol.~54, no.~6,
  pp. 2110--2118, 2006.

\bibitem{Spagolla2019b}
A.~Spagolla, ``An\'{a}lise de estabilidade e s\'{i}ntese de controle para
  sistemas lineares positivos discretos no tempo por meio de desigualdades
  matriciais lineares,'' Master's thesis, Universidade Estuadual de Campinas,
  Campinas, Brazil, 2019.

\bibitem{Vaidyanathan1985}
P.~P. Vaidyanathan, ``The discrete-time bounded-real lemma in digital
  filtering,'' \emph{IEEE Transactions on Circuits and Systems}, vol.~32,
  no.~9, pp. 918--924, 1985.

\bibitem{Uezato1999}
E.~Uezato and M.~Ikeda, ``Strict {LMI} conditions for stability, robust
  stabilization, and $\mathcal{H}_\infty$ control of descriptor systems,'' in
  \emph{Proc. IEEE Conference on Decision and Control}, Phoenix, AZ, 1999, pp.
  4092--4097.

\bibitem{Rehm2002}
A.~Rehm and F.~Allg\"{o}wer, ``An {LMI} approach towards $\mathcal{H}_\infty$
  control of discrete-time descriptor systems,'' in \emph{Proc. American
  Control Conference}, Anchorage, AK, 2002, pp. 614--619.

\bibitem{Wu2006}
A.-G. Wu and G.-R. Duan, ``Enhanced {LMI} representations for $\mathcal{H}_2$
  performance of polytopic uncertaint systsems: Continuous-time case,''
  \emph{International Journal of Automation and Computing}, vol.~3, pp.
  304--308, 2006.

\bibitem{Steentjes2020}
\BIBentryALTinterwordspacing
T.~R.~V. Steentjes, M.~Lazar, and P.~M.~J. {Van den Hof}, ``Distributed
  $\mathcal{H}_2$ control for interconnected discrete-time systems: A
  dissipativity-based approach,'' \emph{arXiv}, 2020. [Online]. Available:
  \url{https://arxiv.org/abs/2001.04875v1}
\BIBentrySTDinterwordspacing

\bibitem{DeCaigny2010}
J.~{De Caigny}, J.~F. Camino, R.~C.~L.~F. Oliveira, P.~L.~D. Peres, and
  J.~Swevers, ``Gain-scheduled $\mathcal{H}_2$ and $\mathcal{H}_\infty$ control
  of discrete-time polytopic time-varying systems,'' \emph{IET Control Theory
  and Applications}, vol.~4, no.~3, pp. 362--380, 2010.

\bibitem{Santos2017}
L.~A.~F. Santos, ``Projeto de controladores e filtros robustos para sistemas
  lineares discretos com enriquecimento de din\^{a}mica,'' Ph.D. dissertation,
  Universidade Estuadual de Campinas, Campinas, Brazil, 2017.

\bibitem{Geromel1993}
J.~C. Geromel, P.~L.~D. Peres, and S.~R. Souza, ``$\mathcal{H}_2$ guaranteed
  cost control for uncertain discrete-time linear systems,''
  \emph{International Journal of Control}, vol.~57, no.~4, pp. 853--864, 1993.

\bibitem{Takaba1997}
K.~Takaba and T.~Katayama, ``Robust $\mathcal{H}_2$ performance of uncertain
  descriptor system,'' in \emph{Proc. European Control Conference}, Brussels,
  Belgium, 1997, pp. 950--955.

\bibitem{Takaba1998}
K.~Takaba, ``Robust $\mathcal{H}_2$ control of descriptor system with
  time-varying uncertainty,'' \emph{International Journal of Control}, vol.~71,
  no.~4, pp. 559--579, 1998.

\bibitem{Ikeda2000}
M.~Ikeda, T.-W. Lee, and E.~Uezato, ``A strict {LMI} condition for
  $\mathcal{H}_2$ control of descriptor systems,'' in \emph{Proc. IEEE
  Conference on Decision and Control}, Sydney, Australia, 2000, pp. 601--604.

\bibitem{Yagoubi2009}
M.~Yagoubi, ``On multiobjective synthesis for parameter-dependent descriptor
  systems,'' \emph{IET Control Theory and Applications}, vol.~4, no.~5, pp.
  817--826, 2010.

\bibitem{Belov2018}
A.~A. Belov, O.~G. Andrianova, and A.~P. Kurdyukov, \emph{Control of
  Discrete-Time Descriptor Systems: An Anisotropy-Based Approach}, ser. Studies
  in Systems, Decision and Control.\hskip 1em plus 0.5em minus 0.4em\relax
  Cham, Switzerland: Springer, 2018, vol. 157.

\bibitem{Yang2002}
D.~M. Yang, Q.~L. Zhang, B.~Yao, and C.~M. Sha, ``$\mathcal{H}_2$ performance
  analysis and control for discrete-time descriptor systems,'' in \emph{Proc.
  World Congress on Intelligent Control and Automation}, Shanghai, China, 2002,
  pp. 3039--3043.

\bibitem{Kang2018}
D.~Kang, S.~Li, and H.-M. Lee, ``Robust $\mathcal{H}_2$ state estimation for
  discrete-time descriptor systems,'' in \emph{Proc. International Conference
  on Information and Communication Technology Convergence}, Jeju, South Korea,
  2018, pp. 1488--1490.

\bibitem{Scherer1997}
C.~Scherer, P.~Gahinet, and M.~Chilali, ``Multiobjective output-feedback
  control via {LMI} optimization,'' \emph{IEEE Transactions on Automatic
  Control}, vol.~42, no.~7, pp. 896--911, 1997.

\bibitem{Rotea1993}
M.~A. Rotea, ``The generalized $\mathcal{H}_2$ control problem,''
  \emph{Automatica}, vol.~29, no.~2, pp. 373--385, 1993.

\bibitem{Kottenstette2014}
N.~Kottenstette, M.~J. McCourt, M.~Xia, V.~Gupta, and P.~J. Antsaklis, ``On
  relationships among passivity, positive realness, and dissipativity in linear
  systems,'' \emph{Automatica}, vol.~50, no.~4, pp. 1003--1016, 2014.

\bibitem{Willems1972}
J.~C. Willems, ``Dissipative dynamical systems - part {I}: General theory,''
  \emph{Archive Rational Mechanics and Analysis}, vol.~45, no.~5, pp. 321--351,
  1972.

\bibitem{Hill1976}
D.~J. Hill and P.~J. Moylan, ``The stability of nonlinear dissipative
  systems,'' \emph{IEEE Transactions on Automatic Control}, vol.~21, no.~5, pp.
  708--711, 1976.

\bibitem{Goodwin1984}
G.~C. Goodwin and K.~S. Sin, \emph{Adaptive Filtering Prediction and
  Control}.\hskip 1em plus 0.5em minus 0.4em\relax Englewood Cliffs, NJ:
  Prentice-Hall, 1984.

\bibitem{Marquez2003}
H.~Marquez, \emph{Nonlinear Control Systems: Analysis and Design}.\hskip 1em
  plus 0.5em minus 0.4em\relax Hoboken, NJ: Wiley, 2003.

\bibitem{Anderson1967}
B.~D.~O. Anderson, ``A system theory criterion for positive real matrices,''
  \emph{SIAM Journal on Control}, vol.~5, no.~2, pp. 171--182, 1967.

\bibitem{Bao2007}
J.~Bao and P.~L. Lee, \emph{Process Control: The Passive Systems
  Approach}.\hskip 1em plus 0.5em minus 0.4em\relax London, UK:
  Springer-Verlag, 2007.

\bibitem{Brogliato2007}
B.~Brogliato, R.~Lozano, B.~Maschke, and O.~Egeland, \emph{Dissipative Systems
  Analysis and Control: Theory and Applications}, 2nd~ed.\hskip 1em plus 0.5em
  minus 0.4em\relax London, UK: Springer, 2007.

\bibitem{Hitz1969}
L.~Hitz and B.~D.~O. Anderson, ``Discrete positive-real functions and their
  application to system stability,'' \emph{Proceedings of the IEEE}, vol. 116,
  no.~1, pp. 153--155, 1969.

\bibitem{Haddad1994}
W.~H. Haddad and D.~S. Bernstein, ``Explicit construction of quadratic
  {Lyapunov} functions for the small gain, positivity, circle, and {Popov}
  theorems and their application to robust stability. {Part II}:
  Discrete‐time theory,'' \emph{International Journal of Robust and Nonlinear
  Control}, vol.~4, no.~2, pp. 249--265, 1994.

\bibitem{Wu1996}
S.-P. Wu, S.~Boyd, and L.~Vandenberghe, ``{FIR} filter design via semidefinite
  programming and spectral factorization,'' in \emph{Proc. IEEE Conference on
  Decision and Control}, Kobe, Japan, 1996, pp. 271--276.

\bibitem{Masubuchi2006}
I.~Masubuchi, ``Dissipativity inequalities for continuous-time descriptor
  systems with applications to synthesis of control gains,'' \emph{Systems \&
  Control Letters}, vol.~55, no.~2, pp. 158--164, 2006.

\bibitem{Freund2004}
R.~W. Freund and F.~Jarre, ``An extension of the positive real lemma to
  descriptor systems,'' \emph{Optimization Methods and Software}, vol.~19,
  no.~1, pp. 69--87, 2004.

\bibitem{Zhang2002}
L.~Zhang, J.~Lam, and S.~Xu, ``On positive realness of descriptor systems,''
  \emph{IEEE Transactions on Circuits and Systems}, vol.~49, no.~3, pp.
  401--407, 2002.

\bibitem{Lee2003}
L.~Lee and J.~L. Chen, ``Strictly positive real lemma and absolute stability
  for discrete-time descriptor systems,'' \emph{IEEE Transactions on Control of
  Network Systems}, vol.~50, no.~6, pp. 788--794, 2003.

\bibitem{Tang2019}
W.~Tang and P.~Daoutidis, ``Input-output data-driven control through
  dissipativity learning,'' in \emph{Proc. American Control Conference},
  Philadelphia, PA, 2019, pp. 4217--4222.

\bibitem{Gupta1994}
S.~Gupta and S.~M. Joshi, ``Some properties and stability results for
  sector-bounded {LTI} systems,'' in \emph{Proc. IEEE Conference on Decision
  and Control}, Lake Buena Vista, FL, 1994, pp. 2973--2978.

\bibitem{Forbes2011}
J.~R. Forbes, ``Extensions of input-output stability theory and the control of
  aerospace systems,'' Ph.D. dissertation, University of Toronto, Toronto,
  Canada, 2011.

\bibitem{Bridgeman2014b}
L.~J. Bridgeman and J.~R. Forbes, ``Conic-sector-based control to circumvent
  passivity violations,'' \emph{International Journal of Control}, vol.~87,
  no.~8, pp. 1467--1477, 2014.

\bibitem{Joshi2002}
S.~M. Joshi and A.~G. Kelkar, ``Design of norm-bounded and sector-bounded {LQG}
  controllers for uncertain systems,'' \emph{Journal of Optimization Theory and
  Applications}, vol. 113, no.~2, pp. 269--282, 2002.

\bibitem{BridgemanThesis}
L.~Bridgeman, ``Methods exploiting and extending the conic sector theorem,''
  Ph.D. dissertation, McGill University, Montreal, Canada, 2016.

\bibitem{Bridgeman2015}
L.~J. Bridgeman and J.~R. Forbes, ``The exterior conic sector lemma,''
  \emph{International Journal of Control}, vol.~88, no.~11, pp. 2250--2263,
  2015.

\bibitem{Iwasaki2003}
T.~Iwasaki, S.~Hara, and H.~Yamauchi, ``Dynamical systems design from a control
  perspective: Finite frequency positive-realness approach,'' \emph{IEEE
  Transactions on Automatic Control}, vol.~48, no.~8, pp. 1337--1354, 2003.

\bibitem{Iwasaki2005b}
T.~Iwasaki, S.~Hara, and A.~L. Fradkov, ``Time domain interpretations of
  frequency domain inequalities on (semi)finite ranges,'' \emph{Systems \&
  Control Letters}, vol.~54, no.~7, pp. 681--691, 2005.

\bibitem{Hara2003}
S.~Hara and T.~Iwasaki, ``Finite frequency characterization of easily
  controllable plant toward structure/control design integration,'' in
  \emph{Control and Modeling of Complex Systems: Cybernetics in the 21st
  Century}, K.~Hashimoto, Y.~Oishi, and Y.~Yamamoto, Eds.\hskip 1em plus 0.5em
  minus 0.4em\relax Boston, MA: Birkhauser, 2003, pp. 183--196.

\bibitem{Iwasaki2005}
T.~Iwasaki and S.~Hara, ``Generalized {KYP} lemma: Unified frequency domain
  inequalities with design applications,'' \emph{IEEE Transactions on Automatic
  Control}, vol.~50, no.~1, pp. 41--59, 2005.

\bibitem{Bridgeman2014}
L.~J. Bridgeman and J.~R. Forbes, ``The minimum gain lemma,''
  \emph{International Journal of Robust and Nonlinear Control}, vol.~25,
  no.~14, pp. 2515--2531, 2015.

\bibitem{Caverly2018}
R.~J. Caverly and J.~R. Forbes, ``$\mathcal{H}_\infty$-optimal parallel
  feedforward control using minimum gain,'' \emph{IEEE Control Systems
  Letters}, vol.~2, no.~4, pp. 677--682, 2018.

\bibitem{Caverly2016}
------, ``Robust controller design using the large gain theorem: The full-state
  feedback case,'' in \emph{Proc. American Control Conference}, Boston, MA,
  July 2016, pp. 3832--3837.

\bibitem{CaverlyPhDThesis}
R.~J. Caverly, ``Optimal output modification and robust control using minimum
  gain and the large gain theorem,'' Ph.D. dissertation, University of
  Michigan, Ann Arbor, MI, 2018.

\bibitem{Lanzon2008}
A.~Lanzon and I.~R. Petersen, ``Stability robustness of a feedback
  interconnection of systems with negative imaginary frequency response,''
  \emph{IEEE Transactions on Automatic Control}, vol.~53, no.~4, pp.
  1042--1046, 2008.

\bibitem{Song2012}
Z.~Song, A.~Lanzon, S.~Pitra, and I.~R. Petersen, ``A negative-imaginary lemma
  without minimality assumptions and robust state-feedback synthesis for
  uncertain negative-imaginary systems,'' \emph{Systems \& Control Letters},
  vol.~61, no.~12, pp. 1269--1276, 2012.

\bibitem{Ferrante2017}
A.~Ferrante, A.~Lanzon, and L.~Ntogramatzidis, ``Discrete-time negative
  imaginary systems,'' \emph{Automatica}, vol.~79, pp. 1--10, May 2017.

\bibitem{Liu2017}
M.~Liu and J.~Xiong, ``Properties and stability analysis of discrete-time
  negative imaginary systems,'' \emph{Automatica}, vol.~83, pp. 58--64,
  September 2017.

\bibitem{Xiong2012}
J.~Xiong, I.~R. Petersen, and A.~Lanzon, ``Finite frequency negative imaginary
  systems,'' \emph{IEEE Transactions on Automatic Control}, vol.~57, no.~11,
  pp. 2917--2922, 2012.

\bibitem{Caverly2019}
R.~J. Caverly and M.~Chakraborty, ``Convex synthesis of strictly negative
  imaginary feedback controllers,'' in \emph{Proc. IEEE Conference on Decision
  and Control}, Nice, France, 2019, pp. 7578--7583.

\bibitem{Lee2019}
K.~Lee and J.~R. Forbes, ``Synthesis of strictly negative imaginary controllers
  using a $\mathcal{H}_\infty$ performance index,'' in \emph{Proc. American
  Control Conference}, Philadelphia, PA, 2019, pp. 497--502.

\bibitem{Lee_MS_thesis}
K.~Lee, ``Synthesis and application of optimal strictly negative imaginary
  controllers,'' Master's thesis, McGill University, Montreal, Canada, 2019.

\bibitem{Hung1998}
Y.~S. Hung and D.~L. Chu, ``Relationships between discrete-time and
  continuous-time algebraic {Riccati} inequalities,'' \emph{Linear Algebra and
  its Applications}, vol. 270, no. 1--3, pp. 287--313, 1998.

\bibitem{Cao1998b}
Y.~Y. Cao, J.~Lam, and Y.~X. Sun, ``Static output feedback stabilization: An
  {ILMI} approach,'' \emph{Automatica}, vol.~34, no.~12, pp. 1641--1645, 1998.

\bibitem{Kucera1995}
V.~Kucera and C.~E. {de Souza}, ``A necessary and sufficient condition for
  output feedback stabilization,'' \emph{Automatica}, vol.~31, no.~9, pp.
  1357--1359, 1995.

\bibitem{Gumussoy2005}
S.~G{\"u}m{\"u}{\c{s}}soy and H.~{\"O}zbay, ``Remarks on strong stabilization
  and stable $\mathcal{H}^\infty$ controller design,'' \emph{IEEE Transactions
  on Automatic Control}, vol.~50, no.~12, pp. 2083--2087, 2005.

\bibitem{Kouvaritakis1976a}
B.~Kouvaritakis and A.~G.~J. MacFarlane, ``Geometric approach to analysis and
  synthesis of system zeros: Part 1. square systems,'' \emph{International
  Journal of Control}, vol.~23, no.~2, pp. 149--160, 1976.

\bibitem{ChilaliGahinet1996}
M.~Chilali and P.~Gahinet, ``$\mathcal{H}_\infty$ design with pole placement
  constraints: An {LMI} approach,'' \emph{IEEE Transactions on Automatic
  Control}, vol.~41, no.~3, pp. 358--367, 1996.

\bibitem{Ohara1991}
A.~Ohara, S.~Nakazumia, and N.~Suda, ``Relations between a paramterization of
  stabilizing state feedback gains and eigenvalue locations,'' \emph{Systems \&
  Control Letters}, vol.~16, no.~4, pp. 261--266, 1991.

\bibitem{Yedavalli1993}
R.~K. Yedavalli, ``Robust root clustering for linear uncertain systems using
  generalized {Lyapunov} theory,'' \emph{Automatica}, vol.~29, no.~1, pp.
  237--240, 1993.

\bibitem{Chadli2013}
M.~Chadli and P.~Borne, \emph{Multiple Models Approach in Automation:
  Takagi-Sugeno Fuzzy Systems}.\hskip 1em plus 0.5em minus 0.4em\relax London,
  UK: John Wiley \& Sons, Inc., 2013.

\bibitem{Xue2019}
X.~Xue, ``Novel robust and adaptive distributed protocol for consensus-based
  control of uncertain multi-agent systems,'' Ph.D. dissertation, North
  Carolina State University, Raleigh, NC, 2019.

\bibitem{Iwasaki2007}
\BIBentryALTinterwordspacing
T.~Iwasaki, \emph{Lecture Notes: Multivariable Control}, December 2007.
  [Online]. Available: \url{https://sites.google.com/g.ucla.edu/cyclab}
\BIBentrySTDinterwordspacing

\bibitem{Kuo2004}
C.-H. Kuo and L.~Lee, ``Robust $\mathcal{D}$-admissibility in generalized {LMI}
  regions for descriptor systems,'' in \emph{Proc. Asian Control Conference},
  Melbourne, Australia, 2004, pp. 1058--1065.

\bibitem{Whidborne2007}
J.~F. Whidborne and J.~Mckernan, ``On the minimization of maximum transient
  energy growth,'' \emph{IEEE Transactions on Automatic Control}, vol.~52,
  no.~9, pp. 1762--1767, 2007.

\bibitem{Polyak2015}
B.~T. Polyak, A.~A. Tremba, M.~V. Khlebnikov, P.~S. Shcherbakov, and G.~V.
  Smirnov, ``Large deviations in linear control systems with nonzero initial
  conditions,'' \emph{Automation and Remote Control}, vol.~76, no.~6, pp.
  957--976, 2015.

\bibitem{Hayes2020}
A.~Hayes, I.~Nompelis, R.~J. Caverly, J.~Mueller, and D.~Gebre-Egziabher,
  ``Dynamic stability analysis of a hypersonic entry vehicle with a non-linear
  aerodynamic model,'' in \emph{Proc. Modeling and Simulation Technologies
  Conference, AIAA Aviation}, Virtual Event, 2020, {AIAA 2020-3201}.

\bibitem{Bernstein1992}
D.~S. Bernstein and W.~M. Haddad, ``Robust controller synthesis using
  {Kharitonov's} theorem,'' \emph{IEEE Transactions on Automatic Control},
  vol.~37, no.~1, pp. 129--132, Jan. 1992.

\bibitem{Dey2018}
R.~Dey, G.~Roy, and V.~E. Balas, \emph{Stability and Stabilization of Linear
  and Fuzzy Time-Delay Systems}, ser. Intelligent Systems Reference
  Library.\hskip 1em plus 0.5em minus 0.4em\relax Cham, Switzerland: Springer,
  2018, vol. 141.

\bibitem{Doyle1991}
J.~Doyle, A.~Packard, and K.~Zhou, ``Review of {LFT}s, {LMI}s, and $\mu$,'' in
  \emph{Conference on Decision and Control}, Brighton, England, 1991, pp.
  1227--1232.

\bibitem{Green2012}
M.~Green and D.~J.~N. Limebeer, \emph{Linear Robust Control}.\hskip 1em plus
  0.5em minus 0.4em\relax Mineaola, NY: Dover, 2012.

\bibitem{Francis1987}
B.~A. Francis, \emph{A Course in $\mathcal{H}_\infty$ Control Theory}, ser.
  Lecture Notes in Control and Information Sciences, M.~Thomas and A.~Wyner,
  Eds.\hskip 1em plus 0.5em minus 0.4em\relax Berlin, Germany: Springer-Verlag,
  1987, vol.~88.

\bibitem{Ogata2010}
K.~Ogata, \emph{Modern Control Engineering}, 5th~ed.\hskip 1em plus 0.5em minus
  0.4em\relax Upper Saddle River, NJ: Prentice Hall, 2010.

\bibitem{DSB580Notes}
D.~S. Bernstein, ``Lecture notes for {AEROSP} 580 - linear feedback control
  system,'' 2014.

\bibitem{Zhou1998}
K.~Zhou and J.~C. Doyle, \emph{Essentials of Robust Control}.\hskip 1em plus
  0.5em minus 0.4em\relax Upper Saddle River, NJ: Prentice-Hall, 1998.

\bibitem{PeetOptNotes22}
\BIBentryALTinterwordspacing
M.~M. Peet, ``{Modern Optimal Control} lecture 22: $\mathcal{H}_2$, {LQR} and
  {LQG},'' 2011. [Online]. Available:
  \url{http://control.asu.edu/Classes/MAE507/507Lecture22.pdf}
\BIBentrySTDinterwordspacing

\bibitem{PeetOptNotes21}
\BIBentryALTinterwordspacing
------, ``{Modern Optimal Control} lecture 21: Optimal output feedback
  control,'' 2011. [Online]. Available:
  \url{http://control.asu.edu/Classes/MAE507/507Lecture21.pdf}
\BIBentrySTDinterwordspacing

\bibitem{LallNotes16}
\BIBentryALTinterwordspacing
S.~Lall, ``Engr210a lecture 16: $\mathcal{H}_\infty$ synthesis,'' November
  2001. [Online]. Available:
  \url{https://lall.stanford.edu/engr210a/lectures/lecture16_2001_11_25_04.pdf}
\BIBentrySTDinterwordspacing

\bibitem{PeetOptRobNotes11}
\BIBentryALTinterwordspacing
M.~M. Peet, ``{LMI Methods in Optimal and Robust Control} lecture 11:
  Relationship between $\mathcal{H}_2$, {LQG} and {LGR} and {LMI}s for state
  and output feedback $\mathcal{H}_2$ synthesis,'' 2016. [Online]. Available:
  \url{http://control.asu.edu/Classes/MAE598/598Lecture11.pdf}
\BIBentrySTDinterwordspacing

\bibitem{Geromel2000}
J.~C. Geromel, J.~Bernussou, G.~Garcia, and M.~C. {de Oliveira},
  ``$\mathcal{H}_2$ and $\mathcal{H}_\infty$ robust filtering for discrete-time
  linear systems,'' \emph{SIAM Journal on Control and Optimization}, vol.~38,
  no.~5, pp. 1353--1368, 2000.

\end{thebibliography}

\printindex

\end{document}